**Czech Technical University in Prague**

**Faculty of Biomedical Engineering**

# Doctoral Thesis

*June 2010* *Michel Kana*

Czech Technical University in Prague
Faculty of Biomedical Engineering
Department of Biomedical Informatics

# MATHEMATICAL MODELS OF CARDIOVASCULAR CONTROL BY THE AUTONOMIC NERVOUS SYSTEM

**Doctoral Thesis**

*Michel Kana*

**Prague**, *June 2010*

Ph.D. Program: Biomedical and Clinical Technology

**Supervisor:** ***Prof. Ing. Jiři Holčík, CSc.***

# Abstract

This thesis develops an integrated mathematical model for autonomic nervous system control on cardiovascular activity. As a working control system, with diverse feedback and feedforward loops interfering with each other, our integrative model is key to understand paradoxal phenomenon such as vagally-mediated tachycardia, fluctuation of sinoatrial rhythms on denervated heart, Mayer waves and tonic activity of sympathetic premotor pacemaker neurons. The model extensively covers cardiovascular neural pathways including a wide range of afferent sensory neurons, central processing by autonomic premotor neurons, efferent outputs via preganglionic and postganglionic autonomic neurons and dynamics of neurotransmitters at cardiovascular effectors organs.

The results achieved in this work are challenging some established methods for assessing autonomic activity on the cardiovascular system, e.g. heart rate variability, blood pressure variability, baroreflex sensitivity and conventional mathematical models based on control theory. Most of these methods picture a reciprocal control of cardiac vagal and sympathetic nervous activity, and neglect simultaneous co-activation of both autonomic efferent branches. Furthermore they lack to integrate cardiovascular reflexes across their many levels of organization and therefore miss to exhibit emerging properties of the regulatory processes.

We performed over 500 cardiovascular experiments using clinical autonomic tests on 72 subjects ranging from 11 to 82 years old and collected typical cardiovascular signals such as electrocardiogram, arterial pulse, arterial blood pressure, respiration pattern, galvanic skin response and skin temperature. Next to a statistical evaluation in the time and frequency domains, the data were especially used to validate the mathematical model by resolving a constrained optimization task. Results bring evidences supporting the hypothesis that Mayer waves result from a rhythmic sympathetic discharge of pacemaker-like sympathetic premotor neurons. Simulation also show that vagally-mediated tachycardia, observed during vagal maneuvers on some subjects could be related to the secretion of vasoactive neurotransmitters by the vagal nerve. We additionally identified model parameters for estimating the resting sympathetic and parasympathetic tone which are believed to be linked to some pathological states. Results show higher vagal tone on young subjects with a decreasing trend with aging, what agrees with the data from heart rate variability studies. Tonic sympathetic activity was found to possibly emerge from pacemaker premotor neurons, but also from activation of chemoreceptors to a lesser extent.

We include a software package as practical work product of this thesis for clinical applications. Our web-based telemedicine platform offers features for connecting doctors with remote patients, including signal processing, statistical evaluation and messaging. We extended the infrastructure with the design of biofeedback solution that upgrades a common mobile phone with low-cost sensors for skin temperature and finger pulse measurements, as well as a software module for uploading and downloading medical information and cardiovascular signals through internet.

In summary this thesis offers a software and hardware application that could be useful in a clinical environment and proposes an integrative model of cardiovascular control that might help for educational and research purposes. The thesis also opens perspectives for future work including validating the markers of autonomic tone provided by our model against data from experiments with pharmacological blockers and invasive neural activity recordings.

## Abstract in Czech

Tato práce pojednává o vývoji integrovaného matematického modelu řízení kardiovaskulární aktivity autonomním nervovým systémem. Vytvořený funkční model, využívající navzájem interferující řídicí mechanismy jak se zpětnou, tak i přímou vazbou, je klíčovým prostředkem k porozumění takových paradoxních jevů, jako jsou vagem zprostředkovaná tachykardie, kolísání sinoatriálního rytmu na denervovaném srdci, Mayerovy oscilace, nebo tonická aktivita sympatických premotorických pacemakerových neuronů. Model zahrnuje vliv mnohých neurokardiovaskulárních vedení, včetně velkého počtu aferentních senzorických nervových cest, centrálního zpracování pomocí autonomních premotorických neuronů, eferentních výstupů zprostředkovaných pregangliovými a postgangliovými nervovými vlákny i vliv dynamiky neurotransmiterů v kardiovaskulárních efektorech.

Dosažené výsledky mohou být užitečné v diskuzi o některých klasických metodách hodnocení autonomní aktivity kardiovaskulární soustavy, jako jsou variabilita srdečního rytmu, variabilita krevního tlaku, baroreflexní senzitivita a konvenční matematické modely založené na teorii řízení. Nabízejí prostor pro další výzkumné aktivity, zejména při hodnocení markerů tonu autonomního nervového systému, stanovených pomocí vytvořeného modelu pro data pořízená při experimentech s farmakologickými blokátory, příp. s invazivně zaznamenávanou nervovou aktivitou.

## Abstract in French

Cette thèse traite de l'élaboration d'un modèle mathématique intégré du contrôle exercé par le système nerveux autonome sur le système cardiovasculaire. Ce modèle représente le fonctionnement de divers mécanismes de régulation neurologique, interférant les uns avec les autres et est une clé pour comprendre des phénomènes paradoxaux, comme la tachycardie parasympathique, les fluctuations à haute-fréquence observées sur le nœud sino-auriculaire du cœur dénervé, les oscillations de Mayer, ou encore l'activité tonique des neurones sympathiques prémoteurs. Le modèle permet de simuler un grand nombre de voies neurales autonomes connues, y compris les voies neurales sensorielles afférentes, les voies neurales centrales autonomes, les sorties efférentes incluant les neurones pré-ganglionnaires et post-ganglionnaires, ainsi que la concentration des neurotransmetteurs au niveau des organes cardiovasculaires tels que ventricules, oreillettes, nœud sino-auriculaire, nœud auriculo-ventriculaire et vaisseaux sanguins.

Les résultats obtenus mettent certaines méthodes classiques d'évaluation de l'activité du système nerveux autonome sur le système cardiovasculaire à l'épreuve, tels que la variabilité du rythme cardiaque, la variabilité de la tension artérielle, la sensibilité baroréflexe, et les modèles mathématiques conventionnelles basés sur la théorie du contrôle.

Ce travail offre des possibilités de poursuivre les recherches, en particulier dans l'évaluation des marqueurs du système nerveux autonome à l'aide des données acquises dans des expériences utilisant des inhibiteurs pharmacologiques.

## Acknowledgment

This work would have not been possible without Lucie, my lovely wife and Simon, my fantastic son. Lucie had so much patience with me along these years and never stopped encouraging me. Her affection was a real source of energy empowering my work.

I am very thankful towards my dear dad and mum, Paul and Elise Kana who set the educational basis for scientific work since my childhood. My achievements are fruits of their constant support and prayers. This work fulfills one of their deepest wishes. I am paying one of the debts a Bamileke descendant has towards his parents.

Special thanks to Helena Simonova and Petr Simon, my parents-in-law for their constant support as well as my sisters Rosette Detsi, Lydie Biandu, Hermine and Laurece Kana who were always giving me courage. I thank my whole big family, grand-pa, grand-ma, aunts, uncles, cousins from the Cameroonian side as well as my family from the Czech side, especially Jaroslava Dvorakova and Anna Tresnakova. I have special thoughts for my cousin, Sonia Zebaze, who will make an excellent medical doctor one day; and my grand-ma M'a Konza who always looks so beautiful and young. As I was still a baby, my uncle Dr. Michel Ngueti dedicated his PhD thesis to me, and wished to see me achieving greater challenges. I am very grateful for his great influence on my scientific thinking.

This thesis wouldn't be without an exceptional PhD supervisor. That's the one who used to bring me back to the main road of success. I sincerely thank my PhD father, Prof. Ing. Jiri Holcik.

This work has been officially reviewed by Prof. Richard Reilly and Doc. Ing. Milan Tysler. I thank them for the constructive feedback. I am very grateful for the strong scientific feedback provided by Doc. MUDr. RNDr. Petr Marsalek. I also thank Dr. Esther Tamm, Prof. Pavel Kucera and MUDr. Jan Wichterle for support and advices during my studies.

The Department of Biomedical Informatics was hosting my research. I felt there at home thanks to the head of department, Dr. Zoltan Szabo. Thanks a lot to Dr. Jan Kauler for the good mood and help with biomedical sensors. I would also thank my colleague and friend Mgr. Radim Krupicka who accompanied me along my studies. I thank doc. Marcel Jirina for the good advices, especially about the artificial neural parts of this thesis. Special regards to Mrs. Lucie Kulhankova for helping out with administrative matters. Financial support has been provided by the Interni Grantova Soutez CVUT, Nadacni Fond Stanislava Hanzla CVUT, Vyzkumni Zamer MSMT grants.

I extremely thank all those people who voluntary participated to my experiments. Without them I would not have enough data to support my research. I would like to address a special thanks to Dr. Norbert Tsopze from the University of Yaounde I in Cameroon, and Dr. J. Donfack from the District Hospital of Dschang for the good collaboration. Similarly I thank Prof. J. Hanak from the Equine Clinic of the Veterinary University in Brno and his staff for helping out with horse experiments. I thank MSc. Serge Dongmo, BSc. Petr Slajchtr and BSc. Andrea Svenkrtova for actively helping me with experiments. Serge and Petr contributed to software development within my thesis. I thank Mrs Jana Ocenaskova for their help with biofeedback sensors, what has improved this work.

One page is not much to mention all those people, who really deserve my regards because without them, I would not be here. They recognize themselves and know that I am grateful for their support.

To the leaving light, O’oo

To the coming light, Simon

Matthew 5.1-12

The Beatitudes

[1]Now when he saw the crowds, he went up on a mountainside and sat down. His disciples came to him,

[2]and he began to teach them saying:

[3]Blessed are the poor in spirit, for theirs is the kingdom of heaven.

[4]Blessed are those who mourn, for they will be comforted.

[5]Blessed are the meek, for they will inherit the earth.

[6]Blessed are those who hunger and thirst for righteousness, for they will be filled.

[7]Blessed are the merciful, for they will be shown mercy.

[8]Blessed are the pure in heart, for they will see God.

[9]Blessed are the peacemakers, for they will be called sons of God.

[10]Blessed are those who are persecuted because of righteousness, for theirs is the kingdom of heaven.

[11]Blessed are you when people insult you, persecute you and falsely say all kinds of evil against you because of
me. [12]Rejoice and be glad, because great is your reward in heaven, for in the same way they persecuted the
prophets who were before you.

Matthieu 5.1-12

Les Béatitudes

[1]Voyant la foule, Jésus monta sur la montagne; et, après qu'il se fut assis, ses disciples s'approchèrent de lui.

[2]Puis, ayant ouvert la bouche, il les enseigna, et dit:

[3]Heureux les pauvres en esprit, car le royaume des cieux est à eux.

[4]Heureux les affligés, car ils seront consolés.

[5]Heureux les débonnaires, car ils hériteront la terre.

[6]Heureux ceux qui ont faim et soif de la justice, car ils seront rassasiés.

[7]Heureux les miséricordieux, car ils obtiendront miséricorde.

[8]Heureux ceux qui ont le cœur pur, car ils verront Dieu.

[9]Heureux ceux qui procurent la paix, car ils seront appelés fils de Dieu.

[10]Heureux ceux qui sont persécutés pour la justice, car le royaume des cieux est à eux.

[11]Heureux serez-vous, lorsqu'on vous outragera, qu'on vous persécutera et qu'on dira faussement de vous toute sorte de mal, à cause de moi.

[12]Réjouissez-vous et soyez dans l'allégresse, parce que votre récompense sera grande dans les cieux; car c'est ainsi qu'on a persécuté les prophètes qui ont été avant vous.

# Table of Contents

# Preface

## Historical Review

The first known written description of the human heart and blood vessels appears around 1600 BC in the Edwin Smith Papyrus. It is also the world's oldest surgical document, credited to have been written by the Egyptian Imhotep, the first engineer, architect and physician in history known by name [1]. Around 200 BC the Greek physician Galen identified venous (dark red) and arterial (brighter and thinner) blood, each with distinct and separate functions. The first known description of the pulmonary, capillary and coronary circulations and pulse is attributed to the Arabian physician Ibn al-Nafis in 1242 [2]. The English physician William Harvey improved this concept experimentally and described in details the systemic circulation in 1628 [3]. Giovanni Borelli first postulated the action of the heart to simulate a piston in the middle of 17$^{th}$ century and the idea that the arteries have to be elastic [4]. Richard Lower first described the muscular fibers of the heart in 1669. Thomas Willis discovered the internal brain carotids, and their communications with the branches of the basilar artery in the middle of 17$^{th}$ century. Stephen Hales described the effect of the elastic arteries in dampening the arterial pulse and performed the first measurements of in vivo blood pressure in horse in 1733 [5]. Jan Evangelista Purkyně discovers the Purkinje fibers in 1839 [6]. Claude Bernard established the scientific method in experimental medicine, defined the underlying principle of homeostasis, set the stage for cardiac catheterization and discovered the existence of vaso-dilator and vaso-constrictor nerves in 1851 [7]. DeCyon and Ludwig further demonstrated the ability of afferent nerves to influence the behavior of the cardiovascular system in 1867 [8]. Records of heart beat using electrodes attached to the chest were first reported by Alexander Muirhead in 1872. The first measurement of monophasic action potential with an extracellular electrode placed on the surface of the heart was performed by Burdon-Sanderson and Page in 1883 [9]. Wilhelm His, Jr. discovered the bundle of His in 1893 and later recognized that the heartbeat has its origin in the individual cells of heart muscle [10]. The atrioventricular node is discovered later by Sunao Tawara in 1906 [11]. In 1913, Willem Einthoven designed a device for more precise measurements of surface body potentials originating from the heart, what will be the origin of modern electrocardiography [12]. Wiggers provided the first descriptions of the simultaneous electrical and mechanical events of a cardiac cycle in 1921 [13]. In 1921 Loewi proved experimentally that signaling across synapses is chemical and discovered to first neurotransmitter to be identified: acetylcholine [14]. He also clarified the blockade and the augmentation of nerve action by certain drugs. Hering discovered the carotid baroreceptors in 1923 [15]. The relationship between baroreceptor firing rate and distension of the vessel wall was put in evidence by Bronk and Stella in 1932 [16]. The carotid sinus and aortic arch chemoreceptors and their influence on the chemical composition of arterial blood were revealed by Heymans in 1933 [17]. The relationship between kidneys and vasoconstriction was advanced by Goldblatt in 1934 [18]. In 1946, Alexander demonstrated that cardiac reflex functions originating from the medullary cardiovascular center [19]. Artrial mechanoreceptors were discovered in 1956 by Henry and Gauer [20]. Randall demonstrated the cardioaugmentor and cardioaccelerator pathways projecting to the heart in 1957 [21]. In 1961 Folkow and Mellander differentiated the aorta, arterioles, capillaries, small venules, larger venules and veins [22]. The fundamental basis of the Frank-Starling mechanism was revealed in 1966 [23]. In 1972

Guyton developed the first model of overall long-term circulation regulation including interactions of the heart, blood vessels, kidney, baroreceptor and chemoreceptor reflexes [24]. In 1979 Korner extended cardiovascular integration from the medulla to hypothalamic, limbic, and other forebrain regions, as well as to the spinal cord [25]. In the last quarter of 20$^{th}$ century, the emphasis has been on the molecular and cellular bases of cardiovascular function [26].

## Motivation

The autonomous nervous system (ANS) together with the hormone and central nervous system form the main regulators in human and animal organism. The autonomic division of the efferent nervous system influences a great number of life situations at rest, exercise or illness. This so-called vegetative nervous system or visceral nervous system controls internal organs, which are not under voluntary control (especially smooth muscles). Generally, we can say that the sympathetic part of vegetative nervous system controls catabolic processes and its parasympathetic part controls anabolic processes. Catabolism is activated by physical exercise and illness whereas anabolic processes are activated by rest or recovery from exercise or illness.

The ANS controls the cardiovascular system. Both its sympathetic and parasympathetic branches influence the heart rate via mostly antagonistic control. Also the heart chambers contractility and the vessels resistance are modulated by the ANS. Assessing the autonomic control over the cardiovascular system is possible directly through its mediators, by specific drugs (sympathomimetic, sympatholytic, vagomimetic, vagolytic drugs). Cardiovascular activity can also be influenced by non-invasive cardiac autonomic tests. Their clinical importance is described in [27]. Those tests are currently used in human medicine with the heart rate variability (HRV) being the most often measured parameter, because it is very sensitive to stress and other stimuli. For example, in [28] there was spectral analysis applied for measuring the HRV, just to evaluate stress conditions in humans. So far, the methods used for assessing level of the ANS activity usually utilize heuristically determined characteristics of the HRV frequency spectra [29] or time-dependency of the spectrum [30]. Unfortunately, these frequency changes are ambiguous [31] what significantly reduces their range of application. Additionally they do not provide a detailed insight view on the neural control processes involved, which are not directly visible on measured data. These facts evolve an effort to look for more reliable alternative algorithms. Rough assessing of the ANS activity based on statistical characteristics of interbeat interval sequences [29] can be one of possible alternative methods. Other approaches use various tools of nonlinear dynamics theory [32] or so called Weighted Principal Component Regression [33]. While such methods provide interesting information, the contribution of neural pathways involved is not covered. We believe better insight into the non-linear regulation processes during cardiovascular control would be beneficial both to research and health services. An approach utilizing mathematical models of cardiovascular control processes (e.g. [34], [35]) looks as a promising tool for the analysis of cardiovascular regulation processes. Application of this approach enabled us to measure levels of the ANS activity in both its branches. Its disadvantages are that it uses parameters of electrical activity of myocardium only and its results were not verified using quantified ANS stimulation.

Following from accessible references and our own experimental work we can claim that sufficiently unambiguous and evincible methods for quantization of the ANS activity does not exist, yet. Such a method would certainly find important applications in human and veterinary medicine.

## Problem Statement

The global aim of the PhD thesis is to develop new methods for assessing and quantifying the level of Autonomic Nervous System (ANS) activity based on mathematical models of cardiovascular control. In order to fulfill this aim we will solve following work packages:

*Experimental Setups*

This work package consists of the definition, arrangement of experiments for appropriate quantifiable stimulation of both branches of the ANS in human. The experiments should utilize non-invasive physical stimuli and make use of existing cardiac autonomic tests.

*Signal Processing Framework*

This work package consists of developing algorithms for processing cardiovascular signals. Adequate measuring equipment should be selected and methodology for noninvasive measuring of basic quantities describing performance and control of the cardiovascular system (e.g. electrocardiogram, blood pressure, respiration signals) should be defined, so that we might be able to assess the reaction to the ANS stimulation. It is necessary to determine and define a set of appropriate signal parameters of the noninvasively recorded signals which are suitable to assess and quantify response of the cardiovascular system.

*Statistical Evaluation Framework*

This work package consists of choosing appropriate time-frequency transforms to assess effectiveness of the stimulation for the given selection of the measured signal parameters. An evaluation framework should be defined including statistical parameters having physiological meaning and a clinical application.

*Mathematical Model of Cardiovascular Control*

This work package consists of the development of the mathematical models of the cardiovascular control by ANS. The models should capture the non-linear processes that influence cardiovascular response to external stimulus. It is necessary to cover neural pathways and quantify the level of modulation of the efferent part of the ANS on the heart and blood vessels. Especially the proposed physiologically-based models should quantify the level of sympathetic and parasympathetic discharge as response to an external stimulus applied during a cardiac autonomic test. The models should be verified by simulation fitting the experimental data.

*Prototypes for Clinical Application*

This work package consists of showing perspectives for a possible application of the results in research or in clinical environments. A software or hardware with build-in modules for simulation and parameters estimation could be a possible outcome.

## Outline for the Dissertation

The dissertation is made of six chapters. While the first chapter presents the state of art, the next four chapters mostly contain author contributions and propose novel approaches, which are experimentally verified and practically realized as preparation for clinical applications.

Chapter 1 gives a short overview on the current research on quantitative assessment of cardiovascular control. It presents prominent methods for quantifying autonomic nervous activity such as heart rate variability. Other time series analysis methods are introduced as well as current methods based on mathematical models. Benefits and limitations of each method are highlighted.

Chapter 2 presents our first result achieved. We described an integrated physiologically-based mathematical model of regulatory processes performed by the autonomic nervous system on the heart and vascular beds. The model extensively covers neural pathways between sensory neurons, autonomic premotor neurons, preganglionic neurons, neurotransmitter dynamics at autonomic ganglionic sites, postganglionic neurons and cardiovascular effectors (sinoatrial node, atrioventricular node, atria, ventricles, vascular beds in splanchnic, extrasplanchnic, skeletal muscles, brain and heart). The model includes variables for simulating baroreflex, respiratory sinus arrhythmias, Mayer waves as well as diverse cardiovascular reflexes such as the diving reflex and the oculocardiac reflex.

Chapter 3 deals with the material and methods we have used in our experimental setups on humans. Subjects are presented in details. Each cardiac autonomic test is explained including its physiology, its materials, and signals recorded. We present a generic signal processing framework developed in the frame of this thesis for batch processing cardiovascular signal and for extracting statistical parameters. The chapter also shows results obtained during animal experiments.

Chapter 4 uses the experimental data to fit the model presented in Chapter 2. Optimization algorithms are developed for parameters estimation. Estimated values are presented and features related to age, gender and race are exhibited. This chapter shows the capability of the model to estimate e.g. the parasympathetic tone, sympathetic tone, level of respiratory sinus arrhythmias, level of Mayer waves and vagally-mediated tachycardia.

Chapter 5 demonstrates perspectives for application of our results in a clinical environment. We developed a web-based software package offering features for patients and cardiac autonomic tests. Signals can be processed and statistical evaluation can be performed online. The package is extended as telemedicine platform where doctors can remotely examine cardiovascular signals uploaded by patients using built-in features for modeling and simulation. Both doctor and patients can communicate using social network features including email and forums. The chapter also presents the design of a biofeedback device which can upgrade a common mobile phone with sensors for skin temperature and arterial pulse measurements. The design includes embedded software capable of processing measured signals by solving a built-in light weight mathematical model of cardiovascular control for estimating parameters related to autonomic nervous activity and display it to the patient in the context of a biofeedback therapy.

Chapter 6 concludes the thesis and gives insight into future developments.

# Chapter 1. Existing Methods for Assessing Cardiovascular Control Quantitatively

This chapter gives a very brief overview on the cardiovascular and respiratory physiology. It includes an overview on state-of-art in quantitative assessment of cardiovascular control.

## 1.1 Background

### 1.1.1 Introduction to Cardiovascular Physiology

The heart is the central organ of the cardiovascular or circulatory system. As illustrated in Figure 1.1, this system is a series of tubes (the blood vessels) filled with fluid (blood) and connected to a pump (the heart). Pressure generated in the heart propels blood through the system continuously. The blood picks up oxygen at the lungs and nutrients in the intestine and then delivers these substances to the body’s cells while simultaneously removing cellular wastes for excretion. In addition, the cardiovascular system provides cell-to-cell communication and immune function.

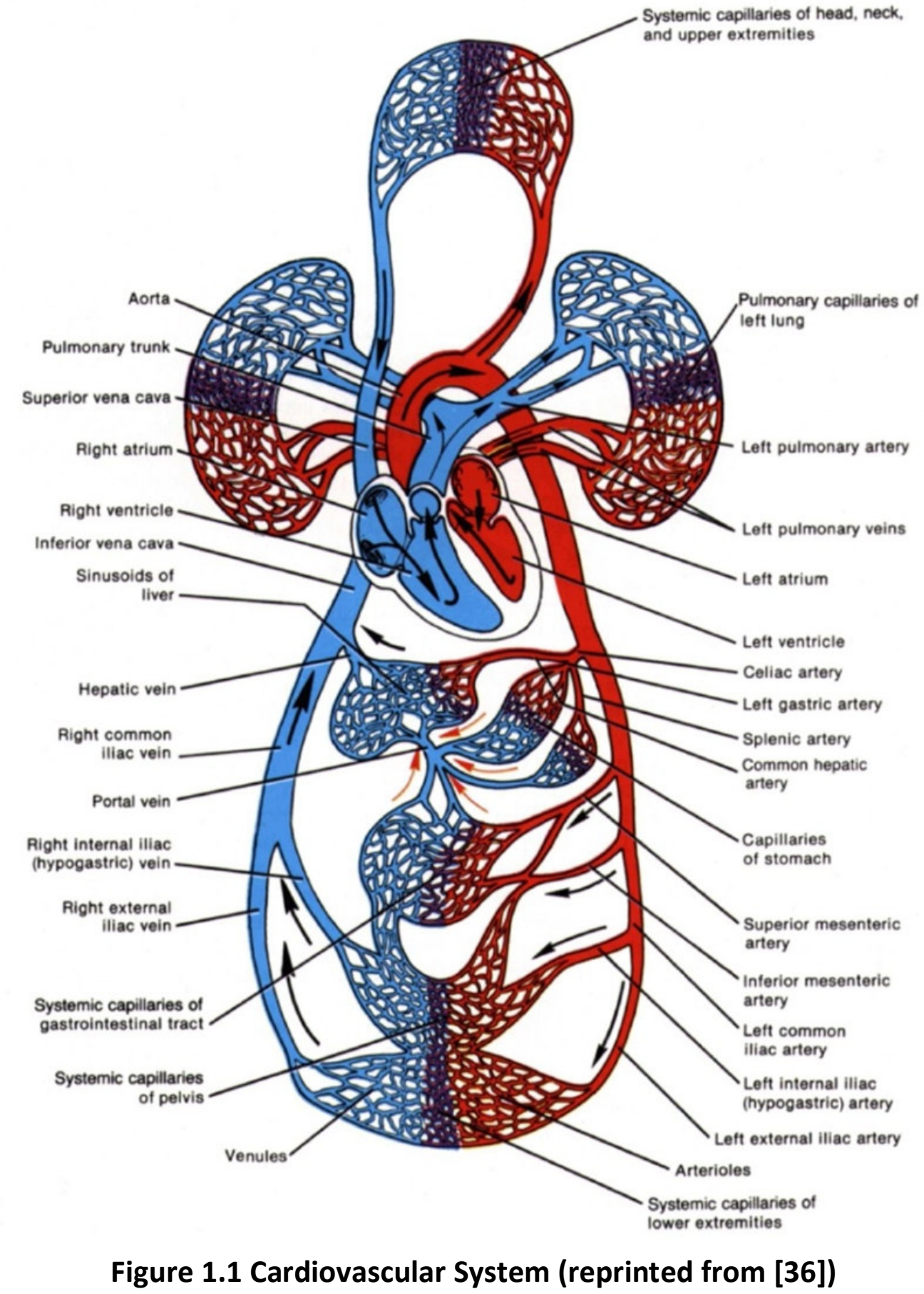

**Figure 1.1 Cardiovascular System (reprinted from [36])**

The myocardium muscle has a few pacemaker cells (1%) and the rest are contractile cells (99%). Heart contraction is normally originated by cardiac pacemaker cells which are located in the sinoatrial node (SA node) and are able to generate electrical signals without nervous simulation. The electrical and mechanical events of the cardiac cycle can be summarized into five phases: atrial and ventricular diastole, atrial systole, ventricular contraction, ventricular ejection, and ventricular relaxation.

Autonomic regulatory mechanisms and circulating hormones modulate cardiovascular function in order to ensure cardiovascular homeostasis by regulating e.g. blood pressure, heart rate, force of contraction of the heart chambers and size of blood vessels. A range of sensory neurons monitor blood pressure (baroreceptors), blood volume (volume receptors), blood chemistry (chemoreceptors) and plasma osmolarity (osmoreceptors). Many afferent neurons terminate in the nucleus of the solitary tract, each type of afferent input having its specific terminal field within the tract. The nucleus of the solitary tract projects to various regions of spinal cord, lower brain stem, midbrain, forebrain, especially to the rostral ventrolateral medulla. These control centers integrate sensory input and initiate an appropriate response via efferent sympathetic and parasympathetic preganglionic neurons which project to an autonomic ganglion outside the central nervous system, where they synapse with the postganglionic neurons. These second neurons have their cell bodies in the ganglion and project their axons to the target cardiovascular effectors. Parasympathetic postganglionic autonomic neurons innervate the sinoatrial and atrioventricular nodes. Sympathetic neurons innervate most of the heart and peripheral vessels additionally.

Several reflexes contribute to the short-term regulation of normal cardiovascular function. The baroreflex functions as a negative feedback system controlling arterial blood pressure. Respiratory sinus arrhythmia is characterized by increase heart rate during inspiration and decrease heart rate during expiration. Thermoreceptors on the face response to cold and activate the diving reflex. Increased intracranial pressure elicits the Cushing reflex, a strong sympathetic pressor response followed by a baroreflex-mediated bradycardia.

### 1.1.2 Overview of Respiratory Physiology

The primary function of the respiratory system is the exchange of oxygen and carbon dioxide between the atmosphere and the blood. Additionally lungs can alter body pH, protect from inhaled pathogens, help with vocalization, and promote water and heat loss from the body. The respiratory system consists of three structures: a conduction system (pharynx, larynx, trachea, bronchi, bronchioles), alveoli (exchange surface) and the chest cavity (diaphragm, bones and muscles of the thorax). Breathing is an active process which involves diaphragm and muscles contraction resulting in an increase of thoracic volume during inspiration and decrease of thoracic volume during expiration. According to Boyle’s law of gases, change of lung volume during ventilation creates pressure gradient and air flows passively from areas of higher pressure to areas of lower pressure. After air has reached the alveolar, gas exchange muss occur across the alveolar-capillary interface. Since the partial pressure of alveolar oxygen is about 100 mmHg and the partial pressure of oxygen in systemic venous blood is only 40 mmHg, oxygen moves down its partial pressure gradient from the alveoli into the capillaries until equilibrium is reached. Conversely systemic venous blood has a carbon dioxide partial pressure of 46 mmHg, whereas pressure in alveoli is 40 mmHg, carbon dioxide moves from the capillaries into the alveoli until equilibrium is reached. Muscles involved in the ventilation are not able to contract spontaneously like cardiac muscle; instead periodic nervous input is required from a group of neurons located mainly within the pons and medulla oblongata. This regulatory system can change its behavior according to inputs from peripheral chemoreceptors in the carotid and aortic bodies, central chemoreceptors in medulla which respond to low oxygen or high carbon dioxide levels in blood. High level of carbon dioxide modulates respiration by increasing the rate and depth of ventilation. Higher brain centers in hypothalamus and cerebrum can also affect the basic breathing pattern via voluntary control and emotional stress.

## 1.2 Assessing Autonomic Function with Mean Signal Values

Cardiovascular variables such as blood pressure and heart rate are maintained within a narrow range although a multitude of physiological perturbations occur on short and long-term basis. For example respiration affects venous return and heart rate; vessels diameter in vascular bed adapts to local metabolism demand; standing and exercise affect blood pressure; low environmental temperature elicits changes in heart rate. Complex regulatory reflexes compensate the perturbation in order to maintain cardiovascular homeostasis. A quantitative evaluation of autonomic nervous function and level of influence of both autonomic nervous branches to cardiovascular system are often desirable in clinical practice to monitor disease development and guide therapy; and in research to understand mechanisms of physiological regulation such as baroreflex and pathophysiological processes such as orthostatic hypotension. A first attempt to assess autonomic function is to apply simple statistical methods using mean values of cardiovascular signals. Resting heart rate might provide rough measure of autonomic nervous system function only. Subjects with high resting parasympathetic activity such as athletes have lower resting heart rate. Conditions characterized by increased resting heart rate such as congestive heart failure, anemia or hypovolemia are often associated with higher resting sympathetic activity.

Cardiovascular reaction to input stimulus can provide much more information about characteristics of the cardiovascular system and its control. The inputs can be realized either by chemical stimuli using drugs or physical provocative maneuvers. The heart rate response to cold face test and oculocardiac reflex test for example is often used to assess parasympathetic activity while sympathetic activity is typically evaluated by measuring the blood pressure change during postural change, cold pressor test, sustained handgrip or isometric exercise. In the deep breath test subject is asked to breathe at a paced rate of 6 breaths per minute and the mean difference or ratio between the maximal RR interval value recorded during expiration and the minimal RR interval value recorded during expiration is calculated as a marker of vagal activity. In postural change test subject is asked to actively stand up from a supine position and the so-called 30:15 ratio between the change in RR interval occurring 15 and 30 seconds from assumption of the upright posture is calculated as a marker of overall autonomic activity [37]. Exercise heart rate recovery has been proposed for assessing parasympathetic tone. With normal parasympathetic reactivation in the post-phase of exercise, heart rate rapidly decreases in the first 30 seconds followed by a more gradual decrease. Such a biexponential pattern of heart rate decrease during early recovery is not observed in subjects with heart failure [38].

## 1.3 Heart Rate Variability

Mean values of physiological variables provide only little insight into regulatory processes. Therefore quantitative assessment of autonomic nervous function is preferably performed via analysis of fluctuation of measurable physiological variable such as instantaneous heart rate, blood pressure or instantaneous lung volume.

The first link between heart rate and respiration was reported in the 18$^{th}$ century by Stephen Hales. Since then different animal and human experiments have put attention on the existence of physiological rhythms in the beat-to-beat heart rate signal. Additional studies have drawn the relationship between heart rate variability (HRV) and autonomic nervous tone on the heart.

### 1.3.1 Time Domain Estimates of HRV

In a continuous ECG signal the so-called normal-to-normal (NN) interval is the interval between adjacent QRS complexes. Following statistics can be calculated and evaluated from a series of NN intervals [29].

Common time domain parameters include: the root mean squared differences of successive NN intervals (RMSSD) as a measure of parasympathetic activity; the standard deviation of the NN intervals (SDNN) and the coefficient of variation of the NN intervals (CV), both reflecting combined sympathetic and parasympathetic heart rate modulation; the standard deviation of the average NN interval calculated over short periods of e.g. 5 minutes (SDANN); the mean of the 5-minutes standard deviation of the NN intervals calculated over 24 hours (SDNN index); the number of interval differences of successive NN intervals greater than 50 ms (NN50); the proportion obtained by dividing NN50 by the total number of NN intervals (pNN50); the histogram of NN intervals.

SDNN reflects the overall variability, but strongly depends on the duration of the signal. SDNN increases with the length of recording. Although RMSSD and SDNN provide a measure of the degree of autonomic modulation, they do not indicate the level of autonomic tone regulating heart rate. Moreover RMSSD and SDNN can be subject-specific; values for single data sets do not always correspond to the expected increase or decrease of autonomic nervous activity.

The NN interval series can be converted into a geometrical pattern such as sample density distribution of NN interval duration and sample density distribution of differences between adjacent NN intervals. The following geometrical parameters can be calculated: the integral of the density distribution divided by the maximum of the density distribution (HRV triangular index) and the base of a triangle that approximates the NN interval distribution in space (also called triangular interpolation of NN interval histogram TINN). Both indexes reflect overall variability measured over 24 hours but are more influenced by low frequency components. Despite of the fact that geometrical methods have the advantage to be relatively insensitive to the analytical quality of NN intervals, they need at least 20 minutes recording, what makes them inapplicable to assess short-term control of cardiovascular system [29].

### 1.3.2 Frequency Domain Estimates of HRV

Time-domain estimates of HRV are single-time statistics which are calculated by sampling the variable over the time and by computing quantities that do not depend on the time when the samples were taken. The temporal pattern of variation is better captured using two-time statistics based on the relative time duration between consecutive pairs of samples. Common two-time statistics are the autocorrelation function and power spectral analysis.

Analysis of HRV is usually studied in frequency domain by converting heart rate signal to frequency components. Studies have shown that sympathetic and parasympathetic nervous activities make frequency-specific contributions to heart rate signal. Power spectral density analysis of heart rate fluctuations provides a quantitative noninvasive mean of assessing functioning of cardiovascular control and tells how power (i.e. variance) is distributed as function of frequency. [39]

### Spectral Analysis Methods

The most commonly used methods for calculating power spectral density can be classified as non-parametric and parametric. Non-parametric methods use algorithms based on discrete Fourier transform (DFT) usually implemented by its effective fast algorithm fast Fourier transform (FFT). Fourier (1768–1830) described in his *Theorie Analytique de la Chaleur (The Analytical Theory of Heat)* in 1822 that periodic functions can be expressed as the superposition of an infinite series of sinusoidal functions. This observation is true for piecewise regular functions with a finite number of discontinuities and extrema; conditions that are met by heart rate and blood pressure signals. Therefore the property can be used for power spectral analysis of those signals. Parametric methods are based on modeling signals by linear system, often represented by autoregressive (AR) frequency response. The spectrum resulting from the FFT is derived from all the data present in the recorded signal whereas the AR procedure uses the raw data to identify a best-fitting model from which the final spectrum is then calculated. AR approach has the ability to identify the central frequency of the oscillation in an analytic way and is adequate when attention is focused only on rhythmic fluctuations driven by a fixed-rate oscillator. Furthermore AR methods are particularly appropriate when the number of samples available for the analysis is low. On the other hand, when the analysis is focused on broadband powers, FFT methods are preferable [40]. Figure 1.2 shows the same heart rate spectra obtained by means of different analysis methods: A, data are plotted with an unsmoothed fast Fourier transform (FFT) algorithm; B, data are plotted with an autoregressive (AR) model, the order (=13) of which was determined by Akaike criteria; C, data are plotted with a FFT algorithm smoothed with a Gaussian window and appear like those obtained with the AR model having an order of 13 (B); D, data are plotted with an AR model with a high order (=30) and appear like those obtained with the unsmoothed FFT algorithm (A).

Power spectral density can also be calculated using Least-Squares Spectral Analysis (LSSA) instead of fast Fourier transform. LSSA is based on least squares fit of sinusoidal functions [41], thus is preferable when a straight-forward implementation is needed, for example it can be implemented in Matlab in less than a page of code. The magnitudes in the spectrum determine the contribution of each frequency band to the power (i.e. variance) of the signal.

### Interpretation of spectral components

The spectral values for blood pressure and heart rate variability can be calculated for very low (VLF: 0-0.04 Hz), low (LF: 0.04–0.15 Hz) and high (HF: 0.15–0.5 Hz) frequency bands. Figure 1.3 shows recordings of preganglionic thoracic sympathetic and vagus nerve discharges on one artificially ventilated decerebrate cat. A clear correlation between firing rate and power spectral density can be seen. Low frequency heart rate oscillations are considered to be mediated by combined sympathetic and parasympathetic activity at rest while there is a predominance of sympathetic activity during stressful conditions. Some studies demonstrate that human cardiovascular rhythms occurring at 0.1 Hz are related to activity of the sympathetic nervous system since strong fluctuations of muscle sympathetic nerve activity appear about every 10 seconds, a period that is sufficiently long for vascular smooth muscle and sinoatrial node effectors to respond to released norepinephrine and to modulate arterial pressure and heart rate [42].

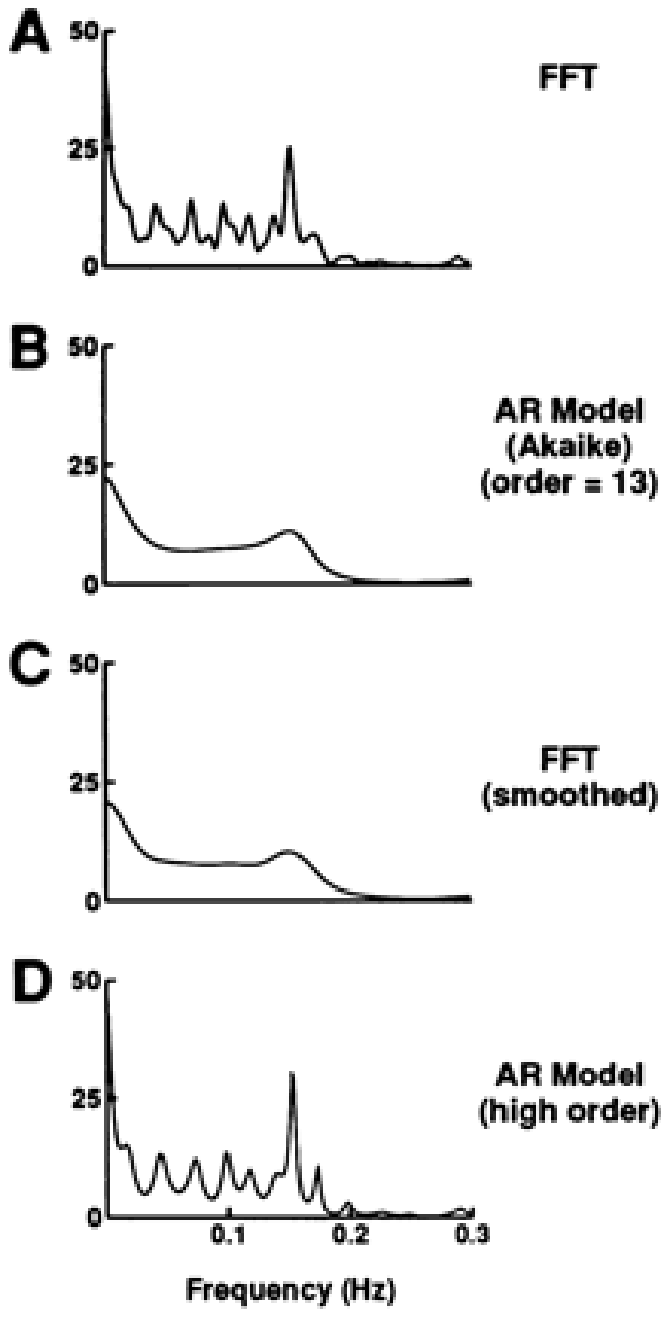

**Figure 1.2 Heart rate spectrum using FFT and AR techniques (reprinted from [40])**

A - unsmoothed fast Fourier transform (FFT) algorithm;
B - autoregressive (AR) model with order 13;
C - fast Fourier transform (FFT) algorithm smoothed with a Gaussian window;
D - autoregressive (AR) model with high order 30

High frequency heart rate oscillations are associated with respiratory sinus arrhythmia and reflect parasympathetic activity. Vagal cardiac control operates like a low-pass filter with a relatively high cutoff frequency, effectively modulating heart rate up to 0.5 Hz, while sympathetic cardiac control operates as a low-pass filter with a much lower cutoff frequency, capable of significantly modulating heart rate only at frequencies below 0.15 Hz [40]. Parasympathetic blockade by atropine eliminates most heart rate fluctuations above 0.15 Hz, while leaving those below 0.15 Hz partly unaffected; whereas cardiac sympathetic blockade with propranolol reduces heart rate fluctuations below 0.15 Hz, while leaving those above 0.15 Hz largely unaffected [43]. At frequencies between 0.025 and 0.07 Hz, the factors involved in heart rate and blood pressure modulation have been regarded as being the renin-angiotensin system, endothelial factors and local influences related to thermoregulation [39] [43].

Measurement of VLF, LF and HF power components is usually made in absolute values of power (unit $ms^2$). Sympathovagal balance is the ratio of the absolute value of power at LF components over the absolute value of power at HF components [42]. LF and HF can also be measured in normalized (also called relative or fractional) units (nu) which represent the relative value of each power component centered at the frequency of interest (LF or HF) divided by total power without power at VLFs. Thus LF and HF in normalized units, denoted by $LF_n$ and $HF_n$ respectively, are defined as follows:

$$LF_n = \frac{LF}{LF + HF}$$
$$HF_n = \frac{HF}{LF + HF} \qquad (1.3.1)$$

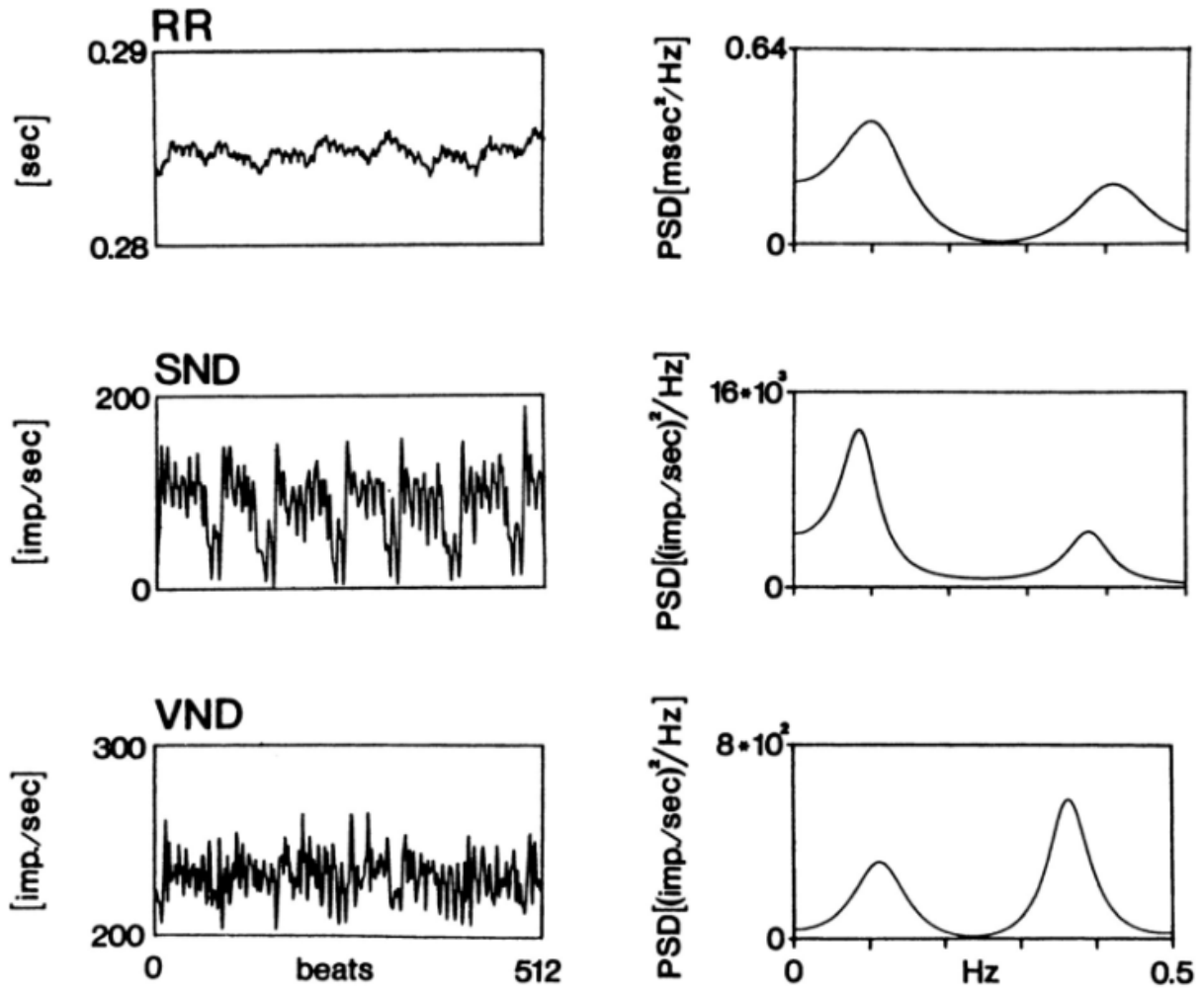

**Figure 1.3 Direct vagal and sympathetic nerves recordings with corresponding HRV (reprinted from [44])**

Spectral analysis of RR interval, preganglionic sympathetic neural discharge (SND) recorded from third left thoracic sympathetic ramus communicans, and efferent vagal neural discharge (VND) simultaneously recorded from left cervical vagus in an artificially ventilated decerebrate cat. Time series of the three signals are illustrated on left panels, whereas their spectra are represented on right panels. A predominant low frequency characterizes RR and SND spectra, whereas a greater respiratory high-frequency component is present in VND variability. PSD, power spectral density.

**Effects of disease, age, race and gender on HRV**

In most autonomic disorders, parasympathetic function is affected before sympathetic function. Depressed HRV after myocardial infarction is characterized by prevalence of sympathetic activity and decrease in vagal tone. Power spectral density of HRV has been used as measure of risk factor after acute myocardial infarction. Reduction of absolute values of power (unit $ms^2$) of LF and HF components has been seen to precede the clinical expression of neuropathy associated with diabetes mellitus. [29]

Absolute values of power (unit $ms^2$) of LF and HF components are associated with age and sex. With increasing age, the high frequency components, which are related to parasympathetic activity, decrease. Women have a lower LF and higher HF/LF ratio than men [45]. The effect of race on heart rate variability is not yet established. Whereas studies in [46] exhibits lower heart rate variability in Black Americans, the opposite has been documented in [45] [47] where black subjects were found to have lower LF, higher HF and higher HF/LF ration than white subjects.

### 1.3.3 Limitations of Heart Rate Variability

Findings in [48] suggest that heart rate variability parameters traditionally associated with parasympathetic tone do not always reliably measure parasympathetic tone, since they decrease with baroreflex parasympathetic stimulation produced by phenylephrine infusion. Furthermore simulation of the right vagus nerve with three different stimulation patterns in anaesthetized, vagotomized and spinal anaesthetized dogs show that the high frequency component of heart rate variability reflects the magnitude of fluctuation in the cardiac parasympathetic input rather than parasympathetic firing rate [49]. Constant frequency vagal stimulation increased cardiac interval, but did not change HRV markedly. Instantaneous stimulation frequency oscillation between 4 and 17 Hz during 5 s period produced a marked heart rate variation. With an increased magnitude of modulation, i.e. the difference between minimum and maximum instantaneous frequency, the high frequency component increased with greater extend. Additional studies put in evidence small respiratory sinus arrhythmia caused by mechanical modulation of sinus rate by stretch after combined pharmacological sympathetic and parasympathetic blockade and after cardiac transplantation [43].

Sympathetic nerve traffic measurement directly by microneurography and intravascular resistance measured by forearm venous occlusion plethysmography during intravenous infusion of nitroprusside, intravenous infusion of phenylephrine or lower body negative pressure have shown that the level of sympathetic cardiovascular modulation cannot always be specifically reflected by the power of heart rate and blood pressure spectral components around 0.1 Hz (LF) [43].

Studies have emphasized the possible occurrence of a high random variability varying from behavioral conditions in components of power spectral density of blood pressure and heart rate, especially in long-term recordings. Neural modulation is influenced by a large number of input signals and a multitude of interactions of central command and reflexes at various brain levels, therefore describing these regulatory processes with narrow frequency regions around 0.1 Hz and 0.3 Hz might be too simplistic. Furthermore the current interpretation of spectral data relies on the assumption that the responses of the system are approximately linear, however neural regulation of the cardiovascular system is characterized by at least two orders of nonlinearity [40].

Although several studies display a quantitative relation between power spectral density of HF components and vagal-cardiac nerve activity, moderate changes of vagal-cardiac nerve activity from baseline do not alter HF power components. Additionally huge changes of power spectral density of HF components provoked by changes in breathing may not reflect changes of vagal-cardiac nerve traffic at all. Therefore it is unclear how sympathovagal balance should be interpreted when changes of breathing frequency occur during recording. Moreover LF power components measured in normalized units did not show significant correlation to levels of myocardial norepinephrine spillover, muscle sympathetic nerve activity or antecubital vein plasma norepinephrine concentrations. [42]

## 1.4 Blood Pressure Variability

High frequency fluctuations of blood pressure are mainly caused by the mechanical effects of respiration on the pressure gradients, size, and functions of the heart and large thoracic vessels. Vagally mediated changes in heart rate and cardiac output play a role in determining the power spectral density of HF components of blood pressure. [43]

Low frequency blood pressure oscillations are considered to be due to sympathetic modulation of vasomotor tone and systemic vascular resistance and also to reflect baroreflex influences on rhythmic fluctuations of arterial blood pressure. Therefore, the power spectral density of LF oscillations of blood pressures are likely to represent useful estimates of sympathetic vascular control [50]. However studies on rats have emphasized that low frequency fluctuations of blood pressure depends not only on sympathetic but also to a substantial extent on non-sympathetic influences [51]. Moreover, sympathetic influences probably do not contribute to low-frequency blood pressure power, having instead a restraining effect. The low frequency components of blood pressure variability should therefore be regarded as a result of combined sympathetic and vagal influences.

Because the use of power spectral density of low frequency components of blood pressure variability is still under debate, nonlinear indexes were investigated in [52]. Authors apply a method of nonlinear time series analysis called recurrence plot. The recurrence plot method produces nonlinear indexes of blood pressure variability and can also be applied to study heart rate variability. It looks for repeated sequences in the data. Figure 1.4 shows a series of 20 systolic blood pressure values *SBP(i)*, where *i* = 1, 2, ..., 20. Two SBP values are the same if their absolute difference is less than a small number *r*, say 2 mmHg. Starting from *SBP(1)*, the algorithm looks whether the same value (105±2 mmHg) occurs later in the series (i.e., for some *j*, with *j* > 1). In Figure 1.4A, the same values are found at *j* = 4, 7, 10, 13, 19. These recurrences can be plotted as points (1, 4), (1, 7), (1, 10), (1, 13) and (1, 19) in a 20 x 20 square the points as shown in Figure 1.4B. Similarly, recurrences of *SBP(2)* can be plotted as points (2, 8) and (2, 16). After plotting recurrent points for all 20 systolic blood pressure values, the next step is to identify the diagonals in the figure. One example is the line that link points (1, 7), (2, 8) and (3, 9). The histograms of the diagonal lengths can be displayed in semilog scale as shown in Figure 1.4C. A decreasing line linking the summits of the histograms can be observed in the figure. The length where this line cuts the abscissa is called $L_{max}$ with value 4 in this case. If $n$ is the total number of observed series and $k$ is the number of the recurrence points, then the percentage of recurrences is then defined as follows:

$$\%rec = \frac{100 \cdot k}{n \cdot (\frac{n-1}{2})} \tag{1.4.1}$$

If $q$ is the count of recurrent points in diagonal lines of length 2 or more, then the percentage of determinism is defined as follows:

$$\%det = \frac{100 \cdot q}{k} \tag{1.4.2}$$

Sympathetic blockade by hexamethonium and in particular α1-sympathetic blockade by prazosin significantly increased $\%rec$, $\%det$ and $L_{max}$ in studies conducted in [52]. Parasympathetic blockade by atropine and β-sympathetic blockade by atenolol did not change the indexes. These results suggest that, in contrast to the LF components of spectral analysis, nonlinear indexes may be more specific markers of sympathetic tones.

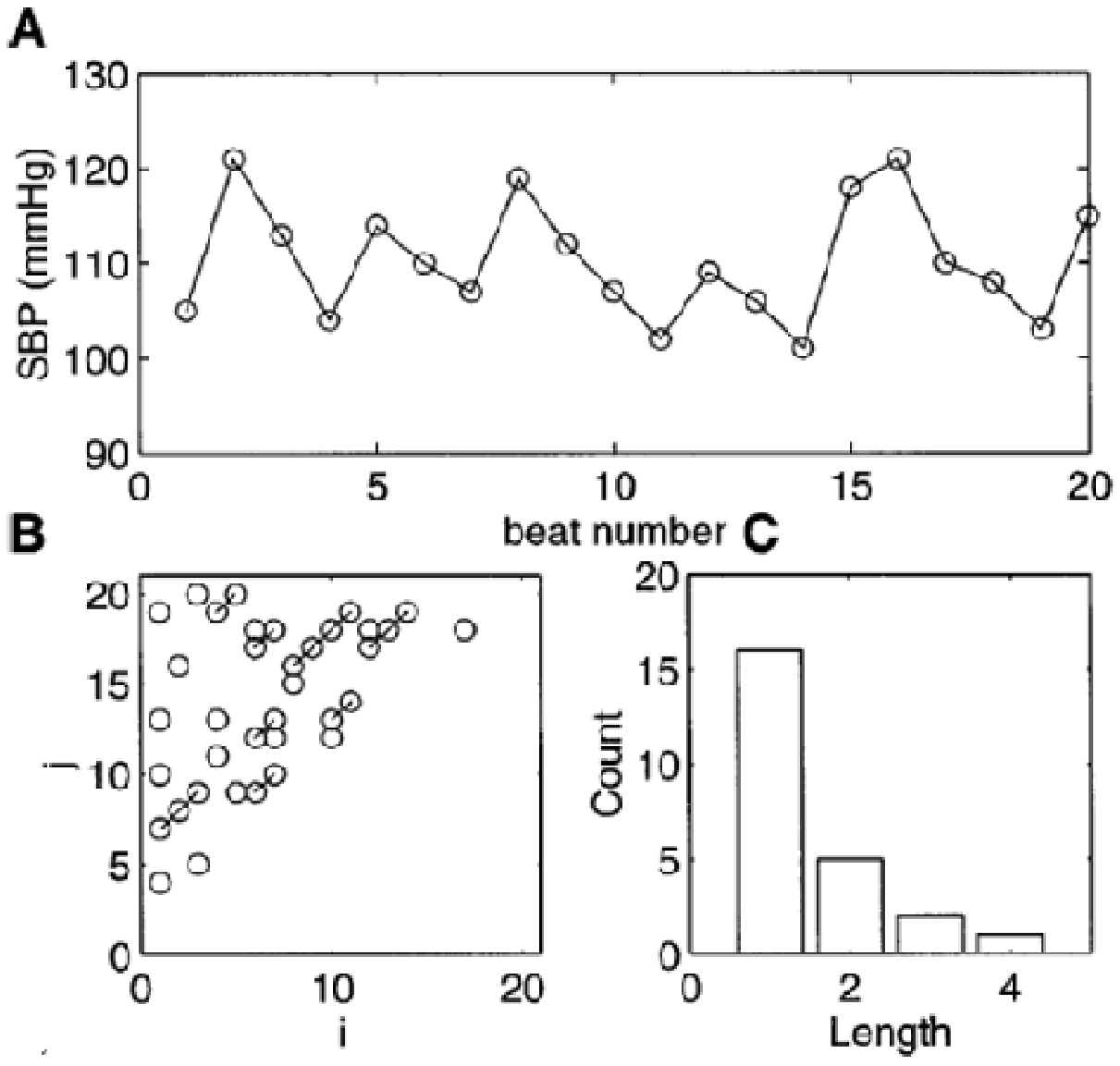

**Figure 1.4 Recurrence plot method for systolic blood pressure (SBP) variability (reprinted from [52])**

## 1.5 Multivariate Fluctuation Analysis

All univariate statistical variables presented in the previous sections fail to capture dependencies between different physiological phenomenon such as heart rate regulation and blood pressure regulation. Multivariable statistics capture such interrelationships and can be used to compute cross-correlation functions, cross-spectra or transfer functions [53]. Furthermore parametric system identification extends transfer function methods and takes into account causality conditions and noises in variables [54].

### 1.5.1 Baroreflex Sensitivity

A time domain method for computing baroreflex sensitivity is based on the identification of sequences of three or more consecutive increasing/decreasing systolic blood pressure and lengthening/shortening of RR interval. The slope of the regression line between systolic blood pressure and RR interval changes is taken as an index of baroreflex sensitivity (unit ms/mmHg).

A more standard assessment of baroreflex sensitivity is made in frequency domain by computation of the squared ratio between the spectral powers of RR interval and systolic blood pressure or of the modulus of the cross-spectrum between systolic blood pressure and RR interval in the frequency regions (0.04 to 0.35 Hz). Baroreflex sensitivity can also be evaluated pharmacologically by relating changes in heart rate to phenylephrine-induced increases and nitroglycerin-induced decreases in arterial pressure [55].

While above methods provide summary information, continuous baroreflex sensitivity can be calculated using Complex Demodulation technique (CDM), which is a nonlinear time domain method of time series analysis. It involves shifting the frequency band of interest to zero by multiplying the original signal with a complex sinusoid at the center frequency of the spectral region of interest (e.g. 0.09 Hz). The resultant complex signal is then low-pass filtered and converted to a polar form to produce amplitude and phase, as a function of time. CDM can work with less than 15 seconds data and provides a continuous assessment of baroreflex sensitivity in the time domain. Other methods presented in the previous paragraphs provide only discrete values over a fix time range. CDM shows values similar to those obtained by the power spectral analysis. A comparison between both methods in supine and standing can be found in [56].

### 1.5.2 Parametric Transfer Function

Heart rate variability does not provide direct information about how heart rate changes in response to varying arterial blood pressure. Pharmacological measurement of baroreflex sensitivity provides information about the coupling between blood pressure and heart rate; it requires however intravenous administration of a vasoactive agent that disrupts normal operating conditions. While the method provides ways for characterizing physiological couplings, it cannot impose causality, which is essential for consistent least-squares estimation when the data are obtained in a closed loop.

Parametric system identification using an autoregressive moving average set of equations has been applied in [54] to impose causality and quantitatively characterize the transfer functions and noise sources associated with closed-loop cardiovascular regulation of heart rate ($HR$), arterial blood pressure ($ABP$) and instantaneous lung volume ($ILV$). The causal structure allows to computationally opening the loop of the closed-loop system, therefore separating the feedforward from the feedback components. A schematic description of the proposed model is shown in Figure 1.5. *Circulatory Mechanics* represents the feedforward coupling between heart rate and arterial blood pressure and is determined by stroke volume and peripheral vascular resistance. *HR Baroreflex* represents feedback baroreflex mediated fluctuations in efferent autonomic activity and *SA Node* represents the coupling between autonomic activity and heart rate. $IVL \rightarrow HR$ represents respiratory sinus arrhythmias. $IVL \rightarrow ABP$ represents mechanical effects of respiration on arterial blood pressure. $N_{HR}$ and $N_{ABP}$ represents perturbing noises on $HR$ which are not caused by fluctuations in $ABP$ or $IVL$ and perturbing noises on $ABP$ which are not caused by fluctuations in $HR$ or $IVL$. The model includes parameters which were determined by minimizing the residual errors using a least squares fit to experimental data obtained during orthostatic stress. According to the authors, results indicate an incremental diminution in the magnitude of the impulse response functions characterizing the two autonomically mediated couplings (*HR Baroreflex* and $IVL \rightarrow HR$) and the power spectral density of the autonomically mediated perturbing noise source ($N_{HR}$) as one progresses from healthy control subjects to minimal, moderate, and severe diabetic autonomic neuropathy groups [54].

A similar approach was adopted in [57] for delineating which major physiological mechanisms contributing to RR Interval and systolic blood pressure variability are affected as a consequence of chronic exposure to sleep disorder breathing. Fluctuations in RR Intervals were assumed to be caused by systolic blood pressure fluctuations via the baroreflex and through direct coupling between respiration and heart rate. Fluctuations in systolic blood pressure were considered to result from changes in intrathoracic pressure during respiration and circulatory dynamics derived from Frank-Starling mechanism and Windkessel runoffs effects. Another application of transfer functions was described in [58] and estimated the residual heart period variability which is neither related to baroreflex nor to direct effects of respiration or low frequency influences independent of baroreflex. This residual fluctuation was interpreted as the effect of the dynamic properties of the sinus node.

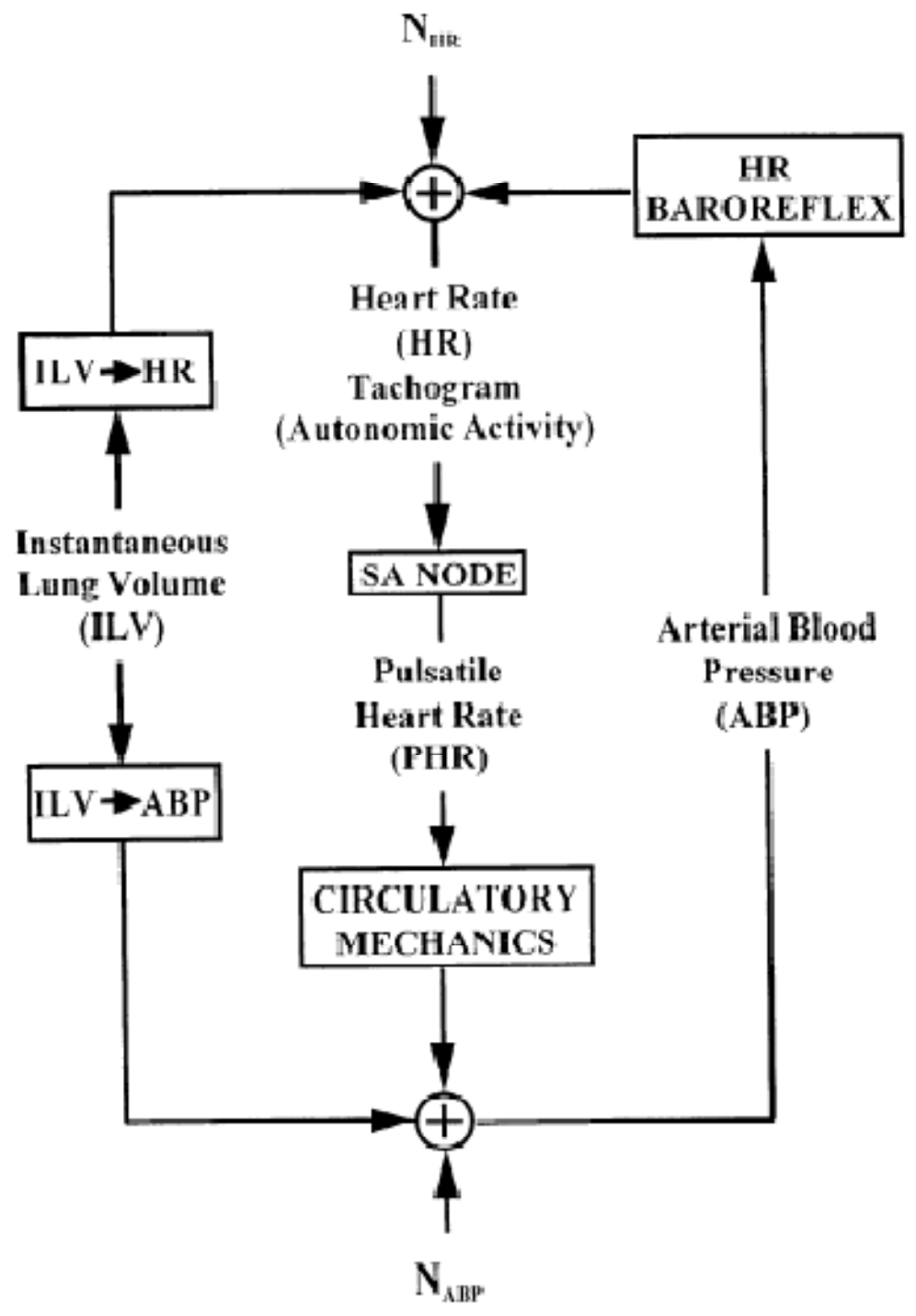

**Figure 1.5 Parametric system identification model (reprinted from [59])**
heart rate (HR), arterial blood pressure (ABP), instantaneous lung volume (IVL), sinoatria node (SA), perturbing noise source ($N_{HR}$) , pulsatile heart rate (PHR)

## 1.6 Plasma Concentration of Cardiac Neurotransmitters

When evaluating the cardiovascular regulation using time-frequency methods, time series analysis and multivariate fluctuation analysis applied to signals recorded in an experimental setup using a provocative maneuver, only baseline values ending approx. 30 seconds before the actual autonomic test are usually considered to assure that baseline values are not biased by preparatory activities, for example, the subject might be afraid of a cold compress approaching the face during the cold face test, what would be reflected by an elevated heart rate just before the autonomic test itself and would corrupt the assessment of autonomic nervous response to the stimuli [50]. Furthermore fluctuation analysis assumes that the regulatory process is stationary, i.e. the statistical properties of the physiological variables do not change over time (stationarity can defined as a difference of less than 5% in the spectral components calculated in two successive 256-beat series). Such requirements are not always fulfilled by the signals and emotional factors make stationary an even unreachable goal.

Parametric functions and minimal kernel approaches introduced in previous sections possibly characterize the dynamics of individual physiological coupling mechanisms involved in cardiovascular regulation. This methodology however assumes that the system is linear and time invariant. This assumption is generally not valid for cardiovascular autonomic control.

Myocardial norepinephrine spillover has shown good capability for measuring sympathetic traffic to the human heart. This measure correlates well with muscle sympathetic nerve activity and plasma norepinephrine levels, at rest and during isometric exercise and mental arithmetic.

The cardiac norepinephrine spillover rate is a biochemical index of the global sympathetic nerve firing rate in the heart (sinoatrial node, atrioventricular node, atria, ventricles). The method requires a continuous intravenous infusion of a radiotracer dose of tritiated norepinephrine to reach steady-state concentration in plasma. Norepinephrine concentration is measured in the coronary sinus via catheterization. Coronary sinus plasma flows is derived from thermodilution-determined blood flows. Endogenous norepinephrine plasma concentration is measured by high-performance liquid chromatography. All concentration values are then used to calculate norepinephrine spillover rate according to the Fick principle. In patients with cardiac failure, the rate was two to three times the value found in age-matched healthy controls (59±14 versus 18+3 ng/min). Compared with young healthy subjects aged 18 to 35 years, older subjects aged 60 to 75 years had an elevated cardiac norepinephrine spillover rate (14.3±2.5 versus 20.1±3.0 ng/min). [60]

Muscle sympathetic nerve activity (MSNA) can be recorded with a tungsten microelectrode, with a tip diameter of a few micrometres, inserted in a muscle fascicle of the peroneal nerve at the fibular head. A reference electrode is inserted subcutaneously a few centimeters away from the recording electrode. When a muscle nerve fascicle has been identified, small electrode adjustments are made until a site is found in which sympathetic impulses can be recorded. MSNA was found to have values ranging from 30 to 50 bursts/min. The mean value of total MSA increased by 81±12% during the last 2 minutes of isometric handgrip. A correlated increase of norepinephrine spillover rate was also simultaneously measured. [61]

## 1.7 Model-based Methods for Assessing Cardiovascular Control

Different methods presented in the previous sections, such as baroreflex sensitivity, plasma or urinary catecholamine levels, and heart rate variability analysis have been employed to quantitatively evaluate autonomic function. However, these methods are both invasive and expensive (e.g. myocardial norepinephrine spillover) or are affected by factors other than alterations in autonomic tone like respiration and they cannot accurately measure sympathetic and parasympathetic tone separately. For example studies show that HRV power in the HF band, above 0.15 Hz, is a satisfactory but incomplete measure of vagal cardiac control [43].

An attempt for improving the quantitative assessment of autonomic cardiovascular control by the spectral analysis approach is to combine the information obtained from the recorded physiological signal with the information derived from mathematical models. Mathematical models help understanding the nonlinear interactions among the different mechanisms and the effect of their individual variability to be accounted for in rigorous quantitative terms [62]. Seminal mathematical models to understand how cardiac output is determined and regulated in response to metabolic demand of the body have been studied for many years by Guyton [24, 63] and Grodins [64]. Over the last decade there has been a large body of research on mathematical models covering features from the molecular to the organ system level. A comprehensive summary of architectures of cardiovascular models is given in [65] where a distinction is made between distributed and lumped models. Distributed models represent the arterial tree in segments where viscous and inertial effects are represented by a resistance and an inductance. Both variables depend on parameters such as segment radius, viscosity, density, length, compliance. Lumped models do not go into such anatomical details, they rather lump vessels into compartments and describe the cardiovascular system as a network of compliances, resistances and inductances. Lumped models can be non-pulsatile or pulsatile. Pulsatility is usually included using a time-varying compliance function. Lumped models can be coupled with mathematical formulation of autonomic nervous control.

### 1.7.1 A Lumped Model of Cardiovascular System

Batzel and coworkers extended the original mathematical model of Grodins with a submodel for local metabolic control [66]. The lumped model does not consider blood flow in the arterial tree and do not distinguish between individual vessels in the venous and arterial part of the systemic or pulmonary circuit. Instead, it lumps them together into four single compliant compartments: *Venous Systemic Compartment* and *Arterial Systemic Compartment; Venous Pulmonary Compartment* and *Arterial Pulmonary Compartment.* The venous systemic and the arterial systemic compartments are connected by the *Systemic Peripheral Compartment.* The venous pulmonary and arterial pulmonary compartments are connected by the *Pulmonary Peripheral Compartment.* Both peripheral compartments represent capillaries, arterioles and venules; and are only characterized by the resistance of their walls to blood flow. The venous systemic and the arterial pulmonary compartments are connected by the *Right Ventricle Compartment.* The venous pulmonary and the arterial systemic compartments are connected by the *Left Ventricle Compartment.* The right and left atrium are assumed to be part of the venous systemic and venous pulmonary compartment respectively.

Since the venous systemic, arterial pulmonary, venous pulmonary and arterial systemic compartments are compliance compartments, we have the following relation, where $V$ is the blood volume, $P$ is the blood pressure and $c$ is the compliance constant of the given compartment.

$$V = c \cdot P \tag{1.7.1}$$

The left and right ventricle are characterized by the generated blood flow, i.e. the cardiac output, which is given by the equation below, where $H$ is the heart rate and $V_{str}$ is the stroke volume, i.e., the volume of blood ejected by one beat of the ventricle.

$$Q = H \cdot V_{str} \tag{1.7.2}$$

The blood flow $F$ through the pulmonary and systemic peripheral compartments depends on the blood pressure in the adjacent compartments and the resistance $R$ against blood in the peripheral regions. This is given by the following relation, where $a$ stands for arterial and $v$ stands for venous.

$$F = \frac{1}{R} \cdot (P_a - P_v) \tag{1.7.3}$$

The following equations describe the filling process and the ventricle stroke volume using the formula of Frank-Starling; where $S$ is the contractility of the ventricle; $P_v$ is the venous filling pressure; $P_a$ is the arterial pressure against which the ventricle has to eject; $R$ is the total resistance to the blood inflow into the ventricle; $t_d$ is the ventricle filling time.

$$\begin{aligned} V_{str} &= \frac{c \cdot (1-k) \cdot P_v \cdot S}{(1-k) \cdot P_a + k.S} \\ k &= e^{-(c \cdot R)^{-1} \cdot t_d} \\ t_d &= \frac{1}{H^{\frac{1}{2}}} \cdot \left( \frac{1}{H^{\frac{1}{2}}} - 0.0387 \right) \end{aligned} \tag{1.7.4}$$

If the subscripts $l, r, as, vs, ap, vp, s, p$ are used to characterize the left heart and the right heart; the arterial systemic, venous systemic, arterial peripheral and venous peripheral compartment; the systemic peripheral and pulmonary peripheral compartment respectively; then volume change over time can be described using the following system of equations:

$$\begin{aligned} C_{as} \cdot \frac{dP_{as}}{dt} &= Q_l - F_s \\ C_{vs} \cdot \frac{dP_{vs}}{dt} &= F_s - Q_r \\ C_{ap} \cdot \frac{dP_{ap}}{dt} &= Q_r - F_p \\ C_{vp} \cdot \frac{dP_{vp}}{dt} &= F_p - Q_l \\ F_s &= \frac{1}{R_s} \cdot (P_{as} - P_{vs}) \\ F_p &= \frac{1}{R_p} \cdot (P_{ap} - P_{vp}) \end{aligned} \tag{1.7.5}$$

The local metabolic control in tissues is achieved by increasing or decreasing the diameter of arterioles, which results in a change of vessel resistance to blood supply with oxygen.

$$\frac{dR_s}{dt} = \frac{1}{K}\left( A_{pesk} \cdot \left( \frac{P_{as} - P_{vs}}{R_s} \right) \cdot C_{a,O_2} - W \right) \tag{1.7.6}$$

where $R_s$ is the peripheral resistance; $A_{pesk}$ is a positive constant; and $C_{a,O_2}$ is the concentration of oxygen in the arterial blood in the capillary region; $K$ is a positive constant. $W$ represents a constant ergonomic workload imposed on muscles in the peripheral region.

This model has been extended with a receding horizon control for the baroreceptor loop in [67]. The controller has to minimize the cost function defined as the error between the current and desired state. The output of the controller is the change in heart rate. This approach to represent the baroreflex control of heart rate using a stabilizing non-linear control derived from optimal control theory however does not quantify the level of activity of efferent sympathetic and parasympathetic fibers.

### 1.7.2 An Open-Loop Model of Cardiovascular Control

Key to the short-term control of blood pressure is the baroreflex or baroreceptor loop, which stabilizes blood pressure by the mean of neural negative feedback. Ottesen and coworkers developed a model of heart rate regulation and divided baroreceptor mechanism into three components: an afferent or sensory component, the central nervous system, and an efferent component [65].

The *Sensory Component* consists of mechanoreceptor cells located in the carotid sinus and aorta arch walls responding to pressure by stretching. Their firing rate $n$ exhibits a number of non-linear phenomena which is modeled using the following system of non-linear differential equations.

$$\begin{aligned}
\frac{dn_1}{dt} &= k_1 \cdot \frac{dP_{as}}{dt} \cdot \frac{n \cdot (M - n)}{(M/2)^2} - \frac{n_1}{\tau_1} \\
\frac{dn_2}{dt} &= k_2 \cdot \frac{dP_{as}}{dt} \cdot \frac{n \cdot (M - n)}{(M/2)^2} - \frac{n_2}{\tau_2} \\
\frac{dn_3}{dt} &= k_3 \cdot \frac{dP_{as}}{dt} \cdot \frac{n \cdot (M - n)}{(M/2)^2} - \frac{n_3}{\tau_3} \\
N &= \frac{M}{2} + \frac{\eta^2}{1+\eta^2} \cdot (M - \frac{M}{2}) \\
n &= n_1 + n_2 + n_3 - N
\end{aligned} \tag{1.7.7}$$

where $k_1, k_2, k_3$ are unknown weighting factors; $\tau_1, \tau_2, \tau_3$ are unknown time constant describing the resetting phenomenon of baroreceptors; $M$ is the high saturation level of the firing rates of baroreceptors with constant value 120 Hz and $N$ is the threshold value of the firing rates depending on an unknown parameter $\eta$ .

Information processing in the *Central Nervous System* is complex and the explicit interaction between firing rate $n$, the efferent sympathetic activity $T_{sym}$, and the efferent parasympathetic activity $T_{par}$ were modeled by Ottesen as follows [68].

$$T_{par} = \frac{n(t)}{M}$$
$$T_{sym} = \frac{1 - \frac{n(t-\tau_d)}{M} + u(t)}{1 + \beta \cdot T_{par}}$$
$$u(t) = -\left[b \cdot (t - t_m)\right]^2 + u_0 \qquad (1.7.8)$$
$$b = \sqrt{\frac{4u_0}{t^2_{per}}}$$
$$t_m = t_{start} + \frac{t_{per}}{2}$$

where $\tau_d$ is the delay of sympathetic response; $\beta$ is the dampening factor of the parasympathetic response; $u(t)$ is the impulse function accounting for the regulation of the sympathetic nerve activity by the vestibulo-sympathetic system; this impulse response is modeled as a parabolic function where $t_{start}, t_{per}$ are the start time and duration of the response, $u_0$ is the amplitude of the response.

The *Efferent Component* models the neural response on the chemical concentration of norepinephrine $C_{nor}$ and acetylcholine $C_{ach}$. The heart rate potential was modeled by Ottesen using an integrate and fire model; when $\phi=1$ the heart beats and the $RR$ interval is taken as the time from $\phi=0$ to $\phi=1$ in milliseconds; the instantenous heart rate $H$ is derived from the $RR$ interval duration in beats/min; $H_0$ is intrinsic heart rate when denervated depending on the age; $M_s, M_p$ represent the strength of the response to changes in the concentrations.

$$\frac{dC_{nor}}{dt} = \frac{-C_{nor} + T_{sym}}{\tau_{nor}}$$
$$\frac{dC_{ach}}{dt} = \frac{-C_{ach} + T_{par}}{\tau_{ach}}$$
$$\frac{d\phi}{dt} = H_0 \cdot (1 + M_s \cdot C_{nor} - M_p \cdot C_{ach}) \qquad (1.7.9)$$
$$H_0 = 1.97 - 9.50 \cdot 10^{-3} \cdot age$$
$$RR = t_{\phi=1} - t_{\phi=0}$$
$$H = \frac{60000}{RR}$$

The model was extended with a nonpulsatile lumped-parameter model of the systemic loop, consisting of the left ventricle, which pumps blood to the arteries, the capillaries, the veins, and back to the right ventricle [69]. Equations for the conservation of average non-pulsatile blood volume are given below, where $P_{as}, P_{vs}$ are the mean arterial and venous pressure respectively; $C_{as}, C_{vs}$ are the compliance of arterial and venous system respectively;

$R_{as}, R_{vs}$ are the resistance to flow through the arterial and venous system respectively; $V_{str}$ is the stroke volume.

$$C_{as} \cdot \frac{dP_{as}}{dt} = -\frac{P_{as} - P_{vs}}{R_{as}} + H \cdot V_{str}$$
$$C_{vs} \cdot \frac{dP_{vs}}{dt} = \frac{P_{as} - P_{vs}}{R_{as}} - \frac{P_{vs}}{R_{vs}} \quad (1.7.10)$$

The model of Ottesen described above provides a quantification of sympathetic $T_{sym}$ and parasympathetic level $T_{par}$ in dimensionless units. It however represents the system as an open-loop and lacks various delays which are known to be present at diverse points of neural pathways. Cardiovascular control system includes some transport and time delays which should be included in mathematical models. The time required for the signals to travel from baroreceptors to the medulla including the processing time in the control centers is about 100 to 300 ms. Typically there is ca. 2.5 s delay before heart rate changes are seen after β-sympathetic nerve stimulation, and a further 7.5 s before the effect is complete. There is ca. 5 s delay before peripheral vasoconstriction begins, after stimulation of α-sympathetic nerves, and a further 15 s delay before vasoconstriction is complete. An excellent review of mathematical models of cardiovascular control including delays is given in [70].

### 1.7.3 A Closed-Loop Model of Cardiovascular Control

Ursino and coworkers developed a complex distributed closed-loop model of the cardiovascular system and applied it to study exercise physiology. Model equations describing pressure and volume in systemic arteries, splanchnic peripheral circulation, splanchnic venous circulation, extrasplanchnic peripheral circulation, extrasplanchnic venous circulation, right atrium, right ventricle, pulmonary arteries, pulmonary peripheral circulation, pulmonary veins, left atrium and left ventricle can be found in [71].

The afferent baroreflex pathway was modeled by Ursino using a linear derivative first-order dynamic block and a sigmoidal static block [62] as follows.

$$f_{ab} = \frac{f_{ab,min} + f_{ab,max} \cdot e^{\frac{\tilde{P} - P_n}{k_{ab}}}}{1 + e^{\frac{\tilde{P} - P_n}{k_{ab}}}}$$
$$k_{ab} = \frac{f_{ab,max} - f_{ab,min}}{4 \cdot G_b} \quad (1.7.11)$$
$$\tau_{pb} \cdot \frac{d\tilde{P}}{dt} = P_{as} + \tau_{zb} \cdot \frac{dP_{as}}{dt} - \tilde{P}$$

where $f_{ab}$ is the firing rate of the carotid baroreceptors in response to the pressure sensed $\tilde{P}$ which depends on the arterial pressure $P_{as}$; $f_{ab,max}, f_{ab,min}$ are the upper and lower saturation levels of firing rate; $\tau_{pb}, \tau_{zb}$ are the time constants for the real pole and the real zero in the linear dynamic block; $P_n$ is the value of baroreceptor pressure at the central point of the sigmoidal function; $k_{ab}$ is a parameter, with the dimension of pressure, related to the slope of the static function at the central point; $G_b$ is the maximum baroreceptor gain.

The afferent chemoreflex pathway was modeled by Ursino using a linear derivative first-order dynamic block and a sigmoidal static block as follows.

$$\frac{df_{ac}}{dt} = \frac{1}{\tau_c} \cdot \left(-f_{ac} + \varphi_{ac}\right)$$
$$\varphi_{ac} = \frac{f_{ac,max} + f_{ac,min} \cdot e^{\frac{P_{as_{O_2}} - P_{as_{O_2 n}}}{k_{ac}}}}{1 + e^{\frac{P_{as_{O_2}} - P_{as_{O_2 n}}}{k_{ac}}}} \tag{1.7.12}$$

where $f_{ac}$ is the firing rate of chemoreceptors as sigmoidal response to changes in partial pressure of oxygen in the arterial systemic compartment $P_{as_{O_2}}$ ; $f_{ac,max}, f_{ac,min}$ are the upper and lower saturation levels of firing rate; $\tau_c$ is a time constant of the chemoreceptors dynamics; $P_{as_{O_{2n}}}$ is the oxygen level at the central point of the sigmoidal function; $k_{ac}$ is a parameter, with the dimension of pressure, related to the slope of the static function at the central point.

The afferent activity of pulmonary stretch receptors was modeled by Ursino using a linear derivative first-order dynamic block and a sigmoidal static block as follows.

$$\frac{df_{ap}}{dt} = \frac{1}{\tau_p} \cdot \left(-f_{ap} + \varphi_{ap}\right)$$
$$\varphi_{ap} = G_{ap} \cdot V_T \tag{1.7.13}$$

where $f_{ap}$ is the firing rate of the pulmonary stretch receptors in response to changes in tidal volume $V_T$ ; $\tau_c$ is a time constant of the lung inflation; $G_{ap}$ is a constant gain.

Efferent sympathetic responses to vessels and heart were modeled by Ursino as exponential trends using weighted sum of afferent inputs from baroreceptors, chemoreceptors and pulmonary stretch receptors [62] as follows.

$$f_{sp} = \begin{cases} f_{es,\infty} + (f_{es,0} - f_{es,\infty}) \cdot e^{k_{es} \cdot (-W_{b,sp} \cdot f_{ab} + W_{c,sp} \cdot f_{ac} - W_{p,sp} \cdot f_{ap} - \theta_{sp})} & if\ f_{sp} < f_{es,max} \\ f_{es,max} & if\ f_{sp} \geq f_{es,max} \end{cases}$$
$$f_{sh} = \begin{cases} f_{es,\infty} + (f_{es,0} - f_{es,\infty}) \cdot e^{k_{es} \cdot (-W_{b,sh} \cdot f_{ab} + W_{c,sh} \cdot f_{ac} - W_{p,sh} \cdot f_{ap} - \theta_{sh})} & if\ f_{sh} < f_{es,max} \\ f_{es,max} & if\ f_{sh} \geq f_{es,max} \end{cases} \tag{1.7.14}$$

where $f_{sp}, f_{sh}$ are the firing rates of efferent sympathetic fibers to vessels and heart; $k_{es}, f_{es,max}, f_{es,\infty}, f_{es,0}$ are constants; $W_{b,sp}, W_{c,sp}, W_{p,sp}, W_{b,sh}, W_{c,sh}, W_{p,sh}$ are synaptic weights; $\theta_{sp}, \theta_{sh}$ are offset terms for sympathetic neural activation.

Efferent parasympathetic response to heart was modeled by Ursino as follows.

$$f_v = \frac{f_{ev,0} + f_{ev,\infty} \cdot e^{\frac{f_{ab} - f_{ab,0}}{k_{ev}}}}{1 + e^{\frac{f_{ab} - f_{ab,0}}{k_{ev}}}} + W_{c,v} \cdot f_{ac} - W_{p,v} \cdot f_{ap} - \theta_v \tag{1.7.15}$$

where $f_v$ is the firing rate of efferent vagal fibers to heart; $k_{ev}, f_{ev,\infty}, f_{ev,0}$ are constants; $W_{c,v}, W_{p,v}$ are synaptic weights; $\theta_v$ is an offset term for parasympathetic neural activation.

The response of vessel resistances was modeled by Ursino to include a pure latency, a monotonic logarithmic static function, and low-pass first order dynamics [62].

$$\begin{aligned} & R_s(t) = \Delta R_s(t) + R_s(0) \\ & \frac{d\Delta R_s}{dt} = \frac{1}{\tau_{R_S}} \cdot (-\Delta R_s + \sigma_{R_s}) \\ & \sigma_{R_s} = \begin{cases} G_{R_S} \cdot ln[f_{sp}(t - D_{R_S}) - f_{es,min} + 1] & if\ f_{sp} \geq f_{es,min} \\ 0 & if\ f_{sp} < f_{es,min} \end{cases} \end{aligned} \tag{1.7.16}$$

where $R_s$ is the resistance of peripheral vessels with $R_s(0)$ being its baseline value; $\tau_{R_S}$ is a time constant; $G_{R_S}$ is a constant gain factor; $D_{R_S}$ is the time delay for sympathetic response to take effect; $f_{es,\min}$ is a threshold for sympathetic stimulation.

The response of cardiac elastances to the sympathetic drive was modeled by Ursino as follows.

$$\begin{aligned} & E_H(t) = \Delta E_H(t) + E_H(0) \\ & \frac{d\Delta E_H}{dt} = \frac{1}{\tau_{E_H}} \cdot (-\Delta E_H + \sigma_{E_H}) \\ & \sigma_{E_H} = \begin{cases} G_{E_H} \cdot ln[f_{sh}(t - D_{E_H}) - f_{es,min} + 1] & if\ f_{sh} \geq f_{es,min} \\ 0 & if\ f_{sh} < f_{es,min} \end{cases} \end{aligned} \tag{1.7.17}$$

where $E_H$ is the elastance of heart chambers with $E_H(0)$ being its baseline value; $\tau_{E_H}$ is a time constant; $G_{E_H}$ is a constant gain factor; $D_{E_H}$ is the time delay for sympathetic response to take effect.

The response of heart period to the sympathetic and parasympathetic drive was modeled by Ursino as follows.

$$\begin{aligned} & T(t) = \Delta T_S(t) + \Delta T_V(t) + T_0 \\ & \frac{d\Delta T_S}{dt} = \frac{1}{\tau_{T_S}} \cdot (-\Delta T_S + \sigma_{T_S}) \\ & \frac{d\Delta T_V}{dt} = \frac{1}{\tau_{T_V}} \cdot (-\Delta T_V + \sigma_{T_V}) \\ & \sigma_{T_S} = \begin{cases} G_{T_S} \cdot ln[f_{sh}(t - D_{T_S}) - f_{es,min} + 1] & if\ f_{sh} \geq f_{es,min} \\ 0 & if\ f_{sh} < f_{es,min} \end{cases} \\ & \sigma_{T_V} = G_{T_S} \cdot f_v(t - D_{T_V}) \end{aligned} \tag{1.7.18}$$

where $T$ is the heart period with $T_0$ being its baseline value; $\tau_{T_S}, \tau_{T_V}$ are a time constants; $G_{T_S}, G_{T_V}$ are constant gain factors; $D_{T_S}, D_{T_V}$ are the time delays for sympathetic and parasympathetic responses to take effect.

The model of Ursino and coworkers described in this section includes analytical description of efferent sympathetic and parasympathetic activity as response to exercise. However additional stimuli on the cardiovascular system are not covered, such as cold stimulation inducing bradycardia and cogbitive stress inducing pressor response.

### 1.7.4 Modeling External Stimuli on Cardiovascular System

The transition from sitting to standing was modeled by Olufsen by including gravitational forces to arterial systemic pressure [72] .

$$
\begin{aligned}
&[P_{as}]_{stand} = [P_{as}]_{sit} + \rho \cdot g \cdot h(t) \\
&h(t) = \begin{cases} 0, & t < t_{st} \\ h_M \cdot (t - t_{st}) / \tau, & t_{st} \le t \le t_{st} + \tau \\ h_M, & t \ge t_{st} + \tau \end{cases}
\end{aligned}
\tag{1.7.19}
$$

where $P_{as}$ is the pressure in arterial systemic compartment in legs; $\rho = 1.055 g / cm^3$ is the density, $g = 981 cm / s^2$ is the gravitational acceleration, $h$ is the change in height from sitting to standing, $t_{st}$ is the timestamp when subject stands up, $\tau$ is the duration of the transition from sit to stand, $h_M$ is the maximum height needed for the mean arterial pressure to drop (e.g. 25.16 cm).

Physical exercise can put some challenges to the cardiovascular system. Elstad and coworkers developed a non-distributed model which demonstrates that a rapid increase in baroreflex setpoint and locally induced vasodilation of vessels in exercising muscles account for almost all cardiovascular changes seen during moderated exercise [73]. Local metabolic demand in exercising muscles modulates peripheral vessel resistance. Vasodilation induced by lower oxygen level in blood was modeled by Magosso using a static gain and first-order low pass dynamics [74].

$$
\begin{aligned}
&R_{am}(t) = \frac{R_s(t)}{1 + x_{am}} \\
&\frac{dx_{am}}{dt} = \frac{1}{\tau_{am}} \cdot [-x_{am} - G_{am} \cdot (C_{v,O_2} - C_{v,O_2 n})] \\
&C_{v,O_2} = C_{a,O_2} - \frac{M_{am}}{F_{am}} \\
&M_{am} = M_{am,n} \cdot (1 + x_M) \\
&\frac{dx_M}{dt} = \frac{1}{\tau_M} \cdot (-x_M - G_M \cdot I)
\end{aligned}
\tag{1.7.20}
$$

where $R_{am}$ is the peripheral resistance in the active muscles; $\tau_{am}$ is a time constant; $G_{am}$ is a static gain constant; $C_{v,O_2}$ is the oxygen concentration in venous blood leaving the active muscles with baseline value $C_{v,O_2 n}$; $C_{a,O_2}$ is oxygen concentration in arterial blood entering the active muscles; $F_{am}$ is blood flow through the active muscles; $M_{am}$ is the oxygen consumption rate with baseline value $M_{am,n}$ and depends on exercise intensity $I$ ; $\tau_M$ is a time constant and $G_M$ is a static gain.

The respiration response to exercise was empirically described by Magosso as a function of relative exercise intensity as follows.

$$
\begin{aligned}
&V_T = V_{T,n} + \Delta V_T \\
&\Delta V_T = \frac{G_{V_T} \cdot I}{\Delta V_{steady}} \cdot \Delta V \\
&V = V_n + \Delta V \cdot I(t - D_V) \\
&\Delta V = \Delta V_{fast} + \Delta V_{slow} \\
&\Delta V_{fast} = 0.45 \cdot \Delta V_{steady} \\
&\frac{d\Delta V_{slow}}{dt} = \frac{1}{\tau_V} \cdot (-\Delta V_{slow} + 0.55 \cdot \Delta V_{steady}) \\
&\Delta V_{steady} = A \cdot I + B \cdot I^2
\end{aligned}
\qquad (1.7.21)
$$

where $V_T$ is the tidal volume with baseline value $V_{T,n}$; $G_{V_T}$ is a static gain; $I$ is exercise intensity; $V$ is the minute ventilation with baseline value $V_n$; $D_V$ is the time delay between onset of exercise and onset of ventilation response; $\tau_V$ is a time constant; $A, B$ are constant parameters.

The direct action of central command on sympathetic efferent activity on heart and vessels as well as on parasympathetic efferent activity on heart were modeled by Magosso and Ursino by modifying equations (1.7.14) and (1.7.15) to include additive terms $\gamma_{sp}, \gamma_{sh}, \gamma_v$ which are sigmoidal functions, defined as follows.

$$
\gamma_i = \frac{\gamma_{i,min} + \gamma_{i,max} \cdot e^{\frac{I - I_{0,i}}{k_{cc,i}}}}{1 + e^{\frac{I - I_{0,i}}{k_{cc,i}}}} \qquad i = sp, sh, s \qquad (1.7.22)
$$

where $\gamma_{i,min}, \gamma_{i,max}$ are the minimum and maximum saturation level of central command; $I_{0,i}$ is the value of exercise intensity at the central point of the sigmoid; $k_{cc,i}$ is a parameter related to the slope of the sigmoid at the central point.

## 1.8 Conclusion

Autonomic neural regulation of the cardiovascular system is characterized by at least two orders of nonlinearity. Examples include the nonadditive nature of the interactions between cardiac sympathetic and parasympathetic responses, the cardiac phase–dependent response to vagal stimuli and the nonlinear modulation of vagal and sympathetic neural outflow by respiration. Additional nonlinearities may originate in specific behavioral and experimental conditions. Such system nonlinearities are present regardless the operating point and will drive the cardiovascular system control mechanisms to operate out of their linear range. Although physiological control systems appear to have steady-state responses that are sigmoidal and include a threshold, a linear operating regime and a saturation point, the dynamic response can be thought of as continuously moving up and down the sigmoid curve. An increase in the mean operating point of the input may be associated with an increase or decrease in the dynamic gain of the system. Therefore changes in the activity of cardiovascular control mechanisms may not be linearly related to changes in heart rate variability (HRV) or blood pressure variability (BPV). Thus, a measure of HRV or BPV may fail to quantify alterations of autonomic cardiovascular influences in several instances [40]. Interrelating both variables to compute baroreflex sensitivity index overcomes the limitations of univariate cardiovascular signal analysis but its frequency variant does not take into account the fact that heart rate and blood pressure are physiologically coupled in a closed-loop control system while the sequence variant of baroreflex sensitivity does not consider the influence of respiration on heart rate.

Mathematical models, such as those proposed by Ursino and coworkers in the previous section 1.7 aim to solve such limitations of signal-based methods for assessing cardiovascular control. Whereas these conventional mathematical models of neural regulation picture a reciprocal control of cardiac vagal and sympathetic nervous activity, as seen during a baroreflex, many other reflexes involve simultaneous co-activation of both autonomic efferent branches. Autonomic co-activation occurs during peripheral chemoreceptor, diving, oculocardiac, somatic nociceptor reflex responses as well as being evoked from structures within higher brain centers. Vagal activity can by itself produce a paradoxical vagally mediated tachycardia. [75]

Another factor which is missing in most mathematical models is the fact that, even at rest, the heart receives tonic drives from both sympathetic and parasympathetic cardiac nerves. Presympathetic cardiovascular neurons in the rostral part of the ventrolateral medulla have the capacity to generate action potentials in the absence of synaptic inputs. These neurons are also chemosensitive and this may contribute to their tonic activation under conditions of hypoxia or hypercapnia. [76]

As emphasized in the work by Eckberg [77], many mathematical models involve prediction of cardiovascular changes that occur when a subject is perturbed by diverse provocative maneuvers. They however fail to include some index of the intensity of the perturbation that elicits the response which is linearly or nonlinearity related to the perturbation; and the factor in subjects' abilities to meet the challenges posed by the perturbations.

All these missing features constitute the challenges which this thesis aims to solve.

# Chapter 2. Our Integrative Physiologically-based Model of Cardiovascular Control

In the previous chapter we put existing methods for quantitative assessment of autonomic nervous activity on the cardiovascular system under loop and identified some unfulfilled requirements. In this chapter we adopt a physiological approach and develop an integrative model of cardiovascular control that quantifies the activity of both sympathetic and parasympathetic branches which determine normal cardiovascular behavior. The model should be further used for quantifying the activity of autonomic nervous system on the cardiovascular system.

## 2.1 Modeling Notation and Requirements

An excellent introduction to the modeling concept is given in [78]. Modeling is one of the basic methods of science used to describe and understand complex systems. A physiological model is an abstract representation that enables us to answer questions about the underlying physiological system. The model simulates those aspects of the physiological system that are relevant to the problem under consideration. The application domain is modeled as a set of objects and relationships.

A modeling notation system is a graphical, mathematical and textual set of rules for representing the model. We will borrow the UML notation from the field of Software Engineering that provides a standard for modeling complex software systems. UML stands for Unified Modeling Language. [78]

The structure of the physiological systems, i.e. cardiovascular system, respiratory system and autonomic nervous system will be described using UML Class Diagrams. Class diagrams represent the anatomical features of the system in terms of classes, objects, attributes, operations and their associations. An object concept represents a real-world object, e.g. a neuron; a class is a set of objects which share common properties, e.g. the class of all baroreceptors; an attribute is a property of an object, e.g. threshold of a neuron; an operation is a concrete activity that the object can perform, e.g. firing of a neuron; an association represents a link between related objects, e.g. the baroreceptor is linked to the rostral ventrolateral medulla. The dynamics of the physiological systems will be modeled using differential equations.

It is necessary to start the modeling activity by defining which parts of the physiological system should be topic of the modeling effort. We consider only the cardio-respiratory control system. Following anatomical and physiological structures are **in scope** of our model:

- The model should focus on autonomic nervous control on the cardiovascular system and makes use of well-established models of its mechanical part.
- The model should include a quantification of various external stimuli influencing the cardiovascular system.
- The model should include a representation of ongoing sympathetic and parasympathetic activity in the absence of any inputs. This pacemaker-like activity is referred as autonomic tone.
- The mechanical part of the model should include pulmonary and systemic circulation. The model should distinguish between the splanchnic and extrasplanchnic circulations since they have different roles in the baroreflex control, causing volume redistribution in response to hemodynamic perturbations [71].
- Arterial pressure pulsatility should be considered since it has a considerable effect on the carotid baroreflex. Pulsatile perfusion of the carotid sinuses modifies the frequency in the sinus nerve and alters baroreflex gain [71].
- Interaction between carotid baroreflex and Frank-Starling relationship should be considered since this can strongly modulate cardiac output and venous return.
- The model should include the effect of chemoreceptors, lung and atria stretch receptors and cold receptors on the heart and blood vessels.

Following anatomical and physiological structures are **out of scope** of our model:

- We do not distinguish between individual vessels in the arterial tree since we are only interest in global control of peripheral vascular resistance (i.e. mostly in arterioles).
- Long-term regulation is not considered because our experimental setups include stimulations which concern a time period of only 1 to 5 min after hemodynamic perturbations. During this period hormonal regulation have negligible effects on the regulatory response.
- Gas exchange at lung and tissues level is not included in the model. We focus on cardiovascular control with little insight in respiration control.
- We ignore inertial effect of blood in venous compartments since they generally function as blood reservoir and the major inertance is already covered in the arterial compartments.

## 2.2 Model of the Cardiovascular System

### 2.2.1 Anatomical Model of the Cardiovascular System

Our anatomical model of the cardiovascular system (without its control by the autonomic nervous system) is a modified representation of the model by Ursino [71]. We schematically described his model using the UML class diagram depicted in Figure 2.1 where classes represent cardiovascular compartments and associations represent blood flow between compartments. The minus sign indicates a model parameter and the plus sign indicates a model variable. Both notations are derived from the UML convention.

The heart is composed of cardiac muscle called myocardium, covered by a thin layer of epithelium and connective tissues, the whole encased in a though membranous sac, the pericardium. The heart is divided by a central wall, or septum, into left and right halves. Each half consists of an atrium and a ventricle, represented here respectively by the classes *Left Atrium ($la$), Left Ventricle ($lv$), Right Atrium ($ra$)* and *Right Ventricle ($rv$)*. Atriums receive blood into the heart and ventricles pump blood out of the heart. The right atrium receives deoxygenated blood coming from body's organs through the venae cavae and pumps it into the right ventricle. From there it is pumped through the pulmonary arteries to the lungs, where it is oxygenated. Arteries, alveolar circulation and veins in the lungs are represented by the classes *Pulmonary Arteries ($pa$), Pulmonary Peripheral ($pp$)* and *Pulmonary Venous ($pv$)* respectively. Blood travels from the lungs to the left atria through the pulmonary veins. It is then passed to the left ventricle. Blood pumped out of the left ventricle enters the large artery aorta. The aorta branches into smaller and smaller arteries that finally lead into a network of capillaries. The arterial tree is lumped into the class *Systemic Arteries ($sa$)*. The network of capillaries and venules is represented by the class *Systemic Peripheral ($sp$)*. After leaving capillaries, blood moves to small veins, then into larger and larger veins until it reaches the superior and inferior venae cavae, which empty into the right atria again. Venous circulation is represented by the class *Systemic Venous*. Heart valves ensure one-way blood flow in the heart: the atrioventricular valves and the semilunar valves.

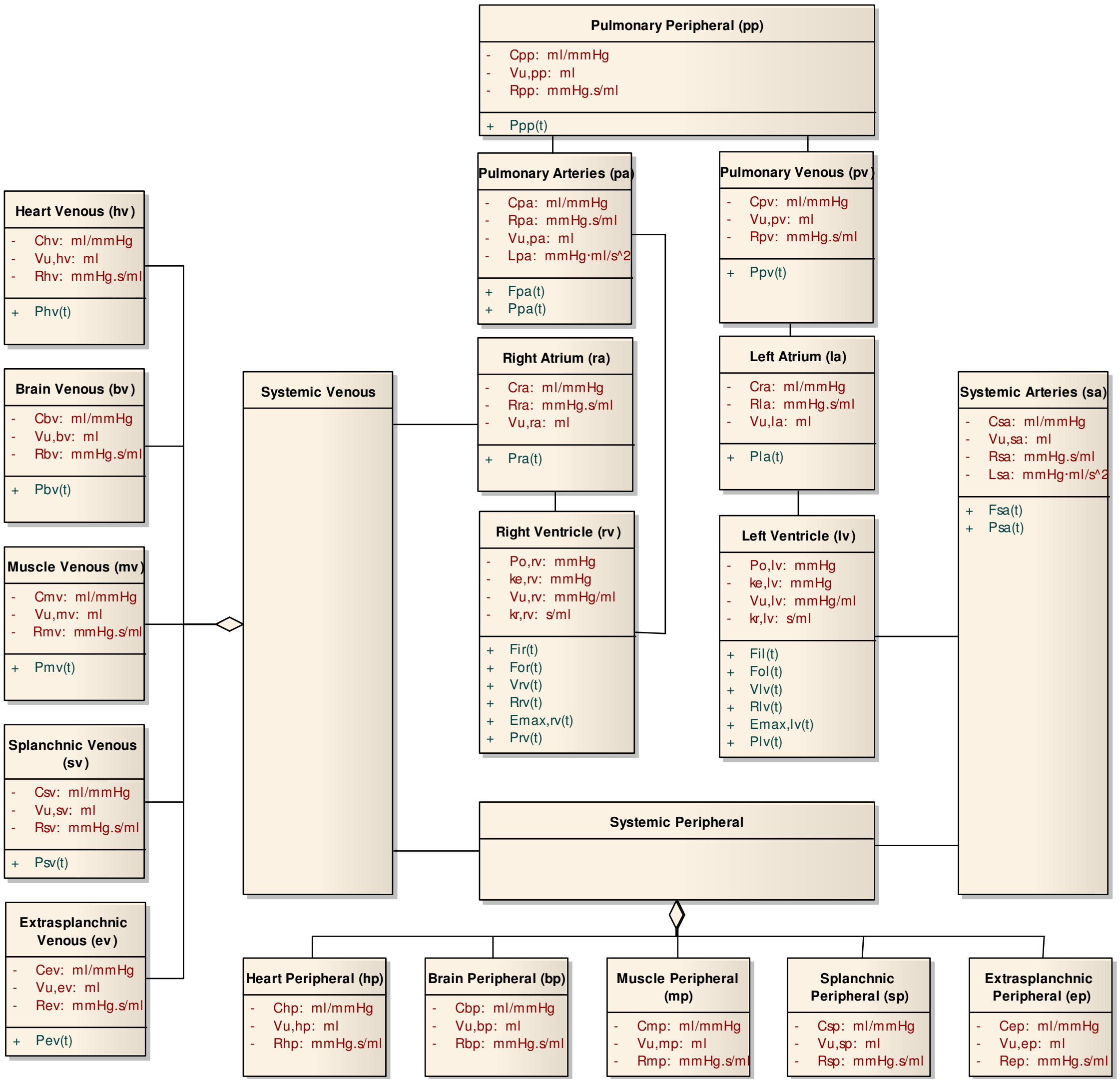

**Figure 2.1 Anatomical Model of Cardiovascular System**

Each class represents an anatomical component. Parameters are preceded by the sign "-" and variables are preceded with by the sign "+"

The systemic peripheral and venous systemic circulations are subdivided into distinct compartments, which exhibit different baroreflex response [71] and different metabolic demand [62]. The corresponding peripheral and venous sub-compartments are coronary, brain, skeletal muscles, splanchnic (intestine, stomach, liver, pancreas, spleen) and extrasplanchnic (kidneys). The corresponding classes are designated using subscripts $hp, bp, mp, sp, ep$ for the systemic peripheral circulation and $hv, bv, mv, sv, ev$ for the systemic venous circulation, respectively.

Venous and arterial compartments are characterized by their compliance, unstressed volume and resistance. These properties are modeled as class attributes using corresponding symbols $C, V_u, R$. The inertial effect of blood in arteries is modeled as class attribute using the inertance symbol $L$.

The capability of circulatory components to generate fluid dynamics is modeled by defining the operations which are available in the corresponding class. Operations related to flow, volume, resistance, elastance and pressure dynamics have the symbols $F, V, R, E, P$ respectively.

### 2.2.2 Physiological Model of the Cardiovascular System

The equations for the conservation of mass and balance of flow forces describing the mechanical activities resulting in a cardiac cycle have been adapted from [62, 71]. We will not describe them in details here since we are only interested in modeling neural control. The equations can be found in Appendix A1 and give the relationship between pressure, volume, flow, resistance, compliance, inertance and elastance during a cardiac cycle which can be resumed in five phases.

In the **first phase**, atrial and ventricular diastole, atria and ventricles are relaxing. The atria are filling with blood from veins.  The ventricles have just completed a contraction and as they relax, the AV valves open and blood flows by gravity from the atria into the ventricles.

In the **second phase**, atrial systole, although most blood enters the ventricles in the phase 1, the last 20% of filling is accomplished during atrial contraction. A small amount of blood is forced backward into the veins.

In the **third phase**, ventricular contraction, blood pushing the underside of AV valves forces them to close. Vibrations following closure of the valves create the first heart sound. With both AV valves closed and ventricles still contracting, high pressure develops. This phase is called isovolumic ventricular contraction. While ventricles are contracting, atria are relaxing. When atrial pressure falls below that in the veins, blood flows from the veins into the atria again.

In the **fourth phase**, ventricular ejection, ventricles have contracted and generated enough pressure to open the semilunar valves. High-pressured blood is pushed into arteries. During this phase, the AV valves remain closed and the atria continue to fill. The stroke volume is approx. 70 mL. Cardiac output is obtained by multiplying the stroke volume with the heart rate (e.g. 70 mL X 72 beats/min ≈ 5 L/min, this means that, at rest, one side of the heart pumps all the blood in the body through it in only one minute). During exercise cardiac output can increase to 35 L/min.

In the **fifth phase**, ventricular relaxation, pressure in ventricles falls below the pressure in arteries and blood starts to flow backward into the heart. This backflow forces the semilunar valves to close. This period is called isovolumic ventricular relaxation because the volume of blood in the ventricles is not changing. When ventricular relaxation causes ventricular pressure to become less than atrial pressure, the AV valves open and blood rushes from the atria into the ventricles. The cardiac cycle has begun again.

## 2.3 Anatomical Model of Cardiovascular Control

Most of the anatomical and physiological background information in this section can be found in any basic physiological textbook [79, 80, 81]. Autonomic control centers in the brain monitor and regulate important functions. The hypothalamus is the center for homeostasis, it activates sympathetic nervous system, maintains body temperature, controls body osmolarity, controls reproductive function, controls food intake, influences behavior, emotions and the cardiovascular control center. The pons makes the relay station between cerebrum and cerebellum and coordinates breathing. The cardiovascular integration centers which receive sensory inputs, the interconnection between nervous control regions, and the sources of efferent signals are gradually being identified but the map is far from complete.

The following five classes of neurons constitute the central neural pathways for cardiovascular regulation as depicted in Figure 2.2:

- Primary afferent neurons that bring sensory information to the brain.
- Interneurons that forward sensory information and inputs from higher brain centers to the premotor neurons.
- Autonomic premotor neurons that control the activities of preganglionic neurons.
- Preganglionic autonomic motor neurons that project to the autonomic ganglion.
- Postganglionic autonomic motor neurons that innervate the effectors organ and implement the cardiovascular response.

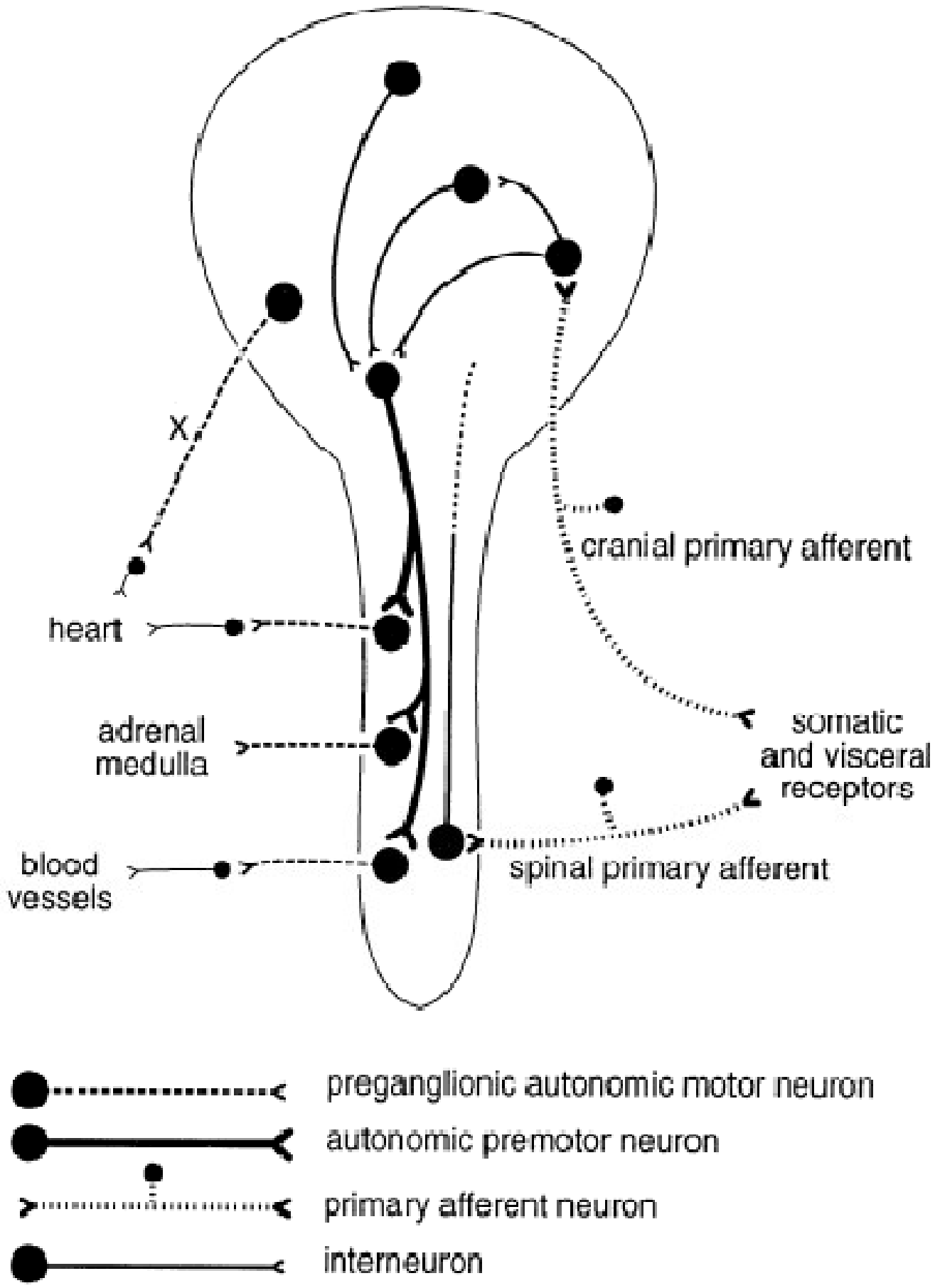

**Figure 2.2 Central Cardiovascular Neural Pathways (reprinted from [81])**

### 2.3.1 Primary Afferent Neurons

A wide range of sensory neurons monitor cardiovascular state. Their organization is modeled using the UML Class Diagram depicted in Figure 2.3 where each class represents a group of receptors sharing the same functionalities. Main classes are baroreceptors, chemoreceptors, stretch receptors, thermoreceptors and nociceptors. They are further subdivided in subclasses using the UML notation for specialization (arrow with triangle ending).

Baroreceptors are stretch-sensitive mechanoreceptors located in the walls of the carotid arteries and aorta, where they monitor the pressure of blood flowing to the brain (carotid baroreceptors) and to the body (aortic baroreceptors). When increased blood pressure in the arteries stretches the baroreceptor membrane, firing rate of the receptor increases. If the blood pressure falls, the firing rate decreases.

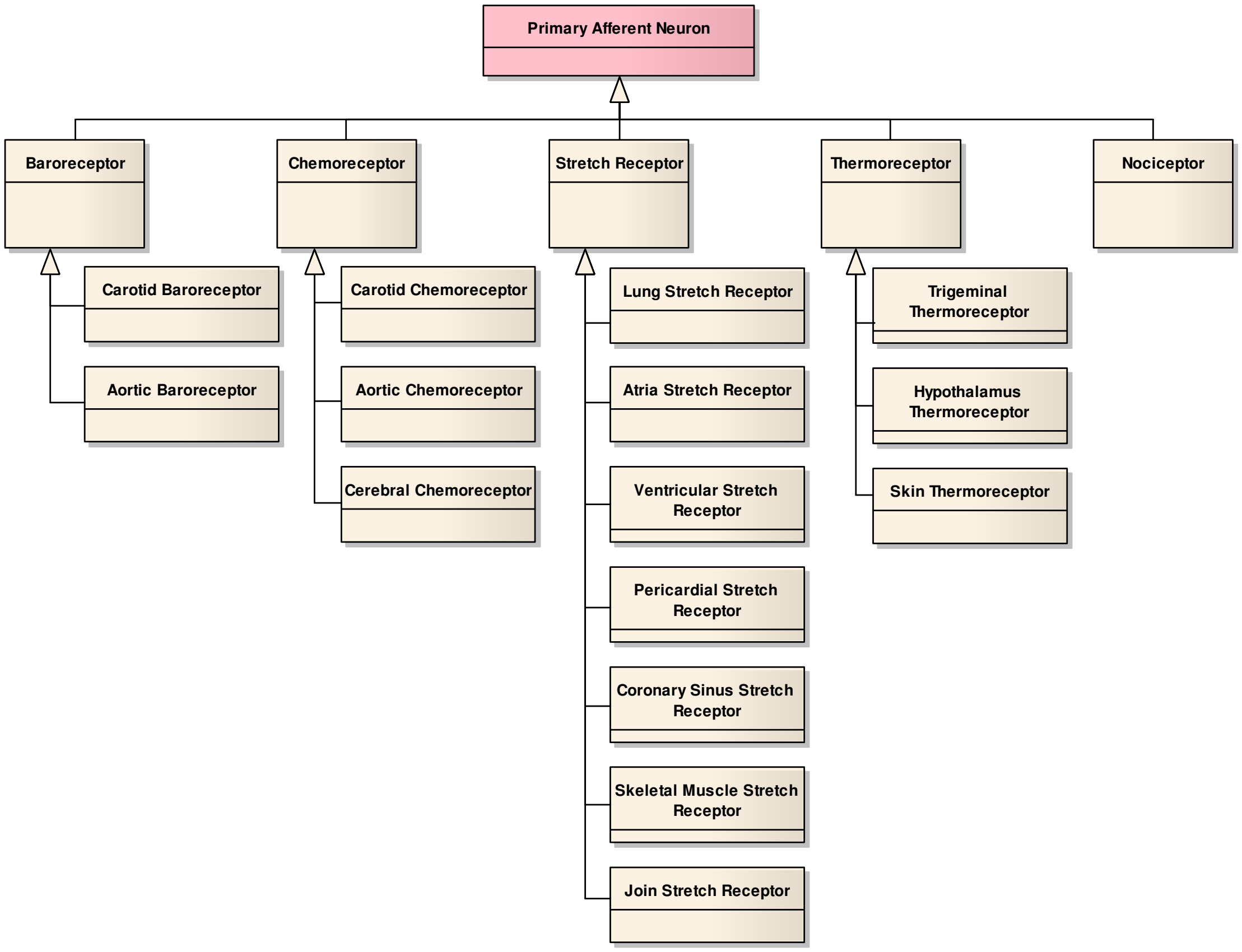

**Figure 2.3 UML Class Diagram for Primary Afferent Neurons**

Cardiovascular function can be modulated by input from peripheral receptors other than the baroreceptors. For example, low blood flow causes pulmonary ventilation to increase. The cardiovascular control center also has reciprocal communication with centers controlling respiration. Increase in breathing rate is usually accompanied by increase in cardiac output.

Additional receptors deliver afferent information: pulmonary artery pressure sensors project to the medulla; right and left atria receptors sense intravascular volume; ventricular, pericardial and coronary sinus stretch sensors causes a reflex fall in pressure. Thermoreceptors in the face sense low temperature and elicit bradycardia. Cerebral or muscle chemoreceptors sense oxygen and carbondioxyde level in brain tissues or skeletal muscles. Thermoreceptors in the hypothalamus monitor blood temperature. Peripheral sensory nerves (pain fibers from viscera and large blood vessels) simulation may elicit a very strong depressor response and causes fainting (vasovagal syncope). Muscles and joints stretch receptors sense tension and length of change and can stimulate sympathetic activity to help reinforcing cardiovascular response to exercise.

Many afferent neurons terminate in the nucleus of the solitary tract, each type of afferent input having its specific terminal field within the tract. Some studies suggest that the solitary tract is capable of modulating or even inhibit afferent input information.

### 2.3.2 Autonomic Premotor Neurons

Sensory neurons projects to autonomic premotor neurons. Their organization is modeled using the UML Class Diagram depicted in Figure 2.4 where each class represents a group of premotor neurons sharing the same functionalities. Reflex reactions are integrated and coordinated through sympathetic and vagal premotor neurons which are further subdivided into inhibitory and excitatory subclasses using the UML notation for specialization (arrow with triangle ending). Premotor neurons can group to form an autonomic nervous center, what is shown using the aggregation notation in UML (arrow with a diamond ending).

Autonomic nervous centers include specific cell groups in the rostral ventrolateral medulla, rostral ventromedial medulla, caudal raphe nuclei, as well as noradrenergic cell group in the caudal ventrolateral pons, and the paraventricular nucleus in the hypothalamus. They innervate the preganglionic outflow to the adrenal medulla and all major sympathetic ganglia.

The CNS (Central Nervous System) coordinates the reflex control of blood pressure mostly from the medulla. Cells in the rostral ventrolateral medulla play a crucial role in the tonic and phasic regulation of blood pressure. The primary goal is to maintain adequate blood flow to the brain and the heart. Action potentials from the baroreceptors travel to the medulla via sensory neurons. The control center integrates sensory input and initiates an appropriate response. Excitation of the rostral ventrolateral medulla cells produces an increase in blood pressure, increase in heart rate and release of adrenomedullary catecholamines. The response is quite rapid and changes cardiac output and peripheral resistance within two heartbeats.  Studies indicate that rostral ventrolateral medulla cells do not affect the sympathetic output to noncardiovascular effectors.

Some rostral ventrolateral medulla neurons display an ongoing synaptic pacemaker activity, which is propagated in the sympathetic preganglionic neurons. These neurons exhibit a spontaneous regular activity even in absence of synaptic inputs. They project to the spinal cord and are inhibited by baroreceptors activation. Blocking all sympathetic activity leads to a drop of arterial pressure of approximately 50%, indicating a continuous basal level of sympathetic firing at normal pressure, called sympathetic tone. The pacemaker neurons within the rostral ventrolateral medulla have much higher firing rates and conduction velocities than non-pacemaker neurons. [82]

Stimulation of rostral ventromedial medulla results in modest increase in blood pressure whereas stimulation of medullary raphe nuclei can elicit either a decrease or an increase of blood pressure depending on the region stimulated. This suggests that medullary raphe nuclei includes both sympathoexcitatory and sympathoinhibitory neurons. Stimulation of the A5 noradrenergic cell group in the pons produces decrease in heart rate and blood pressure.

Stimulation of the caudal ventrolateral medulla cells is accompagnied with inhibition of sympathetic vasomotor activity and a decrease in cardiac contractility. In contrast inhibition of the caudal ventrolateral medulla cells results in the opposite effect. It was suggested that these neurons project to the rostral ventromedial medulla.

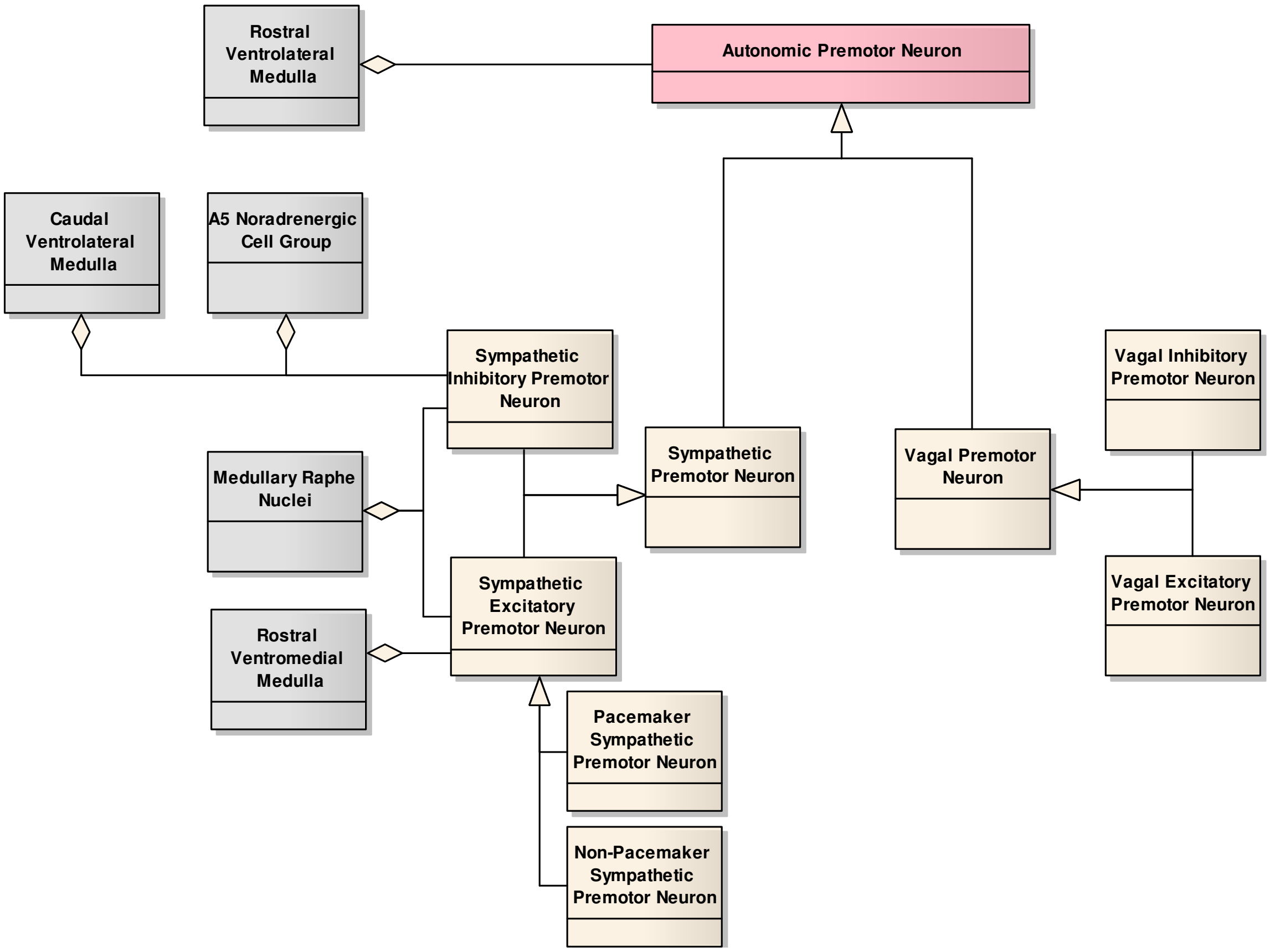

**Figure 2.4 UML Class Diagram for Autonomic Premotor Neurons**

### 2.3.3 Preganglionic Autonomic Motor Neurons

Reflex reactions are carried out by the efferent division of the autonomous nervous system (ANS) whose organization is modeled using the UML Class Diagram depicted in Figure 2.5, where each class represents a group of preganglionic motor neurons sharing the same functionalities. The autonomic division is subdivided into sympathetic and parasympathetic branches which have anatomical differences according to the origin of cell bodies of their preganglionic neurons.

The cell bodies of vagal preganglionic neurons innervating cardiac ganglia are located in the nucleus ambiguous within the ventrolateral medulla, but to a lesser extent also within the dorsal motor nucleus of the vagus. Cardiac vagal preganglionic neurons fire in synchrony with the cardiac cycle, due to an excitatory input arising mainly from baroreceptors. However they also react to inputs from peripheral chemoreceptors, cardiac receptors, and trigeminal cold receptors activated during the diving reflex. Additionally vagal preganglionic nerves are inhibited during the inspiration phase of the respiratory cycle. In general parasympathetic ganglia are located close to their target tissues. The major parasympathetic pathway is the vagus nerve (cranial nerve X). [42]

The cell bodies of sympathetic preganglionic neurons are located in the thoracic (T1) and upper lumbar segments of spinal cord (S2 – S4), especially in the intermediolateral cell column. Groups of nerves with vasomotor and nonvasomotor functions form entirely separate populations and have different locations within the spinal cord. Furthermore sympathetic preganglionic neurons controlling blood vessels in different effectors organs form separate groups, each with their own physiological characteristics. [42]

Particular physiological situations including emotional stress, pain, exercise, blood loss or heart failure activate the adrenal preganglionic sympathetic nerves, which project to both suprarenal glands (adrenal medulla).

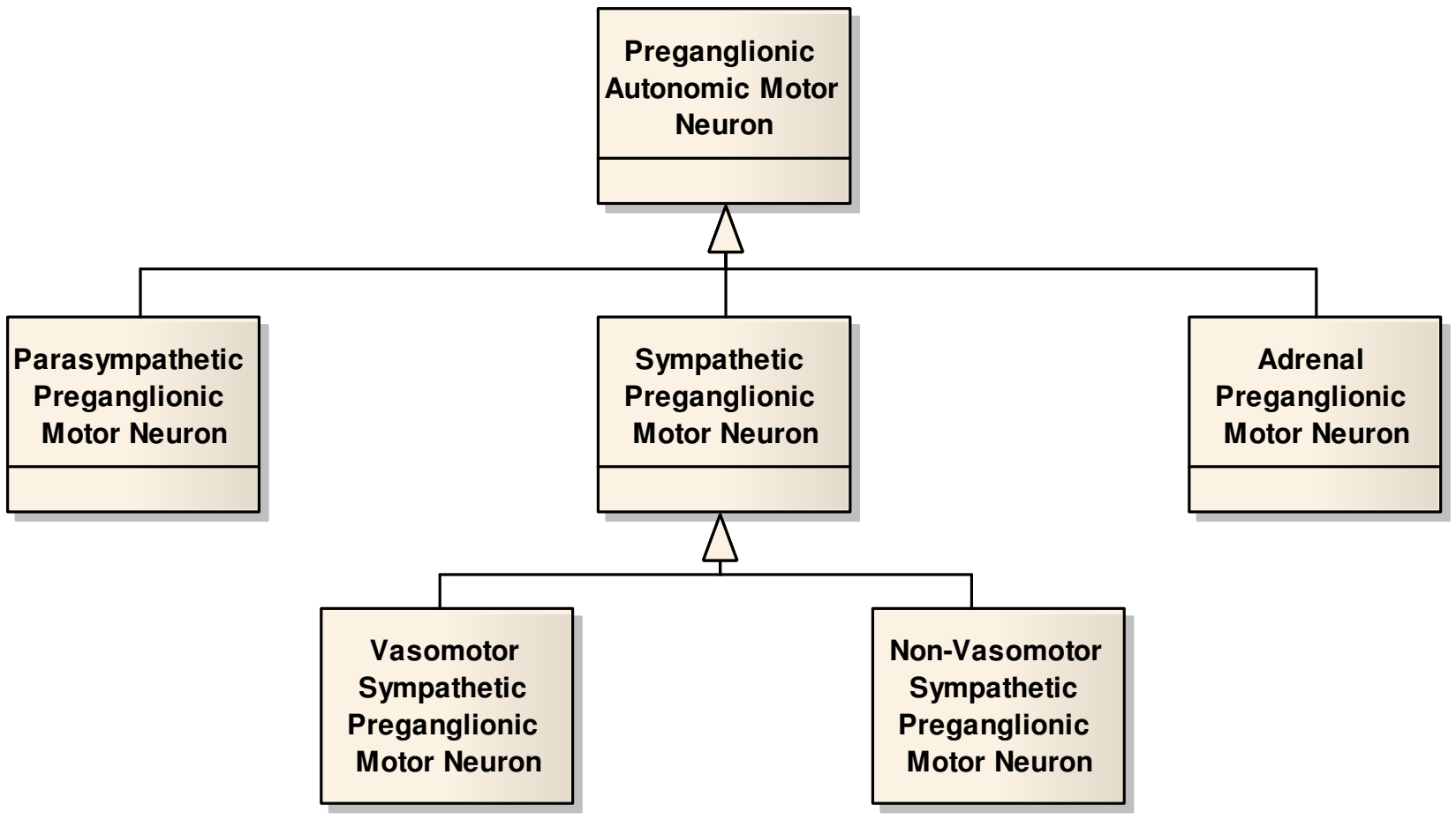

**Figure 2.5 UML Class Diagram for Preganglionic Autonomic Motor Neuron**

### 2.3.4 Postganglionic Autonomic Motor Neurons

Postganglionic autonomic neurons have their cell bodies in the autonomic ganglion and project their axon to the target tissue. Their organization is modeled using the UML Class Diagram depicted in Figure 2.6, where each class represents a group of postganglionic motor neurons sharing the same functionalities. Both sympathetic and parasympathetic postganglionic motor neurons innervate the heart.

The main nerve branches controlling the sympathetic behavior of the heart are bunched together into the cardiopulmonary splanchnic nerves. They approach the base of the heart before splitting into smaller nerves which innervate much of the myocardium, with high concentration around the aortic arch and ventricles.

Parasympathetic postganglionic autonomic neurons originate from ganglia located near the sinoatrial and atrioventricular node. They innervate a large portion of artria. The right vagus nerve primary innervates the sinoatrial node while the left vagus nerve innervates the atrioventricular node. Parasympathetic innervations of ventricles is negligible compared to sympathetic innervations (16-50%), whereas parasympathetic division outcomes sympathetic division by 30-60% in atria. Endocardial and epicardial distribution of sympathetic nerve endings appears to be equal whereas endocardial innervation by parasympathetic nerves surpasses epicardial innervations by some 100%.

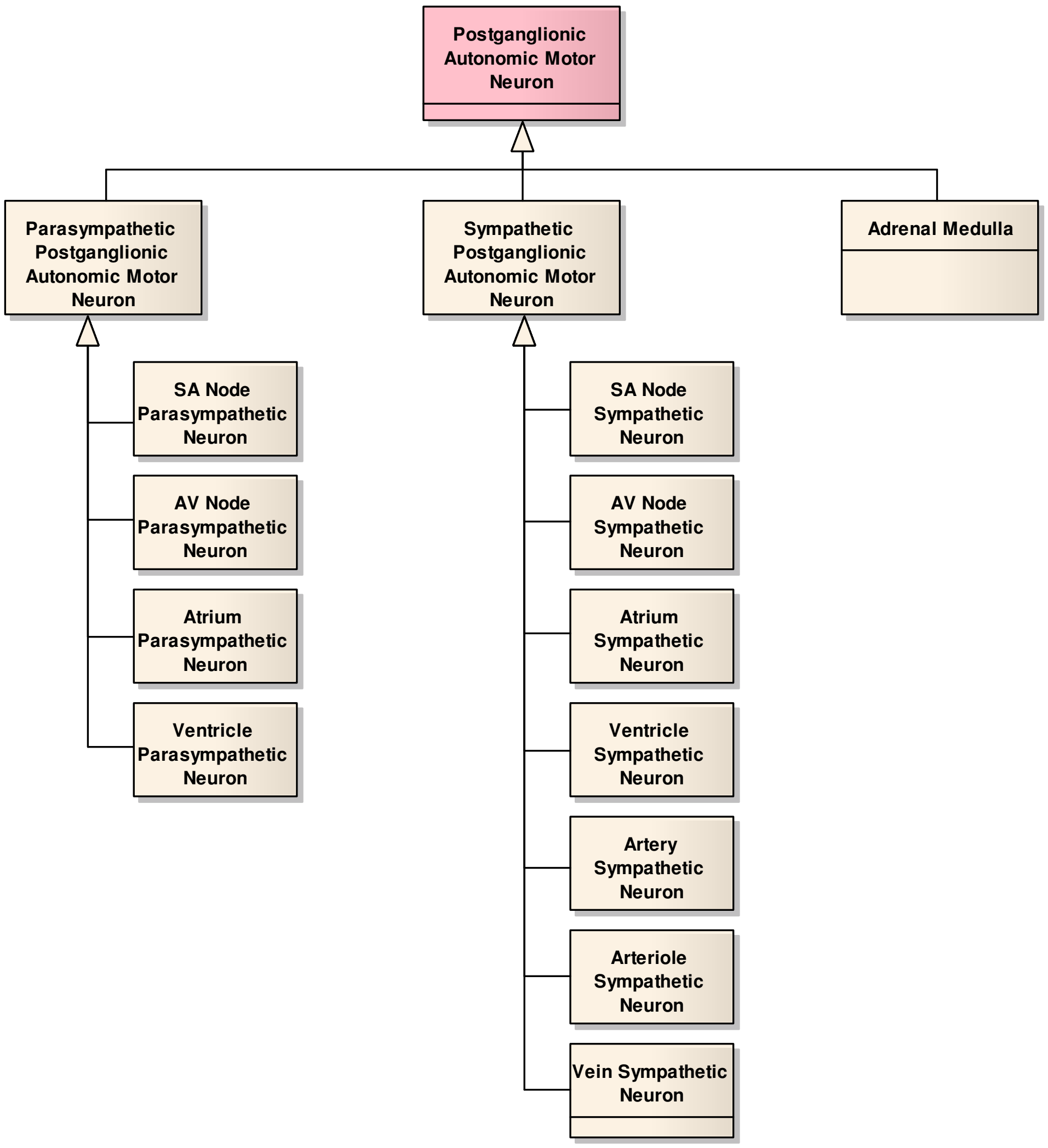

**Figure 2.6 UML Class Diagram for Postganglionic Autonomic Motor Neuron**

Sympathetic postganglionic neurons release norepinephrine at cardiac muscle cells. As result the muscular contraction commences sooner with greater force, the systole duration shortens and the diastole duration increases. With increased relaxation time, the preload increases causing a positive inotropic effect after Frank-Starling law. This results in increased stroke volume.

The adrenal medulla release catecholamines into the circulation which modify heart rate and peripheral vascular resistance. Heart rate is also dependent on changes in circulating thyroxin (thyrotoxicosis causes tachycardia) and by changes in body core temperature (hyperthermia increases heart rate).

Sympathetic postganglionic neurons also innervate blood vessels (arteries, arterioles, veins) which respond with vasoconstriction, vasodilation, venoconstriction, venodilation.

## 2.4 Physiological Model of Primary Afferent Neurons

### 2.4.1 Model of Baroreceptors

Baroreceptors are sensitive to the rate of change of blood pressure and to the mean of pressure. Reduced mean pressure reinforces the baroreflex when pressure falls. Maximal baroreceptors sensitivity occurs when the set point of normal mean arterial pressure is reached (ca. 95 mm Hg in adults). Small deviations from the set point induce large change in firing rate of baroreceptors, however the set point and response shape is not fixed. The resetting of the receptors can occur locally at the level of the baroreceptors themselves or centrally in cardiovascular control centers. Long-term blood pressure regulation cannot be achieved by baroreceptors inputs because of their adaptative nature. We presented a non-linear model of baroreceptors dynamics in section 1.7.2 which has been proposed by Ottesen and co-workers. The model was used in open-loop condition to successfully simulate changes in heart rate induced by orthostatic stress, arterial pressure being provided as input. We recall equation (1.7.7) and describe baroreceptors firing rate $f_{ab}$ as follows:

$$
\begin{aligned}
\frac{dn_{b1}}{dt} &= k_{b1} \cdot \frac{dP_{sa}}{dt} \cdot \frac{f_{ab} \cdot (M_b - f_{ab})}{(M_b / 2)^2} - \frac{n_{b1}}{t_{b1}} \\
\frac{dn_{b2}}{dt} &= k_{b2} \cdot \frac{dP_{sa}}{dt} \cdot \frac{f_{ab} \cdot (M_b - f_{ab})}{(M_b / 2)^2} - \frac{n_{b2}}{t_{b2}} \\
\frac{dn_{b3}}{dt} &= k_{b3} \cdot \frac{dP_{sa}}{dt} \cdot \frac{f_{ab} \cdot (M_b - f_{ab})}{(M_b / 2)^2} - \frac{n_{b3}}{t_{b3}} \\
N_b &= \frac{M_b}{2} + \frac{\eta_b^2}{1+\eta_b^2} \cdot \left( M_b - \frac{M_b}{2} \right) \\
f_{ab} &= n_{b1} + n_{b2} + n_{b3} + N_b
\end{aligned}
\tag{2.4.1}
$$

where $k_{b1}, k_{b2}, k_{b3}$ are constant weighting factors; $\tau_{b1}, \tau_{b2}, \tau_{b3}$ are time constant describing the resetting phenomenon; $M_b$ is the high saturation level of the firing rates and $N_b$ is the threshold value of the firing rates depending on the constant $\eta_b$. Constant values are given in Table 2.4.1 as calculated by Ottesen after fitting the model with heart rate and blood pressure data obtained on healthy young subjects [65]. A simulation of the equations above is given in Figure 2.7 where a decrease of baroreceptor firing rate can be observed with decreased arterial blood pressure from 50$^{th}$ to 60$^{th}$ second.

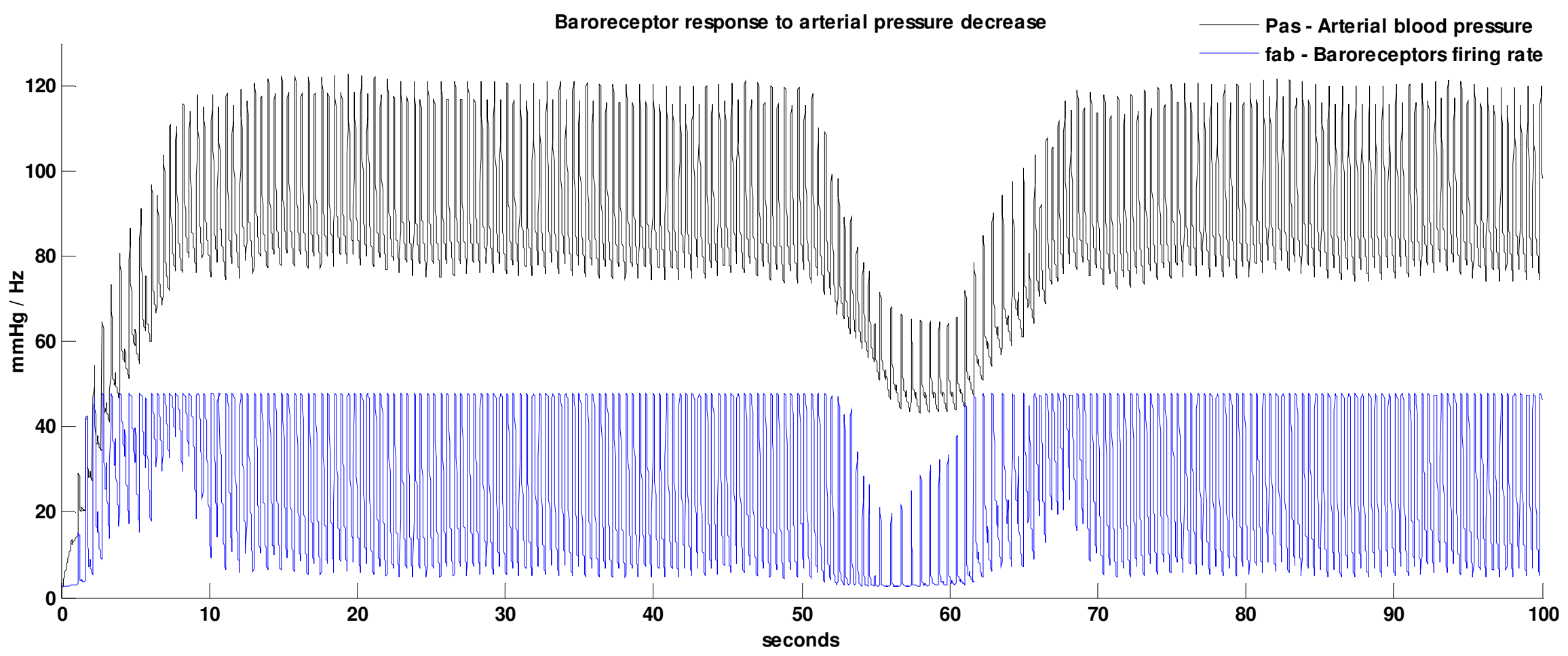

**Figure 2.7 Simulation of baroreceptors**

Receptors firing rate (lower graph) with decreasing arterial blood pressure (upper graph)

Table 2.4.1 Constant values for baroreceptors

| **Constant** | **Value** | **Description** |
|---|---|---|
| $k_{b1}$ | 3.50 ± 4.0 $mmHg^{-1}$ | gain constant for short time response component of baroreceptor |
| $k_{b2}$ | 0.73 ± 0.5 $mmHg^{-1}$ | gain constant for intermediate time response component of baroreceptor |
| $k_{b3}$ | 1.41 ± 1.4 $mmHg^{-1}$ | gain constant for long time response component of baroreceptor |
| $\tau_{b1}$ | 0.82 ± 0.5 s | resetting time for short time response component of baroreceptor |
| $\tau_{b2}$ | 5.58 ± 2.0 s | resetting time for intermediate time response component of baroreceptor |
| $\tau_{b3}$ | 262.03 ± 16 s | resetting time for long time response component of baroreceptor |
| $M_b$ | 120 Hz | maximal firing rate of baroreceptor |
| $\eta_b$ | 1.28 ± 0.3 | model constant |

### 2.4.2 Model of Chemoreceptors

Peripheral chemoreceptors in the carotid and aortic bodies respond to low oxygen or high carbon dioxide levels in blood. Central chemoreceptors lie on the ventral surface of the medulla and sense carbon dioxide level in the cerebrospinal fluid and respond to the resulting pH change. High level of carbon dioxide modulates respiration by increasing the rate and depth of ventilation. A model of chemoreceptors firing rate $f_{ac}$ has been described in section 1.7.3 as sigmoidal response to changes in partial pressure of oxygen $P_{as_{O_2}}$ in the arterial systemic compartment using equation (1.7.12) .

We adopt the equation in our model and use the constant values given in Table 2.4.2 as calculated by Ursino after fitting the model with experimental data [62]. A simulation of the equation is given in Figure 2.8 where an increase of chemoreceptors firing rate can be observed with breath holding, i.e. decreased oxygen partial pressure in arterial blood.

Table 2.4.2 Constant values for chemoreceptors

| Constant | Value | Description |
|---|---|---|
| $\tau_c$ | 2 s | time constant of the chemoreceptor dynamics |
| $k_{ac}$ | 29.27 mmHg | slope of the chemoreceptor sigmoid at the central point |
| $f_{ac,\min}$ | 1.16 Hz | lower saturation of chemoreceptor discharge |
| $f_{ac,\max}$ | 17.07 Hz | upper saturation of chemoreceptor discharge |
| $P_{as_{O_{2n}}}$ | 45mmHg | oxygen partial pressure at the central point of chemoreceptor sigmoid |

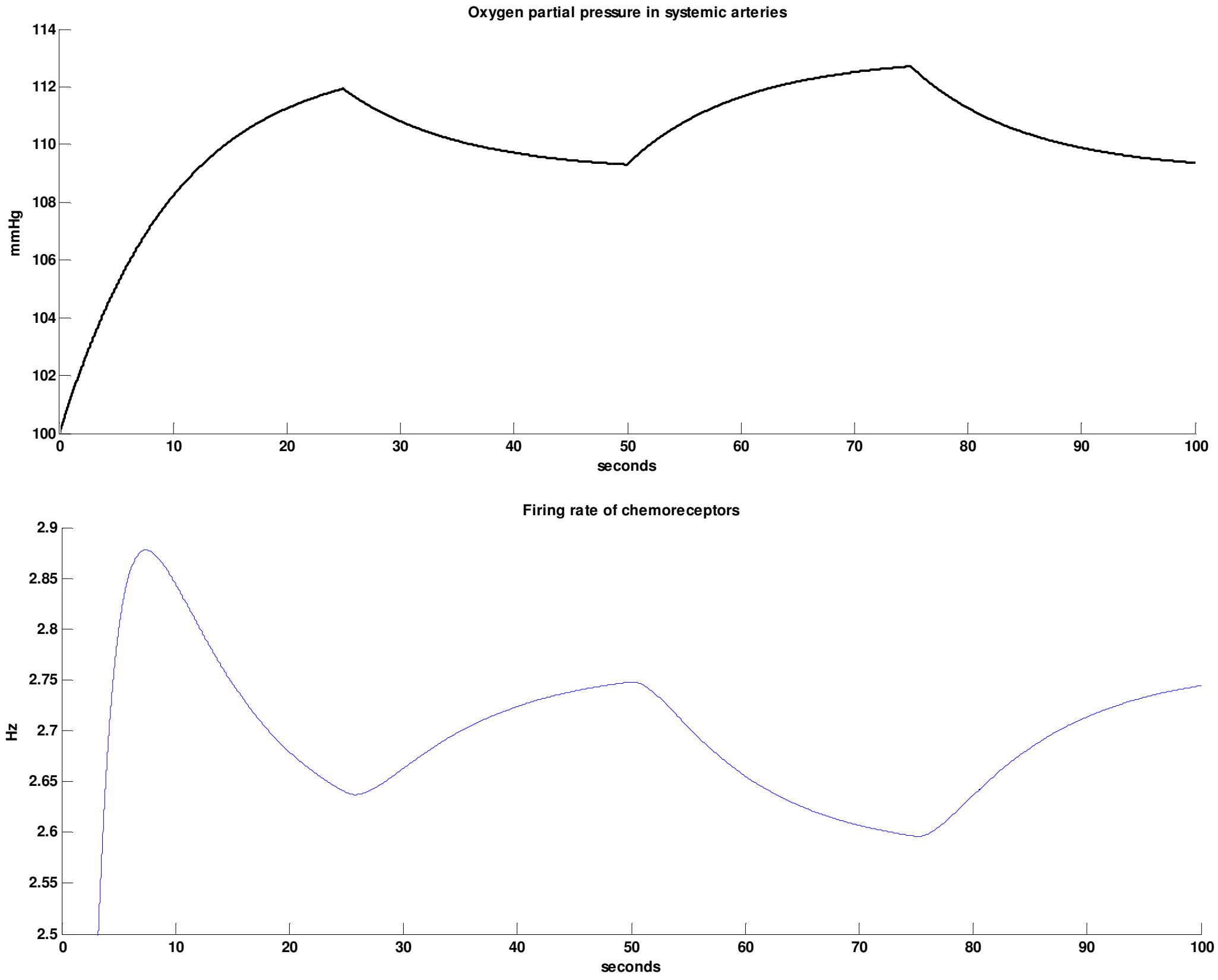

**Figure 2.8 Simulation of chemoreceptors**

Receptors firing rate (bottom panel) in response to oxygen partial pressure changes (top panel)

### 2.4.3 Model of Lung Stretch Receptors

Inflation of lungs at low pressure activates slowly adapting stretch receptors with vagal myelinated A-fibers which project to the medulla. Their activity depends on the tidal volume, which depends on chemoreceptors activity. Chemoreceptors inform about low oxygen level; response is increased tidal volume; lung stretch receptors measure the response and inform medulla, thus creating a closed-loop. We adopt the model of pulmonary stretch receptors firing rate $f_{ap}$ which has been introduced in section 1.7.3, equation (1.7.13). It describes how the firing rate increases with increased tidal volume $V_T$.

Constant values are given in Table 2.4.3 as calculated by Ursino after fitting the model with experimental data [62]. A simulation of the equation is given in Figure 2.9 where a decrease of stretch receptors firing rate can be observed with breathe holding after expiration, i.e. decreased tidal volume.

Table 2.4.3 Constant values for lung stretch receptors

| **Constant** | **Value** | **Description** |
|---|---|---|
| $\tau_c$ | 2 s | time constant of the lung stretch receptors dynamics |
| $G_{ap}$ | 23.29 $L^{-1}$ Hz | constant gain factor |

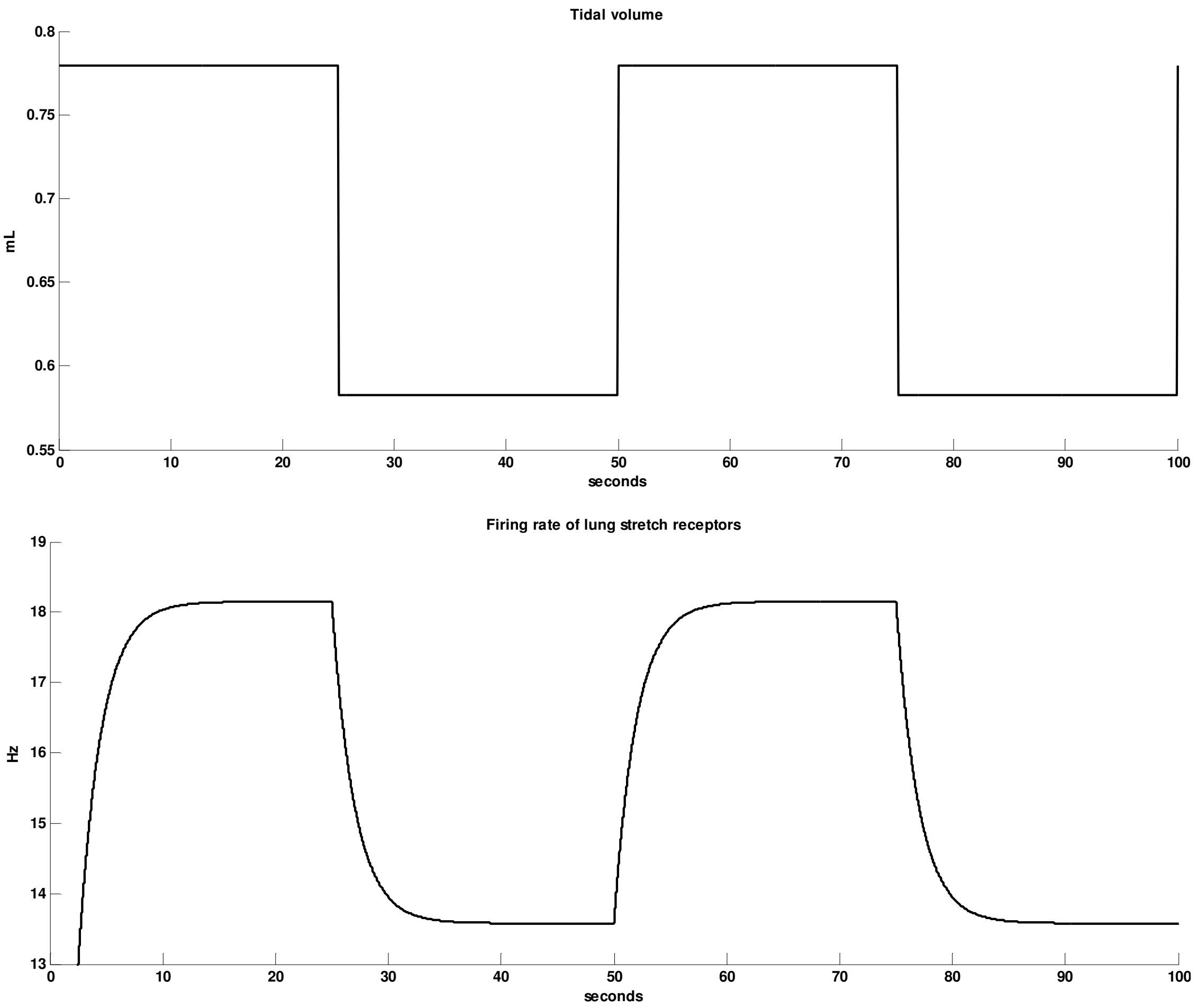

**Figure 2.9 Simulation of lung stretch receptors**

Receptors firing rate (bottom panel) in response to tidal volume change (top panel)

### 2.4.4 Model of Atria Stretch Receptors

Stretch receptors in atria are able to respond to changes in preload, afterload, and contractility of the heart. The afferent activity is modeled using a linear derivative first-order dynamic block and a sigmoidal static block as suggested by Magosso and Liang [83, 84]:

$$f_{aa} = \frac{f_{aa,max} \cdot e^{\frac{\tilde{P}_{pv} - P_{pv,n}}{k_{aa}}}}{1 + e^{\frac{\tilde{P}_{pv} - P_{pv,n}}{k_{aa}}}}$$
$$t_{aa} \cdot \frac{d\tilde{P}_{pv}}{dt} = P_{pv} - \tilde{P}_{pv} \qquad (2.4.2)$$

where $f_{aa}$ is the firing rate of the atria stretch receptor in response to the sensed pressure $\tilde{P}_{pv}$ which depends on the pulmonary venous pressure $P_{pv}$; $f_{aa,\max}$ is the upper saturation level of firing rate; $\tau_{aa}$ is a time constant; $P_{pv,n}$ is the value of venous pressure at the central point of the sigmoidal function; $k_{aa}$ is a parameter, with the dimension of pressure, related to the slope of the static function at the central point.

Constant values are given in Table 2.4.4 as calculated by Magosso after fitting the model with experimental data [62]. A simulation of the equations above is given in Figure 2.10 where a decrease of stretch receptors firing rate can be observed with decreased pulmonary venous pressure.

Table 2.4.4 Constant values for atria stretch receptor

| **Constant** | **Value** | **Description** |
|---|---|---|
| $P_{pv,n}$ | 6.0 mmHg | set-point value of the cardiopulmonary baroreflex |
| $f_{aa,max}$ | 18.0 Hz | upper saturation of atria stretch receptor |
| $\tau_{aa}$ | 6.37 s | time constant of atria stretch receptor afferent response |
| $k_{aa}$ | 3.429 mmHg | slope of the atria stretch receptor sigmoid at the central point |

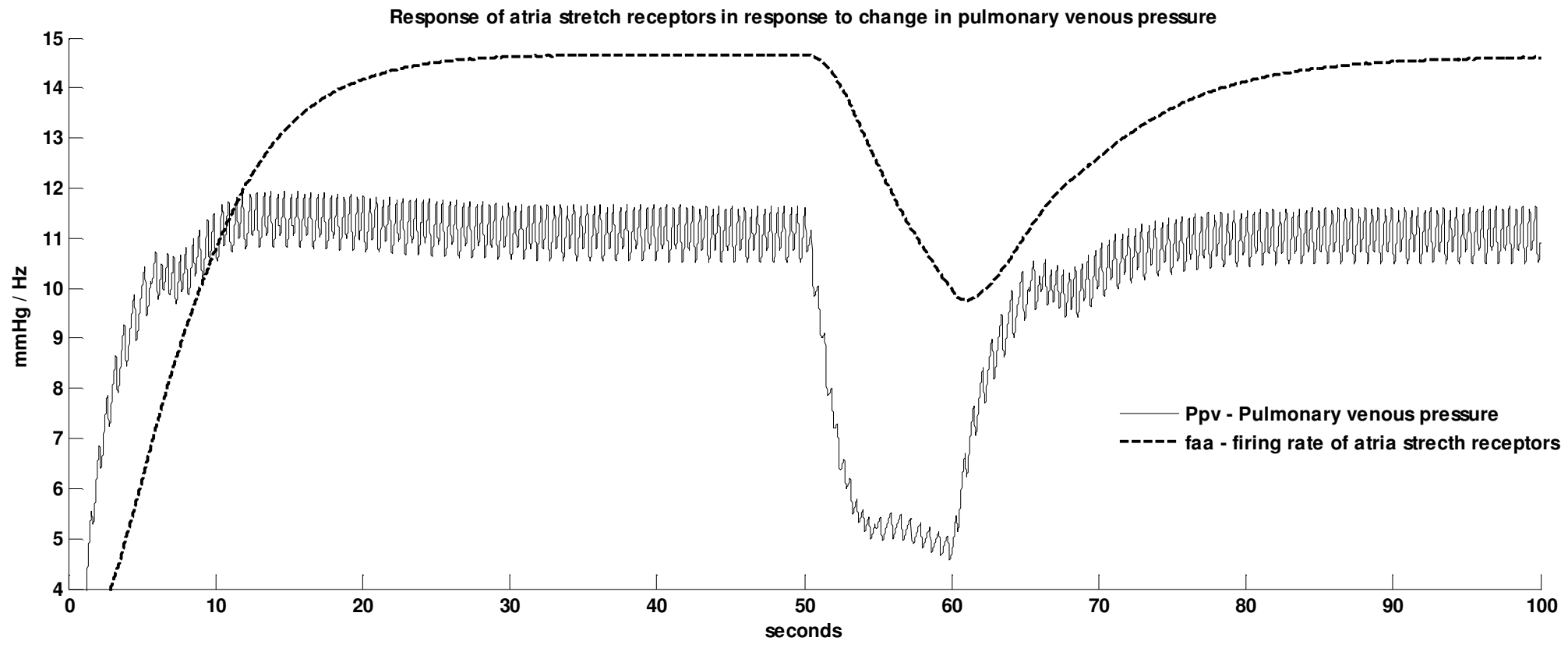

**Figure 2.10 Simulation of atria stretch receptors**

Receptors firing rate (dashed line) in response to changes in pulmonary venous pressure (plain line)

Right and left atria receptors also sense intravascular volume (i.e. atria stretch or preload) and project to the hypothalamus and cortex. Stretch causes decreased antidiuretic hormone release, which results in increased urine output. Stretch causes also release of atrial natiuretic peptide, which acts on the kidneys and promotes body fluid volume reduction. Ventricular, pericardial and coronary sinus stretch sensors causes a reflex fall in pressure. We do not include these processes in our model for simplicity.

### 2.4.5 Model of Cold Receptors

The mechanism of transduction in cold receptors with temperature change is not fully understood and we could not find any mathematical model of their function in the known literature. In the following paragraphs we will develop mathematical equations that relate cold stimuli to the neural activity of cold receptors.

Innocuous cold receptors sense nonpainful cold temperature, whereas temperature nociceptors respond to noxious cold or heat, which can damage organs. At normal skin temperature innocuous cold receptors exhibit a spontaneous, periodic discharge that increases with cooling and decreases with warming. In contrast, cold nocireceptors are quiet at rest and fire only to extreme low temperatures. It has been recently suggested that Transient Receptor Potential Melastatin 8 (TRPM8) is the best molecular candidate to explain the mechanism of transduction of cold in both innocuous thermoreceptors and nociceptors. TRPM8 are activated by the threshold of 25°C ± 1 [85]. Cold receptors usually increase their firing rate when the threshold temperature is reached. Their initial response indicates a change in temperature and their sustained response informs about the ambient temperature. Some cold receptors also briefly respond to high temperature (> 45°C). This is known as a paradoxical response to heat. A repetitive beating activity and burst (grouped) discharges were observed in cold receptor populations at constant temperatures and after rapid cooling [86]. Such studies were conducted by Hutchison who did a quantitative analysis of orofacial thermoreceptive neurons in the superficial medullary dorsal horn of the rat [87]. He classified the receptors after their response to decreasing temperatures (see Figure 2.11). Type 1 had a bell-shaped static stimulus response function. Type 2 had a high maintained or increasing static firing rate as temperature decreased. Type 3 showed highly dynamic components in its response function. Similar response functions were obtained in [85] when investigating the response of neurons located in trigeminal subnucleus caudalis (Vc) during intraoral cooling (via ice water and menthol) of TRPM8 channels on rats.

We distinguish between facial receptors which are located on the forehead and can elicit the diving reflex [88] and receptors which are located on hands and feet. Below a certain threshold the firing rate of facial receptors (also called trigeminal thermoreceptors) promptly rises until saturation is reached, then rapidly decreases to resting values even when stimuli is maintained. We model this behavior using an exponential decay function $f_{at_f}$ as follows.

$$f_{at_f} = \begin{cases} f_{at_f,min} & if \quad t < t_{start} \\ f_{at_f,max} \cdot e^{-k_{at_f} \cdot (t - t_{start})} & if \quad t_{start} \le t \le t_{stop} \\ f_{at_f,min} & if \quad t > t_{stop} \end{cases} \tag{2.4.3}$$

where $t_{start}, t_{stop}$ are model inputs representing the start and stop timestamps of cold stimulation; $f_{at_f,min}$ is the lower saturation level of firing rate with constant value 5 Hz; $f_{at_f,max}$ is a model parameter representing the upper saturation level of firing rate; $k_{at_f}$ is model parameter related to the slope of the exponential decay function.

When hands or feet are immersed in cold water (temperature around 0 $^{o}$C), heart rate response indicates that cold receptors of type 2 are mostly activated compared to other types, since the response is maintained during the whole period of stimulation. We model the firing rate $f_{at_s}$ of cold receptors on hands and feet using a sigmoid function as follows.

$$f_{at_s} = \begin{cases} 0 & if \quad t < t_{start} \\ \dfrac{f_{at_s,max}}{1 + f_{at_s,max} \cdot e^{-k_{at_s} \cdot (t - t_{start})}} & if \quad t_{start} \le t \le t_{stop} \\ 0 & if \quad t > t_{stop} \end{cases} \tag{2.4.4}$$

where $t_{start}, t_{stop}$ are model inputs representing the start and stop timestamps of cold stimulation; $k_{at_s}$ is a parameter related to the slope of the cold response; $f_{at_s,max}$ is the maximal firing rate of cold receptors.

Typical parameters values are summarized in Table 2.4.5 after we have tried several values so that the simulated firing rate will fit the shape of neural spikes measured invasively on rats and published in [85]. A simulation of above equations is given in Figure 2.13 and Figure 2.14 where cold receptors of face and hands/feet increase their firing rate with decreasing temperature, followed by an exponential decay and sigmoidal increase respectively. This response shape agrees with experimental data in Figure 2.12.

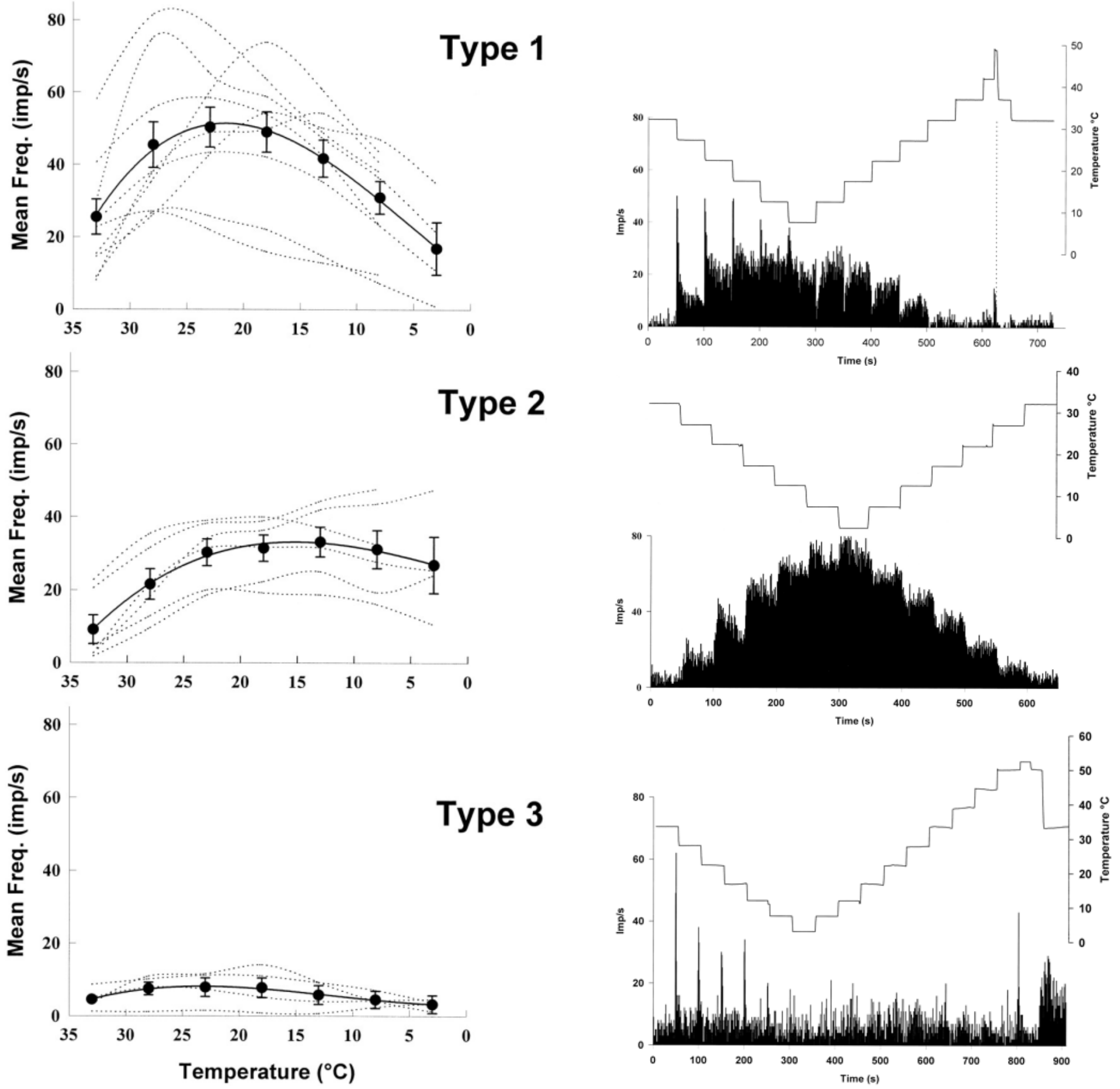

**Figure 2.11 Types of cold receptors (reprinted from [87])**

Left panels – mean firing rates with gradual decreasing temperature
Right panels – mean firing rates with step decrease and increase temperature

Table 2.4.5 Parameters values for cold receptor

| Parameter | Value | Description |
|---|---|---|
| $f_{at_f,max}$ | 50 Hz | upper saturation of facial cold receptor |
| $k_{at_f}$ | 0.2 s$^{-1}$ | slope of the exponential decay of facial cold receptor |
| $f_{at_s,max}$ | 72 Hz | upper saturation of cold receptor on hands and feet |
| $k_{at_s}$ | 1.03 s$^{-1}$ | slope of the sigmoid response of cold receptor on hands and feet |

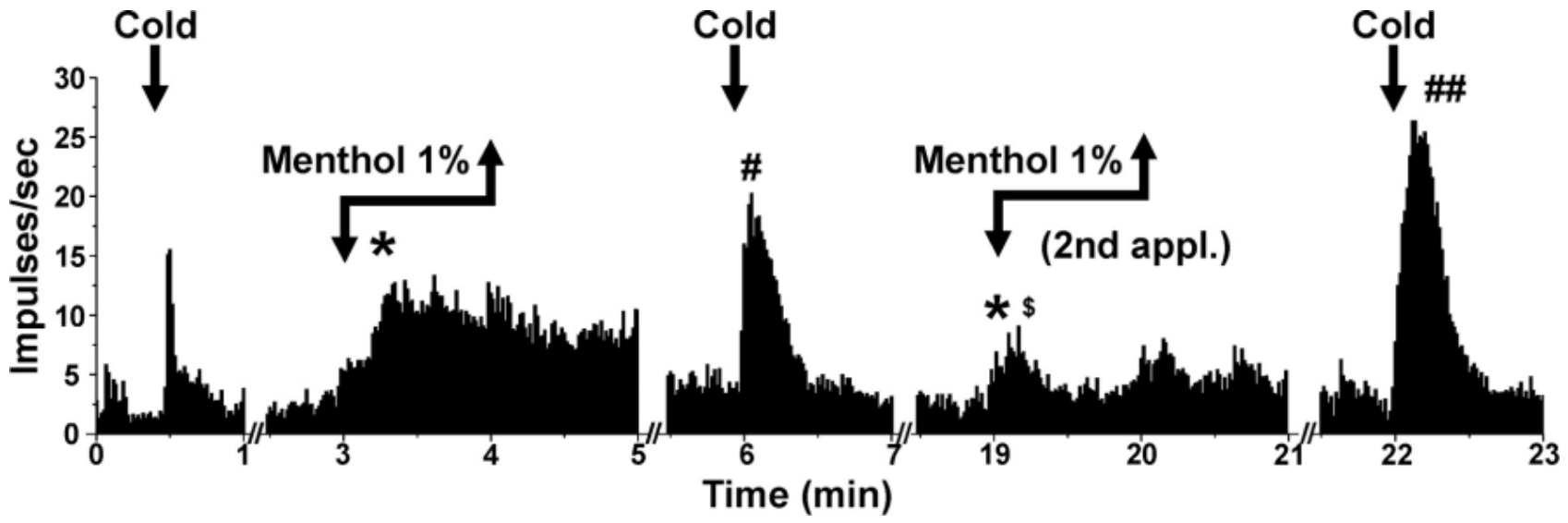

**Figure 2.12 Cold receptor response to sustained cold stimuli and menthol (reprinted from [85])**

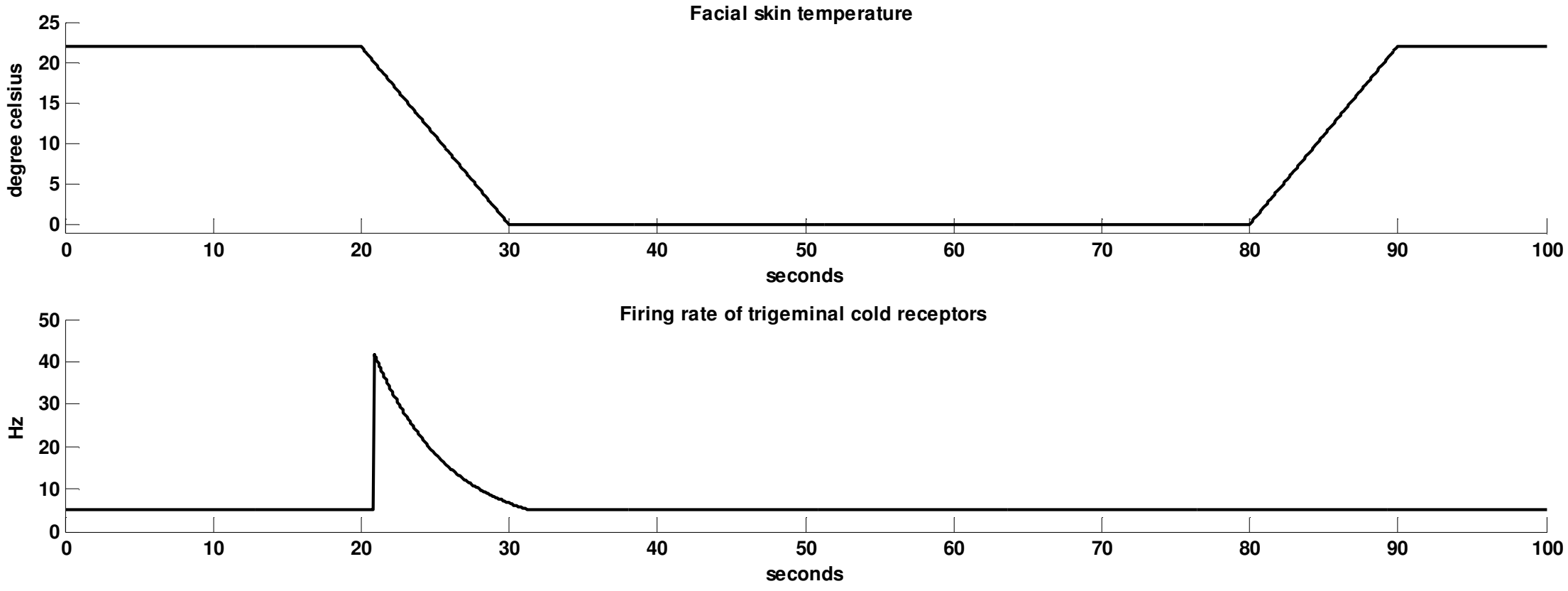

**Figure 2.13 Simulation of trigeminal cold receptors**

Receptors firing rate (bottom graph) as response to low temperature (low graph)

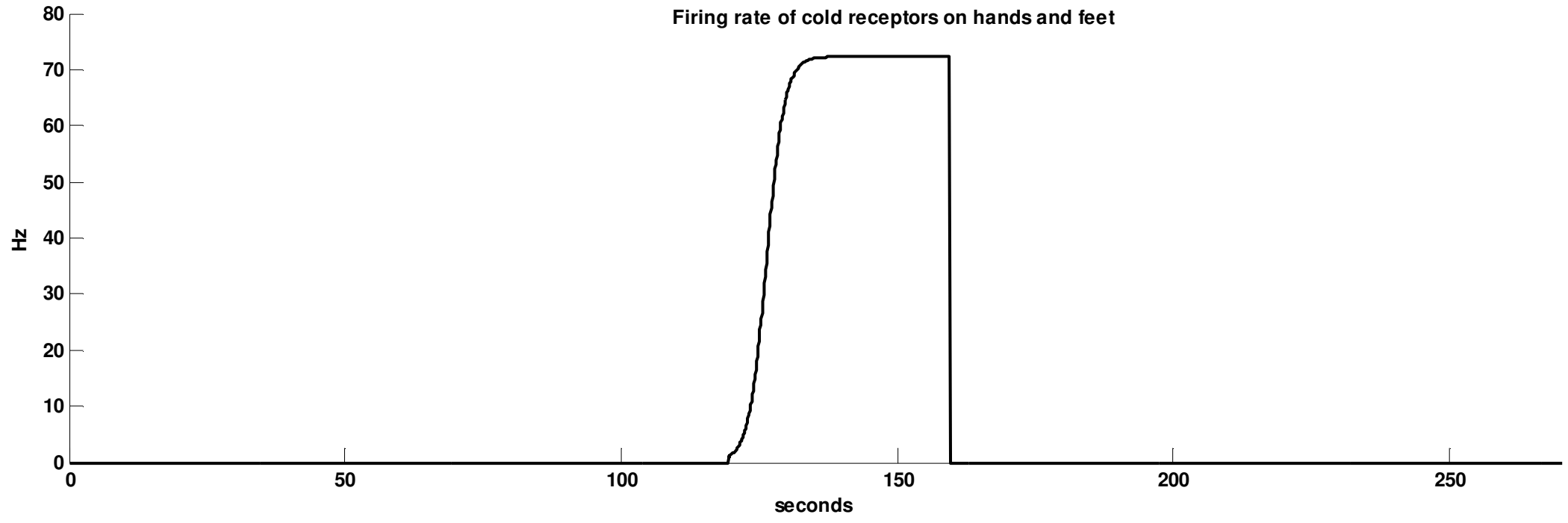

**Figure 2.14 Simulation of cold receptors on hands/feet by low temperature**

Temperature drops from 22°C to 0°C at 120$^{th}$ second until 160$^{th}$ second

### 2.4.6 Model of Eye Pressure Sensors

The Oculocardiac reflex is a trigeminovagal reflex that can produce potent bradycardia and cardiac arrhythmias such as nodal rhythm, ectopic beats, ventricular fibrillation, or asystole. The reflex may be induced by applying pressure on the eyeball or caused by an orbital hematoma, ocular trauma, eye pain, and traction of an extraocular muscle during treatment of poor alignment of the visual axis (strabismus) [89]. We could not find any mathematical model of this reflex in the known literature and will propose such a model in the following paragraphs.

Experimental data recorded in [90] (Figure 2.15) illustrate that receptor activity depends on the pressure or stretch applied on the cornea. When pressure occurs, the firing rate promptly rises until saturation is reached. Such a behavior can be modeled using a sigmoid function as follows.

$$f_{ao} = \begin{cases} 0 & if \quad t < t_{start} \\ \dfrac{f_{ao,max} \cdot e^{-k_{ao} \cdot (t - t_{start})}}{1 + f_{ao,max} \cdot e^{-k_{ao} \cdot (t - t_{start})}} & if \quad t_{start} \le t \le t_{stop} \\ 0 & if \quad t > t_{stop} \end{cases} \tag{2.4.5}$$

where $f_{ao}$ is the firing rate of sensors in eyeball in response to the pressure exerted on eye ball; $f_{ao,max}$ is a model parameter representing the upper saturation level of firing rate; $k_{ao}$ is a model parameter related to the slope of the sigmoid function; $t_{start}, t_{stop}$ are model inputs representing the start and stop timestamps of pressure;

Parameters values are given in Table 2.4.6 after trying several values so that the simulated firing rate fits the shape of electro-oculograms recorded in [90] during oculocardiac reflex. A simulation of the equations above is given in Figure 2.16 where oculo-pressure receptors increase their firing rate with increased pressure on eye balls.

Table 2.4.6 Parameters values for oculo-pressure receptor

| Parameter | Value | Description |
|---|---|---|
| $f_{ao,max}$ | 17 Hz | upper saturation of oculo-pressure receptor |
| $k_{ao}$ | 0.7 $s^{-1}$ | slope of the sigmoid response of oculo-pressure receptor |

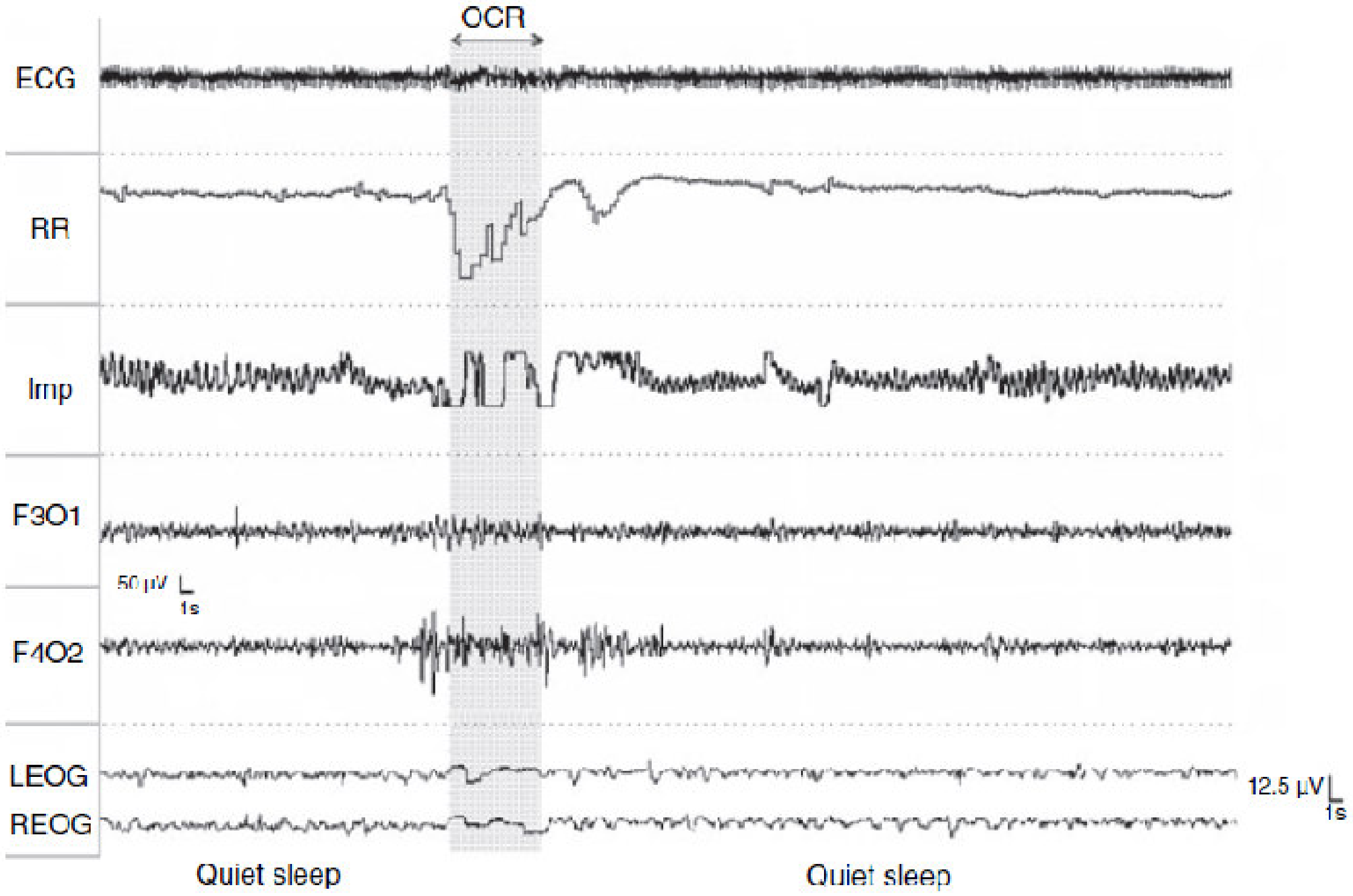

**Figure 2.15 Recorded changes induced by the oculocardiac reflex (reprinted from [90])**
ECG, RR intervals, thoracic respiratory movements by impedancemetry (Imp), right and left electroencephalograms (F3O1 and F4O2), left and right electro-oculograms (LEOG and REOG)

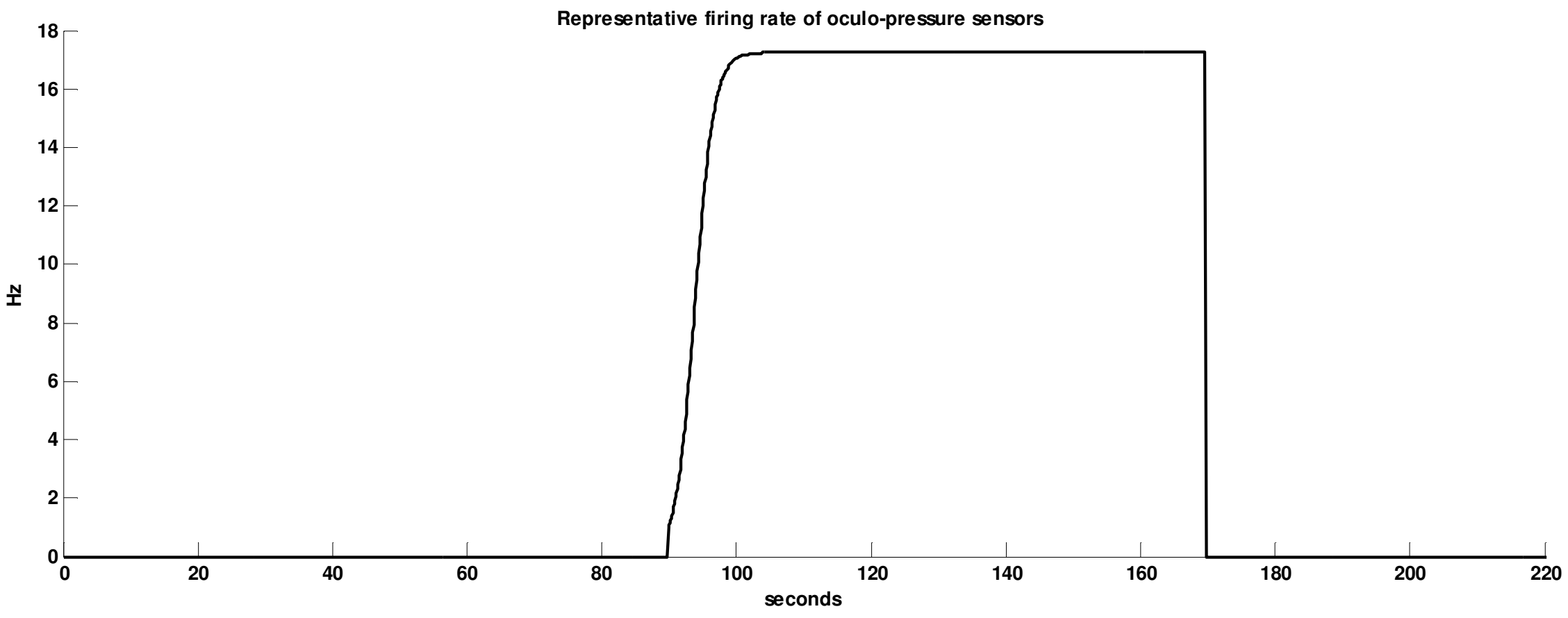

**Figure 2.16 Simulation of oculo-pressure receptors**
pressure on eye balls is applied from 90th to 170th second

### 2.4.7 Model of Central Input during Mental Stress

Invasive studies with direct microneurographically measurement of muscle sympathetic nerve activity (MSNA) have exhibited an increased sympathetic activity during mental stress, induced for example by arithmetic tasks and characterized by an overall increase of heart rate and blood pressure [91]. Figure 2.17 shows MSNA recordings during mental tasks and the corresponding heart rate and blood pressure changes.

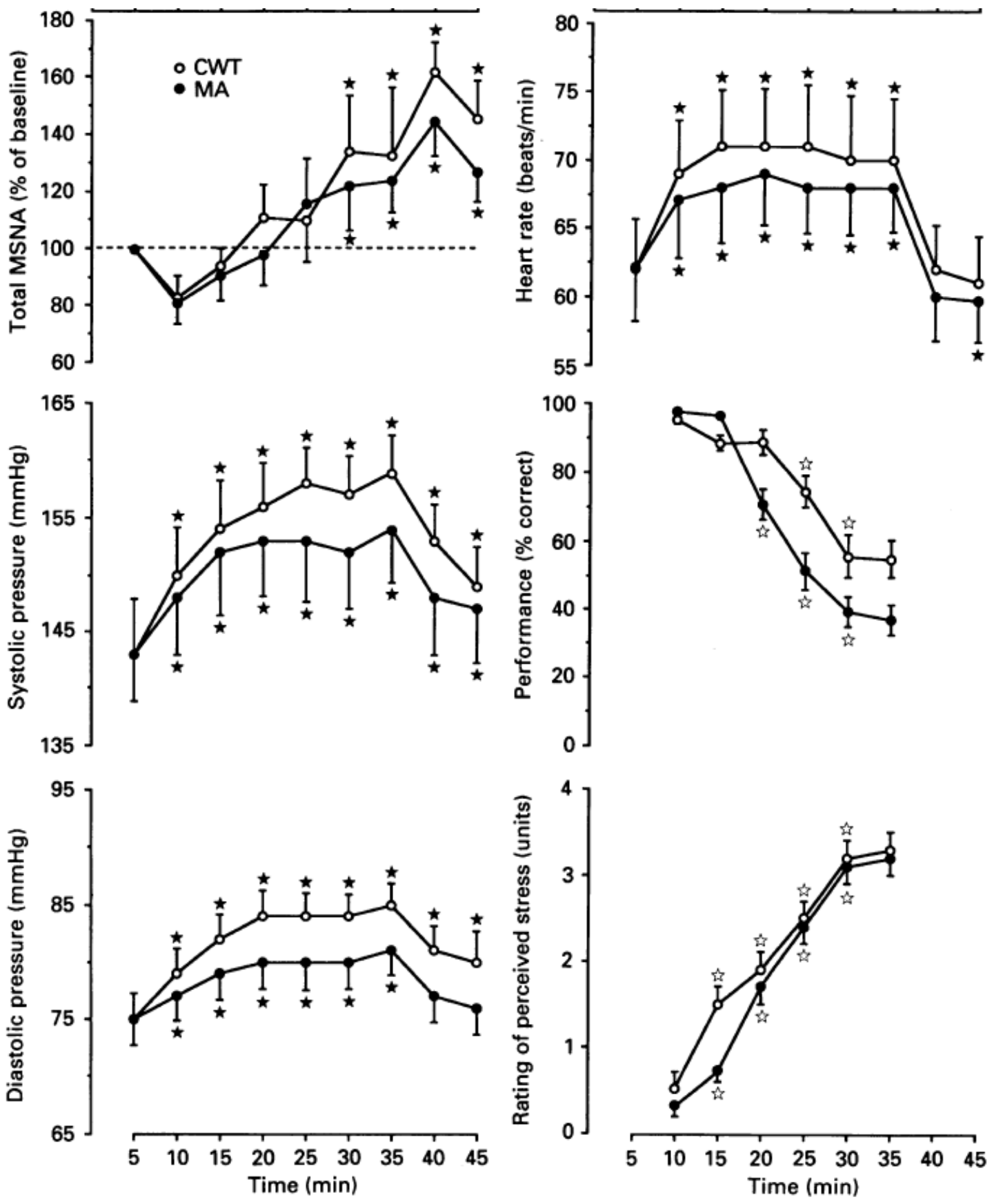

**Figure 2.17 MSNA, heart rate and blood pressure recordings during stress (reprinted from [91])**
MSNA (muscle sympathetic nerve activity), CWT (colour word test), MA (mental arithmetic)
5 min of baseline, 30 min of mental stress, 10 min of recovery

In order to reproduce the observed heart rate changes, which have a plateau-like shape, we assume that there exist some kinds of mental stress sensors, which communicate the level of stress to sympathetic premotor neurons using a logistic function $f_{as}$. We propose the following model with outputs normalized in the interval 0 to 1.

$$f_{as} = \begin{cases} 0 & if \quad t < t_{start} \\ \dfrac{f_{as,max}}{1 + f_{as,max} \cdot e^{-k_{as} \cdot t}} & if \quad t_{start} \le t \le t_{stop} \\ 0 & if \quad t > t_{stop} \end{cases} \qquad (2.4.6)$$

where $t_{start}, t_{stop}$ are model inputs representing the start time and stop time of mental stress; $k_{as}$ is a parameter related to the slope of the mental stress sensor response; $f_{as,max}$ is the maximal value.

Typical parameters values are summarized given in Table 2.4.7 after trying several values so that the simulated firing rate fits the shape of microneurographically measurement of muscle sympathetic nerve activity performed in [91] during cognitive stress. A simulation of above equations is given in Figure 2.18 where increased heart rate and blood pressure is visible as response to mental stress from 150th to 220th second.

Table 2.4.7 Parameter values for mental stress sensors

| **Constant** | **Value** | **Description** |
|---|---|---|
| $f_{as,max}$ | 20 Hz | upper saturation of mental stress sensors |
| $k_{as}$ | 0.2 $s^{-1}$ | slope of the mental stress sensor response |

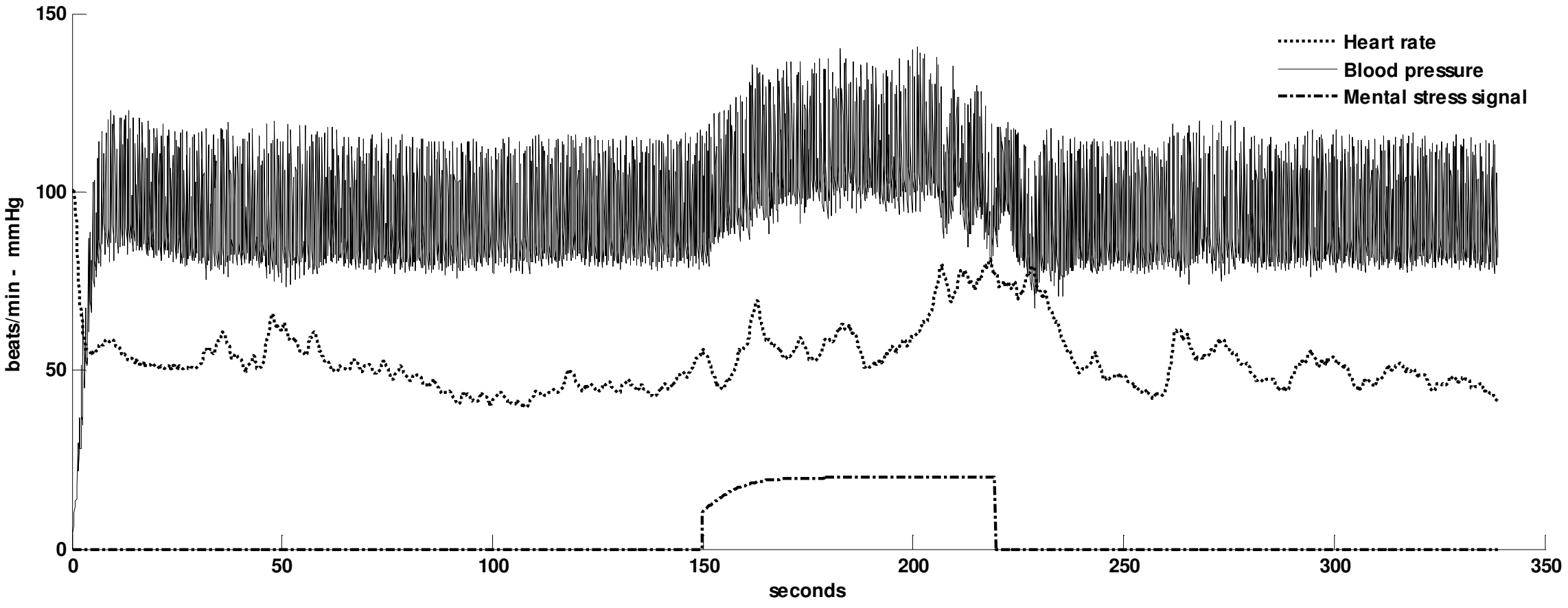

**Figure 2.18 Simulation of mental stress sensors**
Mental arithmetic was performed from 150th to 220th second.

## 2.5 Physiological Model of Autonomic Premotor Neurons

### 2.5.1 Model of Sympathetic Premotor Neurons

There are evidences that rostral ventrolateral medulla cells do not exert a uniform control over the sympathetic vasomotor outflow. Ursino distinguished between premotor neurons which control sympathetic preganglionic neurons projecting to heart, brain, skeletal muscles, splanchnic and extrasplanchnic vascular beds in his model [62]. We further distinguish between SA node, AV node and ventricles. Sympathetic innervation in remaining parts of heart is negligible. We denote the firing rate of sympathetic premotor neurons to both SA node and AV node, ventricles, brain, skeletal muscles, splanchnic and extrasplanchnic vascular beds by $f_{sn_{pm}}, f_{sv_{pm}}, f_{sb_{pm}}, f_{sm_{pm}}, f_{ss_{pm}}, f_{se_{pm}}$ respectively. The rostral ventrolateral medulla receives inputs from interneurons originating in supraspinal nuclei, prefrontal cortex, hypothalamic defense area, fastigial pressor area, fastigial nucleus in the cerebellum and primary afferent nerves from peripheral baroreceptors, chemoreceptors, cardiopulmonary receptors, renal receptors and vestibular receptors as well as inputs from the central respiratory rhythm generator. Furthermore studies suggest that rostral ventrolateral medulla vasomotor cells have ascending projections to brain stem and diencephalic nuclei to inform on the degree of sympathetic motor outflow.

In our modeling effort we consider inputs from afferent neurons including baroreceptors, chemoreceptors, lung stretch receptors, atria stretch receptors, trigeminal cold receptors, cold receptors in hands/feet and mental stress sensors characterized by their firing rate $f_{ab}, f_{ac}, f_{ap}, f_{aa}, f_{at_f}, f_{at_s}, f_{as}$.

We assume that premotor neurons have additive properties of the sensory inputs and model their firing rate using an exponential trend with weighted sensory inputs. The weight ($W$) is applied negatively when increased sensory input inhibit premotor neuron output and positively elsewhere. For example the baroreflex functions as a negative feedback system such that baroreceptors firing normally inhibit sympathetic autonomic premotor neurons in rostral ventrolateral medulla. Decreased blood pressure causes decreased baroreceptors firing, therefore increased sympathetic tone together with decrease vagal tone. Although small increases of arterial pressure provoked reciprocal changes of sympathetic and vagal neural outflows, larger increases of pressure provoked exclusively increases of vagal-cardiac nerve activity (see Figure 2.19). Thus baroreflex stimulation provokes reciprocal changes of sympathetic activity (negative weight) and parasympathetic activity (positive weight) whereas some physiological interventions such as diving reflex provoke parallel changes.

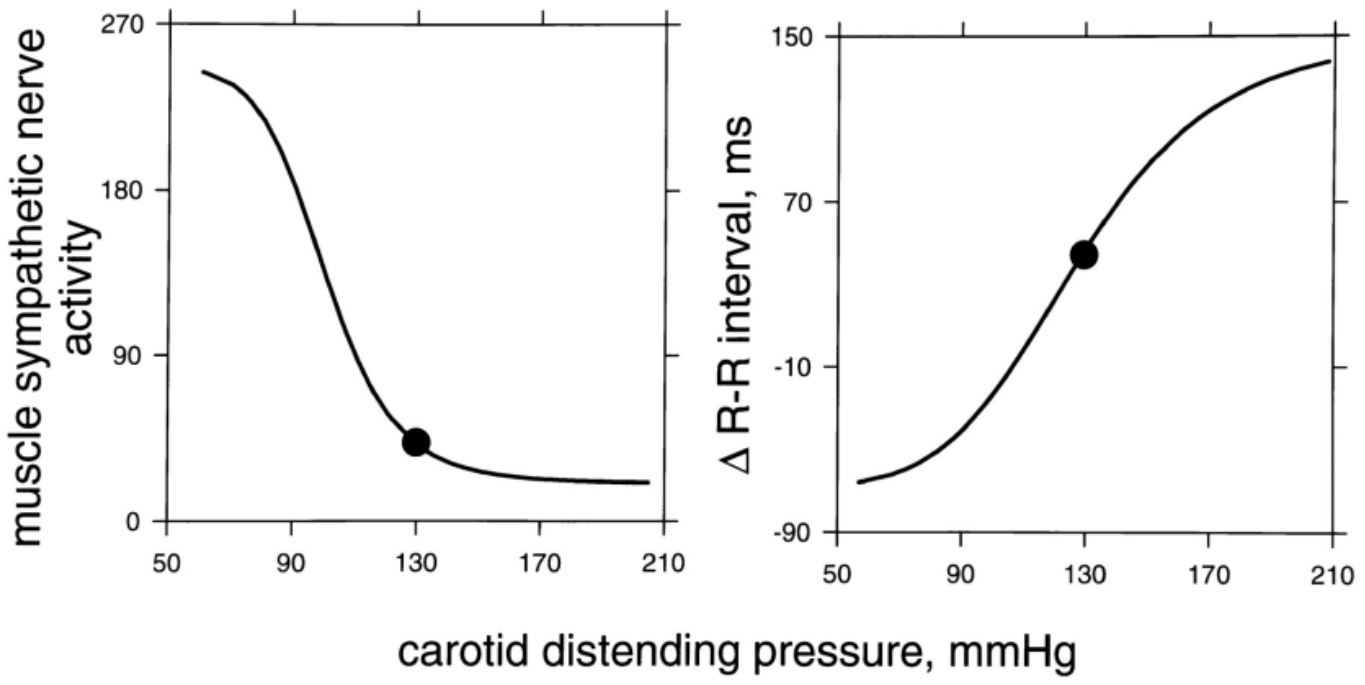

**Figure 2.19 MSNA and RR changes with carotid distending pressure (reprinted from [42])**
MSNA (sympathetic nerve activity), RR (RR interval) - Large increase of arterial pressure causes moderate decrease of sympathetic activity (left panel) and large increase of RR interval durations (right panel)

Some rostral ventrolateral medulla neurons display an ongoing excitatory activity even in absence of sensory inputs. The continuous basal level of sympathetic firing at normal pressure is called sympathetic tone. The power of this activity lies in the 2 to 10 Hz range in anesthetized decebrated animals. The origin of spontaneous activity is still under debate. A theory suggests that some premotor neurons are chemosensitive and are tonically excited at normal level of oxygen or carbon dioxide partial pressure [76]. Another theory suggests the existence of cell groups with electrophysiological pacemaker properties [81]. We developed a model of sympathetic tone $f_{es,0,sn}$ as follows.

$$f_{es,0,sn} = max\left(f_{es,0,high} + W_{es,0,c} \cdot f_{ac}, f_{es,0,min}\right) \tag{2.5.1}$$

where $f_{es,0,high}$ is a model parameter representing the firing rate of cells group exhibiting a pacemaker activity in the frequency range 2 to 10 Hz; $W_{es,0,c}$ is a model parameter representing synaptic weight applied to sensory inputs from chemoreceptors $f_{ac}$; $f_{es,0,min}$ is a model constant representing the minimal basal firing rate of sympathetic premotor neurons.

Mayer waves are arterial pressure oscillations lower than respiration rate [92]. They have a characteristic frequency of approximately 0.1 Hz and result in fluctuations in sympathetic vasomotor tone since they are strongly reduced or abolished with $\alpha$-adrenoceptors blockade. Studies suggest that Mayer waves do not depend on age, gender or posture. The origin of the waves remains under investigation with two prominent theories [93]. The pacemaker theory suggests a central oscillator implemented by a group of medullary or spinal neurons which generate slow sympathetic nervous activity rhythms to vasculature independent of afferent inputs. For example sinoaortic denervated, vagotomized and decerebrated cats present rhythmic discharge of medullary neurons involved in the regulation of the cardiovascular system. This is usually called vascular tone. The baroreflex theory suggests that Mayer waves are an epiphenomenon of delayed baroreflex operation. They are a resonance effect of the baroreflex mediated sympathetic vasoconstriction control loop and have no specific function. We take the pacemaker theory into account at this level and introduce an additional model variable $f_{es,0,p}$ representing the vascular tone which is derived from equation (2.5.1) as follows.

$$f_{es,0,p} = max\left(f_{es,0,high} + W_{es,0,c} \cdot f_{ac} + W_{es,0,low} \cdot f_{es,0,low}, f_{es,0,min}\right) \tag{2.5.2}$$

where $W_{es,0,low}$ is a synaptic weight applied to the variable $f_{es,0,low}$ which represents the rate of slower oscillation in premotor neurons activity as modelled in the equation (2.5.3) below. We assume that the slow oscillator presents a bursting property characterized by brief bursts of oscillatory activity interspersed with quiescent periods. Such phenomenon is usually described using FitzHugh-Nagumo models. Hindmarsh and Rose distil this biophysical mechanism into a simple model containing only polynomials [82]. We adapt his model and describe $f_{es,0,low}$ as follows.

$$\begin{aligned}
\frac{dx_{es}}{dt} &= 2.8 \cdot x_{es}^{\ 2} - x_{es}^{\ 3} - y_{es} - z_{es} + 0.05 \\
\frac{dy_{es}}{dt} &= (2.8 + 1.6) \cdot x_{es}^{\ 2} - y_{es} \\
\frac{dz_{es}}{dt} &= r_{es,0,low} \cdot \left(9 \cdot x_{es} + k_{es,0,low} - z_{es}\right) \\
f_{es,0,low} &= max(x_{es}, 0)
\end{aligned} \tag{2.5.3}$$

where $x_{es}, y_{es}, z_{es}$ are polynomial variables; $r_{es,0,low}, k_{es,0,low}$ are model parameters that determine the bursting frequency; $f_{es,0,low}$ is the instantaneous number of spikes per seconds, i.e the firing rate of the slow oscillator generating Mayer waves.

Parameters of the subsystem generating sympathetic tone and Mayer waves are summarized in Table 2.5.1 after trying different parameters sets so that model output fits the shape of firing rate of the slow sympathetic oscillator as measured on sham-operated and sinoartic baroreceptor denervated rats in [81] [93] [82]. A simulation of the equations above is given in Figure 2.20 where the sympathetic basal tone can be seen, as well as the vascular tone.

Table 2.5.1 Parameters values for rhythmic sympathetic premotor neurons

| **Parameter** | **Value** | **Description** |
|---|---|---|
| $f_{es,0,\min}$ | 16.11 Hz | minimal basal firing rate of sympathetic premotor neurons |
| $f_{es,0,high}$ | [2-10] Hz | firing rate of pacemaker sympathetic premotor neurons exhibiting a pacemaker activity |
| $W_{c,es,0}$ | 2.1 | synaptic weight applied to sensory inputs from chemoreceptors |
| $W_{es,0,low}$ | 2.1 | synaptic weight applied to low frequency oscillators in sympathetic premotor neurons |
| $r_{es,0,low}$ | 0.05 | parameters determining the rate of low frequency oscillators in sympathetic premotor neurons |
| $k_{es,0,low}$ | 3 | |

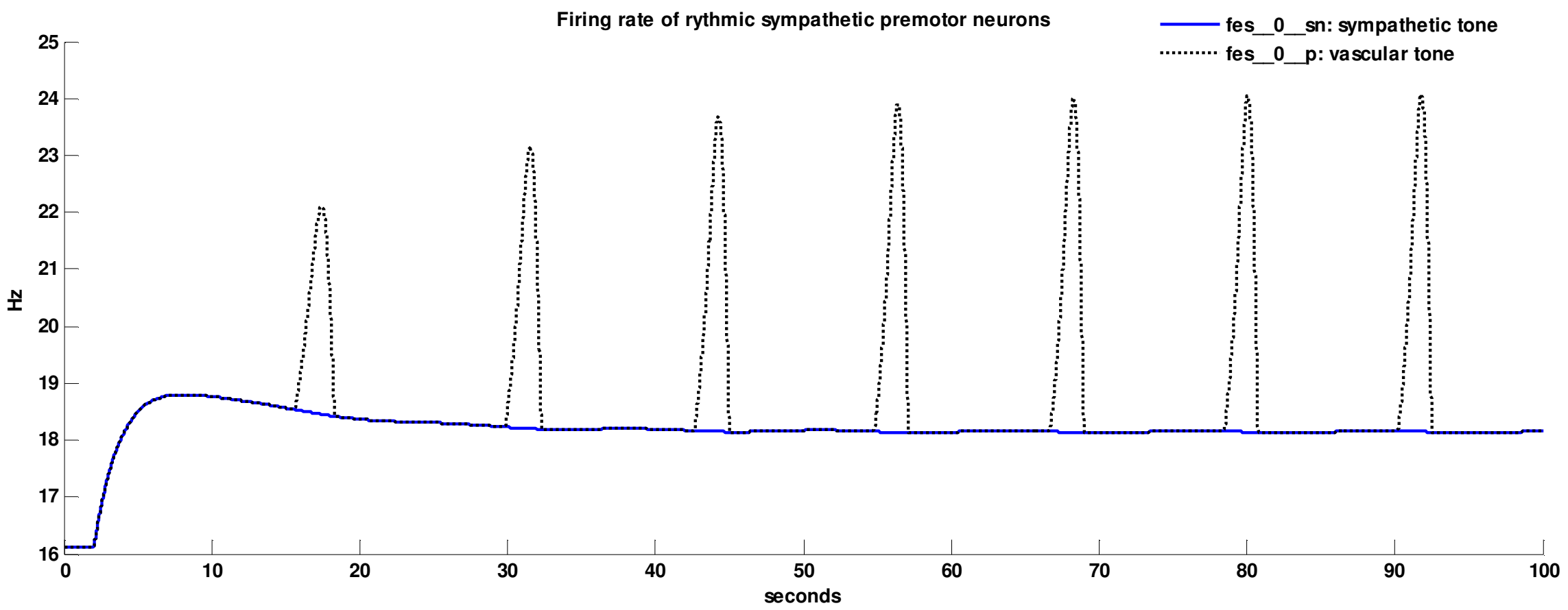

**Figure 2.20 Simulation of rhythmic sympathetic premotor neurons**
Firing rate of neurons generating sympathetic tone (plain line) and Mayer waves (dashed lines)

Sympathetic premotor neurons which control sympathetic preganglionic neurons projecting to the SA Node and AV node are inhibited by increased activity of baroreceptors, lung stretch receptors, atria stretch receptors and trigeminal cold receptors. They are excited by increased activity of chemoreceptors, cold receptors in hands/feet and mental stress sensors. Therefore we model the firing rate of sympathetic premotor neurons controlling SA node and AV node as follows.

$$f_{sn_{pm}} = \begin{cases} f_{es,\infty} + (f_{es,0,sn} - f_{es,\infty}) \cdot e^{k_{es} \cdot (-W_{b,sn} \cdot f_{ab} + W_{c,sn} \cdot f_{ac} - W_{p,sn} \cdot f_{ap} - W_{a,sn} \cdot f_{aa} - W_{t_f,sn} \cdot f_{at_f} + W_{t_s,sn} \cdot f_{at_s} + W_{s,sn} \cdot f_{as} - \theta_{sn})} & if\ f_{sn_{pm}} < f_{es,max} \\ f_{es,max} & if\ f_{sn_{pm}} \geq f_{es,max} \end{cases}$$

(2.5.4)

where $W_{b,sn}, W_{c,sn}, W_{p,sn}, W_{a,sn}, W_{t_f,sn}, W_{t_s,sn}, W_{s,sn}$ are synaptic weights applied to sensory inputs from baroreceptors, chemoreceptors, lung stretch receptors, atria stretch receptors, trigeminal cold receptors, cold receptors in hands/feet and mental stress sensors respectively; $\theta_{sn}$ is an offset term for sympathetic neural activation. Experiments reported in [81] show that maximal inhibitory input will not abolish the activity of inhibitory premotor neurons. Similarly minimal excitatory inputs will not silent excitatory premotor neurons. Therefore we model the residual activity of sympathetic premotor neurons using the parameter $f_{es,\infty}$. Additional unknown physiological factors can modulate the weighted sum of sensory inputs. We include a corresponding constant gain $k_{es}$. The maximal firing rate of sympathetic premotor neurons is captured by model constant $f_{es,max}$.

The overall shape of equation (2.5.4) is similar to the equation developed by Ursino [62] for simulating the activity of efferent sympathetic neurons on heart (see section 1.7.3). Our contribution is the inclusion of components $-W_{a,sn} \cdot f_{aa} - W_{t_f,sn} \cdot f_{at_f} + W_{t_s,sn} \cdot f_{at_s} + W_{s,sn} \cdot f_{as}$ and $f_{es,0,sn}$.

Constant values are given in Table 2.5.2 as calculated by Ursino after fitting their model with in-vitro data from invasive neural measurements on dog [94]. A simulation of the above equation is given in Figure 2.21 where an increase of premotor neurons firing rate can be observed as response to a decrease in arterial blood pressure from 50th to 60th second.

Table 2.5.2 Constant values for sympathetic premotor neurons controlling SA and AV nodes

| Constant | Value | Description |
|---|---|---|
| $f_{es,0,min}$ | 16.11 Hz | minimal basal firing rate of sympathetic premotor neurons |
| $f_{es,\infty}$ | 2.1 Hz | residual activity of sympathetic premotor neurons in absence of inputs |
| $f_{es,max}$ | 60 Hz | maximal firing rate of sympathetic premotor neurons |
| $k_{es}$ | 0.0675 | constant gain of all synaptic inputs to sympathetic premotor neurons |
| $W_{b,sn}$ | 0.3 | synaptic weight applied to sensory inputs from baroreceptors |
| $W_{c,sn}$ | 1 | synaptic weight applied to sensory inputs from chemoreceptors |
| $W_{p,sn}$ | 0 | synaptic weight applied to sensory inputs from lung stretch receptors |
| $W_{a,sn}$ | 0 | synaptic weight applied to sensory inputs from atria stretch receptors |
| $W_{t_f,sn}$ | 1 | synaptic weight applied to sensory inputs from trigeminal cold receptors |
| $W_{t_s,sn}$ | 1 | synaptic weight applied to sensory inputs from cold receptors in hands/feet |
| $W_{s,sn}$ | 1 | synaptic weight applied to sensory inputs from mental stress sensors |
| $\theta_{sn}$ | -49 Hz | offset term for neural activation of sympathetic premotor neuron |

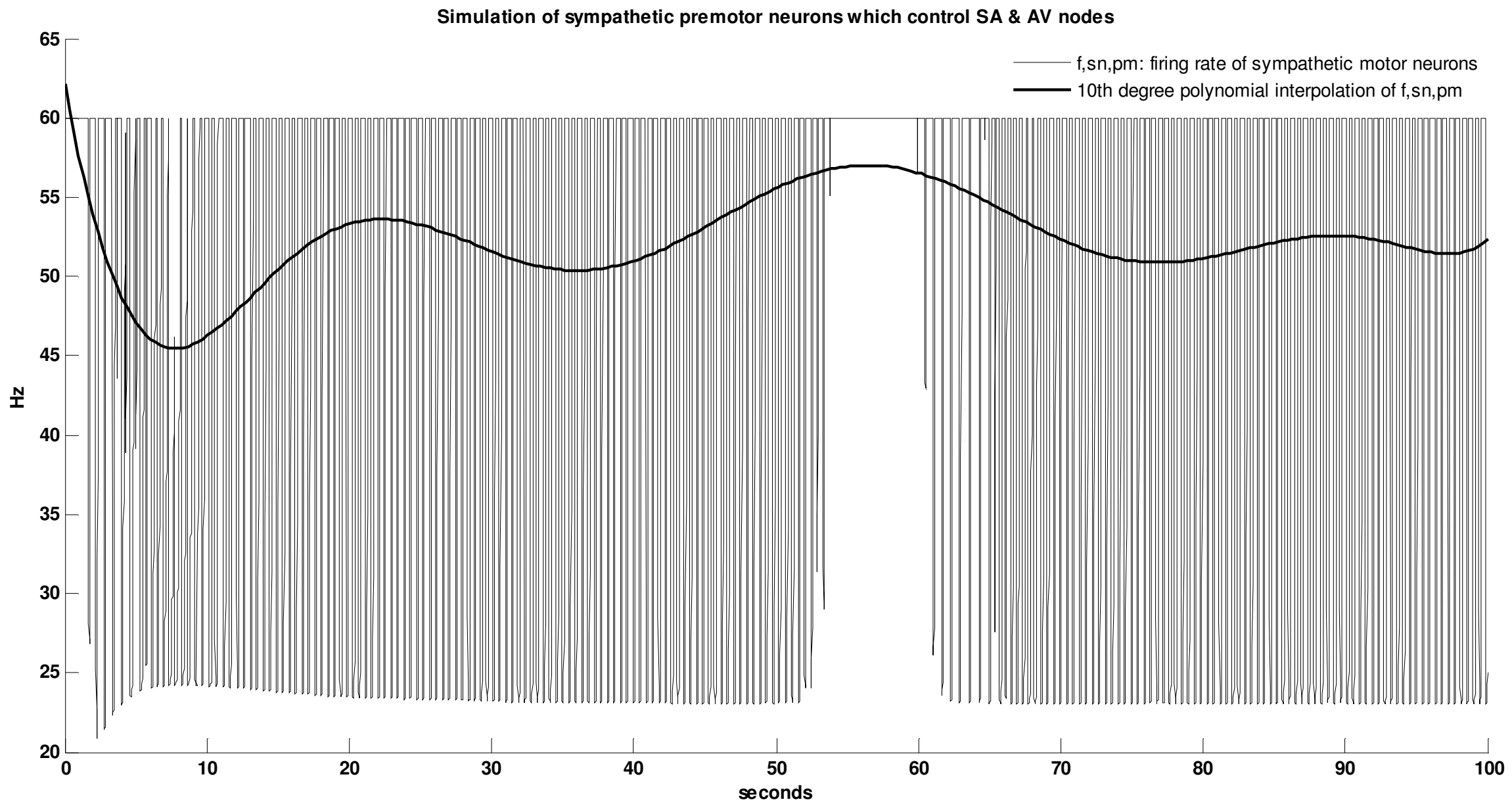

**Figure 2.21 Simulation of sympathetic premotor neurons controlling SA & AV nodes**
Response to arterial blood pressure decreases between 50th and 60th second

Sympathetic premotor neurons which control sympathetic preganglionic neurons projecting to ventricles, brain, skeletal muscles, splanchnic and extrasplanchnic vascular beds are inhibited by increased activity of baroreceptors, lung stretch receptors, atria and stretch receptors. They are excited by increased activity of chemoreceptors, trigeminal cold receptors, cold receptors in hands/feet and mental stress centers. Therefore we model the firing rate of sympathetic premotor neurons controlling ventricles, brain, skeletal muscles, splanchnic and extrasplanchnic effectors as follows.

$$f_{si_{pm}} = \begin{cases} f_{es,\infty} + (f_{es,0,p} - f_{es,\infty}) \cdot e^{k_{es}\cdot(-W_{b,si}\cdot f_{ab} + W_{c,si}\cdot f_{ac} - W_{p,si}\cdot f_{ap} - W_{a,si}\cdot f_{aa} + W_{t_f,si}\cdot f_{at_f} + W_{t_s,si}\cdot f_{at_s} + W_{s,si}\cdot f_{as} - \theta_{si})} & if\ f_{si_{pm}} < f_{es,max} \\ f_{es,max} & if\ f_{si_{pm}} \geq f_{es,max} \end{cases}$$

$$i = v, b, m, s, e \qquad (2.5.5)$$

where $W_{b,si}, W_{c,si}, W_{s,si}, W_{p,si}, W_{t_s,si}, W_{t_f,si}, W_{s,si}$ are synaptic weights applied to sensory inputs from baroreceptors, chemoreceptors, lung stretch receptors, atria stretch receptors, trigeminal cold receptors, cold receptors in hands/feet and mental stress sensors respectively; $\theta_{si}$ is an offset term for sympathetic neural activation. $f_{at_f}$ is positively weighted compared to equation (2.5.4) because increased activity of trigeminal cold receptors causes increased activity of sympathetic premotor neurons that control vascular beds.

The overall shape of equation (2.5.5) is similar to the equation developed by Ursino [62] for simulating the activity of efferent sympathetic neurons on vasculature (see section 1.7.3). Our contribution is the inclusion of components $-W_{a,si} \cdot f_{aa} + W_{t_f,si} \cdot f_{at_f} + W_{t_s,si} \cdot f_{at_s} + W_{s,si} \cdot f_{as}$ and $f_{es,0,p}$.

Constant values are given in Table 2.5.3 as calculated by Ursino after fitting their model with in-vitro data [62]. A simulation of the above equation for ventricles is given in Figure 2.22 where an increase of premotor neurons firing rate can be observed when there is a decrease in arterial blood pressure from 50th to 60th second.

Table 2.5.3 Constant values for sympathetic premotor neurons controlling vascular beds

| Constant | Value | Description |
|---|---|---|
| $W_{b,si}$ | 0.3 | synaptic weight applied to sensory inputs from baroreceptors |
| $W_{c,si}$ | 5 | synaptic weight applied to sensory inputs from chemoreceptors |
| $W_{p,si}$ | 0.34 | synaptic weight applied to sensory inputs from lung stretch receptors |
| $W_{a,si}$ | 1 | synaptic weight applied to sensory inputs from atria stretch receptors |
| $W_{t_f,si}$ | 1 | synaptic weight applied to sensory inputs from trigeminal cold receptors |
| $W_{t_s,si}$ | 1 | synaptic weight applied to sensory inputs from cold receptors in hands/feet |
| $W_{s,si}$ | 0.5 | synaptic weight applied to sensory inputs from mental stress sensors |
| $\theta_{si}$ | 7.34 Hz | offset term for neural activation of sympathetic premotor neuron |

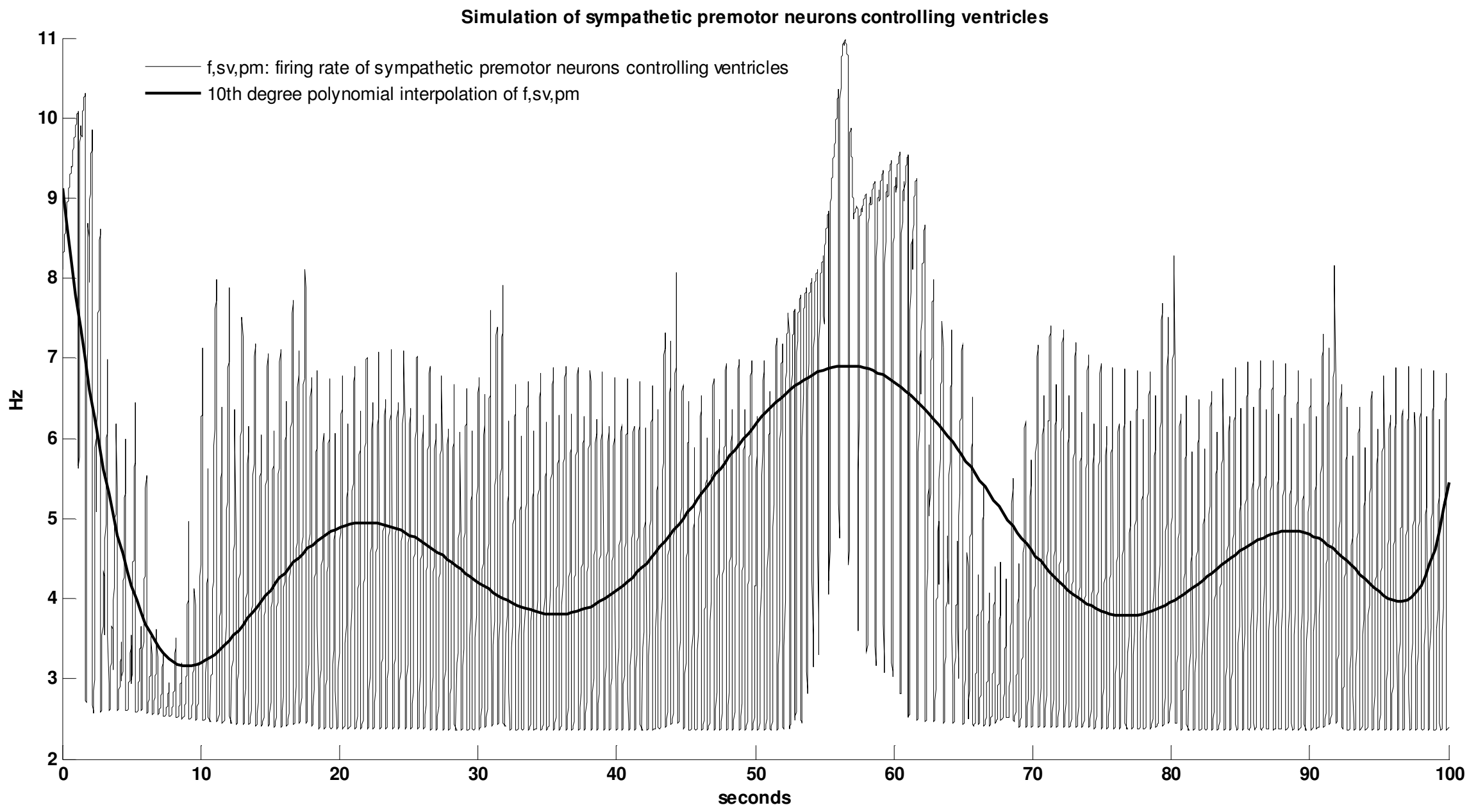

**Figure 2.22 Simulation of sympathetic premotor neurons controlling ventricular vascular beds**
Response to arterial blood pressure decreases between 50th and 60th second

### 2.5.2 Model of Parasympathetic Premotor Neurons

The intrinsic depolarization rate of pacemaker cells in the sinoatrial node is approximately 1.67 Hz despite the resting rate of a normal adult is maintained at a relatively constant value 1.17 Hz. This is achieved via a continuous parasympathetic tonic activity called vagal or parasympathetic tone. It becomes diminished in disease states including hypertension and heart failure. Measuring and increasing vagal tone may be an effective clinical procedure in treating some heart diseases. We propose the following model of resting vagal tone $f_{ev,0}$.

$$f_{ev,0} = f_{ev,0,max} - 1.66 \cdot 10^{-3} \cdot age \tag{2.5.6}$$

where $f_{ev,0,max}$ is the intrinsic firing rate of vagal nerves; $age$ is a model input representing the age of the subject. $1.66 \cdot 10^{-3}$ was chosen using a similar method as the one established by Jose relating age and intrinsic heart rate [95].

Mean value of parasympathetic tone is given in Table 2.5.4 after trying different values to obtain a resting heart rate of 70 beats/min on a 30 years old healthy subject. A simulation of the equation above provided a corresponding parasympathetic basal tone equals to 3.15 Hz.

Table 2.5.4 Parameter value for parasympathetic premotor neurons generating vagal tone

| Parameter | Value | Description |
|---|---|---|
| $f_{ev,0,max}$ | 3.2 Hz | Maximal intrinsic firing rate of vagal nerves |

Parasympathetic premotor neurons which control parasympathetic preganglionic neurons projecting to SA node, AV node, atria and ventricles are excited by increased activity of baroreceptors, chemoreceptors, trigeminal cold receptors and occulo-pressure receptors. They are inhibited by increased activity of lung stretch receptors. Therefore we model the firing rate $f_{ev_{pm}}$ of parasympathetic premotor neurons controlling the heart as follows.

$$f_{ev_{pm}} = \frac{f_{ev,0} + f_{ev,\infty} \cdot e^{\frac{f_{ab}-f_{ab,0}}{k_{ev}}}}{1 + e^{\frac{f_{ab}-f_{ab,0}}{k_{ev}}}} + W_{c,v} \cdot f_{ac} - W_{p,v} \cdot f_{ap} + W_{t_f,v} \cdot f_{at_f} + W_{o,v} \cdot f_{ao} - \theta_v \tag{2.5.7}$$

where $k_{ev}, f_{ev,\infty}$ are model constants; $W_{c,v}, W_{p,v}, W_{t_f,v}, W_{o,v}$ are synaptic weights; $\theta_v$ is an offset term for parasympathetic neural activation.

We will now include Respiratory Sinus Arrhythmia (RSA) into our model. RSA is characterized by increase heart rate during inspiration and decrease heart rate during expiration. It is most pronounced in young subjects and athletes and decreases with age. The neural control mechanism involved in respiratory sinus arrhythmia is complex and still under debate. Following explanations have been raised [96]. One explanation correlates respiratory pump to the baroreflex including cardiopulmonary stretch mechanoreceptors. Intrathoracic pressure decreases during inspiration, allowing more venous blood to return to heart. Venous return to the right atria increases and the right ventricle pumps more blood to the lungs. Therefore blood volume in pulmonary vascular bed increases, but most of it does not return to the left atria, because of low pressure in the lungs. This causes the left ventricle to fill in a lesser extend and stroke volume is reduced. This decreases blood pressure and triggers the baroreflex. Reponse is inhibition of vagal efferent nerves to the heart and increase vascular tone. Another explanation suggests a central mechanism via a direct coupling of the respiratory centers and the vasomotor center in the brain. This second explanation seems to be more plausible because several studies demonstrate that mechanical mechanism in the respiratory pump is unable to account entirely for the presence of RSA. They suggest either a pre-inspiratory decrease in sensitivity to baroreceptor afferent stimuli or a direct inhibition of vagal efferent nerves just prior to inspiration [97]. In order to take RSA into account in our model, we assume a direct coupling between respiratory centers and parasympathetic autonomic premotor neurons and we modify equation (2.5.7) to include an additional weighted factor so that vagal activity is inhibited with increased inspiratory-expiratory somatic motor neurons activity as follows.

$$f_{ev_{pm}} = \frac{f_{ev,0} + f_{ev,\infty} \cdot e^{\frac{f_{ab} - f_{ab,0}}{k_{ev}}}}{1 + e^{\frac{f_{ab} - f_{ab,0}}{k_{ev}}}} + W_{c,v} \cdot f_{ac} - W_{p,v} \cdot f_{ap} + W_{t_f,v} \cdot f_{at_f} + W_{o,v} \cdot f_{ao} - W_{resp,v} \cdot f_{resp} - \theta_v$$

(2.5.8)

where $f_{resp}$ is the firing rate of inspiratory-expiratory somatic motor neurons defined by equation (2.8.6) in section 2.8.4; $W_{resp,v}$ is a model parameter representing the synaptic weight, with initial value 0.105 Hz.

The overall shape of equation (2.5.8) is similar to the equation developed by Ursino [62] for simulating the activity of efferent parasympathetic neurons on heart (see section 1.7.3). Our contribution is the inclusion of components $W_{t_f,v} \cdot f_{at_f} + W_{o,v} \cdot f_{ao} - W_{resp,v} \cdot f_{resp}$ and $f_{ev,0}$.

Constant values are given in Table 2.5.5 as calculated by Ursino after fitting their model with experimental in-vitro data [62]. A simulation of the above equation is given in Figure 2.23 where a decrease of premotor neurons firing rate can be observed when there is a decrease in arterial blood pressure from 50th to 60th second.

Table 2.5.5 Constant values for parasympathetic premotor neurons controlling SA/AV nodes

| Constant | Value | Description |
|---|---|---|
| $f_{ev,\infty}$ | 2.1 Hz | residual activity of parasympathetic premotor neurons in absence of inputs |
| $k_{ev}$ | 7.06 Hz | constant affecting baroreceptor inputs to parasympathetic premotor neurons |
| $W_{c,v}$ | 0.2 | synaptic weight applied to sensory inputs from chemoreceptors |
| $W_{p,v}$ | 0.105 | synaptic weight applied to sensory inputs from lung stretch receptors |
| $W_{t_f,v}$ | 0.105 | synaptic weight applied to sensory inputs from trigeminal cold receptors |
| $W_{o,v}$ | 0.105 | synaptic weight applied to sensory inputs from oculo-pressure receptors |
| $\theta_v$ | -0.68 Hz | offset term for neural activation of sympathetic premotor neuron |

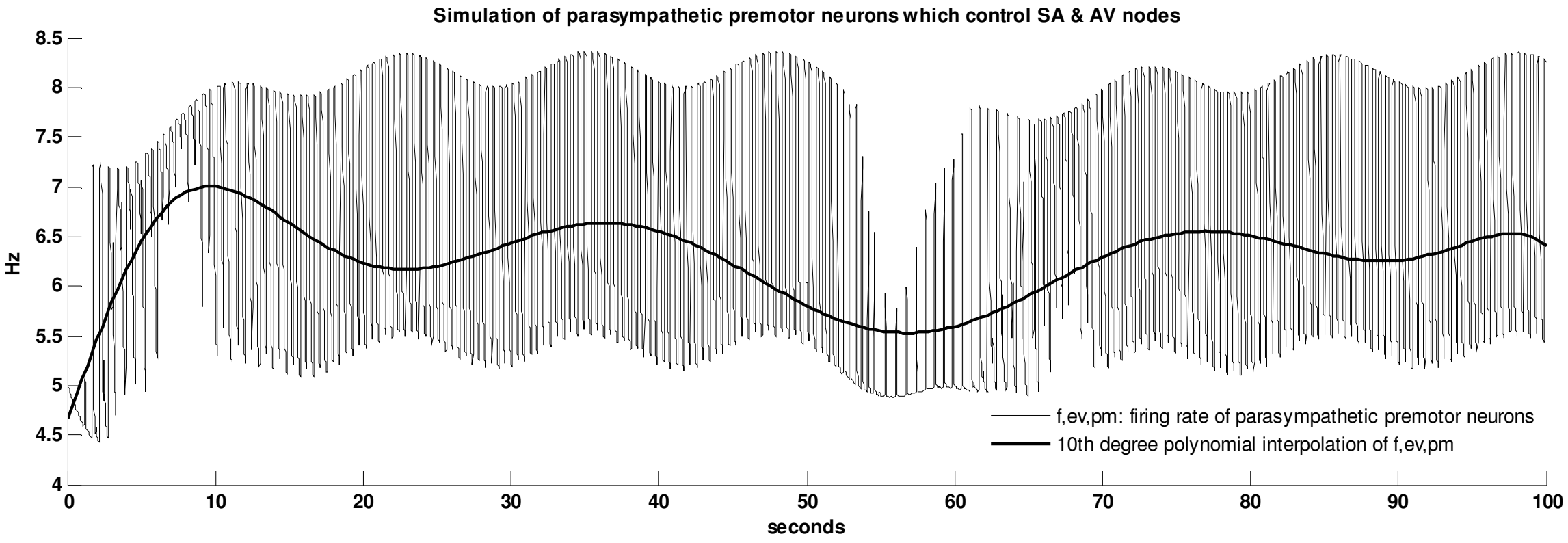

**Figure 2.23 Simulation of parasympathetic motor neurons controlling SA & AV nodes**
Response to arterial blood pressure decreases between 50th and 60th second

## 2.6 Physiological Model of Preganglionic Autonomic Motor Neurons

Autonomic premotor neurons synapse with parasympathetic preganglionic neurons in medulla oblongata. Premotor neurons synapse with sympathetic preganglionic neurons in the thoracic and lumbar region of the spinal cord.

Autonomic preganglionic neurons project to autonomic ganglia where they release acetylcholine (ACh) as neurotransmitter. Neurotransmitters are chemical substance used to carry information between two neurons. Additional neurotransmitters can modulate the activities of vagal preganglionic nerves, such as excitatory amino acids, gamma-aminobutyric acids, serotonin and opioid peptides.[42]

In order to model the relationship between premotor neuron activity and acetylcholine release in autonomic ganglia, Ottesen lumped the long chain of biochemical reactions into a single first order reaction equation and took the accumulated release time to be equal to the average clearance and consumption time for the respective substances. He kept the concentrations in non-dimensional form in order to reduce the number of parameters to estimate. [68] An example is equation (1.7.9). We propose a similar equation to model acetylcholine concentration in parasympathetic autonomic ganglia using the nondimensionalized variable $C_{ach_{ig},v}$ as follows.

$$\frac{dC_{ach_{ig},v}}{dt} = \frac{-C_{ach_{ig},v} + f_{ev_{pm}}}{\tau_{ach_{ig},v}} \tag{2.6.1}$$

where $f_{ev_{pm}}$ is the firing rate of parasympathetic premotor neurons controlling the heart; $\tau_{ach_{ig},v}$ is a characteristic time scale for vagal acetylcholine.

Acetylcholine concentration in sympathetic autonomic ganglia where postganglionic neurons synapse and project to the SA Node and AV node is modeled using the nondimensionalized variable $C_{ach_{ig},sn}$ as follows.

$$\frac{dC_{ach_{ig},sn}}{dt} = \frac{-C_{ach_{ig},sn} + f_{sn_{pm}}}{\tau_{ach_{ig},sn}} \tag{2.6.2}$$

where $f_{sn_{pm}}$ is the firing rate of sympathetic premotor neurons controlling the SA and AV nodes; $\tau_{ach_{ig},sn}$ is a characteristic time scale for sympathetic acetylcholine.

Postganglionic sympathetic neurons which control vascular beds originate in sympathetic autonomic ganglia as well. Their activity also depends on acetylcholine concentration. Acetylcholine release for vascular control is normally controlled by premotor neurons in rostral ventrolateral medulla. However bilateral inhibition of these cells causes a much greater decrease in renal than splenic sympathetic activity; the residual activity might be of spinal origin. This supports the possibility of other source to maintain blood pressure since fall in blood pressure after bilateral lesion of the rostral ventrolateral medulla is not sustained. We model this residual activity by including a new parameter $f_{si_{res}}$ representing the firing rate of spinal neurons capable of releasing acetylcholine in autonomic ganglia without central input.

Therefore changes in acetylcholine concentration is modeled using the nondimensionalized variable $C_{ach_{ig},si}$ (with $i$ equals $v$ for ventricles, $b$ for brain, $m$ for skeletal muscles, $s$ for splanchnic and $e$ for extrasplanchnic) as follows.

$$\frac{dC_{ach_{ig},si}}{dt} = \frac{-C_{ach,si} + max(f_{si_{pm}}, f_{es_{resi}})}{\tau_{ach_{ig},si}} \qquad (2.6.3)$$

$$i = v, b, m, s, e$$

where $f_{si_{pm}}$ is the firing rate of sympathetic premotor neurons controlling vascular beds; $\tau_{ach_{ig},si}$ is a characteristic time scale for sympathetic acetylcholine; $f_{es_{resi}}$ is a model constant representing the firing rate of residual sympathetic activity originating in spinal cord.

Constant values for acetylcholine concentration in autonomic ganglia are given in Table 2.5.6 as calculated by Ottesen after fitting their model with heart rate and blood pressure data [65]. A simulation of the equations above is given in Figure 2.24 where an increase of acetylcholine concentration in sympathetic autonomic ganglia and a decrease of concentration in parasympathetic autonomic ganglia can be observed as response to decreased arterial blood pressure between 50th and 60th second.

Table 2.5.6 Constant values for acetylcholine concentration in autonomic ganglia

| Constant | Value | Description |
|---|---|---|
| $\tau_{ach_{ig},v}$ | 1.32 | characteristic time scale for acetylcholine release by parasympathetic preganglionic neurons |
| $\tau_{ach_{ig},sn}$ | 0.72 | characteristic time scale for acetylcholine release by sympathetic preganglionic neurons for controlling SA and AV nodes |
| $\tau_{ach_{ig},si}$ | 0.72 | characteristic time scale for acetylcholine release by sympathetic preganglionic neurons for controlling vessels (with $i$ equals $v$ for ventricles, $b$ for brain, $m$ for skeletal muscles, $s$ for splanchnic and $e$ for extrasplanchnic) |
| $f_{es_{resi}}$ | 1 Hz | firing rate of residual sympathetic activity originating in spinal cord |

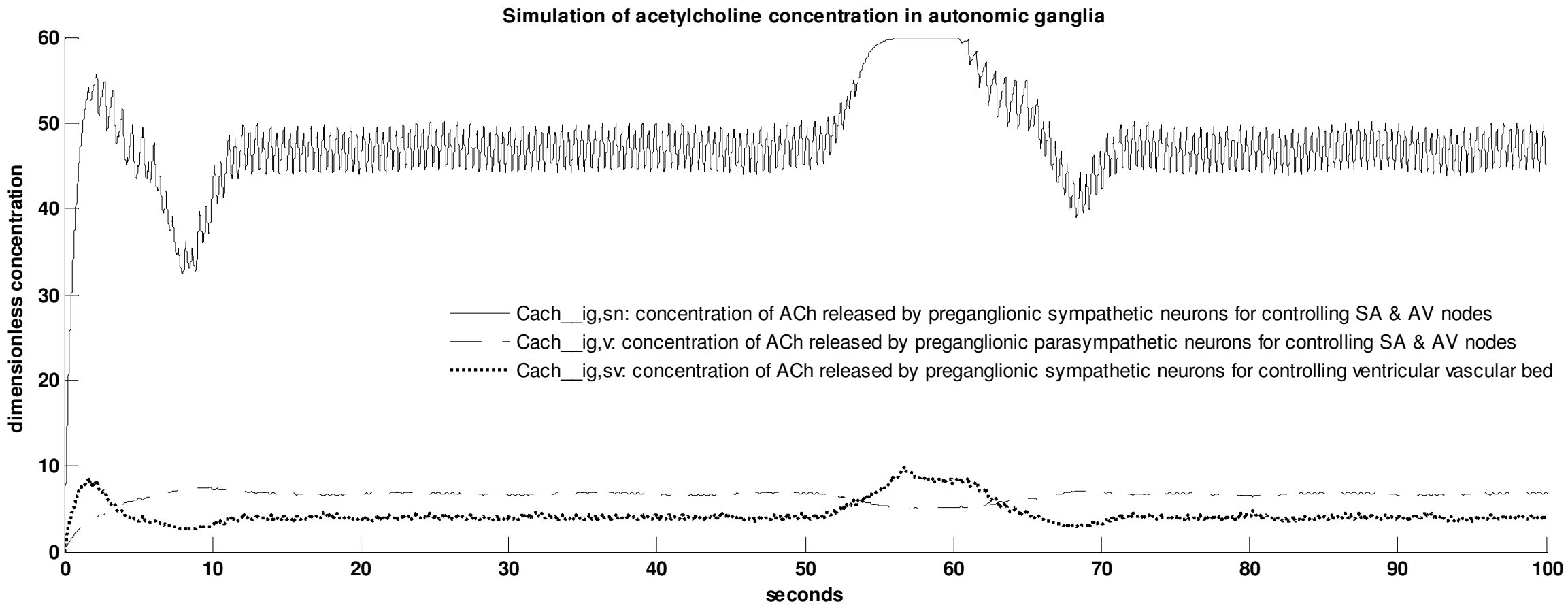

**Figure 2.24 Simulation of acetylcholine concentration in autonomic ganglia**
Response to arterial blood pressure decreases between 50th and 60th second

## 2.7 Physiological Model of Postganglionic Autonomic Motor Neurons

Postganglionic neurons originate from autonomic ganglia and project to cardiovascular effectors where they secrete neurotransmitters which modulate the function of target cells. Autonomic postganglionic axons end with a series of swollen areas called varicosity, which lie across the surface of the target tissue. Neurotransmitters are diffused in the interstitial fluid, so that a single postganglionic neuron can affect a large area of the target tissue. Sympathetic postganglionic neurons release norepinephrine, whereas parasympathetic postganglionic neurons secrete acetylcholine onto the target cell.

We assume that the concentration of acetylcholine and norepinephrine released on target cell by postganglionic neurons linearly depends on the concentration of acetylcholine released by preganglionic neurons in the autonomic ganglia. We model acetylcholine concentration in SA and AV nodes using the nondimensionalized variable $C_{ach_{og},v}$ as follows.

$$C_{ach_{og},v} = k_{ach_{og},v} \cdot C_{ach_{ig},v} \tag{2.7.1}$$

where $C_{ach_{ig},v}$ is the concentration of acetylcholine released by the parasympathetic preganglionic neuron in autonomic ganglia and $k_{ach_{og},v}$ is a model constant representing modulation of acetylcholine concentration by local effects at cell membrane receptors in SA node.

Vagally-mediated atropine-sensitive tachycardia has been observed after somatic nociceptors stimulation suggesting the presence of vasoactive peptides within vagal nerves ending [98]. Therefore norepinephrine concentration in SA and AV nodes might depend on concentration of acetylcholine released by both sympathetic and parasympathetic preganglionic neurons projecting to heart. We model norepinephrine concentration in SA and AV nodes using the nondimensionalized variable $C_{nor_{og},sn}$ as follows.

$$C_{nor_{og},sn} = k_{nor_{og},sn} \cdot C_{ach_{ig},sn} + k_{nor_{og},v} \cdot C_{ach_{ig},v} \tag{2.7.2}$$

where $C_{ach_{ig},sn}, C_{ach_{ig},v}$ are the concentrations of acetylcholine released by the sympathetic and parasympathetic preganglionic neuron in autonomic ganglia respectively; $k_{nor_{og},sn}$ is a model constant and $k_{nor_{og},v}$ is model parameter representing modulation of norepinephrine concentration by local effects at cell membrane receptors in SA node.

We model norepinephrine concentration in vascular beds using the nondimensionalized variable $C_{nor_{og},si}$ (with $i$ equals $v$ for ventricles, $b$ for brain, $m$ for skeletal muscles, $s$ for splanchnic and $e$ for extrasplanchnic) as follows.

$$\begin{gathered} C_{nor_{og},si} = k_{nor_{og},si} \cdot C_{ach_{ig},si} \\ i = v, b, m, s, e \end{gathered} \tag{2.7.3}$$

where $C_{ach_{ig},si}$ is the concentration of acetylcholine released by the sympathetic preganglionic neuron in autonomic ganglia and $k_{nor_{og},si}$ is a model parameter representing modulation factor of norepinephrine by neuropeptides released by vagal nerve endings at SA & AV nodes.

Constant values for acetylcholine and norepinephrine concentrations in SA and AV nodes are given in Table 2.5.7 as well as constants for norepinephrine concentration in vascular beds. For reason of simplicity we assume the values to be equal to 1. Detailed studies of neurotransmitter dynamic at ganglionic sites are required for determining more accurate constant values. The parameter for vagally-mediated tachycardia is given in Table 2.5.8. Simulation of the above equations is given in Figure 2.25 where a decrease of acetylcholine concentration in SA & AV nodes and an increase of norepinephrine concentration in SA & AV nodes and ventricular vascular beds can be observed as response to decreased arterial blood pressure between $50^{th}$ and $60^{th}$ second.

Table 2.5.7 Constant values for concentration of postganglionic neurotransmitters

| Constant | Value | Description |
|---|---|---|
| $k_{ach_{og},v}$ | 1 | modulation factor of acetylcholine concentration by local effects at cell membrane receptors in SA & AV nodes |
| $k_{nor_{og},sn}$ | 1 | modulation factor of norepinephrine concentration by local effects at cell membrane receptors in SA & AV nodes |
| $k_{nor_{og},si}$ | 1 | modulation factor of norepinephrine concentration by local effects at cell membrane receptors in blood vessels wall |

Table 2.5.8 Parameter values for vagally-mediated tachycardia

| Parameter | Value | Description |
|---|---|---|
| $k_{nor_{og},v}$ | 0.1 | modulation factor of norepinephrine concentration by neuropeptides released by vagal nerve endings at SA & AV nodes |

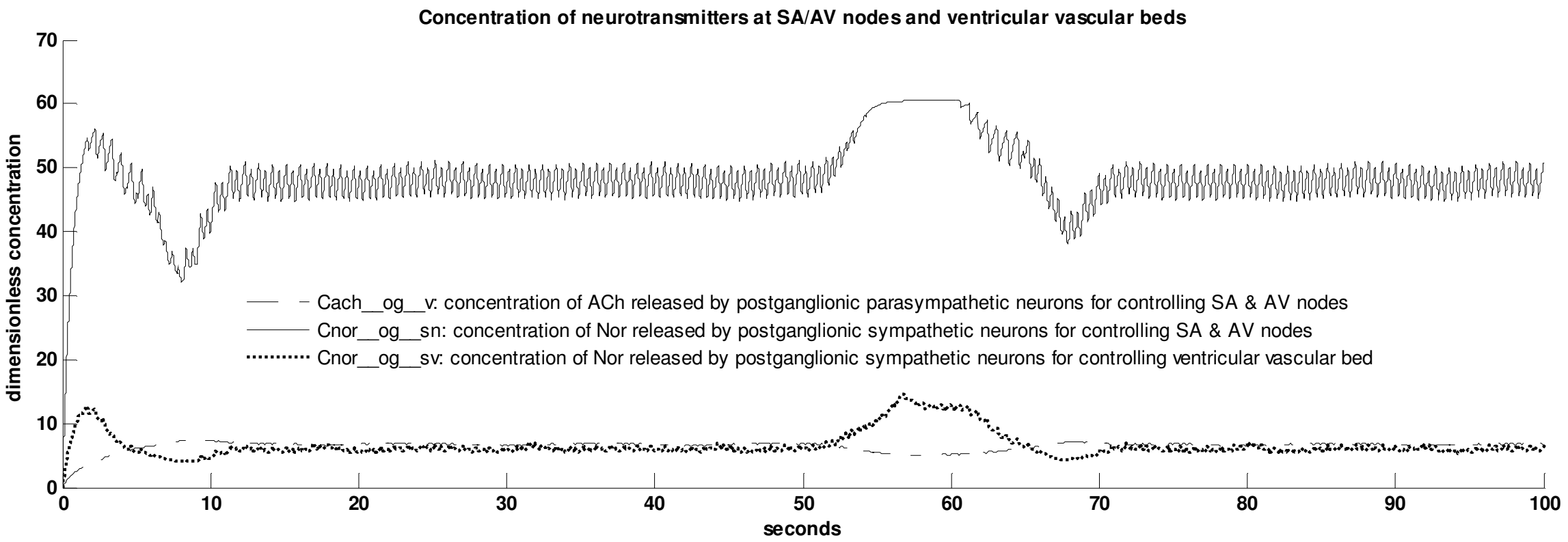

**Figure 2.25 Simulation of acetylcholine and norepinephrine concentrations in heart**
Response to arterial blood pressure decreases between $50^{th}$ and $60^{th}$ second

Sympathetic autonomic preganglionic neurons also project to the adrenal medulla which work as modified sympathetic ganglion. Adrenal medulla include postganglionic neurons which however lack axon that would normally project to target cells. Instead they release catecholamines (approximately 80% epinephrine and 20% norepinephrine) directly into the circulation. We assume that activity in adrenal medulla depends of the weighted sum of premotor inputs. We model secreted epinephrine and norepinephrine concentrations using the nondimensionalized variables $C_{epi,am}, C_{nor,am}$ as follows.

$$\begin{aligned} \frac{dC_{am}}{dt} &= \frac{-C_{am} + W_{ach,am} \cdot (f_{sn_{pm}} + f_{sv_{pm}} + f_{sb_{pm}} + f_{sm_{pm}} + f_{ss_{pm}} + f_{se_{pm}})}{\tau_{am}} \\ C_{epi,am} &= 0.8 \cdot C_{am} \\ C_{nor,am} &= 0.2 \cdot C_{am} \end{aligned} \qquad (2.7.4)$$

where $f_{sn_{pm}}, f_{sv_{pm}}, f_{sb_{pm}}, f_{sm_{pm}}, f_{ss_{pm}}, f_{se_{pm}}$ are the firing rate of sympathetic premotor neurons controlling SA/AV nodes, ventricular, brain, muscles, splanchnic and exptrasplanchnic beds respectively; $W_{ach,am}$ is a synaptic weight. $\tau_{am}$ is a characteristic time scale for neurotransmitters in medulla.

Constant values for adrenal epinephrine and norepinephrine concentrations in blood stream are given in Table 2.5.9. Simulation of the above equations is given in Figure 2.26 where a increased concentration can be observed as response to decreased arterial blood pressure between 50th and 60th second.

Table 2.5.9 Constant values for concentration of adrenal neurotransmitters

| Constant | Value | Description |
|---|---|---|
| $W_{ach,am}$ | 0.166 | synaptic weight applied to sympathetic premotor inputs to adrenal medulla |
| $\tau_{am}$ | 1.32 | characteristic time scale for neurotransmitters in adrenal medulla |

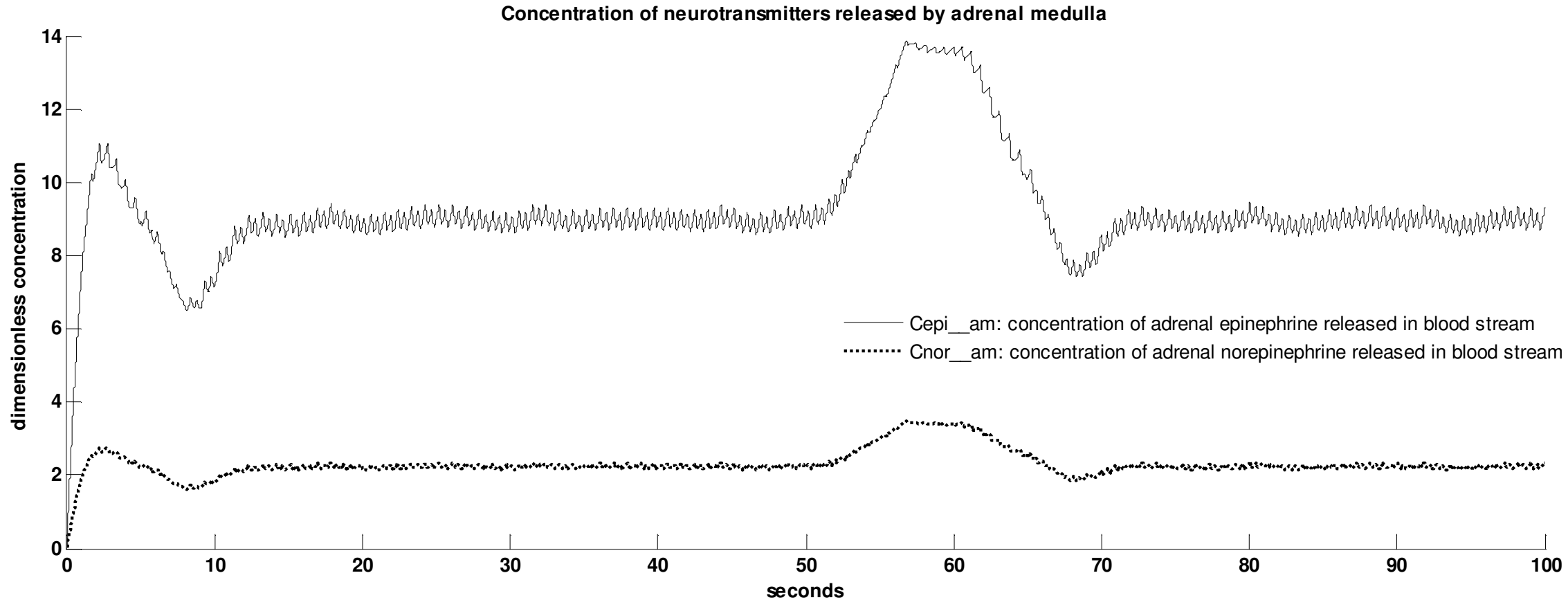

**Figure 2.26 Simulation of adrenal epinephrine and norepinephrine concentrations**
Response to arterial blood pressure decreases between 50th and 60th second

## 2.8 Physiological Model of Cardiovascular Effectors

### 2.8.1 Modulation of Heart Rate by Postganglionic Neurons

The interval between action potentials defines heart rate and can be shortened by altering the permeability of pacemaker cells to specific ions which in turn can be modified by norepinephrine secreted by sympathetic postganglionic neurons and circulating catecholamines (norepinephrine) secreted by adrenal medulla and acting via $\beta_1$-adrenoceptors located on SA nodal cells. Increased ion flow through $I_f$ and $Ca^{2+}$ channels and increased permeability to $Na^+$ and $Ca^{2+}$ during the pacemaker potential phase speed up depolarization and heart rate. Additionally catecholamines norepinephrine and epinephrine enhance conduction of action potentials through the AV node.

Parasympathetic postganglionic neurons release the neurotransmitter acetylcholine (Ach) which increases $K^+$ permeability, hyperpolarizing the cell. At the same time ACh causes $Ca^{2+}$ permeability to decrease, slowing the rate at which the pacemaker potential depolarizes. Both effects cause the cell to delaying the onset of the action potential, slowing the heart rate. ACh also slows the conduction of action potentials through the AV node, thereby increasing the AV node delay.

Existing prominent mathematical models assume that heart period $T$ is obtained as sum of components corresponding to vagal drive $\Delta T_{ach_{og},v}$, sympathetic drive $\Delta T_{nor_{og},sn}$ and intrinsic effects $T_0$. We further include influence from adrenal medulla $\Delta T_{nor,am}$ and high frequency intracardiac fluctuations $\Delta T_{rsa}$

We model the component $\Delta T_{rsa}$ in order to simulate the heart rate variability that has been observed after combined sympatho-vagal blockade as well as in transplanted denervated human heart, suggesting an intracardiac mechanism, which might contribute to respiratory sinus arrhythmias (RSA). $\Delta T_{rsa}$ has been modelled as positive sinusoidal function for simplicity. We model the influence of adrenal medulla ($\Delta T_{nor,am}$) on heart rate using a pure latency, a monotonic logarithmic static function, and low-pass first order dynamics.

Parasympathetic activity dominates in the control of heart rate. Increase in parasympathetic tone evokes a pronounced bradycardia even when there is high level of sympathetic activity. In the presence of high or moderate parasympathetic tone, change in sympathetic activity elicits negligible heart rate modulation. It has been suggested that parasympathetic efferent neurons may inhibit release of norepinephrine. This effect is considered by including the weighting factor $k_{T_{nor_{og},sn}}$ .

The remaining parts of equations for vagal drive $\Delta T_{ach_{og},v}$ and sympathetic drive $\Delta T_{nor_{og},sn}$ are adapted from Ursino [62] with the difference that concentration of neurotransmitters is used instead of firing rate of premotor neurons. The original model of Ursino is given in equation (1.7.18) for comparison. The model of age-related intrinsic effects on heart rate $T_0$ is adapted from Jose [95].

The resulting equations that determine heart period in our model are given below.

$$HR = \frac{60}{T}$$
$$T = T_0 + \Delta T_{rsa} + \Delta T_{ach_{og},v} + \Delta T_{nor,am} + \Delta T_{nor_{og},sn}$$
$$T_0 = \frac{1}{1.97 - 9.50 \cdot 10^{-3} \cdot age}$$
$$\frac{d\Delta T_{ach_{og},v}}{dt} = \frac{1}{\tau_{T_{ach_{og},v}}} \cdot \left( -\Delta T_{ach_{og},v} + G_{T_{ach_{og},v}} \cdot C_{ach_{og},v}(t - D_{T_{ach_{og},v}}) \right)$$
$$\frac{d\Delta T_{nor_{og},sn}}{dt} = \frac{1}{\tau_{T_{nor_{og},sn}}} \cdot \left( -\Delta T_{nor_{og},sn} + G_{T_{nor_{og},sn}} \cdot ln[C_{nor_{og},sn}(t - D_{T_{nor_{og},sn}}) - k_{T_{nor_{og},sn}} \cdot C_{ach_{og},v}(t - D_{T_{ach_{og},v}}) - C_{nor_{og},sn,min} + 1] \right)$$
$$\frac{d\Delta T_{nor,am}}{dt} = \frac{1}{\tau_{T_{nor,am}}} \cdot \left( -\Delta T_{nor,am} + G_{T_{nor,am}} \cdot ln[C_{nor,am}(t - D_{T_{nor,am}}) - C_{nor,am,min} + 1] \right)$$
$$\Delta T_{rsa}(t) = G_{T_{rsa}} \cdot \left( 1 + \sin\left( 2\pi \cdot f_{rsa} \cdot t \right) \right)$$
$$\sigma_{T_{ach_{og},v}} = G_{T_{ach_{og},v}} \cdot C_{ach_{og},v}(t - D_{T_{ach_{og},v}})$$
$$\sigma_{T_{nor_{og},sn}} = \begin{cases} G_{T_{nor_{og},sn}} \cdot ln[C_{nor_{og},sn}(t - D_{T_{nor_{og},sn}}) - k_{T_{nor_{og},sn}} \cdot C_{ach_{og},v}(t - D_{T_{ach_{og},v}}) - C_{nor_{og},sn,min} + 1] & if\ C_{nor_{og},sn} \geq C_{nor_{og},sn,min} \\ 0 & if\ C_{nor_{og},sn} < C_{nor_{og},sn,min} \end{cases}$$
$$\sigma_{T_{nor,am}} = \begin{cases} G_{T_{nor,am}} \cdot ln[C_{nor,am}(t - D_{T_{nor,am}}) - C_{nor,am,min} + 1] & if\ C_{nor,am} \geq C_{nor,am,min} \\ 0 & if\ C_{nor,am} < C_{nor,am,min} \end{cases}$$

(2.8.1)

where $HR$ is the heart rate (in beats/min); $T$ is the heart period (i.e. RR interval in seconds) with $T_0$ being its intrinsic value in the absence of cardiac innervations, depending on the model input $age$; $\tau_{T_{ach_{og},v}}, \tau_{T_{nor_{og},sn}}, \tau_{T_{nor,am}}$ are time constants; $G_{T_{ach_{og},v}}, G_{T_{nor_{og},sn}}, G_{T_{nor,am}}$ are constant gain factors; $D_{T_{ach_{og},v}}, D_{T_{nor_{og},sn}}, D_{T_{nor,am}}$ are the time delays for parasympathetic and sympathetic responses to take effect; $C_{nor_{og},sn,min}, C_{nor,am,min}$ are threshold for sympathetic stimulation; $f_{rsa}$ is the frequency of high frequency intracardiac fluctuations and $G_{T_{rsa}}$ is their amplitude; $k_{T_{nor_{og},sn}}$ is a weighting factor describing the inhibiting effect that vagal efferent has on release of norepinephrine.

Constant values for parasympathetic $\Delta T_{ach_{og},v}$ and sympathetic drive $\Delta T_{nor_{og},sn}$ are given in Table 2.5.10 as calculated by Ursino after fitting the model with data from in-vitro experiments [62]. Constant values for the adrenal medulla component $\Delta T_{nor,am}$ were assumed to be the same as those for the sympathetic component $\Delta T_{nor_{og},sn}$. Table 2.5.11 includes initial values for parameters that determine high frequency intracardiac fluctuations $\Delta T_{rsa}$. They were calculated after trying several values that fit our own experimental data (heart rate signals). A simulation of the above equation is given in Figure 2.27 where increasing heart rate can be observed when there is a decrease in arterial blood pressure from 50$^{th}$ to 60$^{th}$ second.

Table 2.5.10 Constant values for heart rate modulation

| **Constant** | **Value** | **Description** |
|---|---|---|
| $\tau_{T_{ach_{og},v}}$ | 1.5 | time constant of parasympathetic achetychloline responses |
| $\tau_{T_{nor_{og},sn}}$ | 2 | time constant of sympathetic norepinephrine responses |
| $\tau_{T_{nor,am}}$ | 2 | time constant of adrenal norepinephrine sympathetic responses |
| $k_{T_{nor_{og},sn}}$ | 1 | weighting factor describing the inhibiting effect that vagal efferent has on release of norepinephrine |
| $D_{T_{ach_{og},v}}$ | 0.2 s | time delay for parasympathetic achetychloline responses to take effect |
| $D_{T_{nor_{og},sn}}$ | 2 s | time delay for sympathetic norepinephrine responses to take effect |
| $D_{T_{nor,am}}$ | 2 s | time delay for adrenal norepinephrine sympathetic responses to take effect |
| $C_{nor_{og},sn,min}$ | 2.66 | threshold for sympathetic stimulation by sympathetic norepinephrine |
| $C_{nor,am,min}$ | 2.66 | threshold for sympathetic stimulation by adrenal norepinephrine |
| $G_{T_{rsa}}$ | 0.01 | gain of intracardiac high frequency fluctuations |
| $G_{T_{ach_{og},v}}$ | 0.09 | gain of parasympathetic achetychloline responses on heart rate |
| $G_{T_{nor_{og},sn}}$ | -0.13 | gain of sympathetic norepinephrine responses on heart rate |
| $G_{T_{nor,am}}$ | -0.13 | gain of adrenal norepinephrine sympathetic responses on heart rate |

Table 2.5.11 Parameters values for heart rate modulation

| **Parameter** | **Value** | **Description** |
|---|---|---|
| $f_{rsa}$ | 0.15 Hz | frequency of intracardiac high frequency fluctuations |
| $G_{T_{rsa}}$ | 0.01 s | amplitude of intracardiac high frequency fluctuations |

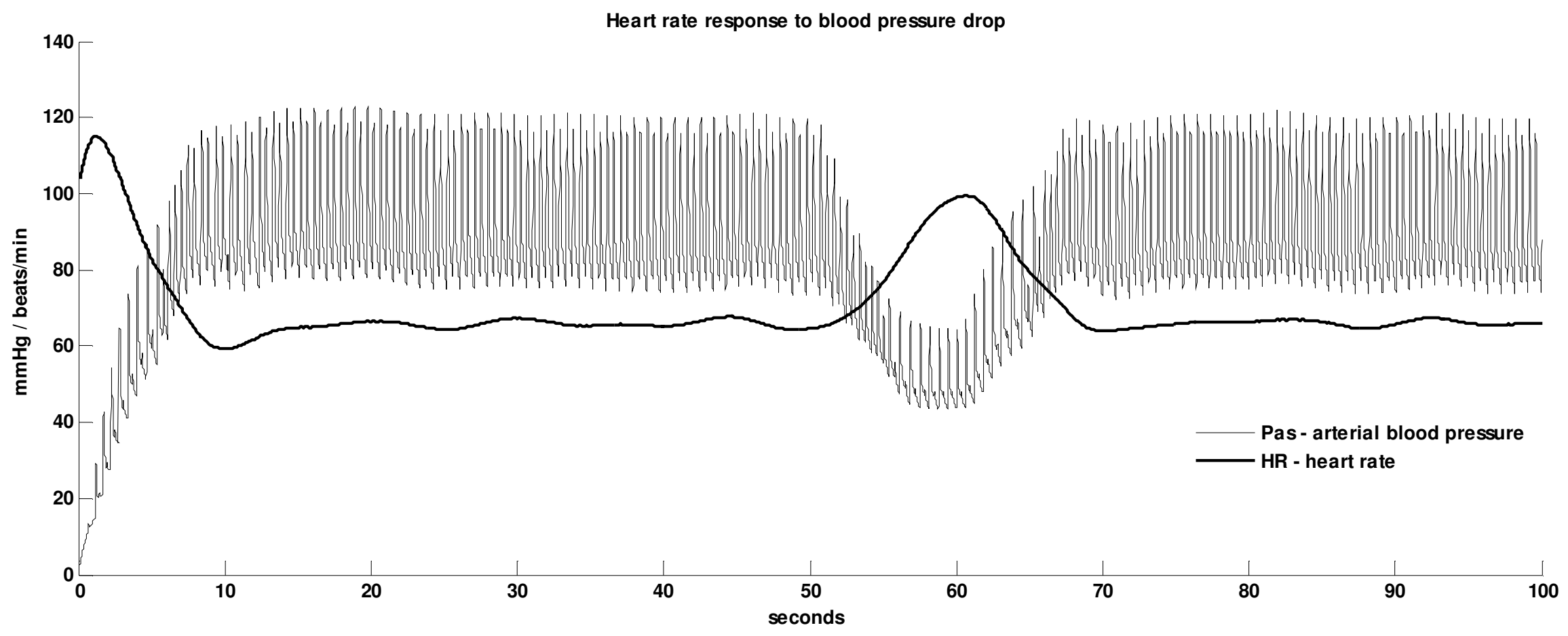

**Figure 2.27 Heart rate response to arterial blood pressure drop**
arterial blood pressure decreases between 50th and 60th second

### 2.8.2 Modulation of Ventricle Elastance by Postganglionic Neurons

Sympathetic postganglionic neurons release norepinephrine at ventricle muscle cells. This promotes influx of $Ca^{2+}$ resulting in greater contraction strength and a shorter duration of contraction. Cardiac contraction is also modified by circulating epinephrine and norepinephrine acting via $\beta_1$-adrenoceptors located on heart muscle. $\beta_2$-adrenoceptors are more sensitive to epinephrine and are typically not innervated by postganglionic neurons. They are located on certain blood vessels and smooth muscles of some organs. They are also present in heart and respond to circulating epinephrine.

Parasympathetic postganglionic neurons release acetylcholine at ventricle muscle cells which bind to M2 muscarinic receptors. This results in a reduction in contraction strength and lengthening of muscular relaxation. The effect is stronger in atria and relative weak in ventricle.

Our goal is to describe the contractile activity of ventricle by means of a Voigt viscoelastic model, i.e., the series arrangement of a time-varying elastance. We model the ventricle elastance at the moment of maximal contraction, i.e. end-systolic elastance. Existing prominent mathematical models assume that ventricle elastance is only controlled by the sum of two components corresponding to sympathetic drive $\Delta E_{nor_{og},sv}$ and intrinsic value $E_0$ . We further include a model of parasympathetic drive $\Delta E_{ach_{og},v}$ , effects of adrenal norepinephrine $\Delta E_{nor,am}$ and effects of adrenal epinephrine $\Delta E_{epi,am}$ .

Although the sympathetic component $\Delta E_{nor_{og},sv}$ is modeled following Ursino [62] as compared to equation (1.7.17), the remaining components are own contributions. We model the changes in ventricle elastance as response to change in norepinephrine, epinephrine and acetylcholine concentrations using a pure latency, a monotonic logarithmic static function, and low-pass first order dynamics.

The resulting equations that determine ventricle elastance in our model are given as follows.

$$
\begin{aligned}
&E(t) = \Delta E_{nor_{og},sv} + \Delta E_{nor,am} + \Delta E_{epi,am} + \Delta E_{ach_{og},v} + E_0 \\
&\frac{d\Delta E_{nor_{og},sv}}{dt} = \frac{1}{\tau_{E_{nor_{og},sn}}} \cdot (-\Delta E_{nor_{og},sv} + \sigma_{E_{nor_{og},sv}}) \\
&\frac{d\Delta E_{nor,am}}{dt} = \frac{1}{\tau_{E_{nor,am}}} \cdot (-\Delta E_{nor,am} + \sigma_{E_{nor,am}}) \\
&\frac{d\Delta E_{epi,am}}{dt} = \frac{1}{\tau_{E_{epi,am}}} \cdot (-\Delta E_{epi,am} + \sigma_{E_{epi,am}}) \\
&\frac{d\Delta E_{ach_{og},v}}{dt} = \frac{1}{\tau_{E_{ach_{og},v}}} \cdot (-\Delta E_{ach_{og},v} + \sigma_{E_{ach_{og},v}}) \\
&\sigma_{E_{nor_{og},sv}} = \begin{cases} G_{E_{nor_{og},sv}} \cdot ln[C_{nor_{og},sv}(t - D_{E_{nor_{og},sv}}) - C_{nor_{og},sv,min} + 1] & if\ C_{nor_{og},sv} \geq C_{nor_{og},sv,min} \\ 0 & if\ C_{nor_{og},sv} < C_{nor_{og},sv,min} \end{cases} \\
&\sigma_{E_{nor,am}} = \begin{cases} G_{E_{nor,am}} \cdot ln[C_{nor,am}(t - D_{E_{nor,am}}) - C_{nor,am,min} + 1] & if\ C_{nor,am} \geq C_{nor,am,min} \\ 0 & if\ C_{nor,am} < C_{nor,am,min} \end{cases} \\
&\sigma_{E_{epi,am}} = \begin{cases} G_{E_{epi,am}} \cdot ln[C_{epi,am}(t - D_{E_{epi,am}}) - C_{epi,am,min} + 1] & if\ C_{epi,am} \geq C_{epi,am,min} \\ 0 & if\ C_{epi,am} < C_{epi,am,min} \end{cases} \\
&\sigma_{E_{ach_{og},v}} = G_{E_{ach_{og},v}} \cdot C_{ach_{og},v}(t - D_{E_{ach_{og},v}})
\end{aligned}
\qquad (2.8.2)
$$

where $E$ is the ventricle elastance with $E_0$ being its intrinsic value in the absence of cardiac innervation; $\tau_{E_{nor_{og},sv}}, \tau_{E_{nor,am}}, \tau_{E_{epi,am}}, \tau_{E_{ach_{og},v}}$ are time constants; $G_{E_{nor_{og},sv}}, G_{E_{nor,am}}, G_{E_{epi,am}}, G_{E_{ach_{og},v}}$ are constant gain factors; $D_{E_{nor_{og},sv}}, D_{E_{nor,am}}, D_{E_{epi,am}}, D_{E_{ach_{og},v}}$ are the time delays for sympathetic and parasympathetic responses to take effect; $C_{nor_{og},sv,min}, C_{nor,am,min}, C_{epi,am,min}$ are threshold for sympathetic stimulation.

We assume that the duration of ventricle contraction depends on norepinephrine, epinephrine and acetylcholine concentrations in ventricle muscles with the same extend as on the elastance. Therefore we model the length of QT intervals (i.e. duration of ventricle contraction) as follows.

$$QT = QT_0 + k_{QT} \times (E_0 - E) \qquad (2.8.3)$$

where $QT_0$ is baseline QT interval duration and $k_{QT}$ is a gain factor.

Constant values for sympathetic drive $\Delta E_{nor_{og},sv}$ and intrinsic value $E_0$ are given in Table 2.5.12 as calculated by Ursino after fitting the model with data from in-vitro experiments [62]. Constant values for effects of adrenal norepinephrine $\Delta E_{nor,am}$ and effects of adrenal epinephrine $\Delta E_{epi,am}$ were assumed to be the same as those for the sympathetic component $\Delta E_{nor_{og},sv}$. Constant values for parasympathetic drive $\Delta E_{ach_{og},v}$ were assumed to be the same as those for parasympathetic drive on heart period (see Table 2.5.10). Table 2.5.13 includes initial values for parameters that determine the duration of QT intervals. They were calculated after trying several values that fit our own experimental data (QT time series). A simulation of the above equation is given in Figure 2.28 where increasing ventricle elastance can be observed when there is a decrease in arterial blood pressure from 50th to 60th second.

Table 2.5.12 Constant values for ventricle elastance modulation

| Constant | Value | Description |
|---|---|---|
| $E_0$ | 2.392 mmHg mL$^{-1}$ | Baseline elastance of ventricles |
| $\tau_{E_{ach_{og},v}}$ | 1.5 s | time constant of parasympathetic achetychloline responses |
| $\tau_{E_{nor_{og},sv}}$ | 8 s | time constant of sympathetic norepinephrine responses |
| $\tau_{E_{nor,am}}$ | 8 s | time constant of adrenal norepinephrine sympathetic responses |
| $\tau_{E_{epi,am}}$ | 8 s | time constant of adrenal epinephrine sympathetic responses |
| $D_{E_{ach_{og},v}}$ | 0.2 s | time delays for parasympathetic achetychloline responses to take effect |
| $D_{E_{nor_{og},sv}}$ | 2 s | time delays for sympathetic norepinephrine responses to take effect |
| $D_{E_{nor,am}}$ | 2 s | time delays for adrenal norepinephrine sympathetic responses to take effect |
| $D_{E_{epi,am}}$ | 2 s | time delays for adrenal epinephrine sympathetic responses to take effect |
| $C_{epi,am}$ | 2.66 | threshold for sympathetic stimulation by adrenal epinephrine |
| $G_{E_{ach_{og},v}}$ | -0.09 mmHg mL$^{-1}$ Hz$^{-1}$ | gain of parasympathetic achetychloline responses on ventricle elastance |
| $G_{E_{nor_{og},sv}}$ | 0.475 mmHg mL$^{-1}$ | gain of sympathetic norepinephrine responses on ventricle elastance |
| $G_{E_{nor,am}}$ | 0.475 mmHg mL$^{-1}$ | gain of adrenal norepinephrine sympathetic responses on ventricle elastance |
| $G_{E_{nor,am}}$ | 0.475 mmHg mL$^{-1}$ | gain of adrenal epinephrine sympathetic responses on ventricle elastance |

Table 2.5.13 Parameters values for ventricle elastance modulation

| Parameter | Value | Description |
|---|---|---|
| $QT_0$ | 300 ms | baseline QT interval duration |
| $k_{QT}$ | 10 mmHg$^{-1}$ mL | gain of elastance on QT interval duration |

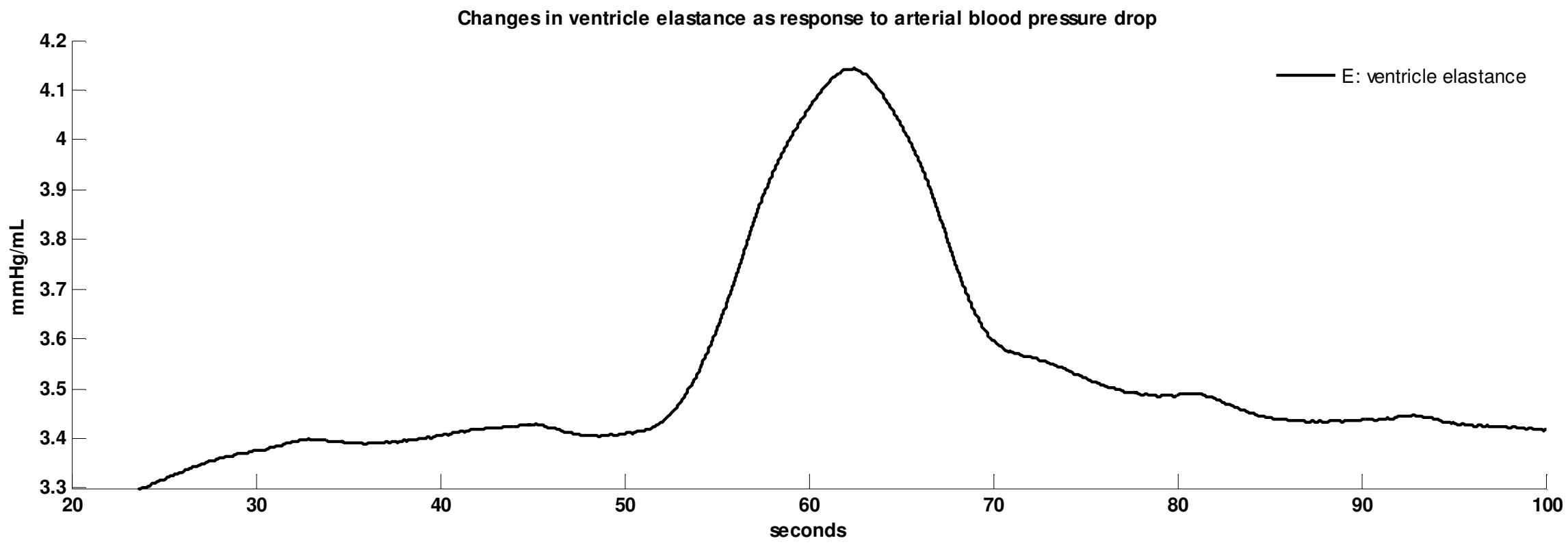

**Figure 2.28 Changes in ventricle elastance as response to arterial blood pressure drop**
arterial blood pressure decreases between 50th and 60th second

### 2.8.3 Modulation of Peripheral Vascular Resistance

The smooth muscle cells within arteriole walls remain constantly slightly contracted via a combination of neuronal, hormonal influences and local effects. The relative degree of contraction within the arterioles is referred to as their basal tone. Veins and venules have a lower basal tone. The arterioles have large amount of smooth muscles in their walls and are the main site of variable resistance. They contribute to more than 60% of the total peripheral resistance. [79]

Sympathetic postganglionic neurons innervate vascular beds. Tonic secretion of norepinephrine on α-receptors causes vasoconstriction. The adrenal medulla also secrete large amount of epinephrine during the flight-or-fight response or strong emotional conditions at arteriolar cell α-receptors enhancing vasoconstriction. However it is noted that α-receptors have a low affinity for epinephrine.

At the opposite of other vascular beds, vessels of heart, skeletal muscle and liver additionally have $\beta_2$ receptors which elicit vasodilatation in the presence of adrenal epinephrine. During the flight-or-fight response while increased sympathetic activity will cause vasoconstriction on other peripheral vessels, vasodilatation will predominate in coronary arteries, skeletal muscles and liver. This diverts blood from nonessential organs, such as gastrointestinal tract, to the skeletal muscles, liver, and heart. However if sympathetic activity is high then more norepinephrine will be released and the more numerous $\alpha_1$ receptors will be activated, causing vasoconstriction.

Blood flow in arterioles is also regulated by local metabolic demand coupled with oxygen consumption. Low oxygen and high carbon dioxide dilate the arterioles. A similar vasodilatation can be observed on vessels of exercising muscles whereas vessels in non-exercising muscles remain constricted.

Existing mathematical models assume that peripheral resistance is determined by the activity of sympathetic neurons and local metabolic demand; an example is equation (1.7.16) from Ursino [62]. We further include the effects of adrenal epinephrine and model peripheral vascular resistance $R_{ip}$ as sum of three components representing effects of sympathetic norepinephrine $\Delta R_{nor_{og},si}$, adrenal epinephrine $\Delta R_{epi,am,i}$ and baseline values $R_{ip,0}$. At the opposite of existing mathematical models, we distinguish between regions which respond to circulating epinephrine with vasoconstriction, these are brain, splanchnic and extrasplanchnic vascular beds; and regions which respond with vasodilation, i.e. heart and muscles. Each component is modeled using a pure latency, a monotonic logarithmic static function, and low-pass first order dynamics.

The resulting equations that determine peripheral vascular resistance in our model are given as follows.

$$
\begin{aligned}
&R_{ip} = \frac{\Delta R_{ip}}{1+M_{ip}} \\
&\Delta R_{ip} = \begin{cases} \Delta R_{nor_{og},si} + \Delta R_{epi,am,i} + R_{ip,0} & if\ i = b,s,e \\ \Delta R_{nor_{og},si} - \Delta R_{epi,am,i} + R_{ip,0} & if\ i = v,m \end{cases} \\
&\frac{d\Delta R_{nor_{og},si}}{dt} = \frac{1}{\tau_{R_{nor_{og}}}} \cdot (-\Delta R_{nor_{og},si} + \sigma_{R_{nor_{og},si}}) \\
&\frac{d\Delta R_{epi,am,i}}{dt} = \frac{1}{\tau_{R_{epi,am}}} \cdot (-\Delta R_{epi,am,i} + \sigma_{R_{epi,am,i}}) \\
&\sigma_{R_{nor_{og},si}} = \begin{cases} G_{R_{nor_{og},si}} \cdot ln[C_{nor_{og},si}(t - D_{R_{nor_{og},si}}) - C_{nor_{og},min} + 1] & if\ C_{nor_{og},si} \geq C_{nor_{og},min} \\ 0 & if\ C_{nor_{og},si} < C_{nor_{og},min} \end{cases} \\
&\sigma_{R_{epi,am,i}} = \begin{cases} G_{R_{epi,am,i}} \cdot ln[C_{epi,am}(t - D_{R_{epi,am,i}}) - C_{epi,am,min} + 1] & if\ C_{epi,am} \geq C_{epi,am,min} \\ 0 & if\ C_{epi,am} < C_{epi,am,min} \end{cases} \\
&i = b,h,m,s,e
\end{aligned}
\tag{2.8.4}
$$

where $R_{ip}$ is the peripheral resistance with basal value $R_{ip,0}$; $\tau_{R_{nor_{og}}}, \tau_{R_{epi,am}}$ are time constants; $G_{R_{nor_{og},si}}, G_{R_{epi,am,i}}$ are constant gain factors; $D_{R_{nor_{og},si}}, D_{R_{epi,am,i}}$ are the time delays for sympathetic responses to take effect; $C_{nor_{og},min}, C_{epi,am,min}$ are threshold for sympathetic stimulation; $M_{ip}$ represents the metabolic demand in the vascular bed as described in equation (2.8.8) in section 2.8.4.

There is ca. 5 s delay before peripheral vasoconstriction begins, after stimulation of α-sympathetic nerves, and a further 15 s delay before vasoconstriction is complete [70]. These delays in baroreflex response implementation on peripheral vessels is a possible explanation to Mayer waves (slow oscillation in vasomotor tone around 0.1 Hz) [92]. They are taken into account in our model using constants $D_{R_{nor_{og},si}}, D_{R_{epi,am,i}}$.

The relationship between peripheral vascular resistance and skin conductance has been investigated in [99], where authors maintained a finger under isothermal conditions while inducing vasoconstriction by cooling the rest of the hand. They found out an artificial relation between peripheral vascular resistance and skin conductance, usually measured as galvanic skin response. We assume this relationship to be linear when pronounced vasoconstriction occurs and model the level of galvanic skin response $G_{mp}$ at finger site as follows.

$$G_{mp} = k_{G_{mp}} \cdot R_{mp} \tag{2.8.5}$$

where $k_{G_{mp}}$ is a gain parameter applied to peripheral vascular resistance in skeletal muscles $R_{mp}$.

Constant values for sympathetic drive $\Delta R_{nor_{og},si}$ and baseline values $R_{ip,0}$ are given in Table 2.5.14 as calculated by Ursino after fitting the model with data from in-vitro experiments [62]. In their studies the baseline vascular resistance in brain was not measured, but deduced with the assumption that normal blood flow entering the brain is 14 to 15% of total cardiac output. Constant values for effects of adrenal epinephrine $\Delta R_{epi,am,i}$ were assumed to be the same as those for the sympathetic component $\Delta R_{nor_{og},si}$. Table 2.5.15 includes initial values for parameters that determine the galvanic skin response $G_{mp}$. They were calculated after trying several values that fit our own experimental data (galvanic skin response signals). A simulation of the above equation is given in Figure 2.29 where increasing peripheral vascular resistance and galvanic skin response can be observed when mental stress is induced from 150$^{th}$ to 220$^{th}$ second.

Table 2.5.14 Constant values for peripheral resistance modulation

| Constant | Value | Description |
|---|---|---|
| $R_{vp,0}$ | 2.392 mmHg s ml$^{-1}$ | baseline resistance of vascular beds in ventricle |
| $R_{bp,0}$ | 6.57 mmHg s ml$^{-1}$ | baseline resistance of vascular beds in brain |
| $R_{mp,0}$ | 2.106 mmHg s ml$^{-1}$ | baseline resistance of vascular beds in skeletal muscles |
| $R_{sp,0}$ | 2.49 mmHg s ml$^{-1}$ | baseline resistance of vascular beds in splanchnic |
| $R_{ep,0}$ | 1.655 mmHg s ml$^{-1}$ | baseline resistance of vascular beds in extrasplanchnic |
| $\tau_{R_{nor_{og},si}}$ | 6 | time constant of sympathetic norepinephrine responses |
| $\tau_{R_{epi,am}}$ | 6 | time constant of adrenal epinephrine sympathetic responses |
| $D_{R_{nor_{og},si}}$ | 5 s | time delays for sympathetic norepinephrine responses to take effect |
| $D_{R_{epi,am,i}}$ | 5 s | time delays for adrenal epinephrine sympathetic responses to take effect |
| $G_{R_{nor_{og},si}}$ | 1.94 mmHg s ml$^{-1}$ | gain of sympathetic norepinephrine responses on vascular beds (with $i$ equals $v$ for ventricles, $b$ for brain, $m$ for skeletal muscles, $s$ for splanchnic and $e$ for extrasplanchnic) |
| $G_{R_{epi,am,i}}$ | 1.94 mmHg s ml$^{-1}$ | gain of adrenal epinephrine sympathetic responses on vascular beds (with $i$ equals $v$ for ventricles, $b$ for brain, $m$ for skeletal muscles, $s$ for splanchnic and $e$ for extrasplanchnic) |

Table 2.5.15 Parameter value for peripheral resistance modulation

| Parameter | Value | Description |
|---|---|---|
| $k_{G_{mp}}$ | 12 µS mmHg$^{-1}$ s$^{-1}$ ml | gain parameter applied to peripheral vascular resistance in skeletal muscles |

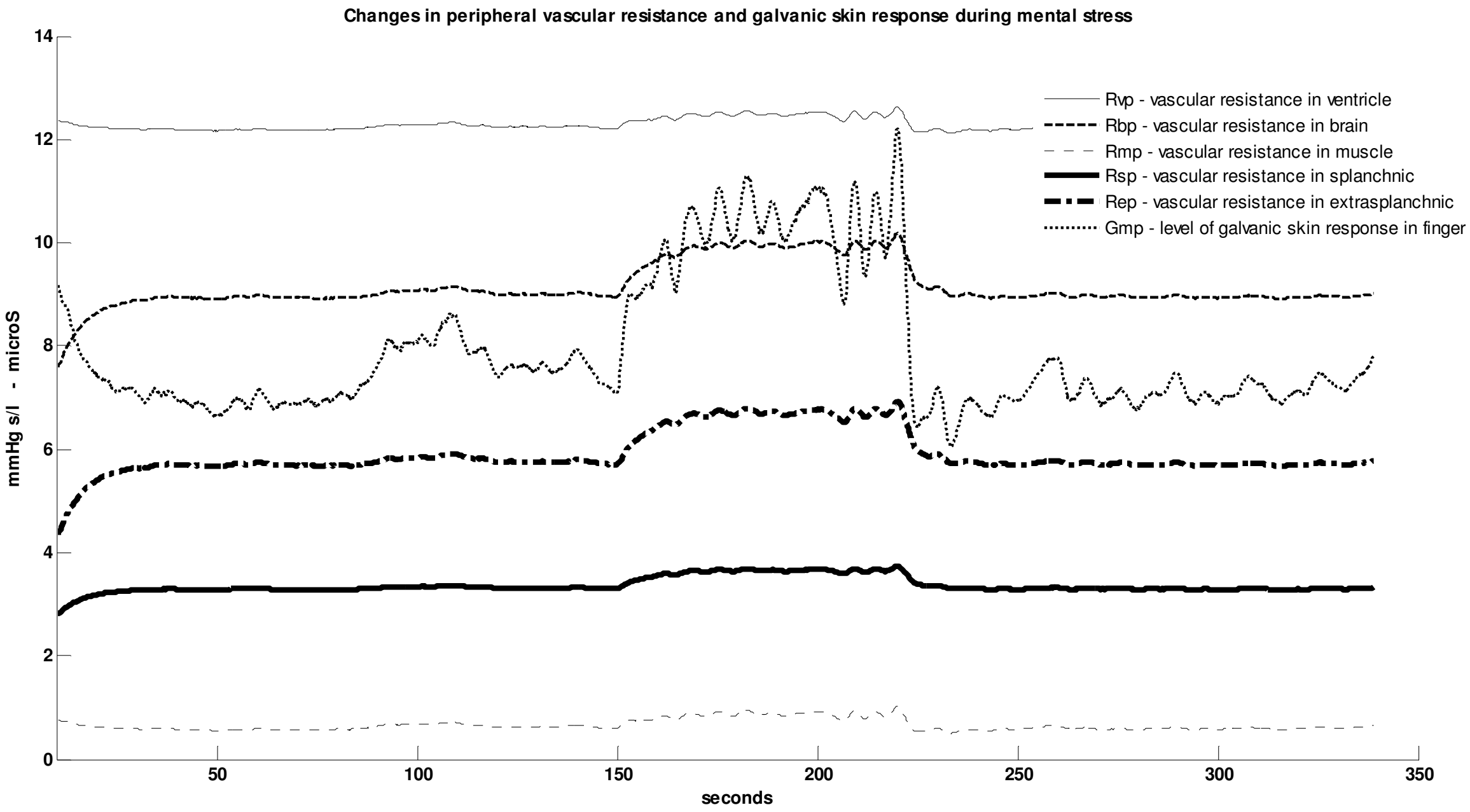

**Figure 2.29 Changes in vascular resistance and galvanic skin response as response to mental stress**

Mental stress between $150^{th}$ and $220^{th}$ second

### 2.8.4 Effects of Respiration and Exercise

Some studies suggest that the intrinsic rhythmic efferent activity to pulmonary muscles is maintained by pacemaker neurons in respiratory control centers. Respiratory neurons are mostly found in two centers in the medulla: the dorsal respiratory group contains inspiratory somatic motor neurons that control the diaphragm, external intercostals, scalene and sternocleidomastoid muscles; the ventral respiratory group contains expiratory somatic motor neurons that control abdominal muscles used for active respiration, internal intercostals, larynx, pharynx and tongue. During quiet respiration at a rate of 12 breaths/min, inspiration is characterized by a gradually increase stimulation of the diaphragm by inspiratory somatic motor neurons for approximately 2 seconds. Over the next 3 seconds passive expiration occurs because of the elastic relaxation of inspiratory muscles and lungs. Expiratory somatic motor neurons are mostly inactive during quiet respiration.

We assume that the inspiratory and expiratory somatic motor neurons present a periodic sine pattern. Respiration can be modulated by exercise intensity and by voluntary control from higher brain centers. We include an arbitrary function representing voluntary respiration pattern $f_{resp,vol}$ and model the resulting respiration pattern $f_{resp}$, respiration rate $f_{resp_{rate}}$ and respiration amplitude $f_{resp_{amp}}$ as follows.

$$
\begin{aligned}
f_{resp_{rate}} &= \begin{cases} \dfrac{1}{T_{resp} - k_{resp} \cdot I} & if \quad f_{resp,vol} = 0 \\ rate(f_{resp,vol}) & if \quad f_{resp,vol} \neq 0 \end{cases} \\
f_{resp_{amp}} &= \begin{cases} A_{resp} + k_{resp} \cdot I & if \quad f_{resp,vol} = 0 \\ amplitude(f_{resp,vol}) & if \quad f_{resp,vol} \neq 0 \end{cases} \\
f_{resp} &= \begin{cases} f_{resp_{amp}} \cdot sin\left(2\pi \cdot f_{resp_{rate}} \cdot t\right) & if \quad f_{resp,vol} = 0 \\ f_{resp,vol} & if \quad f_{resp,vol} \neq 0 \end{cases}
\end{aligned}
\tag{2.8.6}
$$

where $A_{resp}$ is a model input representing the strength and depth of quiet respiration; $T_{resp}$ is a model input representing the period of a quiet respiration cycle; $f_{resp,vol}$ is a model input representing the pattern of voluntary respiration; $I$ is a model input representing exercise intensity; $k_{resp}$ is a gain factor applied to exercise intensity.

The volume of air that moves during a single inspiration or expiration is called tidal volume. The average tidal volume for a 70kg-man during quiet breathing is about 500 mL. The additional air volume which can be inspired to reach maximal inspiration is called inspiratory reserve volume and can reach 3000 mL. The additional air volume which can be exhaled to reach maximal exhalation is called expiratory reserve volume and averages about 1100 mL. The volume of air still remaining in the lungs after maximal exhalation is called residual volume, ca. 1200 mL. We assume that tidal volume depends on respiration signal and exercise intensity and slightly modify the equation (1.7.21) from Magosso [74] (see section 1.7.4) as follows.

$$V_T = V_{T,n} \cdot (f_{resp} + 1) + \frac{G_{V_T} \cdot f_{resp}}{A_{V_T} + B_{V_T} \cdot I} \tag{2.8.7}$$

where $V_T$ is the tidal volume with baseline value $V_{T,n}$; $G_{V_T}$ is a static gain; $I$ is a model input representing exercise intensity; $f_{resp}$ is the respiratory signal; $A_{V_T}, B_{V_T}$ are model constants.

Constant and input values for respiration and tidal volume are given in Table 2.5.16 and Table 2.5.17. Simulation of the above equations is given in Figure 2.30.

Table 2.5.16 Constant values for tidal volume

| **Constant** | **Value** | **Description** |
|---|---|---|
| $A_{V_T}$ | 33 | constant for exercise intensity |
| $B_{V_T}$ | 61 | |
| $k_{resp}$ | 2 | static gain for exercise intensity |
| $V_{T,n}$ | 0.583 L | basal tidal volume |
| $G_{V_T}$ | 1.62 L | static gain for tidal volume |

Table 2.5.17 Input values for respiration and tidal volume

| **Input** | **Value** | **Description** |
|---|---|---|
| $I$ | [0 .. 1] | exercise intensity |
| $A_{resp}$ | 1 | amplitude of respiration signal |
| $T_{resp}$ | 5 s | period of the respiration cycle |
| $f_{resp,vol}$ | [-1 .. 1] | voluntary respiration pattern |

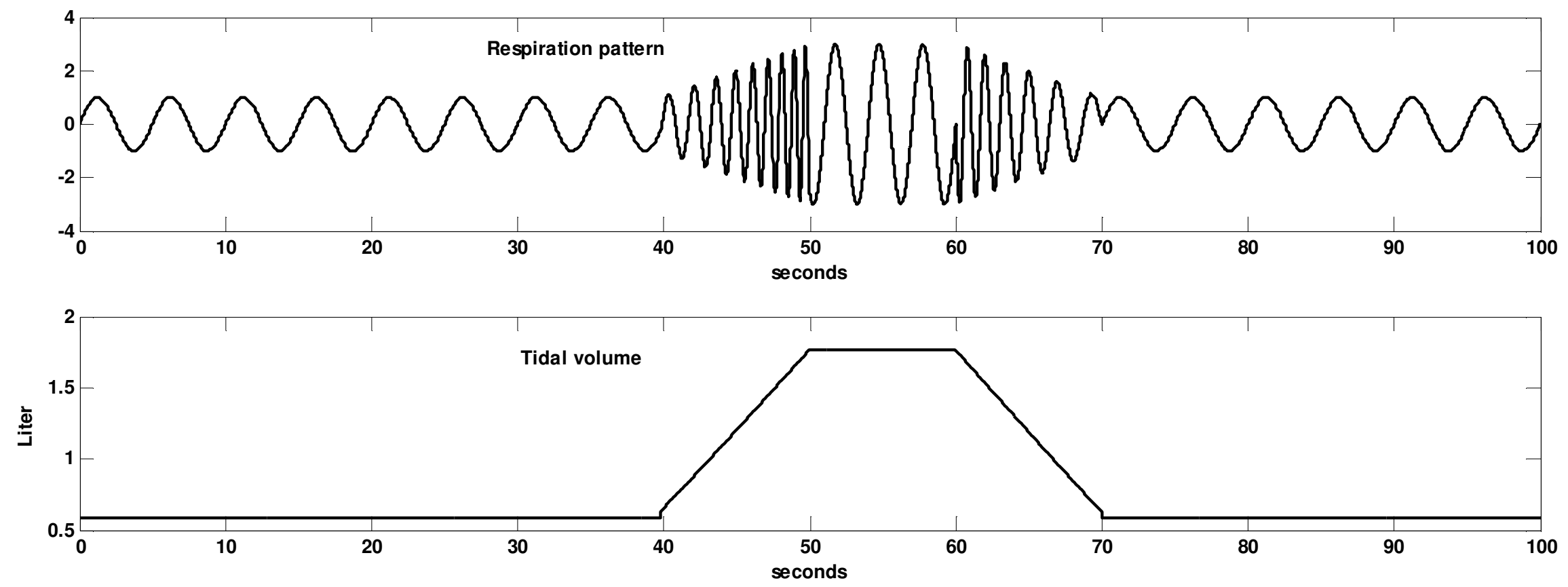

**Figure 2.30 Simulation of respiration signal and corresponding tidal volume**

Local metabolic demand in brain, heart, muscles, splanchnic and extrasplanchnic tissues is adapted from Magosso [74] using a static gain and first-order low pass dynamics as follows.

$$\begin{aligned} & M_{ip} = M_{ip,n} \cdot (1 + x_{M_{ip,n}}) \\ & \frac{dx_{M_{ip,n}}}{dt} = \frac{1}{\tau_{M_{ip,n}}} \cdot (-x_{M_{ip,n}} + G_{M_{ip,n}} \cdot I) \\ & M = \sum M_{ip} \\ & i = b, h, m, s, e \end{aligned} \tag{2.8.8}$$

where $M_{ip}$ is the metabolic demand with baseline value $M_{ip,n}$ ; $I$ is a model input representing exercise intensity; $\tau_{M_{ip}}$ are time constants and $G_{M_{ip}}$ are static gains; $M$ is the total metabolic demand.

Oxygen is diffused into pulmonary blood by ventilation and removed by metabolism in peripheral tissues. We model the dynamics of partial pressure of oxygen in arterial systemic compartment as follows.

$$\begin{aligned} & P_{as_{O_2}} = P_{as_{O_2},n} \cdot (1 + x_{P_{as_{O_2}}}) \\ & \frac{dx_{P_{as_{O_2}}}}{dt} = \frac{1}{\tau_{P_{as_{O_2}}}} \cdot (-x_{P_{as_{O_2}}} + G_{P_{as_{O_2}}} \cdot V_T - k_{P_{as_{O_2}}} \cdot M) \end{aligned} \tag{2.8.9}$$

where $P_{as_{O_2}}$ is oxygen partial pressure with basal value $P_{as_{O_2},n}$ ; $V_T$ is the tidal volume; $\tau_{P_{as_{O_2}}}$ is a time constant and $G_{P_{as_{O_2}}}$ is a static gain; $k_{P_{as_{O_2}}}$ is a static gain representing oxygen consumption to answer metabolic demand $M$ .

Constant values for metabolic demand and oxygen consumption are given in Table 2.5.18. Simulation of the above equations is given in Figure 2.31.

Table 2.5.18 Constant values for metabolic demand and oxygen consumption

| **Constant** | **Value** | **Description** |
|---|---|---|
| $M_{ip,n}$ | 0.516 mL $s^{-1}$ | baseline metabolic demand in vascular beds (with $i$ equals $v$ for ventricles, $b$ for brain, $m$ for skeletal muscles, $s$ for splanchnic and $e$ for extrasplanchnic) |
| $\tau_{M_{ip}}$ | 40 s | time constants of metabolic demand responses to exercise intensity |
| $G_{M_{ip}}$ | 40 s | static gains applied to exercise intensity (with $i$ equals $v$ for ventricles, $b$ for brain, $m$ for skeletal muscles, $s$ for splanchnic and $e$ for extrasplanchnic) |
| $P_{as_{O_2},n}$ | 100 mmHg | baseline value of oxygen partial pressure in arteries |
| $\tau_{P_{as_{O_2}}}$ | 10 s | time constant of oxygen partial pressure |
| $G_{P_{as_{O_2}}}$ | 0.2 s $mL^{-1}$ | gain of tidal volume on oxygen partial pressure |
| $k_{P_{as_{O_2}}}$ | 0.01 $s^2$ $mL^{-1}$ | static gain of metabolic demand |

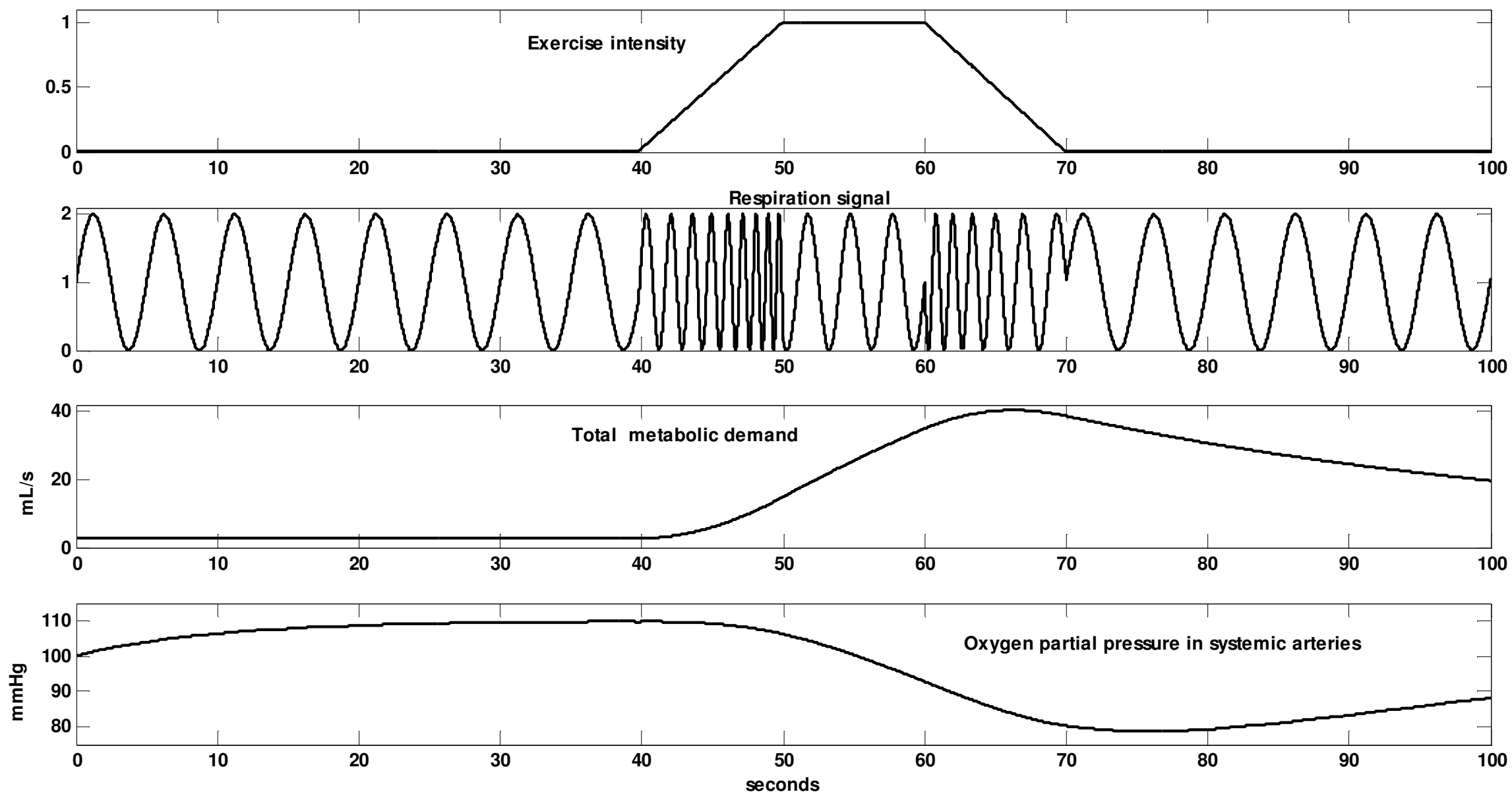

**Figure 2.31 Effects of exercise on the cardiovascular system**
Simulation of respiration signal (2nd graph), metabolic demand (3rd graph) and oxygen partial pressure (4th graph) as response to increased exercise intensity (1st graph)

### 2.8.5 Effects of Gravitational Forces

Change of posture from a sitting or lying position to standing has a reducing effect on venous return and is characterized by reduced arterial blood pressure, activation of baroreflex and increasing heart rate. Studies in [68] with mathematical models exhibit differences in cardiovascular response between head-up tilt and sit-to-stand. During head-up tilt, the head is slowly moved to a location above the heart, while during sit-to-stand no hydrostatic pressure changes are enforced between the head and the heart. The model used is however an open-model of cardiovascular control using the blood pressure as input. In our closed-loop model we were required to model gravitational forces of both types by introducing new variables. Examples of variables have been presented in section 1.7.4 where arterial blood pressure in legs was a function of height. Trial of including these variables in our model was unsuccessful since they exclusively model cerebral blood flow control during postural change from sitting to standing only [72]. Therefore we adopt a different approach based on the observation that heart rate response has a peak increase followed by an exponential decay during sit-to-stand, and a sustained sigmoid-like shape during head-up tilt.

In order to reproduce observed heart rate changes, which have a sigmoid-like shape, we model gravitational forces during head-up tilt as follows.

$$g_{tt} = \begin{cases} 0 & if \quad t < t_{start} \\ \dfrac{g_{tt,max}}{1 + g_{tt,max} \cdot e^{-k_{tt} \cdot t}} & if \quad t_{start} \le t \le t_{stop} \\ 0 & if \quad t > t_{stop} \end{cases} \tag{2.8.10}$$

where $t_{start}, t_{stop}$ are model inputs representing the start time and stop time of tilting; $k_{tt}$ is a parameter related to the slope of the orthostatic response; $g_{tt,max}$ is the maximal value.

Upon standing from sitting, heart rate promptly rises until saturation is reached, then rapidly decreases to resting values even when standing is maintained. Such a behavior can be modeled using an exponential decay function $g_{ss}$ as follows.

$$g_{ss} = \begin{cases} 0 & if \quad t < t_{start} \\ g_{ss,max} \cdot e^{-k_{ss}\cdot(t-t_{start})} & if \quad t_{start} \le t \le t_{stop} \\ 0 & if \quad t > t_{stop} \end{cases} \tag{2.8.11}$$

where $t_{start}, t_{stop}$ are model inputs representing the time when subject starts standing and time when he stands; $k_{ss}$ is a model parameter related to the slope of the exponential decay function; $g_{ss,max}$ is the maximal value.

We rewrite equation (A1.20) from Appendix A1 in order to include the effect of gravitational forces on venous return to right atrium as follows.

$$\frac{dP_{ra}}{dt} = \frac{1}{C_{ra}} \cdot \left( \frac{P_{hv} - P_{ra}}{R_{hv}} + \frac{P_{bv} - P_{ra}}{R_{bv}} + \frac{P_{mv} - P_{ra}}{R_{mv}} + \frac{P_{sv} - P_{ra}}{R_{sv}} + \frac{P_{ev} - P_{ra}}{R_{ev}} - F_{i,r} - g_{tt} - g_{ss} \right) \tag{2.8.12}$$

Parameter values for gravitational effects are given in Table 2.5.19. Simulation of the above equations is given in Figure 2.32 (head-up tilt) and Figure 2.33 (sit-to-stand).

Table 2.5.19 Parameter values for gravitational effects

| **Parameter** | **Value** | **Description** |
|---|---|---|
| $g_{ss,max}$ | 2000 mL | Maximal value of effects due to gravitational forces during active standing from sitting |
| $k_{ss}$ | 0.051 $s^{-1}$ | Slope of the heart rate response during active standing from sitting |
| $g_{tt,max}$ | 1000 mL | Maximal value of effects due to gravitational forces during head-up tilt |
| $k_{tt}$ | 0.2 $s^{-1}$ | Slope of the heart rate response during head-up tilt |

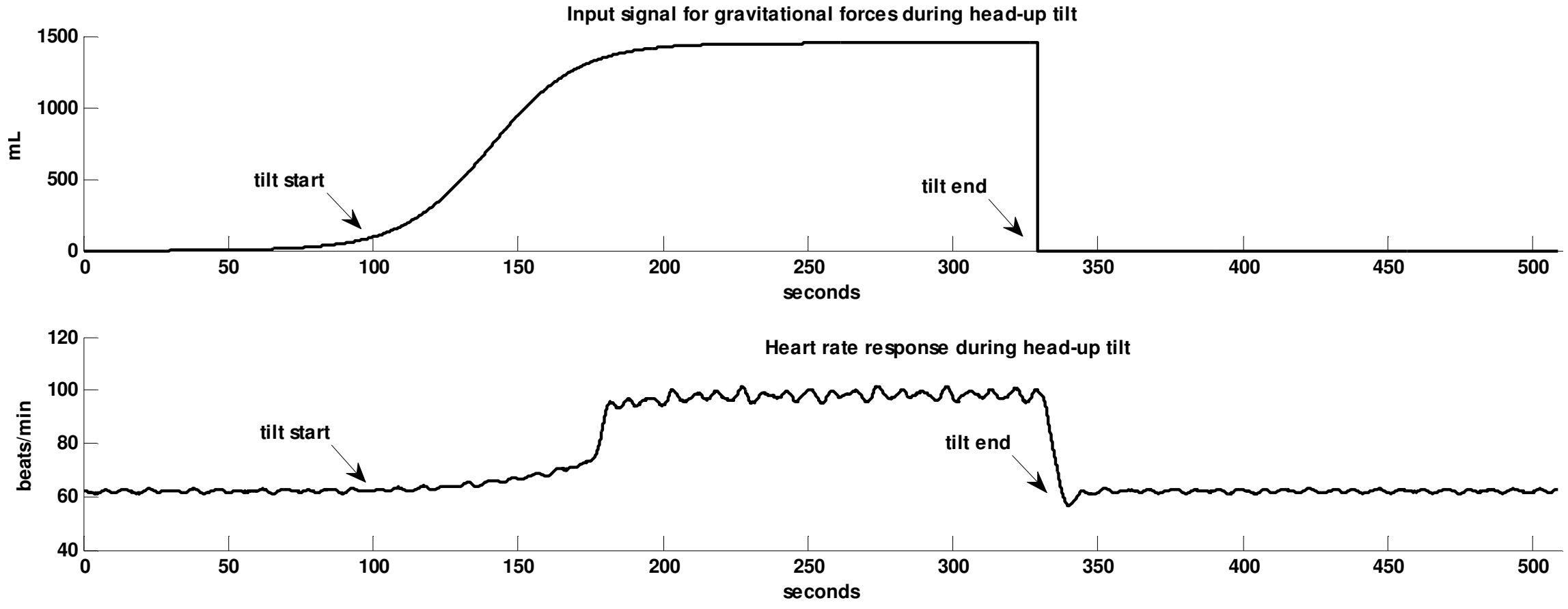

**Figure 2.32 Simulation of heart rate response during head-up tilt**

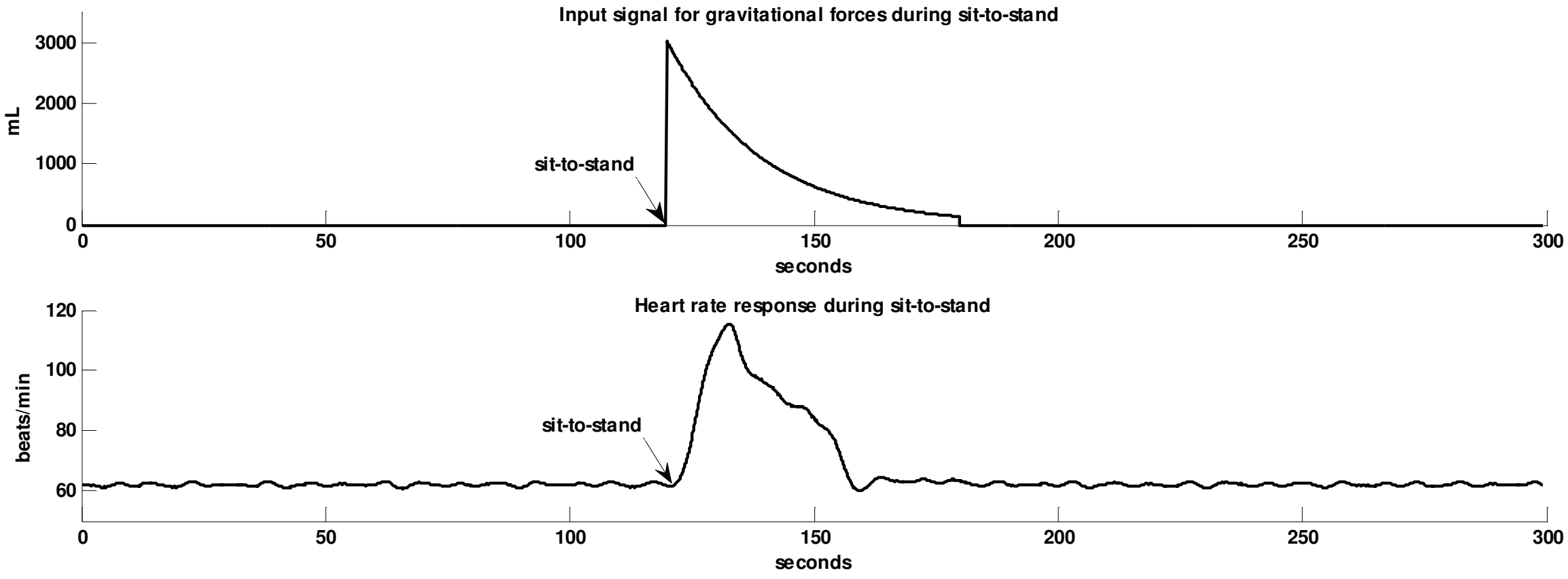

**Figure 2.33 Simulation of heart rate response during sit-to-stand**

### 2.8.6 Effects of Intra-thoracic Pressure Elevation

An abrupt transient increase of intra-thoracic and intra-abdominal pressure provoked by blowing into a resistance of approx. 40 mmHg for about 15 seconds induces complex hemodynamic changes in the cardiovascular system which can be summarized into four phases [100] as modeled in the next paragraphs.

During the Phase 1, the gradual increase in intra-thoracic and intra-abdominal pressure will cause aortic and peripheral vessels compression. This results in increased peripheral resistance. We model this slight increase using a Gaussian function $P_1$:

$$P_1 = P_{1amp} \cdot e^{-\frac{(t - t_{start} - P_{1dur}/8)^2}{2 \cdot (P_{1dur}/8)^2}} \qquad (2.8.13)$$

where $P_{1amp}$ is the amplitude of pressure increase in mmHg; $t_{start}$ is the timestamp in seconds when the expiration starts; $P_{1dur}$ is the duration of the Phase 1 in seconds.

Therefore equation (2.8.4) can be rewritten as:

$$R_{ip} = \frac{\Delta R_{ip}}{1 + M_{ip}} + P_1 \qquad (2.8.14)$$
$$i = v, b, m, s, e$$

where $R_{ip}$ are the resistances in vascular beds.

The sustained high intra-thoracic and intra-abdominal pressure is modeled using a logistic function $P_s$:

$$P_s = \begin{cases} \dfrac{Ps_{max} \cdot 10^{-1} \cdot e^{dP_{rate} \cdot (t - t_{start})}}{Ps_{max} + 10^{-1} \cdot \left(e^{dP_{rate} \cdot (t - t_{start})} + 1\right)} & if \quad t_{start} \le t \le t_{stop} \\ 0 & if \quad t > t_{stop} \end{cases} \qquad (2.8.15)$$

where $Ps_{max}$ is the maximum intra-thoracic pressure with constant value 40 mmHg, $dP_{rate}$ is the growing rate of pressure development at the beginning of the expiration and $t_{stop}$ is the timestamp in seconds when the forced expiration is released, usually 15 seconds after $t_{start}$.

We assume that this intra-thoracic pressure has a linear effect on the pressure in the venous compartments, thus we can rewrite the equations of venous pressures (from Appendix A1) as follows:

$$\begin{aligned} &P_{iv} = P_{iv,0} + k_{ev} \cdot P_s \\ &i = v, b, m, s, e \end{aligned} \tag{2.8.16}$$

where $k_{ev}$ is a weighting factor and $P_{iv,0}$ is the baseline venous pressure.

During the Phase 2, the sustained intra-thoracic and intra-abdominal pressure hinders venous return. We include the resulting decrease of transmural pressure in the venous extrasplanchnic compartment and model stroke volume changes as follows:

$$V_{str} = k_{str} \cdot \frac{P_{ev} - 2 \cdot k_{ev} \cdot P_s}{P_{as}} \tag{2.8.17}$$

where $k_{str}$ is a weighting factor.

During the Phase 3, the forced expiration is released, resulting in a decrease in venous pressure. We model this marked decrease using a negative Gaussian function $P_3$:

$$P_3 = -P_{3amp} \cdot e^{-\frac{\left(t - t_{start} - P_{1dur} - P_{3dur}/4\right)^2}{2 \cdot \left(P_{3dur}/8\right)^2}} \tag{2.8.18}$$

where $P_{3amp}$ is the amplitude of pressure drop in mmHg, $P_{3dur}$ is the duration of the phase 3 in seconds.

Therefore, equation (2.8.17) can be rewritten as follows.

$$V_{str} = k_{str} \cdot \frac{P_{ev} - 2 \cdot k_{ev} \cdot P_s + P_3}{P_{as}} \tag{2.8.19}$$

During Phase 4, the accumulated venous blood returns to heart and is pumped into the constricted arteries causing an overshoot of stroke volume above the normal level. We can rewrite the above equation and include the overshoot effect, as follows:

$$V_{str} = \begin{cases} k_{str} \cdot \dfrac{P_{ev} - 2 \cdot k_{ev} \cdot P_s + P_3}{P_{as}} & if \quad t_{start} \le t \le t_{stop} \\ k_{str} \cdot \dfrac{P_{ev} - 2 \cdot k_{ev} \cdot P_s + P_3 + P_{4amp}}{P_{as}} & if \quad t_{stop} < t \le t_{stop} + P_{3dur} + P_{4dur} \\ k_{str} \cdot \dfrac{P_{ev} - 2 \cdot k_{ev} \cdot P_s + P_3 + \dfrac{P_{4amp}}{5}}{P_{as}} & if \quad t > t_{stop} + P_{3dur} + P_{4dur} \end{cases} \tag{2.8.20}$$

where $P_{4amp}$ is a model parameter representing the overshoot effect on the stroke volume, $P_{4dur}$ is the duration of phase 4 in seconds.

Parameters values are given in Table 2.5.20. Simulation of the above equations is given in Figure 2.34. During phase 1 a brief increase in arterial blood pressure and decrease in heart rate is observed. In phase 2 sustained high intra-thoracic and intra-abdominal pressure hinders venous return, leading to a rapid decrease in arterial blood pressure and increased heart rate. During phase 3, the forced expiration is released resulting in a sudden decrease in arterial blood pressure. During phase 4, the accumulated venous blood returns to heart and is pumped into the constricted arteries causing an “overshoot” of arterial pressure followed by bradycardia.

Table 2.5.20 Parameters values for intra-thoracic pressure elevation

| Parameter | Value | Description |
|---|---|---|
| $k_{ev}$ | 0.29 | weighting factor of intra-thoracic pressure elevation on venous pressure |
| $k_{str}$ | 0.57 liter | weighting factor of intra-thoracic pressure elevation on stroke volume |
| $P_{1dur}$ | 2.49 s | duration of phase 1 of intra-thoracic pressure elevation |
| $P_{1amp}$ | 1.18 mmHg s $ml^{-1}$ | amplitude of phase 1 of intra-thoracic pressure elevation |
| $P_{3dur}$ | 3.94 s | duration of phase 3 of intra-thoracic pressure elevation |
| $P_{3amp}$ | 2.46 mmHg s $ml^{-1}$ | amplitude of phase 3 of intra-thoracic pressure elevation |
| $P_{4dur}$ | 3.66 sec | duration of phase 4 of intra-thoracic pressure elevation |
| $P_{4amp}$ | 3.06 mmHg s $ml^{-1}$ | amplitude of phase 4 of intra-thoracic pressure elevation |
| $dP_{rate}$ | 2.24 | rate of intra-thoracic pressure elevation |

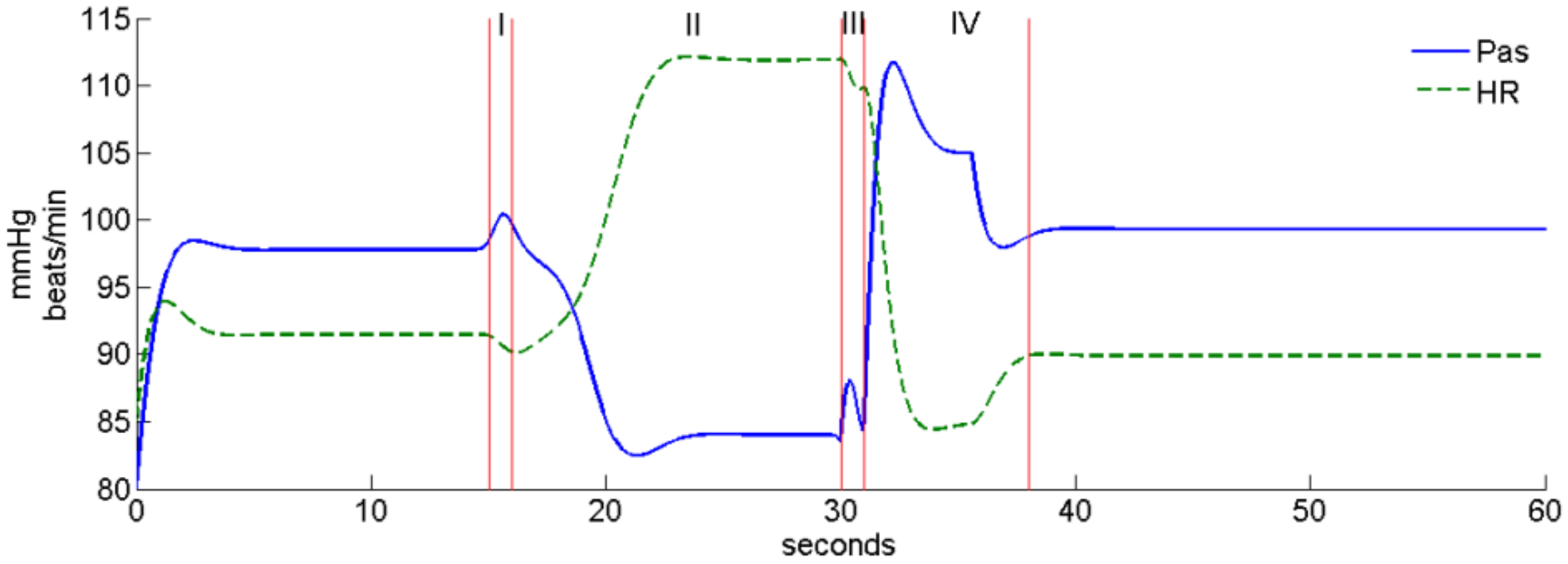

**Figure 2.34 Mean arterial pressure (plain line) and heart rate (dashed line) changes during the VM**

The maneuver covers the timeline from the 15th to the 30th second. Phase 1 goes from 15th to 16th second, phase 2 ends at 30th second, followed by phase 3. The last phase 4 runs from 31st to 38th second.

## 2.9 Summary and Conclusion

In Chapter 1 we presented major limitations of established methods for assessing autonomic nervous activity on the cardiovascular system such as univariate analysis (heart rate variability, blood pressure variability), multivariate analysis (baroreflex sensitivity, transfer functions), biochemical analysis (norepinephrine spillover, muscle sympathetic activity) and model-based analysis (lumped or distributed mathematical models based on control theory). In this Chapter 2 we have proposed a novel integrative approach to mathematical modeling of cardiovascular control and covered single levels of cardiovascular neural pathways from receptors to effectors.

Our major contribution includes novel mathematical models of sympathetic tone, parasympathetic tone, Mayer waves, respiratory sinus arrhythmias, spinal residual activity, vagally-mediated tachycardia and high-frequency intra-cardiac fluctuations. Furthermore we propose unique models of diving reflex, cognitive stress reflex, oculocardiac reflex, reflex reaction to low temperature, reflex reaction to gravitational forces, reflex reaction to intra-thoracic pressure elevation, firing rate of respiratory somatic neurons and concentration of adrenal catecholamine.

An overall overview of the model described in the previous sections is depicted as package diagram in Figure 2.35. The model consists of several negative feedback processes for controlling internal cardiovascular variables such as heart rate, ventricle elastance, arterial blood pressure, and peripheral vascular resistance. The current state of these variables is measured by primary afferent neurons which are lumped into the *Measuring Subsystem*. Their anatomical structure and physiological function were modeled in sections 2.3.1 and 2.4. The *Controller Subsystem* is composed of autonomic premotor neurons as described in sections 2.3.2 and 2.5. The controller subsystem integrates the sensory information and initiates regulatory commands to actuators. The *Actuator Subsystem* consists of preganglionic and postganglionic autonomic neurons as described in sections 2.3.3, 2.3.4, 2.6 and 2.7. They translate the command signals into chemical substances that affect the response of effectors. Cardiovascular effectors build the *Plant Subsystem* including heart and blood vessels. Outputs from effectors are physiological variables which are measured by the measuring subsystem and feedback to the controller, thus starting the loop again. The model also includes feedforward loops especially when the cardiovascular system should face major external disturbances such as cold. Such external disturbances are sensed and feedforward to the controller before they affect the system.

Our physiologically-based model is characterized by input variables (Table 2.9.1), internal state variables (Table 2.9.2), observed output variables (Table 2.9.3) and unknown parameters (Table 2.9.4). The observed output variables can be fit to the measured data in order to provide unique values for the model parameters. Chapter 3 will deal with experimental setups which we have designed in order to gather enough physiological data. Chapter 4 will cover the estimation of values for model parameters and will check the validity of model outputs. The quantitative feature of our model will also be demonstrated and methods for assessing autonomic neural activity of both sympathetic and parasympathetic branches will be developed.

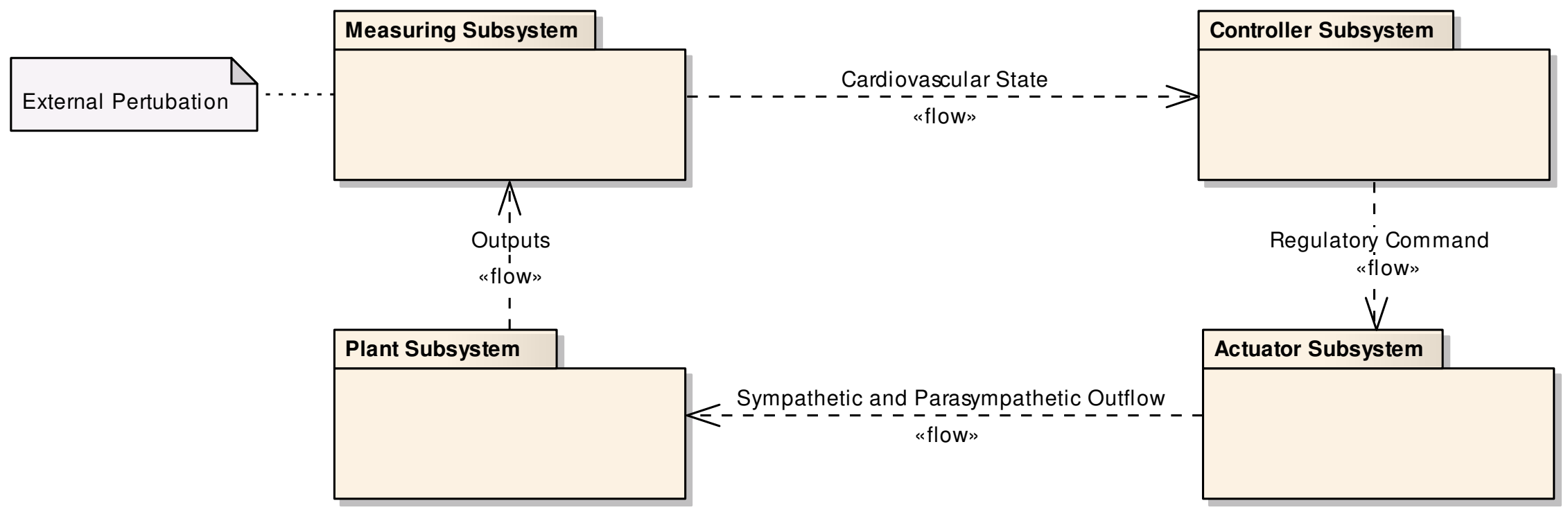

**Figure 2.35 Diagram for Overall Cardiovascular Control**

Table 2.9.1 Model inputs

| **Variable** | **Value Range** | **Description** |
|---|---|---|
| $age$ | [10 .. 100] years | age of the subject |
| $f_{resp,vol}$ | [-1 .. 1] | voluntary respiration pattern |
| $A_{resp}$ | [0 .. 1] | amplitude of the signal representing the strength and depth of quiet respiration |
| $T_{resp}$ | [1 .. 60] s | period of the signal representing the duration of a quiet respiration cycle |
| $t_{start}, t_{stop}$ | [1 .. 300] s | start and end timestamps for external stimulation (e.g. cold stimuli applied to the skin, mental stress, sit to stand, head-up tilt) |
| $I$ | [0 .. 1] | exercise intensity |
| $QT_n$ | [100 .. 500] ms | baseline QT interval duration |

Table 2.9.2 Selected state variables

| Variable | Value Range | Description |
|---|---|---|
| $f_{ab}$ | [0 .. 120] Hz | firing rate of baroreceptor |
| $f_{ac}$ | [1.16 .. 17.07] Hz | firing rate of chemoreceptor |
| $f_{ap}$ | [0 .. 15] Hz | firing rate of pulmonary stretch receptor |
| $f_{aa}$ | [1 .. 18] Hz | firing rate of atria stretch receptor |
| $f_{at_f}$ | [5 .. 50] Hz | firing rate of trigeminal cold receptor |
| $f_{at_s}$ | [5 .. 50] Hz | firing rate of  cold receptor in hands/feet |
| $f_{ao}$ | [5 .. 50] Hz | firing rate of oculo-pressure receptor |
| $f_{as}$ | [5 .. 20] Hz | firing rate of mental stress sensors |
| $f_{es,0,sn}$ | [16.11 .. 20] Hz | sympathetic tone to heart |
| $f_{es,0,p}$ | [16.11 .. 25] Hz | sympathetic tone to vessels |
| $f_{ev,0}$ | [1 .. 3.2] Hz | vagal tone to heart |
| $f_{resp}$ | [0 .. 2] Hz | firing rate of inspiratory and expiratory somatic motor neurons |
| $f_{sn_{pm}}$ | [2 .. 60] Hz | firing rate of sympathetic premotor neurons controlling SA node and AV node |
| $f_{si_{pm}}$ | [2 .. 60] Hz | firing rate of sympathetic premotor neurons controlling vascular beds (with $i$ equals $v$ for ventricles, $b$ for brain, $s$ for splanchnic and $e$ for extrasplanchnic) |
| $f_{ev_{pm}}$ | [2 .. 20] Hz | firing rate of parasympathetic autonomic premotor neurons |
| $R_{ip}$ | [1 .. 50] mmHg s $mL^{-1}$ | peripheral resistance in vascular beds (with $i$ equals $v$ for ventricles, $b$ for brain, $s$ for splanchnic and $e$ for extrasplanchnic) |
| $E$ | [2 .. 10] mmHg $mL^{-1}$ | ventricle elastance |
| $P_{as_{O_2}}$ | [70 .. 110] mmHg | oxygen partial pressure in arteries |
| $f_{resp}$ | [-1 .. 1] | respiration signal |
| $f_{resp_{rate}}$ | [0 .. 60] cycles/min | respiration rate |
| $V_T$ | [0.583 .. 0.75] L | tidal volume |

Table 2.9.3 Observed variables

| Variable | Value Range | Description |
|---|---|---|
| $HR$ | [20 .. 200] beats/min | heart rate |
| $P_{as}$ | [40 .. 200] mmHg | arterial blood pressure |
| $QT$ | [200 .. 500] s | duration of QT intervals |
| $G_{mp}$ | [1 .. 50] µSiemens | level of galvanic skin response in skeletal muscles |

Table 2.9.4 Model parameters

| Parameter | Value | Range | Description |
|---|---|---|---|
| ***Characteristics of sympathetic tone and parasympathetic tone*** | | | |
| $f_{es,0,high}$ | 10 Hz | [2 .. 12] | firing rate of pacemaker sympathetic premotor neurons |
| $W_{c,es,0}$ | 3 | [1 .. 5] | synaptic weight applied to sensory inputs from chemoreceptors |
| $W_{es,0,low}$ | 2.1 | [1 .. 5] | synaptic weight applied to low frequency oscillators in sympathetic premotor neurons |
| $r_{es,0,low}$ | 0.05 | [0.01 .. 0.1] | parameters determining the rate of low frequency oscillators in sympathetic premotor neurons |
| $k_{es,0,low}$ | 3 | [1 .. 5] | |
| $f_{ev,0,max}$ | 3.2 Hz | [1 .. 20] | maximal intrinsic firing rate of vagal nerves |
| ***Characteristics of respiratory sinus arrhythmias & Inspiratory and expiratory somatic motor neurons*** | | | |
| $W_{resp,v}$ | 0.105 Hz | [0.1 .. 5] | synaptic weight applied to inputs from inspiratory-expiratory somatic motor neurons |
| $f_{rsa}$ | 0.15 Hz | [0.1 .. 1] | frequency of intracardiac high frequency fluctuations |
| $G_{T_{rsa}}$ | 0.01 | [0.01 .. 0.1] | gain of intracardiac high frequency fluctuations |
| ***Characteristics of heart period and QT interval duration*** | | | |
| $k_{nor_{og},v}$ | 60 | [0 .. 100] | modulation factor of vagally-mediated tachycardia |
| $k_{QT}$ | 10 $mmHg^{-1}$ mL | [10 .. 20] | gain of elastance on QT interval duration |
| ***Characteristics of peripheral vascular resistance*** | | | |
| $k_{G_{mp}}$ | 12 µS $mmHg^{-1}$ $s^{-1}$ ml | [1 .. 20] | gain parameter applied to peripheral vascular resistance in skeletal muscles |
| ***Characteristics of cold receptors*** | | | |
| $f_{at_f,max}$ | 50 Hz | [0 .. 100] | upper saturation of facial cold receptor |
| $k_{at_f}$ | 0.2 $s^{-1}$ | [0.01 .. 1] | slope of the exponential decay of cold receptor on face |
| $f_{at_s,max}$ | 72 Hz | [0 .. 100] | upper saturation of cold receptor on hands and feet |
| $k_{at_s}$ | 1.03 $s^{-1}$ | [0.01 .. 5] | slope of the sigmoid response of cold receptor on hands and feet |
| ***Characteristics of mental stress sensors*** | | | |
| $f_{as,max}$ | 20 Hz | [1 .. 50] | upper saturation of mental stress sensors |
| $k_{as}$ | 0.2 $s^{-1}$ | [0.1 .. 1] | slope of the mental stress sensor response |
| ***Characteristics of gravitational forces during orthostatic stress*** | | | |
| $g_{ss,max}$ | 2000 mL | [500 .. 5000] | maximal value of effects due to gravitational forces during active standing from sitting |
| $k_{ss}$ | 0.051 $s^{-1}$ | [0.01 .. 1] | slope of the heart rate response during active standing from sitting |
| $g_{tt,max}$ | 500 mL | [100 .. 2000] | maximal value of effects due to gravitational forces during head-up tilt |
| $k_{tt}$ | 0.2 $s^{-1}$ | [0.01 .. 2] | slope of the heart rate response during head-up tilt |
| ***Characteristics of oculo-pressure receptors*** | | | |
| $f_{ao,max}$ | 17 Hz | [0 .. 50] | upper saturation of oculo-pressure receptor |
| $k_{ao}$ | 0.7 $s^{-1}$ | [0.1 .. 2] | slope of the sigmoid response of oculo-pressure receptor |

| Characteristics of intra-thoracic pressure elevation | | | |
|---|---|---|---|
| $k_{ev}$ | 0.29 | [0.1 .. 1] | weighting factor of intra-thoracic pressure elevation on venous pressure |
| $k_{str}$ | 0.57 L | [0.1 .. 1] | weighting factor of intra-thoracic pressure elevation on stroke volume |
| $P_{1dur}$ | 2.49 s | [2 .. 5] | duration of phase 1 of intra-thoracic pressure elevation |
| $P_{1amp}$ | 1.18 mmHg s $ml^{-1}$ | [1 .. 5] | amplitude of phase 1 of intra-thoracic pressure elevation |
| $P_{3dur}$ | 3.94 s | [2 .. 5] | duration of phase 3 of intra-thoracic pressure elevation |
| $P_{3amp}$ | 2.46 mmHg s $ml^{-1}$ | [2 .. 5] | amplitude of phase 3 of intra-thoracic pressure elevation |
| $P_{4dur}$ | 3.66 sec | [2 .. 5] | duration of phase 4 of intra-thoracic pressure elevation |
| $P_{4amp}$ | 3.06 mmHg s $ml^{-1}$ | [3 .. 5] | amplitude of phase 4 of intra-thoracic pressure elevation |
| $dP_{rate}$ | 2.24 | [0.5 .. 5] | rate of intra-thoracic pressure elevation |

# Chapter 3. Material and Methods for Experiments

In the previous chapter we used prior knowledge of physiological processes running in the body and built an integrative model of cardiovascular control. The model includes a set of unknown parameters with physiological meaning. The goal of this chapter is to define experimental setups for qualitative data collection which will be further used to estimate values of those unknown parameters.

There are an important number of patients exhibiting cardiac autonomic disorders. Several studies could link sudden cardiac death and the autonomic nervous system. In the clinical practice a method to investigate cardiac autonomic tone is to perform provocative tests known as cardiac autonomic tests, which are simple, low-cost, non-invasive and can be performed in clinical routine. They trigger cardiovascular reflex arcs with responses, which can be measured using non-invasive techniques (electrocardiography, impedance cardiography, blood pressure, galvanic skin response ...) that can give diagnostic information about cardiac autonomic disorders.

We selected the mental arithmetic test, the Valsalva Maneuver, slow deep breathing, cold face test, cold pressor test, tilt table test, active change of posture and oculocardiac reflex test. We also tried various other experiments such as breathe holding, yawning, aerobic exercise, sustained handgrip, which were finally not included into our framework because of technical and time constraints.

We first experimented on horses in order to verify hypothesis on a denser parasympathetic innervations of equine ventricle causing sudden cardiac death during surgery [101]. Equine experiments accounted only for a small portion of our trials. The majority of experiments were performed on human subjects. Additional species were not included in our research because of technical and administrative boundaries.

This chapter describes all cardiac autonomic tests we have performed; including their physiological background, indications, contraindications, the entry conditions which should be fulfilled before the subject is allowed to take the test, the instrumentation, the activities flow performed during the test, the exceptions which might cause test interruption and examples of normal cardiac signals plots.

## 3.1 Generalities

### 3.1.1 Human Subjects

Experiments were carried out on 72 subjects (ages from 11 to 82 years, mean 36±17 years) in accordance with the Declaration of Helsinki and informal consent was obtained from each participant. Our studies included 25 black women, 17 white women and 13 black men and 16 white men in total (see Figure 3.1). Among those subjects 35 were living in Bafoussam, 1 in Yaoundé, 1 in Douala, Cameroon; 16 live in Kladno, 7 in Prague, 5 in Liberec, Czech Republic; 1 was living in Brussels, Belgium and 1 in Grenoble, France. Details about subjects are given in the Appendix A2.

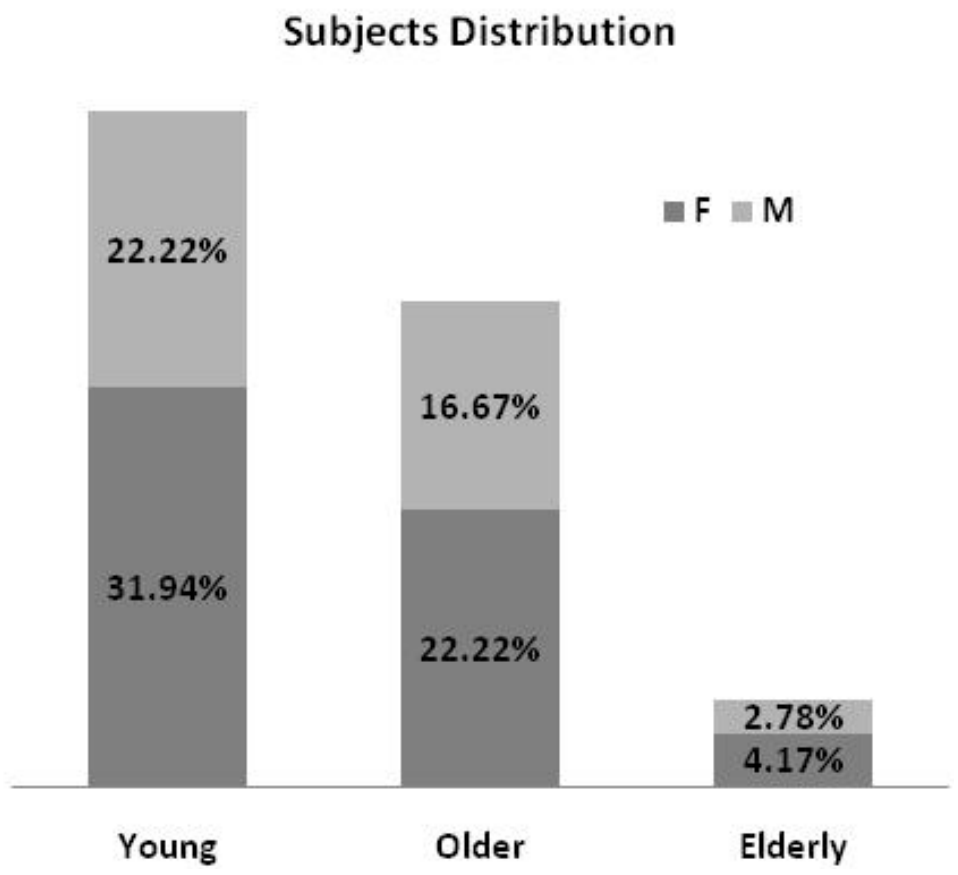

**Figure 3.1 Subjects by age and gender**
Young (less than 29 years); Older (30 to 59 years); Elderly (over 60 years)

3 subjects (n = 8, 10, 26) were presenting palpitation at the moment of the experiment, 2 subjects (n = 8, 20) were suffering of cardiac insufficiency, 4 subjects (n = 16, 20, 45, 55) had hypertension and 1 had aortic stenosis. Some subjects additionally presented non-cardiovascular diseases such as hepatitis B (n = 5), gastric ulcer (n = 8, 10), sleep apnea (n = 38) and diabetes mellitus (n = 41).

Three different experimental sessions were realized in collaboration with the Department of Biomedical Informatics at the Czech Technical University in Prague on February 25, 2009, April 28, 2009 and May 20, 2009. 7 women (ages from 25 to 77 years, mean 42±21 years) and 7 men (ages from 25 to 82 years, mean 48±23 years) underwent the cold face test (CFT) and cold pressor test (CPT) in the first session. The tilt table test (TTT), valsalva maneuver (VM), cold pressor test and oculocardiac reflex test (OCT) were performed on 7 women (ages from 20 to 24 years, mean 21±1 years) and 7 men (ages from 22 to 29 years, mean 24±3 years). The third session consisted of performing the tilt table test, valsalva maneuver and oculocardiac reflex test on 4 women (ages from 15 to 46 years, mean 26±14 years) and 2 men (age 23 years).

Two additional experimental sessions were carried in collaboration with the Department of Computer Science at the University of Yaounde I in Cameroon and District Hospital of Dschang in Cameroon on January 27 and January 28, 2010 where subjects underwent the cold face test, deep breath test (DBT), mental stress test (MST), valsalva maneuver, cold pressor test, active standing and oculocardiac reflex test. The first session included 26 women (ages from 11 to 70 years, mean 37±16 years) and 12 men (ages from 23 to 56 years, mean 47±11 years). The second session included 2 women (ages 17 and 48 years) and 1 man (age 56 years).

### 3.1.2 Instrumentation

Diverse types of instrumentation have been reported in the literature for cardiovascular signal acquisition. Subjects in [102] were instrumented with a Finapres Ohmeda 300 instrument for beat to beat heart rate and blood pressure measurements. Systolic and diastolic blood pressure was continuously measured in [103] from the left radial artery with noninvasive arterial tonometry (Colin Pilot; Colin Medical Instruments). The tonometer consisted of an array of 31 equally spaced piezoresistive pressure transducers, an automated positioning system, signal conditioning and initial calibration by oscillometric cuff measurement of brachial artery blood pressure. Skin blood flow was monitored at the first toe pulp with a Periflux laser Doppler (Perimed). Respiratory frequency was monitored with a two-belt chest-abdomen inductance plethysmograph after calibration (Respitrace calibrator). A Biopac MP35 measurement system was used in [104]. Authors of [105] additionally measured the cardiac output using a Doppler electrocardiographic device (Hewlett-Packard model 77020AC).

We have chosen Biopac MP150 and MP35 measurement systems in our studies (BIOPAC Systems, Inc., Goleta, Figure 3.2). A 3 leads ECG was recorded using Biopac SS2L electrodes placed on the left and right wrists, and on the left part of the abdomen just above the belt (see Figure 3.3). Pulse was measured using a SS4LA Pulse Plethysmograph Transducer placed on the index finger of the right hand. Two Biopac electrodes SS57L were fixed around the index (ground) and middle finger (+) of the left hand for measurement of galvanic skin response. The body skin temperature was measured using a SS6L Temperature Transducer placed closed to the skin. Respiration rate was measured using a Biopac SS5LB transducer to record chest expansion and contraction. The sensors are depicted in Figure 3.4 and Figure 3.5.

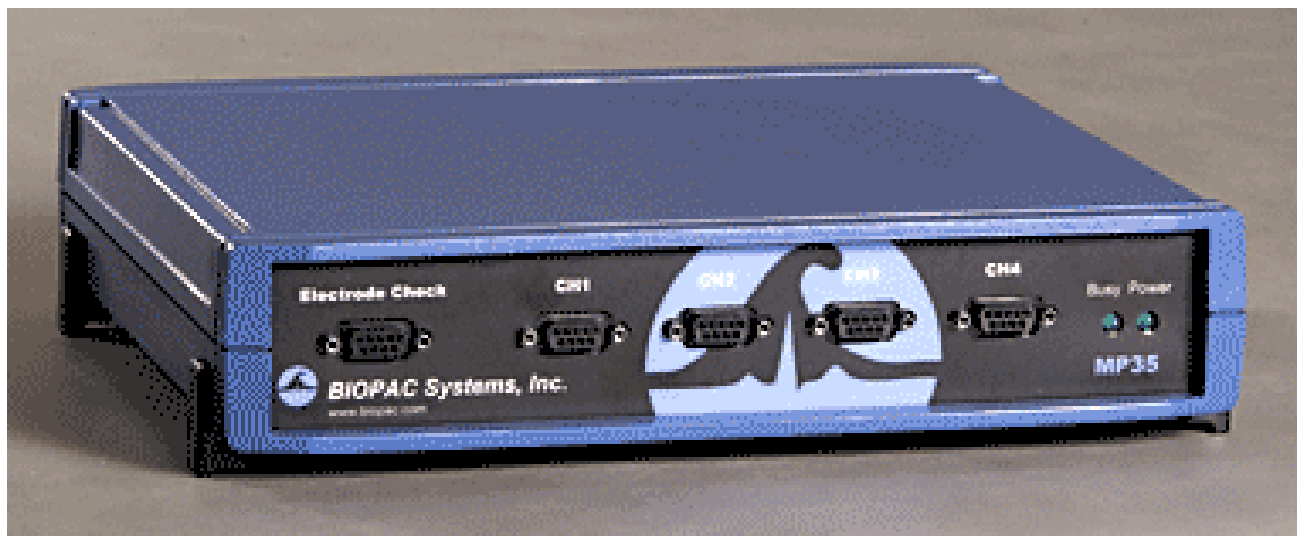

**Figure 3.2 Biopac MP35 Measurement System (reprinted from [106])**

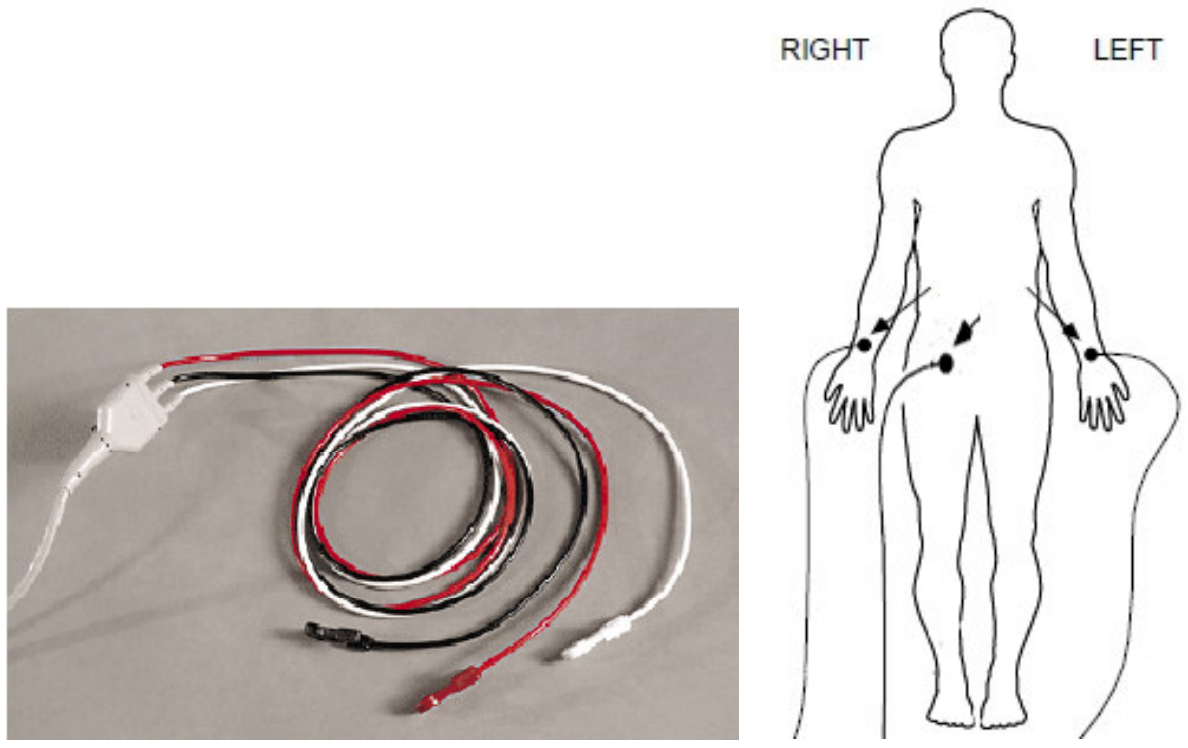

**Figure 3.3 Biopac SS2L Electrodes for ECG measurement (reprinted from [106])**

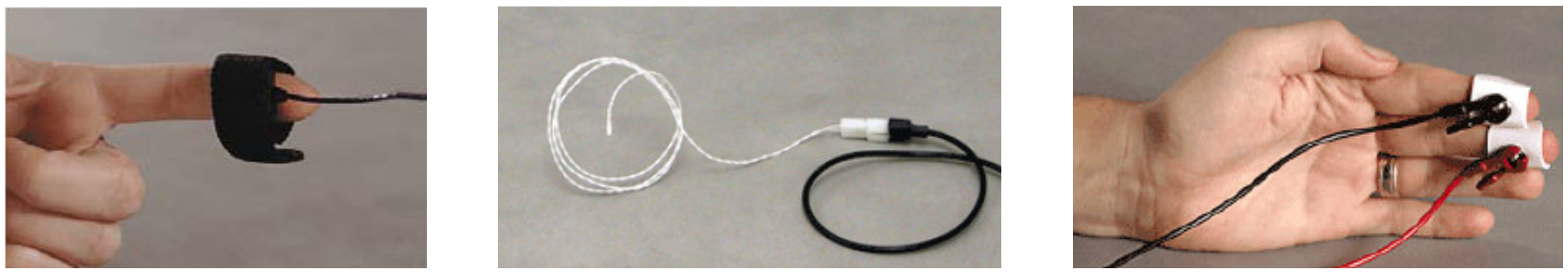

**Figure 3.4 Biopac Sensors (reprinted from [106])**
Pulse measurement (left); skin temperature measurement (center); galvanic skin response measurement (right)

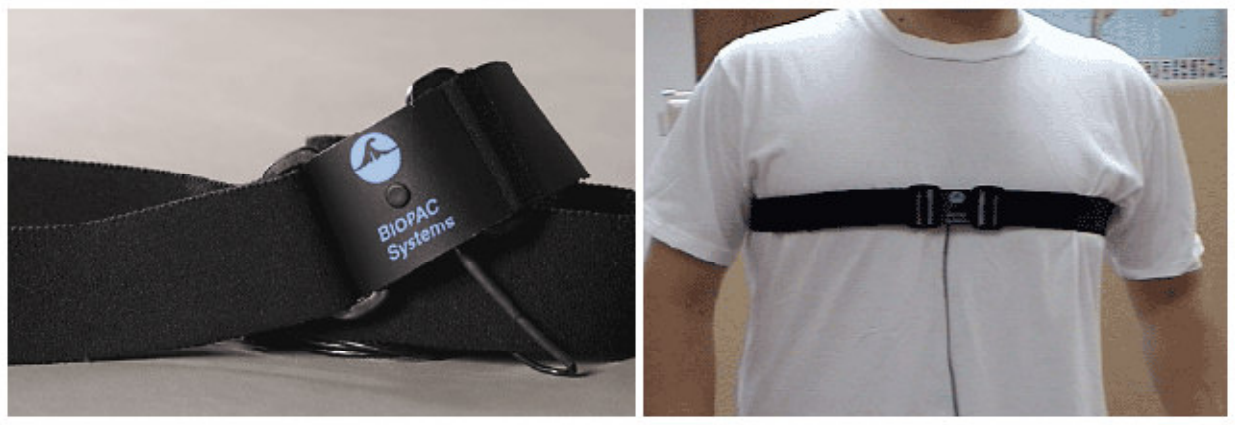

**Figure 3.5 Respiration signal measurement (reprinted from [106])**

The Biopac MP150 or MP35 system was connected to a PC where the Biopac Student Lab PRO software was installed for continuous signal acquisition at a frequency of 500 Hz. A Biopac acquisition file was created with the following settings: the acquisition was set up for *Recording* and *Append into Memory* at a sampling rate of *500 Hz*; the total acquisition length was set to 2 hours; analog channel CH1 had the preset *ECG (.5 - 35 Hz)*; analog channel CH2 had the preset *PPG (Pulse) with* a 50 Hz low pass filter; analog channel CH3 had the preset *EDA (GSR) (0 - 35 Hz)* with a 2 Hz low pass filter; analog channel CH4 had the preset *Respiration (SS5LB)*; calculation channel C1 had the preset *R-R Interval*; calculation channel C2 had the preset *Heart Rate;* calculation channel C3 had the preset *Pulse Rate;* calculation channel C4 had the preset *Respiration Rate.*

Additionally to the Biopac System, a blood pressure device, model OMRON M10-IT, was used for collecting arterial blood pressure samples which were further used to calibrate the pulse waveform. The device was later connected to a PC and the Omron HealthCare Management Software was used to download measured data. Biopac and OMRON systems were synchronized based on the same time.

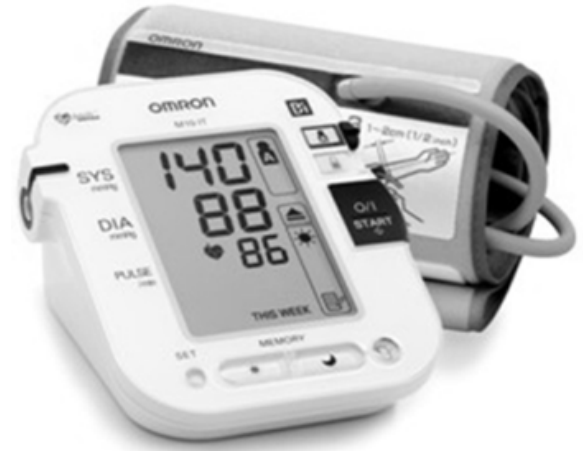

**Figure 3.6 OMRON M10-IT device for cuff blood pressure measurement**
Image reprinted from user manual, http://www.omron.com/, visited June 2010

### 3.1.3 Experimental Protocol

Drugs (blockers) that interfere with parasympathetic or sympathetic transmission were contraindicated before and during the test. Alcohol and smoking was not allowed 24 hours prior to the test. Physical exercise was avoided 24 hours prior to the test in order to avoid long term changes in heart rate variability. Most tests were performed according to following protocol:

***1. Instruction***

The subject was shortly informed about the physiology of the test. He was instructed in details about the activities flow which will be performed. He was asked to report about the degree of discomfort or pain during the test by moving the small finger of his right hand. The faster he moves the finger, the more pain or discomfort he has and the test should be interrupted. The subject was instructed not to talk during the whole measurement.

***2. Demonstration***

The test was first performed on an unwired control person at the presence of the subject, so that he could learn how to behave during the test and be confident on what will be done on him.

***3. Instrumentation***

Subject was wired with a Biopac MP35 measurement system for recording 3 leads ECG, arterial pulse, galvanic skin response, skin temperature and breathing rate signals at 500 Hz. The subject was then stabilized in a supine position on a comfortable table or bed; or in a sitting position on a comfortable chair according to the type of test.

***4. Baseline Measurement***

Baseline measurements were performed using the Biopac Student Lab PRO software until all signals were stable, i.e. until the heart rate did not vary greatly, no tachycardia was present, successive pulse values had a similar amplitude, the galvanic skin response was constant and the breathing frequency was regular. During baseline measurements blood pressure samples were taken using the OMRON device with automatic cuff inflation. Systolic and diastolic values were added to the protocol.

***5. Preparatory Measurement***

A preparatory period of typically 30 seconds was included in the protocol. Recording performed during this period were not part of baseline recordings because during this period the subject was usually influenced by emotional factor induced for example by a cold compress approaching the face during a cold face test.

***6. Measurement during Autonomic Test***

After cardiac signals have stabilized, the external stimulus was applied according to the procedures described in next sections (e.g. cold compress on the face, pressure on eye lids, deep breathing, mental exercise or forced expiration). The current timestamp displayed on the Biopac Student Lab PRO timeline was added to the protocol as starting time of the external stimulus. The subject was observed for pain or discomfort sensation and the test was interrupted if he displayed great discomfort. Otherwise the external stimulus was maintained for the predefined maximal duration depending on the test type (e.g. 15 s for the valsalva maneuver, 180 s for the cold pressor test).

***7. Post Measurement***

After the external stimulus was removed, the subject was instructed to breathe normally and to relax. The current timestamp displayed on the Biopac Student Lab PRO timeline was added to the protocol as end time of the external stimulus. Recordings were resumed until all cardiac signals were stabilized.

***8. Diagnostic Measurement***

After the first autonomic test the subject felt acclimatized with the procedure and the real diagnostic assessment was performed. The *Measurement during Autonomic Test* and *Post Measurement* activities were strictly repeated once more.

***9. Release and Post Evaluation***

After the diagnostic measurement electrodes were released from the subject and he was asked to rate the pain and discomfort factor on a scale of 1 to 5, 1 being full comfort and 5 being strong pain or intense discomfort. He could also report on his feeling during the test, for example if headache, fainting or nausea were present. His feeling after the test at the moment of post evaluation was also included in the protocol.

### 3.1.4 Extracting ECG Parameters

Electrocardiogram (ECG) is used to record the electrical activity in the heart. An electrocardiogram consists of waves (P, Q, R, S, and T), segments between the waves (P-R, S-T) and intervals consisting of a combination of waves and segments (PR, QRS, QT and TP). The first wave is the P wave, which corresponds to the depolarization of the atria. The atria repolarization is too small to be seen but appears as distortion of the PR segment or is incorporated into the QRS complex. The bundle of His and bundle branches are activated during the P-R segment. The PR interval measures the time between atria and ventricle activation. It can reflect vagal control on atrioventricular node. PR interval duration longer than 200 milliseconds suggests conduction impairment. The next trio, the QRS complex, represents the progressive wave of ventricular depolarization. The QRS interval duration is typically 100 milliseconds. The ST segment corresponds to the time between first heart sound (peak of ventricle contraction and opening of aortic valve) and ventricle ejection. The QT interval corresponds to the duration between the onset of ventricle excitation and relaxation. It has a value ranging from 250 to 500 milliseconds. The final T wave represents the repolarization of the ventricles. The amplitude of the T wave has been proposed as measured of sympathetic activity. An illustrative 12-leads ECG signal is displayed in Figure 3.7.

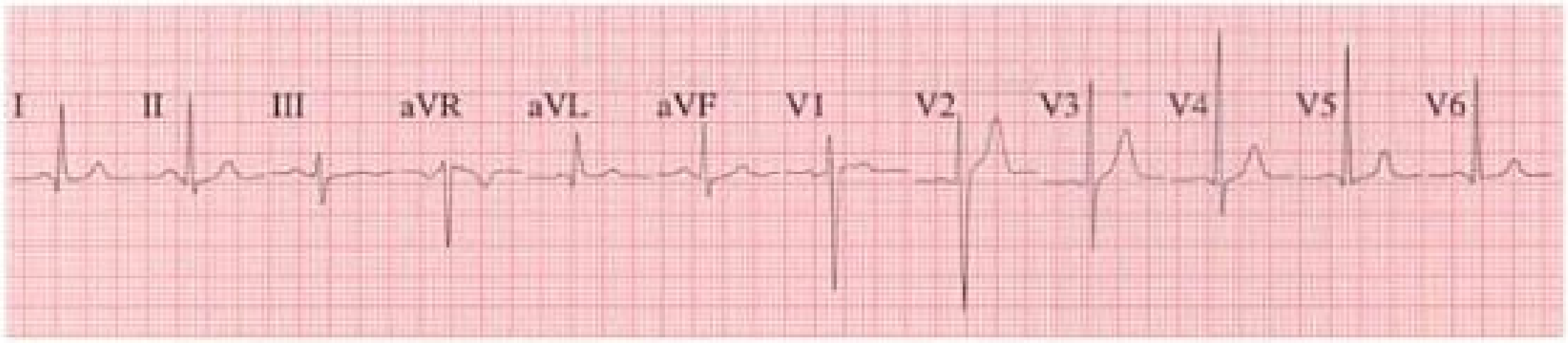

**Figure 3.7 A 12-leads Electrocardiogram (reprinted from [107])**

On a normal ECG signal, the complex QRS has elevated amplitude in relation to the rest of the signal and is therefore usually easy to detect with the help of an algorithm searching for R peaks. Common algorithms store the temporal indication of every R peak in a temporary memory in order to calculate the duration of consecutive R peaks, called RR interval or normal-to-normal (NN) interval. We investigated two algorithms for R peaks detection in an ECG signal because of their relatively easy implementation in our processing environment: non-syntactic QRS Detection [108] and nonlinear high-pass filter R-wave detection [109]. Both algorithms use standards methods of detection by thresholding on the original ECG, or after application of a highpass linear filter, or the methods of detection by derivation and thresholding. They are enough precise in the simplest cases.

The ECG signals were sampled at 500 Hz using the Biopac Student Lab PRO acquisition software. They were exported in time series ASCII files for further processing in Matlab. R-peaks were detected from the ECG signal using an implementation of the non-syntactic QRS detection algorithm offered by the open ANSLAB (Autonomic Nervous System Laboratory) library [110]. For this purpose the ECG signal was sampled down to 400 Hz and filtered to get rid of shifting baseline (baseline wandering noise), muscle artifacts and power line interference using a Butterworth band pass filter. NN intervals durations were calculated as differences of timestamps of consecutive R-peaks in milliseconds.

We developed the following algorithm for automatically removing artifacts from NN intervals series using a smoothing technique. The algorithm plots the series on a graph where the x-axis is the time and the y-axis is the NN interval duration; then calculates a straight line (first order polynomial) that best fits the series and determines the baseline heart rate, i.e. the y-point where the line cuts the y-axis. Next step is to loop through the NN interval durations, to calculate the mean of previous 10 NN interval durations (previous_hr) and smooth up or down the current value if it exceeds or is less than a certain percentage (smoothup_hr or smoothdown_hr) of the previously calculated baseline value. The algorithm is given in Matlab code below, where hr is the original vector of NN interval durations and hrr is the corresponding smoothed vector. An illustrative example is given in Figure 3.8.

```
smoothup_hr = 1.8;
smoothdown_hr = 0.6;
hrr = hr;
t=t(1:length(hr));
baseline_hr = polyval(polyfit(t', hr, 1),0);
i=2;
while i<length(hrr)
   previous_hr = mean(hrr(max(1,i-10):i-1));
   while hrr(i) < smoothdown_hr*baseline_hr && i<length(hrr)
      hrr(max(1,i-2):min(length(hrr),i+2))=previous_hr;
      i = i+1;
   end
   i=i+1;
end;
i=2;
while i<length(hrr)
   previous_hr = mean(hrr(max(1,i-10):i-1));
   while hrr(i) > smoothup_hr*baseline_hr && i<length(hrr)
      hrr(max(1,i-2):min(length(hrr),i+2))=previous_hr;
      i = i+1;
   end
   i=i+1;
end
```

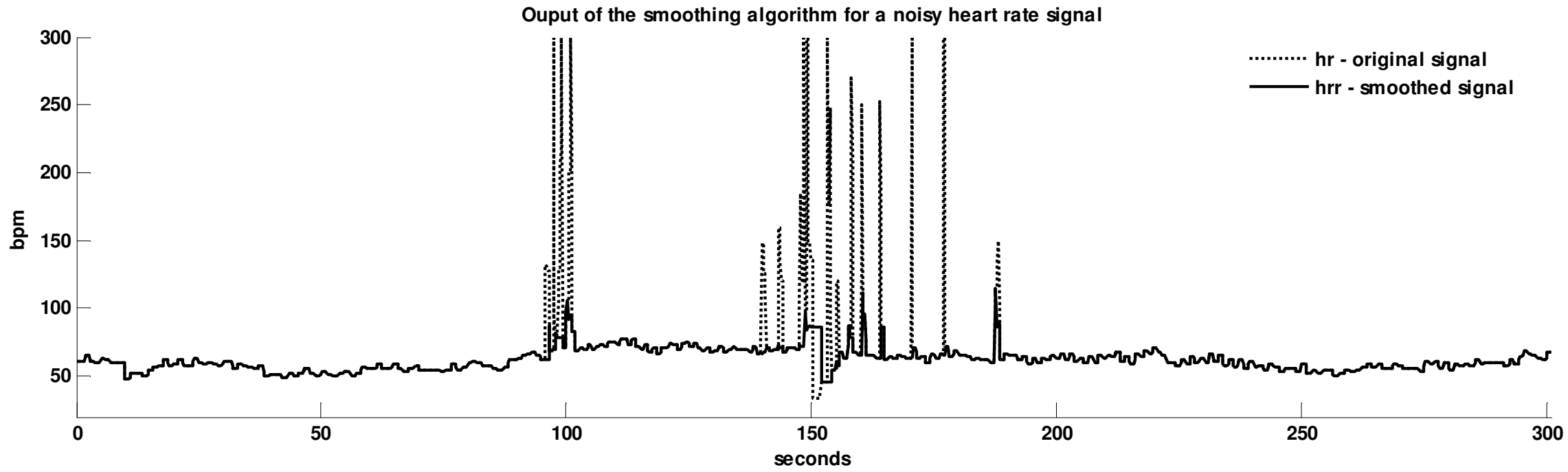

**Figure 3.8 Ouput of the smoothing algorithm for a noisy heart rate signal**
original signal (hr), smoothed signal (hrr)

The smoothed NN time series was manually edited for removing NN values outside physiological range, due to measurement failures such as electrodes displacement. It was then sampled down to 10 Hz and stored in a vector $RR$ (in milliseconds) for further analysis. The series of beat-to-beat heart rate $HR$ (in beats/min) was calculated from the RR intervals as follows:

$$HR = \frac{60000}{RR} \tag{3.1.1}$$

The algorithm for QT interval duration calculation is relatively simple. It first detects the timestamp of a Q point by differentiating the ECG portion shortly before the R peak . After differentiation the maximal slope is computed in this area. The algorithm then goes back to the first flat slope which is the Q point. We used a window of 0.12 second for searching back Q point from R. The next step is to detect the starting time of the T wave using a wave detection algorithm described in [111]. We used the open ANSLAB (Autonomic Nervous System Laboratory) library [110] for processing in Matlab and stored the QT interval durations in a vector $QT$ (unit milliseconds).

### 3.1.5 Estimating Blood Pressure Signal from Pulse Waveform

The arterial blood pressure is the pressure applied on arterial walls as blood traverses the arteries measured in millimeters of mercury (mmHg). The systolic blood pressure $BP_{sys}$ is the highest value reached during systole and diastolic blood pressure $BP_{dias}$ is the lowest value reached during diastole. The pulse pressure $PP$ is the difference between the systolic and diastolic blood pressures and an estimation of mean blood pressure $MAP$ is the time-weighted mean pressure value over one cardiac cycle.

$$\begin{aligned} PP &= BP_{sys} - BP_{dias} \\ MAP &= BP_{dias} + PP/3 \end{aligned} \tag{3.1.2}$$

Blood pressure can be measured noninvasively using palpation, Doppler method, auscultation method, oscillometry, plethysmography or arterial tonometry [80].

The auscultation method consists of placing a cuff on the upper arm and a stethoscope over the brachial artery. The cuff is inflated to a pressure that cuts off blood flow in all arteries until no sound is heard. The cuff is then slowly released until the first so-called Korotkoff sound is heard; the pressure at which this happens is the systolic blood pressure. The cuff is further released and when all sounds disappear, the diastolic blood pressure is obtained. The palpation method is similar to the auscultation method where the stethoscope is replaced by palpation of the artery. The Doppler method is a further modification of the auscultation method where the stethoscope is replaced by a Doppler probe, which detects ultrasounds caused by blood flow. Only systolic blood pressure can be determined by the palpation and Doppler methods. The oscillometry method detects the oscillation caused by blood flow under a cuff.

A tonometry device flattens the artery by applying pressure non-invasively to squeeze the artery against bone using an adjustable air chamber. The applied pressure necessary to maintain the flattened shape is measured by an array of pressure sensors. An algorithm is applied to derive the intraluminal arterial pressure.

Plethysmography method uses a finger plethysmograph to determine the minimum cuff pressure needed to maintain constant finger blood volume. A light-emitting diode and a photoelectric cell detect changes in blood volume and adjust the cuff pressure. The pressure within the transparent inflatable cuff is controlled by an electro-pneumatic system and a PID (Proportional, Integral and Derivator) controller. Any deviation of vascular volume due to changes of intravascular pressures is instantaneously compensated by automatic adjustment of the cuff pressure. This is the principle of FINAPRES and provides a continuous measurement of aortic intravascular blood pressure.

In our studies we had some limitations in measuring blood pressure continuously due to the implied high cost of FINAPRES devices. We rather measured the volume wave form using a Biopac photoplethysmograph as described in section 3.1.2, what gives the relative pulse waveform. The transducer measures the density of blood in the fingertip by shining a light into the finger and measuring how much light gets reflected back. The transducer consists of a matched infrared emitter and photo diode, which transmits changes in blood density, caused by varying blood pressure in finger. A built-in two-stage IR filter and ergonomic housing reduces artifacts from ambient light and subject movement [106]. The signal was sampled at 500 Hz using the Biopac Student Lab PRO acquisition software and was exported in time series ASCII files for further processing in Matlab. The signal was sampled down to 400 Hz, a 20 Hz low-pass filter and 0.30 Hz high-pass filter were applied. The resulting signal was stored in a time series variable $PPG$ (unit millivolts)

Diverse parameters can be extracted from the pulse time series including the beat-to-beat pulse rise time, pulse travel time $PTT$, pulse amplitude $PA$, pulse shape and the variability of each of them [112]. Pulse transit time is the time it takes a pulse wave to travel between two arterial sites. A relatively short $PTT$ is observed with high blood pressure, aging, arteriosclerosis and diabetes mellitus. An estimate of $PTT$ is the interval between the R peak on the ECG and the onset of the corresponding pulse in the finger [113]. The algorithm for calculating $PTT$ simply determines the time of pulse peak as the time when the maximal amplitude is observed between two consecutive R peaks. The difference between that time and the time of R peak is stored for each cardiac cycle in the time series $PTT$ (in milliseconds). The maximal pulse amplitude is stored in the time series $PA$ (Biopac measuring system delivers proportional values of pulse amplitude in millivolts, a transformation to mmHg is performed after calibration as explained in the following paragraph).

Pulse amplitude $PA$ has been suggested as marker of autonomic nervous activity [114]. It is well known that pulse amplitude is related to the underlying mean arterial blood pressure $MAP$. A model of relation between both quantities is proposed in [115] as follows:

$$MAP = a \times e^{b \times PA} + c \times e^{d \times PA} \quad (3.1.3)$$

Using such a model blood pressure can be approximated continuously. Such an approximation does not provide clinically relevant absolute values but is sufficient for our research since we are only interest in the magnitude of change of blood pressure in response to commands from autonomic nervous system. The equation above contains unknown parameters $a, b, c, d$ which can be determined by fitting correct blood pressure values to pulse amplitudes during a calibration phase.

In order to obtain enough data for estimating the four parameters, we performed a separated measurement session for each subject where we measured pulse wave amplitude $PA_{cal}$ and took blood pressure samples at the same time using a digital tonometer device (model OMRON M10-IT) which inflates the cuff automatically and provides ca. 6 samples per minute. The systolic and diastolic values were downloaded from the device using the Omron HealthCare Management Software tool as ASCII file for further processing in Matlab where mean arterial pressure $MAP_{cal}$ was calculated by applying equation (3.1.2). Both $MAP_{cal}$ and $PA_{cal}$ were fit to equation (3.1.3) using Matlab's implementation of the non linear least squares algorithm. The process converged rapidly to unique parameter values for $a, b, c, d$ for each subject. Example values are 104.9 mmHg, -3.8 $mV^{-1}$, 0.2 mmHg and 17.8 $mV^{-1}$ respectively. It was therefore possible to estimate mean arterial pressure $MAP$ from the recorded pulse signal continuously.

### 3.1.6 Processing Respiration Signal

Thoracic contraction and expansion were recorded using a respiration effort transducer. This is useful for determining depth of breath and respiration rate as well as the dispersion index which provides an estimation of the regularity of the breathing activity. The respiration signals were sampled at 500 Hz using the Biopac Student Lab PRO acquisition software. They were exported in time series ASCII files and further processed in Matlab. We used an algorithm based on well-known volume and thorax circumference relationships. The algorithm ideally requires a calibration phase where subjects repeatedly inhale and exhale 750 mL of air in controlled conditions. We did not have materials for controlling respiratory volume or for pacing respiration. Therefore we selected measurements corresponding to quiet regularly breathing and used them for calibration. The algorithm determined calibration weights which were specific to the subject's respiration pattern. Weights were further used to translate measured thoracic circumference into calibrated respiration variables using algorithms offered by the open ANSLAB library [110].

The instantaneous minute respiration rate (in cycles/min) was calculated as the number of respiration cycles per minute. The instantaneous tidal volume (in milliliters) was obtained as volume of lung at the end of inspiration. The instantaneous minute volume (in liter/min) was extracted as the total volume of air flowing into lungs per minute. The instantaneous duty cycle (dimensionless unit) was obtained as ratio of inspiratory duration over total cycle duration. The values were respectively stored in time series $RSR, RST, RSF, RSD$.

In some experimental setups, respiration signal was not recorded because of technical limitations. In such cases we had to estimate the respiration signal from the ECG signal. Pioneer work of Moody and co-workers introduced a signal-processing method which derives respiratory waveforms from ordinary surface electrocardiogram [116]. The so-called ECG-derived respiratory signal (EDR) is calculated from changes in the lead axes of the electrodes on the chest surface using the amplitudes of R-peaks ( $RV$ ). We implemented an algorithm that normalizes the vector of R-peak amplitudes in an interval between -1 and 1 by removing the mean amplitude and dividing the result by the maximal value. Since the R-peak amplitude decreases with inspiration, we switch the signal over the x-axis in order to obtain the ECG-derived respiratory signal. Figure 3.9 shows a Measured Respiration Signal recorded using a respiration effort transducer during deep breathing, the Normalized R-peak

Amplitude signal derived from the ECG and finally the obtained ECG-derived respiration signal, which shows an acceptable match with the measured respiration signal despite of some differences in the amplitude.

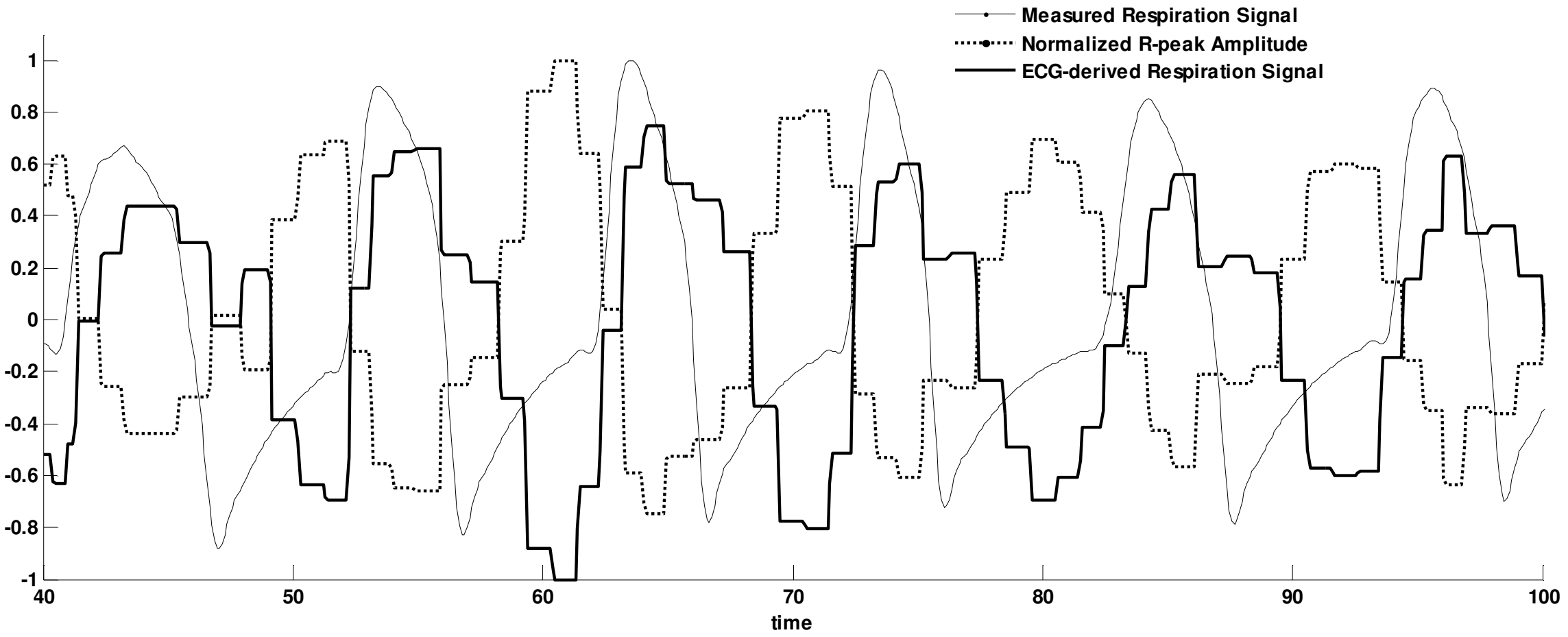

**Figure 3.9 Example of ECG-derived respiration signal**

### 3.1.7 Processing Galvanic Skin Response

A galvanic skin amplifier applies a constant low voltage to the skin through electrodes and measures the current flow. Using Ohm's low the resistance of the underlying conductor can be obtained. The inverse of the resistance is the skin conductance expressed in units called microSiemens (μS). Change in conductance occurs when sweat glands fill or empty sweat. The glands respond to sympathetic nervous stimulation via efferent postganglionic sympathetic nerves releasing acetylcholine [117, 118]. The Galvanic Skin Response (GSR) signals were sampled at 500 Hz using the Biopac Student Lab PRO acquisition software and were exported in time series ASCII files for further processing in Matlab using algorithms offered by the open ANSLAB library [110]. Room temperature was maintained at 20 – 22 $^{o}$C for reproducible GSR response.

Skin conductance level $SCL$ was obtained as baseline tonic level ranging from 10 to 50 μS in absence of discrete environmental event. The non-specific fluctuation rate $NSF$ was extracted as the number of phasic fluctuations per minute, which are not related to experimental events. Skin conductance response amplitude $SCR$ was calculated (in μS) as difference between the peak response and the baseline tonic level. Skin conductance response rise time $SCT$ was obtained (in seconds) as duration between of the onset of the response and peak conductance. Half-response time $SCH$ was calculated (in seconds) as the duration between the peak response and time when conductance returns to a value that is one-half of the amplitude of the peak.

### 3.1.8 Processing Temperature Signal

Temperature signals were used to continuously control skin temperature during experiments with cold stimuli. They were not processed further.

### 3.1.9 Overview of Our Signal Processing Framework in Matlab

We performed over 500 experiments during which up to 6 different signals were recorded simultaneously. Processing all the data in structured way required the development of a signal processing framework. An overview of the workflow is given in Figure 3.10, including loading measured signals into the processing environment, extracting cardiovascular variables, calculating statistics and displaying time course of variables. We implemented the workflow in the Matlab environment and could batch-process all 500 data sets automatically.

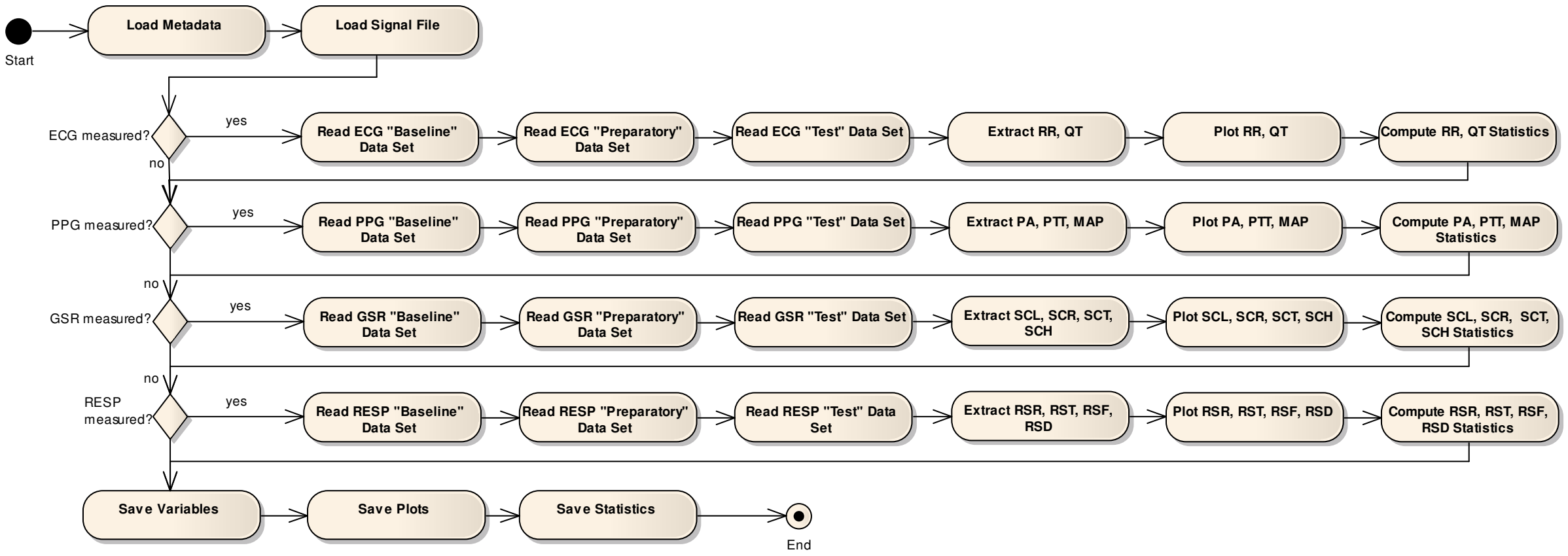

**Figure 3.10 Signal Processing Workflow**

The program first loads descriptive information about experiments and subjects. Figure 3.11 shows how these metadata have been structured. Each experiment is represented by a Cardiac Autonomic Test Metadata object which is characterized by the Abbreviation of the test (e.g. CFT) and it's Name (e.g. Cold Face Test). Each subject who underwent the test is represented as Participating Subject together with some characteristics including his unique identifier n (e.g. n=1), his Name (e.g. Jan Novak), his Sex (e.g. male), his Age (e.g. 21 years) and his Race (e.g. white). A subject can participate to one or many Experiment Sessions which took place on a specific Date (e.g. 20090520) in a given City (e.g. Kladno). Raw signals which were recorded during an experiment session are stored in a Biopac Acquisition File characterized by a Filename (e.g. Jan_Novak.acq) and a Sampling Frequency (e.g. 500 Hz). A Measured Data Set is the portion of a signal that was recorded on a particular Channel (e.g. ECG channel) for a given time interval delimited by a Start Timestamp (e.g. 100$^{th}$ second) and an End Timestamp (e.g. 160$^{th}$ second).

Our signal processing software extracts diverse cardiovascular variables from cardiovascular signals, e.g. RR intervals are extracted from the ECG signal. For each variable $X$ we split the time series into four vectors: $X_{baseline}, X_{preparatory}, X_{test}, X_{post}$ corresponding to values recorded during baseline measurements, values recorded during a preparatory period shortly before applying the stimuli on the subject, values recorded during the cardiac autonomic test and values recorded after the test. The preparatory period of 30 seconds means that baseline measurements end 30 seconds before the cardiac autonomic starts. During this period the subject can be influenced by emotional factor, e.g. fear of a cold compress approaching the eyes during a cold face test. The resulting notch in the heart rate signal is neither part of baseline measurements nor part of the autonomic test itself. That's the reason why we take a preparatory period into account and exclude it from statistical analysis or model fitting.

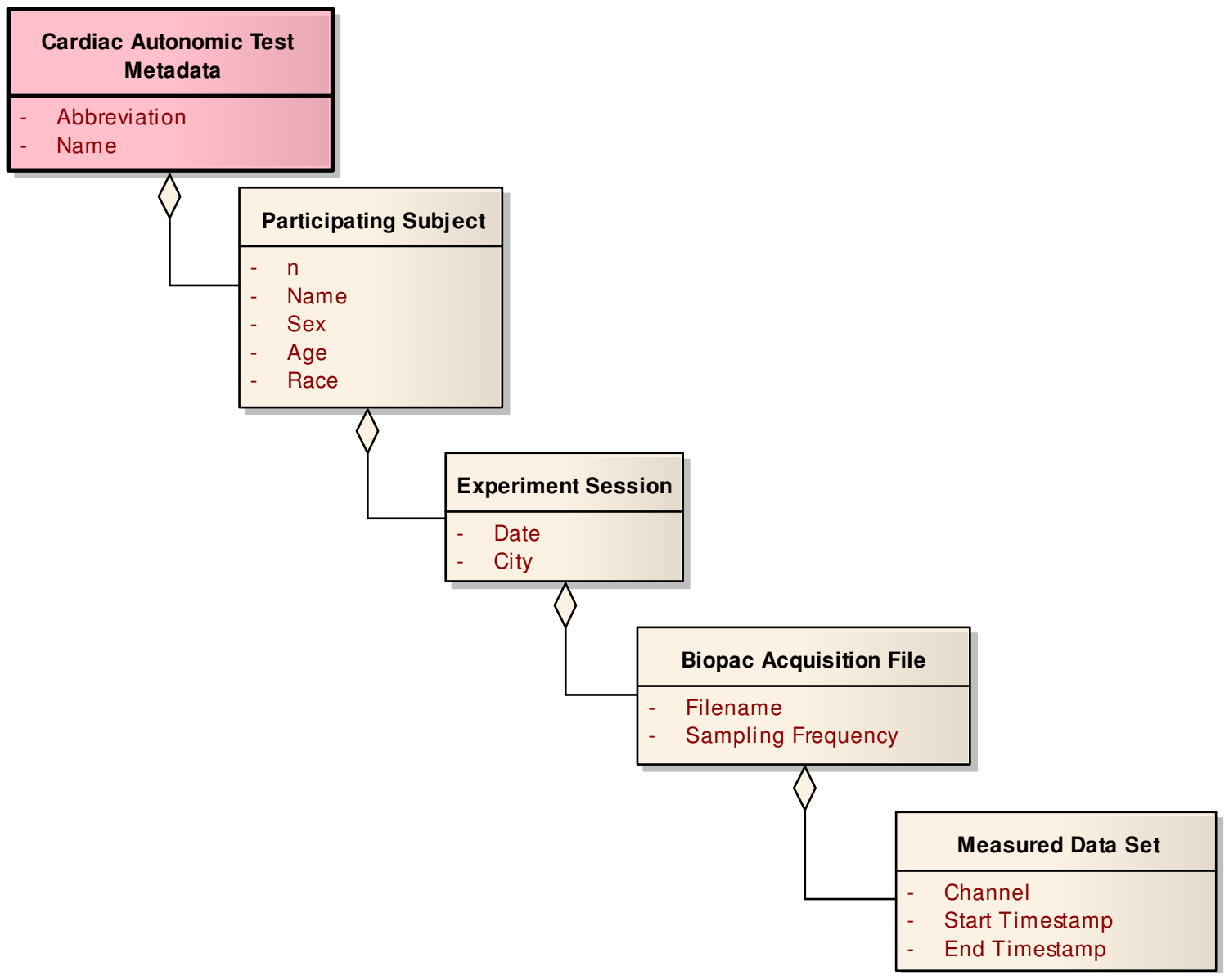

**Figure 3.11 Experimental Metadata Structure**

Subjects displayed various responses to cardiac autonomic tests. A statistical survey was realized in order to organize them into groups according to their level of response. This gave us insights into differences related to age, gender, or race and helped us testing hypothesis using the mathematical model developed in Chapter 2. The following statistical parameters were calculated on the recorded variables.

- $mean(X_{baseline}), max(X_{baseline})$ as the mean and maximum of values recorded during baseline measurements.
- $mean(X_{test}), min(X_{test})$ as the mean and minimum of values recorded during the cardiac autonomic test.
- $mean_{diff}(X)$ as the difference between $mean(X_{test})$ and $mean(X_{baseline})$.
- $mean_{incr}(X)$ as the percentage of increase in mean value between baseline and test phases, i.e. $100 \bullet \left( \frac{mean(X_{test})}{mean(X_{baseline})} - 1 \right)$.
- $std(X_{baseline})$ as the standard deviation of values recorded during baseline measurements.
- $std(X_{test})$ as the standard deviation of values recorded during cardiac autonomic test.
- $std_{diff}(X)$ as the difference between $std(X_{test})$ and $std(X_{baseline})$.
- $std_{incr}(X)$ as the percentage of increase in standard deviation between baseline and test phases, i.e. $100 \bullet \left( \frac{std(X_{test})}{std(X_{baseline})} - 1 \right)$.
- $rmssd(X_{baseline})$ as the root mean square of successive differences of values recorded during baseline measurements.
- $rmssd(X_{test})$ as the root mean square of successive differences of values recorded during during cardiac autonomic test.
- $rmssd_{diff}(X)$ as the difference between $rmssd(X_{baseline})$ and $rmssd(X_{test})$.

- $rmssd_{incr}(X)$ as the percentage of increase in root mean square of successive differences between baseline and test phases, i.e. $100 \bullet \left( \frac{rmssd(X_{test})}{rmssd(X_{baseline})} - 1 \right)$.

For the deep breath test we additionally calculated $EI_{total}(X)$ as the difference between the maximal heart rate during inspiration and minimal heart rate during expiration. We also computed $EI_{ratio}(X)$ as the ratio of minimal heart rate during expiration over maximal heart rate during inspiration.

We also performed a recurrence plot analysis of the heart rate and blood pressure variables and calculated the following statistics using the software KubiosHRV [119] and the Cross Recurrence Plot Toolbox for Matlab [120]. Details of the algorithm have been given in section 1.4.

- $\%rec(X)$ as percentage of reccurence
- $\%det$ as percentage of determinism
- $L_{max}$ as maximal line length

Heart rate and blood pressure variables were additionally analyzed in the frequency domain using an implementation of the Least-Squares Spectral Analysis (LSSA) in Matlab [121]. The magnitudes in the spectrum determine the contribution of each frequency band to the power (i.e. variance) of the variable. We evaluated the low frequency $LF(X)$ and high frequency $HF(X)$ components in the bands [0.015 .. 0.15] Hz and [0.15 .. 0.4] Hz .

## 3.2 Cold Face Test

### 3.2.1 Physiology of the Cold Face Test

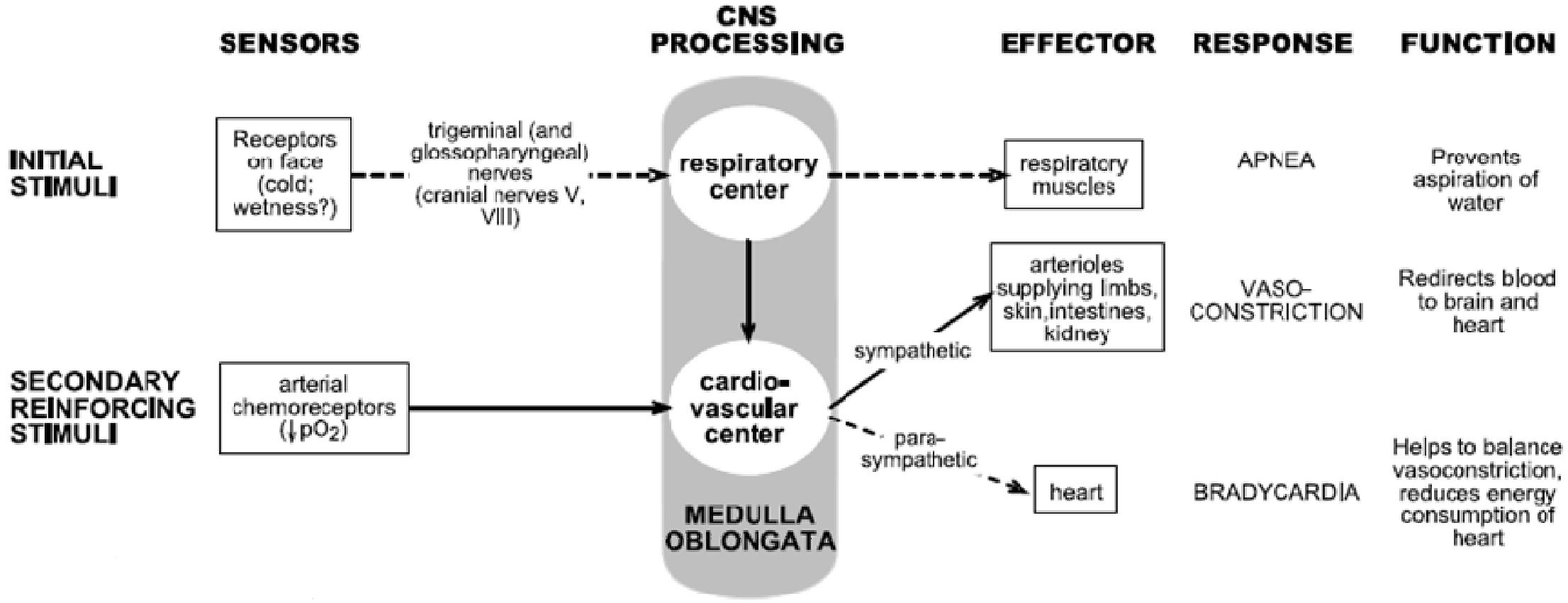

**Figure 3.12 Principal components of the diving reflex (reprinted from [122])**

The Cold Face Test, abbreviated CFT, is a simple clinical test, which activates the diving reflex. A cold stimulus applied to the forehead, usually using a cold compress, activates the afferent temperature sensitive neurons (ophthalmic division of the trigeminal nerve), which leads to the activation of cardiac efferent vagal neuron fibers, resulting in a marked decrease of heart rate. The observed bradycardia results from increased parasympathetic activity on the sinoatrial node and was not influenced by β-adrenoreceptor blockade [88]. Baroreflex activation during CFT has not yet been put in evidence [123]. Increased activity in sympathetic efferent fibers to peripheral arteries in the arms and legs is additionally observed. This results in increases in vascular resistance and eventual rise in blood pressure [124]. Figure 3.12 shows the physiological components involved during cold face test as well as the messages flow between them. The trigeminal nerve or the fifth cranial nerve is mainly to provide sensory information to the brain. Its ophthalmic division is purely sensory. It includes cold temperature receptors, free nerve endings that terminate in the subcutaneous layers of the skin on the forehead. They are sensitive to temperatures lower than the normal body temperature. The receptive field of one thermoreceptor is about 1 mm in diameter. The sensory path consists of a primary sensory neuron, which has its cell body in the trigeminal ganglion. From there a single secondary sensory root enters the pons in the brainstem. The pons relay the cold stimulus down to the medulla, where it is integrated. The vagal response is carried out by the vagus nerve and gives rise to increases in efferent parasympathetic tones on the sinoatrial node of the heart. CFT might simultaneously lead to an increased sympathetic response on the ventricles and vasculature.

Because the cold face test is a vagal maneuver known to inhibit sympathetic efferent activity to the sinoatrial node and enhance parasympathetic activity [125, 126], it offers a suitable experimental setup for gathering experimental data which will fit to our integrated mathematical model of cardiovascular control developed in the previous chapter. This will help quantifying vagal activity.

### 3.2.2 Subjects and Materials

We have performed the cold face test on 50 subjects: 32 women (ages 11 to 77 years, mean 38±17 years) and 18 men (ages 23 to 82 years, mean 48±16 years).

A first experimental session was performed in winter in the Czech Republic on February 25, 2009. A second and third session was performed one year later on January, 27 and 28 in Cameroon during the dry season. In the first experimental session, all 14 subjects were instrumented with a Biopac MP35 system and 3 leads ECG was recorded. Skin temperature was additionally measured on 6 subjects. Galvanic skin response was recorded on 5 subjects following the recommendations in [127].

38 subjects participated in the second experimental sessions (including 2 subjects n = 13, 37 from the first session), 3 of them repeated the test in a third experimental session. In both sessions subjects were instrumented with a Biopac MP35 system and ECG, galvanic skin response and skin temperature were measured.

We had 3 subjects with palpitations (n = 8, 10, 26); 2 with cardiac insufficiency (n = 8, 20); 3 with hypertension (n = 16, 20, 45); 1 with diabetes mellitus type 1 (n = 41); 1 with sleep apnea (n = 38); 1 with gastric ulcer (n = 8, 10) and 1 with hepatitis (n = 5). Literature does not highlight contraindications to the cold face test regarding sex, age or health status. For example the cold face test was performed on swimmers ages 12 to 19 years [88]. Studies in [102] described tests on healthy men and women, age 21 to 54 years. Subjects in [128] were healthy men aged from 24 to 60 years. Subjects in [103] were men and women having familial dysautonomia (Jewish Ashkenazi ancestry, absent deep tendon reflexes, absence of overflow tears, absence of fungiform papillae of the tongue, absent axon flare response after intradermal histamine injection), aged 10 to 44 years. Authors of [123] performed the test on two patients with baroreflex impairment. Studies in [105] support the hypothesis that black males would show greater cardiovascular reactivity to forehead cold stimulation than whites, suggesting taking racial differences into consideration while evaluating the cold face test.

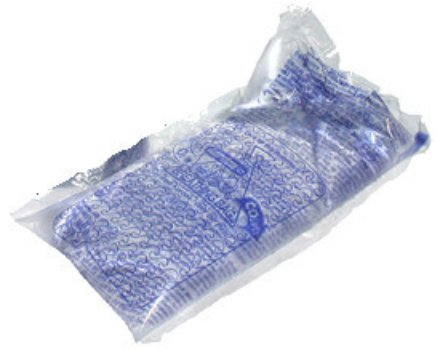

**Figure 3.13 Iced-water filled plastic bag used for the cold face test**

In our studies, after one period of at least 5 min adjustment, baseline measurement was performed for 60 seconds. After baseline measurements a preparatory recording of 30 seconds was included in order to reduce the non-stationary effects induced by experimental conditions. An iced-water filled plastic bag with wet contact surface 15 cm x 5 cm and mass ca. 400 g (Figure 3.13) was then applied bilaterally to the forehead during a period of maximum 60 seconds while measurement was resumed. In general the cold compress had a temperature of 0 degree Celsius and was large enough to cover the forehead, but increasing the mass did not seem to amplify the response. However care was taken to avoid triggering an oculocardiac reflex response. The plastic bag was filled with melting ice-water and was flexible to allow direct contact with the skin. In between tests the plastic bag was be kept in a bucket containing iced-water so that the exterior of the bag remained wet. The test was interrupted when the subject could not sustain the cold pain any more. Subject was encouraged to take a short deep inspiration prior to the test and to hold breathing as long as he could. Most of subjects however disagreed and performed the test while breathing normally.

Studies in [102] recommend performing the cold face test in a setting where the subjects did not continue breathing during the test. Fall in heart rate was larger if the subject was not able to continue breathing. However results in [50] show that bradycardia, blood pressure increase, sympathetic and parasympathetic modulation of the heart rate and blood pressure are independent from the respiratory pattern. The environmental temperature is another factor to be taken into account. In [102] the tests were performed in a quiet room having a constant temperature of 22 degree Celsius. The room was mildly lighted. A swimming pool with water temperature 25 degree Celcius was used in [128]. In our three studies, the subject was sitting on a comfortable chair in a quiet room with constant temperature (21 $^{o}$C ) and lighted with daylight.

### 3.2.3 Results

We performed the cold face test 71 times and responses were evaluated using statistics introduced in section 3.1.9. The $rmssd_{incr}(X)$ was selected as criteria for the overall response since it is a parameter of heart rate variability that should reflect the increase in parasympathetic activity on heart. An increase of 20% or more was rated as *Over Average* response, a lesser increase was rated as *Average* response whereas no increase was evaluated as *Abnormal* response. Subjects were divided into Young, Older and Elderly groups according to their age (less than 30 years for young, between 30 and 60 years for older and more than 60 years for elderly). The distribution of subjects according to their gender and their response to cold is depicted in Figure 3.14 (left panel). Many subjects presented an average or over-average response to the cold stimuli with women being more sensible. We also analyzed the effect of geographical environment on the cardiovascular responses to cold stimuli and noticed e.g. decreasing response intensity with decreasing environmental temperature (see Figure 3.14, right panel), suggesting a certain degree of acclimatization of cold receptors. Figure 3.15 shows how heart rate and RR interval durations were changing during the cold face test on subject ID 4. Mean heart rate dropped from 68 to 62 beats/min while reaching the floor value 54 beats/min at the 40$^{th}$ second during cold stimuli application. RMSSD increased by 54%. A similar but less strong increase of QT intervals duration is observed and maintained even after the release of cold stimulation (see Figure 3.16). CFT induces an increase of systolic blood pressure in the first moment during the application of cold stimuli. This is followed by a drop of systolic pressure which appears to trigger the baroreflex as characterized by some fluctuation in pressure at the end of the test (see Figure 3.17). A test duration of more than 40 seconds seemed to be most appropriate to subjects under 60 years old. The heart rate change diminishes in older subjects and the duration of sustainable cold application, which enhances the fall of heart rate, was also reduced with aging.

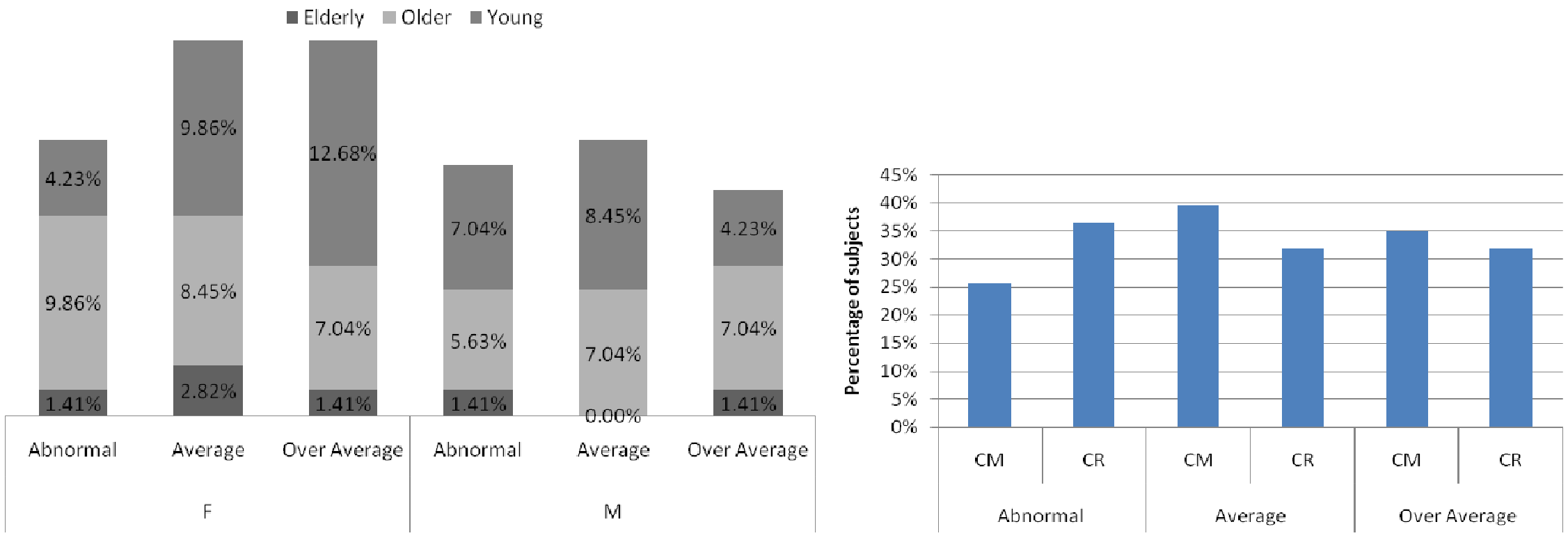

**Figure 3.14 Overall CFT Response**

Young (less than 29 years); Older (30 to 59 years); Elderly (over 60 years)

CM – Cameroon; CR – Czech Republic

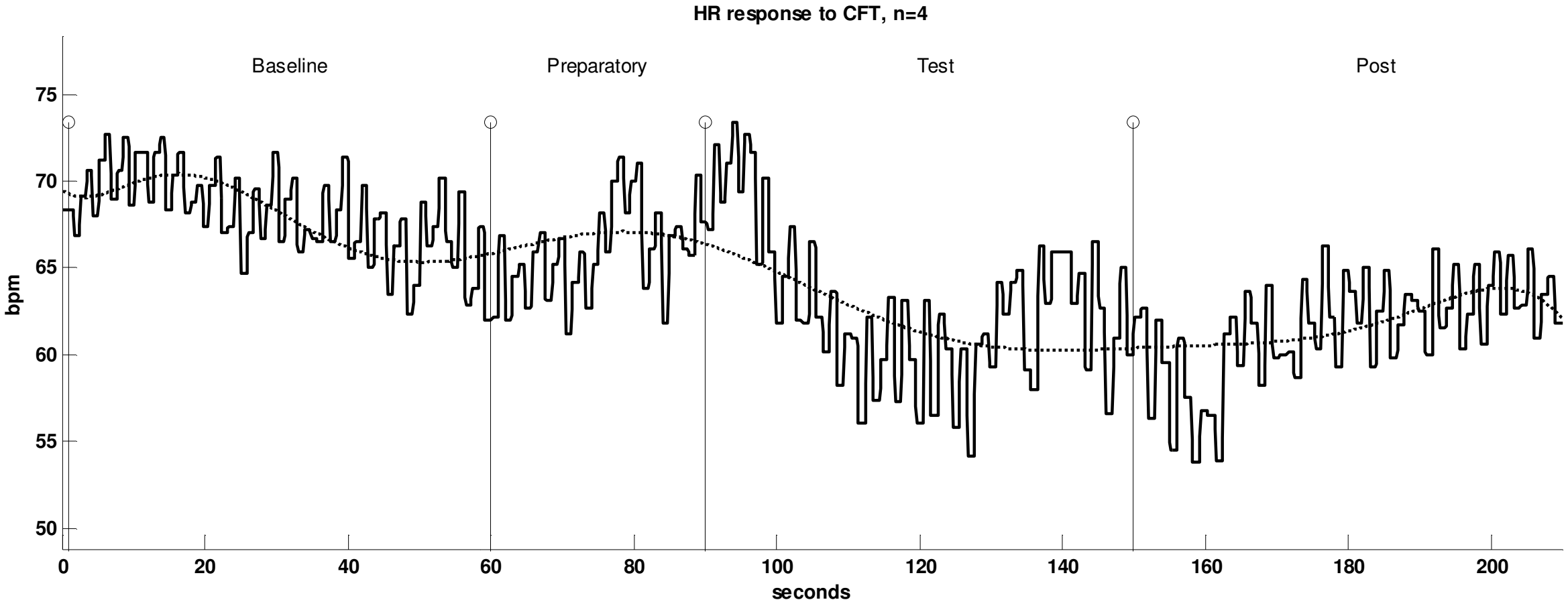

**Figure 3.15 Heart rate response to cold face test**

measurements were performed on a representative subject with ID n=4

dashed line is a $10^{th}$-degree polynomial interpolation of the signal

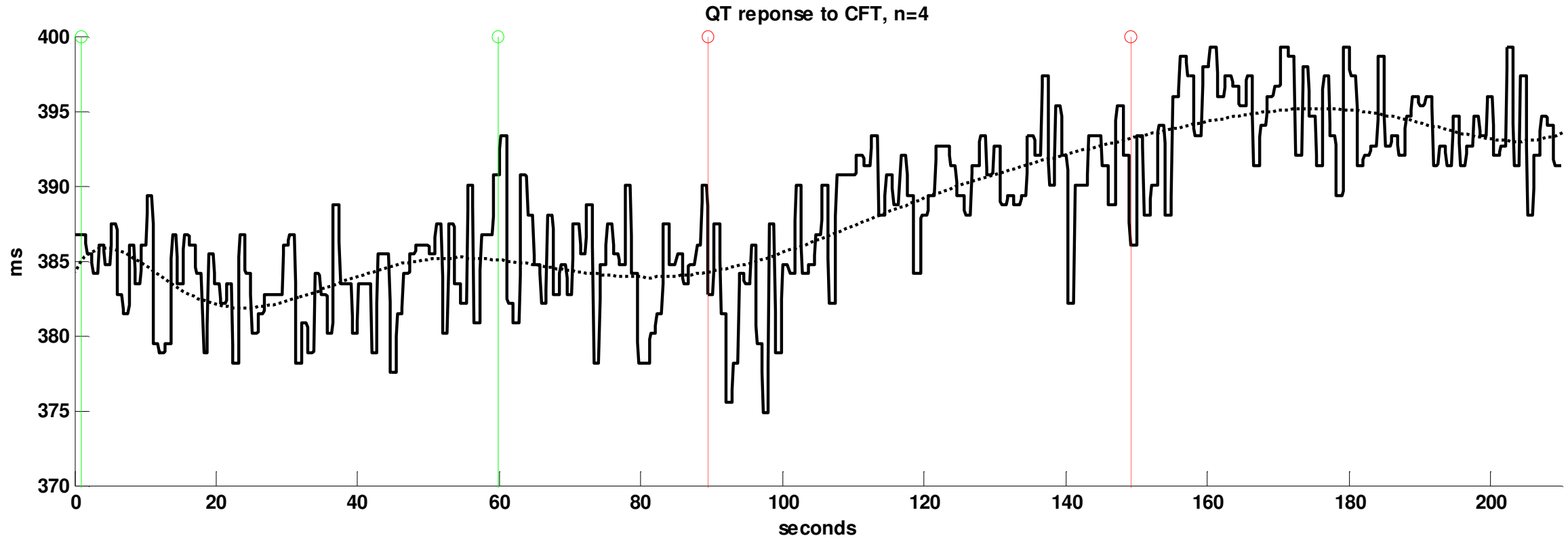

**Figure 3.16 QT response to cold face test**

measurements were performed on a representative subject with ID n=4

dashed line is a $10^{th}$-degree polynomial interpolation of the signal

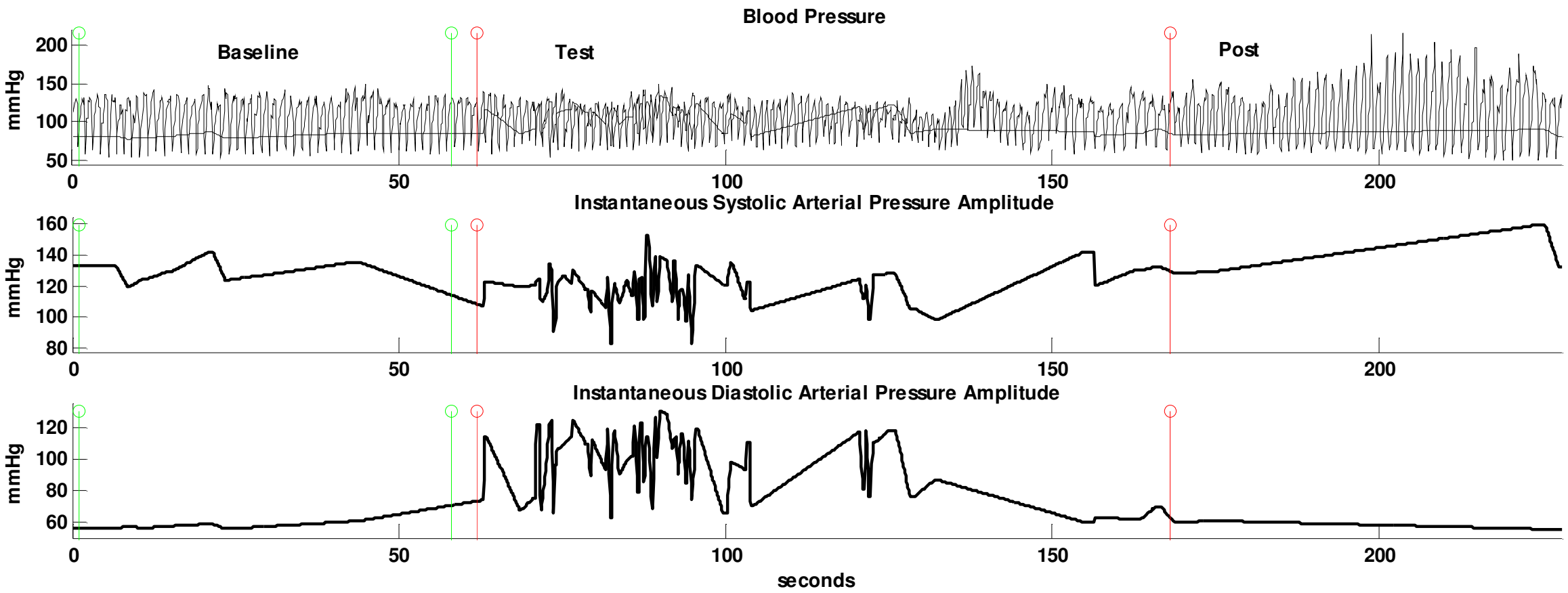

**Figure 3.17 Blood pressure response to cold face test**

measurements were performed on a representative subject with ID n=4

## 3.3 Oculocardiac Reflex Test

### 3.3.1 Physiology of the Oculocardiac Reflex Test

The Oculocardiac Reflex (OCR) is a trigeminovagal reflex that can produce potent bradycardia and cardiac arrhythmias such as nodal rhythm, ectopic beats, ventricular fibrillation, or asystole. The OCR is most often induced during strabismus surgery in children but also occasionally during retinal surgery [89]. The reflex can also be induced by traction on the extrinsic muscles of the eye, intraorbital injections or haematomas, acute glaucoma, and stretching the eyelid's muscles [129]. Positive OCR was considered to have occurred when the heart rate slowed at least 7.2 beats/min in [129]. OCR was defined as decreasing of heart rate by more than 20% from the baseline value in [130].

The oculocardiac reflex is mediated by the ophthalmic part of the trigeminal nerve. It is suggested in [131] that the sensory nerve endings of the trigeminal nerve send neuronal signals via the gasserian ganglion to the sensory nucleus of the trigeminal nerve, forming the afferent pathway of the reflex arc. The afferent pathway continues along the short internuncial fibers in the reticular formation to connect with the efferent pathway in the motor nucleus of the vagus nerve. Cardioinhibitory efferent fibers arising from the motor nucleus of the vagus nerve terminate on the myocardium. These vagal stimuli provoke negative chronotropic and inotropic responses. Consequently, the clinical features of the OCR range from sudden-onset of sinus bradycardia, bradycardia terminating asystole, asystole with no preceding bradycardia, arterial hypotension, apnea, and gastric hypermotility. The pathways are illustrated in Figure 3.18.

### 3.3.2 Subjects and Materials

We have performed the oculocardiac reflex test on 36 subjects: 23 women (age 11 to 51 years, mean 30±14 years) and 13 men (ages 22 to 56 years, mean 36±16 years); among them 1 with aortic stenosis (n = 56), 2 with palpitations (n = 8, 26); 1 with cardiac insufficiency (n = 8). 15 subjects were white and 21 black. The first experimental session was performed in the Czech Republic on April 28, 2009. All 14 subjects were instrumented with a Biopac MP35 student system and following signals were recorded: 3 leads ECG, skin temperature and galvanic skin response. 5 subjects were included in the second experimental sessions with 3 leads ECG and galvanic skin response measurement. The third session included 38 subjects and the fourth included 3 subjects. Both last sessions were realized on January 27-28, 2010 in Cameroon. ECG, galvanic skin response and skin temperature were measured.

The literature provides little information about the oculocardiac reflex test on adults. The study population in [90] comprised 50 asymptomatic, premature neonates (22 females, 28 males) whose mean (±SD) gestational age at birth was 33.5 ± 1.6 weeks. Studies in [129] were conducted on 49 infants and children (53% male, 47% female), aged 6 months to 9 years requiring correction of strabismus. Age limitation or contraindication to the OCR could not be found in the literature. Neither sex nor age differences seemed to cause any statistical difference in the incidence of the OCR.

The OCR was performed in [90] by linking a mercury column to a three-way tap connected to two bags inflated to a pressure between 100 and 140 mmHg. A bag was placed on each closed eye, and a constant pressure of +100 mmHg was then applied for 10 s, or until heart rate decreased by one-third of its baseline value for >2 s. In [132] the subject receives a massage from an instructor, who used his index finger to apply pressure on the eyeballs. Studies in [129] evoked the OCR by traction on the extrinsic muscles of the eye using a mechanical dispositive. The first type of stimulus was the most reflexogenic and consisted of an acute traction of 150 g to 300 g sustained for a minimum of 30 s followed by an acute release. The second stimulus was a slow gradual traction reaching a peak between 150 g and 300 g followed by an acute release. The third stimulus applied a fast progressing traction until a peak between 150 g and 300g followed by an acute release.

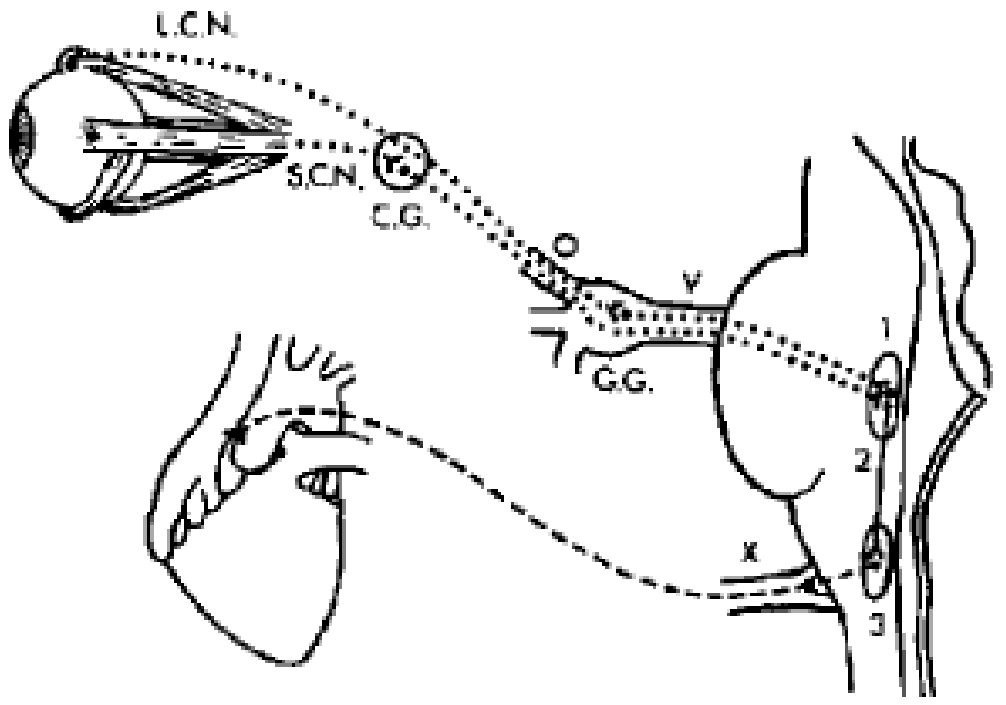

**Figure 3.18 Oculocardiac reflex neural pathways (reprinted from [129])**

LCN (long ciliary nerves), SCN (short ciliary nerve), CG (ciliary ganglion), O (ophthalmic branch of the Vth cranial nerve), GG (gasserian ganglion), 1 (main sensory nucleus of the trigeminal nerve), 2 (short internuncial fibres in the reticular formation), 3 (motor nucleus of the vagus nerve), X (cardiac depressor nerve form the Xth cranial nerve)

We employed a simpler procedure in our studies. The oculocardiac reflex test was performed with the subject in sitting relaxed position with closed eyes. Pressure was applied on the eyes by rotating subjects' eyeballs with closed lids over the outer corner of eyes using the index finger. Subject was asked to concentrate on the light between the eyebrows. This procedure was very delicate because pressure on eyes can be uncomfortable. Subjects were very closely informed about upcoming events and were encouraged to notify any issue. Despite the fact that some subjects reported having blur visibility after the test, no subject requested test interruption. In a different study we lay a small water filled plastic bottle with weight 670 g and contact surface ca. 5 cm$^2$ on each eye. This had applied a pressure of ca. 100 mmHg (1 mm Hg = 0.0013595 g/cm$^2$) for 60 seconds.

### 3.3.3 Results

Figure 3.19 shows the heart rate response to oculocardiac reflex test for one representative subject from our experiments. Heart rate dropped from 71 beats/min to 65 beats/min in average while reaching the floor value 59 beats/min just at the beginning of the eyes massage. RMSSD increased by 8%.

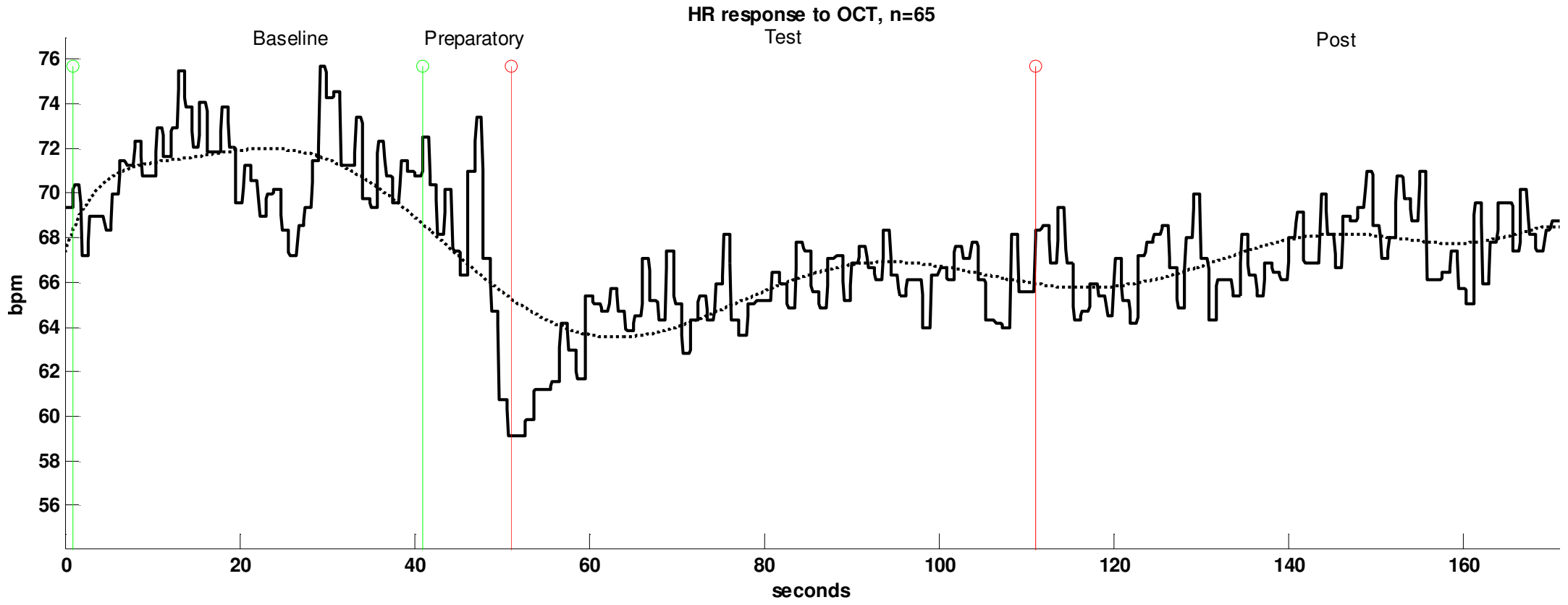

**Figure 3.19 Heart rate response to oculocardiac reflex test**

measurements were performed on a representative subject with ID n=65

dashed line is a 10$^{th}$-degree polynomial interpolation of the signal

## 3.4 Deep Breath Test

### 3.4.1 Physiology of the Deep Breath Test

Deep breathing has proven to be a useful clinical test for assessing cardiovagal dysfunction in autonomic disorders, including diabetic autonomic neuropathy, uremic neuropathy, familial autonomic neuropathies, small fiber neuropathies, pure autonomic failure and multisystem atrophy [133]. It enhanced respiratory sinus arrhythmia (RSA) which is related to the variation in heart rate that occurs during a breathing cycle. Heart rate increases during inspiration and decreases during expiration [97]. During deep breathing at 6 breaths/min (=0.1 Hz), the magnitude of sinus arrhythmia increases profoundly in normal subjects because respiration-mediated and baroreflex-mediated low frequency RR interval oscillations (which normally occur at about 0.1 Hz) synchronize [97] . The depth of breathing for a maximum result requires a tidal volume of approximately 1.2 L for an average adult. Protocols that involve breathing for more than 90 seconds may induce hypocapnea, which can reduce heart rate variability. [133]

### 3.4.2 Subjects and Materials

We have performed the deep breath test on 38 subjects: 26 women (ages from 11 to 70 years, mean 37±16 years) and 12 men (ages from 23 to 56 years, mean 47±11 years); among them 3 with palpitations (n = 8, 10, 26); 2 with cardiac insufficiency (n = 8, 20); 3 with hypertension (n = 16, 20, 45); 1 with sleep apnea (n = 38); 1 with gastric ulcer (n = 8, 10) and 1 with hepatitis (n = 5). The experiments were performed on January 27, 2010 and January 28, 2010 in Cameroon.

The known literature does not exhibit particular distinctions in subjects undergoing the deep breath test. The study population in [134] was composed of 185 patients admitted after a first myocardial infarction 5 days before, 79% male, 21% female, 53% age less than 60 years, 47% age above 60 years. Deep breath tests were conducted in [135] on 10 patients (2 females, 8 males), aged 58±2 years, average height was 173±3 cm, average weight 87±6 kg and BMI 28.8±1.7 kg/m$^2$ who had a confirmed diagnosis of type 2 diabetes mellitus. The study also included 11 healthy controls (four women and seven men). Their average age was 54±3 years, average height 168±2 cm, average weight 76±5 kg and average BMI 26.6±1.3 kg/m$^2$. Studies in [136] included 6 males and 4 females with familial dysautonomia, mean age 25±12 years. Heart rate variability during deep breathing is in general reduced in individuals on cardiac medication, with left ventricular hypertrophy or ECG signs of myocardial infarction. Even in healthy persons the parasympathetic function is inversely associated with age and left ventricular mass [137]. Contra-indications to the deep breath test include avoiding nicotine, caffeine or alcohol for 24 h before the experiments and studies should be performed at least 4 h after consuming a light meal [135]. A history of chronic alcohol abuse, carcinoma, myopathy, hyper- or hypothyroidism, arterial hypertension, cardiac arrhythmia, diagnosed atherosclerosis, previous organ transplantation, renal disease could be a possible contraindication for the deep breath test [135]. In a large population study, patients who had fever, diarrhea, atrial fibrillation, frequent arrhythmias, liver cirrhosis, glaucoma or chronic renal failure, and those who took minor or major tranquilizers were excluded [138].

Regarding experimental setups for deep breath test, it is common to execute the test at a rate of 6 respiration cycles per minute, which corresponds to 0.1 Hz: 5 seconds for each inhalation and 5 seconds for each exhalation [134]. The examiner has a chronometer and raises his hand to signal the start of each inhalation and lowered it to signal the start of each exhalation. Participants can also be asked to breathe according to a continuous graph presented on a computer monitor in front of them [139]. The test usually begins with calibration during spontaneous breathing followed by paced breathing. Next to the pacing frequency, another parameter is the subject position. Tests were conducted with patients in a semi-reclining position of 45° in [139]. The tests in [135] were performed in a quiet, temperature controlled (24°C) room with the participant lying in a supine position; however subjects underwent the test in a sitting position in [137] without measurable impacts in the results. The time of day when the test is performed, seems to be of minor importance. Studies in [134] were performed in the morning; however in [136] the test was conducted between 4 and 6 p.m. in order to avoid the circadian influence on heart rate and autonomic nervous system function.

In our studies, participants were in a relaxed sitting position and were taught to perform three deep breathes in a first experimental setup. Participants of the second experimental setup were instructed to breathe at a rate of 6 respiration cycles per minute. The examiner paced the breathing with a chronometer and gave a vocal signal for the start of each inhalation and exhalation.

### 3.4.3 Results

Simple indexes can be used for the response evaluation to deep breath test [134]. The respiratory stress response (RSR) can be calculated as the ratio of pick area, which is the net peak energy around the breathing frequency (0.1 Hz), by the sum of all the other areas of the power spectrum curve in order to indicate significant coronary disease [139] . In [133] the mean heart rate range (MHRR or $E\text{-}I_{mean}$) was calculated as difference between the maximal heart rate during inspiration and minimal HR during expiration for a single breathing cycle. The expiratory-to-inspiratory ratio (E/I) is calculated as the longest RR interval during expiration divided by the shortest RR interval during inspiration for a single breathing cycle. MHRR and E/I can also be averaged over successive breaths. Response to deep breathing is elevated in children and trained adults; and decreases with age or inactivity, so it is essential to use evaluating methods with well-defined age-stratified normal values [140]. Values of MHRR above $(4.39 - 0.033 \cdot \text{age})^2$ and of E/I above $1 + \exp(-1.12 - 0.0198 \cdot \text{age})$ can be regarded as normal [137].

Additional measures are $E\text{-}I_{total}$, which is the difference between the maximal heart rate during all 6 inspirations and minimal heart rate during all expirations; $E\text{-}I_{max}$, which is the difference between maximum and minimum heart rate for every breathing cycle, with the largest difference being selected; $E\text{-}I_{median}$, which is the calculation of differences between maximum and minimum heart rate for all breathing cycles, thereafter sorted by their magnitude. Averaged difference of the third and fourth value is computed. Similarly $E/I_{median}$ is the median ratio of the longest RR interval during expiration over the shortest RR interval during inspiration within every breathing cycle. $E\text{-}I_{median}$, $E/I_{median}$, and MCR are most resistant to anomalies (in-test variance, premature ventricular contraction, deviation of the respiration rate). $E\text{-}I_{median}$ should be used preferentially, providing the highest precision and independence from heart rate [141]. The peak power at the highest frequencies (> 0.15 Hz) in the spectral analysis of RR intervals can also be calculated if the signal does not contain premature beats.

Figure 3.20 shows heart rate and pulse amplitude responses to deep breath test at 0.1 Hz, which we performed on one representative subject. The response corresponds to a value of $E\text{-}I_{total}$ and E/I equal to 21.2 beats/min and 0.69 respectively.

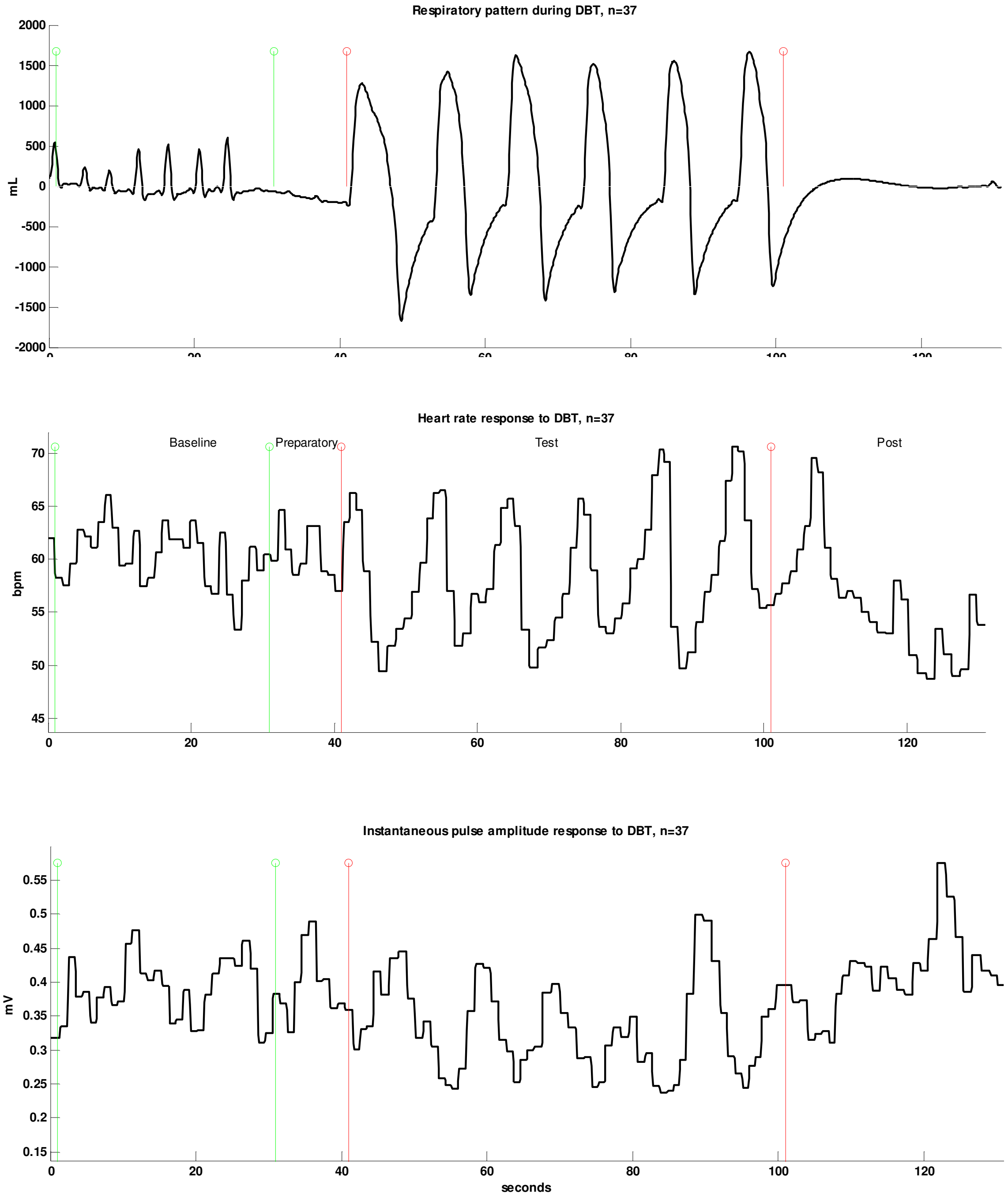

**Figure 3.20 Heart rate and pulse response to deep breath test**

measurements were performed on one representative subject with ID n=37

respiration pattern (top graph), heart rate (middle graph), pulse amplitude (bottom graph)

## 3.5 Mental Stress Test

### 3.5.1 Physiology of the Mental Stress Test

Diverse tests can induce mental stress such as mental arithmetic, computer quiz, delayed auditory feedback task, reaction-time task; what lead to an increase in mean heart rate, systolic and diastolic pressure, and cardiac output. Invasive measurements of muscle sympathetic nerve activity have demonstrated increased sympathetic activity during mental stress. Results with pharmacological blockades demonstrate an increased sympathetic control of heart rate; atropine did not appear to affect the magnitude of the response [142]. Functional magnetic resonance imaging (fMRI) shows that sympathetic activation through mental stress is associated with distinct cerebral regions being immediately involved in task performance (visual, motor, and premotor areas). Other activated regions (right insula, dorsolateral superior frontal gyrus, cerebellar regions) are unrelated to task performance. These latter regions have previously been considered to be involved in mediating different stress responses [143].

Spectral analysis of heart rate variability has shown increased LF power and LF/HF ratio, which are both indices of sympathetic activity [142] and decrease of HF power. Baroreceptor sensitivity decreases as well, suggesting an increased autonomic reactivity. Baroreflex function can be modulated by behavior or mental challenges at relay sites in the medulla, pons and hypothalamus [144]. MST is considered effective if the mean heart rate increased by at least 15 beats/min [142]. Changes in systolic and diastolic blood pressure, cardiac output and ventricular pre-ejection period can also be assessed.

### 3.5.2 Subjects and Materials

In our studies 26 women (ages from 11 to 70 years, mean 37±16 years) and 12 men (ages from 23 to 56 years, mean 47±11 years) were asked to resolve a series of arithmetic operations including additions, subtractions and multiplications. Each operation was given spontaneously and randomly by an instructor. Subject could speak and calculate aloud. The test was conducted for 60 seconds. We had additional experimental setups where a limited number of subjects were asked to subtract the number 17 from 1000, then from the results, and so on.

Alternatively subjects can be asked to subtract continuously the numbers 6 or 7 from a 2 or 3 digit number as fast as possible with each calculation lasting max 5 seconds. A new number is randomly provided every 5 seconds on a computer screen fixed at eye level. Halfway through the mental stress, subjects are asked to answer more correctly, irrespective of the number of correct answers. These procedures are designed to help reduce adaptation to the stress condition [144]. In order to avoid increased HF of the spectrum of HRV, speaking is avoided in some studies and the computer displayed each time four possible solutions and the participants are asked to choose the correct one by pressing a keyboard key placed closed to the finger. They are given feedback on the display that their answer was incorrect, regardless of the accuracy of their response, for a predetermined 60% of the questions in order to ensure that all participants are stressed, independently of individual mental arithmetic abilities. Some studies however recommend to perform the tests aloud in order to improve sympathetic response, however this adds more respiratory sinus arrhythmia influence on the spectral analysis heart rate variability, by increasing its high frequency component, what could be mistaking as increased vagal tone [142].

The Stroop test (Colour Word Interference Test [145]) is another kind of cognitive stress using a card that contains 5 lines and 4 rows. Each line has the names of 4 colors: green, pink, blue, and brown. The sequence of the 4 colors is different in each line but the color names are printed in an ink color different from the color itself. The subject is requested to respond with the colors he is seeing as fast as he could. This task can be initiated using a software tool that displays a four-digit number on a computer screen [146].

Mental stress can be performed by combining diverse kind of tests. For example in [147], all patients underwent serial subtractions, Stroop color word task, public reading, emotionally charged public speaking, and a competitive video computer game.

There are references in the literature where the MST was conducted on special groups of patients. The MST was performed on 15 migraine patients (13 females, 2 males) and 15 healthy control subjects (13 females, 2 males), age 27±9 years [145]. 38 healthy students (19 men, 19 women), ages 26±5 years were recruited for studies in [146]. 35 patients with evidence of ambulatory ischemia during Holter monitoring sessions participated in MST studies in [147]. 36 mild hypertensive men (57%) and women (23%), mean age 50±8 years with 12 or more years of education, underwent a mental arithmetic test in [148]. 12 healthy control subjects were also included, mean age 46±9 years. 16 healthy right-handed male subjects (mean age 23±2 years) with no report of neurological or psychiatric disorders were examined with the MST in [143].

### 3.5.3 Results

Figure 3.21 illustrates heart rate response; we have recorded during mental stress on a representative subject, who was asked to subtract 17 from 1000. Heart rate increased by 14% from a baseline value 52 beats/min. Heart rate variability decreased by 10% indicating vagal withdrawal. Respiration rate increased while tidal volume, minute volume and duty cycle decreased (see Figure 3.22). The stress state is also reflected in the galvanic skin response (Figure 3.23). Tonic skin conductance level (SCL) increased by 13%. Non-specific fluctuations increased as well as their rise time and half-recovery time. All these factors indicate increased sympathetic activity to sweat glands.

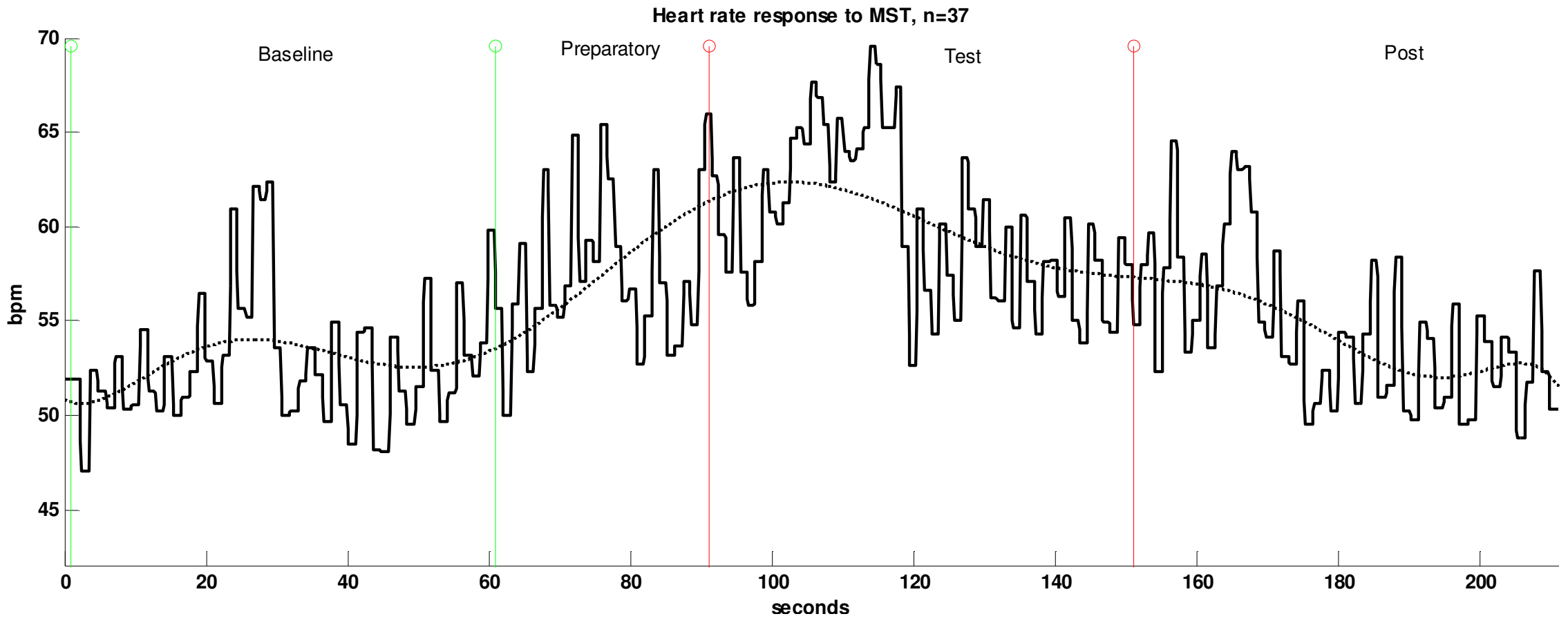

**Figure 3.21 Heart rate response to mental stress test**
measurements were performed on one representative subject with ID n=37
dashed line is a 10th-degree polynomial interpolation of the signal

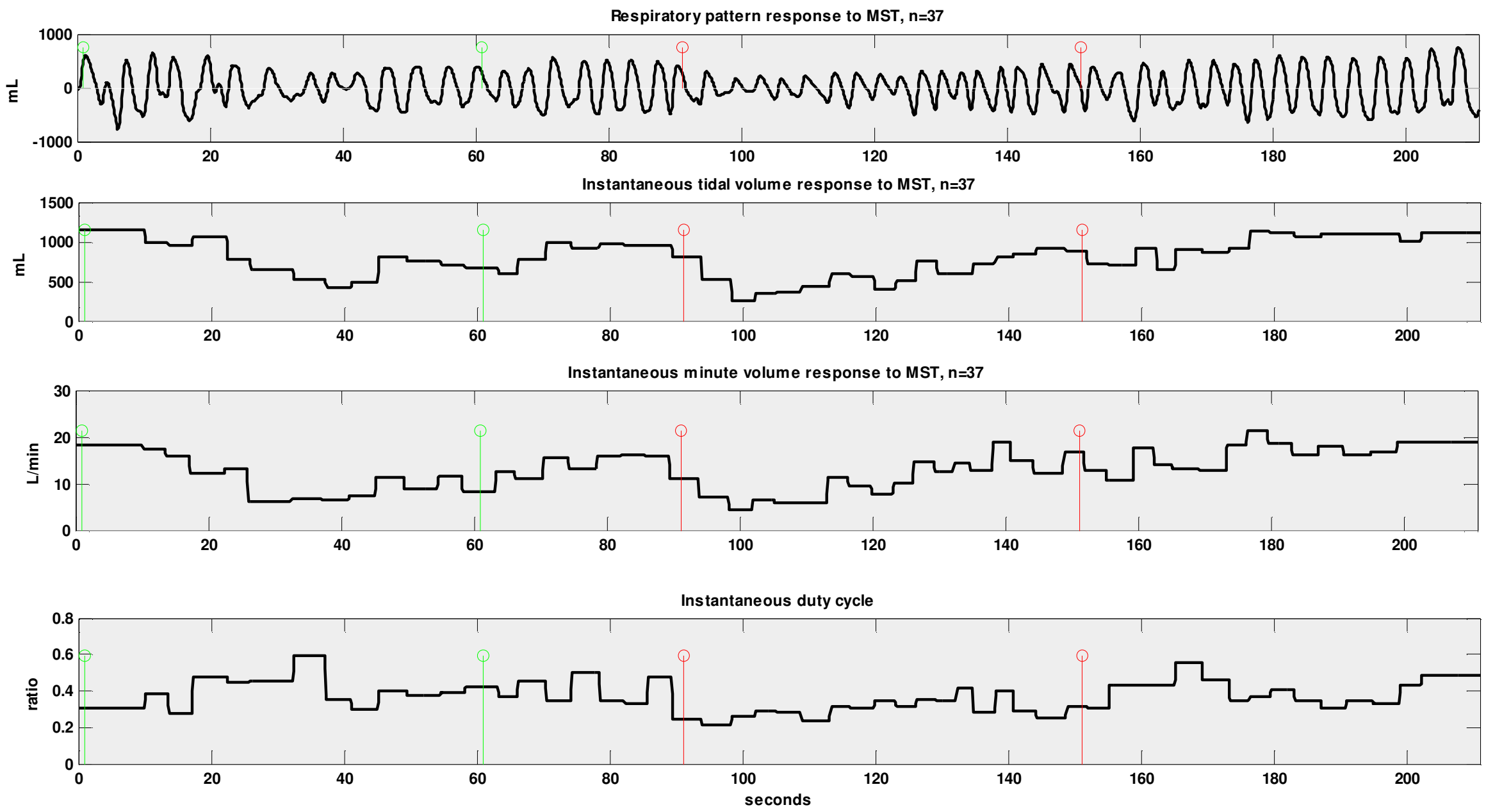

**Figure 3.22 Changes in respiration during mental stress test**

measurements were performed on one representative subject with ID n=37

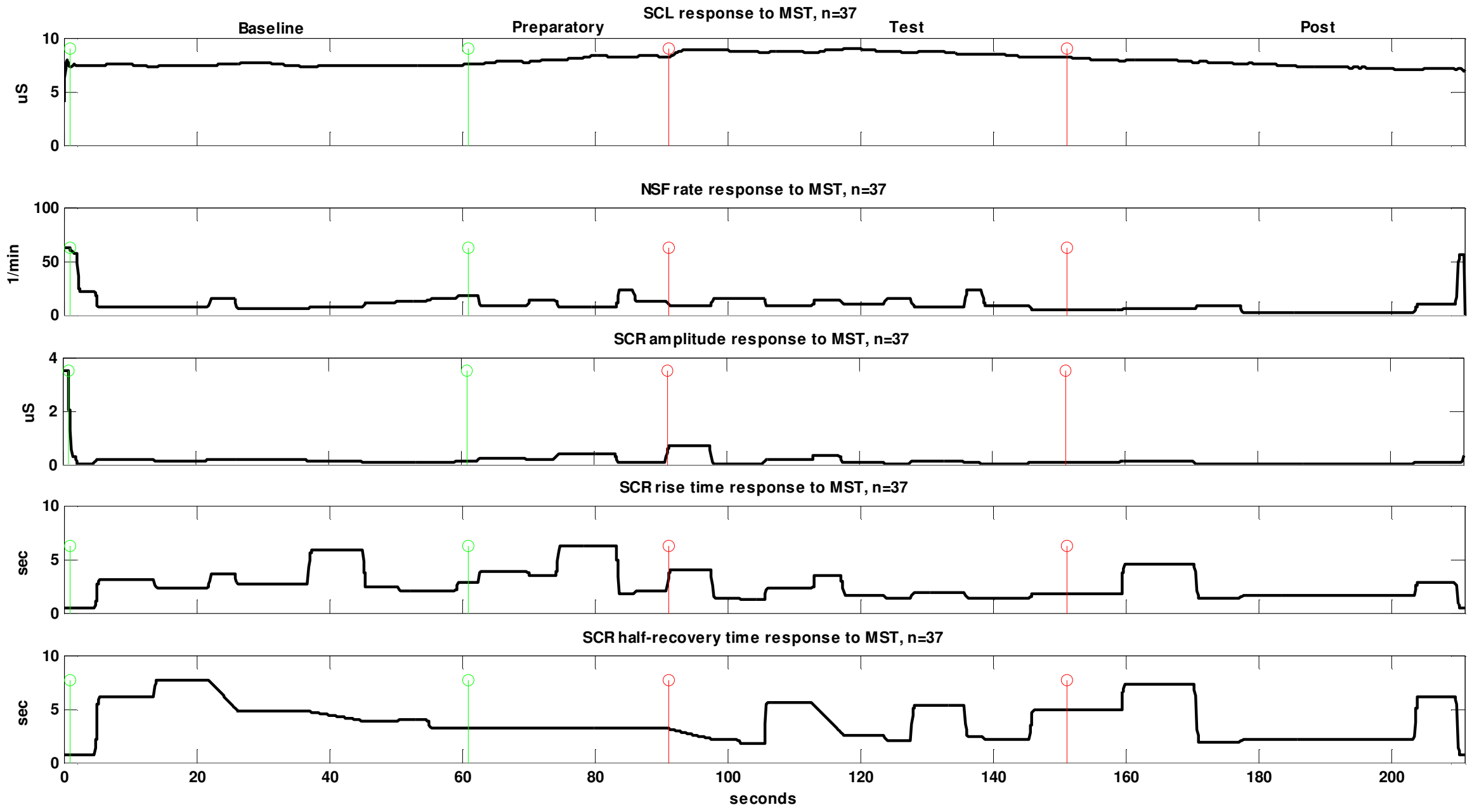

**Figure 3.23 Galvanic skin response during mental stress test**

measurements were performed on one representative subject with ID n=37

## 3.6 Valsalva Maneuver

### 3.6.1 Physiology of the Valsalva Maneuver

The Valsalva Maneuver (VM) is a cardiac autonomic test, named after the physician and anatomist Antonio Maria Valsalva (1677-1723), which consists of taking a deep breath and expiring against a 40 mmHg resistance for approx. 15 seconds. This cardiac autonomic test induces physiological changes and initiates a complex reflex. The reflex pathways involved are both of sympathetic and parasympathetic nature [142]. The reflex arcs involved in the Valsalva maneuver can be summarized in four phases [142]. During phase 1 a brief increase in arterial blood pressure and decrease in heart rate is observed. In phase 2 sustained high intra-thoracic and intra-abdominal pressure hinders venous return, leading to a rapid decrease in arterial blood pressure and increased heart rate. During phase 3, the forced expiration is released resulting in a sudden decrease in arterial blood pressure. During Phase 4, the accumulated venous blood returns to heart and is pumped into the constricted arteries causing an “overshoot” of arterial pressure. Parasympathetic blockade with atropine abolishes the heart rate change, whereas sympathetic blockade does not.

### 3.6.2 Subjects and Methods

We have performed the Valsalva maneuver on 54 subjects: 32 women (age 11 to 70 years, mean 33±16 years) and 22 men (ages 22 to 56 years, mean 33±15 years); among them 1 with multiple sclerosis (n = 60), 3 with palpitations (n = 8, 10, 26); 2 with cardiac insufficiency (n = 8, 20); 3 with hypertension (n = 16, 20, 45); 1 with sleep apnea (n = 38); 1 with gastric ulcer (n = 8, 10) and 1 with hepatitis (n = 5). 17 subjects were white and 37 black.

The first two experimental sessions were performed in the Czech Republic on April 28, 2009 and May 20, 2009. Two additional sessions were performed on January, 27 and 28 in Cameroon. In the first experimental session, all 14 subjects were instrumented with a Biopac MP35 system and following signals were recorded: 3 leads ECG, skin temperature and galvanic skin response. 5 subjects were included in the second experimental sessions with 3 leads ECG and galvanic skin response measurement. The third and fourth sessions included 38 subjects and 3 subjects respectively. ECG, galvanic skin response and skin temperature were measured.

The maneuver was performed in a non-standardized way without galvanometer. A galvanometer and other accessories were not available in our experimental environment. After taking a deep breath, the subject was asked to blow inside an empty plastic bottle (volume 1,5 liters) hold by the instructor at the level of mouth. The subject was encouraged to exhale as strong and as long as he could. The common test duration was 15 seconds. The maneuver was repeated until the heart rate response had a typical VM shape. Subject was sitting during the maneuver as recommended in [149].

The duration of the expiratory straining varies from 10 seconds to 30 seconds in studies but the 15 seconds duration is the most common protocol employed. A 20 seconds protocol is recommended for autonomic assessment. Healthy controls, chronic Chagas’ disease patients have been able to sustain the expiratory strain for 20 seconds very well. In general it depends on the capacity of the subject to sustain a continuous strain, the presence of undesirable effects, and sufficient amplitude to detect autonomic disorders. The maneuver should be considered effective when facial flushing and plethora, neck vein engorgement, and increased muscle tension in the abdominal wall are observed in addition to the full expansion of the thoracic cage. The maximum depth of inspiration before straining, the duration of straining and its magnitude dictated by the magnitude of the intraoral pressure should all be well controlled to generating uniform and persistent elevation of intrathoracic pressure during the straining. The supine position appears to be consensual, but when necessary a sitting or standing position can be used, but these later positions have significant influence on arterial pressure changes and baroreflex sensitivity. It is recommended to repeat the maneuver up to four times for a more precise functional evaluation and the mean of indexes should be obtained. [149]

Subjects who performed the VM were from diverse population groups in the literature. Studies in [150] included 27 diabetic patients: 12 women (2 with diabetes type I and 10 with diabetes type II) and 15 men (4 with diabetes type I and 11 with type II). The age of the patients ranged from 28 to 76 years with a mean of 48 years and the duration of the diabetes range from 1 to 45 years with a mean of 13 years. In [149] the maneuver was performed on 52 healthy subjects (39 men and 13 women) aged 17 to 49 years. Potential complications include retinal hemorrhage, urinary incontinence, syncope, chest pain, arrhythmias, severe hypertensive, hypotensive reactions and cerebral stroke. Subjects with hypertension, hypotension, coronary disease, valve disease, cardiac congenital disease, heart failure, cardiovascular asthenia and previous syncopal attacks should perform the maneuver very cautiously, preferentially using the 15 seconds straining duration. [149]

### 3.6.3 Results

The Valsalva Ratio is determined as the maximum heart rate obtained during the maneuver divided by the lowest heart rate occurring within 30 seconds after the peak. Time change-dependent indexes, such as the time for phase III to IV bradycardia and the velocity for attaining this bradycardia, are more sensitive for the detection of cardiac autonomic dysfunction than amplitude related indexes. An index of the baroreceptor reflex sensitivity can be calculated from the correlation between each systolic pressure peak and the corresponding subsequent R-R interval during the hypertensive overshoot in phase IV. [149]

Figure 3.24 illustrates the dynamics of heart rate changes during the four phases of the maneuver, performed on one representative subject from our experiments.

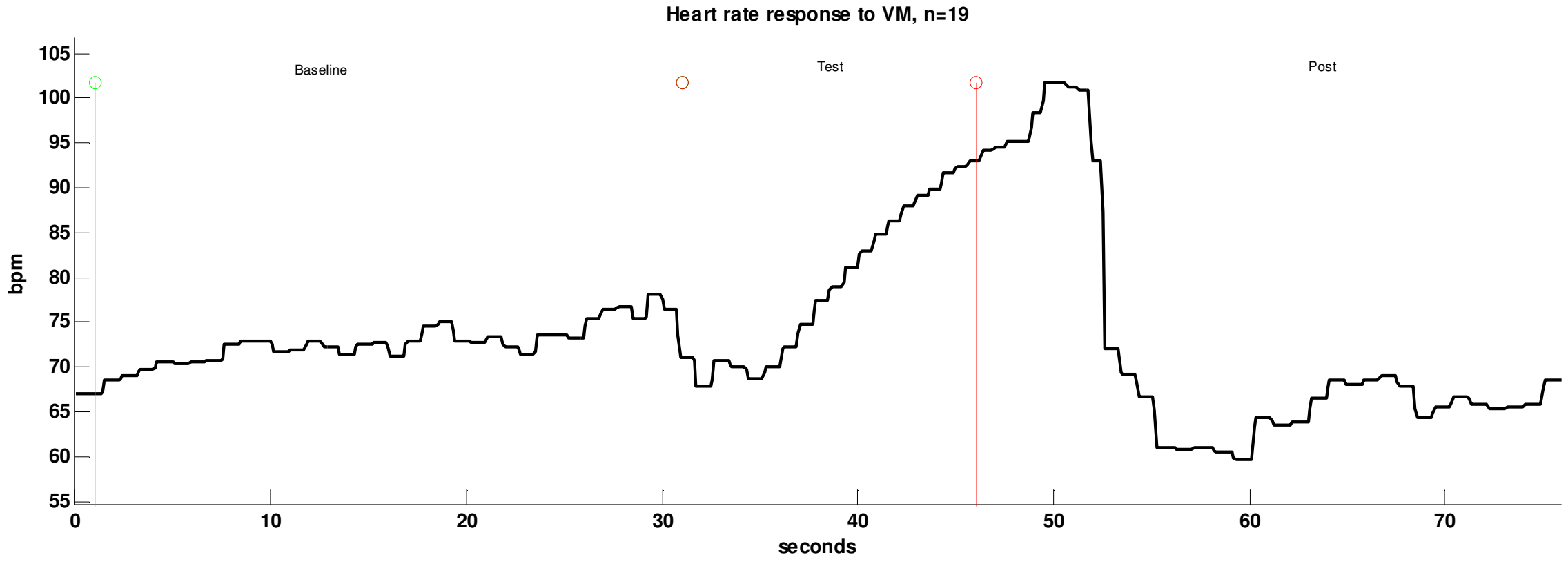

**Figure 3.24 Heart rate response to Valsalva maneuver**
measurements were performed on one representative subject with ID n=19

## 3.7 Other Cardiac Autonomic Tests

### 3.7.1 Active Change of Posture

Change of body posture involves complex interactions between the autonomous nervous system and the cardiovascular system. Postural change from sitting or lying to standing pulls blood to lower parts of the body due to gravitational forces, triggering reflex response to the resulting drop in arterial blood pressure and cerebral perfusion. Subject with orthostatic intolerance disorders manifest signs of syncope, orthostatic hypotension or falls [151]. Orthostatic intolerance is diagnosed on the basis of a drop in arterial blood pressure or an excessive increase in heart rate measured a certain time after the assumption of the upright posture [152]. In [151] subjects were sitting in a straight-backed chair with their legs elevated at 90° in front of them and were then asked to stand. Standing was defined as the moment both feet touched the floor.

In our studies subjects were taught to actively stand up from a sitting position on a chair. They were given a vocal signal for starting standing and care was taken not to displace electrodes while standing. Subject was asked to remain standing with relaxed muscles for 30 seconds. In another experimental setup, electrodes on both hands were hold using an elastic band or gloves, the left arm was stabilized using a sling and subject was asked to stand up from a lying position.

In our studies 16 women (ages between 17 and 54 years, mean 42±10 years) and 7 men (ages between 23 and 56 years, mean 37±12 years) underwent active change of posture. Figure 3.25 shows a typical heart rate response recorded on a subject who moved from a sitting to standing position. Venous blood pooling triggers the baroreflex which responds by increasing heart rate from 52 beats/min to 72 beats/min. Sit from standing causes an initial heart rate increase followed by a stronger decrease (see Figure 3.26).

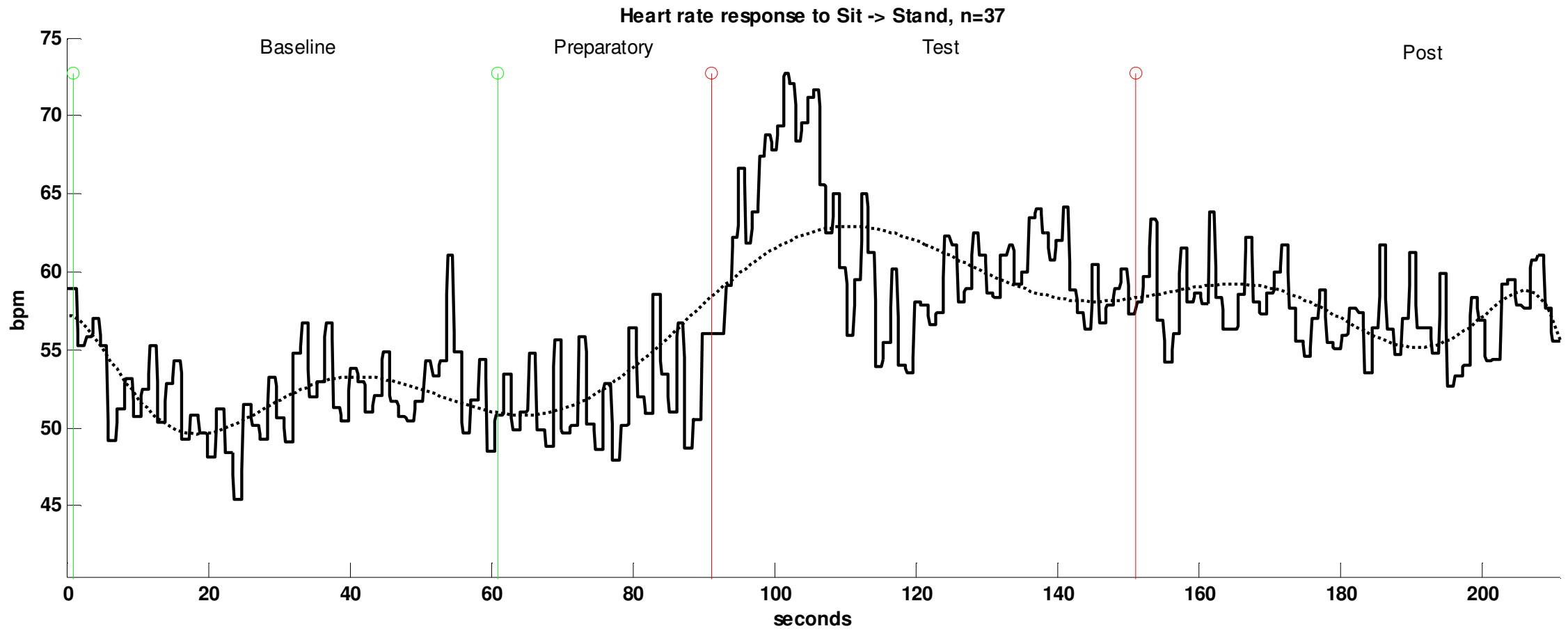

**Figure 3.25 heart rate response to active standing from sitting**
measurements were performed on one representative subject with ID n=37
dashed line is a 10th-degree polynomial interpolation of the signal

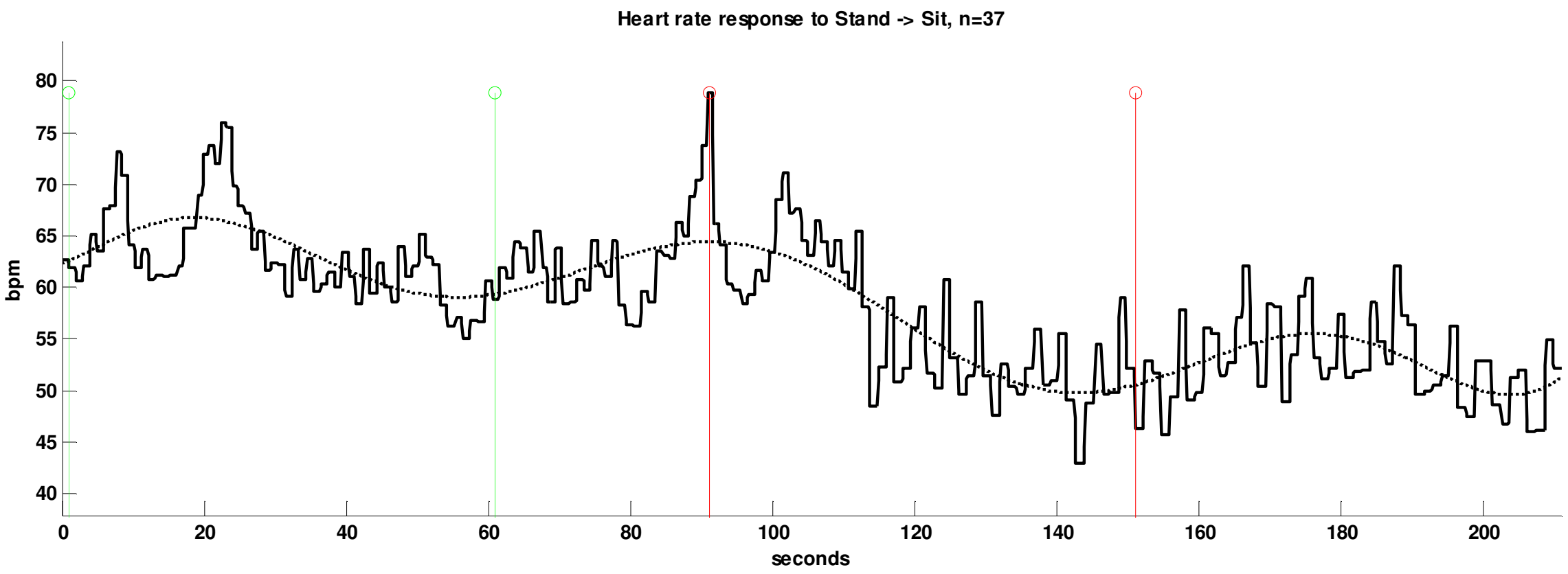

**Figure 3.26 Heart rate response to active sitting from standing**
measurements were performed on one representative subject with ID n=37
dashed line is a 10$^{th}$-degree polynomial interpolation of the signal

### 3.7.2 Tilt Table Test

Tilt table test is another way to induce orthostatic stress. Whereas active standing can be achieved within a couple of seconds, tilt is a slower process where the subject is lying on a tilt table which is then slowly tilted towards the upright position so that regulatory processes are activated even before the subject is fully tilted. Another difference is that limited muscle activity is involved in tilt table test. Furthermore intracranial pressure decrease and venous return from cerebral circulation increases as result of higher gravitational forces between the neck and the torso; meanwhile venous return from lower parts of the body decreases [68].

Our studies included 9 women (ages between 15 and 46 years, mean 26±11 years) and 8 men (ages between 22 and 29 years, mean 23±2 years). After being stabilized on a lying position, the table was manually and slowly tilted up to 70$^{o}$ during approx. 10 seconds. The upright position was kept for approx. 180 seconds. The table was then slowly tilted back to the horizontal position and post-recordings were resumed for another 60 seconds. An example of the corresponding heart rate, recorded in our experiments, is visible on Figure 3.27.

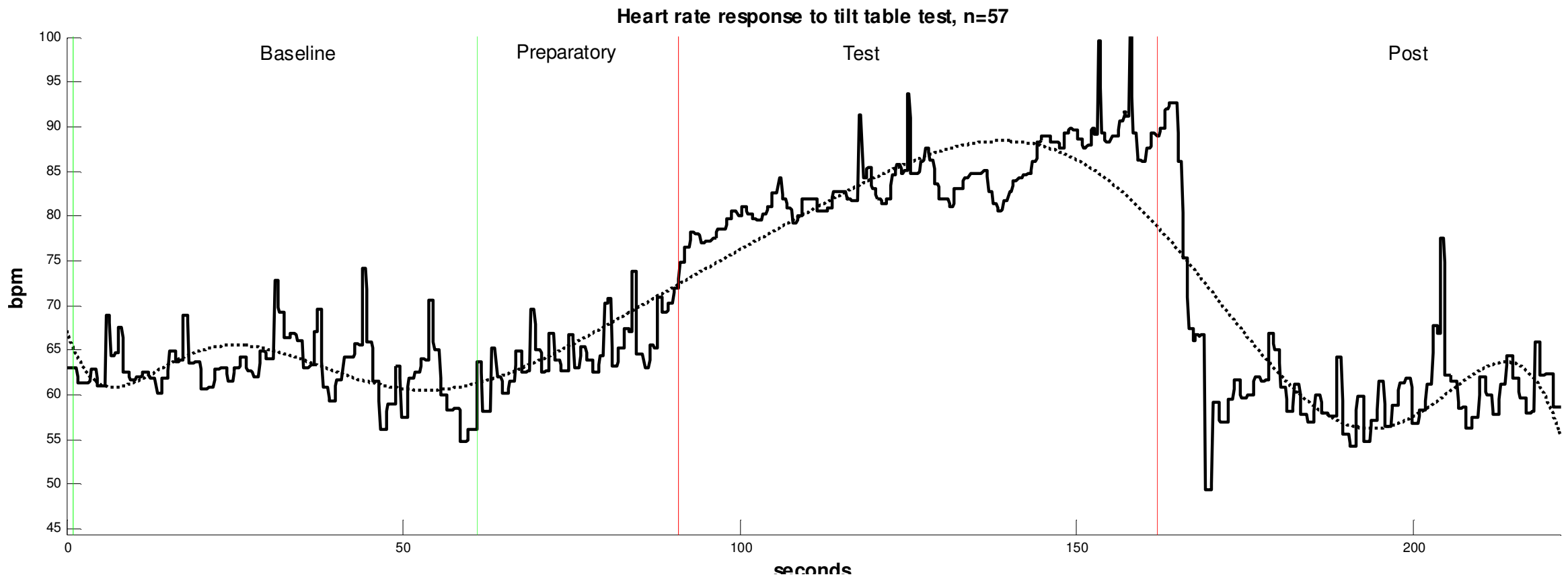

**Figure 3.27 Heart rate response to tilt table test**
measurements were performed on one representative subject with ID n=57
dashed line is a 10$^{th}$-degree polynomial interpolation of the signal

### 3.7.3 Cold Pressor Test

The cold pressor test is performed by immersing hands or feet of the subject into cold water with temperature between 0 and 5 °C. The low temperature excites skin nociceptors which project to brain centers. The response is a strong global sympathetic activation resulting in increased blood pressure, heart rate and in some cases galvanic skin response [153] [154] [155].

In our studies 33 women (ages between 11 and 77 years, mean 36±16 years) and 20 men (ages between 22 and 82 years, mean 42±16 years) underwent the cold pressor test. Both left and right feet were immersed in a bucket filled with iced-water with temperature around 0 °C. Most subjects reported painful sensation but were able to sustain the cold stimuli for 40 to 180 seconds. An example of the corresponding heart rate response, recorded in our experiments, is visible in Figure 3.28.

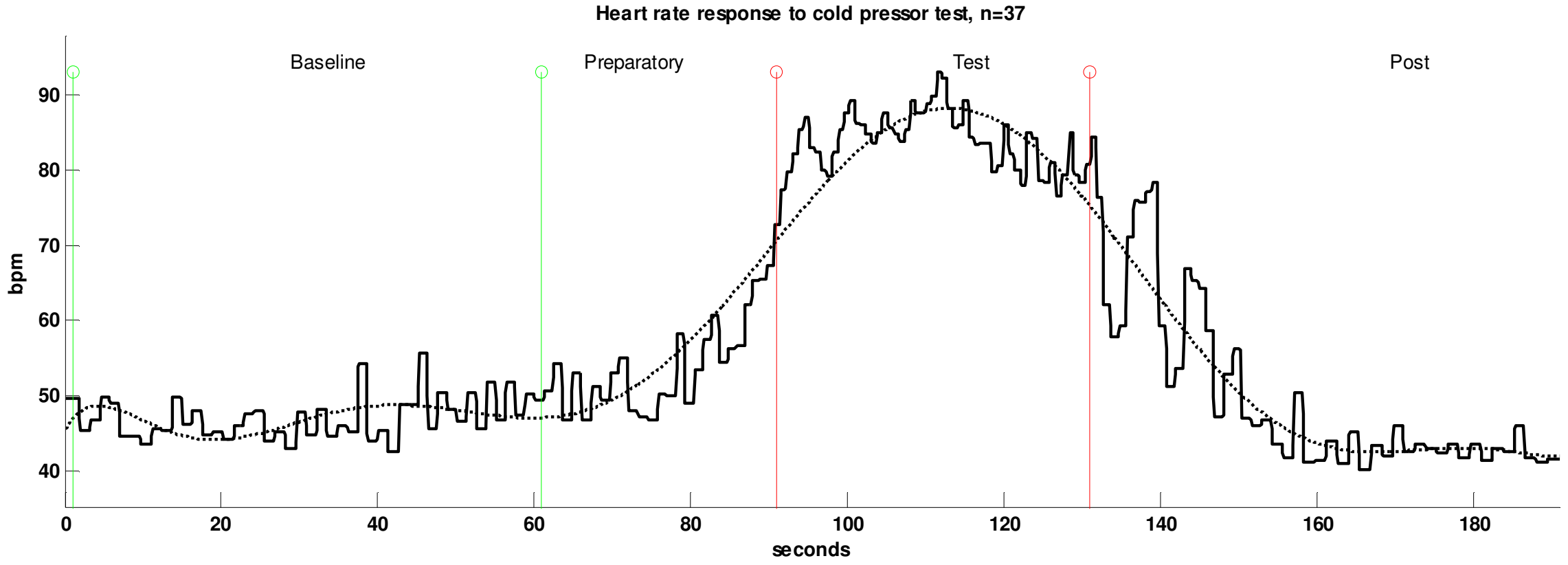

**Figure 3.28 Heart rate response to cold pressor test**
measurements were performed on one representative subject with ID n=37
dashed line is a 10$^{th}$-degree polynomial interpolation of the signal

## 3.8 Experiments on Horses

### 3.8.1 Materials and Methods

In humans, activation of the diving reflex by a cold stimulus to the face (Cold Face Test) results in bradycardia, peripheral vasoconstriction and increased blood pressure (see section 3.2). Immersing hands or feet into cold water during the so-called Cold Pressor Test results in a strong sympathetic response (see section 3.7.3). Responses of horses to Cold Face Test or Cold Pressor Test have not yet been evaluated in the known literature. We undertook this study to assess the effect of both tests on the cardiovascular system of horses and identify potential future development in the application of our model of cardiovascular control in equine organism.

All procedures described were approved at the Equine Clinic of the University of Veterinary and Pharmaceutical Sciences in Brno, Czech Republic. Five healthy clinic owned horses with no evidence of cardiac disautonomia, pulmonary or metabolic diseases were chosen based on physical characteristics such as: age, sex, weight, height, body mass index, degree of acclimatization to cold and level of exercise. We had three stallion and two gelding thoroughbreds, age 5 to 7 years, weight 500 to 600 kg, body score 6 (moderately fleshy). Horses were normotensive and were not taking any medication. They were not feed 3 hours before the test and were not submitted to exercise 24 hours prior to the test.

The tests were performed with the subjects in relaxed, standing position in a quiet room with an ambient temperature of ca. 22°C or in the stage with an ambient temperature of ca. 15°C. The cold stimulus was applied using cold blue gel packs with dimension 20 x 30 cm. Some packs were stored at 4°C in a fridge and the rest at -18°C in a freezer. A Biopac MP150 research system was used together with 3-leads ECG equipment. Biopac Acknowledge software was used to record a 500 Hz signal, which was exported in ASCII files for further processing in Matlab, where RR, QT interval durations, as well as T-wave amplitudes were extracted.

Table 3.8.1. illustrates the experimental setups we performed, using the following factors:

- Stimulus Temperature: *4°C, -18°C* or *20°C*

  Either cold gel packs from the fridge (with a temperature around 4°C) or from the freezer (with a temperature around -18°C) were used. A special setup was realized with warmed packs at 20°C.
- Direct Application: *yes or no*

  In the case of -18°C gel packs, a cotton bandage was used between skin and pack to protect the skin. A direct application means, the cotton bandage was not used and gel packs were directly applied to the skin.
- Body Place: *forehead, eyes, nose, neck, face side, metacarpus or forelimb*

  During the Cold Face Test, the gel packs were applied on the forehead (and maxillary region), between the eyes, on the nose, on the right side of the face or around the neck.

  During the Cold Pressor Test, the gel packs were applied around the distal part of both forelimbs (around the metacarpus, area between the carpus or hock and the fetlock joint), or around the proximal part of the both forelimbs (around the antebrachium, area of the front leg between the carpus and elbow).
- Test Duration: *60 to 300 seconds*

  The cold gel packs were applied for a duration of 60 to 300 seconds.

| Table 3.8.1 Experimental Setups | | | | |
|---|---|---|---|---|
| *Test Name* | *Temp.* | *Direct* | *Body Place* | *Test Duration* |
| Cold Face Test Forehead with 4°C | +4°C | yes | forehead | 60 seconds |
| Cold Face Test Forehead with -18°C | -18°C | no | forehead | 60 seconds |
| Cold Face Test Forehead with -18°C Direct | -18°C | yes | forehead | 60 or 300 seconds |
| Cold Face Test Eyes with -18°C Direct | -18°C | yes | between eyes | 300 seconds |
| Cold Face Test Nose with -18°C Direct | -18°C | yes | nose | 300 seconds |
| Cold Face Test Side with -18°C Direct | -18°C | yes | face side | 300 seconds |
| Cold Face Test Neck with 4°C | +4°C | yes | neck | 60 seconds |
| Cold Face Test Neck with -18°C | -18°C | no | neck | 60 seconds |
| Cold Face Test Neck with -18°C Direct | -18°C | yes | neck | 60 seconds |
| Cold Face Test Head with 20 °C Direct | +20 °C | yes | head | 60 seconds |
| Cold Pressor Test Distal with 4°C | +4°C | yes | metacarpus | 180 seconds |
| Cold Pressor Test Distal with -18 °C | -18°C | no | metacarpus | 180 seconds |
| Cold Pressor Test Distal with -18 °C Direct | -18°C | yes | metacarpus | 180 seconds |
| Cold Pressor Test Proximal with -18 °C Direct | -18°C | yes | forelimbs | 180 seconds |

### 3.8.2 Equine Response to Cold Face Test

In our study horses did not show significant bradycardia as expected when using a cold stimulation of 4°C (see heart rate response in Figure 3.29, cold packs were applied from the 60th to 120th second). Horses were rather afraid of the packs approaching their face and the heart rate did not decrease like it would have been the case in humans. When cold packs at much lower temperature (-18°C) were applied to the forehead with an isolating bandage to protect their skin, the horses also appeared afraid and showed signs of discomfort with increased heart rate (see Figure 3.30, cold packs were applied from the 60th to 120th second).

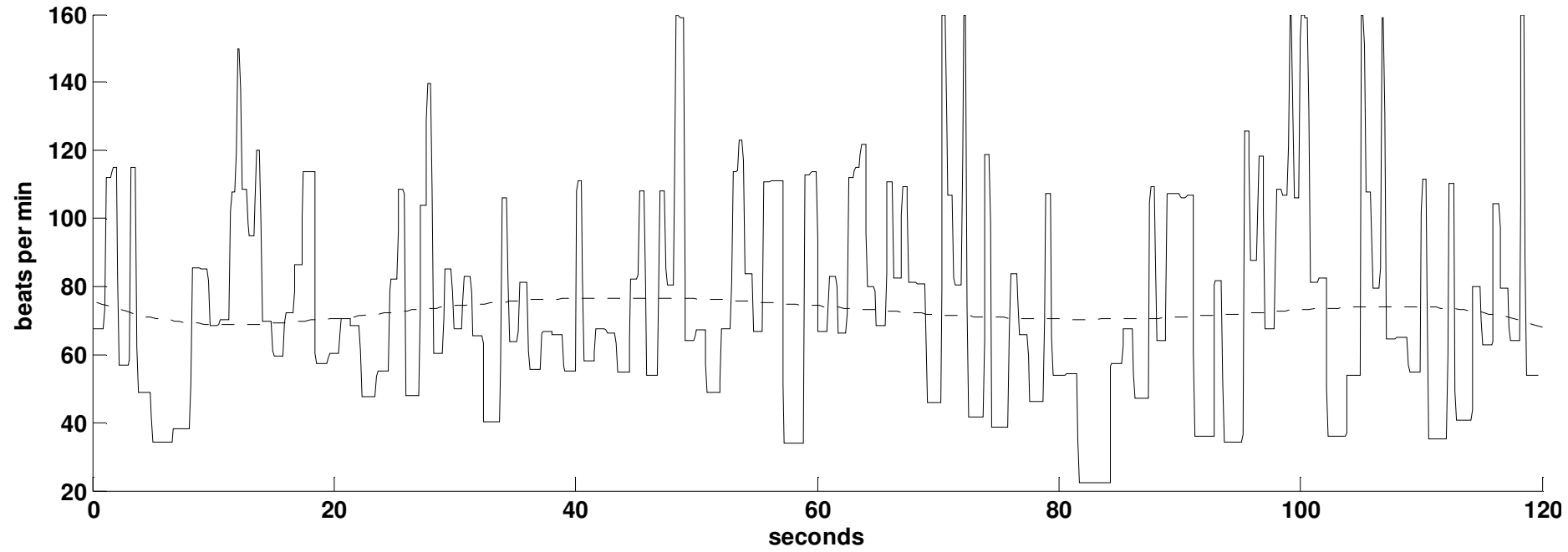

**Figure 3.29 Equine HR Response - Cold Face Test - forehead with 4°C**
cold packs were applied from the 60th to 120th second
dashed line is a 10th-degree polynomial interpolation of the signal

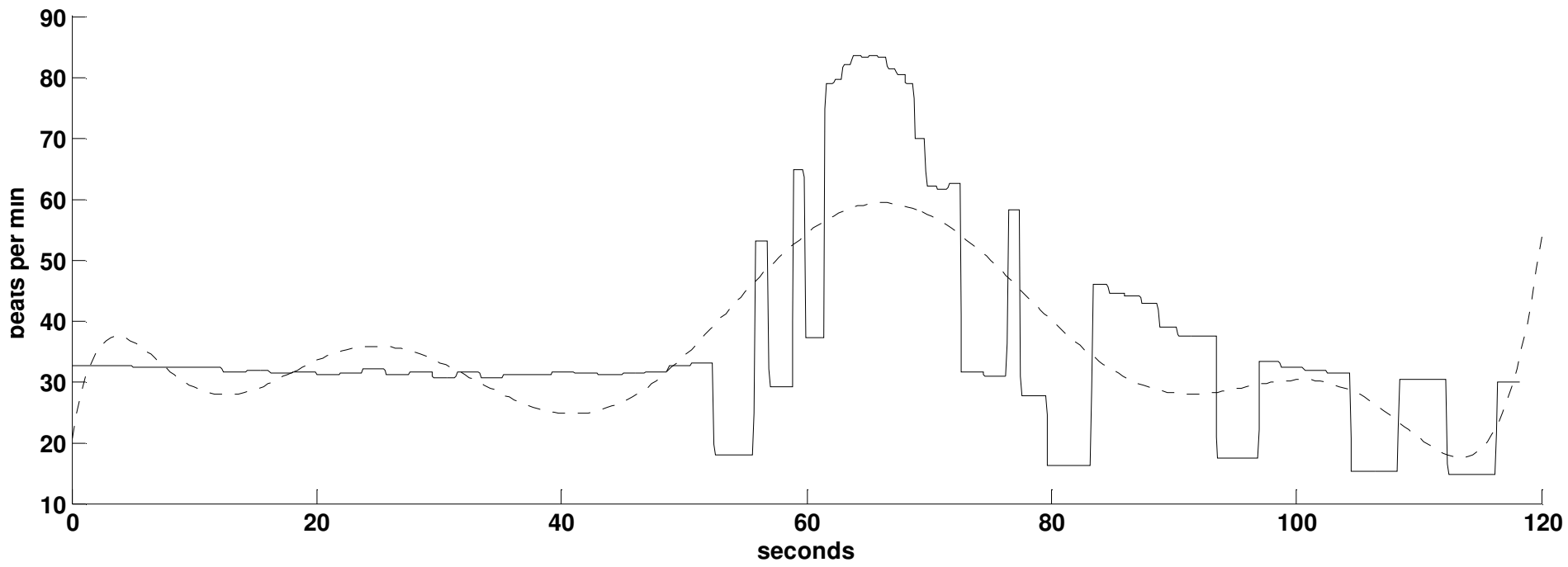

**Figure 3.30 Equine HR Response - Cold Face Test - forehead with -18°C**
cold packs were applied from the 60th to 120th second
dashed line is a 10th-degree polynomial interpolation of the signal

When the same cold packs (-18°C) were directly applied to the face of the horse without isolating bandage, a similar tachycardia response was observed (Figure 3.31, cold packs were applied from the 60th to 120th second). Probably the diving reflex was triggered but was too weak compare to the pain response via skin nociceptors. Noxious cold induces pain and may elicit cardiovascular changes [156]. The QT intervals, which reflect the ventricular depolarization time, also shorten during the test (Figure 3.32). T-wave amplitude of ECG is thought to reflect the sympathetic tone on the heart; it decreases during elevated sympathetic discharge [157]. The T-wave amplitude increase just at the beginning of the response could indicate a surplus of vagal activity. A short time after application of cold stimulus, the T-wave amplitude became lower as shown in Figure 3.33.

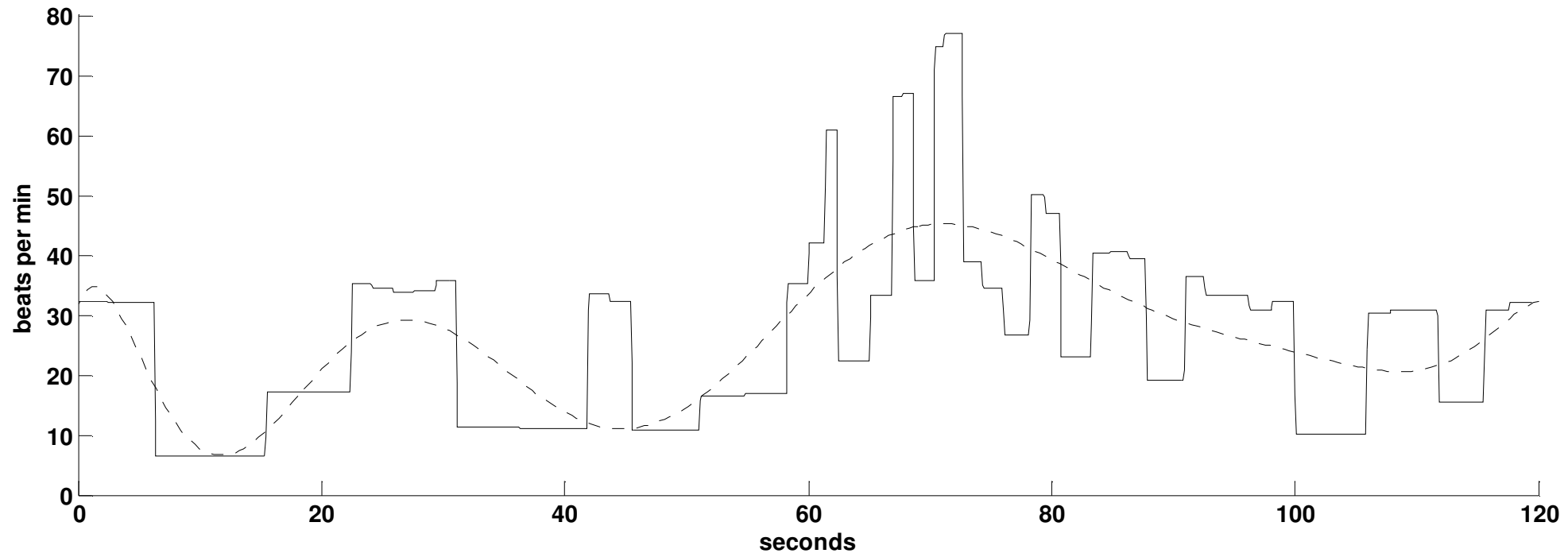

**Figure 3.31 Equine HR Response - Cold Face Test - forehead with -18°C Direct**
cold packs were applied from the 60th to 120th second
dashed line is a 10th-degree polynomial interpolation of the signal

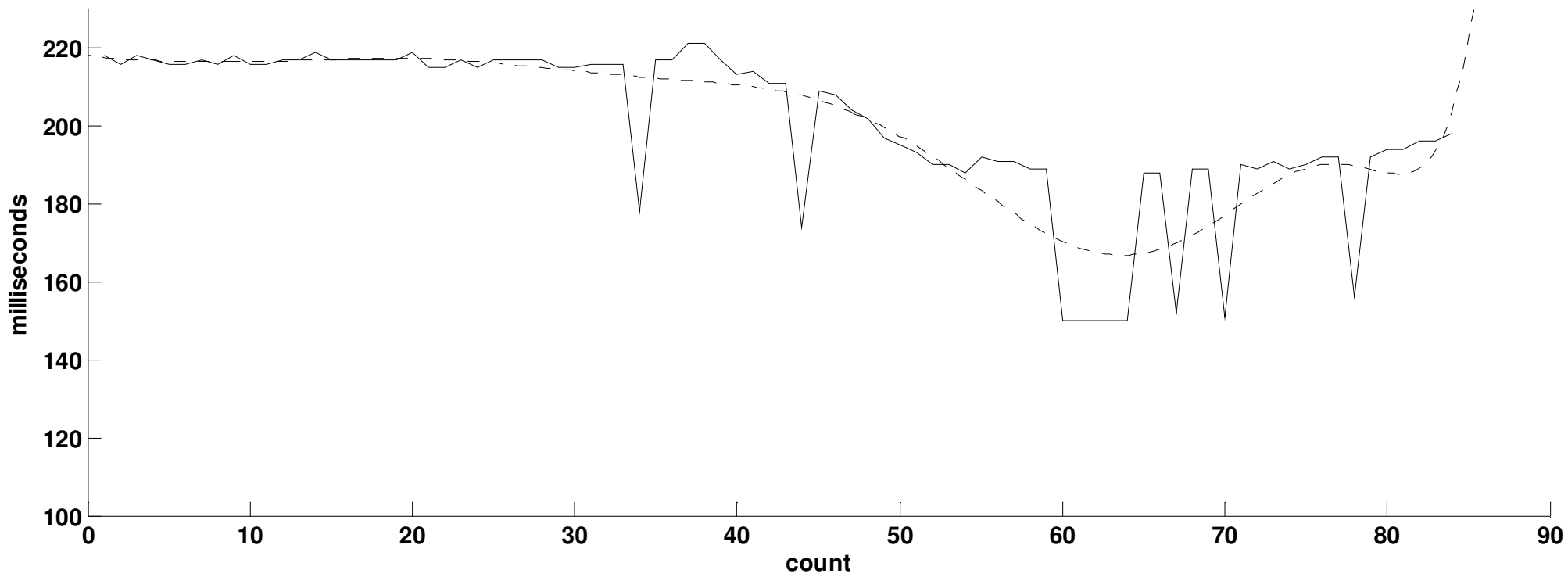

**Figure 3.32 Equine QT Response - Cold Face Test - forehead with -18°C Direct**
first 45 values were recorded before the cold application
last 45 values were recorded during cold application
dashed line is a 10th-degree polynomial interpolation of the signal

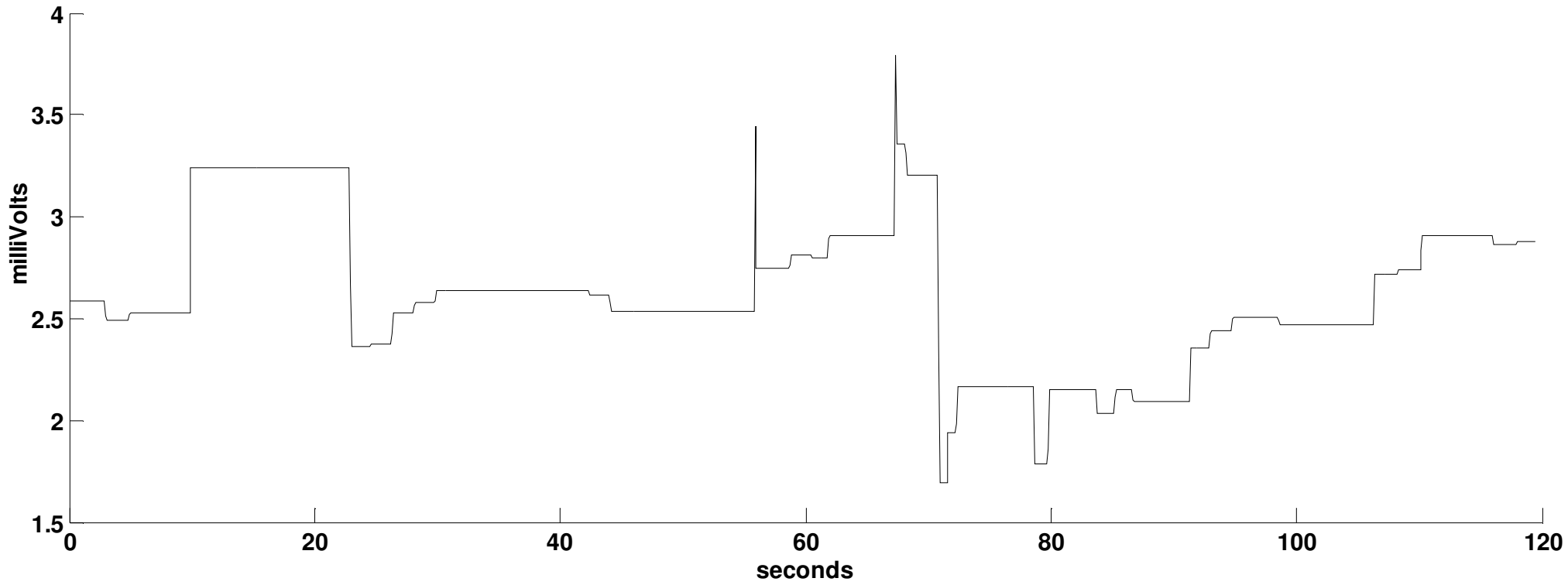

**Figure 3.33 Equine T-wave Amplitude - Cold Face Test - forehead with -18°C Direct**
cold packs were applied from the 60th to 120th second

Since cooling the forehead of horses did not appear to activate the diving reflex as expected, additional experimental setups were arranged to investigate the place of the face and neck where cold receptors are likely to trigger the diving reflex. Applying the cold packs around the neck of the horses did not show significant heart rate change (Figure 3.34, cold packs were applied from the 60th to 120th second). At 4°C the RR interval durations did not change and cold packs at -18°C without skin isolation did not trigger a response either. However weak increase of heart rate was noticed when cold packs at -18°C were directly to the neck with an isolating bandage.

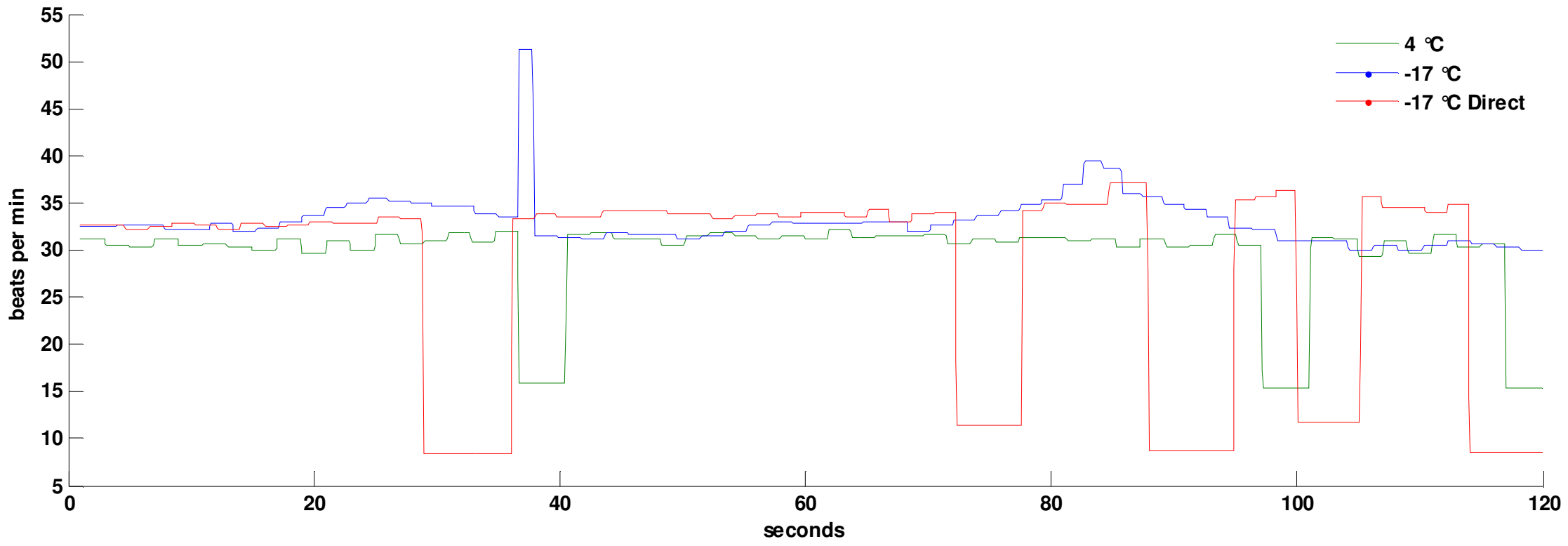

**Figure 3.34 Equine HR Response - Cold Face Test at neck**
cold packs were applied from the 60th to 120th second

We defined additional experimental setups to investigate the heart rate response when cold gel packs were directly applied between the eyes, on the forehead, nose or on the side of the face. Results in Figure 3.35 suggest the eyes and side of face to be the most appropriate body places which might trigger the diving reflex.

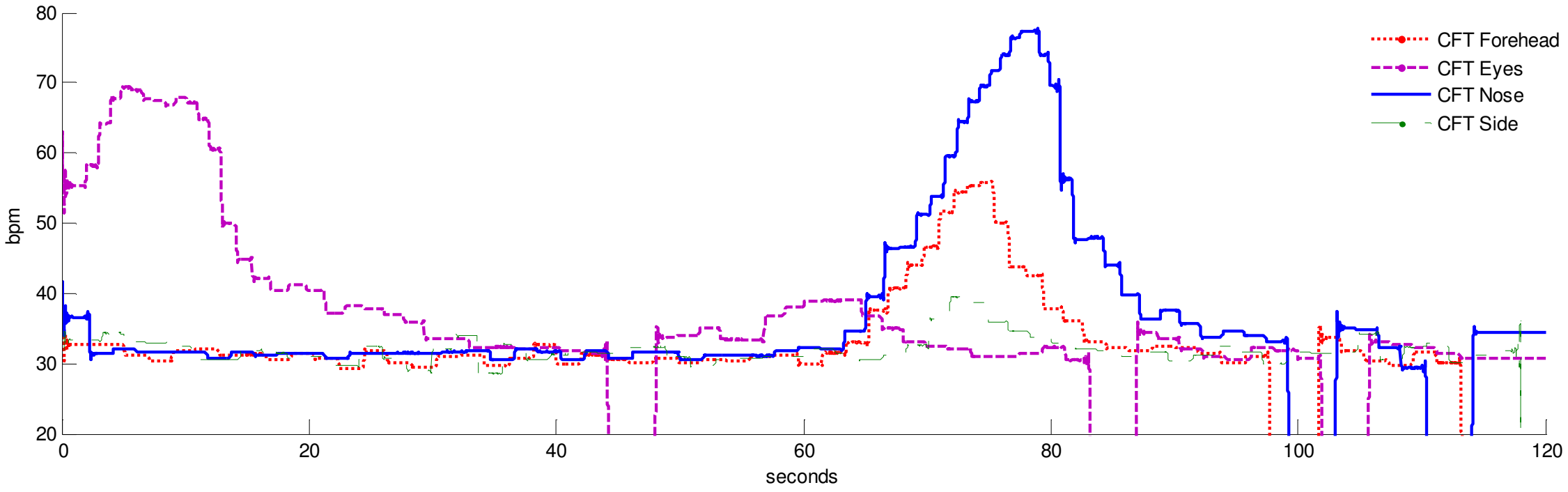

**Figure 3.35 Equine HR Response - Cold Face Test on head**
cold packs were applied from the 60th to 120th second

Horses are very sensible to objects approaching their face. The generally observed tachycardia during various versions of cold face test is likely due to fear or pain. In order to verify this assumption, the blue gel packs were warm up to room temperature (around +20°C) and were applied on different places of the face. Normally most cold receptors and pain receptors will not be activated during such a test. Therefore the observed increase of heart rate is obviously due to fear (see Figure 3.36, warm packs were applied from the 60th to 120th second). The fear factor depends on the location and is more significant on the forehead where the heart rate increased by ca. 60%.

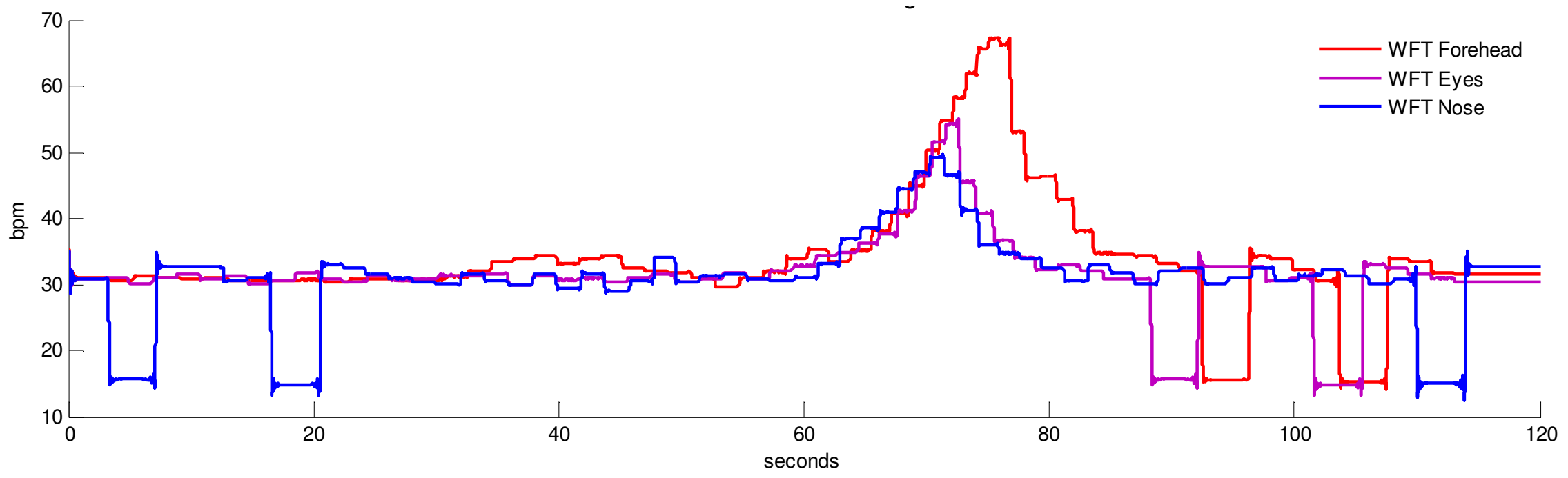

**Figure 3.36 Equine HR Response – Cold Face Test - forehead with +20°C Direct**
cold packs were applied from the 60th to 120th second

We came to the idea of warming the forehead prior to cooling. This placebo effect should reduce the fear factor and develop acclimatization with the experimental conditions. This assumption could be verified on Figure 3.37 and Figure 3.38 (packs were applied from the 300th to 600th second). The subject in Figure 3.37 passed 4 warming tests prior to the cold stimulus and was more familiarized with the blue gel pack than subject in Figure 3.38, who passed only one warm test. Therefore when deducing the increase of heart rate of 60% due to the fear factor from the total heart rate response, it is thinkable that the subjects demonstrated bradycardia via diving reflex but this response was damped by the sympathetically mediated fear.

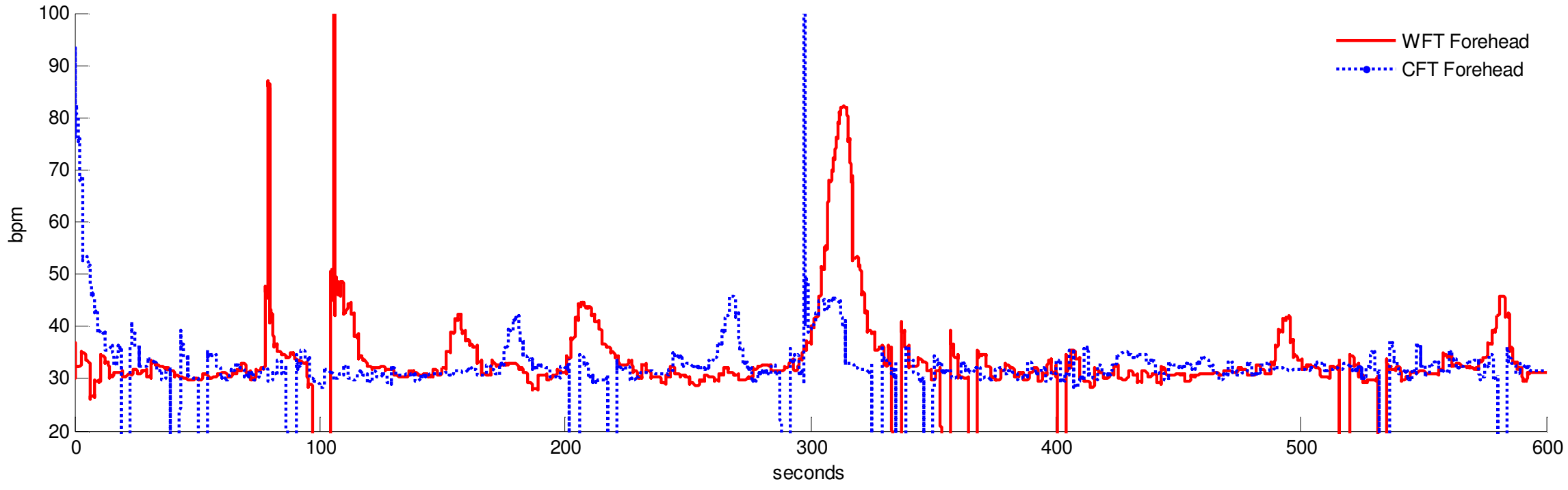

**Figure 3.37 Equine response to Cold Face Test - forehead with -18°C direct in trained subjects**
cold packs were applied from the 300th to 600th second

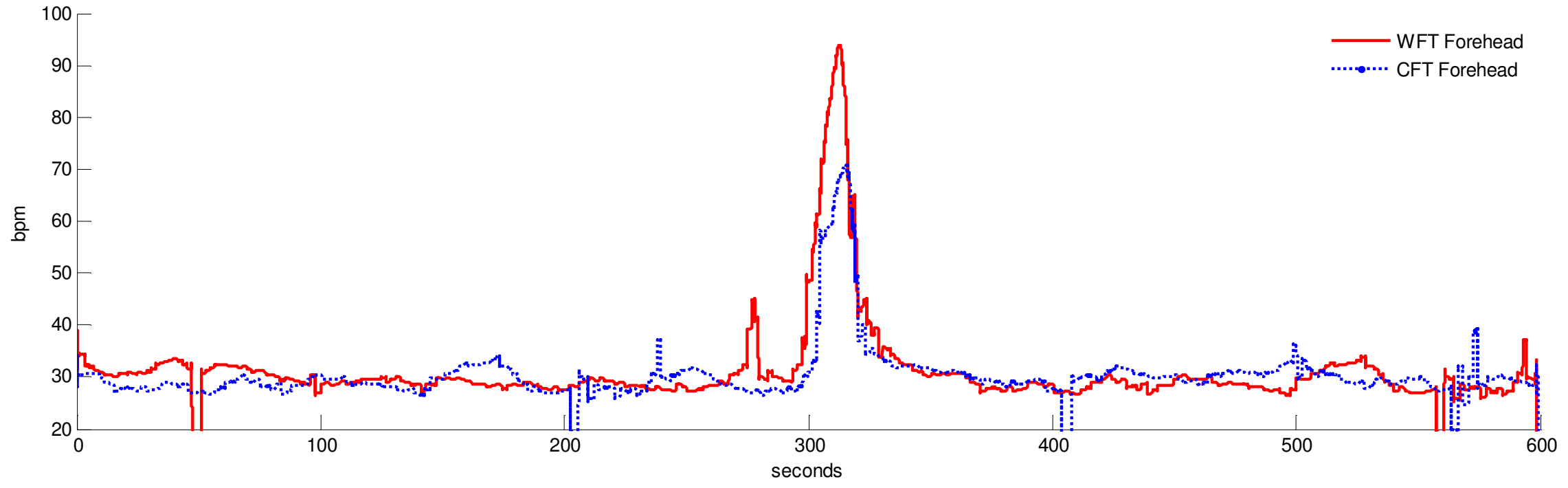

**Figure 3.38 Equine response to Cold Face Test - forehead with -18°C direct in untrained subjects**
cold packs were applied from the 300th to 600th second

### 3.8.3 Equine Response to Cold Pressor Test

Cold pressor test was first performed by applying cold packs at the distal part of forelimbs (cannons). When using an isolating bandage between the skin and the packs, no heart rate change was noticed neither at 4°C nor at -18°C (see Figure 3.39 and Figure 3.40, packs were applied from the 180$^{th}$ to 360$^{th}$ second).

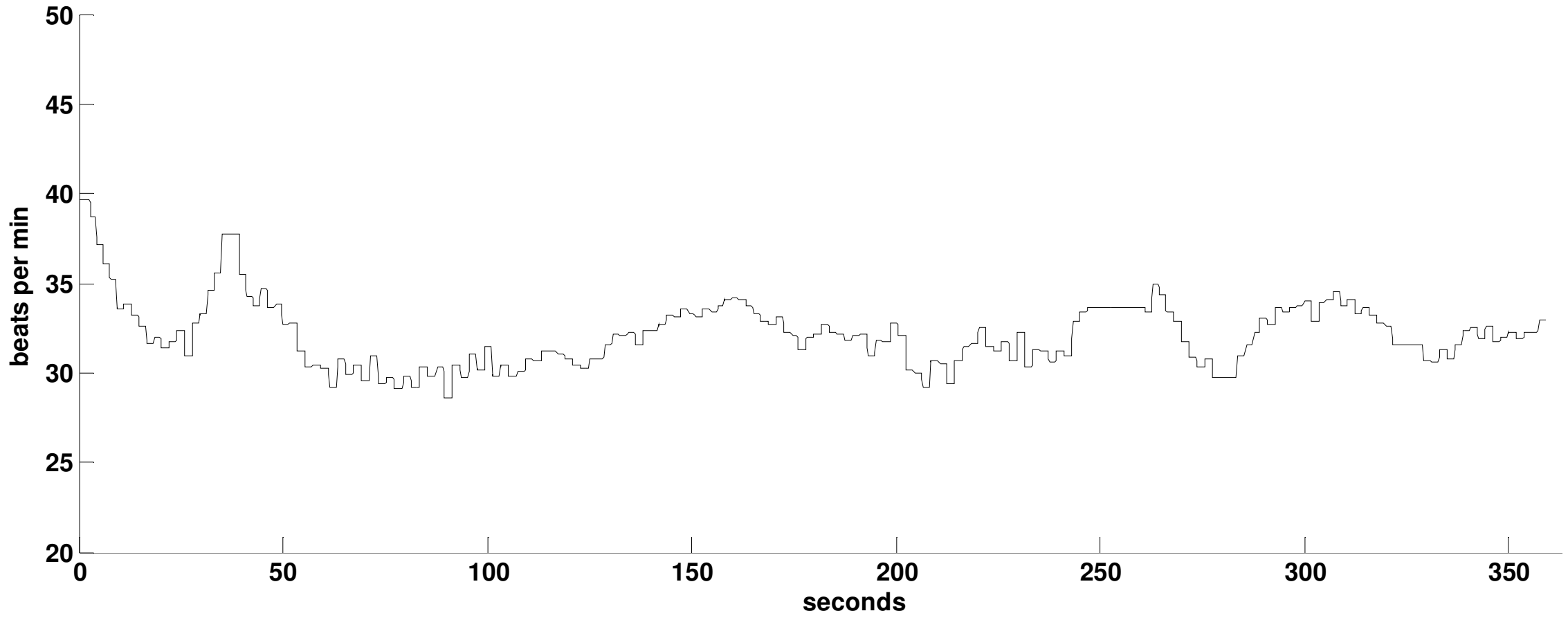

**Figure 3.39 Equine HR Response - CPT at distal position with 4°C**
cold packs were applied from the 180$^{th}$ to 360$^{th}$ second

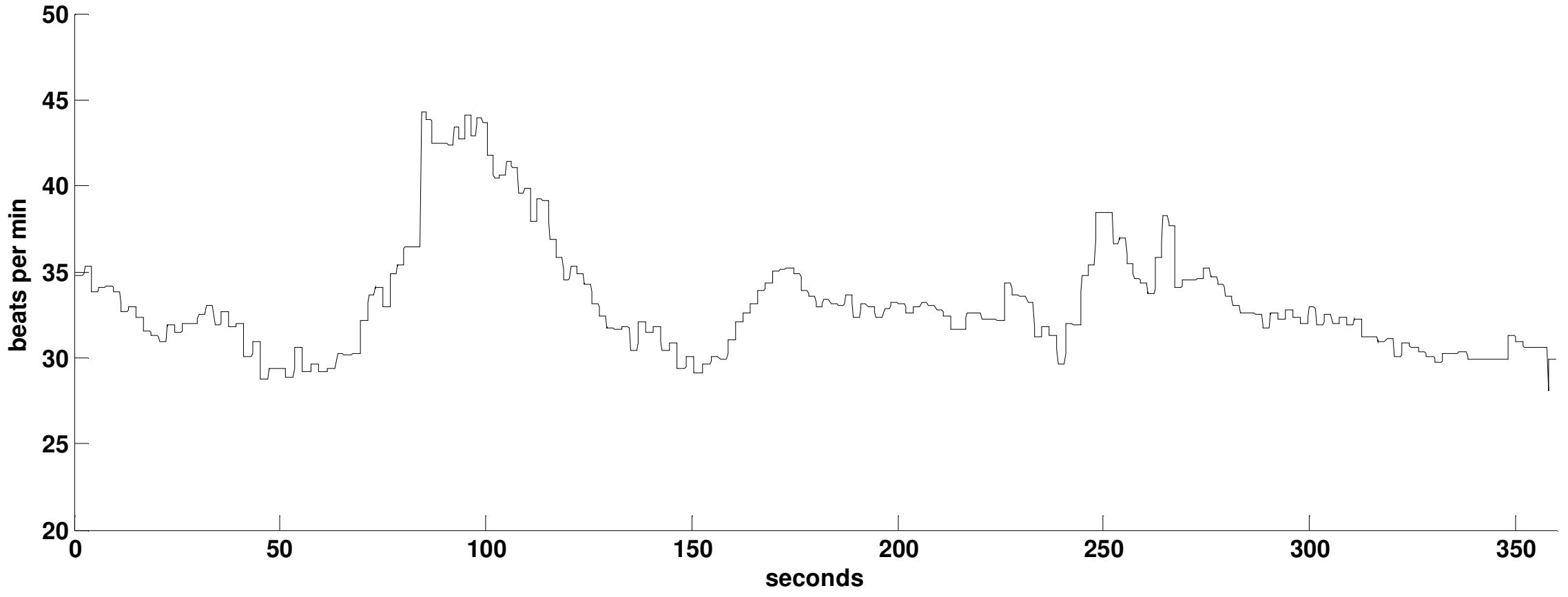

**Figure 3.40 Equine HR Response - CPT at distal position with -18°C**
cold packs were applied from the 180$^{th}$ to 360$^{th}$ second

Applying -18°C directly to the skin without isolation, triggered tachycardia (Figure 3.41, packs were applied from the 180$^{th}$ to 360$^{th}$ second).This increased heart rate was weaker than the one observed in humans. Cold acclimatization could explain the damped response. Horses are used to low ambient temperature on their legs in the stables.

Both Figure 3.42 and Figure 3.43 reveal an interesting observation, QT intervals durations slightly increase whereas RR intervals shorten. This unusual physiological behavior was first investigated in [101] and explained by denser parasympathetic nerve innervations in equine ventricles than in human ventricles. Figure 3.42 illustrates changes in QT intervals durations, the first 90 values were recorded before the application of cold and the last 90 during cold application. Figure 3.43 shows a RR vs. QT plot of the first 60 seconds during cold application.

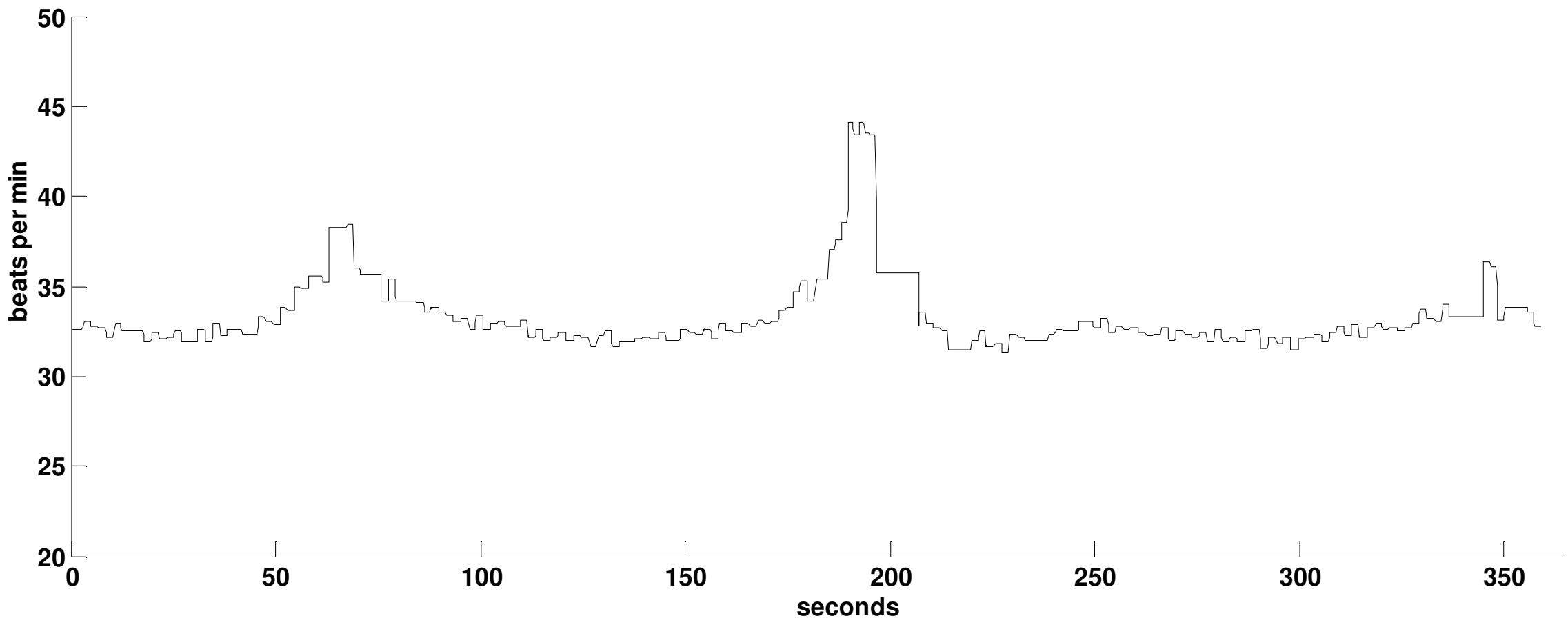

**Figure 3.41 Equine HR Response – CPT at distal position with -18°C direct**
cold packs were applied from the 180th to 360th second

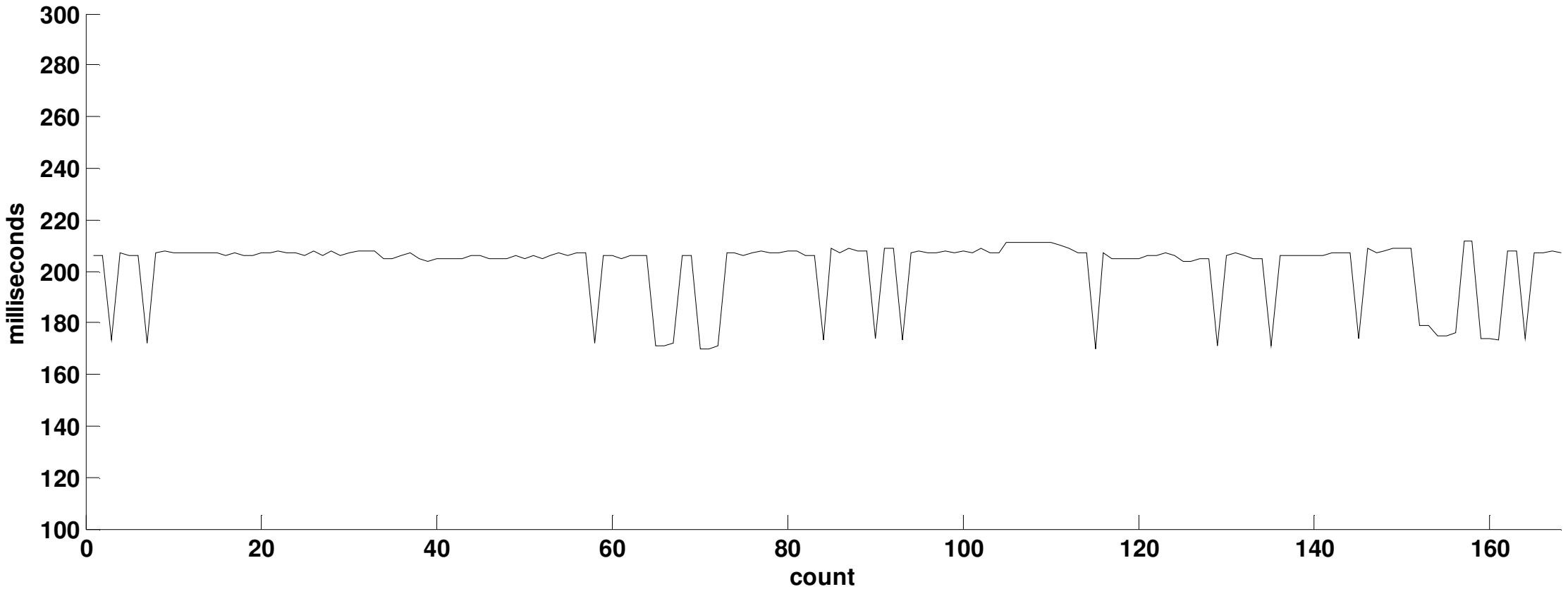

**Figure 3.42 Equine QT Response - CPT at distal position with -18°C direct**
first 90 values were recorded before the cold application
last 90 values were recorded during cold application

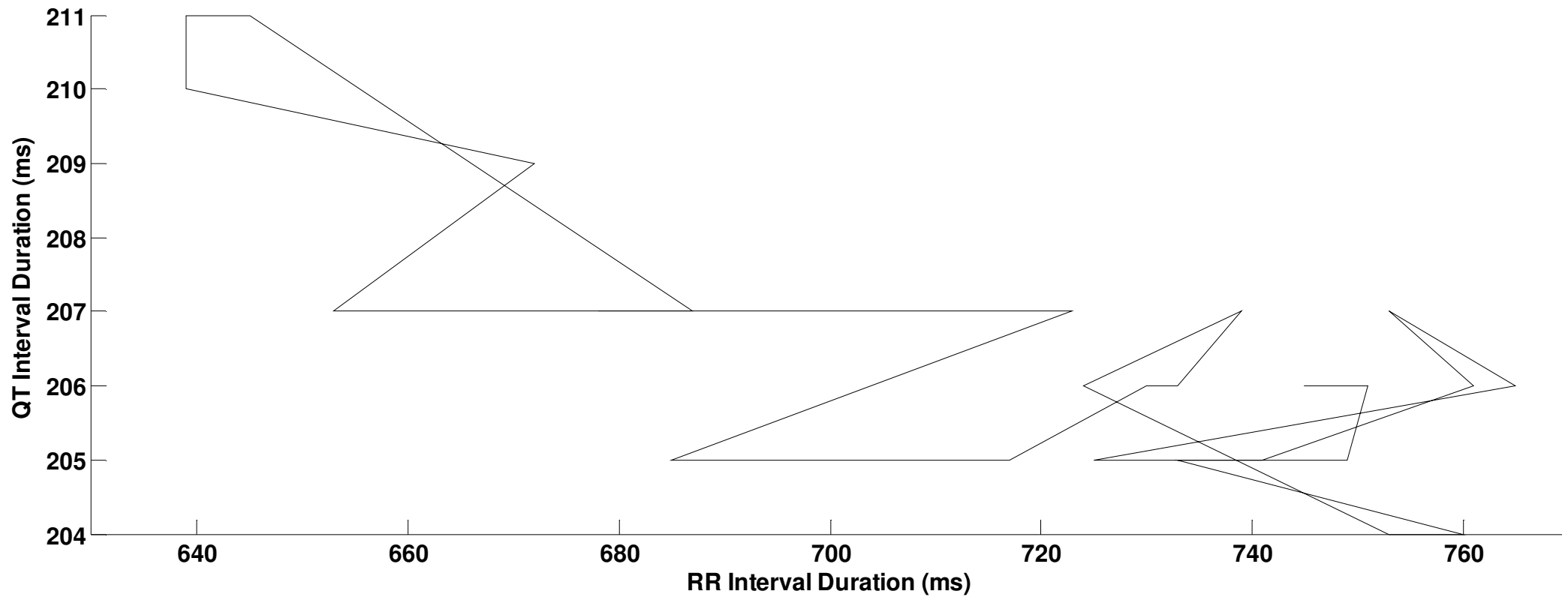

**Figure 3.43 Equine RR-QT Response - Cold Pressor Test Distant with -18°C Direct**

Additional investigations were made with cold packs at -18°C applied directly to the skin of equine forelimbs instead of the metacarpus. The response was less strong than on the metacarpus, however the observed increase of heart rate lasted longer (see Figure 3.44, packs were applied from the 180th to 360th second).

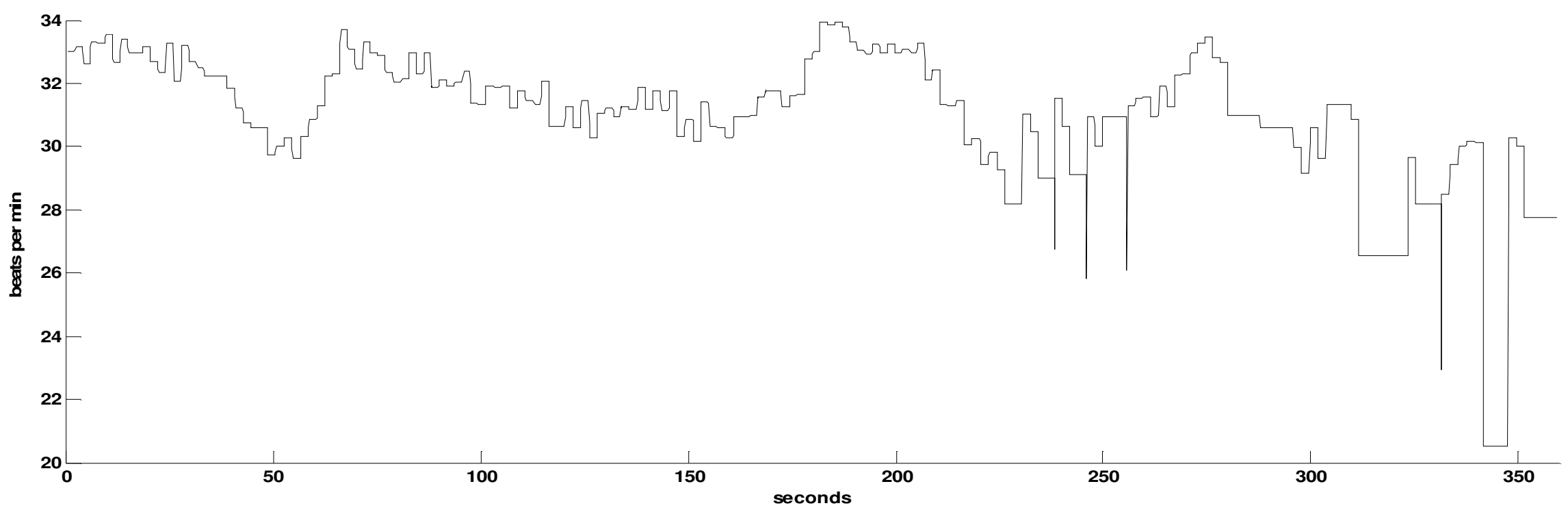

**Figure 3.44 Equine HR Response – CPT at proximal position with -18°C Direct**

As conclusion, equine response to Cold Face Test or Cold Pressor Test is different from the response in humans. Both tests involve emotional factors such as fear and pain. Acclimatization and cold sensibility of the subjects can modulate the results significantly. Horses should be firstly trained with the experimental conditions in order to reduce interferences due to pain and fear. The most appropriate body places where the cold tests could trigger expected cardiovascular response were the forehead, the nose and the forelimbs.

## 3.9 Conclusion

In Chapter 1 we have demonstrated major limitations of current methods for assessing cardiovascular control and proposed novel mathematical models in Chapter 2.

In this Chapter 3 we have undertaken non-invasive experiments in order to gather physiological data for model validation. The experiments were selected from well-known clinical diagnostic tests of cardiovascular function, focusing on acquiring non-invasive cardiovascular signals. Our approach to signal processing was to apply specific known methods, selecting key parameters which are pertinent to the specific problem of cardiovascular control. The intellectual contribution in this chapter is the acquisition of data on human and animal subjects; and the selection of signal processing methods, and specifically on the selection of signal features for inclusion in the integrative models.

In the next Chapter 4 we deal with estimation of values for model parameters for specific experimental data. We will present methods that we believe to be a considerable step towards distinct quantitative assessment of both sympathetic and parasympathetic branches of the autonomic nervous system.

# Chapter 4. Our Approach to Model-based Quantification of Autonomic Nervous Tone

We developed an integrated physiologically-based model of cardiovascular control in Chapter 2. Our model is characterized by input variables, internal state variables, output variables, known constants and unknown parameters. We performed non-invasive experiments in Chapter 3 and collected cardiovascular signals which can now be used to fit the model in order to estimate the values of the unknown parameters for each subject. This is equivalent to resolve an optimization problem. The current chapter will present selected optimization algorithms and develop methods for model-based quantification of autonomic nervous tone.

## 4.1 Optimization Methods

Generally speaking, for some inputs and initial guessed parameter values, the model would produce simulated values of output variables. This simulated behavior of the system can differ from the observed behavior obtained during physical experimentation. Parameter estimation aims to minimize that difference. The goal of this activity is to find the unique set of parameter values for which the difference between simulated and observed behavior is closed to zero. That difference is called objective function. Estimating the parameters is an optimization problem of the objective function. There is a range of methods for parameters estimation including approaches based on non-linear least squares, maximum likehood, Bayesian estimation, simplex method … They can be divided in two groups: unconstrained optimization methods, where the objective function is a function of many variables without any constraints; and constrained optimization, where variables of the objective function are constrained.

In the following paragraphs we will present selected methods for local unconstrained and constraint optimization which will be applied for estimating values of parameters from Table 2.9.4 from section 2.9. Some theoretical background of the methods can be found in more details in the eBook by Frandsen[1].

### 4.1.1 Newton-like Methods

In the unconstrained optimization problem, we look for the smallest, i.e. a local minimizer of a real-valued function $f(x)$ where $x$ is a vector of real variables. In other words, we seek a vector $x^*$, such that $f(x^*) \leq f(x)$ for all $x$ close to $x^*$.

Newton's method gives rise to a wide and important class of algorithms that require computation of the *gradient vector* $f'(x)$ and the *Hessian matrix* $f''(x)$. Although the computation or approximation of the Hessian can be a time-consuming operation, there are many problems for which this computation is justified. The method forms a quadratic model of the objective function around the current iterate $x$: $f(x+h) \approx q(h)$. The model function is defined by: $q(h) = f(x) + h^T f'(x) + \frac{1}{2} h^T f''(x) h$. The idea is to minimize the model $q$ at the current iterate. If $f''(x)$ is positive definite, then $q$ has a unique minimizer at a point where the gradient of $q$ equals zero, i.e. where $f'(x) + f''(x)h = 0$.

[1] Frandsen P. E., Jonasson K., Nielsen H. B., and Tingleff O. (2004). *Unconstrained Optimization (3rd Edition ed.)* [PDF Document]. http://www2.imm.dtu.dk/pubdb/views/edoc_download.php/3217/pdf/imm3217.pdf (Accessed May 18, 2010)

Hence, in Newton's method the new iteration step is obtained as the solution to this last equation. The algorithm is given by:

$$
\begin{aligned}
&begin\\
&\quad x := x_0\\
&\quad repeat\\
&\quad\quad Solve\ f''(x)h_0 = -f'(x)\\
&\quad\quad x := x + h_0\\
&\quad until\ stopping\ criteria\ satisfied\\
&end
\end{aligned}
$$

Convergence is guaranteed if the starting point is sufficiently close to a local minimizer $x^*$ at which the Hessian is positive definite. Moreover, the rate of convergence is quadratic. In most circumstances, basic Newton method has to be modified to achieve convergence.

### 4.1.2 Levenberg-Marquardt Method

Some difficulty is caused in Newton-like methods by non-positive definite Hessian matrices. In this case the underlying quadratic function does not have a unique minimizer and the method is not defined.

A number of methods have been proposed to overcome this problem. We shall concentrate on the so-called damped Newton methods, which are considered to be the most successful in general. The framework for this class of methods is:

$$
\begin{aligned}
&Solve\ (f''(x) + \mu I)h_{dn} = -f'(x)\\
&x := x + \alpha h_{dn}\\
&adjust\ \mu
\end{aligned}
$$

In Levenberg–Marquardt type methods $\mu$ is updated in each iteration step. Given the present value of the parameter, the Cholesky factorization of $f''(x) + \mu I$ is employed to check for positive definiteness, and $\mu$ is increased if the matrix is not significantly positive definite. Otherwise, the solution $h_{dn}$ is easily obtained via the factorization as follows.

$$
\begin{aligned}
&begin\\
&\quad x := x_0\\
&\quad \mu := \mu_0\\
&\quad found := false\\
&\quad k := 0\\
&\quad repeat\\
&\quad\quad while\ f''(x) + \mu I\ not\ positive\ defined\\
&\quad\quad\quad \mu := 2\mu\\
&\quad\quad Solve\ (f''(x) + \mu I)h_{dn} = -f'(x)\\
&\quad\quad Compute\ gain\ factor\ r = \frac{f(x) - f(x + h_{dn})}{q(0) - q(h_{dn})}\\
&\quad\quad if\ r > \delta\\
&\quad\quad\quad x := x + h_{dn}\\
&\quad\quad\quad \mu := \mu * \max\left\{\frac{1}{3}, 1 - (2r-1)^3\right\}\\
&\quad\quad else\\
&\quad\quad\quad \mu := 2\mu\\
&\quad\quad k := k + 1\\
&\quad\quad Update\ found\\
&\quad until\ found\ or\ k > k_{max}\\
&end
\end{aligned}
$$

### 4.1.3 Non-Linear Least Squares Method

The nonlinear least squares problem has the following general form:

$$\begin{gathered}Find\ x^*, a\ local\ minimizer\ of\\ F(x) = \frac{1}{2}\sum_{i=1}^{m}(f_i(x))^2 = \frac{1}{2}\|f(x)\|^2 = \frac{1}{2}f(x)^T f(x)\\ where\ f_i: \mathbb{R}^n \to \mathbb{R}\ are\ given\ functions\end{gathered}$$

Least squares problems often arise in data-fitting applications. Suppose that some physiological process is modeled by a nonlinear function $M(x, t_i)$ that depends on a parameter vector $x$ and time $t_i$. If $b_i$ is the measured output of the process at time $t_i$, then we can compute the discrepancy between the predicted and observed outputs as residuals $f_i(x) = M(x, t_i) - b_i$.

Provided that $f$ has continuous second partial derivatives, we can write its *Taylor expansion* as $f(x+h) = f(x) + J(x)h + O(\|h\|^2)$where $J$ is the Jacobian.

Thus, the gradient is $F'(x) = J(x)^T f(x)$ and the Hessian is $F''(x) = J(x)^T J(x) + \sum_{i=1}^{m} f_i(x) f_i''(x)$. If $x^*$ is a local minimizer, then $F'(x^*) = 0$.

In many practical circumstances, the residuals are small at the solution and the Gauss-Newton method can be used to find that local minimizer. The Gauss–Newton method is based on a linear approximation to the components of $f$ in the neighborhood of $x$: for small $\|h\|$ we see from the Taylor expansion that $f(x+h) \cong f(x) + J(x)h$. Therefore $F(x+h) \cong L(h) = F(x) + h^T J(x)^T f(x) + \frac{1}{2}h^T J(x)^T J(x)h.$

The Gauss-Newton step $h_{gn}$ minimizes $L(h)$ using the algorithm below:

$$\begin{aligned}&begin\\ &\quad x := x_0\\ &\quad found := false\\ &\quad k := 0\\ &\quad repeat\\ &\qquad Solve\ \big(J(x)^T J(x)\big)h_{gn} = -J(x)^T f(x)\\ &\qquad x := x + h_{dn}\\ &\qquad k := k + 1\\ &\qquad Update\ found\\ &\quad until\ found\ or\ k > k_{max}\\ &end\end{aligned}$$

### 4.1.4 Nelder-Mead Method

Physiologically based models usually adopt some simple parameters constraints in form of a finite range (a minimum and a maximum value). This is a bound-constrained optimization problem. More complicated linear and nonlinear constraints between parameters can be defined. The structure of the most common constrained optimization problems is essentially:

$$\begin{aligned}&minimize\ f(x), x \in \mathbb{R}^n\\ &subject\ to\ c_i(x) = 0, i \in E\\ &\qquad\qquad\quad c_i(x) \geq 0, i \in I\end{aligned}$$

where $f(x)$ is the objective function. $c_i(x)$ are the constraints. $E$ is the index set of equality constraints. $I$ is the index set of inequality constraints. The most simple type of unconstrained optimization is obtained when the functions $f(x)$ and $c_i(x)$ are linear functions of $x$. This is the so-called linear programming problem, which can be solved using the downhill simplex method or Nelder-Mead method [158]. This method requires only function evaluations, not derivatives. It approximately finds a locally optimal solution to a problem with N variables when the objective function varies smoothly. In the N-dimensional space, a simplex is a polyhedron with N+1 vertices. The method chooses the N+1 points and defines an initial simplex. The method iteratively updates the worst point by four operations: reflection, expansion, one-dimensional contraction, and multiple contraction. Reflection involves moving the worst point (vertice) of the simplex (where the value of the objective function is the highest) to a point reflected through the remaining N points. If this point is better than the best point, then the method attempts to expand the simplex along this line. This operation is called expansion. On the other hand, if the new point is not much better than the previous point, then the simplex is contracted along one dimension from the highest point. This procedure is called contraction. Moreover, if the new point is worse than the previous points, the simplex is contracted along all dimensions toward the best point and steps down the valley. By repeating this series of operations, the method finds the optimal solution.

## 4.2 Quantification of Resting Autonomic Tone

### 4.2.1 Resting Parasympathetic Tone

Parasympathetic tone is characterized by the model parameter $f_{ev,0,max}$ (maximal intrinsic firing rate of vagal nerves). It has a typical value of 1 Hz at intrinsic heart rate (i.e. 100 beats/min). Subjects with low resting heart rate will have a higher tone (e.g. 20 Hz for 40 beats/min). In order to estimate the individual values for each subject who participated in our studies, a least-squares formulation minimizing the errors between measured heart rate $HR_{meas}$ and simulated heart rate $HR_{sim}$ was defined. We used measured heart rate obtained from baseline recordings at rest during quiet breathing. An algorithm was developed, that goes through all cardiac autonomic tests performed by the subject (using the metadata structure introduced in Figure 3.11) and selects the baseline recording with the lowest mean heart rate. Selecting baseline recording at rest with lowest heart rate might ensure that parasympathetic activity is the main source of control of heart rate. The corresponding cost function, i.e. the discrepancy between measured and simulated values is defined as optimization criteria as follows, where $n$ is the number of data points.

$$J_{f_{ev,0,max}} = \sum_{i=1}^{n} \left[ HR_{meas}(i) - HR_{sim}(i, f_{ev,0,max}) \right]^2 \tag{4.2.1}$$

The model presented in Chapter 2 was implemented in the Matlab Simulink environment. All model inputs (refer to Table 2.9.1) were set to null as well as synaptic weight constants which modulate sensory inputs to parasympathetic premotor neurons (refer to Table 2.5.5). This setup allows us to simulate a physiological condition where parasympathetic efferent activity is determined only by tonic vagal discharge, and not by sensory inputs. In order to calculate the value of parameter $f_{ev,0,\max}$ that minimizes the cost function $J_{f_{ev,0,\max}}$ we implemented the following algorithm in the Matlab environment.

```
Load measured baseline heart rate data for all experiments and Determine HR_meas with the lowest average
Set subject's age as input of model
Set all other inputs to null
Set all synaptic weights that modulate sensory inputs to parasympathetic premotor neurons to null
Loop by varying model parameter f_ev,0,max in the interval [1 .. 20]
    Simulate the model for the current value of f_ev,0,max and calculate the corresponding HR_sim
    Calculate and store the corresponding value of cost function J_f_ev,0,max
End Loop
Return the value of f_ev,0,max for which J_f_ev,0,max was the lowest
```

The algorithm loads measured heart rate data and calculates the best estimates for model parameter $f_{ev,0,\max}$ in the boundary [1 .. 20] Hz. This boundary was chosen because it best agrees with the minimal resting heart rate (ca. 40 beats/min, subject n=37) and maximal resting heart rate (ca. 100 beats/min, subject n=7) that we encountered in our studies. $f_{ev,0,\max}$ was estimated for all 72 subjects after 13 hours of simulations on a 2.1 GHz processor with 1.5 GB RAM.

Results show higher parasympathetic tone on young subjects with a decreasing trend with increasing age as depicted in Figure 4.1, what agrees with heart rate variability measurements in literature [45]. Young female subjects had a mean parasympathetic tone 5.65±3.11 Hz whereas young male subjects had 7.31±3.16 Hz. A possible explanation why young women had a lower tone is that this group included 6 subjects (ca. 26% of all young women) with mean age 16 years. At such young age, the intrinsic heart rate is high causing the resting heart rate to be elevated even with high parasympathetic activity. Parasympathetic tone is a major factor determining resting heart rate as shown in Figure 4.2.

Young black men had the highest parasympathetic tone (11±5.66 Hz), which however decreases with aging faster than by white men. With a parasympathetic tone averaging 7.15±3.29 Hz, young white women surpassed black female subjects by 47%. Subjects with cardiovascular diseases were not distinguished from others according to their parasympathetic tone. We obtained the mean value 4.25±2.55 Hz. The only observation we made, concerns 3 subjects with palpitations, who had lower parasympathetic tone (2±0.5 Hz).

Since the resting parasympathetic tone might change over the time, it is useful to know how an experimental setup might impact the tone. For example tonic changes during cold application on the forehead gives information about the degree of heart rate response, thus provides an evaluation of trigemino-cardiac reflex. As explained in section 2.5.2, our model is able to predict the firing rate of parasympathetic premotor neurons $f_{ev_{pm}}$ for given experimental data. We denote the simulated premotor neurons signal corresponding to maximal parasympathetic tone by denoted by $\left\lfloor f_{ev_{pm}} \right\rfloor_{max}$. The simulated premotor neurons signal corresponding to current parasympathetic tone is denoted by $\left\lfloor f_{ev_{pm}} \right\rfloor_{current}$. We define parasympathetic level $T_{par}$ in a normalized scale between 0 and 1 as the ratio between both signals as follows.

$$T_{par} = \frac{\left\lfloor f_{ev_{pm}} \right\rfloor_{current}}{\left\lfloor f_{ev_{pm}} \right\rfloor_{max}} \qquad (4.2.2)$$

Simulation of $T_{par}$ will be developed in details in the frame of experimental vagal maneuvers in sections 4.3.2, 4.5.2 and 4.5.3.

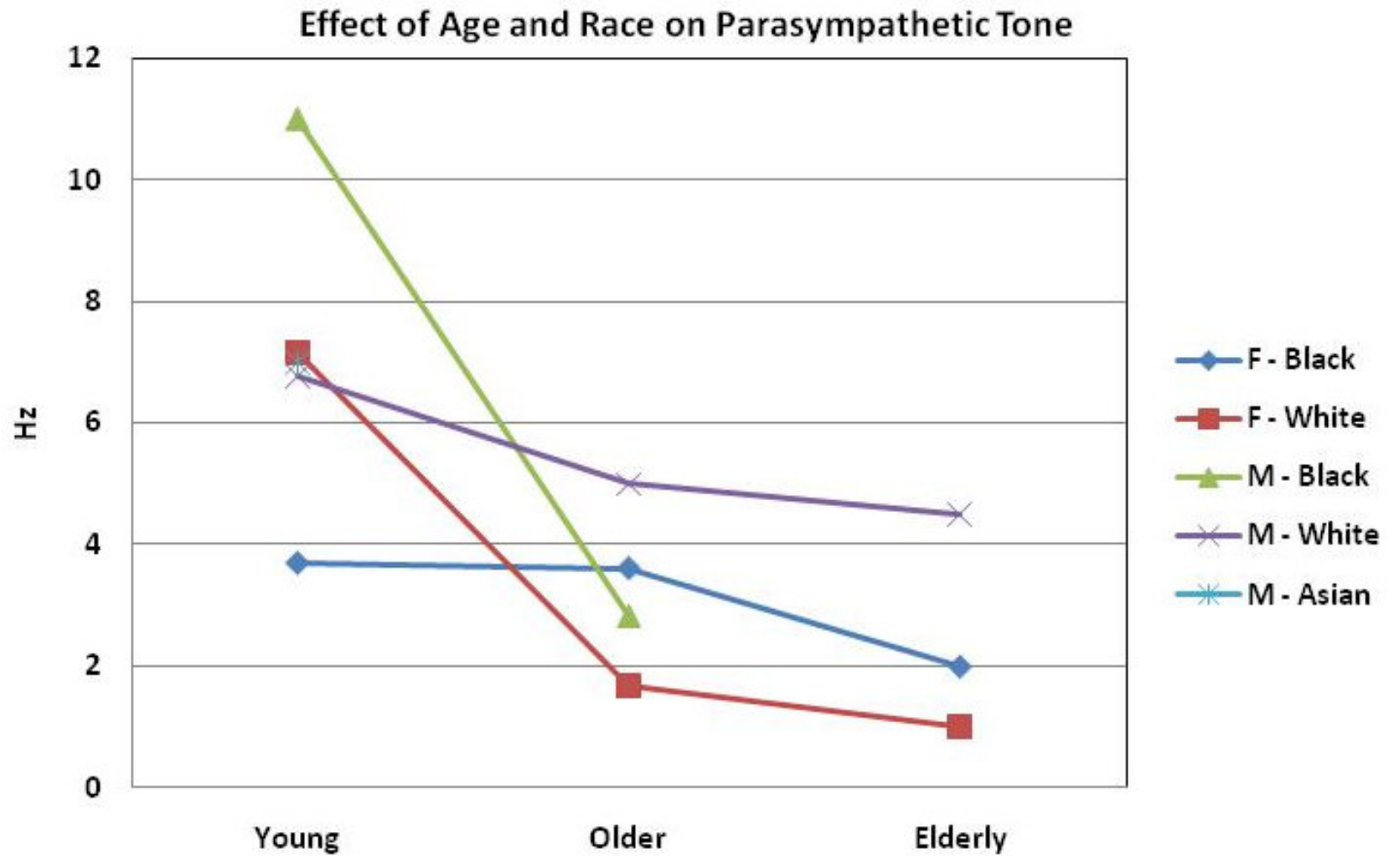

**Figure 4.1 Effect of age and race on parasympathetic tone**
Young (less than 29 years); Older (30 to 59 years); Elderly (over 60 years)

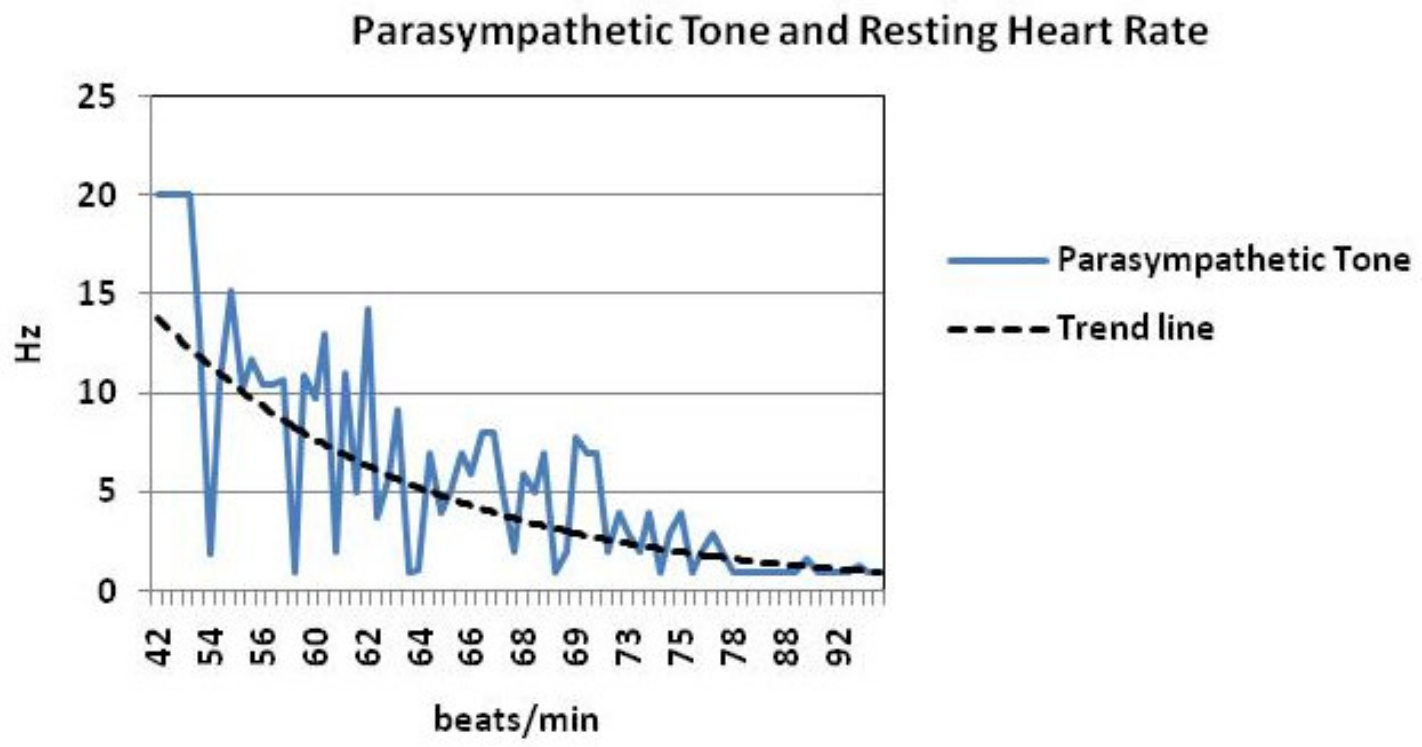

**Figure 4.2 Parasympathetic tone and resting heart rate**

### 4.2.2 Respiratory Sinus Arrhythmias

Respiratory Sinus Arrhythmias (RSA) are characterized by model parameters $W_{resp,v}$ that represents a synaptic weight applied to inputs from inspiratory-expiratory somatic motor neurons. Its value tells in which extend respiration affects heart rate and might provide a measure of high frequency heart rate variability that is correlated to the respiration cycle. We selected the experimental condition provided by the deep breath test (DBT) which is known to induce RSA with high amplitude (see section 3.4 for details). The RSA performance $EI(i)$ was calculated as the difference between the highest heart rate and the lowest heart rate during the $i^{th}$ respiration cycle. We further computed the overall RSA performance ( $EI_{total}$ ) as sum of all individual $EI(i)$ and used it as key performance indicator to select the best data set (i.e. the heart rate recording with the greatest $EI_{total}$ ) for each subject. We developed an algorithm that goes through all cardiac autonomic tests performed by the subject (using the metadata structure depicted in Figure 3.11) and selects the baseline recording with the greatest $EI_{total}$. In order to estimate the optimal value of parameter $W_{resp,v}$ for each subject, we formulated a least square problem which should minimize the sum of squared of the RSA performance $EI(i)$ for all $n$ respiration cycles as follows.

$$J_{W_{resp,v}} = \sum_{i=1}^{n} \left[ EI_{meas}(i) - EI_{sim}(i, W_{resp,v}) \right]^2 \qquad (4.2.3)$$

In some experimental setups, respiration signal was either recorded using a thoracic movement transducer or derived from ECG signal as explained in section 3.1.6. The respiration signal was then used as input to the model as well subject's age. Subject's parasympathetic resting tone was initialized with the value obtained by the algorithm described in section 4.2.1. Intracardiac influences on the estimation of RSA parameter were eliminated by setting the model parameter $G_{T_{rsa}}$ (gain of intracardiac high frequency fluctuations) to null. This setup allows us to only simulate heart rate fluctuations which are matched to respiration.

In order to calculate the value of $W_{resp,v}$ that minimizes the cost function $J_{W_{resp,v}}$, we solved the corresponding nonlinear least-squares problem in the Matlab environment using the lsqnonlin function. Gradient function was calculated by finite-differencing, Hessian was used. The maximal number of iterations was set to 1e5. The trust-region-reflective algorithm was used instead of Levenberg-Marquardt in order to enforce parameter boundaries in the interval [0.1 .. 5] Hz with initial value 0.1 Hz. The minimal and maximal steps were set to 0.1 and 10. They define the steps used by the algorithm within the boundary region. The trust-region-reflective algorithm is based on the interior-reflective Newton method described in [159]. Each iteration involves the approximate solution of a large linear system using the method of preconditioned conjugate gradients (see section 4.1).

$W_{resp,v}$ was estimated for all 72 subjects after 4 hours of simulations on a 2.1 GHz processor with 1.5 GB RAM. Results show a similar trend with aging as the parasympathetic tone, but a more rapid decay of respiratory modulation of heart rate is observed with aging. Figure 4.3 also exhibits a predominant RSA in black subjects (2.37±1.7 Hz) versus 2±1.61 Hz for white subjects although the black population was older (40±15 years) than the white population (32±18 years). Figure 4.4 shows results of model simulation for a young man (n=37) during the deep breath test. The simulated behavior matches the measured behavior on the frequency level, while there is some discrepancy on the amplitude of change in heart rate.

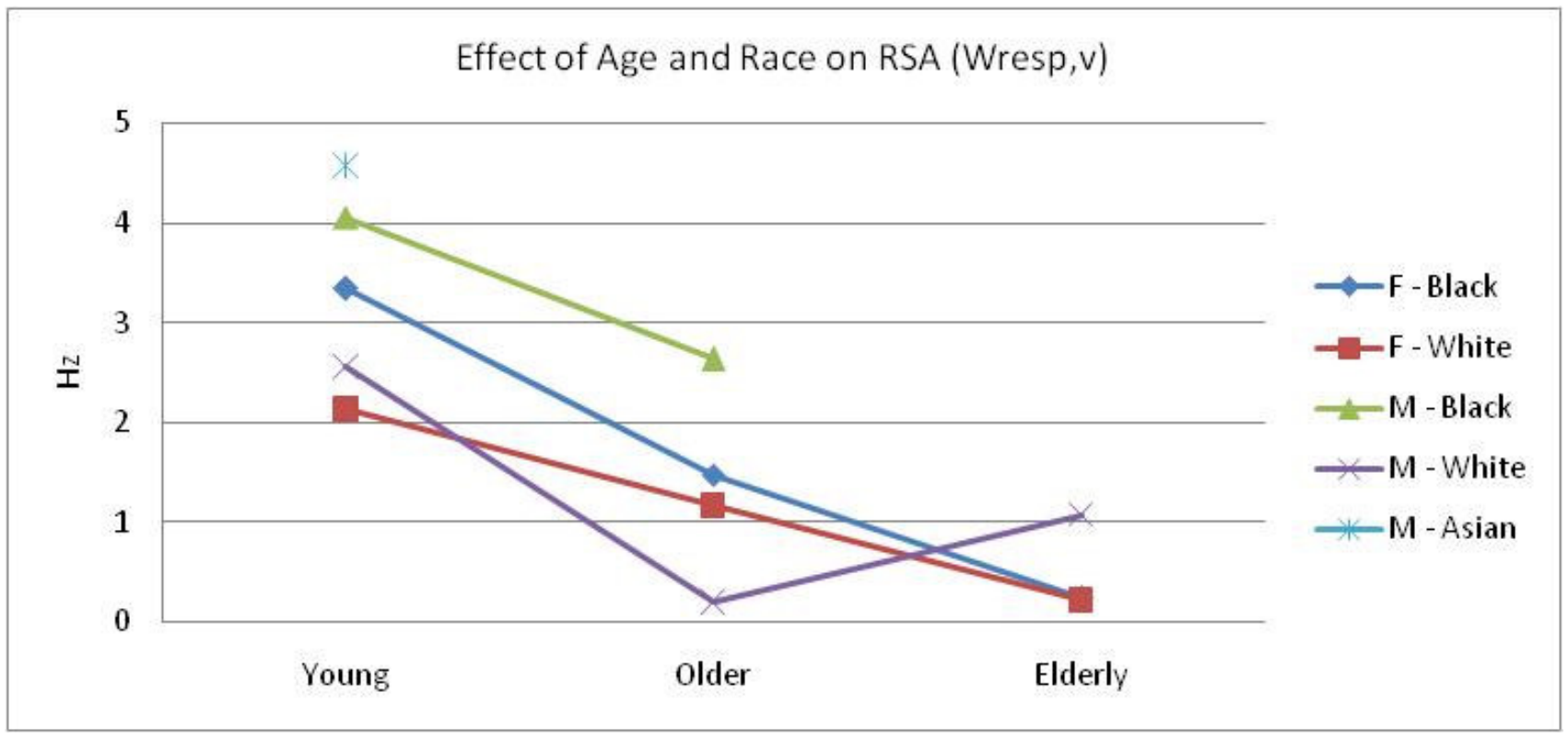

**Figure 4.3 Effect of age and race on respiratory sinus arrhythmias**

Young (up to 29 years); Older (30 to 59 years); Elderly (over 60 years)

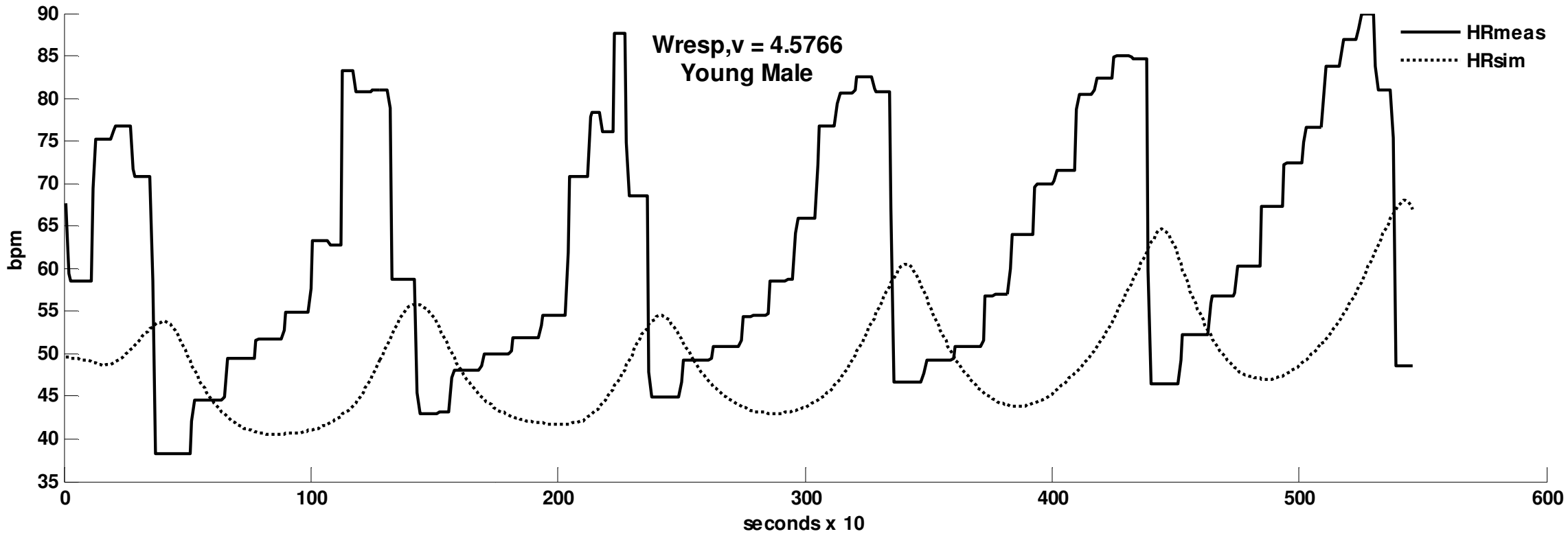

**Figure 4.4 Effect of respiratory sinus arrhythmias on heart rate**

### 4.2.3 Intracardiac High Frequency Fluctuations

The frequency of intracardiac fluctuations is captured by the model parameter $f_{rsa}$ and its amplitude by model parameter $G_{T_{rsa}}$. They should represent residual heart rate variability which is related neither to respiration nor to neural regulation. Their real values could be obtained by fitting the model to recordings performed on a denervated heart, what is rarely possible in human research. Therefore we selected baseline recordings with the greatest $EI_{total}$ during quiet breathing, or if possible during breathe holding. In order to estimate the optimal value of parameter $f_{rsa}$ and $G_{T_{rsa}}$ for each subject, we formulated a least square problem which should minimize the sum of squared of the RSA performance $EI(i)$ for all $n$ respiration cycles present in the measurement as follows.

$$J_{f_{rsa}G_{T_{rsa}}} = \sum_{i=1}^{n} \left[ EI_{meas}(i) - EI_{sim}(i, f_{rsa}, G_{T_{rsa}}) \right]^2 \qquad (4.2.4)$$

The respiration signal was used as input to the model as well subject's age. Subject's parasympathetic resting tone was initialized with the value obtained by the algorithm described in section 4.2.1. RSA parameters were initialized using values estimated in section 4.2.2. In order to calculate the value of $f_{rsa}$ and $G_{T_{rsa}}$ that minimizes the cost function $J_{f_{rsa}G_{T_{rsa}}}$, we solved the corresponding nonlinear least-squares problem in the Matlab environment using the lsqnonlin function where both parameters were constrained in the intervals [0.1 .. 1] Hz and [0.01 .. 0.1] s respectively, with initial guessed values set to 0.1 Hz and 0.01 s. Estimated values clearly show elevated intracardiac fluctuations at a mean frequency of 0.16 Hz on young subjects with amplitude 0.03 s. The frequency decreases down to 0.106 Hz at older ages, and the amplitude to 0.02 s. No big difference was observed between older and elderly subjects. Parameters values are given in Figure 4.5 for each age group and gender. No significant difference was found for subjects with cardiovascular diseases. Hypertensive subjects had 0.101 Hz fluctuations with amplitude 0.016 s, while subjects with palpitation or cardiac insufficiency had 0.11 Hz and 0.019 s.

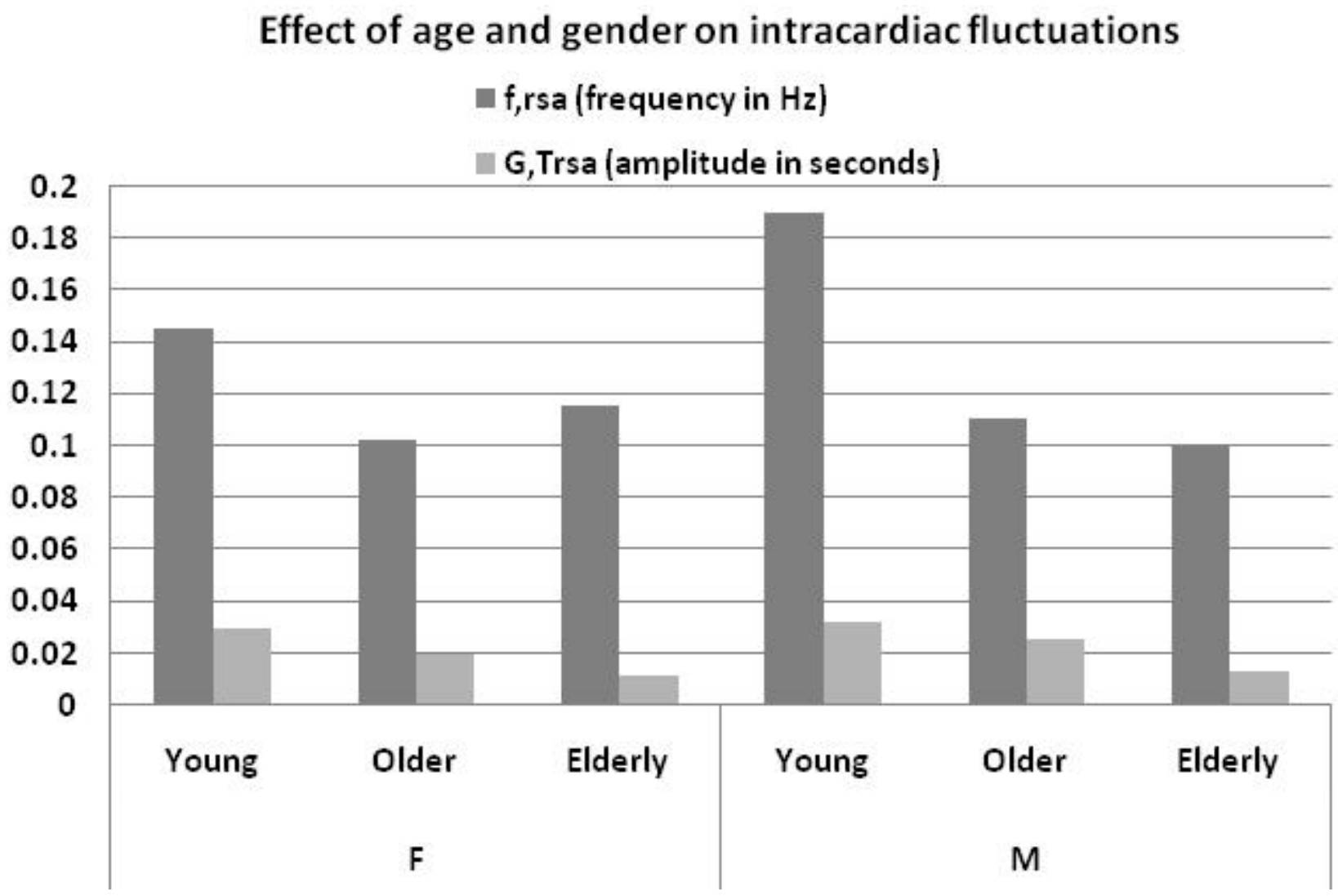

**Figure 4.5 Parameter values for intracardiac fluctuations**
Young (less than 29 years); Older (30 to 59 years); Elderly (over 60 years)

### 4.2.4 Resting Sympathetic Tone in Heart

Efferent sympathetic tone to heart is characterized by the model parameters $f_{es,0,high}$ (firing rate of pacemaker sympathetic premotor neurons), $W_{c,es,0}$ (synaptic weight applied to sensory inputs from chemoreceptors) and $k_{QT}$ (gain factor of elastance on QT interval duration). They quantify the tonic discharge of sympathetic nerves on the cardiovascular system. In order to estimate the individual values for each subject who participated in our studies, a least-squares formulation minimizing the errors between measured $QT_{meas}$ and simulated QT interval durations $QT_{sim}$ was defined. We used measured QT interval durations obtained from baseline recordings at rest based on the assumption that sympathetic modulation of ventricle depolarization duration is controlled by tonic sympathetic discharge at rest, parasympathetic innervations of ventricles being negligible. An algorithm was developed, that goes through all cardiac autonomic tests performed by the subject (using the metadata structure depicted in Figure 3.11) and selects the baseline recording with the longest mean QT interval duration. Selecting baseline recording at rest with longest mean QT interval might ensure that only tonic (basal) sympathetic activity is the main source of control. The corresponding cost function, i.e. the discrepancy between measured and simulated values is defined as optimization criteria as follows, where $n$ is the number of data points.

$$J_{f_{es,0,high}W_{c,es,0}k_{QT}} = \sum_{i=1}^{n} \left[ QT_{meas}(i) - QT_{sim}(i, f_{es,0,high}, W_{c,es,0}, k_{QT}) \right]^2 \tag{4.2.5}$$

After the model has been initialized with subject's age and parameters values estimated in previous sections, we solved a nonlinear least-squares problem aiming to minimize the cost function $J_{f_{es,0,high}W_{c,es,0}k_{QT}}$ in the Matlab environment using the lsqnonlin function where both parameters were constrained in the intervals [2 .. 12] Hz, [1 .. 5] (dimensionless unit) and [10 .. 20] $mmHg^{-1}$ mL respectively, with initial guessed values set to 10 Hz, 3 and 10 $mmHg^{-1}$ mL . The parameters boundaries were chosen according to the minimal and maximal values of physiological variables (RR and QT interval durations).

Parameters were estimated for all 72 subjects after 7 hours of simulation on a 1.8 GHz processor with 4 GB RAM. Results show predominant sympathetic tone for women at all ages. The firing rate $f_{es,0,high}$ of pacemaker sympathetic premotor neurons was found to be 8.84±1.58 Hz for women and 7.92±1.63 Hz for men. The synaptic weight $W_{c,es,0}$ applied to sensory inputs from chemoreceptors contributing to tonic sympathetic activity equaled 1.86±0.55 for women and 1.97±0.57 for men. $k_{QT}$ was found to be ca. 10 $mmHg^{-1}$ mL for all subjects.

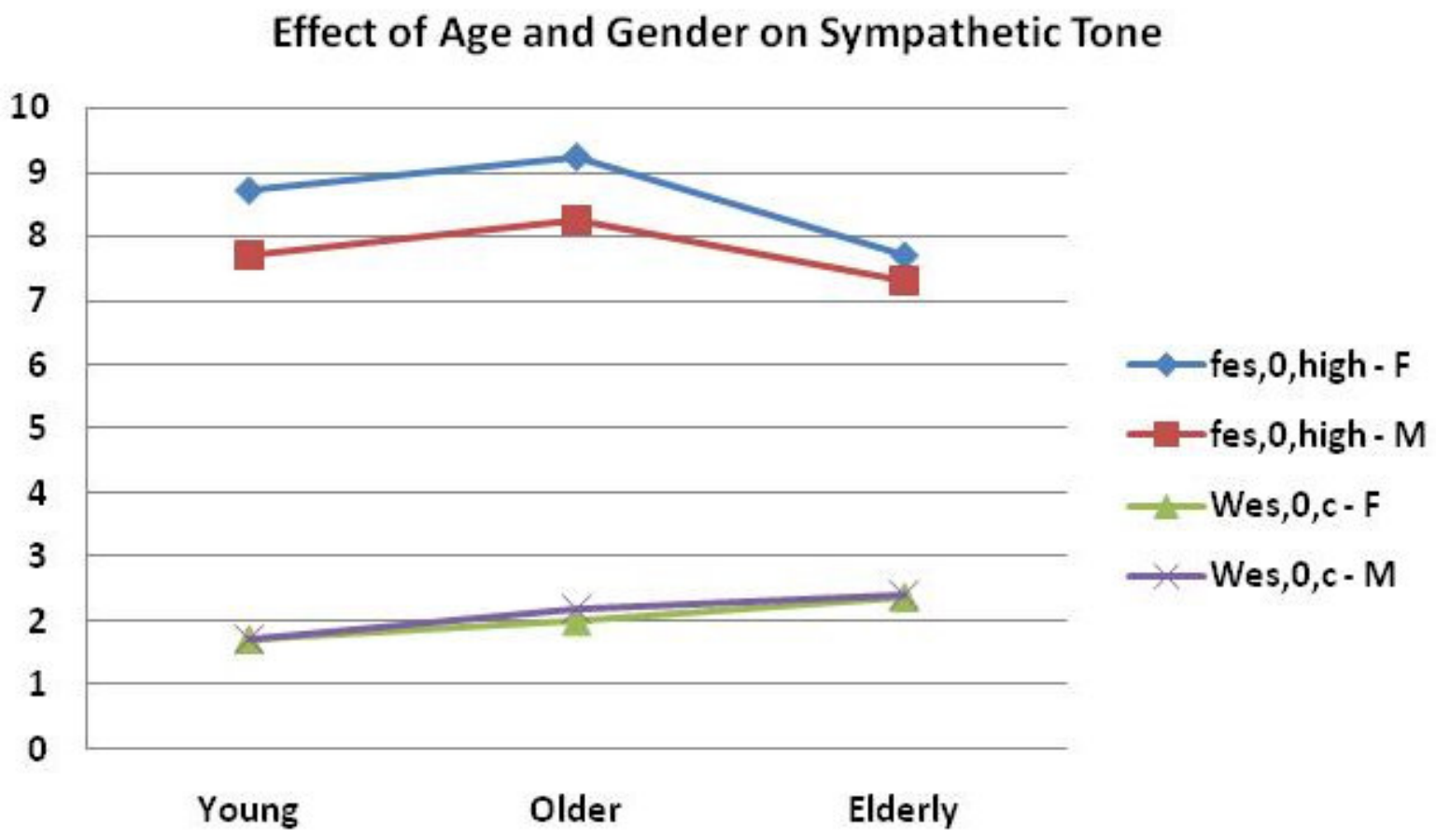

**Figure 4.6 Parameter values for sympathetic tone**
Young (less than 29 years); Older (30 to 59 years); Elderly (over 60 years)

As explained in section 2.5.1, our model is able to predict the firing rate of sympathetic premotor neurons $f_{si_{pm}}$ for given experimental data. The simulated premotor neurons signal corresponding to maximal sympathetic tone is denoted by $\lfloor f_{si_{pm}} \rfloor_{max}$. The simulated premotor neurons signal corresponding to current sympathetic tone is denoted by $\lfloor f_{si_{pm}} \rfloor_{current}$. We define sympathetic level $T_{sym}$ in a normalized scale between 0 and 1 as the ratio between both signals, averaged over all vascular beds (with $i$ equals $v$ for ventricles, $b$ for brain, $m$ for skeletal muscles, $s$ for splanchnic and $e$ for extrasplanchnic) as follows.

$$T_{sym} = \frac{1}{6} \sum_{i=n,v,b,m,s,e} \frac{\lfloor f_{si_{pm}} \rfloor_{current}}{\lfloor f_{si_{pm}} \rfloor_{max}} \qquad (4.2.6)$$

### 4.2.5 Resting Vascular Tone

Low oscillation in cardiovascular variables such as heart rate and blood pressure (so-called Mayer waves) are represented by model parameters $r_{es,0,low}$, $k_{es,0,low}$ which determine the number of burstings per time. The frequency of low oscillation increases with decreasing values of both parameters. The synaptic weight applied to low frequency oscillators in sympathetic premotor neurons is represented by the model parameter $W_{es,0,low}$. We used power spectral density and represented the heart rate signal in the frequency domain using Least-Squares Spectral Analysis (LSSA) and evaluated the low frequency $LF$ in the band [0.015 .. 0.15] Hz.

In order to estimate the values of model parameters $r_{es,0,low}$, $k_{es,0,low}$, $W_{es,0,low}$ for each subject, a least-squares formulation minimizing the errors between magnitude of low frequency components in measured heart rate signal ( $LF_{meas}$ ) and low frequency components in simulated heart rate signal ( $LF_{sim}$ ) was defined. The corresponding cost function, i.e. the discrepancy between measured and simulated values is defined as optimization criteria as follows.

$$J_{r_{es,0,low}k_{es,0,low}W_{es,0,low}} = \left[ LF_{meas}(i) - LF_{sim}(i, r_{es,0,low}, k_{es,0,low}, W_{es,0,low}) \right]^2 \qquad (4.2.7)$$

Model parameters $r_{es,0,low}$, $k_{es,0,low}$, $W_{es,0,low}$ were bounded in the intervals [1 .. 5] (dimensionless unit), [0.01 .. 0.1] (dimensionless unit) and [1 .. 5] (dimensionless unit) respectively, with initial values 3, 0.05 and 3. They were estimated for all 72 subjects after 12 hours of simulation on a 1.8 GHz processor with 4 GB RAM. Values are given in Figure 4.7.

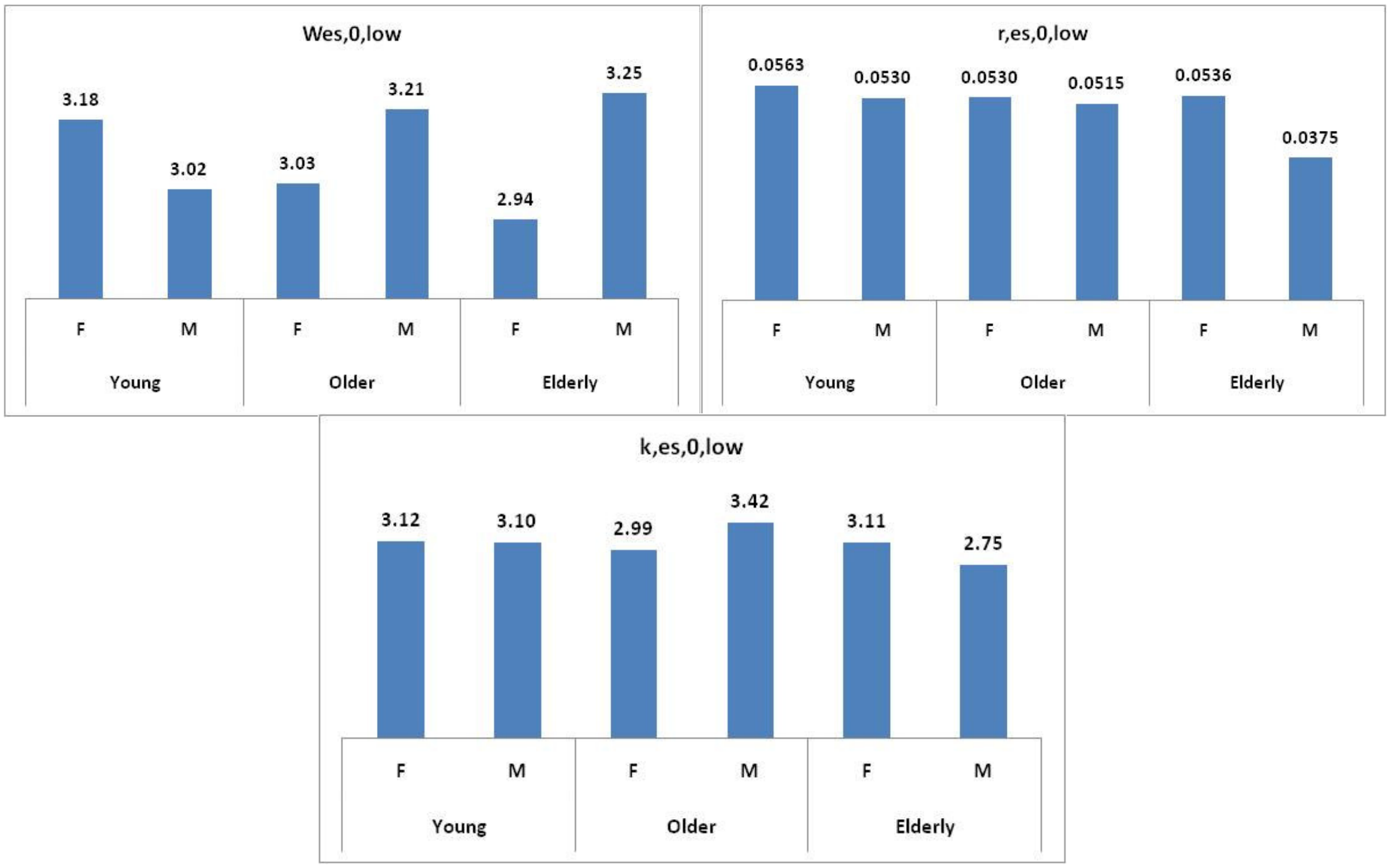

**Figure 4.7 Parameters values for low frequency fluctuations**

## 4.3 Quantification of Autonomic Tone during Vagal Maneuvers

### 4.3.1 Estimation of Parameters of Cold Receptors

The cold face test is a vagal maneuver which has been introduced in section 3.2. A normal response is decreasing heart rate which can be subject-specific. An accurate evaluation of the response requires estimating model parameters related to trigeminal cold receptors located on the forehead. The parameters and corresponding model equations were described in section 2.4.5. Parameters are $f_{at_s,max}$ (upper saturation level of facial cold receptor), which is closely related to the extent of bradycardia, i.e. the maximal number of heart beats decrement observed during the application of cold stimuli; $k_{at_f}$ (slope of the exponential decay of cold receptor), which indicates the duration of the bradycardia response. In order to better understand the meaning of both parameters, model simulations were made when fixing one and varying the second. Table 4.3.1 shows statistics values obtained for $k_{at_f}$ with $f_{at_s,max}$ fixed to 50 Hz. Mean heart rate $mean(HR_{test})$, calculated during cold application, increases with increasing $k_{at_f}$. The difference, in beats, between heart rate before the test and heart rate during the test, $mean_{diff}(HR)$ decreases with increasing $k_{at_f}$. The percentage of decrease, $mean_{incr}(HR)$, what is a measure of the bradycardia during CFT, also decreases with increasing $k_{at_f}$. A second simulation was performed with $k_{at_f}$ fixed to 0.19 $s^{-1}$. Table 4.3.2 shows values obtained when varying $f_{at_s,max}$. Increasing values for $f_{at_s,max}$ cause increasing bradycardia.

Table 4.3.1 Effect of the slope of the exponential decay of facial cold receptor

| $k_{at_f}$ [ $s^{-1}$ ] | $mean(HR_{test})$ beats/min | $mean_{diff}(HR)$ beats/min | $mean_{incr}(HR)$ % |
|---|---|---|---|
| 0.01 | 32.54 | 35.54 | 52.20 |
| 0.1 | 50.54 | 17.54 | 25.76 |
| 0.19 | 54.14 | 13.93 | 20.47 |
| 0.28 | 55.76 | 12.31 | 18.09 |
| 0.37 | 56.63 | 11.45 | 16.81 |
| 0.46 | 56.90 | 11.18 | 16.42 |
| 0.55 | 57.26 | 10.81 | 15.88 |
| 0.64 | 57.37 | 10.71 | 15.73 |
| 0.73 | 57.59 | 10.49 | 15.43 |
| 0.82 | 57.78 | 10.30 | 15.13 |
| 0.91 | 57.95 | 10.13 | 14.88 |
| 1 | 57.96 | 10.11 | 14.86 |

Table 4.3.2 Effect of upper saturation level of facial cold receptor

| $f_{at_s,max}$ [Hz] | $mean(HR_{test})$ beats/min | $mean_{diff}(HR)$ beats/min | $mean_{incr}(HR)$ % |
|---|---|---|---|
| 0 | 59.44 | 8.64 | 12.69 |
| 10 | 58.93 | 9.15 | 13.44 |
| 20 | 57.79 | 10.29 | 15.12 |
| 30 | 56.66 | 11.41 | 16.77 |
| 40 | 55.32 | 12.75 | 18.79 |
| 50 | 54.14 | 13.94 | 20.47 |
| 60 | 53.11 | 14.97 | 21.99 |
| 70 | 52.58 | 15.48 | 22.74 |
| 80 | 52.03 | 16.05 | 23.57 |
| 90 | 51.53 | 16.55 | 24.31 |
| 100 | 51.10 | 16.98 | 24.94 |

In order to estimate the individual parameters values for each subject who underwent the cold face test, we developed an algorithm that goes through all cold face tests performed by the subject (using the metadata structure depicted in Figure 3.11) and selects the recording $HR_{meas}$ with the highest percentage of heart rate fall $HR_{fall}$ between baseline recording $HR_{baseline}$ and recording during cold face test $HR_{test}$, defined as follows.

$$HR_{meas} = HR_{baseline} \cup HR_{test}$$
$$HR_{fall} = 100 \cdot \left( 1 - \frac{min(HR_{test})}{mean(HR_{baseline})} \right) \tag{4.3.1}$$

A least-squares formulation minimizing the errors between selected (measured) heart rate $HR_{meas}$ and simulated heart rate $HR_{sim}$ was defined. The corresponding cost function, i.e. the discrepancy between measured and simulated values is defined as optimization criteria as follows, where $n$ is the number of data points.

$$J_{f_{at_s,max}k_{at_f}} = \sum_{i=1}^{n} \left[ HR_{meas}(i) - HR_{sim}(i, f_{at_s,max}, k_{at_f}) \right]^2 \tag{4.3.2}$$

After the model has been initialized with subject's age and parameters values estimated in previous sections, we solved a nonlinear least-squares problem aiming to minimize the cost function $J_{f_{at_s,max}k_{at_f}}$ in the Matlab environment using the following algorithm.

Load measured heart rate data for all CFT experiments and Determine $HR_{meas}$ with the highest heart rate fall $HR_{fall}$

Load timestamps for baseline, preparatory and test phases; and generate temperature signal

Open Simulink model file

Set subject's age as input of model

Set subject's parameters related to respiratory sinus arrhythmias, sympathetic tone and vascular tone

Estimate parasympathetic tone using baseline recording prior to CFT using the algorithm described in section 4.2.1

Loop by varying model parameter $k_{at_f}$ in the interval [0.01 .. 1] and $f_{at_s,max}$ in the interval [0 .. 100]

  Simulate Simulink model file for the current parameters values and calculate model output $HR_{sim}$

  Calculate and store the corresponding value of cost function $J_{f_{at_s,max}k_{at_f}}$

End Loop

Return the value of $k_{at_f}$ and $f_{at_s,max}$ for which $J_{f_{at_s,max}k_{at_f}}$ was the lowest

$k_{at_f}$ and $f_{at_s,max}$ were estimated after 8 hours of simulation on a 1.8 GHz processor with 4 GB RAM for all 50 subjects who underwent the cold face test. Table 4.3.3 shows mean values obtained per gender and age group. Women exhibit a clear linear trend with aging; the response duration (reflected by $k_{at_f}$) and response intensity (reflected by $f_{at_s,max}$) decrease with aging. This trend is not visible for men because only 4 from 18 were young subjects. They demonstrated a weaker response to cold stimuli than older men.

Table 4.3.3 Parameters values for cold receptors

| **Sex** | **Age Group** | $k_{at_f}$ **[s$^{-1}$]** | $f_{at_s,max}$ **[Hz]** |
|---|---|---|---|
| F | Young | 0.25 | 14.17 |
| | Older | 0.20 | 10.33 |
| | Elderly | 0.09 | 6.67 |
| F Total | | 0.21 | 11.50 |
| M | Young | 0.08 | 10.00 |
| | Older | 0.26 | 13.33 |
| | Elderly | 0.08 | 7.50 |
| M Total | | 0.20 | 11.94 |
| Grand Total | | 0.20 | 11.67 |

Figure 4.8 shows heart rate change during cold application for a young female (n=32) and older (n=4) female. Parameters values for $k_{at_f}$ and $f_{at_s,max}$ are plotted, as well as the corresponding baseline parasympathetic tone $f_{ev,0,\max}$ and the difference between mean heart rate $mean_{diff}(X)$. Our mathematical model responded correctly from the view of an order of values of the output values but incorrectly from the viewpoint of absolute values. Even so the model proved importance to give correct trends and interpretable results.

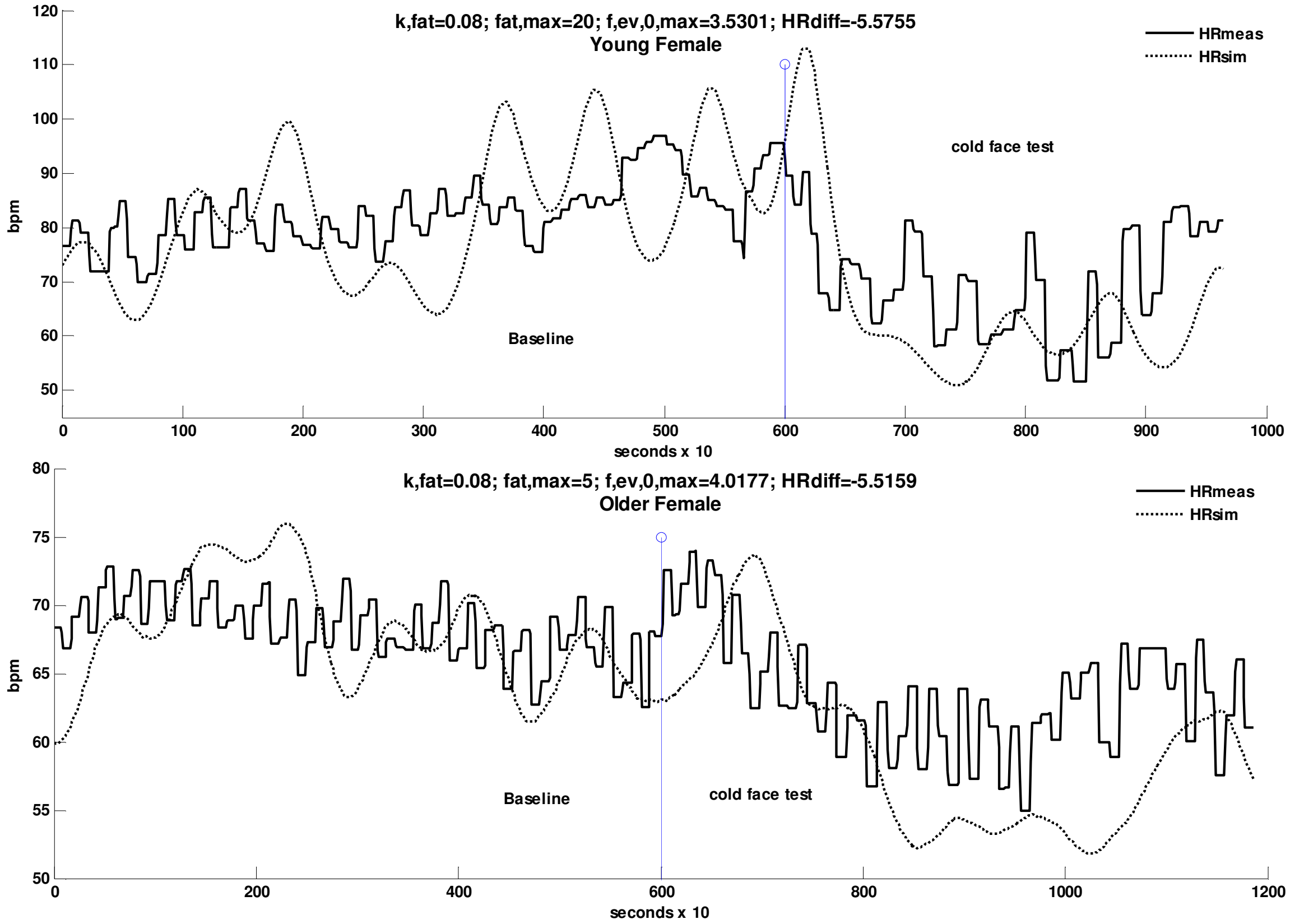

**Figure 4.8 Heart rate response to cold face test**

data are from a young female (top graph) and an older female (bottom graph)

### 4.3.2 Parasympathetic and Sympathetic Response to Cold Face Test

The feedback-feedforward loop regulating temperature and cardiovascular response during the cold face test can be summarized as depicted in Figure 4.9. The model consists of a negative feedback part for controlling the core temperature. Its current value is measured by the hypothalamus central thermoreceptors. The set point of the controlled variable is the target temperature (we use the average body temperature 37 °C, which is however subject to considerable variation [79]). The feedback controller is the hypothalamus thermoregulatory center. It receives temperature as input and generates a command signal in form of sympathetic neuron tone to the actuator (medulla cardiovascular control center). The sympathetic command originated from the hypothalamus reaches the medulla where it is integrated in the cardiovascular control vasocenter. An actuating signal is generated to peripheral vessels for vasoconstriction. Combined feedback and feedforward significantly improves the model of thermoregulation. With feedforward control, major external disturbance can be measured before it affects the system output [160]. The cold stimulus (CS) when applied to the skin induces heat loss by conduction from the blood in skin vessels. This disturbance is sensed by skin thermoreceptors. The mechanism of transduction of temperature change is not fully understood [86]. For simplification we assumed that receptors sense the average of the cold stimulus temperature, room temperature and blood temperature, which make together 22 °C. This measured disturbance is relayed by the Pons into the medulla vagal cardiovascular control center that generates a parasympathetic output signals affecting heart rate. Changes in heart rate and peripheral resistance affect blood flow, and temperature in the vasculature. The new temperature is feedback to the controller and the regulatory process starts again.

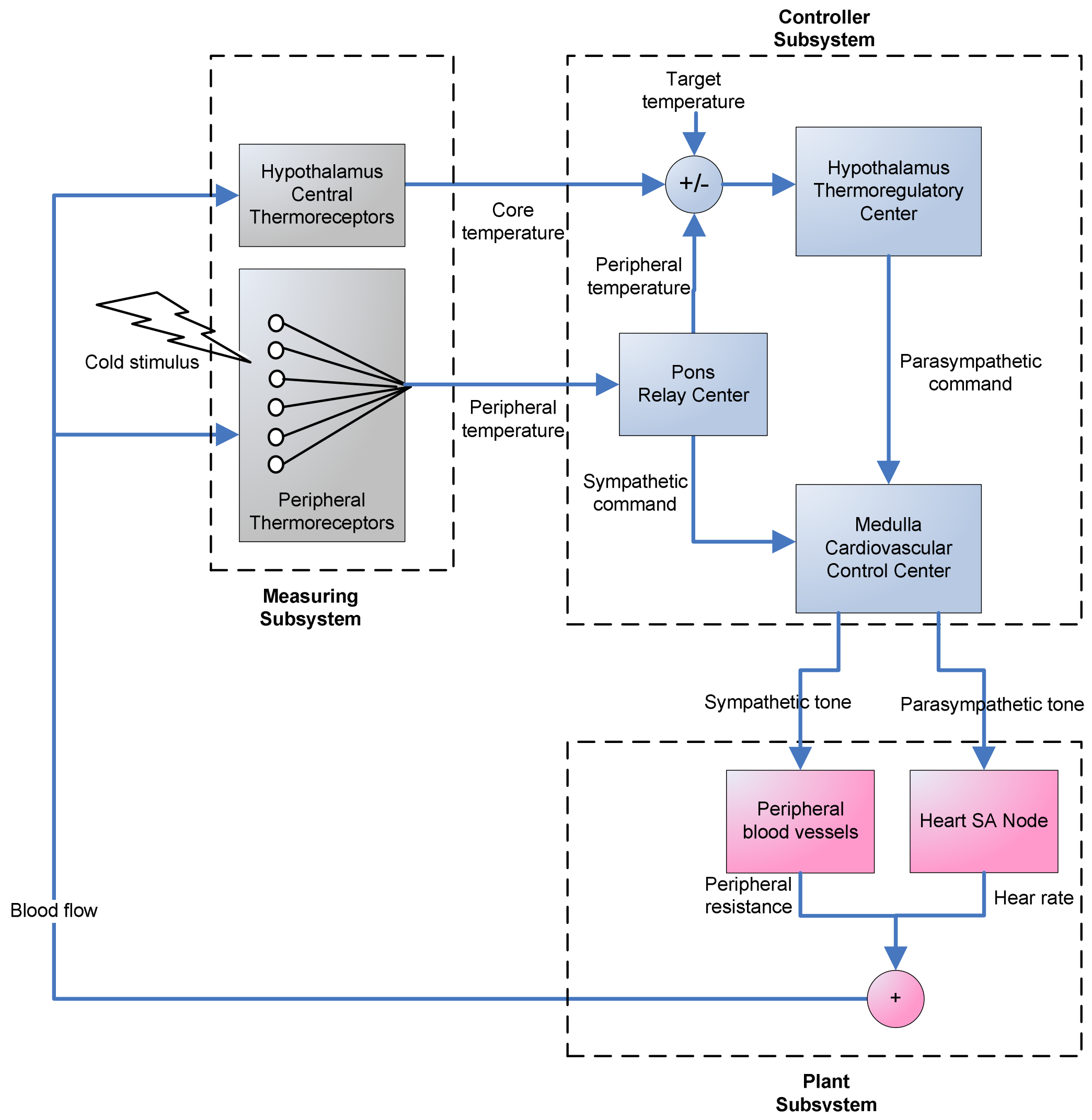

**Figure 4.9 Block diagram of regulatory processes during CFT**

The level of resting parasympathetic tone on the heart determines how strong bradycardia response to CFT will be. In the previous optimization activities, we estimated the resting parasympathetic using the baseline heart rate signals with the lowest mean heart rate ever recorded on the subject. The algorithm has been explained in section 4.2.1. Both resting parasympathetic tone and baseline parasympathetic tone might differ, for example, while subject 32 had a resting parasympathetic tone equals to 4.44 Hz; the tone was 3.53 Hz just before the cold face test. It is necessary to transform the absolute parasympathetic tone into parasympathetic level $T_{par}$ and sympathetic level $T_{sym}$ in a range between 0 and 1 as indicated in section 4.2.1 and 4.2.4. Graphs in Figure 4.10 illustrate the obtained minimum and maximum tone of both branches of the autonomous nervous system during CFT. Young subjects had higher tone of sympathetic activity to the vasculature at rest, whereas older subjects had the predisposition to experience a stronger sympathetic discharge during CFT. The former group had higher parasympathetic level at rest, but it increased to a greater extend for younger subjects. We published these results in [104].

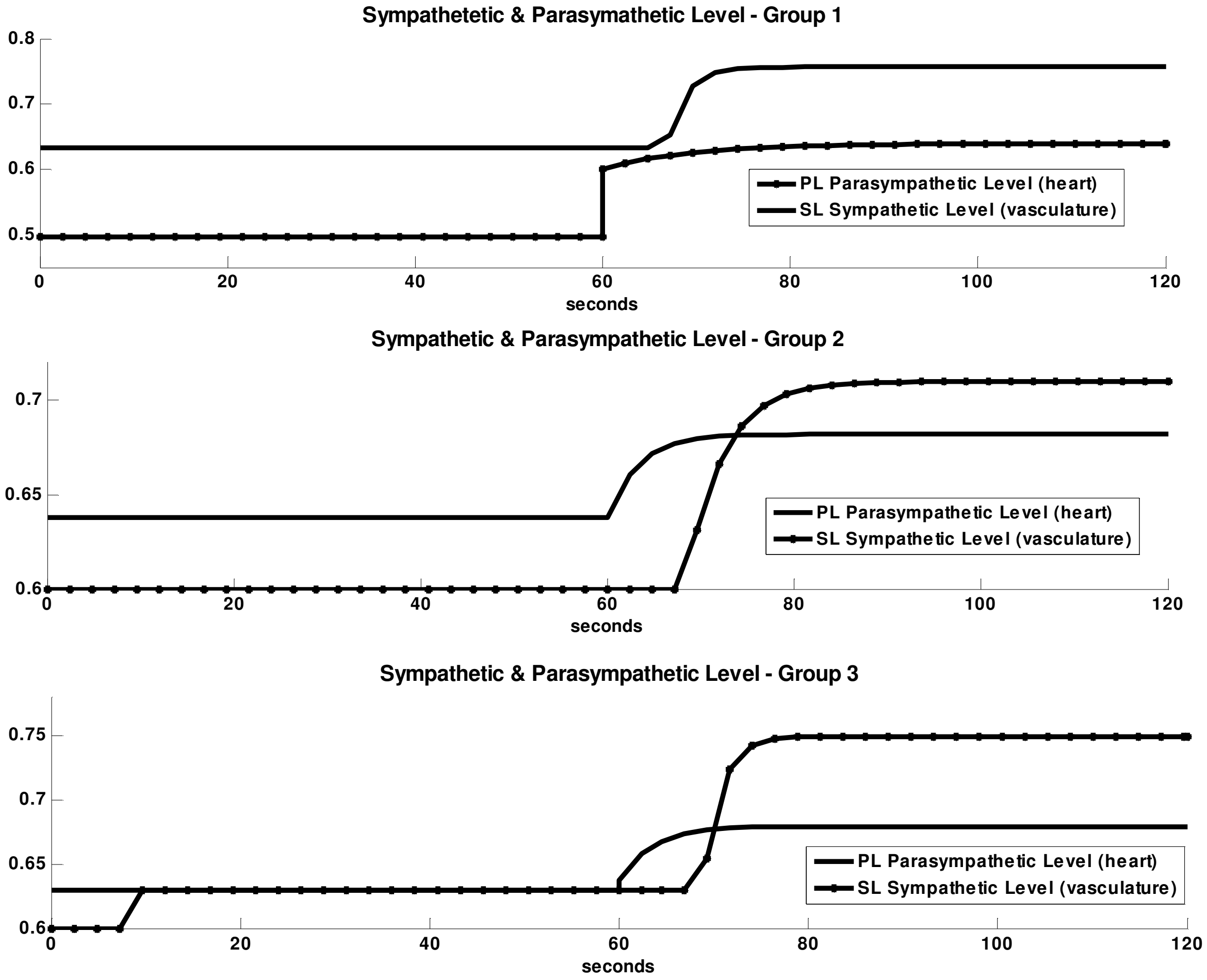

**Figure 4.10 Sympathetic and parasympathetic level during CFT**

Y-axis is normalized in the interval [0 .. 1] in dimensionless unit

Group 1 - Young (less than 29 years); Group 2 - Older (30 to 59 years); Group 3 - Elderly (over 60 years)

### 4.3.3 Vagally-mediated Tachycardia

It is expected that vagally maneuvers such as cold face test (section 3.2) and oculocardiac reflex test (section 3.3) will produce a decrease in heart rate. However some subjects in our studies demonstrated a paradoxal reflex tachycardia. Instead of bradycardia, they had a visible increasing heart rate above baseline value when pressure was applied on eye balls as illustrated in Figure 4.11 (subject n=37, healthy young man).

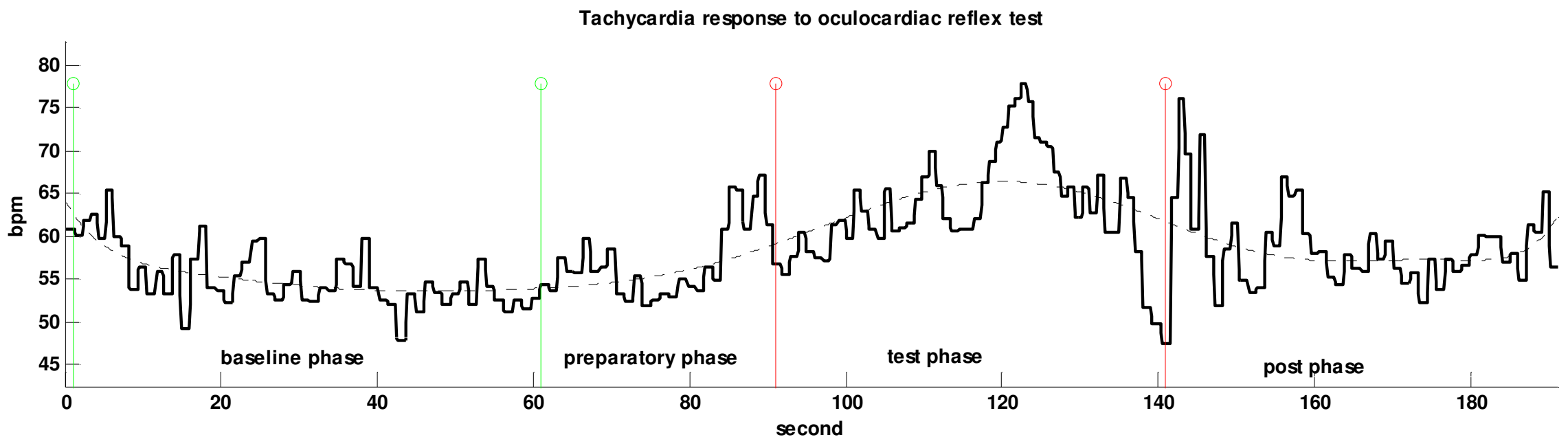

**Figure 4.11 Tachycardia during oculocardiac reflex test**

During the oculocardiac reflex test, nociceptors sense painful sensation and depending on their threshold and operating range, they might elicit a reflex tachycardia which could mask the diving reflex. In such a case, the heart receives not only parasympathetic discharge but also a pain-related pressor sympathetic discharge. Studies in [161] indicate that simultaneous activation of sympathetic nervous system and parasympathetic nervous system was observed in conditioned fear rats. Evidences suggested that simultaneous co-activation may lead to a more efficient cardiac output than activation of the sympathetic limb alone, allowing the system to overcome constricted vascular tree as it is the case during the diving response [98]. It has been found that postganglionic vagal fibers and vasoactive intestinal polypeptide play a role in mediating the vagal tachycardia in anaesthetised dogs [162]. This hypothesis has been included in our model using the parameter $k_{nor_{og},v}$ which tells to which extend the concentration of acetylcholine released by the parasympathetic preganglionic neuron in autonomic ganglia would modulate the release of norepinephrine in SA and AV nodes.

While evaluating a tachycardia response to oculocardiac reflex test, we are interested in determining the values of model parameter $k_{nor_{og},v}$ (see section 2.7 for more details about the model equations). For simplicity we assume that this parameter is described by a linear function of the pressure signal generated by pressure sensors in eyes. The analytic form of oculo-pressure sensor response $f_{ao}$ has been given in section 2.4.6. We obtain the following relationship.

$$k_{nor_{og},v}(t) = k_{vt} \cdot f_{ao}(t) \tag{4.3.3}$$

where $k_{vt}$ is a scalar parameter.

$f_{ao}$ is characterized by the model parameters $f_{ao,max}$ (upper saturation of oculo-pressure receptor) and $k_{ao}$ ( slope of the sigmoid response of oculo-pressure receptor). In order to estimate their value as well as the value of the scalar parameter $k_{vt}$, we formulated a least-squares optimization problem minimizing the errors between measured heart rate $HR_{meas}$ and simulated heart rate $HR_{sim}$.

We found the values 14.98 Hz, 0.43 $s^{-1}$ and 4 for $f_{ao,max}$, $k_{ao}$ and $k_{vt}$ respectively for the subject whose tachycardia response was illustrated in Figure 4.11. The corresponding simulated firing rate of oculo-pressure receptors can be seen in the upper panel of Figure 4.12 when pressure is applied from 60$^{th}$ to 110$^{th}$ second. The lower panel shows that our model could successfully predict the paradoxal tachycardia response. Heart rate increases despite of more acetylcholine being produced at ganglionic site, as it can be seen in Figure 4.13, pre-ganglionic parasympathetic neurons increase their activity during oculocardiac reflex test by 11.97% while pre-ganglionic sympathetic neurons decrease their activity by 8.07%. One should excpect a bradycardia response, but this is not the case because post-ganglionic parasympathetic neurons enhance the release of norepineprine at SA node site, as shown in Figure 4.14. Vagally-mediated norepinephrine production increases by over 1000% which completely masks the increase in acetylcholine concentration (only 11.97%). Out of these results, our model simulation brings additional evidence that a possible explanation why some subjects demonstrate tachycardia during vagal maneuver could be a vagally-mediated production of sympathetic neurotransmitters at heart site.

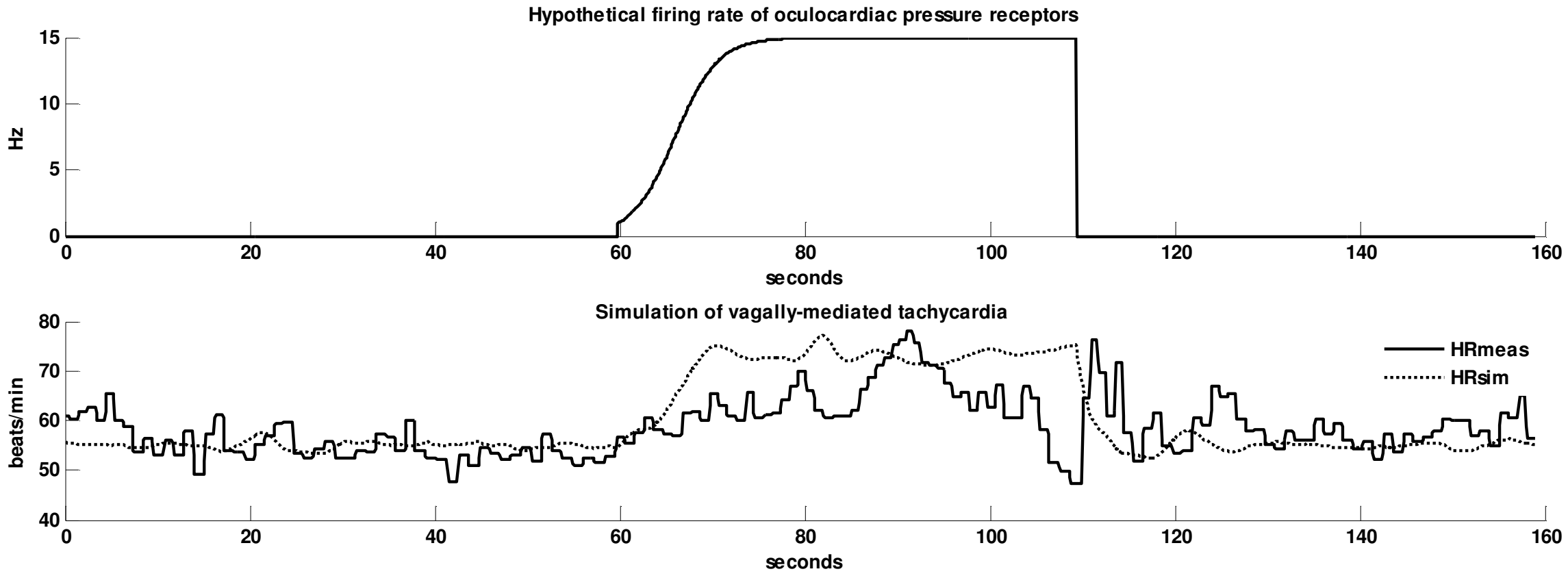

**Figure 4.12 Vagally-mediated tachycardia during oculocardiac reflex test**

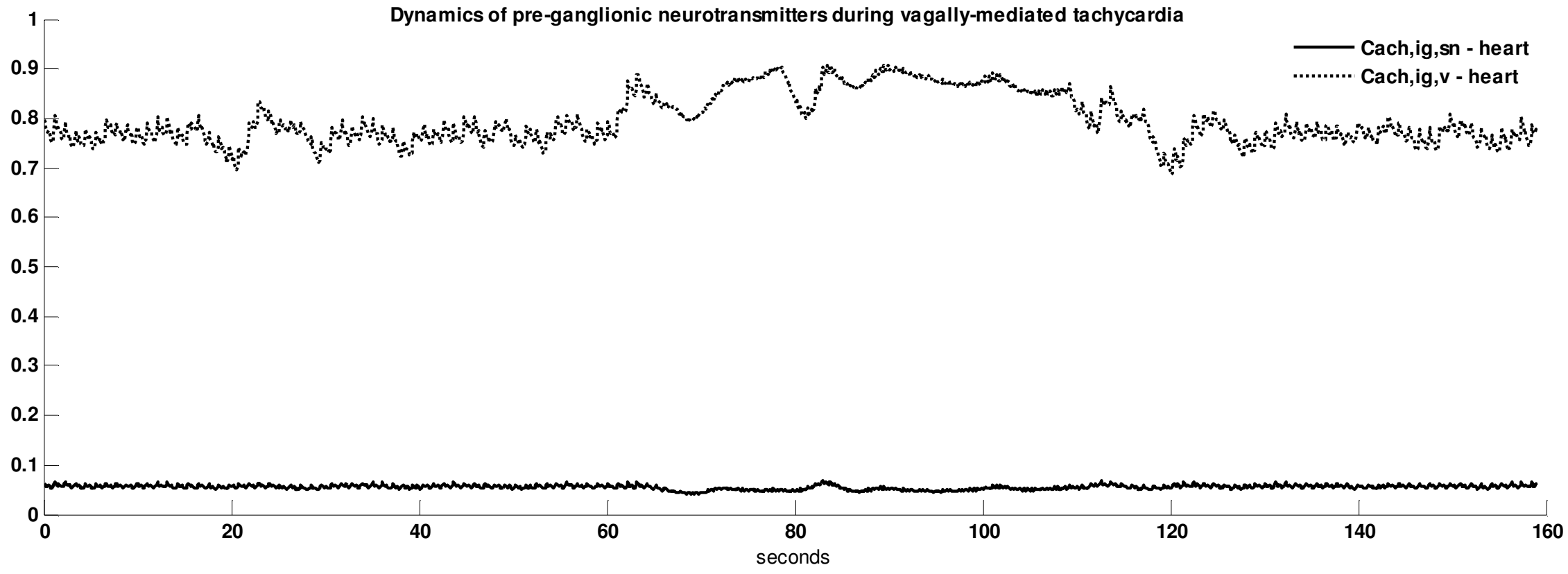

**Figure 4.13 Preganglionic neurotransmitters on heart during vagally-mediated tachycardia**

Y-axis is normalized in the interval [0 .. 1] in dimensionless unit

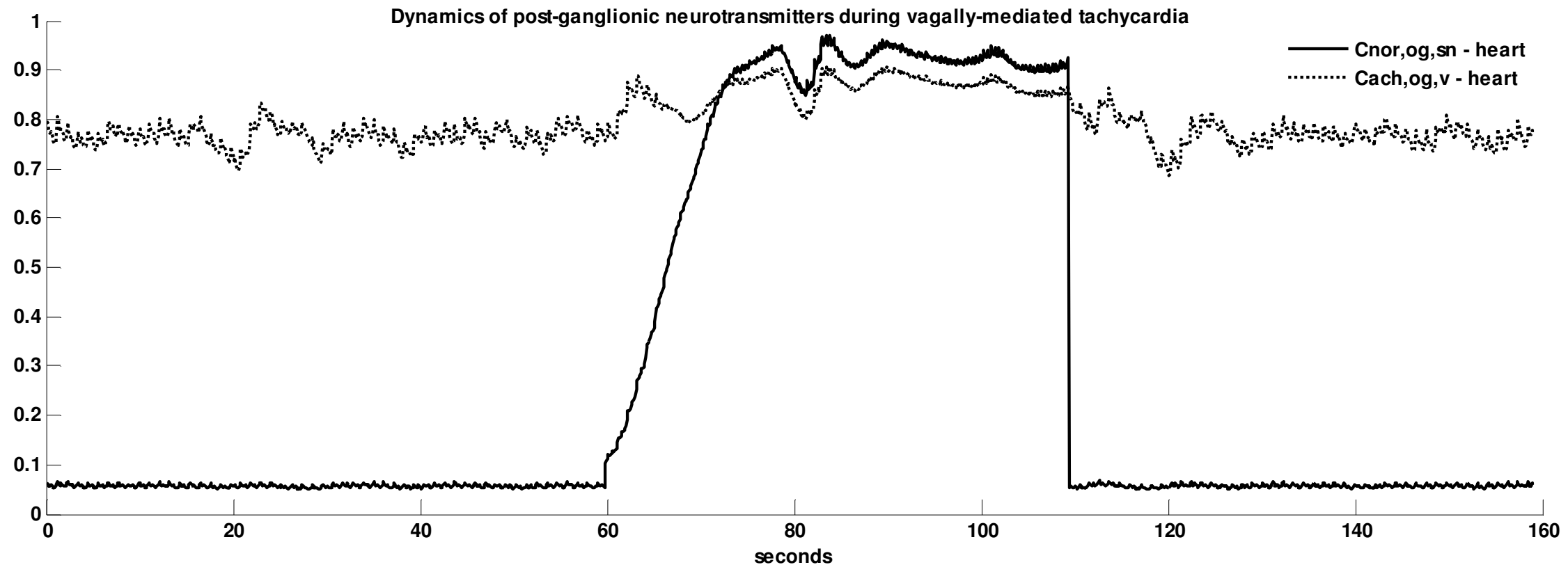

**Figure 4.14 Postganglionic neurotransmitters on heart during vagally-mediated tachycardia**

Y-axis is normalized in the interval [0 .. 1] in dimensionless unit

## 4.4 Quantification of Autonomic Tone during Sympathetic Maneuvers

### 4.4.1 Estimation of Parameters related to Mental Stress

When submitting the subject to mental arithmetic stress we are interested in determining the values of model parameters $f_{as,max}$ and $k_{as}$ which are related to upper saturation of mental stress sensors and the slope of the mental stress sensor response respectively, as described in section 2.4.7. In order to estimate the individual parameters values for each subject who underwent the mental stress test, we formulate a least-squares optimization problem minimizing the errors between measured heart rate $HR_{meas}$ and simulated heart rate $HR_{sim}$; between measured mean arterial blood pressure $MAP_{meas}$ and simulated mean arterial blood pressure $MAP_{sim}$; between galvanic skin level $G_{mp,meas}$ measured at finger site and simulated galvanic skin level $G_{mp,sim}$. The corresponding cost function, i.e. the discrepancy between measured and simulated values is defined as optimization criteria as follows, where $n$ is the number of data points.

$$J_{f_{as,max}k_{as}} = \sum_{i=1}^{n}\left[HR_{meas}(i) - HR_{sim}(i)\right]^2 + \sum_{i=1}^{n}\left[MAP_{meas}(i) - MAP_{sim}(i)\right]^2 + \sum_{i=1}^{n}\left[G_{mp,meas}(i) - G_{mp,sim}(i)\right]^2 \tag{4.4.1}$$

After the model has been initialized with subject's age and parameters values related to sympathetic tone and vascular tone as estimated in previous sections, we solved the nonlinear least-squares problem in the Matlab environment as follows.

Load measured heart rate, blood pressure and galvanic skin level data
If respiration was measured, then load respiration signal, otherwise generate respiration signal from ECG
Load timestamps for baseline, preparatory and test phases
Open Simulink model file
Set subject's age as input of model
Set subject's parameters related to respiratory sinus arrhythmias, sympathetic tone and vascular tone
Set subject's respiration signal
Estimate parasympathetic tone using baseline recording by applying the algorithm described in section 4.2.1
Loop by varying model parameter $f_{as,max}$ in the interval [1 .. 50] and $k_{as}$ in the interval [0.1 .. 1]
    Simulate Simulink model for current parameters values and return model outputs $HR_{meas}, MAP_{sim}, G_{mp,sim}$
    Calculate and store the corresponding value of cost function $J_{f_{as,max}k_{as}}$
End Loop
Return the value of $f_{as,max}$ and $k_{as}$ for which $J_{f_{as,max}k_{as}}$ was the lowest

The parameters $f_{as,max}$ and $k_{as}$ were estimated to 14.07 Hz and 13.22 $s^{-1}$ for a young healthy man (subject n=37). Results are depicted in Figure 4.15 when subject was asked to perform a mental arithmetic task from 120$^{th}$ to 190$^{th}$ second. Heart rate increased by 12% and galvanic skin level increased by 0.42 microSiemens. The results show an acceptable fit between simulated and measured data. We computed the parasympathetic level $T_{par}$ and sympathetic level $T_{sym}$ as indicated in section 4.2.1 and 4.2.4 in a range between 0 and 1. For the normalization, the maximal firing rate of sympathetic premotor neurons simulated for young subjects was used; it was equal to 60 Hz. The maximal firing rate of parasympathetic premotor neurons was found to be 10 Hz. Results in Figure 4.16 show increasing sympathetic level to heart by 24.31% from baseline level 0.6852 to the value 0.8518 during stress. Vagal level decreases by 7.76% from 0.8058 to 0.7431.

In order to compare the obtained results with heart rate variability, we evaluated the low frequency (LF) and high frequency (HF) components of the heart rate signal before and during mental stress. Figure 4.17 shows a comparison with model-based estimates of autonomic tone (sympathetic and parasympathetic level). Spectral power density values for low frequency components (LF) increased by 27.94%, what agrees with the increased in model-based estimate of sympathetic tone (24.31%). However there is no agreement between values of high frequency (HF) components and model-based estimate of parasympathetic tone. While HF increases by 2.69%, the former decreases by 7.76%. We would rather expect a decrease in HF components during pressor response. Obviously our model-based estimates of parasympathetic tone give more plausible values.

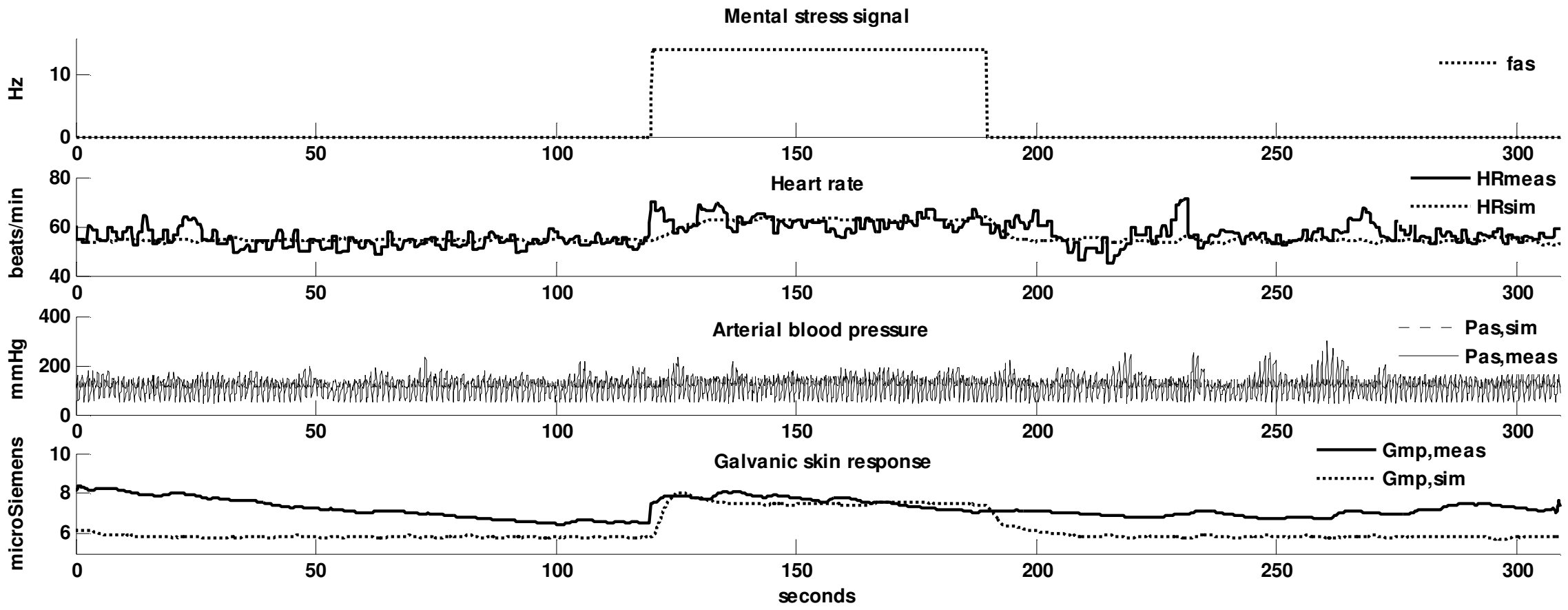

**Figure 4.15 Simulation response to mental stress**

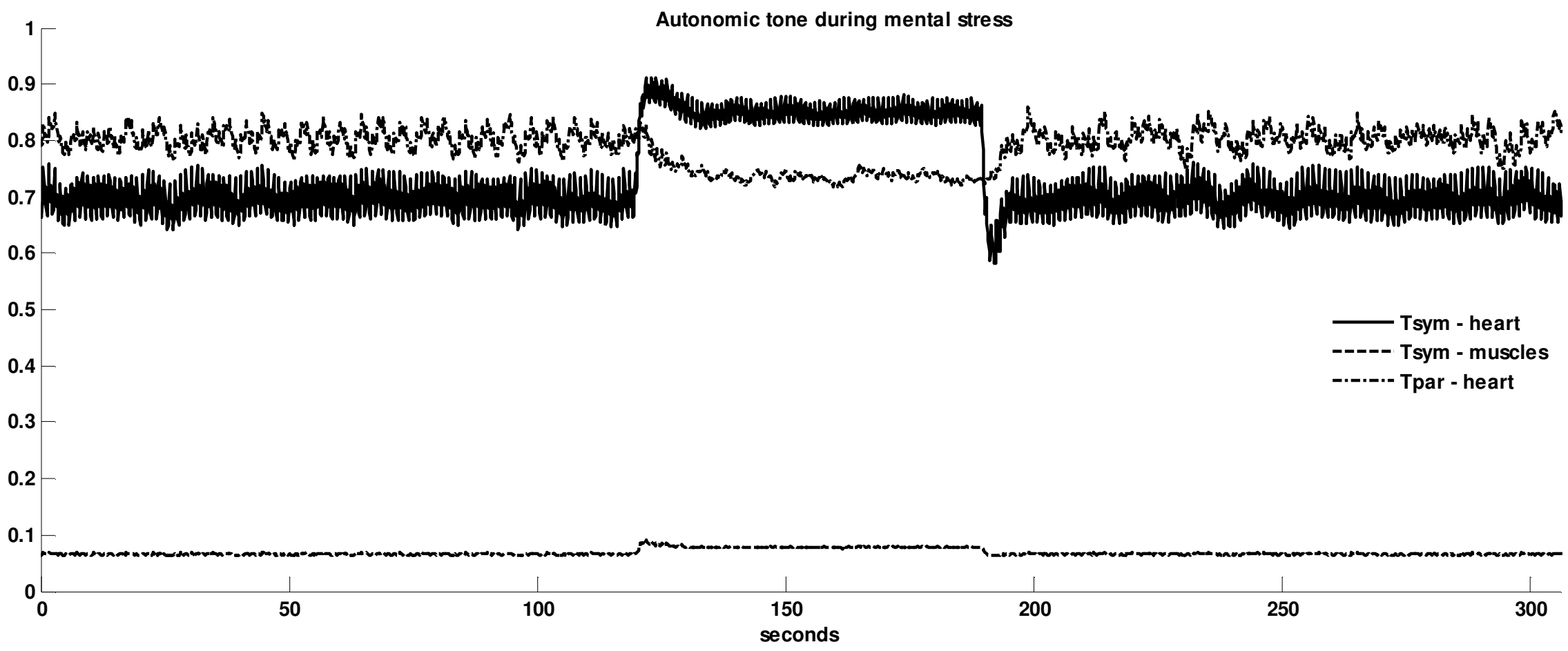

**Figure 4.16 Sympathetic and parasympathetic level during mental stress**

Y-axis is normalized in the interval [0 .. 1] in dimensionless unit

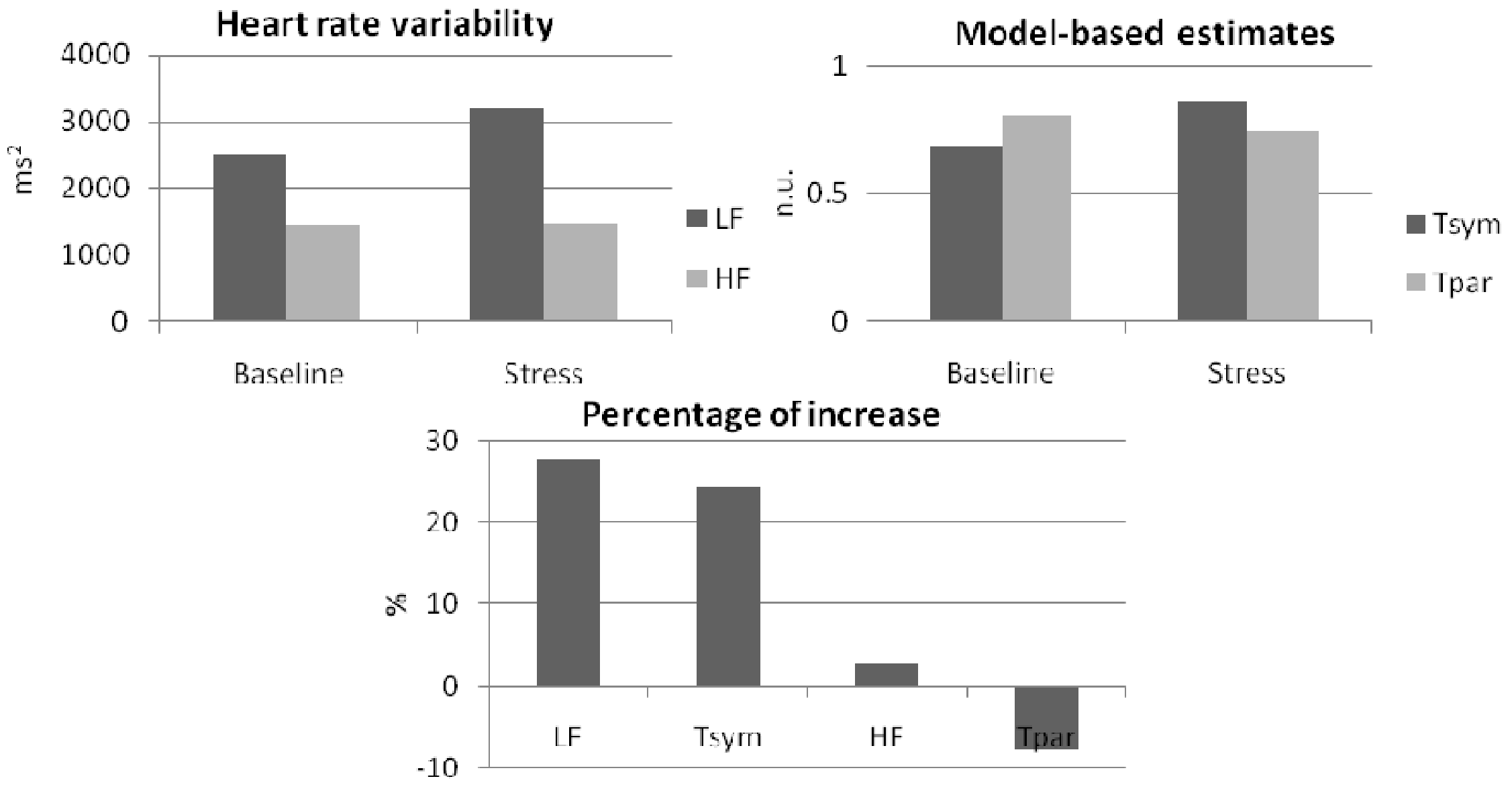

**Figure 4.17 HRV parameters and model-based estimates**

Model-based estimates are normalized in the interval [0 .. 1] in dimensionless unit (n.u.)

### 4.4.2 Estimation of Parameters related to Cold Pressor Test

Immersing subject's hand or foot in cold water induces a global pressor response with increased heart rate, peripheral vascular resistance and blood pressure (see section 3.7.3). In order to quantify autonomic response during pressor test, we need to estimate the model parameters related to cold receptors at hands and feet level. These are $f_{at_s,max}$ ( upper saturation of cold receptor on hands and feet) and $k_{at_s}$ ( slope of the sigmoid response of cold receptor on hands and feet). We defined an optimization task aiming to minimize the error between model output and measured data similar to the one presented in the previous section 4.4.1. Both parameters were found to be equal to 72.32 Hz and 1.0391 $s^{-1}$ respectively for a healthy young man (subject n=37). As shown in Figure 4.18, heart rate increases by 72% from 49.55 beats/min to 83.68 beats/min. Cold stimuli was applied from $120^{th}$ to $160^{th}$ second. The global pressor response is clearly reflected by model-based estimates of autonomic tone, as depicted in Figure 4.19. Sympathetic level to heart increases by 80% whereas sympathetic level to muscles increases by more than 1000%. Parasympathetic level decreases by 10%.

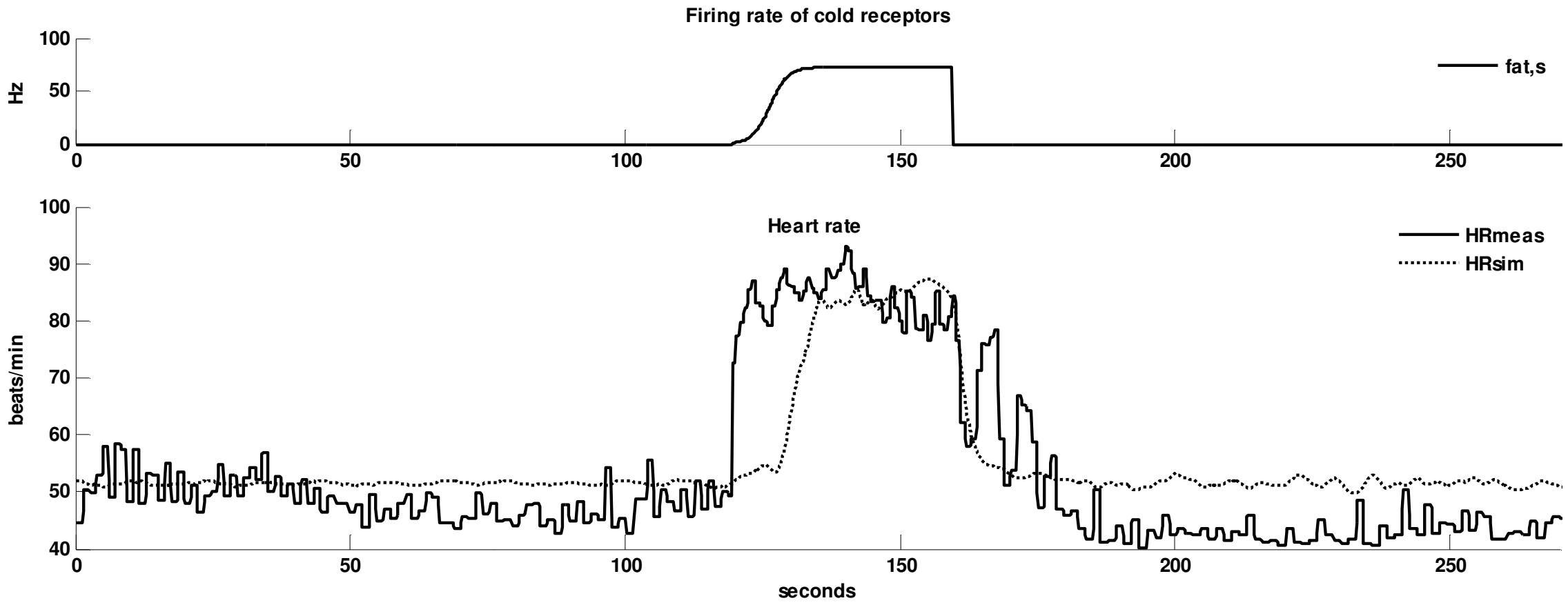

**Figure 4.18 Heart rate response to cold pressor test**

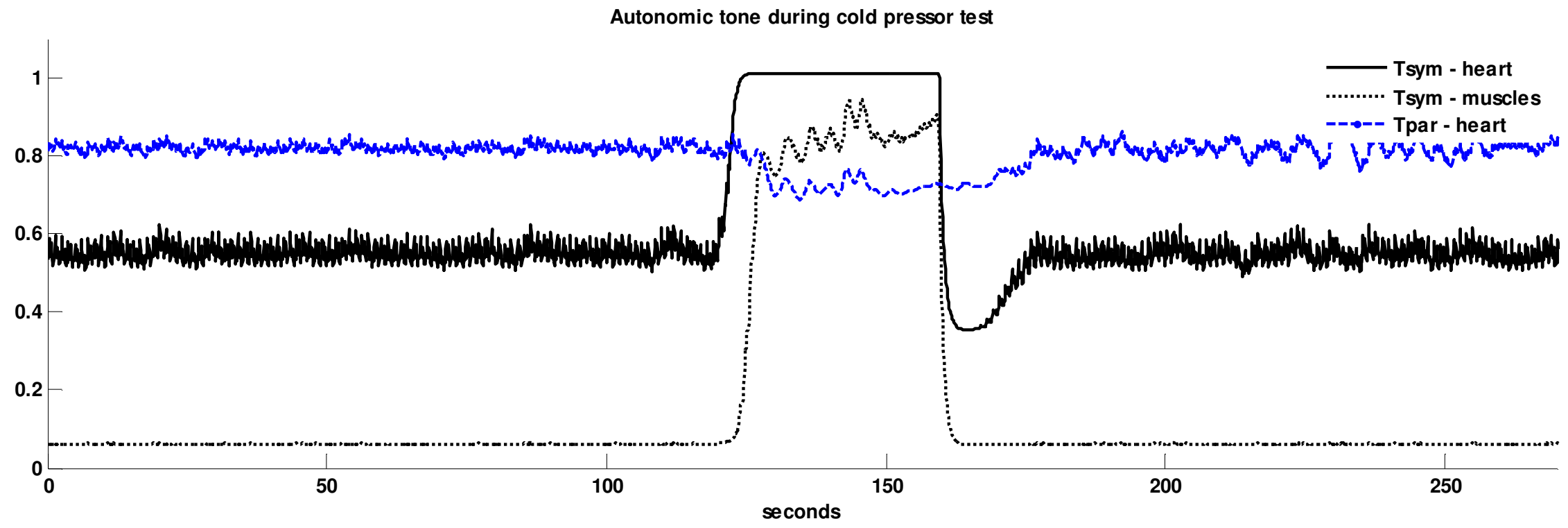

**Figure 4.19 Sympathetic and parasympathetic level during cold pressor test**
Y-axis is normalized in the interval [0 .. 1] in dimensionless unit

## 4.5 Quantification of Autonomic Tone during Mixed Maneuvers

We selected the Valsalva Maneuver (VM) as a cardiac autonomic test which triggers a combined sympathetic and parasympathetic response; and is therefore suitable for assessing the integrity of autonomic nervous control on the cardiovascular system [100]. Its physiological background and experimental setup was introduced in section 3.6. In the following sections we will present some limitation of heart rate variability (HRV) as tool for assessing autonomic nervous control during the VM and show how our mathematical model produces better results. We published results presented in this section as conference contribution [163]. An extended version was submitted as a journal publication [164].

### 4.5.1 Limitations of HRV as Measure of Autonomic Activity

There exist signal analysis methods for evaluating cardiovascular response during the Valsalva Maneuver including wavelets transform [165] and T-wave width alteration [150]. Although such methods provide insight information, the contribution of neural pathways involved was not covered. An attempt using frequency domain analysis of heart rate variability (HRV) was made in [166] and [167]. HRV analysis of autonomic balance during the Valsalva Maneuver can held erroneous results because of the very short duration of the test (15 to 30 seconds). Recording of approximately 1 minute is needed to access the high frequency components of HRV; while 2 minutes are needed to evaluate the low frequency components. Very low frequency assessment from short-term recordings (less than 5 minutes) is a dubious measurement and should be avoided [29]. The spectrum of the HRV signal is usually calculated using non-parametric methods based on Fast Fourier Transforms. A rigorous evaluation requires at least 60 seconds ECG. This basic condition is not met by the Valsalva Maneuver, which is usually performed for less than 30 seconds. Therefore time-domain analysis is more suitable in the evaluation of HRV during VM. The standard deviation of NN intervals (SDNN) reflects the overall variability, but strongly depends on the duration of the signal. SDNN increases with the length of recording. The square root of the mean squared differences of successive NN intervals (RMSSD) estimates high frequency variations of heart rate and is proposed as a measure of parasympathetic activity.

We calculated time-domain estimates of HRV for heart rate signals recorded during our own experiments, as described in section 3.6. Selected data sets and corresponding HRV values are presented in Table 4.5.1. The SDNN and RMSSD for each single phase were calculated on the portion of NN intervals of that phase. Changes in SDNN are visible from one phase to the next phase of the maneuver. Greatest heart rate variability occurs in the phase 2, where SDNN equals 39.23 ms compare to 20.97 ms before the maneuver. Phase 2 is also characterized by a decreased RMSSD from 29.75 ms before the maneuver to 19.17 ms, which suggests a withdrawal of vagal activity and increased sympathetic activity as response to the transient decrease in blood pressure. Phase 4 bradycardia is characterized by an increase of RMSSD from 19.17 ms in phase 3 to 37.57 ms in phase 4. Table 4.5.1 additionally includes the Valsalva Ratio (VR) computed as the maximal heart rate recorded in phase 2 divided by the minimal heart rate recorded in the bradycardia phase 4.

Although RMSSD and SDNN provide a measure of the degree of autonomic modulation, they do not indicate the level of autonomic tone regulating heart rate for each single phase of the VM. Moreover RMSSD and SDNN can be subject-specific; values for single data sets do not always correspond to the expected increase or decrease of autonomic nervous activity. For example, in data set 7, RMSSD surprisingly increases from 13.51 ms in phase 1 to 28.86 ms in phase 2 although phase 2 should be characterized by a decrease of parasympathetic activity. Such inconsistency motivated us to investigate the usage of estimates of autonomic tone based on our mathematical model.

Table 4.5.1 Time-domain estimates of heart rate variability (in milliseconds) during the Valsalva Maneuver for 12 data sets obtained on 5 healthy subjects

| | | Mean | Data Set 1 | Data Set 2 | Data Set 3 | Data Set 4 | Data Set 5 | Data Set 6 | Data Set 7 | Data Set 8 | Data Set 9 | Data Set 10 | Data Set 11 | Data Set 12 |
|---|---|---|---|---|---|---|---|---|---|---|---|---|---|---|
| SDNN | *Baseline* | **20.97±5.29 ms** | 17.12 | 21.30 | 22.11 | 17.03 | 25.38 | 30.74 | 19.03 | 19.88 | 14.23 | 13.13 | 24.97 | 26.68 |
| | *Phase 1* | **13.49±4.79 ms** | 19.27 | 18.90 | 11.85 | 12.98 | 19.31 | 10.01 | 11.16 | 18.09 | 7.60 | 6.11 | 9.71 | 16.83 |
| | *Phase 2* | **39.23±3.59 ms** | 39.93 | 38.21 | 41.13 | 44.47 | 34.66 | 43.33 | 42.60 | 41.58 | 38.18 | 33.98 | 38.50 | 34.19 |
| | *Phase 3* | **28.21±7.6 ms** | 17.23 | 28.26 | 37.10 | 28.07 | 23.68 | 34.04 | 22.68 | 27.93 | 14.13 | 34.31 | 37.61 | 33.44 |
| | *Phase 4* | **23.13±10.34 ms** | 38.59 | 11.24 | 18.51 | 14.01 | 31.68 | 16.42 | 38.45 | 12.96 | 35.27 | 13.82 | 23.90 | 22.72 |
| | *Post VM* | **21.13±1.12 ms** | 21.39 | 21.16 | 21.53 | 20.63 | 18.44 | 21.43 | 23.36 | 20.74 | 20.49 | 21.45 | 21.66 | 21.34 |
| RMSSD | *Baseline* | **29.75±7.4 ms** | 24.44 | 30.41 | 31.57 | 24.31 | 34.04 | 43.87 | 27.16 | 28.38 | 20.32 | 18.74 | 35.64 | 38.08 |
| | *Phase 1* | **17.96±9.96 ms** | 27.11 | 28.80 | 13.92 | 16.95 | 33.29 | 9.68 | 13.51 | 29.57 | 7.00 | 3.52 | 9.85 | 22.32 |
| | *Phase 2* | **24.8±7.34 ms** | 23.76 | 15.55 | 25.89 | 19.53 | 24.10 | 22.83 | 28.86 | 18.59 | 18.04 | 18.59 | 34.40 | 35.45 |
| | *Phase 3* | **19.17±6.27 ms** | 21.26 | 11.44 | 19.13 | 13.43 | 25.29 | 17.13 | 31.11 | 12.85 | 14.82 | 13.95 | 25.28 | 24.38 |
| | *Phase 4* | **37.57±7.01 ms** | 38.15 | 35.38 | 41.19 | 46.52 | 37.03 | 44.45 | 43.10 | 44.66 | 34.53 | 25.36 | 35.19 | 25.32 |
| | *Post VM* | **30.5±1.74 ms** | 29.79 | 29.93 | 30.44 | 28.98 | 34.87 | 30.28 | 32.91 | 29.35 | 28.55 | 30.36 | 30.57 | 29.98 |
| VR | | **1.41±0.04** | 1.413 | 1.405 | 1.425 | 1.45 | 1.361 | 1.46 | 1.45 | 1.44 | 1.4 | 1.36 | 1.4 | 1.36 |

### 4.5.2 Parameters Estimation for Intra-thoracic Pressure Elevation

Mathematical models look as a promising tool for evaluating autonomic nervous control of blood pressure, heart rate and vascular resistance during each single phase of the Valsalva Maneuver. A simple mathematical model for estimating the baroreflex gain during the VM is presented in [168], which however does not cover the pulmonary aspect. Authors of [169] and [84] proposed integrated cardiovascular and respiratory models for the VM. Both models include multiplicity of mechanisms for simulating hemodynamic response; atrial, ventricular and lung mechanics. We adopted a different approach focusing on neural control during the model development activities we have performed in Chapter 2. The section 2.8.6 of that chapter included a model of hemodynamic changes due to intra-thoracic pressure elevation. The model parts involved during VM are schematically depicted in Figure 4.20.

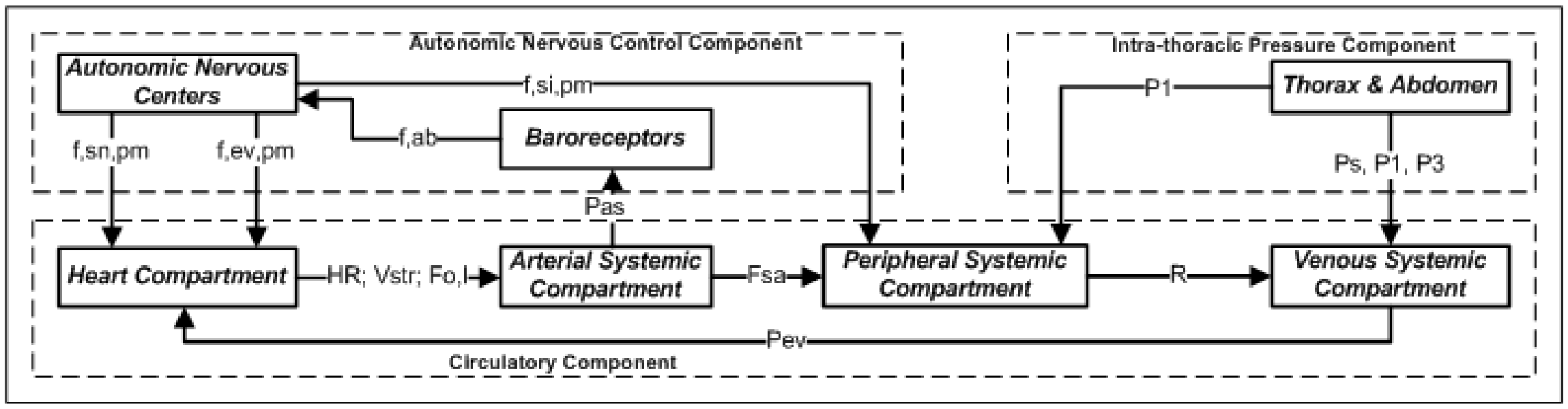

**Figure 4.20 Block diagram describing cardiovascular control during the Valsalva Maneuver**

Measured galvanic skin response signal ($GSR_{meas}$, refer to section 3.1.7) was used to approximate measured peripheral vascular resistance at finger site as follows:

$$R_{mp,meas} = \frac{1}{GSR_{meas}} \tag{4.5.1}$$

The model equations were implemented using the Systems Biology Toolbox 2 for Matlab [170]. The toolbox calls Matlab *ode23s* solver for resolving the ordinary differential equations. *ode23s* is a one-step differential equation solver based on a modified Rosenbrock formula of order 2 [171].

We formulated a weighted least-squares problem minimizing the errors between measured data ($HR_{meas}, R_{mp,meas}$) and model outputs ($HR_{sim}, R_{mp,sim}$). The corresponding cost function is given as follows.

$$J = \frac{1}{n \cdot \overline{HR_{meas}}} \sum_{i=1}^{n} \left[ HR_{meas}(i) - HR_{sim}(i) \right]^2 + \frac{1}{n \cdot \overline{R_{mp,meas}}} \sum_{i=1}^{n} \left[ R_{mp,meas}(i) - R_{mp,sim}(i) \right]^2 \quad (4.5.2)$$

where $n$ is the number of data points in the time series (value is 600 data points). $\overline{HR_{meas}}$, $\overline{R_{mp,meas}}$ are mean of values in vectors containing individual heart rate and peripheral vascular resistance values measured during the experiments on a beat-to-beat basis. $HR_{sim}, R_{mp,sim}$ are their corresponding simulated values.

In order to find the parameters values that minimize the cost function $J$, the Nelder–Mead algorithm [158] was used for bound-constrained optimization. It is based on cost functional evaluations of sequences of downhill simplexes (see section 4.1.4). The model includes unknown parameters, which are related to intra-thoracic pressure elevation during the VM (see Table 2.5.20 in section 2.8.6). They were estimated for each data set after an average of 2983 iterations and 38468 functions evaluations with a final optimal cost function value in order of 0.000228002. Mean parameters values are presented in Table 4.5.2 for 12 data sets obtained from 5 subjects.

Figure 4.22 and Figure 4.23 show a good fit of model outputs to data obtained from individual measurements on one representative 23 years old healthy man. 97.16% and 97.44% of residuals for heart rate and peripheral vascular resistance resp. are within the confidence interval with 95% significance level. The corresponding parameter estimates are included in Table 2 under data set 5.

During Phase I of the maneuver, the gradual increase in intra-thoracic and intra-abdominal pressure causes aortic and peripheral vessels compression. It follows an increase in peripheral resistance resulting in a mild increase in blood pressure from 15$^{th}$ to 16$^{th}$ second (see Figure 4.21).

During Phase II, the sustained high intra-thoracic and intra-abdominal pressure hinders venous return, leading to a decrease in cardiac output and consequently a marked decrease in arterial systolic pressure from the 16$^{th}$ to the 20$^{th}$ second (see Figure 4.21). This fall in arterial pressure is detected by baroreceptors in the carotid artery sinus and aorta. The parasympathetic level $T_{par}$ and sympathetic level $T_{sym}$ were computed as indicated in section 4.2.1 and 4.2.4 in a range between 0 and 1. The autonomic response includes an increase of sympathetic tone (see Figure 4.24-b) to the heart and vessels as well as a decrease of parasympathetic tone (see Figure 4.24-a) in the heart. This causes a reflex tachycardia and peripheral vasoconstriction (see increased peripheral vascular resistance in Figure 4.23). The sympathetically mediated response results in some rise of arterial blood pressure from 20$^{th}$ to 30$^{th}$ second (see Figure 4.21), which however does not overcome the elevated intra-thoracic pressure. Since the blood pressure is still below the normal level, the baroreflex remains active in the whole phase II, resulting in a continuous rize of heart rate [172].

During Phase III, the forced expiration is released, resulting in a decrease in intra-thoracic pressure, thus the pressure around the aorta decreases and consequently arterial blood pressure slightly decreases from 30$^{th}$ to 31$^{st}$ second (see Figure 4.21). Meanwhile heart rate is still elevated.

During Phase IV, the accumulated venous blood returns to the heart and is pumped into the constricted arteries causing an “overshoot” of arterial pressure above the normal level from 31st to 38th second (Figure 4.21). This is detected by the baroreceptors and results in a reflex bradycardia. Sympathetic discharge to the heart decreases (see Figure 4.24-b) whereas parasympathetic discharge increases (see Figure 4.24-a), causing a decrease of heart rate. As result arterial blood pressure returns to normal level.

Table 4.5.2 Estimated values of model parameters for VM

| Mean ±STD | Data Set 1 | Data Set 2 | Data Set 3 | Data Set 4 | Data Set 5 | Data Set 6 | Data Set 7 | Data Set 8 | Data Set 9 | Data Set 10 | Data Set 11 | Data Set 12 |
|---|---|---|---|---|---|---|---|---|---|---|---|---|
| **0.29±0.08** | 0.36 | 0.26 | 0.33 | 0.29 | 0.12 | 0.30 | 0.42 | 0.28 | 0.23 | 0.26 | 0.36 | 0.32 |
| **0.57±0.12 liter** | 0.58 | 0.62 | 0.46 | 0.47 | 0.77 | 0.49 | 0.75 | 0.49 | 0.73 | 0.47 | 0.57 | 0.44 |
| **2.49±0.55 sec** | 2.30 | 2.91 | 2.55 | 2.22 | 2.82 | 2.24 | 2.16 | 4.01 | 2.21 | 2.15 | 2.25 | 2.09 |
| **1.18±0.24 $\mu Mho^{-1}$** | 1.05 | 1.06 | 1.61 | 1.36 | 1.02 | 1.15 | 1.01 | 1.71 | 1.08 | 1.02 | 1.04 | 1.03 |
| **3.94±1.06 sec** | 2.40 | 4.52 | 4.99 | 4.43 | 2.92 | 4.79 | 2.41 | 4.42 | 2.41 | 4.90 | 4.50 | 4.63 |
| **2.46±0.6 $\mu Mho^{-1}$** | 2.05 | 2.01 | 2.03 | 2.47 | 3.00 | 2.26 | 4.05 | 2.94 | 2.21 | 2.06 | 2.16 | 2.32 |
| **3.66±0.91 sec** | 3.08 | 4.59 | 4.73 | 2.52 | 2.65 | 4.67 | 2.49 | 3.61 | 3.21 | 4.76 | 4.36 | 3.22 |
| **3.06±0.08 $\mu Mho^{-1}$** | 3.03 | 3.00 | 3.02 | 3.00 | 3.01 | 3.05 | 3.23 | 3.01 | 3.05 | 3.02 | 3.22 | 3.05 |
| **2.24±0.21** | 2.29 | 2.30 | 2.31 | 2.25 | 1.57 | 2.31 | 2.31 | 2.29 | 2.35 | 2.31 | 2.30 | 2.24 |

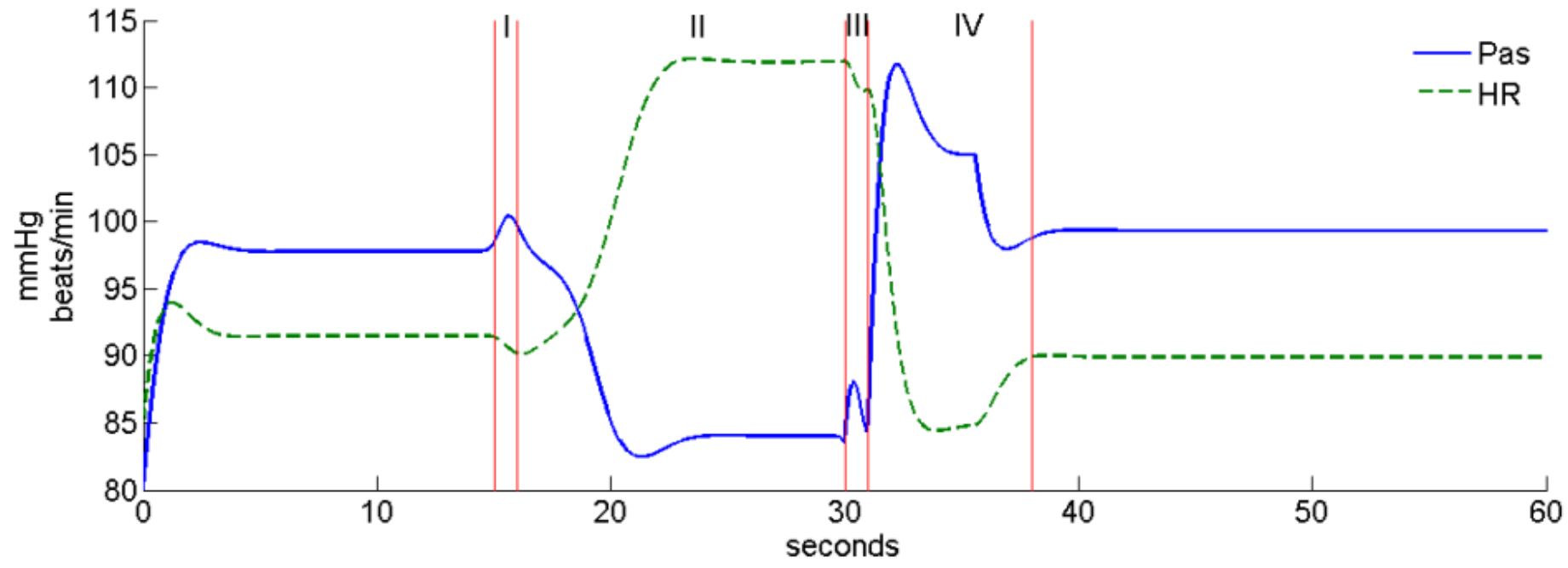

**Figure 4.21 Mean arterial pressure (plain line) and heart rate (dashed line) changes during the VM**

The maneuver covers the timeline from the 15th to the 30th second. Phase I goes from 15th to 16th second, phase II ends at 30th second, followed by phase III. The last phase IV runs from 31st to 38th second.

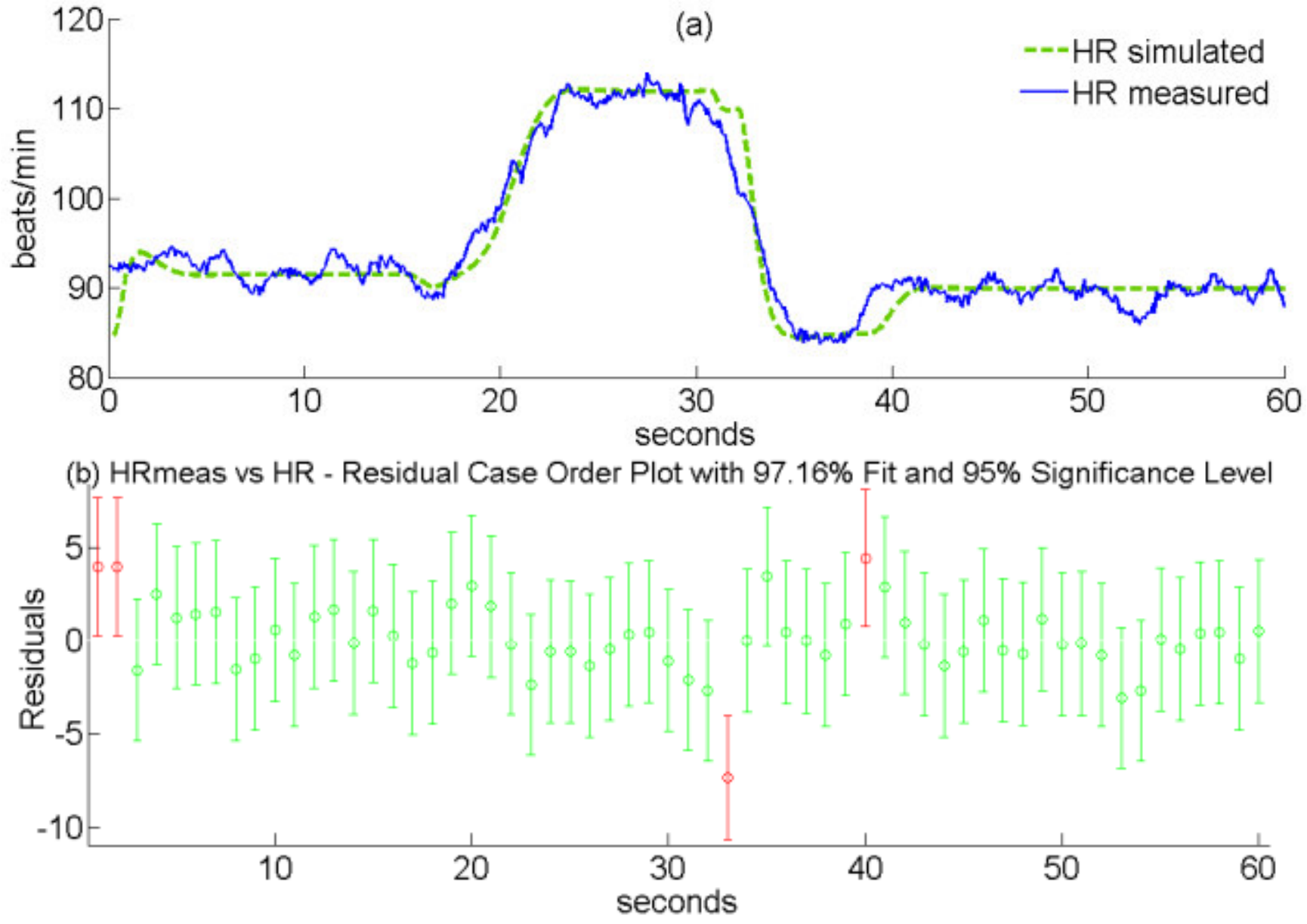

**Figure 4.22 Heart rate change during VM**

(a) predicted value (dashed line) and experimental data (plain line) shows a 97.16% good fit (b)

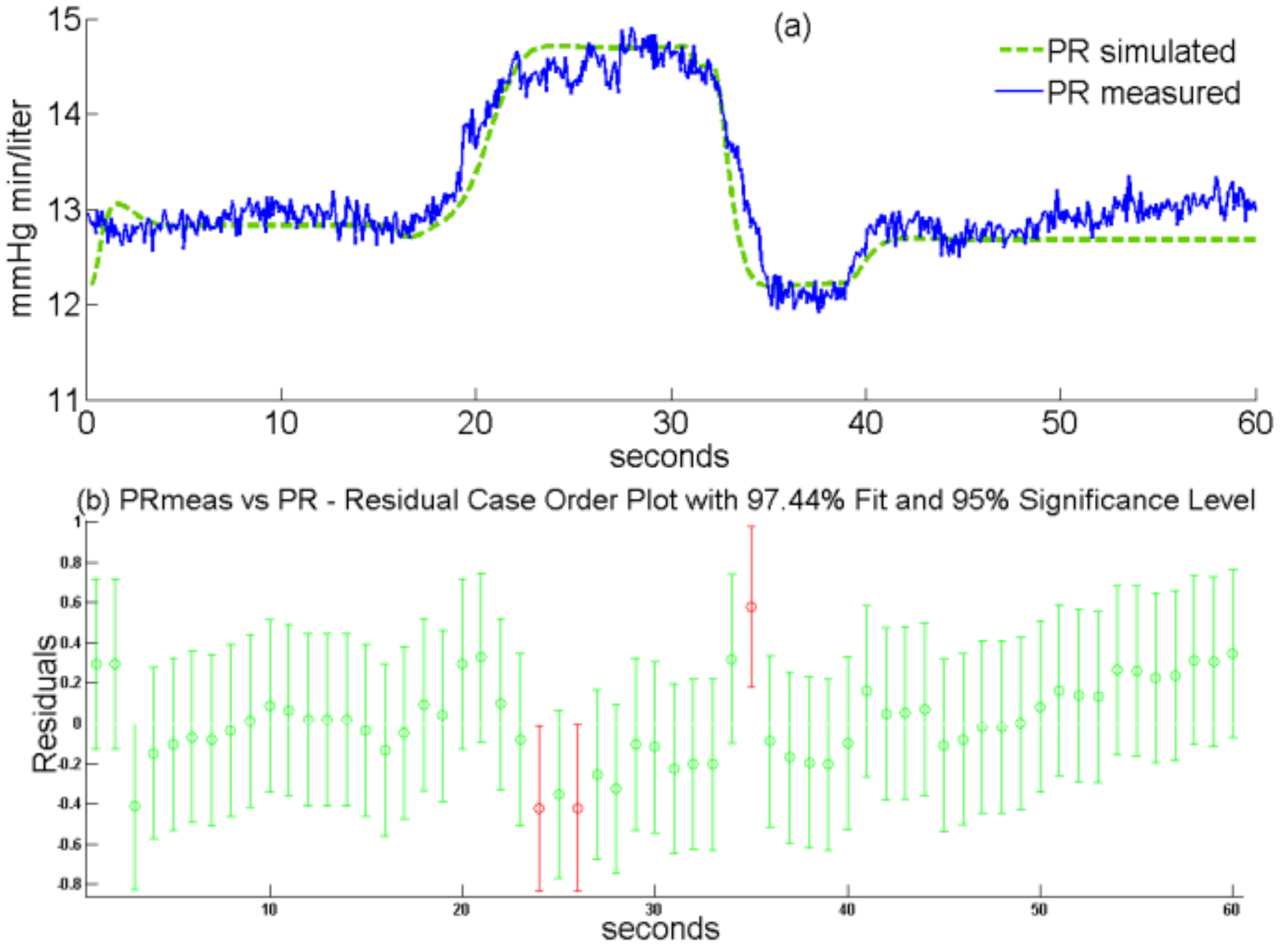

**Figure 4.23 Changes in peripheral vascular resistance during VM**

(a) predicted value (dashed line) and experimental data (plain line) shows a 97.44% good fit (b)

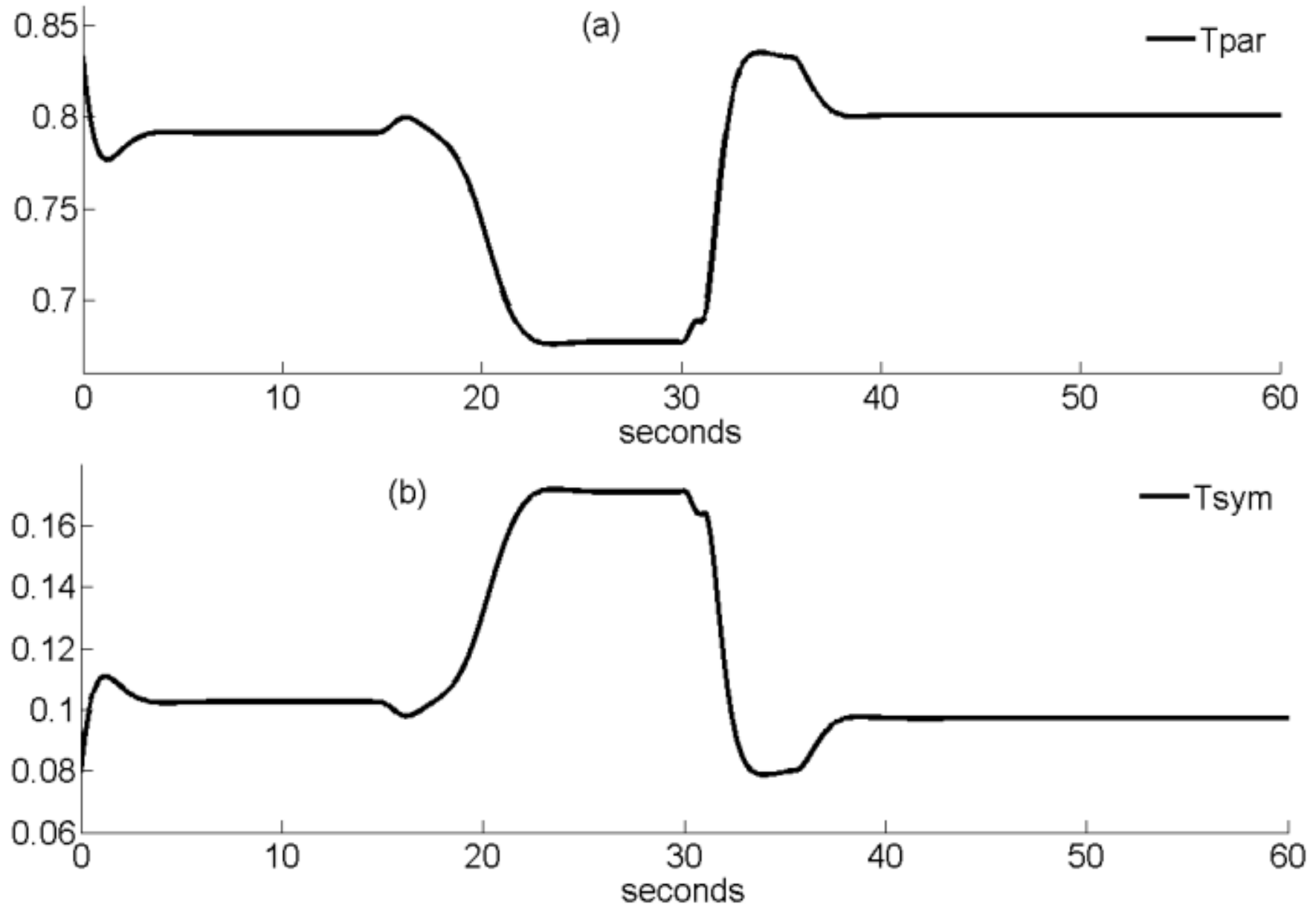

**Figure 4.24 Predicted parasympathetic (a) and sympathetic (b) activity during VM**

Y-axis is normalized in the interval [0 .. 1] in dimensionless unit

The maneuver was performed on a 23 years old healthy man and covers the timeline from the 15$^{th}$ to the 30$^{th}$ second. Phase I is characterized by slight increase of parasympathetic activity from 15$^{th}$ to 18$^{th}$ second; there is baroreflex mediated parasympathetic withdrawal and increased sympathetic tone in phase II until 30$^{th}$ second. Bradycardia characterizes phase IV which runs from 33$^{rd}$ to 36$^{th}$ second. Sympathetic activity is reduced in phase IV in order to compensate arterial pressure overshot.

### 4.5.3 Model-based Estimates of Autonomic Tones

As an alternative to heart rate variability we defined some markers of autonomic activity during the Valsalva Maneuver. The first class of model-based estimate is the so-called **Mean Autonomic Tone (MAT)** which is calculated as the average parasympathetic or sympathetic tone for a given phase of the Valsalva Maneuver using the formulas below. The index $i$ represents the phase of the maneuver which can be *baseline* (i.e. before the maneuver), *phase 1*, *phase 2*, *phase 3*, *phase 4* and *post VM* (i.e. after the maneuver). $T_{par}\{i\}$ is the portion of the parasympathetic signal corresponding to the phase $i$. $T_{sym}\{i\}$ is the portion of the sympathetic signal corresponding to the phase $i$. $MAT_{par}$ and $MAT_{sym}$ are the average parasympathetic tone and average sympathetic tone in the phase $i$ respectively.

$$\begin{aligned} MAT_{par}(i) &= mean\left(T_{par}\{i\}\right) \\ MAT_{sym}(i) &= mean\left(T_{sym}\{i\}\right) \end{aligned} \qquad (4.5.3)$$

A second class of model-based estimate is the so-called **Total Autonomic Tone (TAT)** which is calculated as the sum of all individual tones within the portion of parasympathetic or sympathetic signal for a given phase using the following formulas.

$$\begin{aligned} TAT_{par}(i) &= sum\left(T_{par}\{i\}\right) \\ TAT_{sym}(i) &= sum\left(T_{sym}\{i\}\right) \end{aligned} \qquad (4.5.4)$$

The third and fourth classes of model-based estimates are the so-called **Minimal Autonomic Tone (MIAT)** and **Maximal Autonomic Tone (MAAT)** which are calculated respectively as the minimum and maximum tone within the portion of parasympathetic or sympathetic signal for a given phase using the following formulas.

$$\begin{aligned} MIAT_{par}(i) &= min\left(T_{par}\{i\}\right) \\ MIAT_{sym}(i) &= min\left(T_{sym}\{i\}\right) \\ MAAT_{par}(i) &= max\left(T_{par}\{i\}\right) \\ MAAT_{sym}(i) &= max\left(T_{sym}\{i\}\right) \end{aligned} \qquad (4.5.5)$$

The last class of model-based estimate is the so-called **Autonomic Tone Balance (ATB)** which is calculated as the ratio of total sympathetic tone over total parasympathetic tone for a given phase using the following formulas.

$$ATB(i) = \frac{TAT_{sym}(i)}{TAT_{par}(i)} \qquad (4.5.6)$$

The model-based estimates of autonomic nervous activity were calculated for each data set and each phase of the Valsalva Maneuver. We therefore obtained 12 values per phase for each model-based estimate.

The mean values of $MAT_{par}$, $MAT_{sym}$, $TAT_{par}$, $TAT_{sym}$ and $ATB$ for all subjects are reported in Table 4.5.3 in dimensionless units. The Mean Autonomic Tone for parasympathetic activity $MAT_{par}$ decreases from its baseline value 0.872±0.032 down to 0.386±0.227 in phase 3, indicating a baroreflex-mediated withdrawal of parasympathetic activity. In the bradycardia phase 4, $MAT_{par}$ increases to the average value 0.674±0.296. Heart rate increase during phase 2 is mostly due to higher balance between parasympathetic and sympathetic tone. This is reflected in the value of Autonomic Tone Balance $ATB$, which increases by 500% from its baseline value 0.065 to 0.378 in phase 2. Mean sympathetic activity is the highest in the phase 2 when baroreflex attempts to compensate arterial pressure fall. This is reflected in the increase of Mean Autonomic Tone for sympathetic activity $MAT_{sym}$ from 0.057±0.018 to 0.256±0.087.

Table 4.5.4 shows model-based estimates for maximal and minimal autonomic tones which were calculated for a 23 years old healthy man based on data set 5. The Maximal Autonomic Tone of parasympathetic activity $MAAT_{par}$ obtained in phase 4 was found to be equal 0.835 whereas Maximal Autonomic Tone of sympathetic activity equals only 0.128 in that phase. Sympathetic activation in phase 2 was promoted by a decrease of parasympathetic activity down to a Minimal Autonomic Tone $MIAT_{par}$ equals 0.102.

We investigated possible correlations between time-domain indexes of heart rate variability and the model-based estimates proposed in this work. Results are presented in Figure 4.25. Some correlations were found between the standard deviation of NN intervals (SDNN) and the total autonomic tone ($TAT$, defined as sum of $TAT_{par}$ and $TAT_{sym}$). When moving from phase 1 to phase 2, SDNN increases by 79% while $TAT$ increases by 400% (see Figure 4.25-a). Both quantities are correlated in the same direction, but with different extent. The amplitude of SDNN change is small compared to $TAT$ because the later is a cumulative variable depending on the duration of the phase while SDNN is a cumulative mean value. A similar correlation is observed when moving from phase 3 to phase 4; where SDNN decreases by 33% while $TAT$ decreases by 80%. Both SDNN and $TAT$ reflect the overall heart rate variability; they however have the disadvantage to increase with the length of recording. $TAT$ can be used as alternative to SDNN when the length of recording is less than one minute.

A stronger correlation was found between SDNN and the autonomic tone balance ( $ATB$ ). As shown in Figure 4.25-b SDNN and $ATB$ increase by resp. 79% and 84% from phase 1 to phase 2. Both quantities respectively decrease by 33% and 23% from phase 3 to phase 4. $ATB$ does not depend on the duration of recording because it is a ratio between $TAT_{par}$ and $TAT_{sym}$. Since $ATB$ shows a stronger correlation with SDNN and is moreover independent of the length of recording, it is a good candidate for duration-independent measure of overall heart rate variability during the Valsalva Maneuver.

Another correlation was found between the square root of the mean squared differences of successive NN intervals (RMSSD) and mean autonomic tone of parasympathetic activity ( $MAT_{par}$ ).Figure 4.25-c shows 28% RMSSD decrease corresponding to 12% decrease for $MAT_{par}$ from phase 1 to phase 2. RMSSD increases by 48% from phase 3 to phase 4 whereas $MAT_{par}$ increases by 19%. This correlation suggests that $MAT_{par}$ might be used as a quantitative marker of parasympathetic activity as an alternative to RMSSD.

These results do not to emphasize on correlations between our model-based estimates and HRV parameters, they rather demonstrate that our method is more reliable in the special case of very short ECG recordings (e.g. 15 seconds).

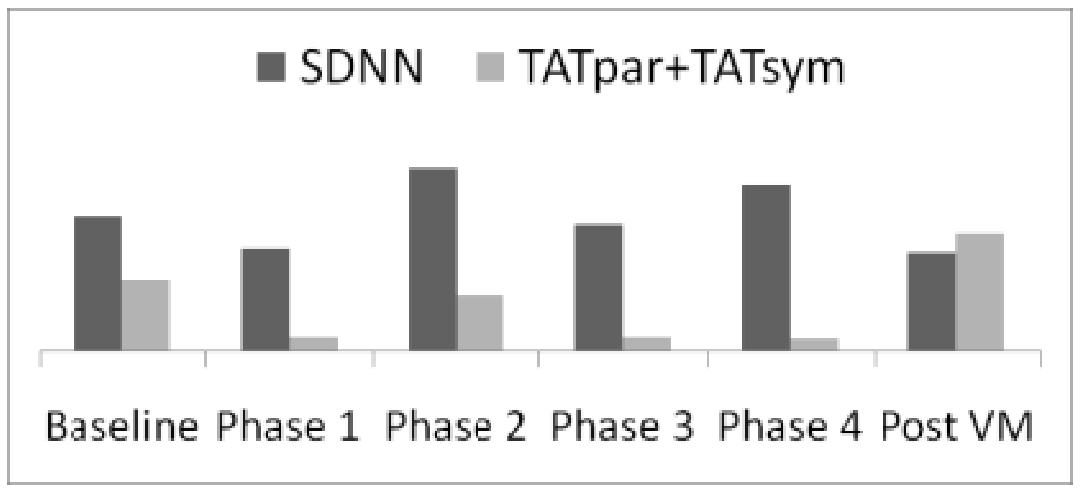

(a)

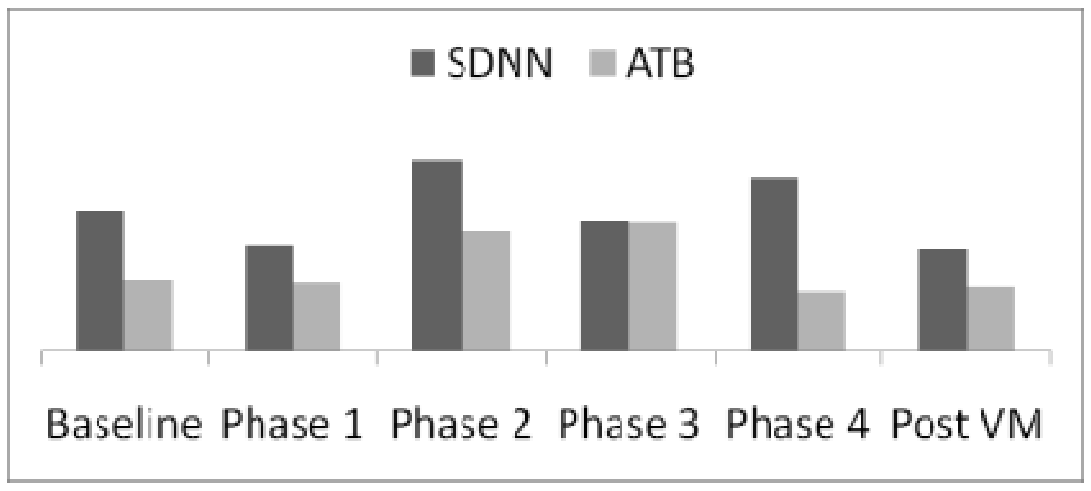

(b)

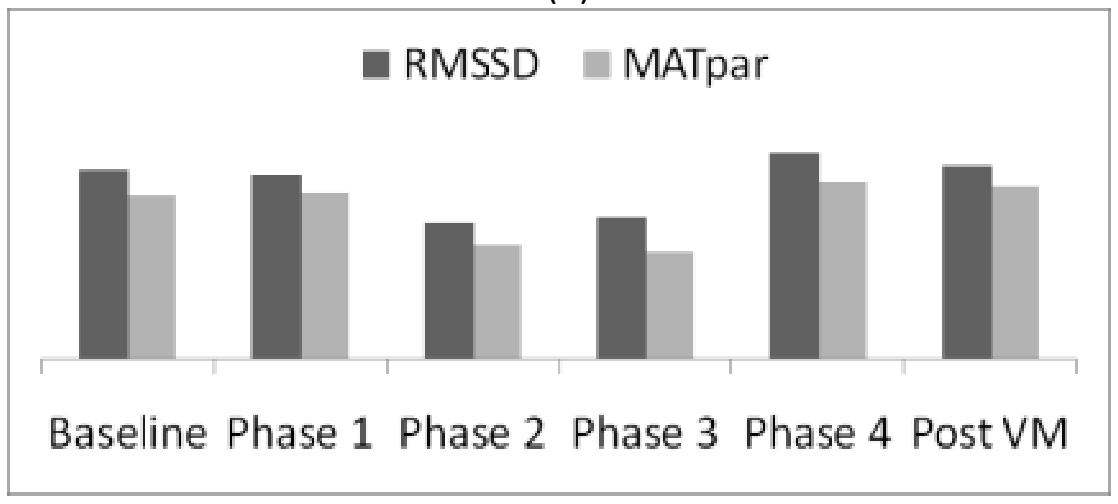

(c)

**Figure 4.25 Correlation between time-domain indexes of HRV and model-based estimates**
Y-axis for model-based estimates is normalized in the interval [0 .. 1] in dimensionless unit.
(a) SDNN (standard deviation of NN intervals in milliseconds) is correlated with $TAT$ (total autonomic tone) from phase 1 to phase 2 and from phase 3 to phase 4. (b) SDNN is additionally correlated with $ATB$ (autonomic tone balance) from phase 1 to phase 2 and from phase 2 to phase 4. (c) RMSSD (mean squared differences of successive NN intervals in milliseconds) is correlated with $MAT_{par}$ (mean autonomic tone for parasympathetic activity). $ATB$ and $MAT_{par}$ were amplified by a factor of 100 for graphical presentation.

Table 4.5.3 Mean values of model-based estimates of autonomic tone (normalized in the interval [0 .. 1] in dimensionless unit) obtained from five healthy subjects during VM

| | **Mean Autonomic Tone** | | **Total Autonomic Tone** | | **Autonomic Tone Balance** |
|---|---|---|---|---|---|
| | $MAT_{par}$ | $MAT_{sym}$ | $TAT_{par}$ | $TAT_{sym}$ | $ATB$ |
| **Baseline** | 0.872±0.032 | 0.057±0.018 | 13.086±0.475 | 0.856±0.275 | 0.065 |
| **Phase 1** | 0.880±0.033 | 0.054±0.018 | 2.270±0.479 | 0.137±0.054 | 0.060 |
| **Phase 2** | 0.595±0.052 | 0.256±0.087 | 7.501±0.645 | 2.833±0.744 | 0.378 |
| **Phase 3** | 0.386±0.227 | 0.254±0.146 | 1.451±0.945 | 0.961±0.624 | 0.662 |
| **Phase 4** | 0.674±0.296 | 0.137±0.136 | 2.913±0.796 | 0.437±0.334 | 0.150 |
| **Post VM** | 0.877±0.036 | 0.050±0.027 | 20.25±1.425 | 1.171±0.662 | 0.058 |

Table 4.5.4 Model-based estimates for minimal and maximal autonomic tone (normalized in the interval [0 .. 1] in dimensionless unit) obtained on a 25 years old healthy man

| | **Minimum Autonomic Tone** | | **Maximum Autonomic Tone** | |
|---|---|---|---|---|
| | $MIAT_{par}$ | $MIAT_{sym}$ | $MAAT_{par}$ | $MAAT_{sym}$ |
| **Baseline** | 0.777 | 0.08 | 0.833 | 0.111 |
| **Phase 1** | 0.791 | 0.098 | 0.800 | 0.103 |
| **Phase 2** | 0.676 | 0.102 | 0.793 | 0.172 |
| **Phase 3** | 0.677 | 0.128 | 0.747 | 0.171 |
| **Phase 4** | 0.747 | 0.079 | 0.835 | 0.128 |
| **Post VM** | 0.800 | 0.079 | 0.835 | 0.098 |

## 4.6 Sensitivity Analysis

While parameter estimation assumes the availability of reliable experimental data, this is not the case in reality. Experiments are not perfect and measurements are always deteriorated by noise. In further steps the following question should be answered: is the quality of the measurement data sufficient for identifying the model parameters? This includes computing the confidence region of the parameter estimates, computing how sensitive model outputs are when inputs change or computing how sensitive outputs are when parameters change. This information gives feedback on the experimental condition and helps adjusting elements such as frequency of measurements and duration of experiments. This is generally called Optimal Experimental Design problem. Local sensitivity analysis can be performed by direct differentiation of model output with respect to the parameter of interest. The direct differentiation method is only possible when the model equations are available in an analytical form. Otherwise it is preferable to use the numerical integration method. [173]

Our model is rather complex and includes a large amount of differential equations relating 33 model parameters to 4 model outputs. Direct differentiation would imply a lot of error prone manual calculations since the Matlab Simulink environment does not offer ways to partially derivate model outputs automatically. The same difficulty is encountered with numerical integration. Under the assumption that all parameters behave like a Gaussian random variable, sensitivity analysis can be performed by varying the values of a given parameter with a known variance; and calculating the variance of the mean value of each model output. A similar approach has held good results in [174]. We compute the sensitivity of each model output $Y$ with regards to a given parameter $k$ by determining how much variance the parameter produces in a significant statistic $Y_{k_j,stat}$ of model output $Y$. We choose the following four statistics and subdivide the parameters in four groups accordingly: parameters that affect mean values of model outputs form group 1, parameters that affect low frequency fluctuations in model outputs are in group 2, parameters that affect high frequency fluctuations in model outputs form group 3 and those that affect the difference between mean value of model output before and during experimental stimulation are in group 4. An overview of the parameter groups and model outputs is given in Table 4.6.1. Parameters related to intra-thoracic pressure elevation were not included for simplicity.

For group 1, sensibility is calculated from the variance of the mean values of model outputs. For group 2, sensibility is calculated from the variance of low frequency components of model outputs, measured in absolute value of power. For group 3, sensibility is calculated from the variance of high frequency components of model outputs, measured in absolute value of power. For group 4, sensibility is calculated from the variance of the differences between mean values of model outputs before the stimuli and after the stimuli.

Let's consider $k$ a model parameter bounded in the interval $\left[k_{min}...k_{max}\right]$, $Y$ a model output and $f$ a function representing our model. The sensitivity of model output $Y$ with regards to model parameter $k$ is denoted $S_{Y,k}$ and is calculated as follows:

```
For each V_{k,i} in the interval [k_min ...2 · k_max]
    Generate a random set of 10 real numbers in the interval [k_min ...k_max] with variance V_{k,i}
    For each k_j in the random set
        Calculate model output Y_{k_j} = f(k_j)
        If k is member of Parameter Group 1
            Calculate Y_{k_j,stat} as mean value of model output
        End If
        If k is member of Parameter Group 2
            Calculate Y_{k_j,stat} as low frequency components measured in absolute values of power
        End If
        If k is member of Parameter Group 3
            Calculate Y_{k_j,stat} as high frequency components measured in absolute values of power
        End If
        If k is member of Parameter Group 4
            Calculate Y_{k_j,stat} as difference between mean value of model output before and during the
        stimuli
        End If
    End For
    Determine the variance of model outputs V_{Y,i} = variance(Y_{k_j,stat})_{j=1...10}
    Store the variances S_{Y,k}(i,1) = V_{k,i} and S_{Y,k}(i,2) = V_{Y,i}
End For
Return the sensitivity matrix S_{Y,k} where V_{k,i} is the first column and V_{Y,i} is the second column
```

In order to illustrate how the algorithm works, let's consider $k = f_{ev,0,max}$, the model parameter representing parasympathetic tone, bounded in the interval $\left[k_{min}...k_{max}\right] = \left[1...20\right]$. We are interested in calculating the sensitivity $S_{HR,f_{ev,0,max}}$ of model output $HR$ (heart rate) with regards to $f_{ev,0,max}$. For each $V_{f_{ev,0,max},i}$ in the interval $\left[k_{min}...2 \cdot k_{max}\right] = \left[1...40\right]$, the algorithm chooses a random set of 10 real numbers in the interval $\left[1...20\right]$ with variance $V_{f_{ev,0,max},i}$. For example if $V_{f_{ev,0,max},i} = 4.53$, then the random set could be $\{7.54,6.76,9.24,9.83,8.58,6.69,7.14,10.06,11.87,4.4551\}$. For each $f_{ev,0,max_j}$ in the random set, the algorithm simulates the model and calculates the mean value of simulated heart rate $HR_{f_{ev,0,max_j},stat}$. After the mean value of heart rate has been obtained for all $f_{ev,0,max_j}$ in the random set, the algorithm computes the variance of the mean values as $V_{HR,i}$, which is equal to $1.85$ in the case of $V_{f_{ev,0,max},i} = 4.53$. By repeating the process, we obtain an output variance $V_{HR,i}$ for each parameter variance $V_{f_{ev,0,max},i}$. The set of such couples forms the sensitivity of $HR$ (heart rate) with regards to $f_{ev,0,max}$ (parasympathetic tone), as shown in the table attached to Figure 4.26.

Table 4.6.1 Sensitivity degrees of model outputs with regards to parameters (≥ 1 means high sensitivity, 0 means no sensitivity)

| Group | Name | Description | $HR$ | $QT$ | $P_{as}$ | $G_{mp}$ | Overall Sensitivity |
|---|---|---|---|---|---|---|---|
| Group 1 – Parameters with mean effect | $f_{es,0,high}$ | firing rate of pacemaker sympathetic premotor neurons | 0.0035 | 0.2312 | 0.0494 | 0.1618 | **0.4459** |
| | $W_{c,es,0}$ | synaptic weight applied to sensory inputs from chemoreceptors | 0.0065 | 1 | 1 | 1 | **3.0065** |
| | $f_{ev,0,max}$ | parasympathetic tone (maximal intrinsic firing rate of vagal nerves) | 0.2255 | 0.1444 | 0.1839 | 0.0011 | **0.5549** |
| | $k_{nor_{og},v}$ | modulation factor of vagally-mediated tachycardia | 1 | 0.0342 | 0.6975 | 0.0037 | **1.7354** |
| | $k_{QT}$ | gain of elastance on QT interval duration | 0 | 0.1592 | 0 | 0 | **0.1592** |
| | $k_{G_{mp}}$ | gain parameter applied to peripheral vascular resistance in skeletal muscles | 0 | 0 | 0 | 0.7191 | **0.7191** |
| Group 2 - Parameters with low frequency fluctuation effect | $W_{es,0,low}$ | synaptic weight applied to low frequency oscillators in sympathetic premotor neurons | 0.5875 | 0.2396 | 0.9856 | 0.1077 | **1.9204** |
| | $r_{es,0,low}$ | parameters determining the rate of low frequency oscillators in sympathetic premotor neurons | 0.6905 | 0.4148 | 0 | 0.6456 | **1.7509** |
| | $k_{es,0,low}$ | | 0 | 0 | 0.0001 | 0 | **0.0001** |
| Group 3 - Parameters with high frequency fluctuation effect | $W_{resp,v}$ | synaptic weight applied to inputs from inspiratory-expiratory somatic motor neurons | 0 | 0.996 | 0.0037 | 0.0014 | **1.0011** |
| | $f_{rsa}$ | frequency of intracardiac high frequency fluctuations | 1.2189 | 0.0046 | 0.0013 | 0.0037 | **1.2285** |
| | $G_{T_{rsa}}$ | gain of intracardiac high frequency fluctuations | 0.0002 | 0.0203 | 0.3696 | 0.4297 | **0.8198** |
| Group 4 - Parameters related to external stimuli | $f_{at_f,max}$ | upper saturation of facial cold receptor | 0.0959 | 0.139 | 0.6185 | 0.2418 | **1.0952** |
| | $k_{at_f}$ | slope of the exponential decay of cold receptor on face | 0.046 | 0.0009 | 0.0808 | 0.0762 | **0.2039** |
| | $f_{at_s,max}$ | upper saturation of cold receptor on hands and feet | 0.528 | 0.7791 | 0.4236 | 0.7146 | **2.4453** |
| | $k_{at_s}$ | slope of the sigmoid response of cold receptor on hands and feet | 0.0022 | 0.0283 | 0.0264 | 0.0307 | **0.0876** |
| | $f_{as,max}$ | upper saturation of mental stress sensors | 0.3079 | 0.2631 | 0.2149 | 0.1815 | **0.9674** |
| | $k_{as}$ | slope of the mental stress sensor response | 0.1263 | 0.1333 | 0.1417 | 0.1403 | **0.5416** |
| | $g_{ss,max}$ | maximal value of effects due to gravitational forces during active standing from sitting | 0.0004 | 0 | 0.0016 | 0 | **0.002** |
| | $k_{ss}$ | slope of the heart rate response during active standing from sitting | 0.0202 | 0.0001 | 0.2554 | 0.0003 | **0.276** |
| | $g_{tt,max}$ | maximal value of effects due to gravitational forces during head-up tilt | 0.0056 | 0.0001 | 0.0622 | 0 | **0.0679** |
| | $k_{tt}$ | slope of the heart rate response during head-up tilt | 0.0012 | 0 | 0.2329 | 0 | **0.2341** |
| | $f_{ao,max}$ | upper saturation of oculo-pressure receptor | 0.0517 | 0.0055 | 0.0067 | 0 | **0.0639** |
| | $k_{ao}$ | slope of the sigmoid response of oculo-pressure receptor | 0 | 0.0033 | 0.006 | 0.0002 | **0.0094** |

| $V_{f_{ev,0,max},i}$ | $V_{HR,i}$ |
|---|---|
| 0.59 | 0.51 |
| 4.53 | 1.85 |
| 6.35 | 2.70 |
| 6.58 | 3.50 |
| 7.63 | 3.49 |
| 9.74 | 4.85 |
| 10.82 | 6.68 |
| 12.21 | 5.91 |
| 21.86 | 10.29 |
| 27.95 | 14.47 |
| 37.63 | 18.35 |

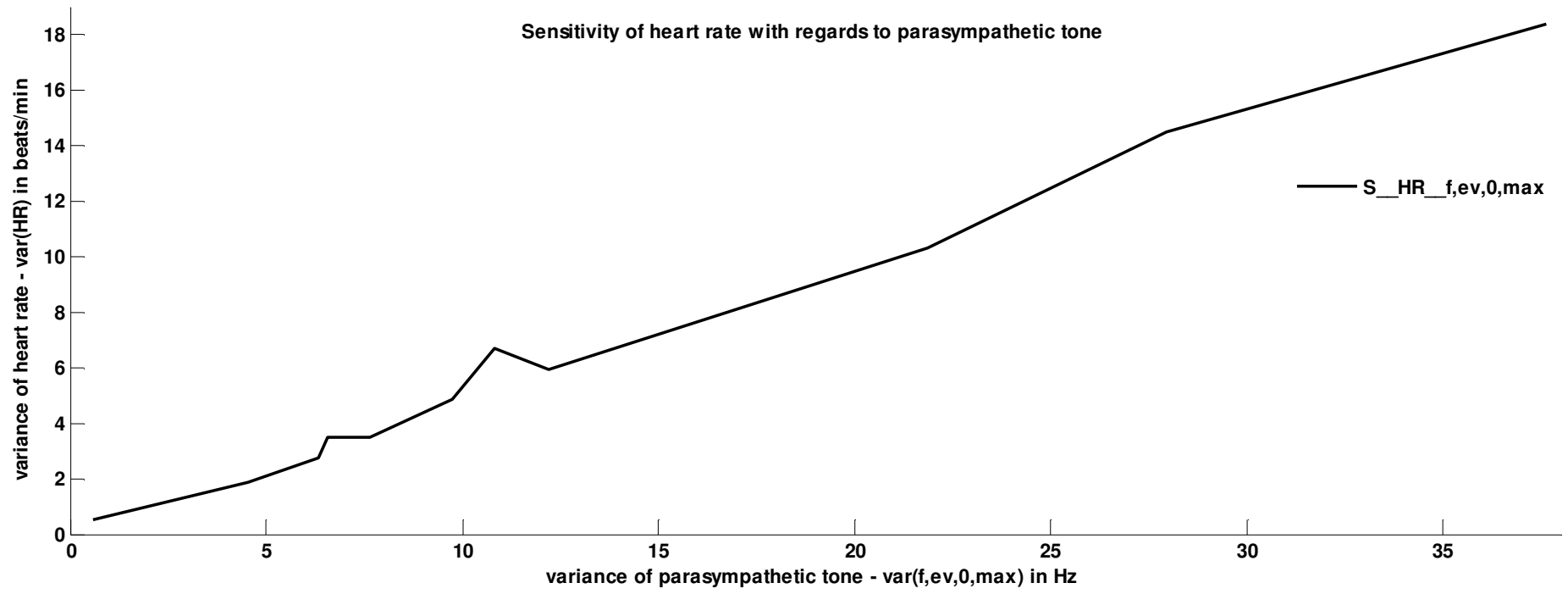

**Figure 4.26 Sensitivity of heart rate with regards to parasympathetic tone**

The plot in Figure 4.26 shows that the more we vary model parameter $f_{ev,0,max}$ , the more model output $HR$ varies in the same direction. In other terms, heart rate is sensitive to changes in model parameter representing parasympathetic tone.

In order to assess the degree of sensitivity for each of the four groups of parameters, we normalized each axis of plots like the one obtained in Figure 4.26 in a scale between 0 and 1 by dividing the values over the maximal variance. After fitting the resulting plot to a polynomial of order 1, we calculate the slope of the line as **degree of sensitivity**, which tells us in which extent the model output is sensitive to the given model parameter. Figure 4.27 shows results obtained for heart rate and parameters which affect its mean value (i.e. Group 1). While mean heart rate is very sensitive to modulation factor of vagally-mediated tachycardia ($k_{nor_{og},v}$) and parasympathetic tone ($f_{ev,0,max}$) with degree of sensitivity equal to 1 and 0.2255 respectively; varying the parameters $f_{es,0,high}$, $W_{c,es,0}$, $k_{QT}$, $k_{G_{mp}}$ has almost no effect on mean heart rate; degree of sensitivity was 0.0035, 0.0065, 0 and 0 respectively. The same procedure was applied for calculating the degree of sensitivity of remaining model outputs $QT$ (QT interval duration), $P_{as}$ (arterial blood pressure) and $G_{mp}$ (level of galvanic skin response). The obtained degrees of sensitivity are reported in Table 4.6.1. While model parameter $W_{c,es,0}$ does not affect heart rate; $QT$ , $P_{as}$ and $G_{mp}$ are highly sensitive to its values (degree of sensitivity was 1). For each parameter we additionally calculate the **overall sensitivity** as sum of individual degrees of sensitivity obtained from all model outputs. This allowed us to rank parameters globally from the most sensitive to the less sensitive: $W_{c,es,0}$ was the parameter with the greatest overall sensitivity (=3.0065) whereas $k_{es,0,low}$ was almost not sensitive (overall sensitivity = 0.0001) to any model output and we might consider transforming it into a model constant. The overall sensitivity is reported in Table 4.6.1 as well. Details about how model outputs vary when parameters vary can be verified in Figure 4.27 to Figure 4.42, for each parameters group.

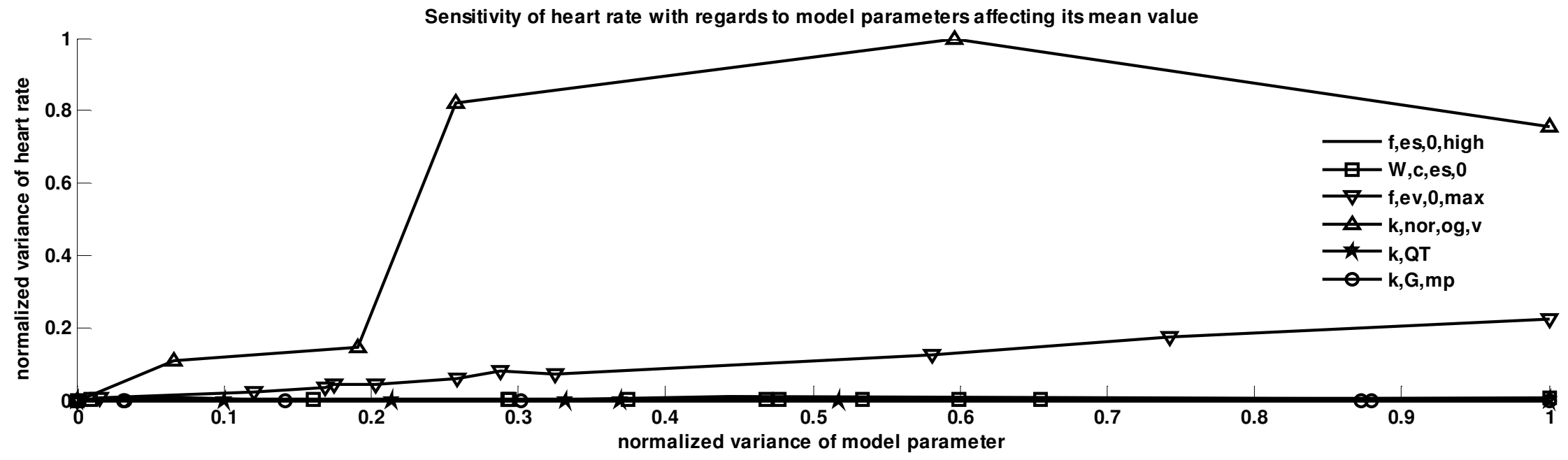

Figure 4.27 Sensitivity of heart rate with regards to parameters from Group 1

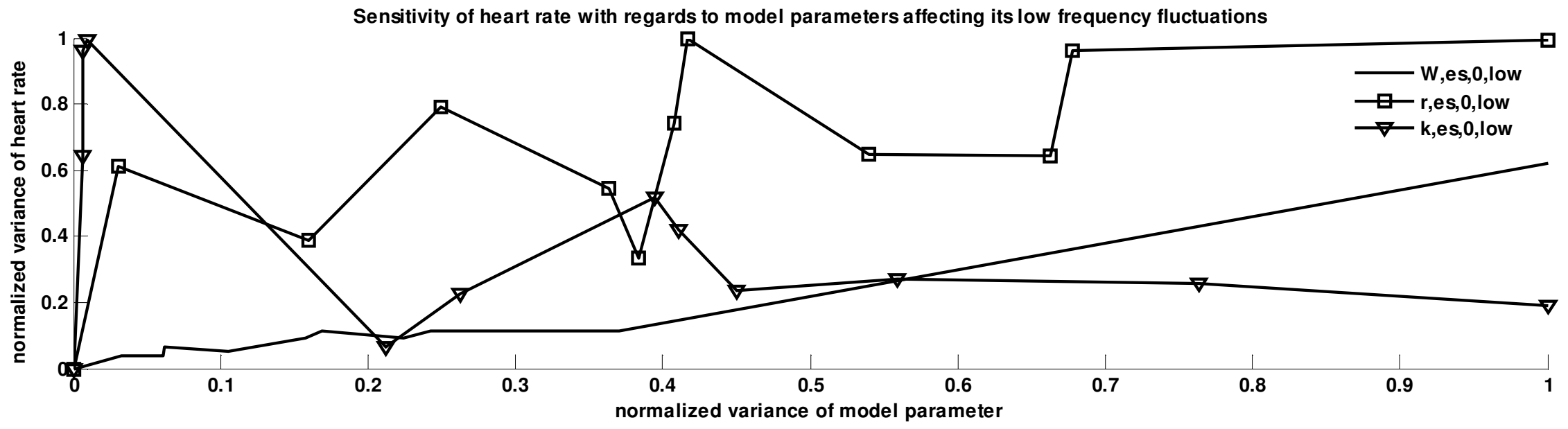

Figure 4.28 Sensitivity of heart rate with regards to parameters from Group 2

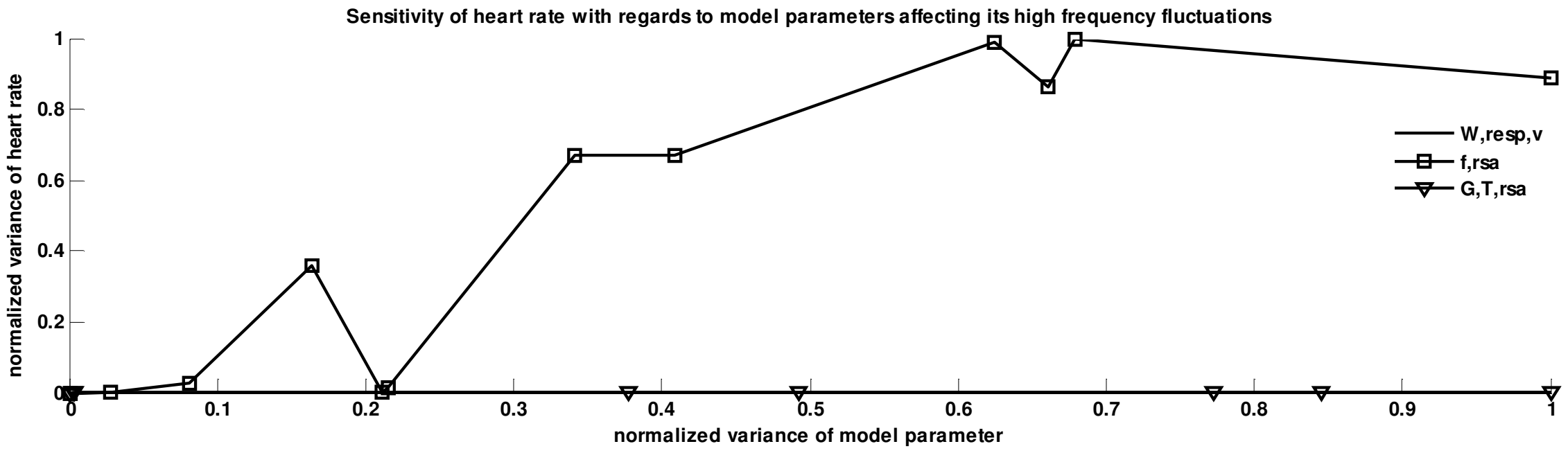

Figure 4.29 Sensitivity of heart rate with regards to parameters from Group 3

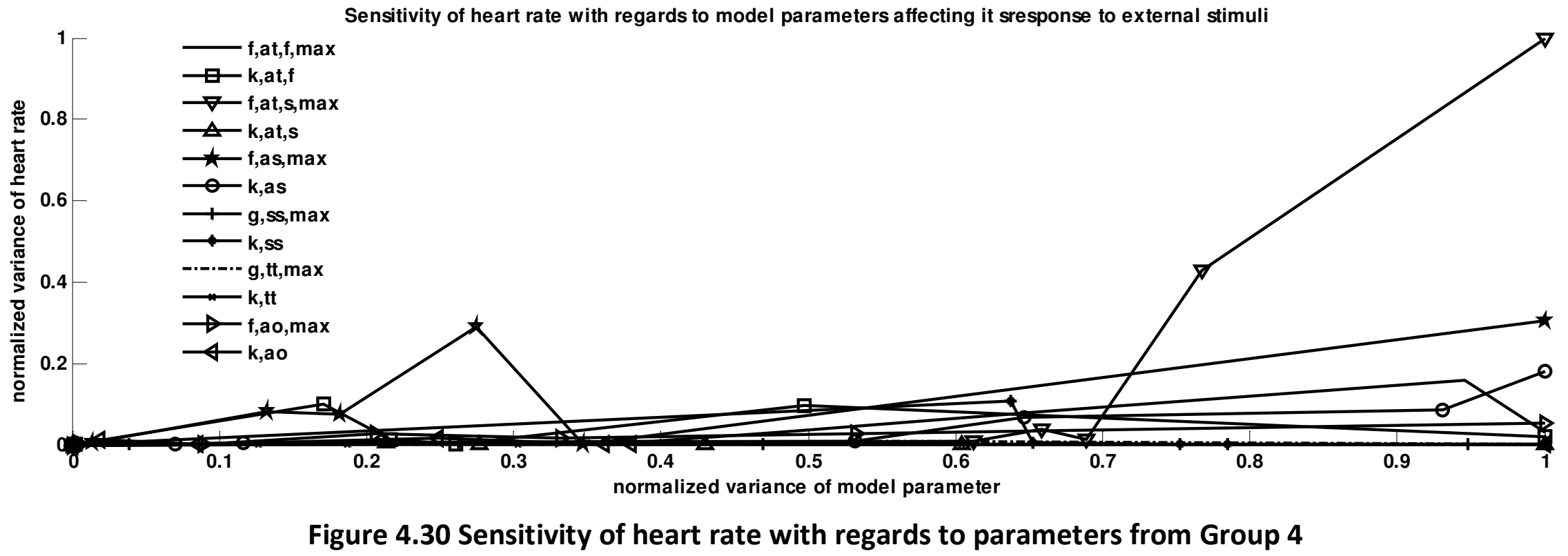

Figure 4.30 Sensitivity of heart rate with regards to parameters from Group 4

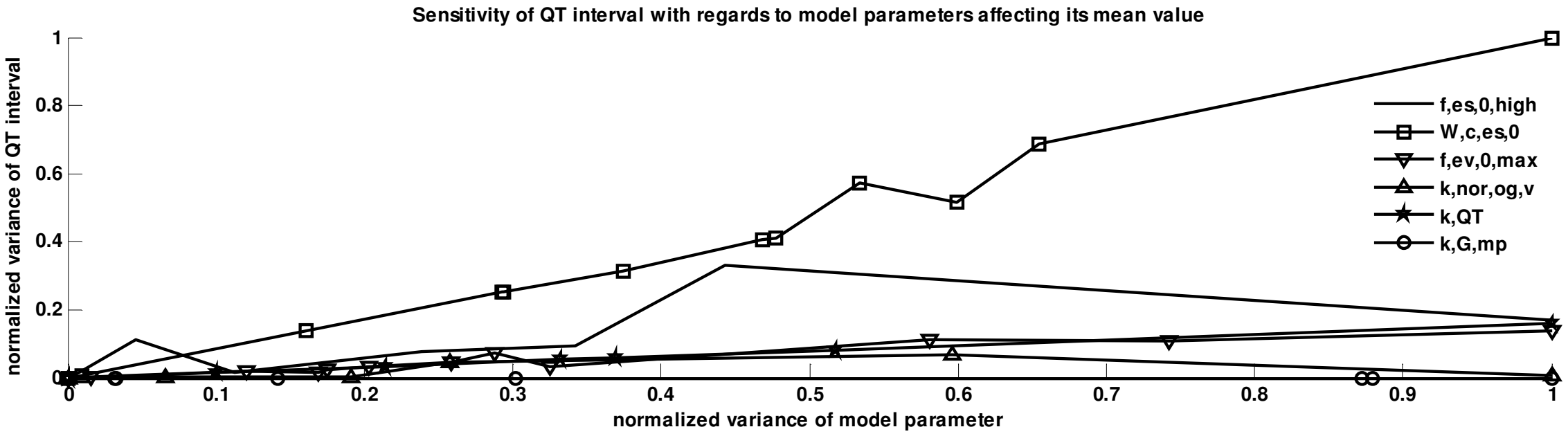

**Figure 4.31 Sensitivity of QT interval with regards to parameters from Group 1**

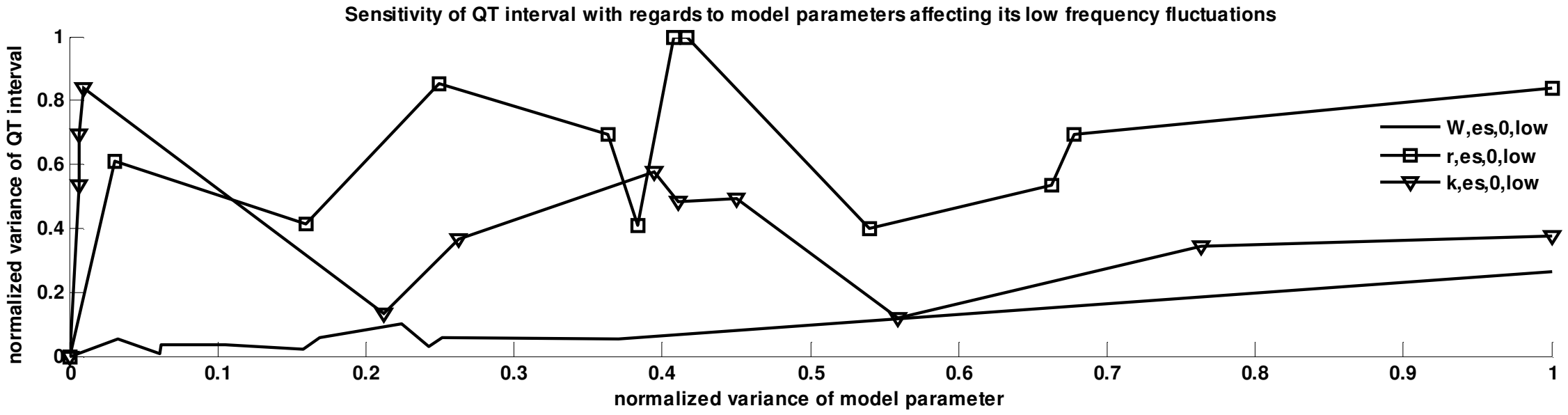

**Figure 4.32 Sensitivity of QT interval with regards to parameters from Group 2**

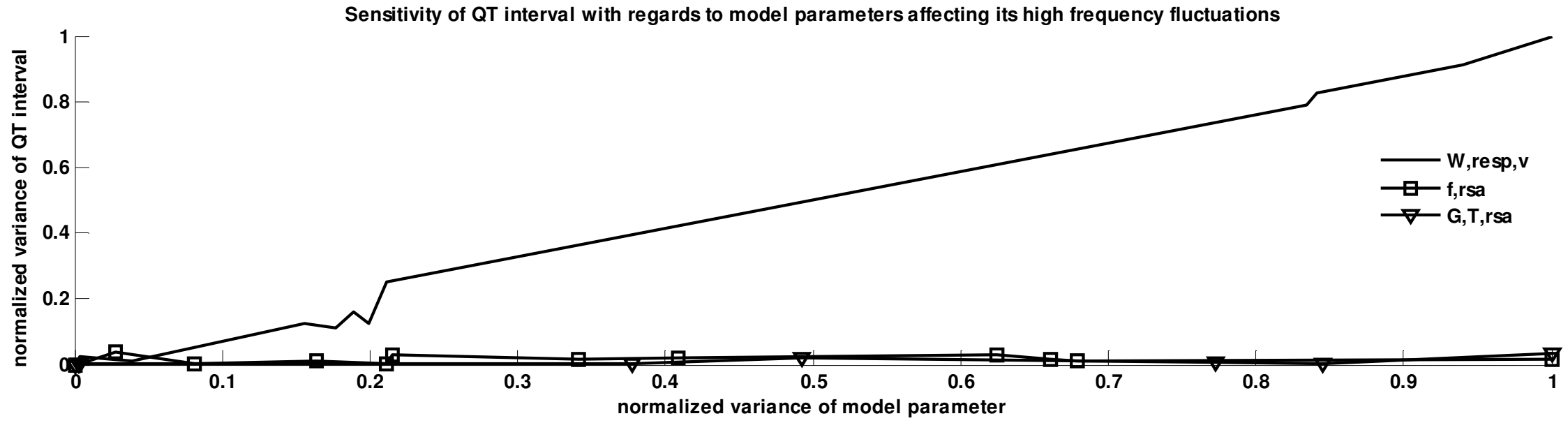

**Figure 4.33 Sensitivity of QT interval with regards to parameters from Group 3**

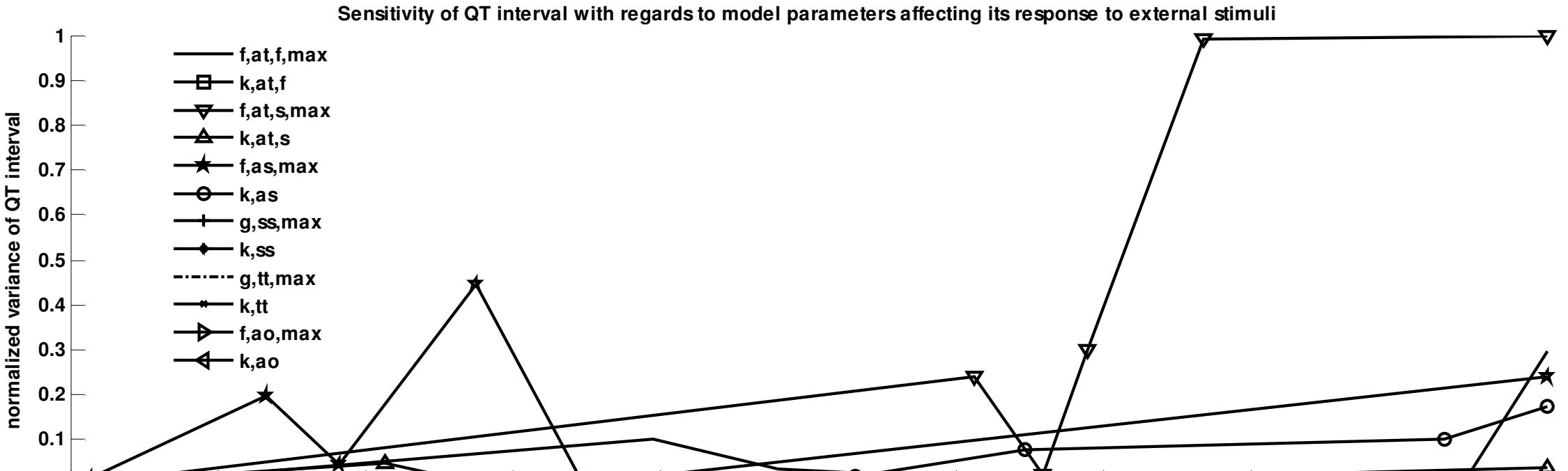

**Figure 4.34 Sensitivity of QT interval with regards to parameters from Group 4**

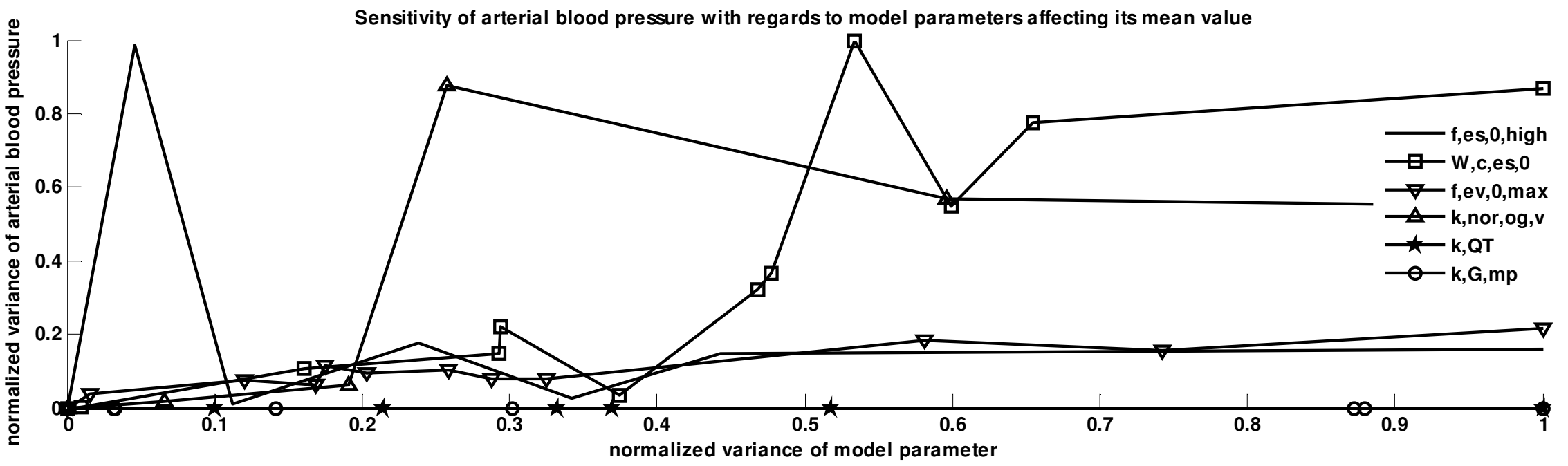

**Figure 4.35 Sensitivity of arterial blood pressure with regards to parameters from Group 1**

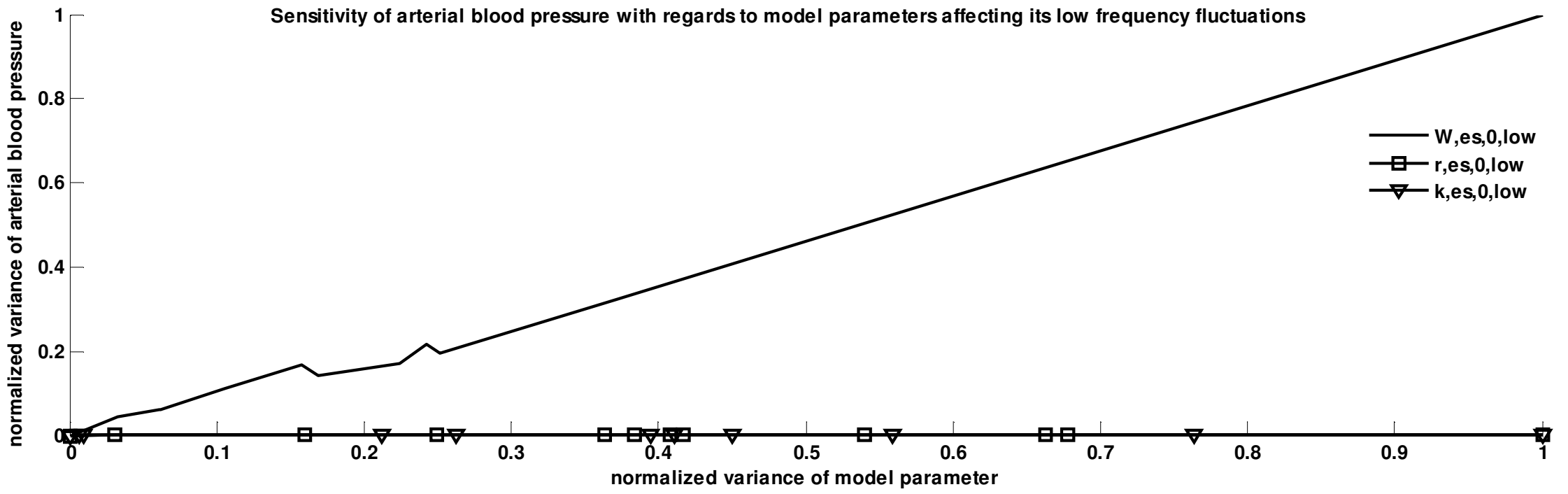

**Figure 4.36 Sensitivity of arterial blood pressure with regards to parameters from Group 2**

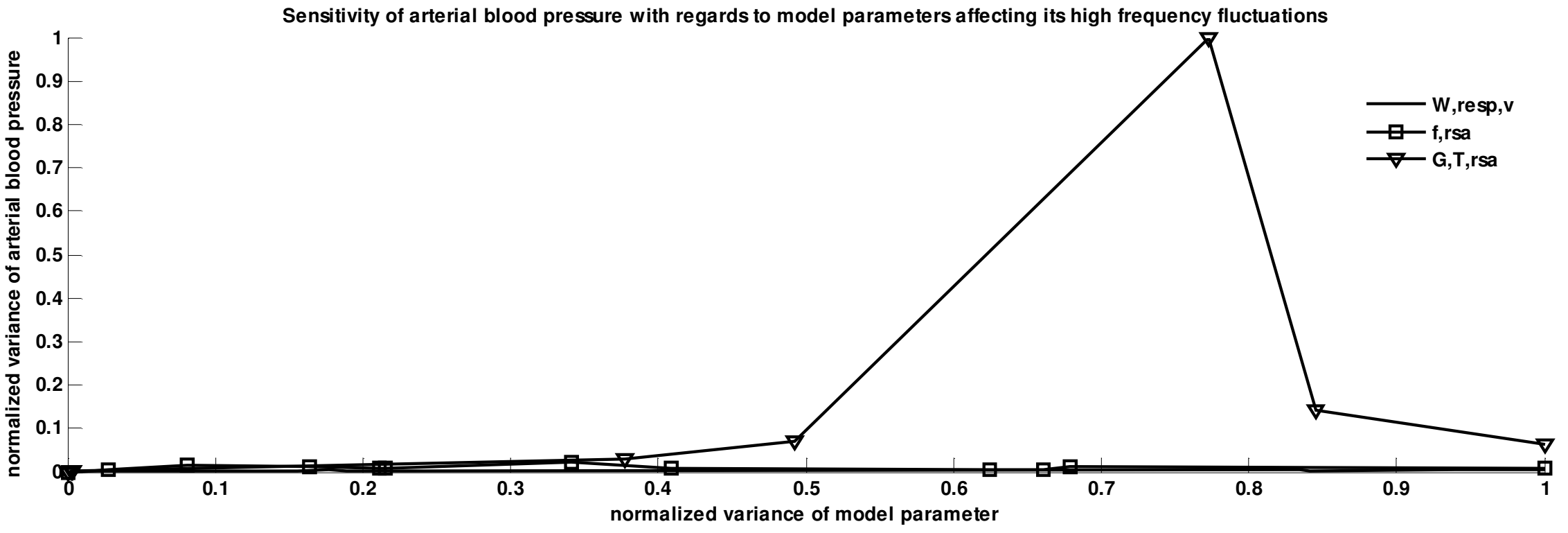

**Figure 4.37 Sensitivity of arterial blood pressure with regards to parameters from Group 3**

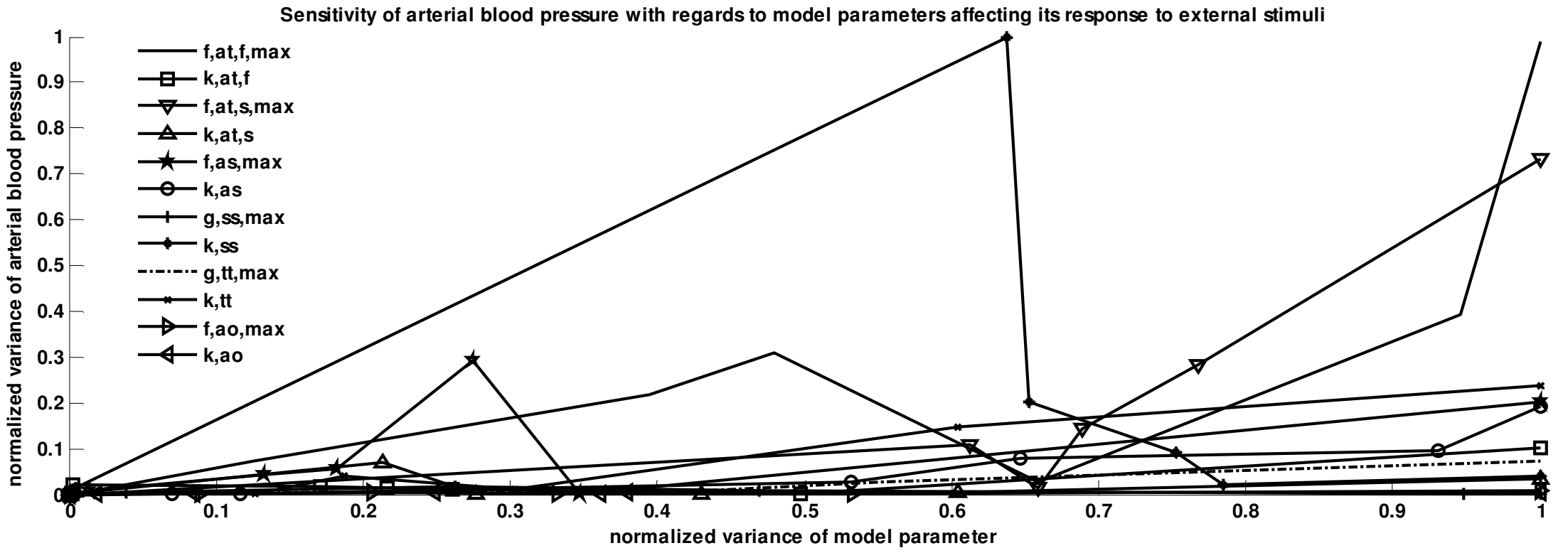

**Figure 4.38 Sensitivity of arterial blood pressure with regards to parameters from Group 4**

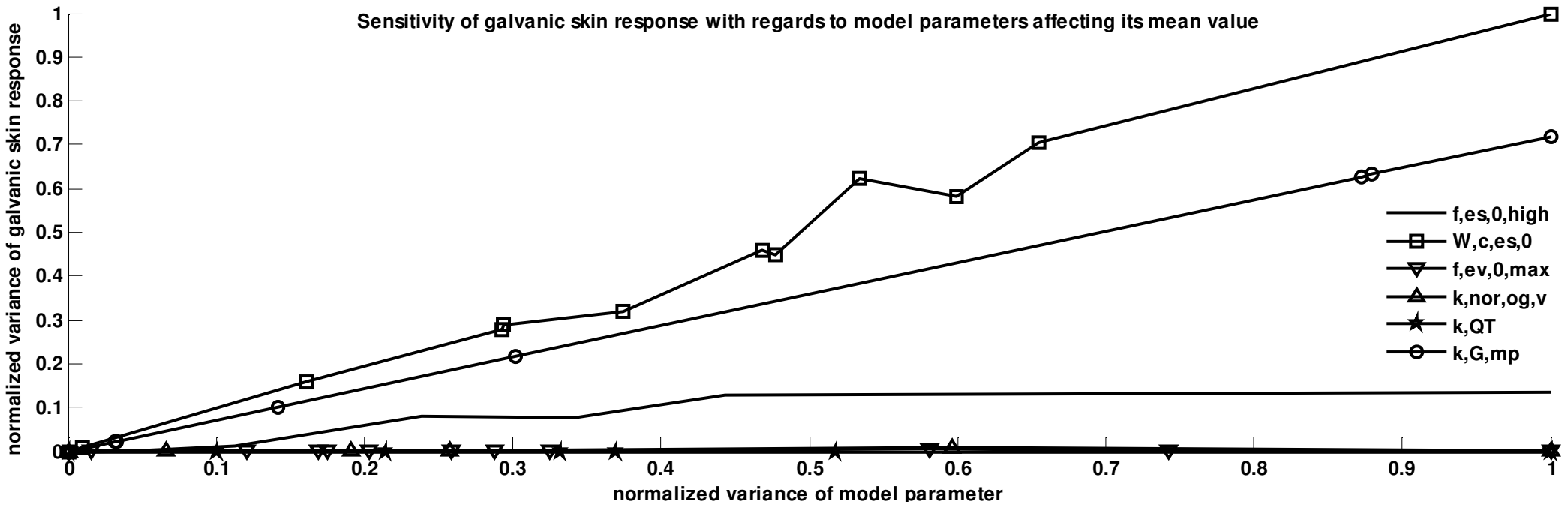

**Figure 4.39 Sensitivity of galvanic skin response with regards to parameters from Group 1**

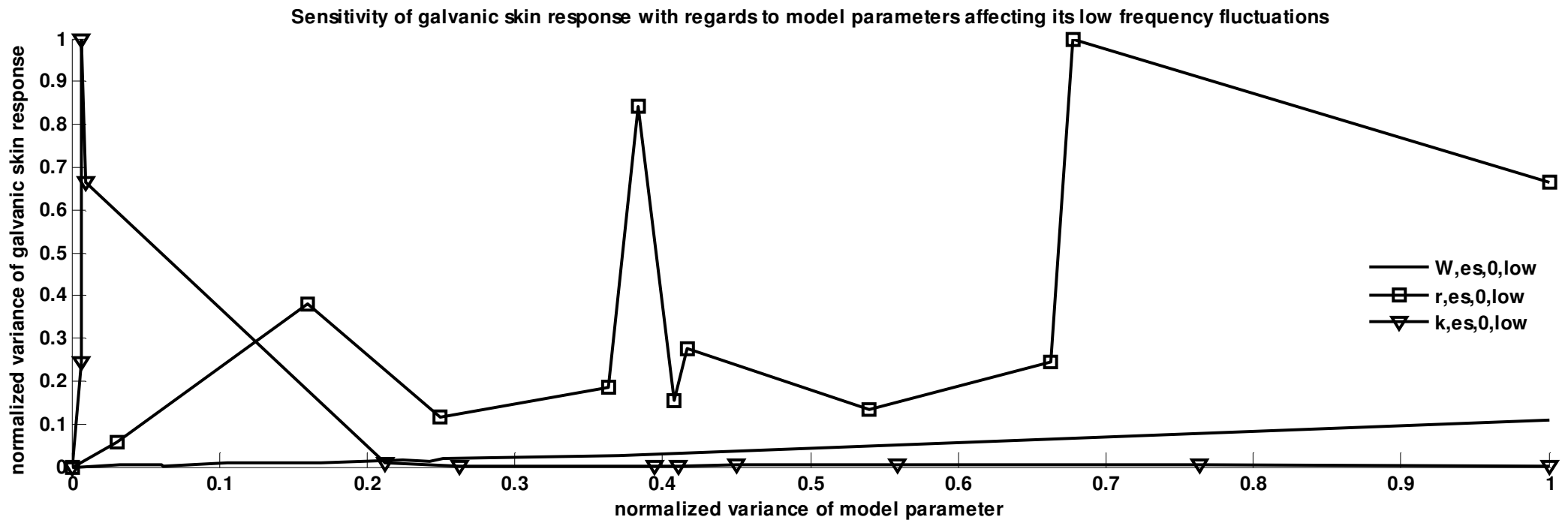

**Figure 4.40 Sensitivity of galvanic skin response with regards to parameters from Group 2**

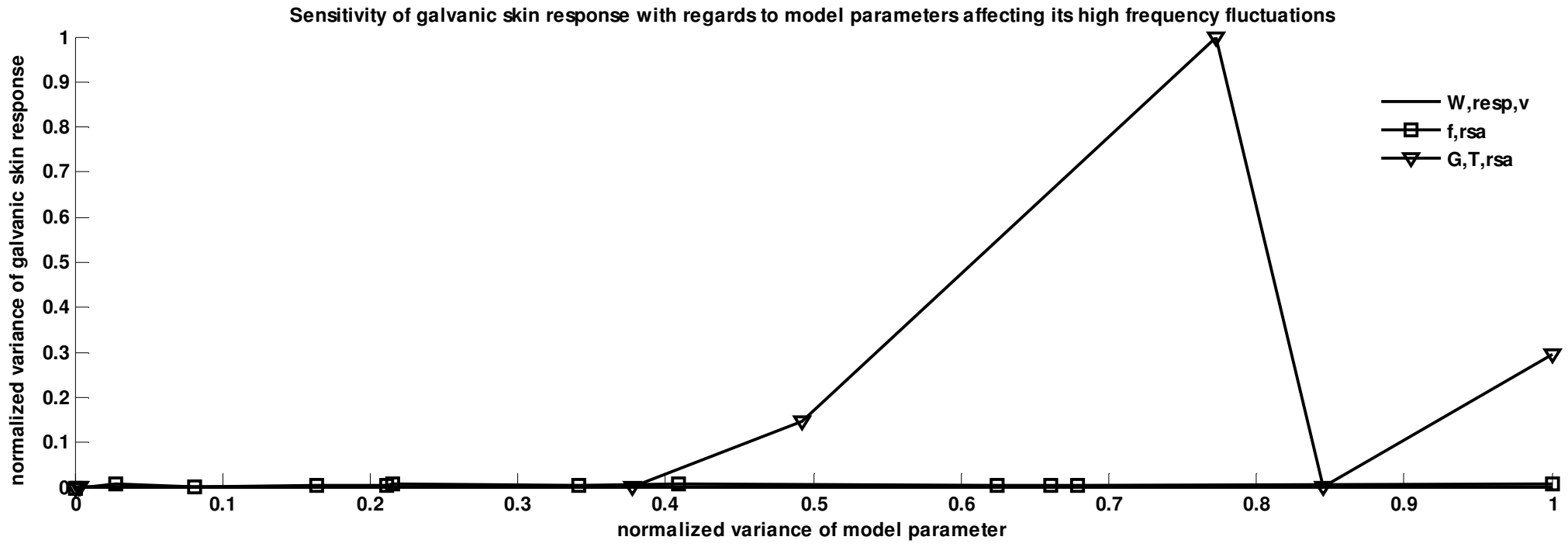

**Figure 4.41 Sensitivity of galvanic skin response with regards to parameters from Group 3**

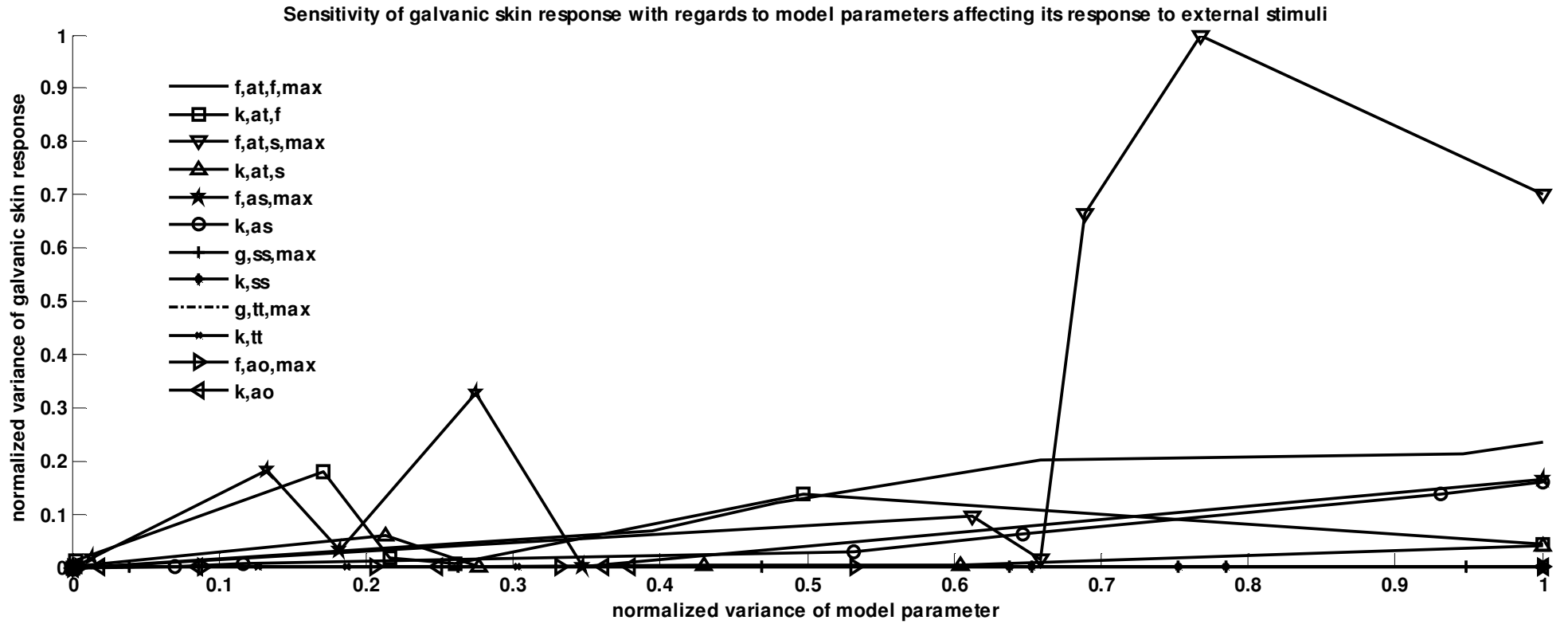

**Figure 4.42 Sensitivity of galvanic skin response with regards to parameters from Group 4**

## 4.7 Conclusion

Our main contribution lies in the development of an integrative model for cardiovascular control activity in Chapter 2, and its validation with non-invasive in-vivo data (human and equine, see Chapter 3) and optimization methods as outlined in this Chapter 4.

Results bring supporting evidences that Mayer waves result from a rhythmic sympathetic discharge of pacemaker-like sympathetic premotor neurons. We show that vagally-mediated tachycardia could be related to the secretion of vasoactive neurotransmitters by the vagal nerve. We could quantify the resting sympathetic and parasympathetic tone separately; and demonstrate correlation with aging and gender. Our method appeared to provide more consistent results than heart rate variability during short-term cardiovascular control in a timeframe up to 30 seconds.

Conclusions include the selection of specific integrative model parameters for their ability to provide quantitative analysis of cardiovascular control. Conclusions from the in-vivo experimentation show that the models do allow trend and behavioral effects to be modeled.

A subsidiary contribution with a more practical bent of pattern recognition and feature extraction for telemedicine and biofeedback will be presented in the next Chapter 5.

# Chapter 5. Perspectives for Application in Research and Clinical Environments

We have proposed methods for quantitatively assess autonomic nervous activity on cardiovascular system in the previous chapters. Experimental arrangements were performed and obtained data were used to fit our integrative model of cardiovascular control. This allowed us to estimate sympathetic and parasympathetic tones of individual subjects. Features related to gender, age and disease state were highlighted. In this chapter we demonstrate how the model could be used to simulate and generate cardiovascular signals when common experimental setups do not allow them to be measured noninvasively. We will also open some perspectives for clinical application of the results and present a software package that was developed in the frame of this research.

## 5.1 Generation of Training Data for Artificial Neural Networks

Measuring the tonic level of parasympathetic activity delivered by the brain to the heart is a very delicate operation in laboratory on animals, involving identifying the vagus nerve and placing electrodes. Muscle sympathetic nerve activity might be easier to record using microneurographic technique, which involves inserting tungsten microelectrodes pericutaneously into sympathetic fascicles in the peroneal nerve and recording multiunit sympathetic nerve discharge. A prominent study was made in [44] where preganglionic sympathetic neural discharge was recorded from third left thoracic sympathetic ramus communicans, and efferent vagal neural discharge was simultaneously recorded from left cervical vagus in an artificially ventilated decerebrate cat. Such recordings are very rare on animals and difficult to realize on humans.

The lack of experimental data has put the emphasis on developing methods for predicting the behavior of sympathetic and parasympathetic nerves using data which are available non-invasively such as the Electrocardiogram (ECG). We developed such methods based on the model of cardiovascular control presented in Chapter 2 and used optimization methods to fit experimental data and estimated the level of activity of both sympathetic and parasympathetic branches in Chapter 4. In this section, both estimated signals are further used to train artificial neural networks that are capable of combining the information obtained from the recorded physiological signal with the information derived from mathematical models. Artificial Neural Networks (ANN) have shown great performance in identifying hidden relationships between data [175] [176] and approximation of functions [177]. Biologically inspired neural nets have been applied to model the baroreceptor reflex responsible for the short-term blood pressure regulation [178]. Classifiers were applied to risk stratification in hypertension [179].

### 5.1.1 ANN for Estimating Autonomic Tone during Orthostatic Stress

We published results presented in this section in [180]. Postural change from a lying position to an upright posture causes decrease of arterial blood pressure and triggers the baroreflex. Experiments were performed using a tilt table during heart rate, pulse and galvanic skin response recording, as described in section 3.7.2.

We used the software Statistica Neural Networks to investigate multi-layer perceptrons (MLP). The selected MLP neural network (see Figure 5.1) has two hidden layers with 27 and 9 nodes, three inputs and three outputs. The inputs to the network are measured heart rate in beats/min., measured mean arterial blood pressure in mmHg and measured galvanic skin response in μMho. The outputs from the MLP are simulated heart rate in beats/min., simulated parasympathetic level (obtained as described in section 4.2.1) and simulated sympathetic level (obtained as described in section 4.2.4). Both the inputs and outputs have been normalized to the interval 0-1. Product of inputs and weights has been used as a postsynaptic function in both hidden layers. Hyperbolic tangent has been used an activation function in both layers as well. The MLP has been trained using training, validation and testing sets. The original data has been split into these sets randomly in the ratio 2:1:1. Two-phase training has been used. In the first run well-know backpropagation has been employed. In the consequent run the conjugate gradient training algorithm has been applied to precisely tune the network. The overall training scheme is given in Figure 5.2.

The second task we were facing was a problem of classification patients according to the following characteristics: their sex (male or female), their age (15-46 years), their weight (50-110 kg) and their health status (1 for young healthy, 2 for elderly healthy, 3 for hypertensive). We have designed a similar network using MLP as well. The inputs to the classifier were values of the model parameters estimated in Chapter 4; and the outputs were the corresponding patient characteristics.

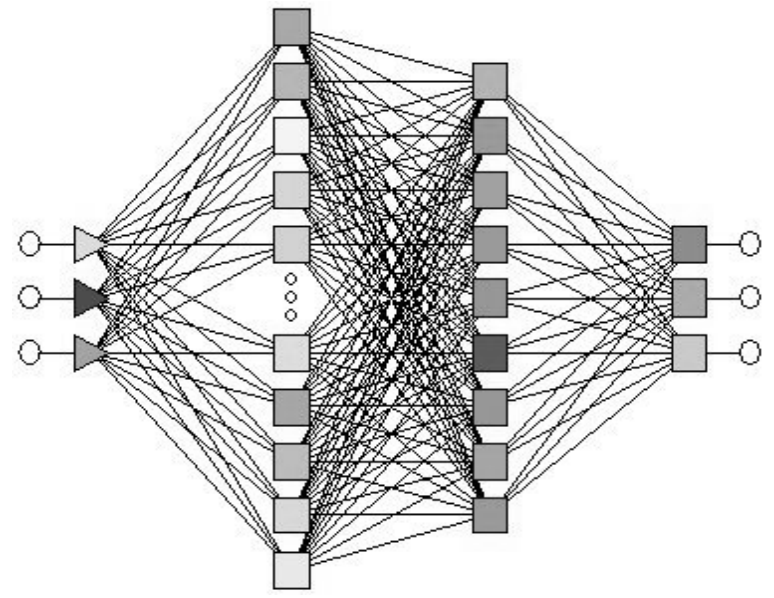

**Figure 5.1 Multilayer perceptron neural network architecture**

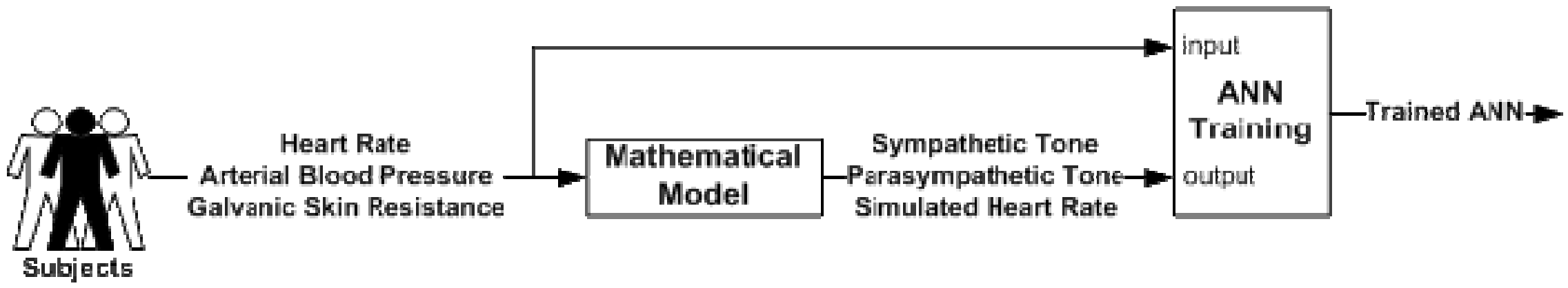

**Figure 5.2 Training of Multilayer perceptron for assessing autonomic nervous activity**

During training, the least-square error for the ANN and the classifier varies within range 0.11 -0.06. More important is that the error signal was approximately the same for training, validation and testing set, what is a good measure of ANNs quality training. The mathematical model produced an error within range 0.0022- 0.088.

After the ANN and the classifier were trained, their efficiency was tested on data from testing set. These data were applied as input to the ANN on one side and to the mathematical model on the other side. Both models output a simulated galvanic skin reponse (Figure 5.3) and a heart rate signal showing a good fit to the measured data (see

Figure 5.4). However the ANN appeared to include more non-linearity than the mathematical model. The ANN estimates sympathetic and parasympathetic tones as well. This output was compared to the one generated by the mathematical model. Figure 5.5 shows the corresponding decreasing parasympathetic and increasing sympathetic tones with a good fit to the mathematical model.

The estimated values of the mathematical model parameters (from section Chapter 4) were applied as input to the classifier. We found out that the parameters contain useful information for predicting the health status and sex. The classifier was able to assign the right health status to unknown patients based on the corresponding estimated values of model parameters. More difficult task was the prediction of age and weight. In both cases the MLP has been employed as a regression model. After training the ANN responded correctly from the view of an order of values of the output values but incorrectly from the viewpoint of absolute values. Even so the network proved importance to give correct trends and interpretable results.

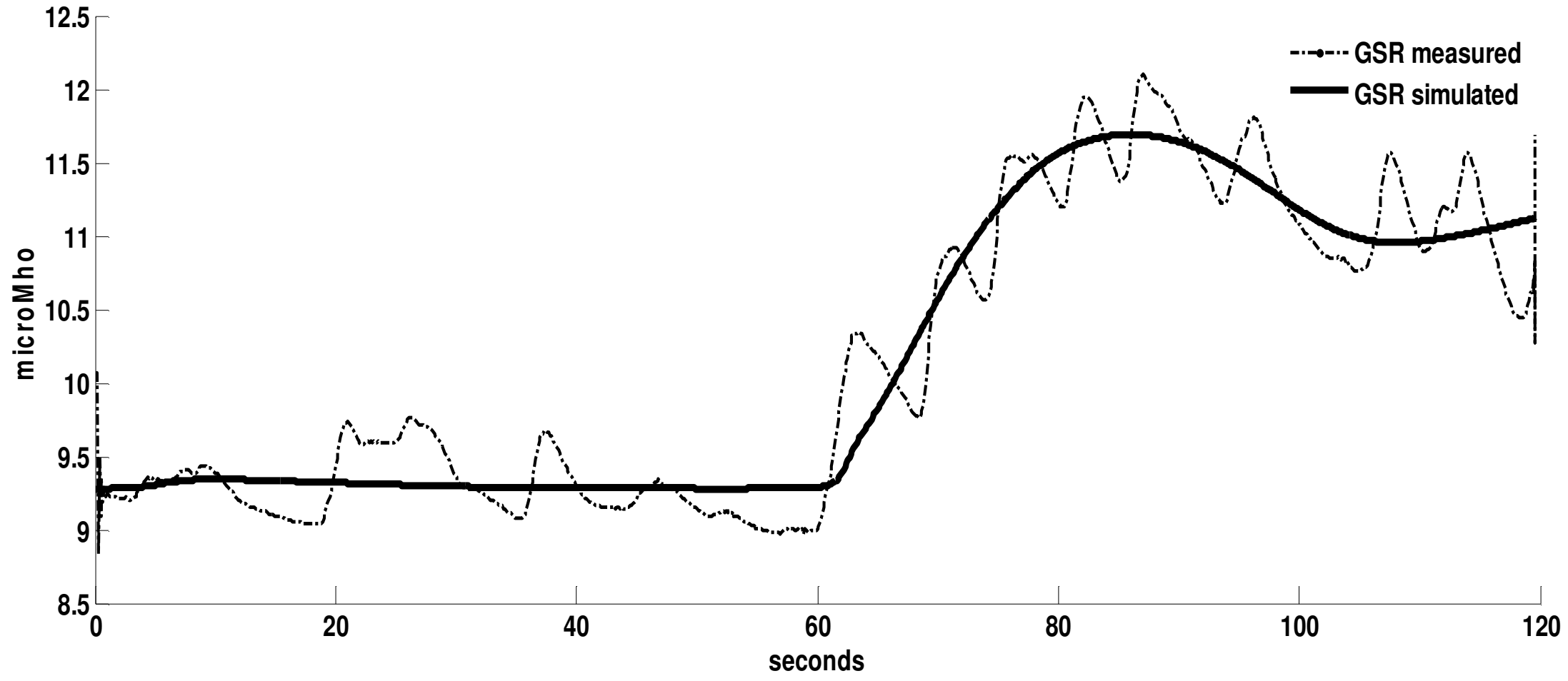

**Figure 5.3 Galvanic skin response during orthostatic stress**

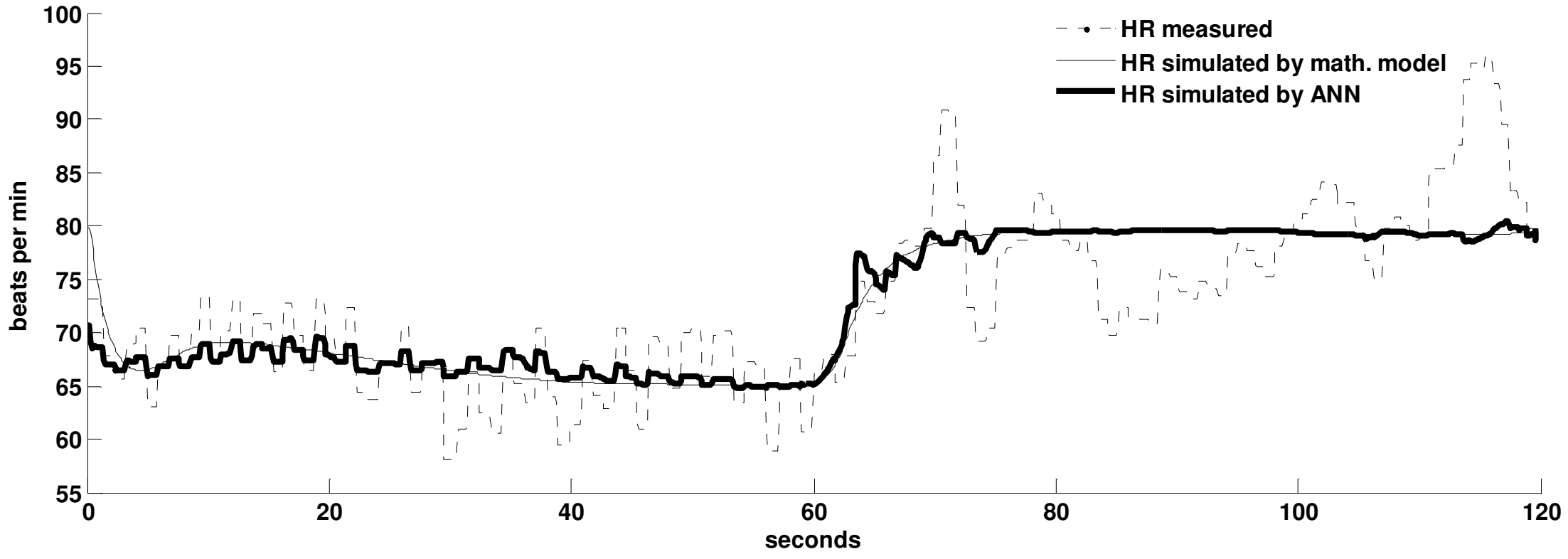

**Figure 5.4 Mean heart rate response during orthostatic stress**

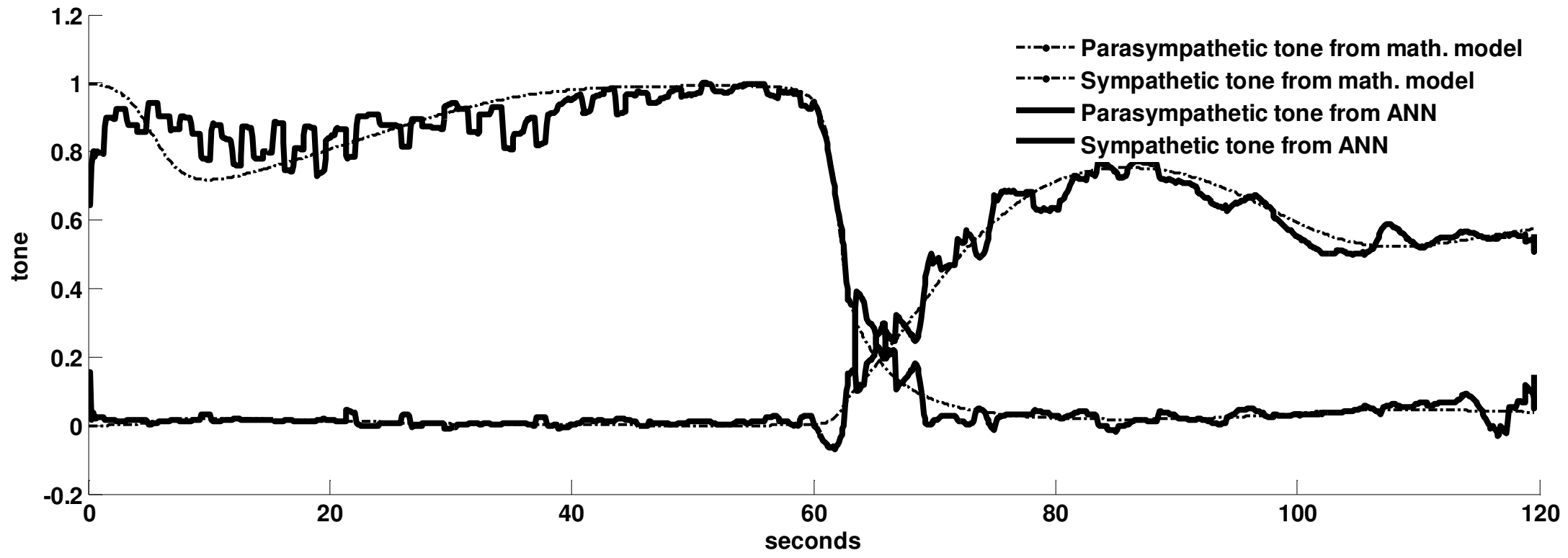

**Figure 5.5 Mean parasympathetic and sympathetic tone**

Y-axis is normalized in the interval [0 .. 1] in dimensionless unit

### 5.1.2 ANN for Estimating Parasympathetic Level during Cold Face Test

We submitted results presented in this section as journal publication in [181]. Most results were published within a Master thesis [182] under our supervision. The intellectual contribution of the thesis was the architecture of the ANNs and the implementation of the software tool in Matlab.

The estimated parasympathetic level (PL) obtained via model fitting to data recorded in experiments with cold stimuli (as described in section 4.2.1) was used as training set for multi-layer perceptrons (MLP). Experimental data were recorded during the cold face test (CFT), which is a non-invasive clinical test known to inhibit sympathetic efferent activity to the sinoatrial node and enhance parasympathetic activity [183] [125]. Because the response to CFT on heart rate is exclusively parasympathetic, it offers a suitable experimental setup for gathering training data for our MLPs. Results include a software tool for simulating the networks and displaying the PL together with an estimate of cardiac health status in response to the CFT.

The cold face test was performed as described in section 3.2. Additionally a standard clinical check of cardiovascular autonomic function was performed by physicians on each subject. Cuff blood pressure was measured; heart and thorax sounds were consulted; clinical signs were observed. Furthermore subjects underwent basic cardiac autonomic tests including Valsalva Maneuver, cold pressor test, active standing, deep breathing test and oculocardiac reflex test. The physician evaluated all results and assigned a health status score to the subject, 10 being perfect health of cardiac autonomic control and 1 being total insufficiency. The data was saved in a table with two columns, the first showing the ID of the subject and the second his health status score.

The response to cold face test can be strongly modulated by factors related to the emotional state of the subject. The emotional perception or level of discomfort was provided by the subject and added as input for the network on a range from 0 to 5. Strong pain (e.g. headache) causing abrupt interruption of the experiment corresponds to score 5, localized intense but sustainable pain corresponds to score 4, localized mild pain corresponds to score 3, fear sensation corresponds to score 2, calmness corresponds to score 1 and total comfort corresponds to score 0. The data was saved in a table with two columns, the first showing the ID of the subject and the second his emotional factor.

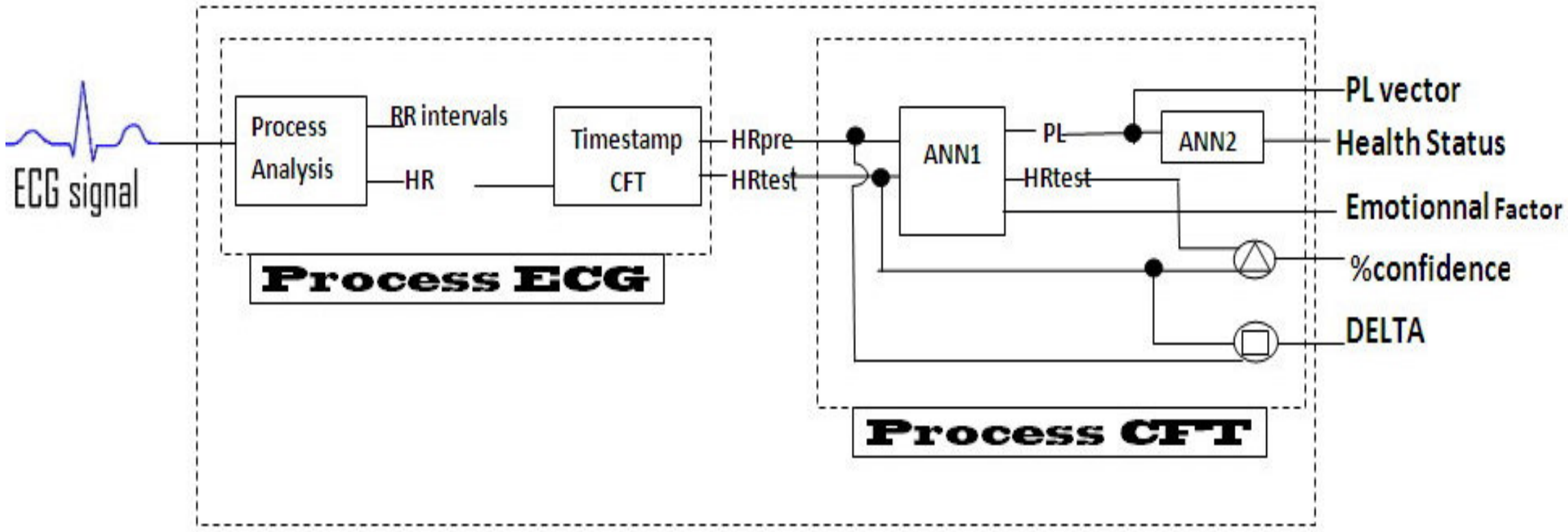

**Figure 5.6 Architecture of the ANN-based evaluation of CFT response**

In order to make use of features of artificial neural networks for quantitative assessment of parasympathetic level we developed a solution whose architecture is depicted in Figure 5.6. We distinguish two main components, a first one for signal processing (*Process ECG*) and a second one for ANN simulation (*Process CFT*).

The entry of *Process ECG* is the ECG signal recorded during the cold face test. This component is responsible for signal processing and extraction of the time series $NN_{pre}$, $NN_{test}$, $HR_{pre}$ and $HR_{test}$. $HR_{pre}$ and $HR_{test}$ are further used as input of the inputs of *Process CFT* component. A MLP artificial network ANN1 uses the inputs and estimates the corresponding parasympathetic time series $PL$. ANN1 also returns a prediction of heart rate time course during CFT which is further compared to the measured values in order to compute a percentage of confidence ($\%confidence$) telling how much error was made by $ANN1$ when estimating $HR_{test}$. ANN1 additionally predicts the emotional factor ($Emotional_Factor$) of the subject according to modulations of his response to CFT by emotional factors such as fear and pain. A second MLP artificial neural network ANN2 is designed to predict the patient's health status. It takes as input, the parasympathetic level and predicts the health status ($Health_Status$) in a normalized on a scale of 1 to 10. Units for both ANN1 and ANN2 are threshold units which use a sigmoid as activation function. It is non polynomial and is therefore suitable to our case since Multi Layer Perceptrons with non polynomial activation function are theoretically able to approximation any function. [22]. *Process CFT* also produces the rate of decrease of the heart rate ($DELTA$, also called $HR_{fall}$) during the cold face test, calculated according to equation (4.3.1) from section 4.3.1.

After processing experimental data we obtained time series for $HR_{pre}$ and $HR_{test}$ as well two tables containing the $Health_Status$ and $Emotional_Factor$. Our model of cardiovascular control (presented in Chapter 2) was fit to each heart rate signal and the estimated level of parasympathetic activity $PL_{pre}$ and $PL_{test}$ was also used as training data for the networks.

We investigated Multi-Layer Perceptrons using the Neural Networks Toolbox version 5.1 of Matlab version 7.1. For ANN1 several tests have been achieved on different MLP with three layers, 120 inputs and 181 outputs and the results were not satisfactory due to the big sizes of ANN. We therefore decided to normalize the durations of cold stimulus to 20 seconds, 30 seconds, 40 seconds and 60 seconds. Thus we obtained 4 groups of training data for precisely 4 ANNs which were mounted into the final network. The structure of networks for ANN1 was (20-35-31), i.e. 20 neurons in the input layer, 35 neurons in the hidden layer and 31 neurons in the output layer. The sizes of input and output layers were determined by the number of variables from which the model is expected to establish. The number of the hidden neurons was randomly chosen, as we cannot explain, but this number relatively provides satisfaction. Similiarly the structures of ANN2 having presented the best performance had 2 neurons in the input layer, 27 neurons in the hidden layer and 1 neuron in the output layer. The inputs were reduced to $mean(PL_{pre})$ and $mean(PL_{test})$.

The evaluation of the ANN was done by Holdout [184]. The holdout consists of randomly dividing the original pattern set into three disjoint sets: training (60%), validation (20%) and test set (20%). The validation of the ANN was done during each iteration of the training. It told if the ANN was converging to minimal error. The test set was used to estimate the confidence of the network ($\%confidence$). Performance of the network measured by cross-validation [184] using the coefficient of correlation of Pearson (~ 0.98) , the number of iterations (~ 38), the retained gradient error (~ 5.9e−4) and the absolute mean square error (~ 4.31e −5) shows a rapid convergence to the solution.

A ratio between the parasympathetic level returned by the artificial neural networks for the period before and during the cold face test was calculated as indicator of parasympathetic response in percentage as follows.

$$PL_{ratio} = 100 \cdot \left( \frac{mean(PL_{test})}{mean(PL_{pre})} - 1 \right) \quad (5.1.1)$$

Overall results from trained ANN show a mean increase of PL during the cold face test by 43% ($PL_{ratio}$) from baseline value 0.472±0.275, which agrees with a mean increase of RMSSD by 45% ($RMSSD_{ratio}$) from baseline value 8.75±5.10. Table 5.1.1 shows the results obtained for 8 representative subjects. PL increased only by 8% for hypertensive subjects (from 0.76 to 0.83) obviously because this condition is characterized by higher sympathetic activity. HRV was surprisingly higher for those subjects (RMSSD increased by 46% from 15.01 to 21.97) suggesting that RMSSD unfortunately also includes fluctuations due to sympathetic activity. Subjects with cardiac insufficiency had the lowest increase of PL (only 4%) suggesting a parasympathetic impairment. A similar response was also observed on subjects with palpitation. Subject with hepatitis had low resting PL (0.15), but the response to CFT was significant (PL increased by 48%). Young subjects had a higher parasympathetic response (heart rate dropped by 33% from 72±8 bpm) than older subjects (heart rate dropped by 18% from 76±13 bpm). Figure 5.7 shows how parasympathetic level was changing during CFT for a 17 years old female subject (n=17).

Table 5.1.1 Statistical analysis of HRV parameter and parasympathetic level obtained from ANNs

| **Subject** | | | | **Baseline** | | | **Cold Face Test** | | | **Ratio** | | |
|---|---|---|---|---|---|---|---|---|---|---|---|---|
| ***n*** | ***Sex*** | ***Age*** *years* | ***Health Status*** | ***HR*** *beats/min* | ***RMSSD*** *ms* | ***PL*** *n.u* | ***HR*** *beats/min* | ***RMSSD*** *ms* | ***PL*** *n.u* | ***$HR_{ratio}$*** *%* | ***$RMSSD_{ratio}$*** *%* | ***$PL_{ratio}$*** *%* |
| 4 | F | 50 | Healthy | 68 | 11.45 | 0.64 | 54 | 17.67 | 0.92 | 20% | 54% | 44% |
| 5 | M | 51 | Hepatitis B | 91 | 2.15 | 0.15 | 83 | 3.40 | 0.22 | 10% | 58% | 48% |
| 6 | M | 54 | Healthy | 89 | 2.59 | 0.18 | 57 | 20.12 | 0.35 | 36% | 676% | 94% |
| 16 | F | 50 | Hypertension | 66 | 15.01 | 0.76 | 54 | 21.97 | 0.83 | 18% | 46% | 8% |
| 20 | F | 70 | Cardiac insufficiency | 61 | 7.34 | 0.84 | 54 | 8.43 | 0.87 | 12% | 14% | 4% |
| 26 | F | 43 | Palpitation | 86 | 3.94 | 0.37 | 75 | 4.29 | 0.47 | 12% | 9% | 28% |
| 31 | M | 23 | Healthy | 66 | 12.64 | 0.26 | 48 | 23.80 | 0.81 | 29% | 88% | 206% |
| 32 | F | 17 | Healthy | 78 | 14.18 | 0.37 | 50 | 34.42 | 0.64 | 36% | 142% | 73% |

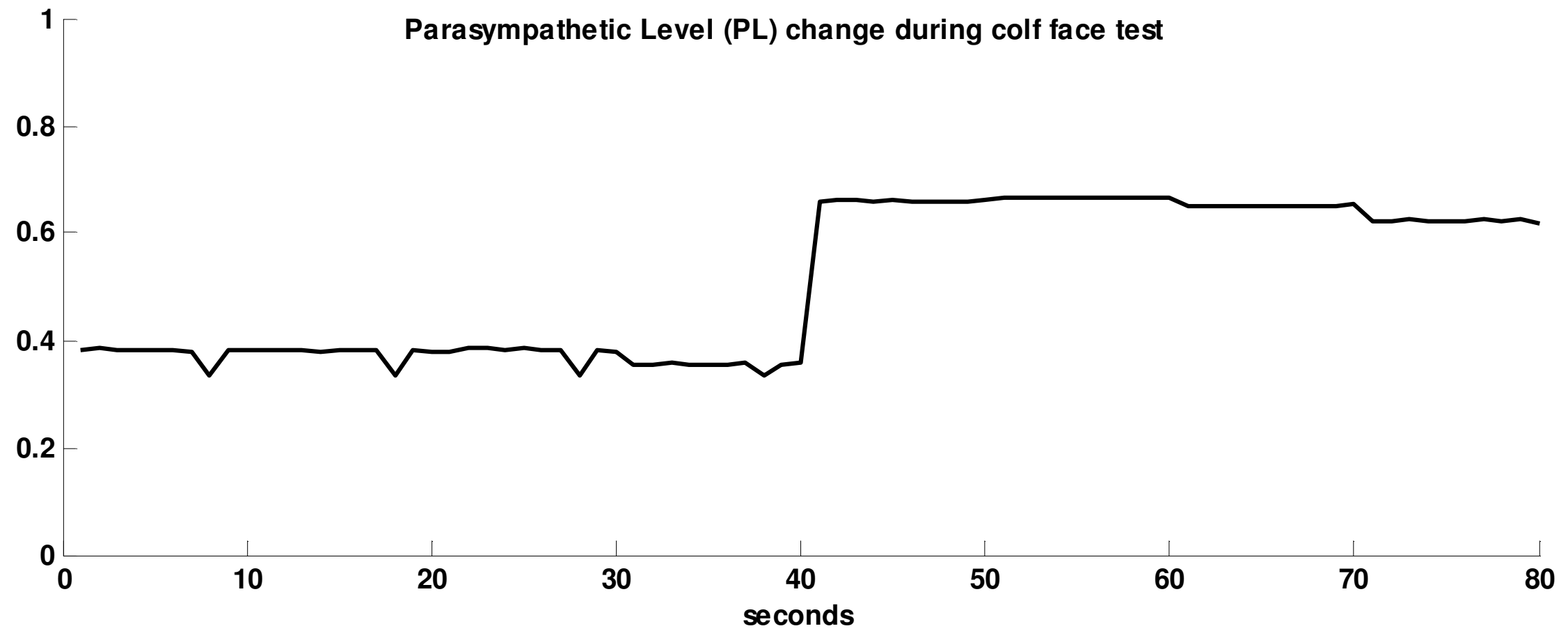

**Figure 5.7 Estimated parasympathetic level during CFT**
measurements were performed on subject with ID n=32
cold stimuli was applied from $40^{th}$ to $80^{th}$ sec
Y-axis is normalized in the interval [0 .. 1] in dimensionless unit

We developed a Matlab software tool offering a user interface (see Figure 5.8) for loading an ECG signal which was recorded during the cold face test in ASCII format and entering the sampling frequency. The user can additionally enter the starting timestamp of the cold face test as well as its duration. The tool automatically detects QRS complexes, performs smoothing, filtering and generates NN intervals HR time series. The software tool simulates the artificial neural networks and estimates the parasympathetic level, ratio of heart rate decrease and health status of the subject. The software offers some plotting tools for the visualization of heart rate and parasympathetic level time course.

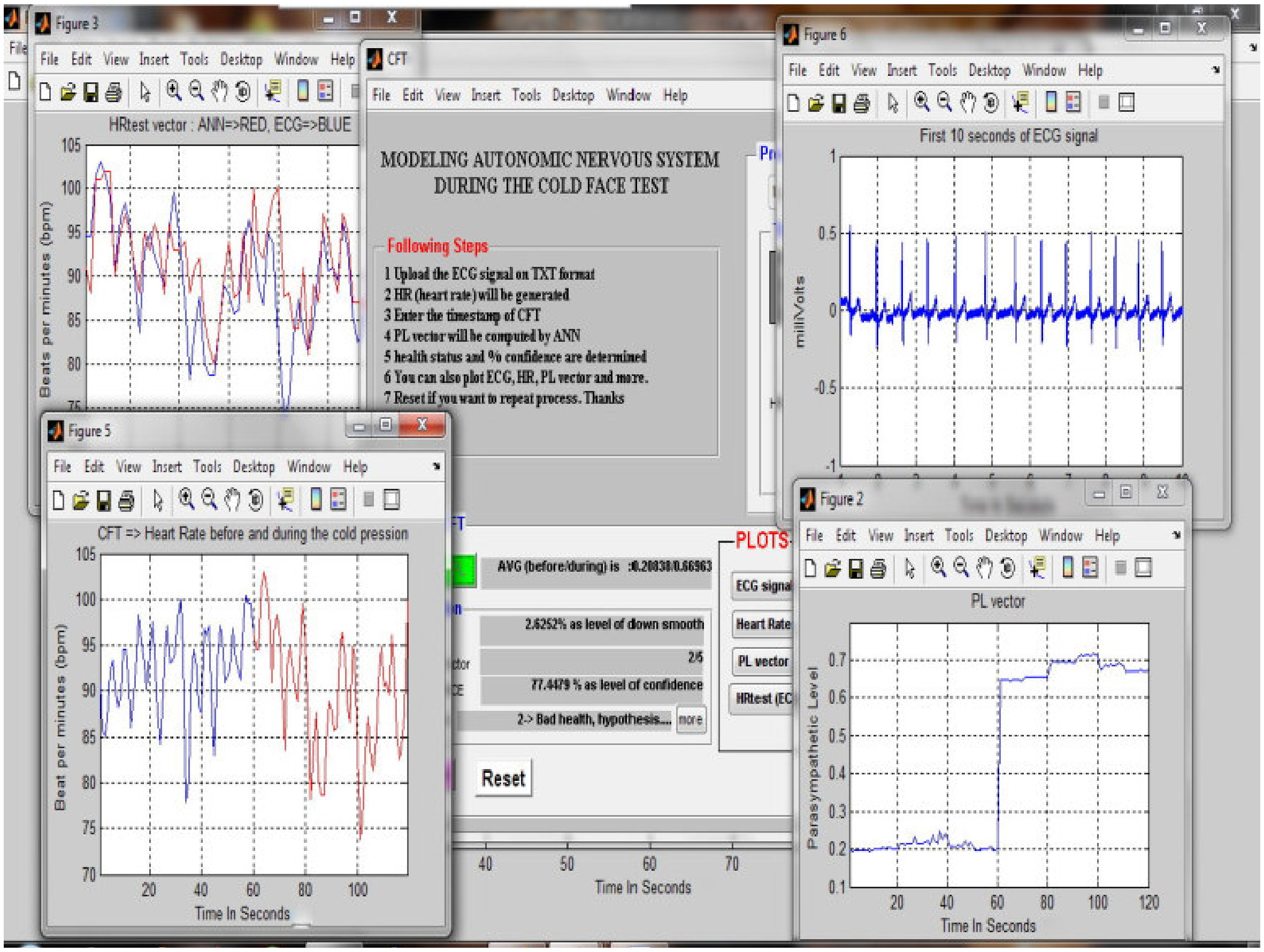

**Figure 5.8 User interface of the Matlab software tool for CFT evaluation using ANNs**

## 5.2 A Telemedicine Platform for Cardiology

### 5.2.1 Background

We published parts of results presented in this section in [185] [186]. The software was partly developed within two Bachelor Theses [187] [188] under our supervision. The intellectual contribution of the theses was within software architecture and coding.

Several studies have linked some patterns of variability to pathological states of the cardiovascular system. Cardiovascular autonomic dysfunction may result in substantial disability, morbidity and even death [189]. The features include orthostatic hypotension, baroreflex failure, cardiac arrhythmias, poor exercise tolerance, infarction and sudden death. Diabetic autonomic neuropathy is known to be the most common autonomic neuropathy in the majority of industrialized countries. Myocardium infarction is another leading cause of death worldwide.

Monitoring patients is a very important part of disease management, especially after a heart attack. When patients leave the hospital it might be necessary to access cardiovascular function remotely in some cases. Blood pressure monitoring of hypertensive patients is a crucial part of the treatment. Risk stratification is another major reason of preventive cardiovascular assessment. Evaluation of heart rate variability has been proven to help with prognosis for low vagal tone and diagnostic of lethal arrhythmias. Cardiac autonomic tests such as Valsalva Maneuver, Cold Face Test, Tilt-Table Test or Deep Breath Test offer means to trigger reflex arcs with a cardiovascular response reflected in cardiac signals (see Chapter 3). Physicians and researchers usually extract time-frequency parameters which could provide emerging physiological properties of the subject. Despite of diverse evaluation methods of tests there is a growing need for a framework for tests management, signal processing and parameters estimation which could be applied to families of tests.

The goal of this project is to realize a web-based telemedicine platform allowing cardiovascular signal uploading and remote assessment of cardiac function. Methods for cardiovascular signal acquisition, storage, processing, and evaluation should be implemented. Results should include a web-based software package for managing patients, experiments and signals. Additionally features for simulation and statistical evaluation can be provided. The software should provide means for manually uploading e.g. an Electrocardiogram (ECG) signal using a web interface and automatically analyzing the signals in order to extract time-frequency parameters. An important feature of the software is to include messaging tools. Patients can share ECG signals, links, photos, videos, status messages and comments with other patients or physicians. This work should make use of available open-source software components for signal processing, modeling and simulation. Conversation tools and features for social networks should be included, such as Emails and Forums.

We will present the work products in the next sections following Software Engineering guidelines [78]. The requirements analysis phase consisted of gathering requirements for the software to be developed and investigating what the system should ideally do. We retrieved functional, access, security, performance and other relevant requirements in a requirements document. In the system design phase, we proposed system architecture for more flexibility and better user experience. In the implementation phase code was developed, a deployment infrastructure was chosen and a prototype of the software was finalized.

### 5.2.2 Requirements Analysis

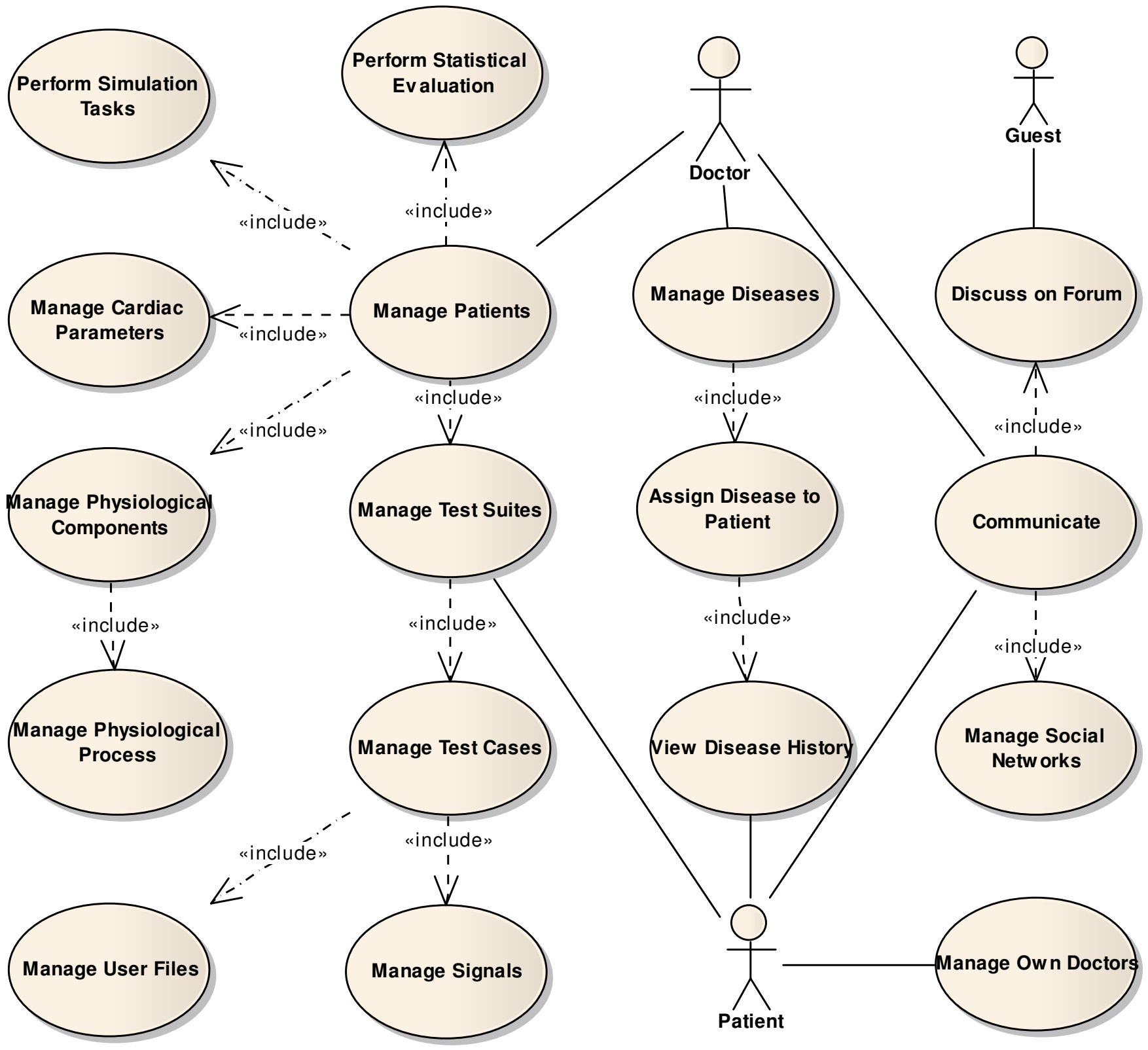

**Figure 5.9 Use cases for Telemedicine Platform**

As depicted in the UML Use Cases Diagram in Figure 5.9, the software should offer a set of functionalities to 1) manage patients, test suites, test cases, physiological components and processes; 2) manage cardiovascular signals and other patient files; 3) perform statistical evaluation; 4) simulate mathematical models of cardiovascular control and estimate cardiac parameters 5) discuss with doctors and other patients using forums and social networks.

The use case *Manage Patients* provides the doctor with features to view, update, add or delete data of patients which are assigned to him. Each patient can choose one or several doctor who can have access to private files in the user case *Manage Own Doctors*.

In the use case *Manage Test Suites*, doctor is able to define a test suite and assign it to a patient. The patient can also create own test suites and choose which doctor will have access to its data. A test suite is a set of test cases aiming to diagnose particular disease stages, e.g a battery of cardiac autonomic tests including Valsalva maneuver, active standing and deep breath test might be used to diagnose autonomic neuropathy in diabetes.

In the use case *Manage Signals* the doctor or patient can upload measurements performed during a test case. The system automatically extracts relevant information, e.g. RR interval durations are extracted from a raw Electrocardiogram (ECG) signal. The doctor can compute statistical parameters (e.g. mean heart rate) in the use case *Perform Statistical Evaluation*. User can also upload other files than signal files, such as scanned images of a paper ECG or print outs of an echocardiogram in the use case *Manage User Files*.

When performing autonomic tests, the measured cardiac signals are usually used to determine hidden cardiac parameters, such as sympathetic tone, parasympathetic tone or ventricle contractility. The system includes mathematical models for given cardiac autonomic tests, which can be used by the doctor to estimate the values of cardiac parameters in the use cases *Perform Simulation Tasks* and *Manage Cardiac Parameters*. Doctor can analyse both cardiac parameters and statistical parameters in order to make a diagnostic and assign a disease or disease stage to the patient in the use case *Assign Disease to Patient*. Disease information can be managed by doctors in the user case *Manage Diseases*. A disease is characterized by some properties such as symptoms, clinical signs, causes, prognoses features and management.

In the Telemedicine social network a conversation is a stream of text, images, audio files, but also signal files, cardiac parameters, statistical parameters. Within a conversation System offers shortcuts for the user to upload new signals or other objects. System displays a step by step wizard, where user can select existing test suites/test case or create new ones, enter signal attributes, upload signal file, and choose doctors who should have access to the files. The user can also insert existing objects to a conversation. System displays snapshots of conversation objects such as images and signals using thumbnails. When the user clicks on the object, System offers a detailed view of the object, e.g. using a signal viewer applet. Furthermore doctors have shortcuts for performing statistical evaluation, cardiac parameters estimation and disease assignment. These are all features available in the use cases *Communicate* and *Manage Social Networks*. Guest users can only make use of the Forum.

### 5.2.3 Software Design

In order to fulfill the requirements elicited in section 5.2.2 the system has been designed in a modular way using three-tier architecture (Figure 5.10). The GUI components are grouped in a Client Tier that presents an interactive and mostly web-based user interface and relays user requests to a remote Application Tier, where the processing logic is implemented. The application tier is built on the top of existing software grouped together in the External Software Package Tier. Our system benefits from its interface to the Systems Biology Toolbox 2 [170] in order to estimate values of the cardiovascular model parameters and evaluate their sensitivity in the Matlab environment. It makes use of the WaveForm DataBase software package from the Physionet project [190] for signals processing. Signals are visualized using the open-source JFreeChart API[2][191].

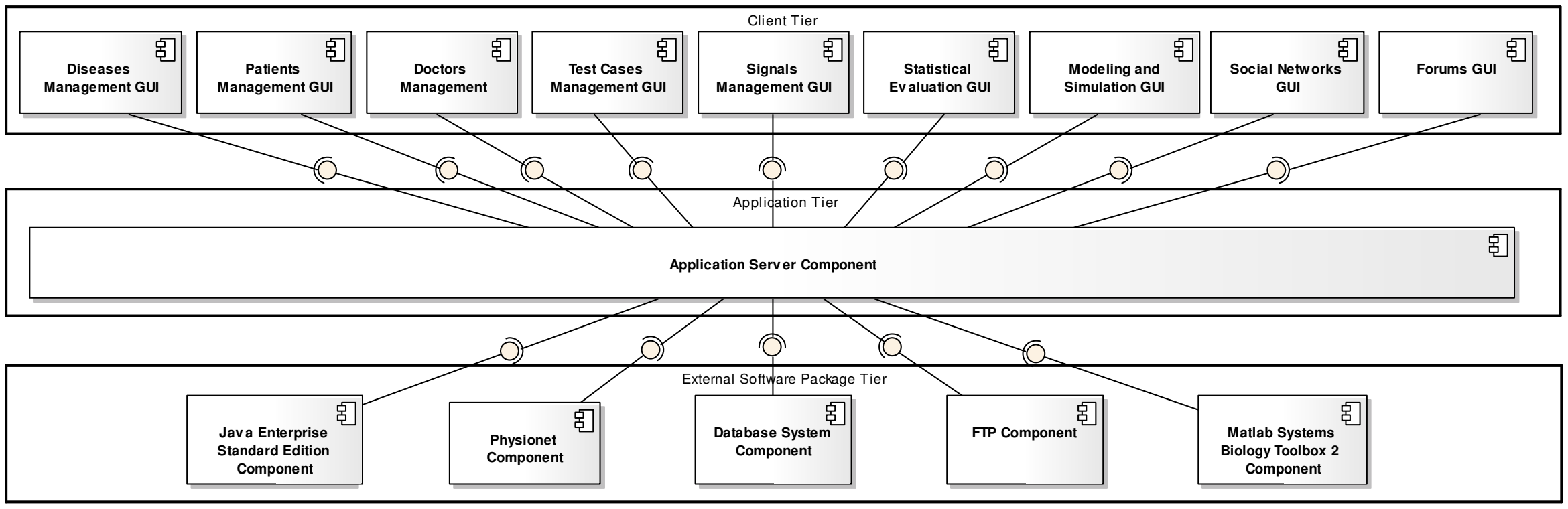

**Figure 5.10 System Architecture of the Telemedicine Platform**

Persistent application objects are stored in a relational database with structure depicted in Figure 5.11 as UML Class Diagram. A Test Suite is a set of Test Cases, each having components to be tested, a description of the dynamic of these components, measured cardiac signals, evaluated statistical parameters and simulated cardiac parameters. The components to be tested are physiological cardiac components that can be tissues (e.g. sinoatrial node), organs (e.g. heart), single neurons (e.g. ophthalmic trigeminal thermoreceptors) or autonomic nervous centers (e.g. medulla). The Test Dynamic describes the messages passed between physiological components during an autonomic test. For example during the orthostatic stress test the baroreceptors send a message (decreased blood pressure) to the vasomotor center, which sends a message (increased sympathetic tonus) to the sinoatrial node.

[2] Gilbert D. (2009). *JFreeChart*. Available: http://www.jfree.org/jfreechart (Accessed June 10, 2010)

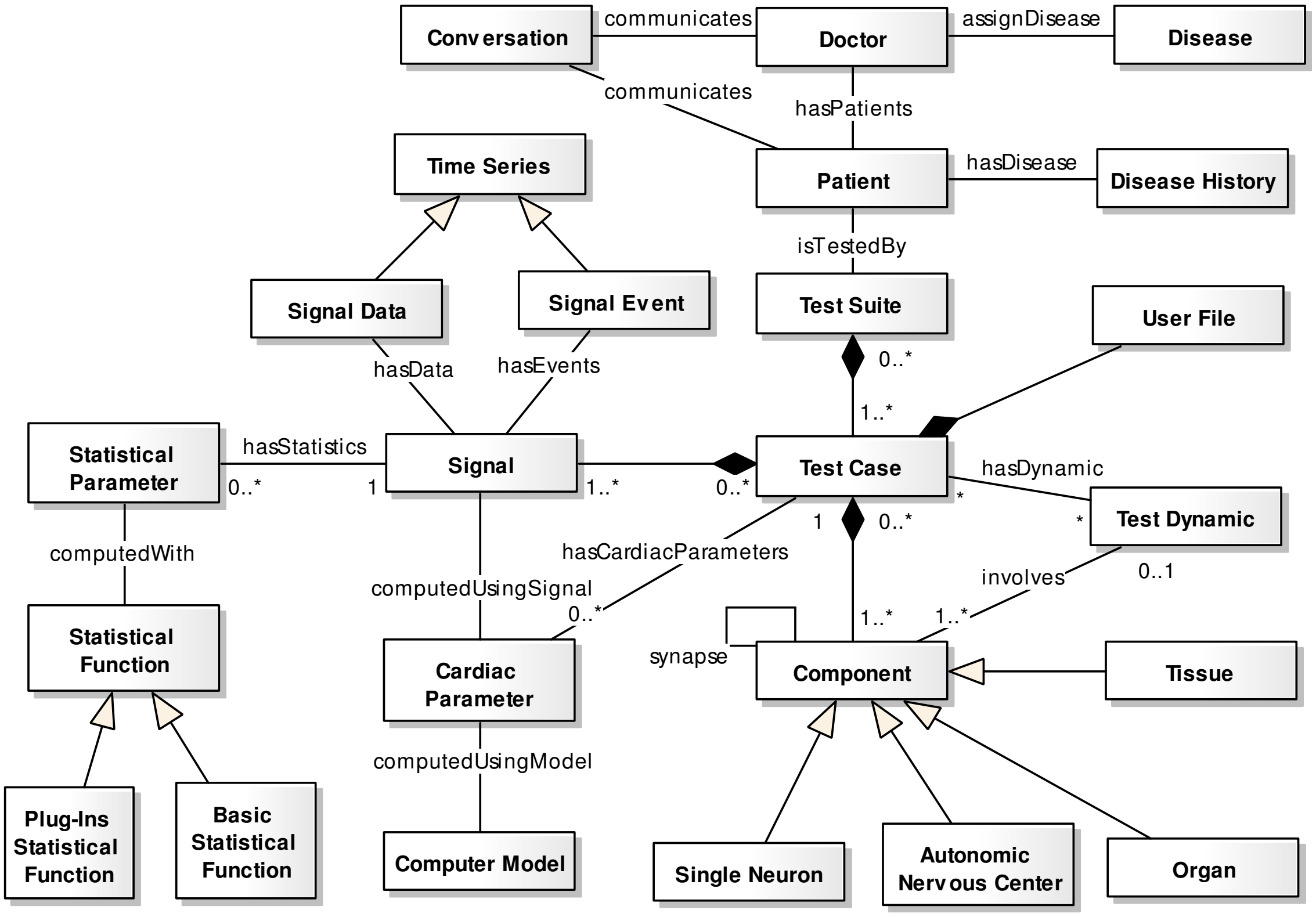

**Figure 5.11 UML class diagram of the Telemedicine platform**

Implemented mathematical models of cardiovascular control are stored in the Computer Model entity of the database. They are used to fit the measured data and estimate model parameters, which are further stored in the Cardiac Parameter entity. The Statistical Parameter entity stores results of statistical evaluation performed using Basic Statistical Functions or Plug-Ins, which are external pieces of software uploaded by the user to perform special statistical tasks.

Cardiac signal data are managed in three layers. In the Raw Signal Layer, signal data are represented as time series of real numbers with properties such as name, sampling frequency, data unit and signal type (e.g. electrocardiogram, impedance cardiogram, blood pressure, galvanic skin response, skin temperature or respiration rate). A signal has a data matrix, with the first row showing the timestamps in seconds and the second row showing the signal values. It also has an events matrix, first row showing the timestamps and second row showing the events text (e.g. ‘cold compress applied to the forehead during the cold face test’). The Parameterized Signal Layer provides tools to extract information from raw signals. For example RR interval durations are extracted from the ECG signal; Left Ventricular Ejection Times are extracted from the impedance cardiogram signal. In the Statistical Signal

Layer, statistical calculations are performed on the parameterized signals using statistical functions and formulas. Signal data can be exported or imported in the ASCII format.

### 5.2.4 Implementation

Our software package is object-oriented and developed in PHP, Java, C, SQL and Matlab programming environments. It delivers a powerful web based interface with a set of portable HTML pages and Java-Scripts. The main menu for patients management is depicted in Figure 5.12. Doctors can log in; define some autonomic tests (see Figure 5.13 and Figure 5.14); process and evaluate measured signals; and establish diagnosis; all this from their Internet browser without additional software. The database is initialized with diseases information selected from the Oxford handbook of cardiology [192], see Figure 5.15.

The Web-based user interface relays user requests to an Ubuntu 8.10 Linux Server including an Apache 2.2 web server and a PHP 5 engine, where the processing logic for patients and test cases management is implemented. A special PHP module makes use of command-line tools from the Wave Form DataBase (WFDB) library for signal processing. The system can be easily extended with additional signal processing tools. Signals are visualized with a WebStart Java applet, making use of the JFreeChart API and running on the client's browser (see Figure 5.16).

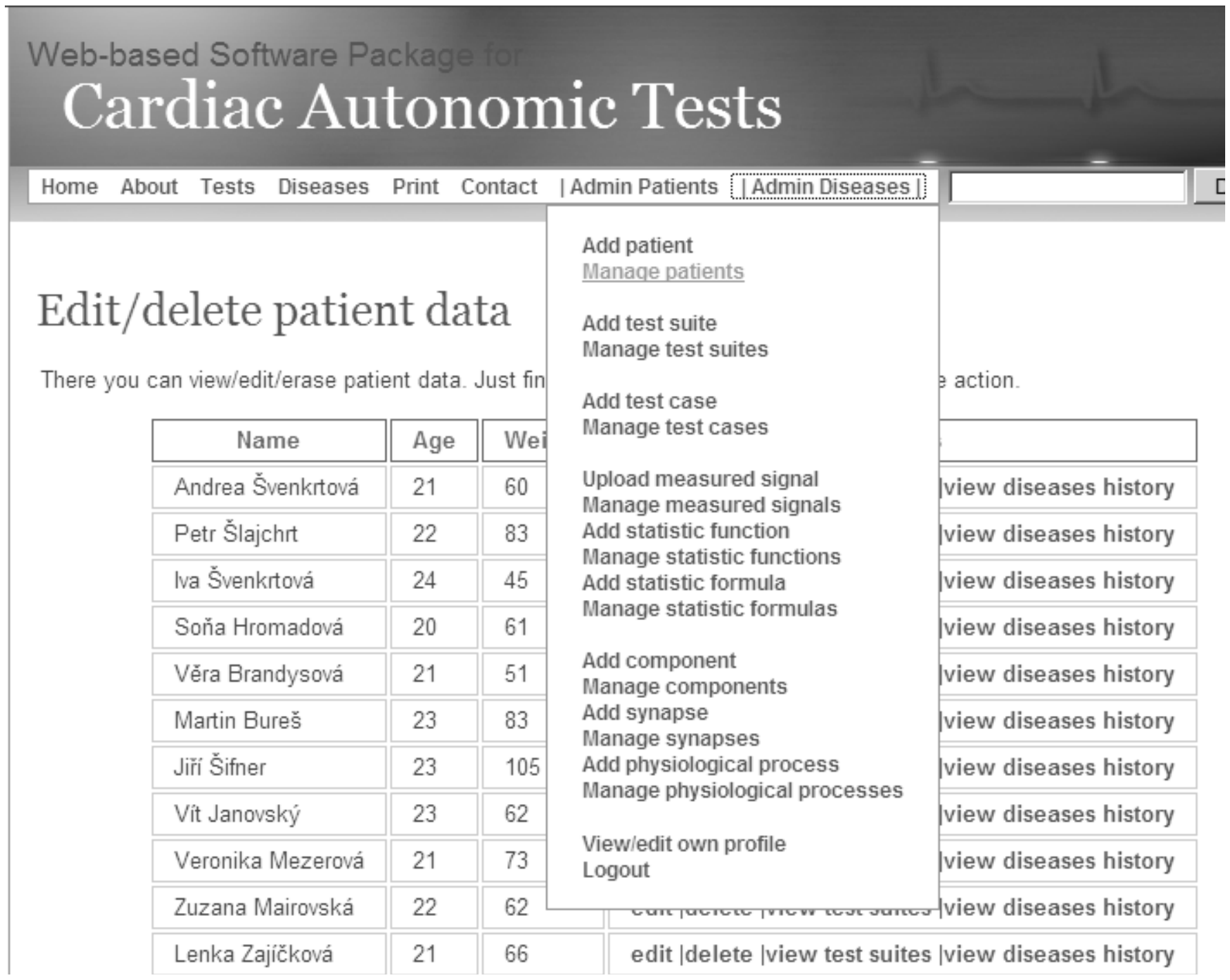

**Figure 5.12 Telemedicine platform - menu for patients management**

## Edit/delete test suite

There you can view/edit/erase test suites. Just find desired suite and click on respective action.

| Name | Date ctreated | Patient | Description | actions |
|---|---|---|---|---|
| Experiments session 1 | 2009-04-28 | Jiří Šifner | This test suite contains test results on the valsalva manoeuvre, tilt-table test, cold pressor test and oculocardiac reflex test | edit delete test cases |
| Experiments session 1 | 2009-04-28 | Petr Šlajchrt | This test suite contains test results on the valsalva manoeuvre, tilt-table test, cold pressor test and oculocardiac reflex test | edit delete test cases |
| Experiments session 1 | 2009-04-28 | Iva Švenkrtová | This test suite contains test results on the valsalva manoeuvre, tilt-table test, cold pressor test and oculocardiac reflex test | edit delete test cases |
| Experiments session 1 | 2009-04-28 | Soňa Hromadová | This test suite contains test results on the valsalva manoeuvre, tilt-table test, cold pressor test and oculocardiac reflex test | edit delete test cases |

**Figure 5.13 Telemedicine platform - menu for test suites management**

## Edit/delete test cases

There you can view/edit/erase test cases. Just find desired component and click on respective action.

| Test Suite | Name & Description | Type | Components | Phys Proc. | actions |
|---|---|---|---|---|---|
| **Experiments session 1** Petr Šlajchrt | **Cold Pressor Test** The Cold Pressor Test (CPT) is usually performed by immersing a subject's hand into cold water (temperature usually under 5 C°) for 1 to 5 minutes. In healthy human subjects, a global vascular sympathetic response with increased skin vascular | Cold Pressor Test | **Baroreceptor SA node Vagal Center Vena** | **Cold pressor test** | **edit delete signals cardiac parameters** |

**Figure 5.14 Telemedicine platform - menu for test cases management**

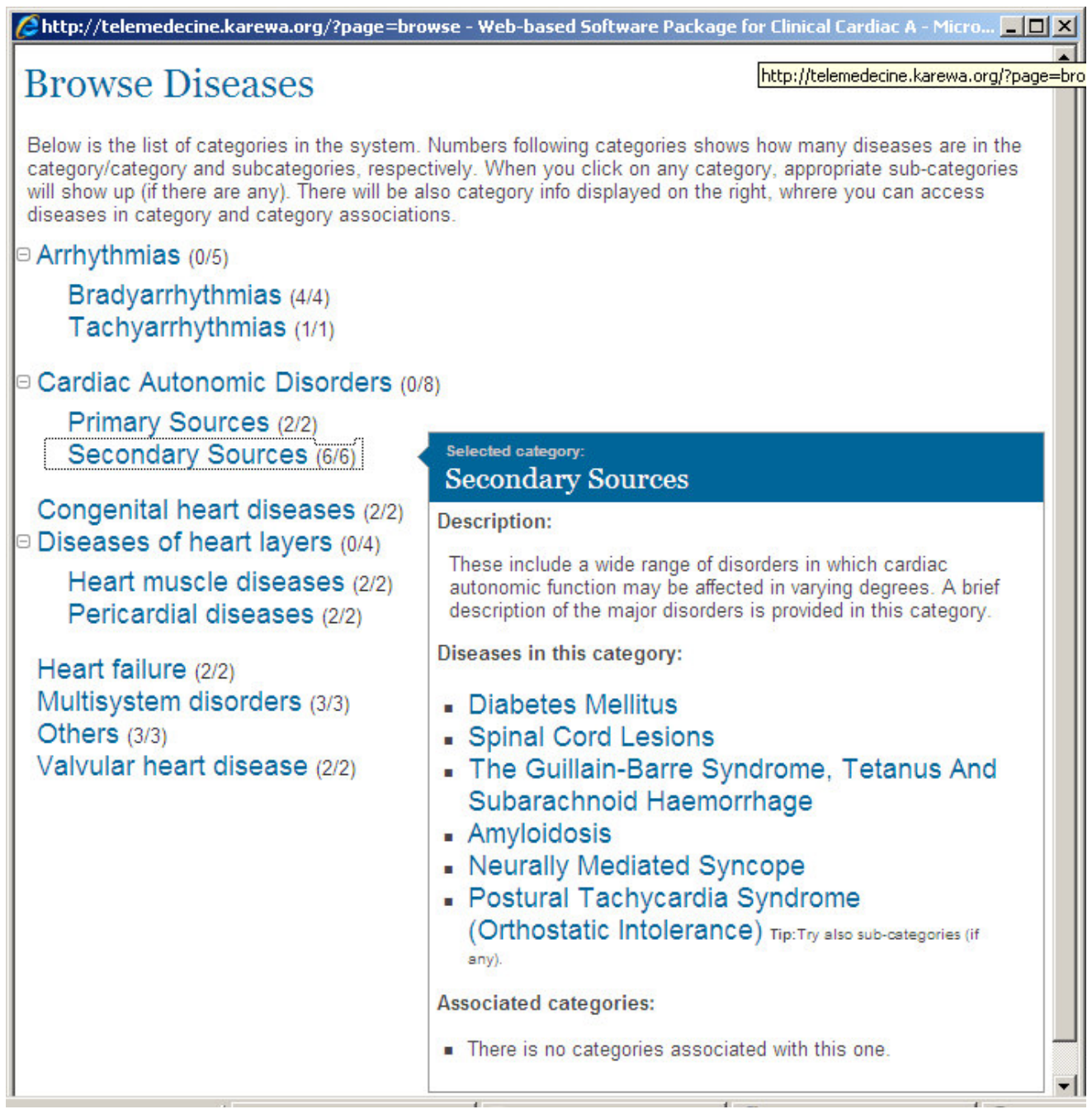

**Figure 5.15 Telemedecine platform - disease database**

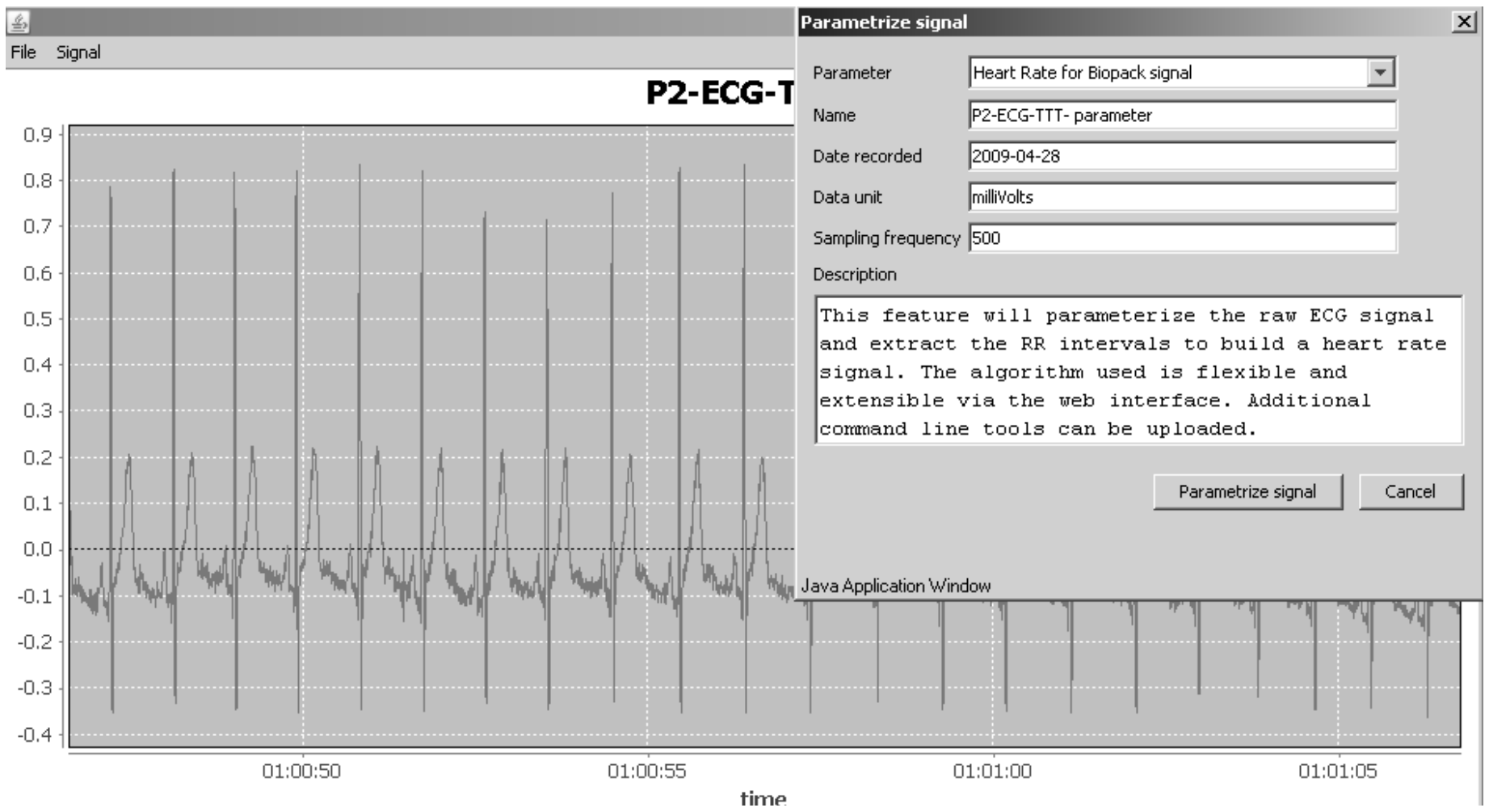

**Figure 5.16 Telemedicine platform – signals visualization**

A special PHP module implements features for statistical evaluation. Statistical parameters are stored as formulas containing the statistical function used; as well as the time range of the signal to which it applies. Basic statistical functions are functions which are available in PHP (e.g. min, max, mean, standard deviation, variance, and covariance). Support for plug-ins is also enabled. These are statistical functions, which are not available in our system, but the user can implement and upload executables to the system in order to perform special statistical calculation, such as total spectral power of a heart rate signal. A parser has been implemented using the PHP Tokenizer, in order to parse and evaluate statistical formulas such as (longest R-R interval) / (shortest R-R interval) for the Valsalva ratio, or mean[1:60] for the mean of signal values from the 1$^{st}$ to the 60$^{th}$ second. The corresponding screenshot is given in Figure 5.17.

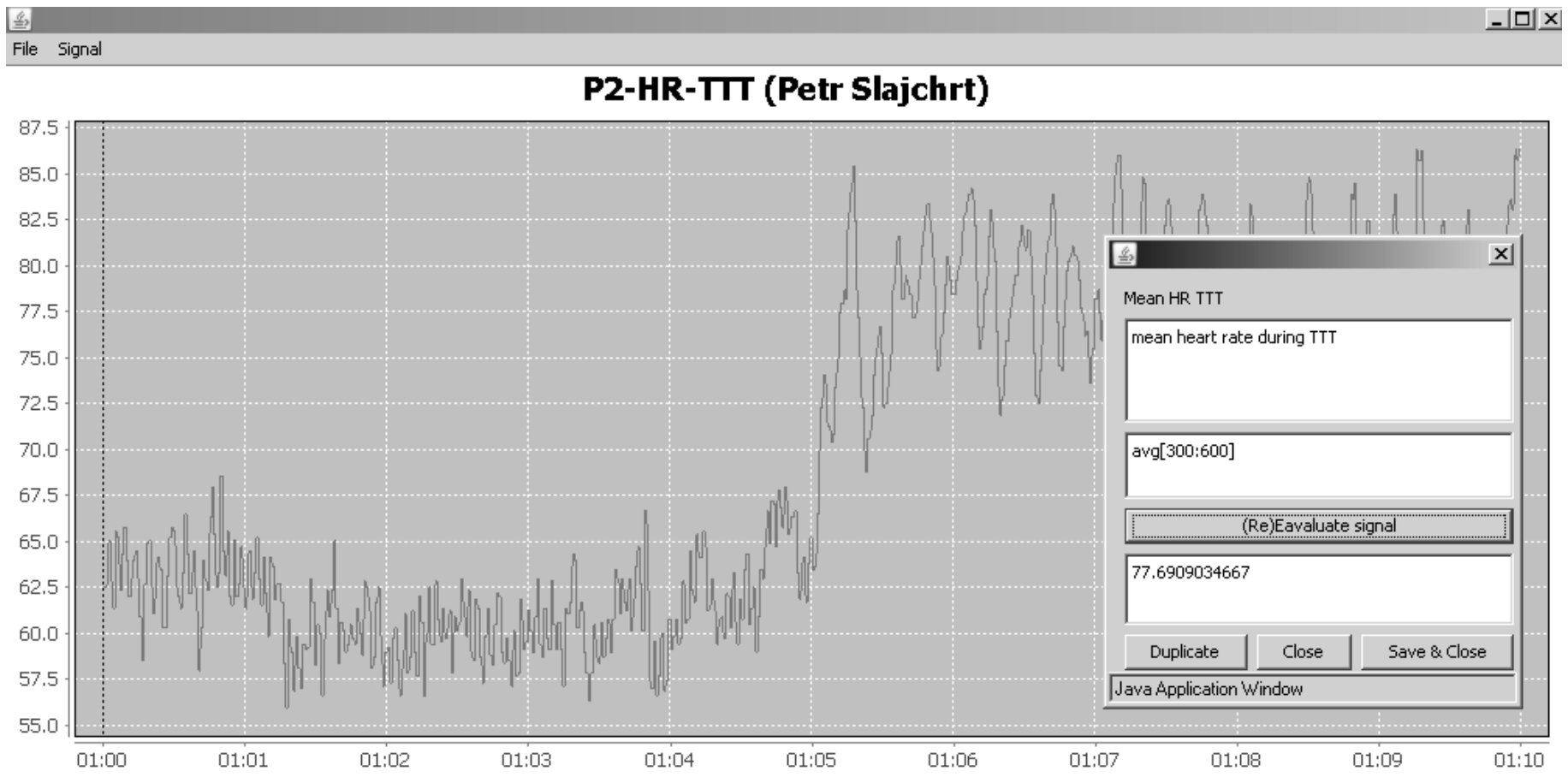

**Figure 5.17 Telemedicine platform - statistical evaluation**

Data are stored in a MySQL 5 database server running on the same server as the web server for fast data access. Signal data are stored in the database, however big signals are automatically stored on a FTP server. A special PHP module streams the signal file and makes its storage type transparent to the rest of the system.

Mathematical models are stored in the database or on the FTP server. Systems Biology Toolbox 2 models are small text files which are stored in the database directly. Simulink models, Matlab M-files, Systems Biology Toolbox 2 projects are stored on the FTP server. We implemented a special Matlab bridge (see Figure 5.18) to download or upload models from or to the FTP server. The Matlab bridge makes also use of the Matlab Database Toolbox to connect to the MySQL server and process patient data and signals for parameters estimation tasks (see Figure 5.19).

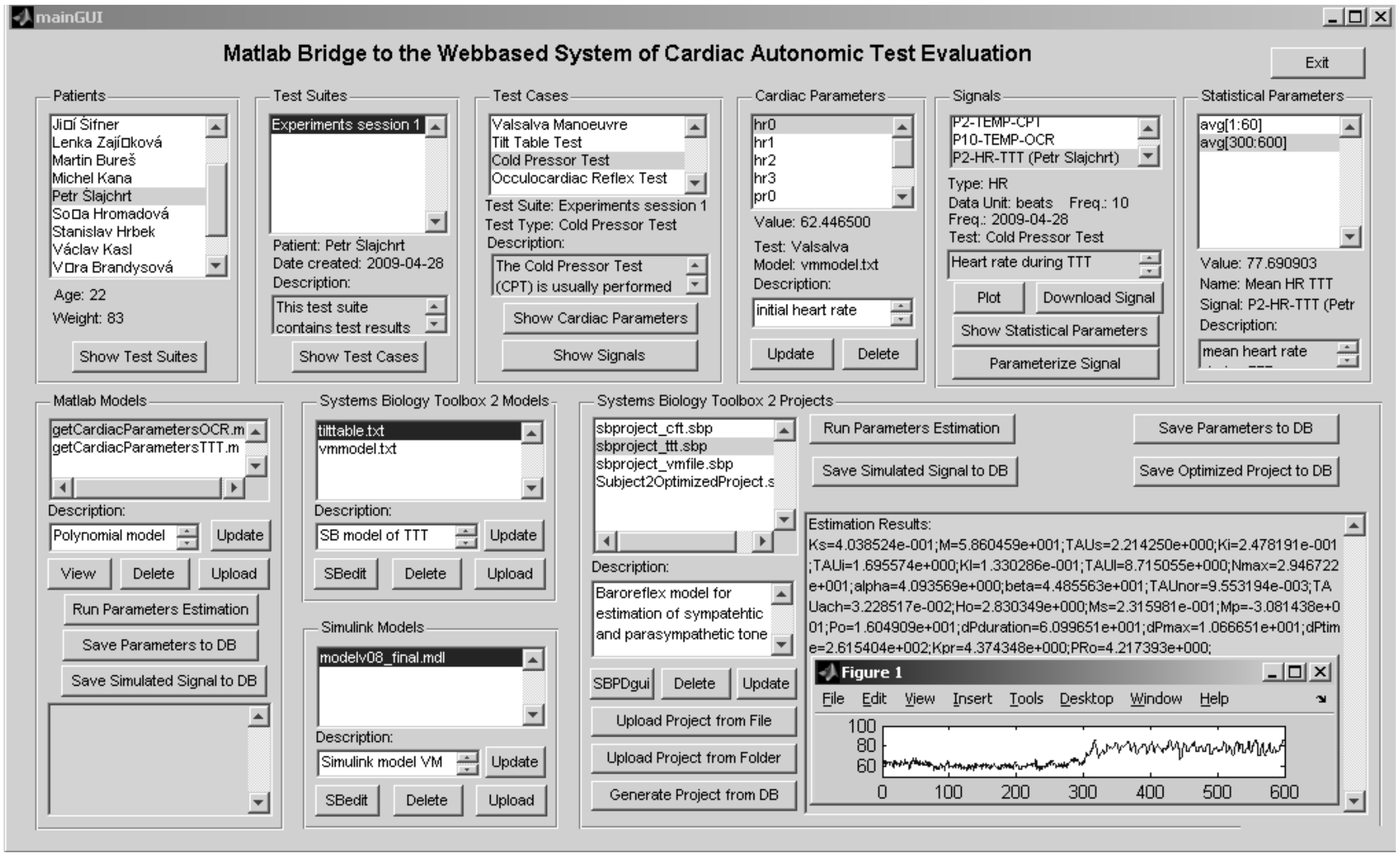

**Figure 5.18 Telemedicine platform - Matlab bridge GUI**

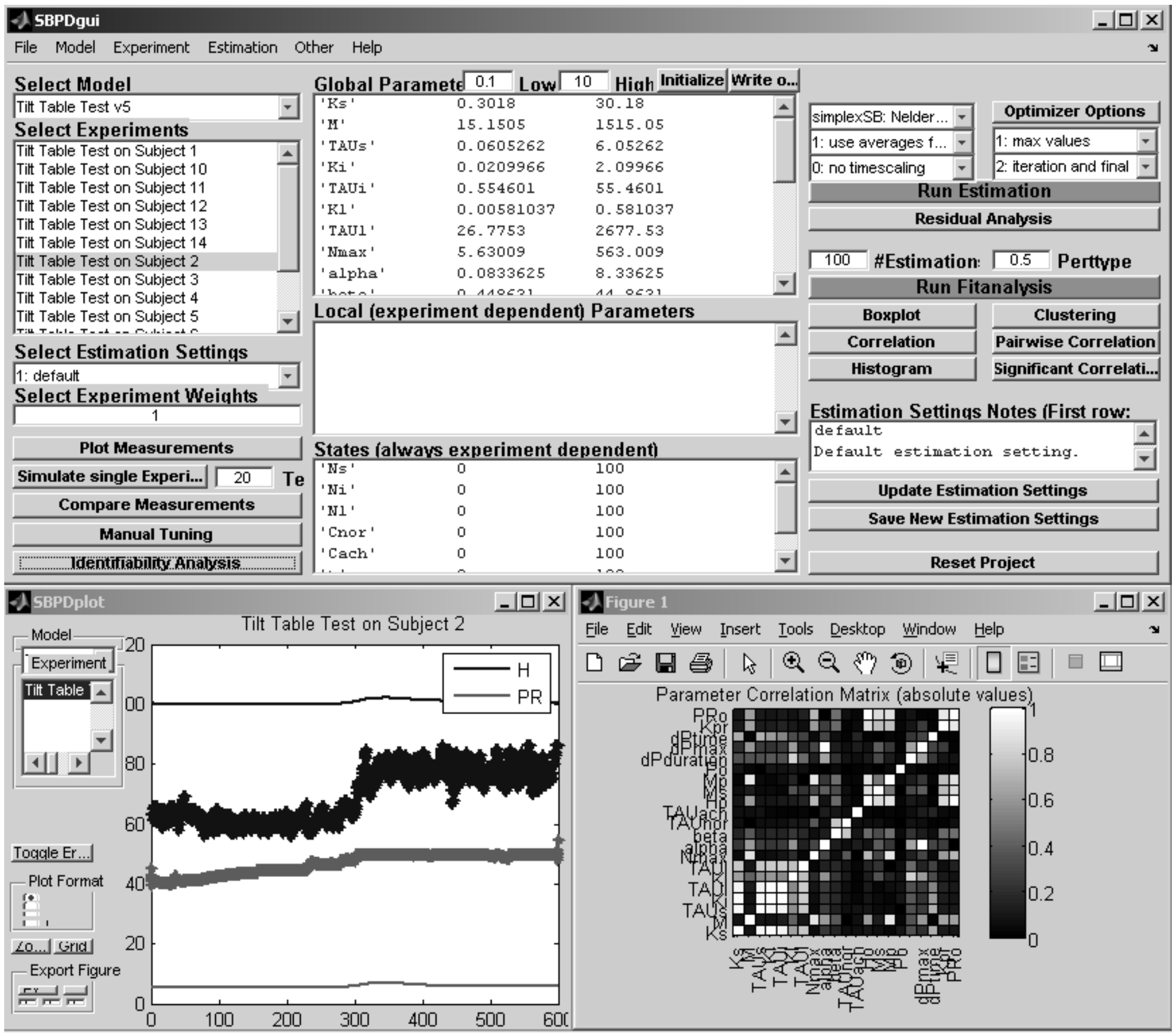

**Figure 5.19 Telemedicine plattform - parameters estimation GUI**

## 5.3 A Portable Biofeedback Device for Autonomic Tone Assessment

### 5.3.1 Background

Several devices have appeared on the market to train body's natural relaxation response against stress, chronic pain, depression, anxiety or asthma. They usually measure finger pulse, galvanic skin response or skin temperature and display the measured signal on a small display. For example the StressEraser[3] is a device that measures real-time beat-to-beat heart rate via an infrared finger photoplethysmograph sensor. It breaks the heart rate signal into waves and analyzes the amplitude and smoothness of each wave in order to estimate respiratory sinus arrhythmias. The StressEraser is proprietary, rather expensive and does not provide a programmable interface. Other existing devices include simple signal processing software for band pass filtering and features extraction. Although they offer interesting services for remote assessment of cardiovascular function; they are not yet per default shipped with integrated software for modeling and simulation. Some attempts have been made by systems such as the Task Force Monitor[4] which offers continuous blood pressure , electrocardiogram and impedance cardiogram measurement together with a computer station for performing signals analysis and heart rate variability all-in-one. Such systems are being included in the clinical practice to a higher extend but they are expensive and rather inaccessible for mobile users. Mobility has become an important requirement for high quality medical care. The concept is being applied for monitoring elderly including fall risk assessment, breathing and pulse monitors. In sport mobile devices measure the effectiveness of the heart as a pump compared to the exercise load, thus giving a feedback on the next level of training. Ambulatory services benefit more and more from mobile devices, easy to wear, which can give adhoc information on the cardiovascular state of the patient. Telemedicine is offered by private clinics together with some private health insurances, the target group being patients with chronic heart diseases, highly mobile patients such as business travelers, who want to be coupled with their doctors back home. These services are highly professional but very expensive and demand a complex network of infrastructure.

Our goal is to develop solutions for intelligent multi-purpose configurable portable devices for continuous biofeedback and assessment of cardiovascular function. This includes a series of mobile hardware devices with embedded software for cardiovascular signal acquisition and local evaluation using biofeedback techniques. The devices should have common interfaces for low-cost sensors plug-and-play. For example an infrared sensor might provide finger pulse measurement into a microprocessor unit. An embedded mathematical model fits the heart rate and estimates the level auf autonomic tone continuously. The level is displayed by a led system or a digital display in a three-color scale (red, orange, green).

User groups spam from mobile patients requiring continuous monitoring by private doctors, e.g. diabetic, hypertension patients; to elderly requiring continuous monitoring by doctors and family members. Athletes requiring mobile performance evaluation, monitoring of resting heart rate, exercise recovery time, sympathetic activation time, and parasympathetic recovery time also belong to the target group. Subjects, psychotherapists, physiotherapists requiring holistic therapy, stress management, psychological self-therapy, self-perception of biological signals can find some usage of these solutions. Furthermore

---

[3] Direct W. C. (2010). *StressEraser*. Available: http://stresseraser.com (Accessed May 26, 2010)

[4] CNSystems. (2010). *Task Force Monitor*. Available: http://www.cnsystems.at (Accessed May 26, 2010)

subjects under diet or fitness program who needs calorie and cardiac health control; fat cells estimator; energy consumption balance are also part of the target group.

### 5.3.2 Requirements analysis

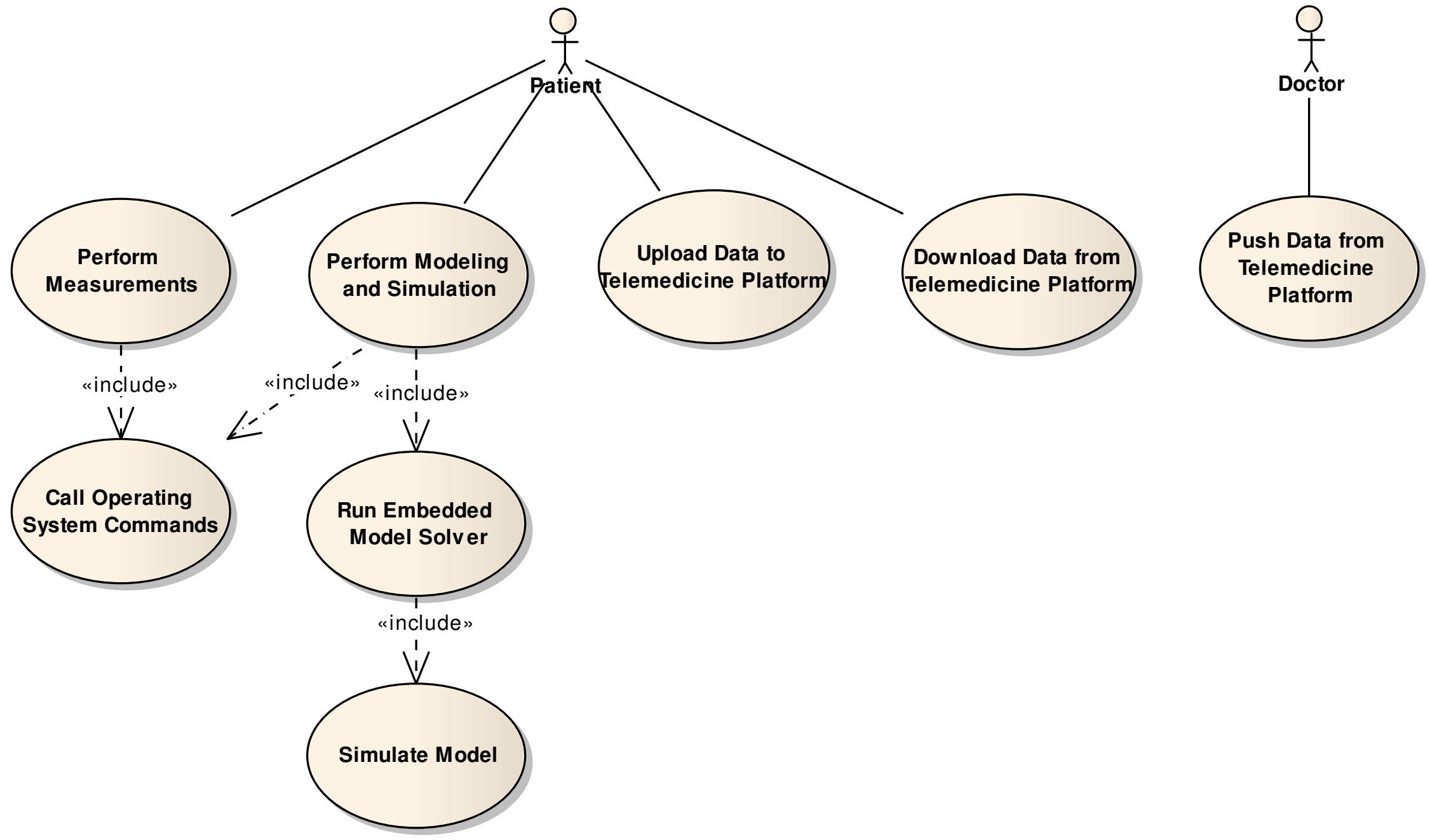

**Figure 5.20 Use cases diagram for biofeedback solution**

The Biofeedback solution should be designed to fulfill the requirements of mobility, portability, low cost, efficiency, and usability as depicted in Figure 5.20. The device should be as small as a mobile phone with interfaces for USB, Wifi, infrared or Bluetooth sensors plug-ins. Light-weight sensors for pulse, skin temperature, electromyography, galvanic skin response measurement should be designed for providing cardiovascular information in the user case *Perform Measurements*. The device should have a built-in digital display for providing feedback to the user in an understandable format. The portable device needs an operating system which interfaces with the underlying hardware and offers means to access the sensors, the memory, the processor and the display in the use case *Call Operating System Commands*.

An API should be developed on the top of the operating system to enable customized software units for modeling and simulation in the use case *Perform Modeling and Simulation*. The device should make use of the model of cardiovascular control proposed in Chapter 2 making it portable, small and efficient enough to run on smaller hardware environments. Portable environment such as mobile phones, personal data assistants have limitations in processing capabilities, working memory and display interface. Running a complex cardiovascular model might be more than challenging. Modeling and simulation is a resource consuming task demanding high processing capacity and memory. Paradigms for embedded software development for modeling and simulation purposes should be investigated. Efficient algorithm should be developed for differential equations solving in a light processing environment in the use case *Run Embedded Model Solver*. The solver will be embedded in a mobile device with limited CPU and RAM, therefore should be optimized to run in real-time and provide continuous results for a continuous input.

Mathematical models of cardiovascular control are usually resolved in computational environments such as Matlab using built-in functions. Since our devices will have their own model solvers we need to define a language for writing models that can be read by the solver. The constructs of such a language should be defined to support the translation of mathematical formulas into computer-like instructions. A parser and interpreter for the language should be realized and load into the operating system environment of the device in the use case *Simulate Model*.

A communication module should be developed on the top of the operating system to interface the mobile device with the telemedicine platforms presented in section 5.2 via the internet for updates, user profiling, cardiovascular signals download in the use case *Download Data from Telemedicine Platform*. The module should provide means for automatically uploading signals from the device to the server for evaluation by doctors in the use case *Upload Data to Telemedicine Platform*. Doctors can also push configuration elements towards the device to set up some parameters and influence signals measurement in the use case *Push Data to Telemedicine Platform*. Doctors can also push messages or signals to the patient via this channel.

### 5.3.3 Design of the Biofeedback Solution

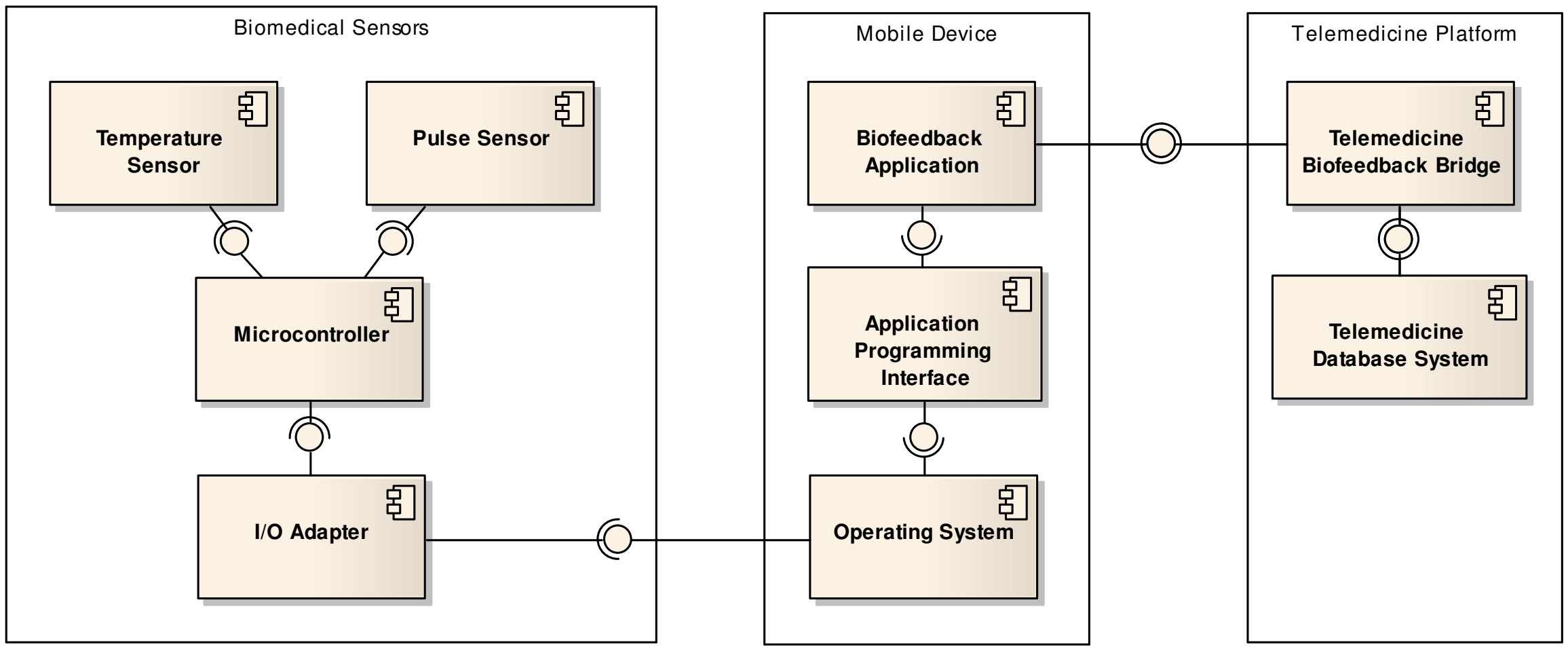

**Figure 5.21 Architecture of the biofeedback solution**

As depicted in Figure 5.21 the system is designed as three components including *Biomedical Sensors,* a *Mobile Device* and the *Telemedicine Platform*. A *Temperature Sensor* provides sensory information about skin temperature, while a *Pulse Sensor* provides a signal describing the pulse waveform at finger level. Both information are processed in a Microcontroller and converted by an *I/O Adapter* into a ready-to-use format. The *Mobile Device* component implements a higher processing logic for signal processing, simulation and remote communication. Such features are realized by a *Biofeedback Application* that makes used of an *Application Programming Interface* offered by the mobile device to call commands of the *Mobile Operating System*. This later component controls hardware components such as input ports from the Biomedical Sensors, keyboard and display for user interaction. The *Biofeedback Application* also acts as client and communicates with the remote *Telemedicine Platform* using a network technology. The *Telemedicine Platform* has been described in section 5.2 and is augmented with a *Telemedicine Biofeedback Bridge* for pushing information to the mobile device or downloading information which is further saved in the *Telemedicine Database System*.

### 5.3.4 Realization of the Biofeedback Solution

We chose to build own sensors and opted for an Object-Oriented Programmable Interface Controller (ooPic) which is a Microchip PIC microcontroller (Figure 5.22) that is programmable in Object Oriented Basic, C, or Java programming languages.

A feedback thermometer detects skin temperature with a thermocouple that is usually taped to a finger and connected to an ooPIC-R board via I/O connector. The thermocouple works following the principle that, when two wires with dissimilar metals are joined and one of the ends is heated, there is a continuous current that flows in the circuit. The voltage in the circuit is read and the temperature is derived using an algorithm programmed in the ooPic controller.

A Led emitting infrared light and a photoresistor are also taped to the finger and are connected to the ooPIC-R board via I/O connector as well. The photoresistor is a resistor whose resistance changes depending on the amount of light incident upon it. The voltage across the circuit is measured and the corresponding finger pulse peaks are approximated in the ooPic controller.

ooPic program for controlling the thermocouple, infrared Led and photoresistor uses objects from the ooPic library. The oLED Object is a hardware object that uses a single digital I/O lines to control the brightness of the LED in order to illuminate the skin. The oA2D10 object is used to read the voltage on the input lines corresponding to the thermocouple and photoresistor with a 10-Bit resolution. The oSerialPort provides a high-speed asynchronous serial I/O port with a 4-byte buffer and is used to write the temperature and pulse peaks information to the RS232 lines. An adapter is used to convert RS232 serial signals to USB.

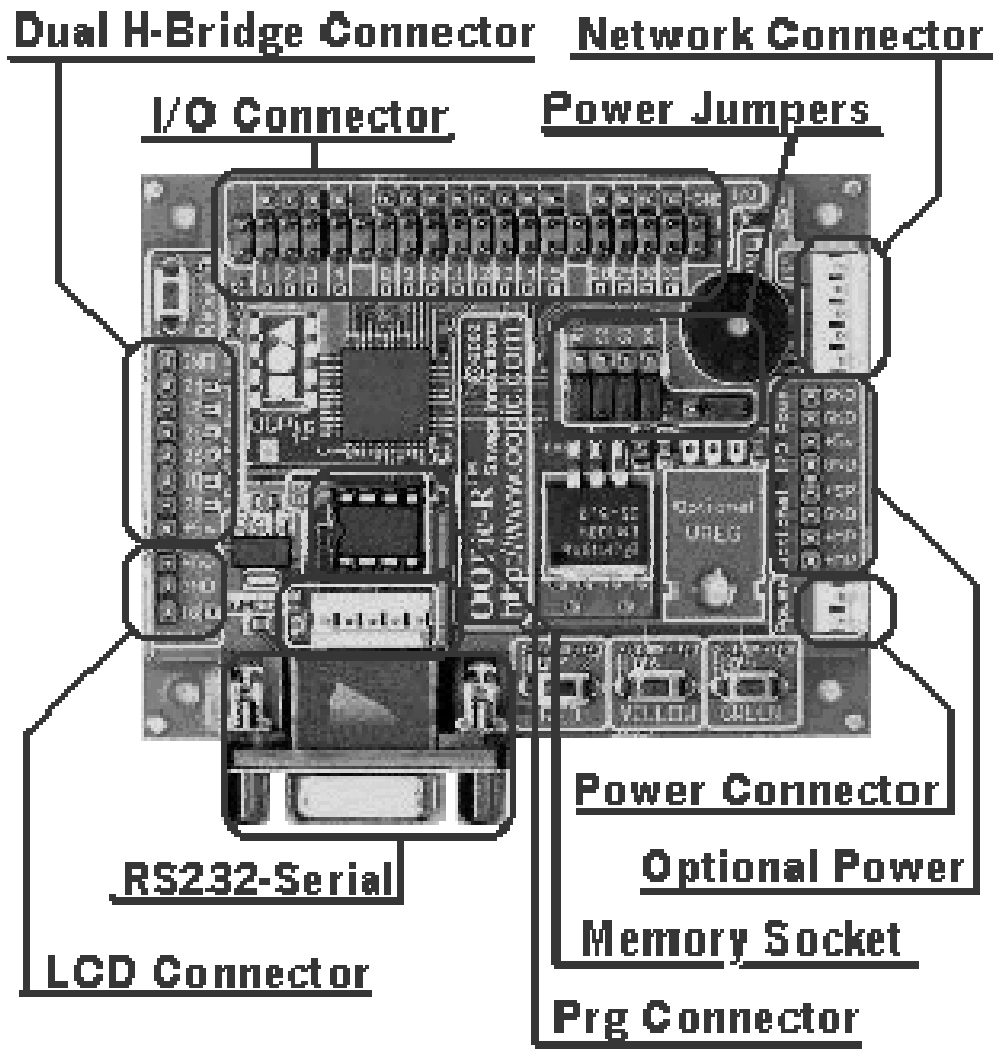

**Figure 5.22 OOPic board** [5]

The ooPic board is connected to a mobile phone via USB. We chose the Nokia E51 model [6] because it is programmable, low cost and offers high performances. 130 MB memory is available for user applications and can be extended up to 4GB with a microSD card. It has a MiniUSB v2.0 slot for transferring data and supports various types of network connectivity including GPRS, EDGE, UMTS and HSDPA.

---

[5] SavageInnovations. (2008). *Object-Oriented Programmable Interface Controller (OOPic)*. Available: http://www.oopic.com/ (Accessed May 26, 2010)

[6] NokiaCooporation. (2010). *Nokia E51*. Available: http://www.nokiausa.com/find-products/phones/nokia-e51/specifications (Accessed May 27, 2010)

Nokia E51 runs on the Symbian 9.2 operating system with Series 60 3rd Edition user interface. Symbian is the most popular open source operating system designed for mobile devices and the so-called smart phones [7]. We use its subsystems responsible for networking via sockets and serial communication via USB in order to access the ooPic sensors. Symbian also offers a Java Micro Edition virtual machine and a software development kit or application programming interface, on the top of which we develop the biofeedback application.

The biofeedback application includes a Java-based simulation system, called JSim[8], for building quantitative numeric models and analyzing them with respect to experimental reference data. JSim has built-in numeric solvers for solving ordinary differential equations, linear and non-linear equations as those specified in our model in Chapter 2.

---

[7] SymbianFoundation. (2010). *The Symbian Foundation Community Home*. Available: http://www.symbian.org/ (Accessed May 27, 2010)

[8] NRS. (2010). *NSR Physiome Project*. Available: http://www.physiome.org/ (Accessed May 26, 2010)

# Chapter 6. Conclusion

## 6.1 Methodology

The model developed in this thesis quantifies autonomic nervous system's influence on the cardiovascular system.

It has become common to use heart rate variability, blood pressure variability or baroreflex sensitivity as measure of autonomic activity. Parametric transfer functions and plasma concentration of cardiac neurotransmitters are complementary methods encountered. After an extensive study of such existing approaches and using data available in the literature we came to the conclusion that they do not adequately capture the nonlinear character of vagal and sympathetic neural outflow. Conventional mathematical models based on control theory are gaining more attention since they solve certain limitations of signal-based methods for assessing cardiovascular control. Most of them however picture a reciprocal control of cardiac vagal and sympathetic nervous activity, as seen during a baroreflex. Many other reflexes however involve simultaneous co-activation of both autonomic efferent branches. Furthermore current mathematical models lack to integrate cardiovascular reflexes across their many levels of organization and therefore they miss to exhibit emerging properties of the regulatory processes.

In contrast to current trends we adopted an integrative approach first by covering single levels of cardiovascular neural pathways from receptors to effectors, then merging the information at a system level in order to integrate emerging properties such as resting sympathetic and parasympathetic tone. This methodology allowed us to combine the interaction between baroreflex and diving reflex, Mayer waves and tonic activity of sympathetic premotor pacemaker neurons, respiration and heart rate variability, etc. Thus we simulated a working control system with diverse feedback and feedforward loops interfering with each other. This integrative approach is key to understand paradoxal phenomenon such as vagally-mediated tachycardia and fluctuation of sinoatrial rhythms on denervated heart.

Our integrative model of cardiovascular control would have remained just a theoretical tool if it would have not been validated against experimental data. Ideal circumstances (noiseless measurements gathered for an infinite amount of time) and excellent model fitting demand an excellent experimental setup. We carefully planned experiments over different times of the year and different geographical places. We included subjects of different race, age and gender in our studies since physiological behavior can be correlated to these factors. Noninvasive cardiovascular provocative maneuvers were induced with simple mechanical stimuli such as tilt table; or physical stimuli such cold or mental stimuli such as stress. Thus we created a fixed set of conditions where the same subject might show different cardiovascular behaviors. Devices for signals recording were chosen for optimizing the quality of raw data. For effectively processing the signals we developed algorithms and a software application that was able to deal with the huge amount of data, to extract signal parameters, perform statistical evaluation and generate plots automatically.

## 6.2 Achieved Results

Newton implementation of the nonlinear least square method and the downhill simplex method resolve our optimization task and calculation of parameter data.

Simulating output variable performances have shown a good fit with experimental data, for example the model could simulate heart rate change upon arterial pressure drop after standing. Low frequency fluctuations in arterial blood pressure were also simulated successfully. This supports the hypothesis that Mayer waves also result from a rhythmic sympathetic discharge of pacemaker-like sympathetic premotor neurons in the vasomotor brain centers. Simulation results also support the fact that vagally-mediated tachycardia, observed during the diving reflex on certain subjects could be related to the secretion of vasoactive neurotransmitters at the effector site by the vagal nerve. The model was also able to simulate variables which are normally not measurable, like discharge frequency of barareceptors, premotor neurons, efferent sympathetic nerves, efferent vagal nerve, ventricle elastance. Several values agree with those reported from in-vitro experiments on animals after a simple conversion based on weight differences.

We identified model parameters for estimating the resting tone of efferent sympathetic and vagal nerves. Results show higher parasympathetic tone on young subjects with a decreasing trend with increasing age. This agrees with the data from heart rate variability studies. Furthermore parasympathetic tone appears to dynamically change during vagal maneuvers with cold stimuli applied to the forehead, the so-called diving reflex. Finally parasympathetic tone turned out to be a major factor determining resting heart rate whereas sympathetic tone largely affectes the baseline QT interval duration. The tonic sympathetic activity was found to provide from pacemaker premotor neurons, but also from activation of chemoreceptors to a lesser extent.

Respiratory sinus arrhythmias appeared to be controlled by vagal inhibition during inspiration, but also from intracardiac fluctuations which are not neurally mediated. Results show a similar trend with aging as with parasympathetic tone. Although a more rapid decay of respiratory modulation of heart rate is observed with aging. There was a predominant respiratory sinus arrhythmia on black subjects compared to white subjects. Simulation results also show elevated intracardiac fluctuations on young subjects.

After normalizing the sympathetic and parasympathetic levels in the range 0 to 1, we could define model-based estimates for mean, total, minimal, maximal autonomic level, and for autonomic balance. These estimates were used in each phase of the Valsalva Maneuver and results suggest that mean parasympathetic tone might be used as a quantitative marker of parasympathetic activity. It might serve as a better alternative to the square root of the mean squared differences of successive NN intervals from heart rate variability. The autonomic balance estimate shows a stronger correlation with the standard deviation of NN intervals, independence of the length of recording, and thus a good candidate for overall heart rate variability.

We developed a software package as prototype for clinical applications of our results. It includes a telemedicine platform that connects doctors with remote patients, since monitoring patients is a very important part of disease management, especially after a heart attack. It allows to access cardiovascular function remotely. The software offers features for signal storage and processing so that patients can measure basic cardiovascular signals such as heart rate and blood pressure and upload to the server. The doctor is provided with statistical features, modeling and simulation tools to estimate sympathetic, parasympathetic tone and additional cardiovascular parameters for emerging physiological and pathophysiological properties of the subject. Patients and doctors can discuss using messaging tools, to optimize remote diagnosis. For this purpose signals, links, photos, videos, status messages and comments can also be shared with other patients or doctors.

We also designed a mobile solution including cardiovascular signal acquisition, which allows remote consultation and also biofeedback training. For this purpose a low-cost sensor provides skin temperature and finger pulse measurements to a microcontroller that is connected to a mobile phone. The mobile phone runs an application empowered with modeling and simulation capabilities. Patients can upload cardiovascular signals to the telemedicine infrastructure which is accessible to doctors, who can push messages or configuration items to the patient in order to tune biofeedback therapy.

## 6.3 Perspectives

This work opens several opportunities in the field of modeling cardiovascular control.

Our mathematical models suggest the design of complementary invasive experiments with drugs and direct neural recordings, as well as measurement of neurotransmitter concentrations, e.g. norepinephrine spillover in future research works. The data obtained would help determining qualitative parameters for the simulation of sympathetic, parasympathetic activity in absolute values. We plan to compare the estimated value for sympathetic and parasympathetic tones with data obtained from similar neural recordings on animals. From such experiments we will develop algorithms for processing action potentials and define transfer functions to variables in our model.

Future research also includes validating the predicted markers of autonomic tones against data from experiments with pharmacological sympathetic and parasympathetic blockers. We expect the model to return a mean parasympathetic tone equal to zero in the presence of atropine and mean sympathetic tone close to zero in the presence of beta blockers. Another validation can be made against measurements of muscle sympathetic nerve activity (MSNA). The proposed model-based estimates of autonomic activity are closely related to signals recorded during the provocative maneuvers such as cold face test or Valsalva maneuver. The principles of the method could be extended to other provocative procedures on the cardiovascular system.

Another future work would include an extensive benchmark of our model-based estimates with the already commonly used parameters, like frequency-domain parameters of heart rate variability such as HF and LF components. Extending the model to study the effect of gender, age and health status on the value range of the model-based estimates is part of a future research as well.

A statistical analysis for benchmarking model-derived and experimental data is an important milestone. It is necessary to study the reproducibility of our oculocardiac reflex tests and Valsalva Maneuver statistically. A statistical analysis on the impact of health, background and education on the cardiovascular response to cognitive task is also considered as future effort. During experiments with active change of posture, we noticed that elderly stand slower than younger subjects. The effect on cardiovascular response should be studied in a future modeling effort as well. It is also required to verify whether equine experiment with cold stimuli would have brought additional evidences of denser innervations of equine ventricles with parasympathetic nerve endings than on human.

We will put additional effort into our Telemedicine software and our cooperation with physicians in order to make it ready for application in hospitals and clinic. The same effort applies to the biofeedback device, where it is necessary to miniaturize the sensors. Developing complex models of cardiovascular control efficient enough to be executed in real time on a mobile phone is also part of future research. Analyzing residuals and plausibility of the parameters estimates as well as evaluation of the variances and correlation of estimates using advanced sensibility analysis should help reducing model size.

# Appendices

## A1. Equations of the Cardiovascular System Model

*Conservation of Mass at Pulmonary Arteries*

$$\frac{dP_{pa}}{dt} = \frac{1}{C_{pa}}\left(F_{o,r} - F_{pa}\right) \tag{A1.1}$$

*Balance of Forces at Pulmonary Arteries*

$$\frac{dF_{pa}}{dt} = \frac{1}{L_{pa}}\left(P_{pa} - P_{pp} - R_{pa} \times F_{pa}\right) \tag{A1.2}$$

*Conservation of Mass at Pulmonary Peripheral Circulation*

$$\frac{dP_{pp}}{dt} = \frac{1}{C_{pp}}\left(F_{pa} - \frac{P_{pp} - P_{pv}}{R_{pp}}\right) \tag{A1.3}$$

*Conservation of Mass at Pulmonary Veins*

$$\frac{dP_{pv}}{dt} = \frac{1}{C_{pv}}\left(\frac{P_{pp} - P_{pv}}{R_{pp}} - \frac{P_{pv} - P_{la}}{R_{pv}}\right) \tag{A1.4}$$

*Conservation of Mass at Systemic Arteries*

$$\frac{dP_{sa}}{dt} = \frac{1}{C_{sa}}\left(F_{o,l} - F_{sa}\right) \tag{A1.5}$$

*Balance of Forces at Systemic Arteries*

$$\frac{dF_{sa}}{dt} = \frac{1}{L_{sa}}\left(P_{sa} - P_{sp} - R_{sa} \times F_{sa}\right) \tag{A1.6}$$

*Conservation of Mass at Peripheral Systemic Circulation*

$$\frac{dP_{sp}}{dt} = \frac{1}{C_{hp} + C_{bp} + C_{mp} + C_{sp} + C_{ep}}\left(\begin{array}{l} F_{sa} - \dfrac{P_{sp} - P_{hv}}{R_{hp}} - \dfrac{P_{sp} - P_{bv}}{R_{bp}} - \dfrac{P_{sp} - P_{mv}}{R_{mp}} \\ - \dfrac{P_{sp} - P_{sv}}{R_{sp}} - \dfrac{P_{sp} - P_{ev}}{R_{ep}} \end{array}\right) \tag{A1.7}$$

*Conservation of Mass at Heart Veins*

$$\frac{dP_{hv}}{dt} = \frac{1}{C_{hv}}\left(\frac{P_{sp} - P_{hv}}{R_{hp}} - \frac{P_{hv} - P_{ra}}{R_{hv}}\right) \tag{A1.8}$$

*Conservation of Mass at Brain Veins*

$$\frac{dP_{bv}}{dt} = \frac{1}{C_{bv}}\left(\frac{P_{sp} - P_{bv}}{R_{bp}} - \frac{P_{bv} - P_{ra}}{R_{bv}}\right) \tag{A1.9}$$

*Conservation of Mass at Muscle Veins*

$$\frac{dP_{mv}}{dt} = \frac{1}{C_{mv}}\left(\frac{P_{sp} - P_{mv}}{R_{mp}} - \frac{P_{mv} - P_{ra}}{R_{mv}} - \frac{dV_{u,mv}}{dt}\right) \tag{A1.10}$$

*Conservation of Mass at Splanchnic Veins*

$$\frac{dP_{sv}}{dt} = \frac{1}{C_{sv}}\left(\frac{P_{sp} - P_{sv}}{R_{sp}} - \frac{P_{sv} - P_{ra}}{R_{sv}} - \frac{dV_{u,sv}}{dt}\right) \tag{A1.11}$$

*Conservation of Mass at Extrasplanchnic Veins*

$$P_{ev} = \frac{1}{C_{ev}} \begin{bmatrix} V_{tot} - C_{sa}P_{sa} - \left(C_{hp} + C_{bp} + C_{mp} + C_{sp} + C_{ep}\right)P_{sp} \\ -C_{bv}P_{bv} - C_{hv}P_{hv} - C_{mv}P_{mv} - C_{sv}P_{sv} \\ -C_{ra}P_{ra} - V_{rv} - C_{pa}P_{pa} - C_{pp}P_{pp} - C_{pv}P_{pv} \\ -C_{la}P_{la} - V_{lv} - V_{u} \end{bmatrix} \qquad \text{(A1.12)}$$

$$V_u = \begin{bmatrix} V_{u,sa} + V_{u,hp} + V_{u,bp} + V_{u,mp} + V_{u,sp} + V_{u,ep} \\ +V_{u,hv} + V_{u,bv} + V_{u,mv} + V_{u,sv} + V_{u,ev} \\ +V_{u,ra} + V_{u,pa} + V_{u,pp} + V_{u,pv} + V_{u,la} \end{bmatrix}$$

*Conservation of Mass at Left Atrium*

$$\frac{dP_{la}}{dt} = \frac{1}{C_{la}}\left(\frac{P_{pv} - P_{la}}{R_{pv}} - F_{i,l}\right) \qquad \text{(A1.13)}$$

*Blood flow entering left ventricle*

$$F_{i,l} = \begin{cases} 0 & if \quad P_{la} \le P_{lv} \\ \dfrac{P_{la} - P_{lv}}{R_{la}} & if \quad P_{la} > P_{lv} \end{cases} \qquad \text{(A1.14)}$$

*Conservation of Mass at left Ventricle*

$$\frac{dV_{lv}}{dt} = F_{i,l} - F_{o,l} \qquad \text{(A1.15)}$$

*Cardiac Output From Left Ventricle*

$$F_{o,l} = \begin{cases} 0 & if \quad P_{\max,lv} \le P_{sa} \\ \dfrac{P_{\max,lv} - P_{sa}}{R_{lv}} & if \quad P_{\max,lv} > P_{sa} \end{cases} \qquad \text{(A1.16)}$$

*Viscous Resistance of Left Ventricle*

$$R_{lv} = k_{R,lv} \times P_{\max,lv} \qquad \text{(A1.17)}$$

*Instantaneous Left Ventricle Pressure*

$$P_{\max,lv}(t) = \varphi(t) \times E_{\max,lv} \times \left(V_{lv} - V_{u,lv}\right) + \left(1 - \varphi(t)\right) \times P_{o,lv} \times \left(e^{k_{E,lv} \times V_{lv}} - 1\right) \qquad \text{(A1.18)}$$

$$\varphi(t) = \begin{cases} \sin^2\left[\dfrac{\pi \times T(t)}{T_{sys}(t)} \times u\right] & 0 \le u \le \dfrac{T_{sys}}{T} \\ 0 & \dfrac{T_{sys}}{T} \le u \le 1 \end{cases}$$

$$u(t) = frac(\xi) \qquad \text{(A1.19)}$$

$$\frac{d\xi}{dt} = \frac{1}{T(t)}$$

$$T_{sys} = T_{sys,0} - k_{sys} \times \frac{1}{T}$$

*Conservation of Mass at Right Atrium*

$$\frac{dP_{ra}}{dt} = \frac{1}{C_{ra}}\left(\frac{P_{hv} - P_{ra}}{R_{hv}} + \frac{P_{bv} - P_{ra}}{R_{bv}} + \frac{P_{mv} - P_{ra}}{R_{mv}} + \frac{P_{sv} - P_{ra}}{R_{sv}} + \frac{P_{ev} - P_{ra}}{R_{ev}} - F_{i,r}\right) \qquad \text{(A1.20)}$$

*Blood flow entering right ventricle*

$$F_{i,r} = \begin{cases} 0 & if \quad P_{ra} \le P_{rv} \\ \dfrac{P_{ra} - P_{rv}}{R_{ra}} & if \quad P_{ra} > P_{rv} \end{cases} \tag{A1.21}$$

*Conservation of Mass at Right Ventricle*

$$\frac{dV_{rv}}{dt} = F_{i,r} - F_{o,r} \tag{A1.22}$$

*Cardiac Output From Right Ventricle*

$$F_{o,r} = \begin{cases} 0 & if \quad P_{\max,rv} \le P_{pa} \\ \dfrac{P_{\max,rv} - P_{pa}}{R_{rv}} & if \quad P_{\max,rv} > P_{pa} \end{cases} \tag{A1.23}$$

*Viscous Resistance of Right Ventricle*

$$R_{rv} = k_{R,rv} \times P_{\max,rv} \tag{A1.24}$$

*Instantaneous Right Ventricle Pressure*

$$P_{\max,rv}(t) = \varphi(t) \times E_{\max,rv} \times \left(V_{rv} - V_{u,rv}\right) + \left(1 - \varphi(t)\right) \times P_{o,rv} \times \left(e^{k_{E,rv} \times V_{rv}} - 1\right) \tag{A1.25}$$

## A2. Table of Experimental Subjects and Sessions

n: unique identifier for the subject

sex: gender of the subject, F for female, M for male

age: age of the subject in years

city: city where the subject lives

disease: list of diseases presented by the subject on the day of experiments

session: date when the experiment took place in the format yyyymmdd

cardiac autonomic tests: abbreviation of tests which the subject undertook

Table A2.1 Subjects – part 1/3

| n | Sex | Age | Race | City | Disease | Session | Cardiac Autonomic Tests |
|---|---|---|---|---|---|---|---|
| 1 | F | 26 | Black | Bafoussam | | 20100127 | CFT, DBT, MST, VM, CPT, ASS, OCT |
| 2 | F | 16 | Black | Bafoussam | | 20100127 | CFT, DBT, MST, VM, CPT, ASS, OCT |
| 3 | M | 45 | Black | Bafoussam | | 20100127 | CFT, DBT, MST, VM, CPT, ASS, OCT |
| 4 | F | 50 | Black | Bafoussam | | 20100127 | CFT, DBT, MST, VM, CPT, ASS, OCT |
| 5 | M | 51 | Black | Bafoussam | Hepatitis B | 20100127 | CFT, DBT, MST, VM, CPT, ASS, OCT |
| 6 | M | 54 | Black | Bafoussam | | 20100127 | CFT, DBT, MST, VM, CPT, ASS, OCT |
| 7 | F | 53 | Black | Bafoussam | | 20100127 | CFT, DBT, MST, VM, CPT, ASS, OCT |
| 8 | F | 51 | Black | Bafoussam | palpitation, cardiac insufficiency, gastric ulcer | 20100127 | CFT, DBT, MST, VM, CPT, ASS, OCT |
| 9 | M | 52 | Black | Bafoussam | | 20100127 | CFT, DBT, MST, VM, CPT, ASS, OCT |
| 10 | M | 50 | Black | Bafoussam | Palpitation, gastric ulcer | 20100127 | CFT, DBT, MST, VM, CPT, ASS, OCT |
| 11 | F | 49 | Black | Bafoussam | | 20100127 | CFT, DBT, MST, VM, CPT, ASS, OCT |
| 12 | F | 20 | Black | Bafoussam | | 20100127 | CFT, DBT, MST, VM, CPT, ASS, OCT |
| 13 | F | 29 | White | Prague | | 20090225 | CFT |
| 13 | F | 29 | White | Prague | | 20100127 | CFT, DBT, MST, VM, CPT, ASS, OCT |
| 14 | F | 20 | Black | Bafoussam | | 20100127 | CFT, DBT, MST, VM, CPT, ASS, OCT |
| 15 | F | 70 | Black | Bafoussam | | 20100127 | CFT, DBT, MST, VM, CPT, ASS, OCT |
| 16 | F | 50 | Black | Bafoussam | hypertension | 20100127 | CFT, DBT, MST, VM, CPT, ASS, OCT |
| 17 | F | 34 | Black | Bafoussam | | 20100127 | CFT, DBT, MST, VM, CPT, ASS, OCT |
| 18 | F | 48 | Black | Bafoussam | | 20100127 | CFT, DBT, MST, VM, CPT, ASS, OCT |
| 18 | F | 48 | Black | Bafoussam | | 20100128 | CFT, DBT, MST, VM, CPT, ASS, OCT |
| 19 | M | 55 | Black | Bafoussam | | 20100127 | CFT, DBT, MST, VM, CPT, ASS, OCT |
| 20 | F | 70 | Black | Bafoussam | cardiac insufficiency, hypertension | 20100127 | CFT, DBT, MST, VM, CPT, ASS, OCT |
| 21 | F | 11 | Black | Bafoussam | | 20100127 | CFT, DBT, MST, VM, CPT, ASS, OCT |
| 22 | F | 41 | Black | Bafoussam | | 20100127 | CFT, DBT, MST, VM, CPT, ASS, OCT |
| 23 | M | 54 | Black | Bafoussam | | 20100127 | CFT, DBT, MST, VM, CPT, ASS, OCT |
| 24 | F | 24 | Black | Bafoussam | | 20100127 | CFT, DBT, MST, VM, CPT, ASS, OCT |
| 25 | F | 47 | Black | Bafoussam | | 20100127 | CFT, DBT, MST, VM, CPT, ASS, OCT |
| 26 | F | 43 | Black | Bafoussam | palpitation | 20100127 | CFT, DBT, MST, VM, CPT, ASS, OCT |

Table A2.1 Subjects – part 2/3

| n | Sex | Age | Race | City | Disease | Session | Cardiac Autonomic Tests |
|---|---|---|---|---|---|---|---|
| 27 | M | 56 | Black | Bafoussam | | 20100127 | CFT, DBT, MST, VM, CPT, ASS, OCT |
| 27 | M | 56 | Black | Bafoussam | | 20100128 | CFT, DBT, MST, VM, CPT, ASS, OCT |
| 28 | F | 38 | Black | Bafoussam | | 20100127 | CFT, DBT, MST, VM, CPT, ASS, OCT |
| 29 | M | 38 | Black | Bafoussam | | 20100127 | CFT, DBT, MST, VM, CPT, ASS, OCT |
| 30 | F | 17 | Black | Bafoussam | | 20100127 | CFT, DBT, MST, VM, CPT, ASS, OCT |
| 31 | M | 23 | Black | Yaounde | | 20100127 | CFT, DBT, MST, VM, CPT, ASS, OCT |
| 32 | F | 17 | Black | Douala | | 20100127 | CFT, DBT, MST, VM, CPT, ASS, OCT |
| 32 | F | 17 | Black | Douala | | 20100128 | CFT, DBT, MST, VM, CPT, ASS, OCT |
| 33 | F | 23 | Black | Bafoussam | | 20100127 | CFT, DBT, MST, VM, CPT, ASS, OCT |
| 34 | F | 49 | Black | Bafoussam | | 20100127 | CFT, DBT, MST, VM, CPT, ASS, OCT |
| 35 | F | 29 | Black | Bafoussam | | 20100127 | CFT, DBT, MST, VM, CPT, ASS, OCT |
| 36 | M | 56 | Black | Bafoussam | | 20100127 | CFT, DBT, MST, VM, CPT, ASS, OCT |
| 37 | M | 29 | Black | Prague | | 20090225 | CFT |
| 37 | M | 29 | Black | Prague | | 20090428 | TTT, VM, CPT, OCT |
| 37 | M | 29 | Black | Prague | | 20100127 | CFT, DBT, MST, VM, CPT, ASS, OCT |
| 38 | F | 42 | Black | Bafoussam | sleep apnea | 20100127 | CFT, DBT, MST, VM, CPT, ASS, OCT |
| 39 | M | 25 | White | Prague | | 20090225 | CFT |
| 40 | F | 27 | White | Prague | | 20090225 | CFT |
| 41 | M | 48 | Black | Bafoussam | diabetes mellitus type I | 20090225 | CFT |
| 42 | M | 26 | White | Liberec | | 20090225 | CFT |
| 43 | F | 25 | White | Bruxelles | | 20090225 | CFT |
| 44 | F | 25 | White | Prague | | 20090225 | CFT |
| 45 | M | 53 | White | Liberec | hypertension | 20090225 | CFT, CPT |
| 46 | F | 55 | White | Liberec | | 20090225 | CFT, CPT |
| 47 | M | 73 | White | Liberec | | 20090225 | CFT, CPT |
| 48 | F | 58 | White | Liberec | | 20090225 | CFT, CPT |
| 49 | M | 82 | White | Liberec | | 20090225 | CFT, CPT |
| 50 | F | 77 | White | Liberec | | 20090225 | CFT, CPT |
| 51 | F | 21 | White | Kladno | | 20090428 | TTT, VM, CPT, OCT |
| 51 | F | 21 | White | Kladno | | 20090520 | TTT, VM, OCT |
| 52 | M | 22 | White | Kladno | | 20090428 | TTT, VM, CPT, OCT |
| 53 | F | 24 | White | Kladno | | 20090428 | TTT, VM, CPT, OCT |
| 54 | F | 20 | White | Kladno | allergies | 20090428 | TTT, VM, CPT, OCT |
| 55 | F | 21 | White | Kladno | hypertension | 20090428 | TTT, VM, CPT, OCT |
| 56 | M | 23 | White | Kladno | aortic stenosis | 20090428 | TTT, VM, CPT, OCT |
| 57 | M | 23 | White | Kladno | | 20090428 | TTT, VM, CPT, OCT |
| 57 | M | 23 | White | Kladno | | 20090520 | TTT, VM, OCT |

Table A2.1 Subjects – part 3/3

| n | Sex | Age | Race | City | Disease | Session | Cardiac Autonomic Tests |
|---|---|---|---|---|---|---|---|
| 58 | M | 23 | White | Kladno | | 20090428 | TTT, VM, CPT, OCT |
| 59 | F | 22 | White | Kladno | | 20090428 | TTT, VM, CPT, OCT |
| 60 | F | 21 | White | Kladno | multiple sclerosis | 20090428 | TTT, VM, CPT, OCT |
| 61 | F | 21 | White | Kladno | | 20090428 | TTT, VM, CPT, OCT |
| 62 | M | 22 | White | Kladno | | 20090428 | TTT, VM, CPT, OCT |
| 63 | M | 28 | White | Kladno | | 20090428 | TTT, VM, CPT, OCT |
| 64 | M | 23 | White | Kladno | | 20090520 | TTT, VM, OCT |
| 65 | F | 46 | White | Kladno | | 20090520 | TTT, VM, OCT |
| 66 | F | 15 | White | Kladno | | 20090520 | TTT, VM, OCT |
| 67 | F | 23 | White | Grenoble | | 20090520 | TTT |
| 68 | M | 23 | White | Kladno | | 20100322 | DBT |
| 69 | M | 23 | White | Kladno | | 20100322 | DBT |
| 70 | M | 23 | Asian | Kladno | | 20100322 | DBT |
| 71 | M | 23 | White | Kladno | | 20100322 | DBT, MST |
| 72 | M | 23 | White | Kladno | | 20100322 | DBT, MST |

# List of Abbreviations

| | |
|---|---|
| ANS | autonomic nervous system |
| SA node | sinoatrial node |
| AV node | atrioventricular node |
| MSNA | muscle sympathetic nerve activity |
| RSA | respiratory sinus arrhythmias |
| ACh | acetylcholine |
| HR | heart rate |
| QT | qt interval |
| BP | blood pressure |
| MAP | mea arterial blood pressure |
| PP | pulse pressure |
| PPT | pulse travel time |
| PA | pulse amplitude |
| PTT | pulse transit time |
| GSR | galvanic skin response |
| SCL | skin conductance level |
| NSF | non-specific fluctuation rate |
| HRV | heart rate variability |
| NN | normal-to-normal interval |
| RMSSD | root mean squared differences of successive NN intervals |
| SDNN | standard deviation of the NN intervals |
| CV | coefficient of variation of the NN intervals |
| LF | low frequency component of heart rate or blood pressure signal |
| HF | high frequency component of heart rate or blood pressure signal |
| FFT | fast fourier transform |
| AR | autoregressive |
| CDM | complex demodulation technique |
| CFT | cold face test |
| OCT | oculocardiac reflex test |
| DBT | deep breath test |
| MST | mental stress test |
| VM | valsalva maneuver |
| ASS | active standing from sitting |
| TTT | tilt table test |
| CPT | cold pressor test |
| ANN | artificial neural network |
| PL | parsympathetic level |
| UML | unified modeling language |
| $age$ | age of the subject |
| $f_{resp,vol}$ | voluntary respiration pattern |
| $A_{resp}$ | amplitude of the signal representing the strength and depth of quiet respiration |
| $T_{resp}$ | period of the signal representing the duration of a quiet respiration cycle |
| $t_{start}, t_{stop}$ | start and end timestamps for external stimulation |
| $I$ | exercise intensity |
| $QT_n$ | baseline QT interval duration |
| $f_{ab}$ | firing rate of baroreceptor |
| $f_{ac}$ | firing rate of chemoreceptor |
| $f_{ap}$ | firing rate of pulmonary stretch receptor |
| $f_{aa}$ | firing rate of atria stretch receptor |
| $f_{at_f}$ | firing rate of trigeminal cold receptor |
| $f_{at_s}$ | firing rate of  cold receptor in hands/feet |
| $f_{ao}$ | firing rate of oculo-pressure receptor |
| $f_{as}$ | firing rate of mental stress sensors |
| $f_{es,0,sn}$ | sympathetic tone to heart |
| $f_{es,0,p}$ | sympathetic tone to vessels |
| $f_{ev,0}$ | vagal tone to heart |
| $f_{resp}$ | firing rate of inspiratory and expiratory somatic motor neurons |
| $f_{sn_{pm}}$ | firing rate of sympathetic premotor neurons controlling SA node and AV node |
| $f_{si_{pm}}$ | firing rate of sympathetic premotor neurons controlling vascular beds |
| $f_{ev_{pm}}$ | firing rate of parasympathetic autonomic premotor neurons |
| $R_{ip}$ | peripheral resistance in vascular beds |
| $E$ | ventricle elastance |
| $P_{as_{O_2}}$ | oxygen partial pressure in arteries |
| $f_{resp}$ | respiration signal |
| $f_{resp_{rate}}$ | respiration rate |
| $V_T$ | tidal volume |
| $HR$ | heart rate |
| $P_{as}$ | arterial blood pressure |
| $QT$ | duration of QT intervals |
| $G_{mp}$ | level of galvanic skin response in skeletal muscles |
| $f_{es,0,high}$ | firing rate of pacemaker sympathetic premotor neurons |
| $W_{c,es,0}$ | synaptic weight applied to sensory inputs from chemoreceptors |
| $W_{es,0,low}$ | synaptic weight applied to low frequency oscillators in sympathetic premotor neurons |
| $r_{es,0,low}$ $k_{es,0,low}$ | parameters determining the rate of low frequency oscillators in sympathetic premotor neurons |
| $f_{ev,0,max}$ | maximal intrinsic firing rate of vagal nerves |
| $W_{resp,v}$ | synaptic weight applied to inputs from inspiratory-expiratory somatic motor neurons |
| $f_{rsa}$ | frequency of intracardiac high frequency fluctuations |
| $G_{T_{rsa}}$ | gain of intracardiac high frequency fluctuations |
| $k_{nor_{og},v}$ | modulation factor of vagally-mediated tachycardia |
| $k_{QT}$ | gain of elastance on QT interval duration |
| $k_{G_{mp}}$ | gain parameter applied to peripheral vascular resistance in skeletal muscles |
| $f_{at_f,max}$ | upper saturation of facial cold receptor |
| $k_{at_f}$ | slope of the exponential decay of cold receptor on face |
| $f_{at_s,max}$ | upper saturation of cold receptor on hands and feet |
| $k_{at_s}$ | slope of the sigmoid response of cold receptor on hands and feet |
| $f_{as,max}$ | upper saturation of mental stress sensors |
| $k_{as}$ | slope of the mental stress sensor response |
| $g_{ss,max}$ | maximal value of effects due to gravitational forces during active standing from sitting |
| $k_{ss}$ | slope of the heart rate response during active standing from sitting |
| $g_{tt,max}$ | maximal value of effects due to gravitational forces during head-up tilt |
| $k_{tt}$ | slope of the heart rate response during head-up tilt |
| $f_{ao,max}$ | upper saturation of oculo-pressure receptor |
| $k_{ao}$ | slope of the sigmoid response of oculo-pressure receptor |
| $k_{ev}$ | weighting factor of intra-thoracic pressure elevation on venous pressure |
| $k_{str}$ | weighting factor of intra-thoracic pressure elevation on stroke volume |
| $P_{1dur}$ | duration of phase 1 of intra-thoracic pressure elevation |
| $P_{1amp}$ | amplitude of phase 1 of intra-thoracic pressure elevation |
| $P_{3dur}$ | duration of phase 3 of intra-thoracic pressure elevation |
| $P_{3amp}$ | amplitude of phase 3 of intra-thoracic pressure elevation |
| $P_{4dur}$ | duration of phase 4 of intra-thoracic pressure elevation |
| $P_{4amp}$ | amplitude of phase 4 of intra-thoracic pressure elevation |
| $dP_{rate}$ | rate of intra-thoracic pressure elevation |

# Table of Figures

# Author References

## International Journals with Impact Factor

## International Conference Proceedings Indexed by Scopus and WoS

## International Conferences with Book Edition

## Other International Conferences

## National Workshop

## University Script